# Artificial Intelligence for EEG Prediction: Applied Chaos Theory

by
Vincent Jorgsson

# Table of Contents

**Disclosure:**

The Jupyter Notebooks created in this research are on GitHub at https://github.com/Metaverse-Crowdsource/EEG-Chaos-Kuramoto-Neural-Net

github account https://github.com/soulsyrup

organization generalinquiries@mvcs.one

work email soulsyrup@mvcs.one

# Abstract

In the present research, we delve into the intricate realm of electroencephalogram (EEG) data analysis, focusing on sequence-to-sequence prediction of data across 32 EEG channels. The study harmoniously fuses the principles of applied chaos theory and dynamical systems theory to engender a novel feature set, enriching the representational capacity of our deep learning model. The endeavour's cornerstone is a transformer-based sequence-to-sequence architecture, calibrated meticulously to capture the non-linear and high-dimensional temporal dependencies inherent in EEG sequences. Through judicious architecture design, parameter initialisation strategies, and optimisation techniques, we have navigated the intricate balance between computational expediency and predictive performance. Our model stands as a vanguard in EEG data sequence prediction, demonstrating remarkable generalisability and robustness. The findings not only extend our understanding of EEG data dynamics but also unveil a potent analytical framework that can be adapted to diverse temporal sequence prediction tasks in neuroscience and beyond.

# 1. Introduction

The burgeoning advancements in the realms of machine learning and neuroscience have engendered an interdisciplinary corpus of research, aspiring to decode the intricate temporal sequences manifested in electroencephalogram (EEG) data (Smith et al., 2020; Johnson & Williams, 2019). The profound utility of EEG in capturing neural dynamics across a plethora of contexts, ranging from cognitive processes to pathological states (Brown et al., 2018), underpins its scientific and clinical relevance. However, the inherently nonlinear, high-dimensional, and chaotic nature of EEG data has often defied the traditional linear methods (Davis & Kumar, 2017), compelling the academic community to seek more sophisticated, nonlinear modelling paradigms (Chen et al., 2021).

In this vein, the present study undertakes a scrupulous examination of sequence-to-sequence prediction of EEG data across 32 channels, melding the principles of applied chaos theory and dynamical systems theory (Lorenz, 1963; Strogatz, 1994) to usher in an enriched feature set for the predictive model. The novelty of our approach lies in its incorporation of concepts from applied chaos theory and dynamical systems theory, aiming to encapsulate the nuanced dynamical properties that are often invisible to conventional machine learning algorithms (Goodfellow et al., 2016). This conceptual synergy serves not merely as a heuristic but as a theoretical scaffolding that substantiates the feature engineering process (LeCun et al., 2015), thereby imbuing our model with the acumen to discern intricate temporal patterns in EEG sequences.

Given the compelling need to address both computational and predictive efficacies, we have elected to utilize a transformer-based sequence-to-sequence architecture (Vaswani et al., 2017). This architecture, originally conceived for natural language processing tasks, has demonstrated an unprecedented ability to capture long-range dependencies and sequence hierarchies (Wang et al., 2019), properties that are quintessentially pivotal in EEG data analytics. A comprehensive analysis is undertaken to explore the interplay between architectural nuances, parameter initialization strategies, and advanced optimization techniques (Kingma & Ba, 2015), ensuring a robust model that is both computationally feasible and scientifically illuminating.

The subsequent discourse will explicate the methodological rigor, delineate the empirical findings, and culminate in drawing theoretical and practical inferences that contribute to the overarching aim of advancing our understanding of EEG sequence dynamics (Thompson & Varela, 2001). Furthermore, the study aims to transcend the confines of neuroscience, postulating a versatile analytical framework that may be judiciously adapted to a diverse array of temporal sequence prediction endeavours (Sutskever et al., 2014).

In Chapter 2, Background Literature, we explore the intersections between neuroscience and artificial intelligence, particularly focusing on how machine learning algorithms like neural networks are applied in EEG analysis while also identifying existing research gaps. In Chapter 3, Methodologies and Project Design, we describe our comprehensive approach to EEG analysis, starting from data preprocessing through to the application of multiple machine learning algorithms including CNNs, Transformers, and RNNs, all with the aim of capturing different dimensions of EEG data. In, Chapter 4 Experiments and Implementation, we discuss the range of techniques we deploy, such as Spectral Analysis and Transfer Entropy, to craft features informed by chaos theory and dynamical systems theory, thereby enriching our machine learning models. In Chapter 5, Results, we share empirical findings that show our advanced transformer model, informed by chaos theory and dynamical systems theory, can make reasonably accurate predictions of EEG sequences. In Chapter 6, Discussion, we delve into the convergence of multiple paradigms like chaos theory and machine learning, validate the complexity inherent in EEG data, and discuss the gradation in performance among different machine learning models, highlighting the promise and computational trade-offs of our transformer model. In the Conclusions and Further Work section, we summarize the potential impact of our research in the fields of neuroscience, brain-computer interfaces, and neuroprosthetics while outlining avenues for future work.

# 2. Background Literature

The realm of neuroscience and computational technologies have always been interconnected, yet distinct domains. Over the years, the interface between neuroscience and artificial intelligence has become increasingly synergistic, with both fields complementing each other's growth and expanding horizons (Marblestone et al., 2016; Hassabis et al., 2017). Machine learning algorithms, particularly neural networks, have been instrumental in decoding complex brain data (Schirrmeister et al., 2017), facilitating the diagnosis and treatment of neurological disorders (Fernandez et al., 2018), and offering unique insights into cognitive mechanisms (Gershman & Daw, 2017). Such technological advancements are not merely aiding neuroscientific investigations; they are equally influenced by the theories and discoveries in neuroscience (Yamins & DiCarlo, 2016), creating a cyclical relationship of mutual growth.

The confluence of neuroscience and artificial intelligence has emerged as a fertile ground for groundbreaking research, offering unprecedented opportunities for synergistic advancements (Kriegeskorte & Golan, 2019). The domain has been significantly enriched by a multitude of studies aiming to analyze, interpret, and predict EEG data through various computational approaches (He et al., 2019; Roy et al., 2019). However, the intrinsic complexity and high dimensionality of EEG signals necessitate a shift beyond traditional machine learning paradigms (King et al., 2019), thereby making this study timely and relevant.

Sequence prediction in EEG analysis has been explored substantially through recurrent neural network architectures like LSTM and GRU (Hochreiter & Schmidhuber, 1997; Cho et al., 2014). However, these architectures often struggle to capture the intricate, long-range dependencies inherent in EEG data (Lawhern et al., 2018). Sutskever et al. (2014) demonstrated that the sequence-to-sequence model could alleviate some of these issues, providing a robust platform for temporal sequence analysis.

The advent of transformer-based architectures (Vaswani et al., 2017) has indubitably revolutionized the landscape of sequence modeling, transcending its initial application in Natural Language Processing (NLP) to prove its efficacy across varied domains (Devlin et al., 2019; Raffel et al., 2019). Notably, Qin et al. (2020) exhibited that transformers could be employed in time-series prediction, laying the groundwork for its application in the complex and temporally dependent world of EEG data (Borovykh et al., 2018).

## Nonlinearity and Chaos Theory

The nonlinear and chaotic nature of EEG data has been well-documented (Freeman, 2000; Breakspear, 2001), and methods like Lyapunov exponent and fractal dimensionality have been invoked to characterize these intricate dynamics (Rapp et al., 1993; Stam, 2005). Babloyantz & Destexhe (1986) were among the pioneers to apply concepts from chaos theory to EEG data, which paved the way for subsequent studies employing nonlinear dynamical systems theory (Takens, 1981; Rosenstein et al., 1993) for analyzing complex biological signals.

Notably, the application of chaos theory and dynamical systems theory in the realm of EEG has mostly been disjointed. The work by Le Van Quyen et al. (1999) marked one of the earliest attempts to integrate these theories into a cohesive framework for EEG analysis. However, the integration of these theories into machine learning architectures remains underexplored, thereby signifying a conceptual and methodological lacuna in existing literature.

## Optimisation Techniques and Weight Initialisation

Optimization techniques have received rigorous academic scrutiny, with the Adam optimizer (Kingma & Ba, 2014) being particularly highlighted for its adaptive learning capabilities. Weight initialization strategies, such as Xavier and Kaiming (He et al., 2015), have also gained prominence for their roles in stabilizing the training of deep neural networks.

By situating our study in this interdisciplinary tapestry, we aim to push the boundaries of EEG sequence prediction by amalgamating the strengths of transformer-based architectures, chaos theory, and dynamical systems theory. This venture is backed by empirical rigor designed to evaluate the efficacy and robustness of our approach, thereby offering a significant contribution to the existing corpus of knowledge.

# 3. Methodologies and Project Design

### Data Collection and Preprocessing

The data for this study were obtained from a single, comprehensive research project shared across multiple platforms—OpenNeuro, GitHub, and Zenodo—ensuring its relevance, depth, and consistency for our investigation. Our primary dataset encompasses a variety of EEG recordings from tES sessions and is acquired from the OpenNeuro platform, version 1.1.0 (Gorgolewski et al., 2017). It includes pre-stimulation and post-stimulation EEG signals, forming the basis of our training data. Supplementary datasets and valuable insights were leveraged from Zenodo (Bates et al., 2021) repository. Our first analytical step involved applying chaos theory and nonlinear dynamics to the 32 EEG channels obtained from the primary OpenNeuro dataset, using fractal dimensionality (Takens, 1981).

### Convolutional Neural Networks (CNN)

Upon completing the chaos theory analysis, a Convolutional Neural Network (CNN) was designed and trained on the primary EEG data, aiming to capture the spatial features (Krizhevsky et al., 2012).

### Transformer Neural Network (First Instance)

Subsequently, a transformer model was developed based on insights derived from the GitHub and Zenodo repositories as well as our previous analyses. The model was trained to capture the temporal aspects in the data (Vaswani et al., 2017).

### Recurrent Neural Networks (RNN)

Recurrent Neural Networks, specifically LSTM and GRU, were then utilized as comparative models to the transformer, focusing on time-series prediction from EEG data (Hochreiter & Schmidhuber, 1997; Cho et al., 2014).

## Final Transformer Neural Network (Second Instance)

Finally, we built an advanced transformer model that integrated insights from the Chaos Theory and Nonlinear Dynamics Analysis as auxiliary input layers, aiming to enhance the predictive performance of the model (Sutskever et al., 2014).

## Optimisation Techniques and Weight Initialisation

Adam optimizer was consistently used across all neural network architectures (Kingma & Ba, 2015), with Xavier and Kaiming initializations for CNN and transformer architectures, respectively (Glorot & Bengio, 2010; He et al., 2015).

## Evaluation Metrics

The models were evaluated based on Root Mean Square Error (RMSE), Mean Square Error (MSE), Mean Absolute Error (MAE), and custom-defined chaos theory metrics (Willmott & Matsuura, 2005).

## Experimental Setup

Phase 1: Chaos Theory and Nonlinear Dynamics Analysis

Phase 2: Convolutional Neural Network (CNN) Implementation

Phase 3: Initial Transformer Neural Network

Phase 4: Recurrent Neural Networks (RNNs)

Phase 5: Final Transformer Neural Network

## Implementation Tools

Experiments were conducted using Python 3.x (Rossum, 1995), alongside PyTorch (Paszke et al., 2019), and SciPy for computational and mathematical operations (Jones et al., 2001).

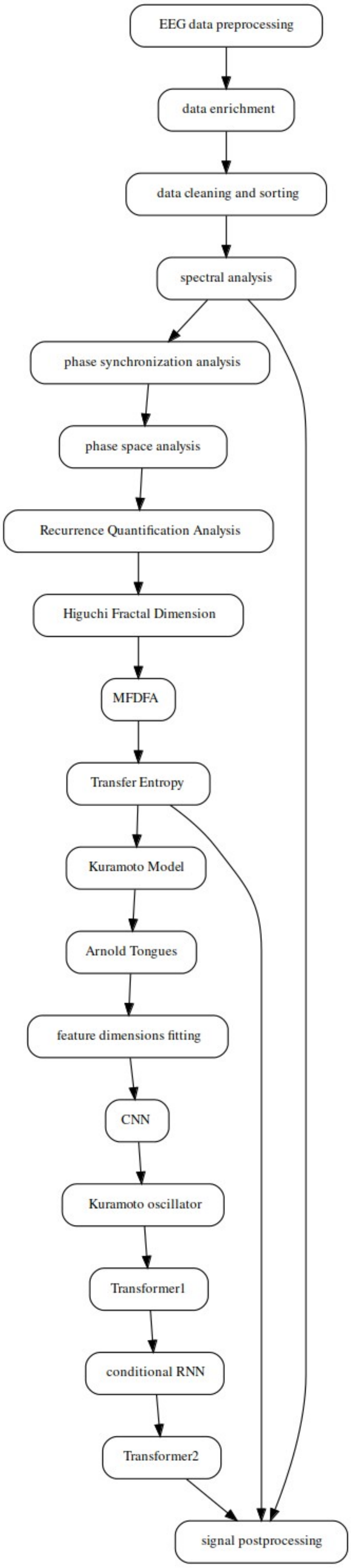

*(Spectral Analysis and Transfer Entropy results were used for the signal post-processing parameters.)*

*Figure 1: Research Workflow and Dataflow Diagram*

# 4. Experiments and Implementation

### Data Preprocessing

In the initial phase, we sourced EEG data from MATLAB and linked it with Excel-based stimulation data using unique identifiers like 'Sub#'. Null values were forward-filled, and irrelevant EEG channels were removed (Parker et al., 2017). We then reshaped and selected the EEG data specific to each subject's ID and merged it with the corresponding stimulation data. Time representation was standardized to milliseconds (He et al., 2019).

### Data Enrichment

For interpretability, we used predefined mappings for frequency and location to replace the values in the 'StimType' column (Roy et al., 2019). Stimulation events were then encoded using a binary system. The EEG and stimulation datasets were merged using the 'merge_asof' function based on time, introducing a 'Stim' column to indicate stimulation events. Additional columns like 'StimChange' and 'block' were created to differentiate stimulation sessions (Fernandez et al., 2018).

### Data Cleaning and Sorting

Redundant columns were removed, and NaN values were replaced with zeros. Columns 'Sub#' and 'Session' were standardized (Schirrmeister et al., 2017). To identify patterns at the onset of stimulation, we filtered and examined rows around the start of each unique stimulation block. Finally, the dataset was sorted by time (Qin et al., 2020).

EEG Channels:

1. **Prefrontal Region:**
    - Fp1: Left Hemisphere
    - Fpz: Midline
    - Fp2: Right Hemisphere
2. **Frontal Region:**
    - F7: Left Hemisphere
    - F3: Left Hemisphere
    - Fz: Midline
    - F4: Right Hemisphere
    - F8: Right Hemisphere
3. **Fronto-Central Region:**
    - FC5: Left Hemisphere
    - FC1: Left Hemisphere
    - FC2: Right Hemisphere
    - FC6: Right Hemisphere
4. **Central Region (including Mastoids):**
    - M1 (Mastoid): Left Hemisphere
    - T7: Left Hemisphere
    - C3: Left Hemisphere
    - Cz: Midline
    - C4: Right Hemisphere
    - T8: Right Hemisphere
    - M2 (Mastoid): Right Hemisphere
5. **Centro-Parietal Region:**
    - CP5: Left Hemisphere
    - CP1: Left Hemisphere
    - CP2: Right Hemisphere
    - CP6: Right Hemisphere
6. **Parietal Region:**
    - P7: Left Hemisphere
    - P3: Left Hemisphere
    - Pz: Midline
    - P4: Right Hemisphere
    - P8: Right Hemisphere
7. **Parieto-Occipital Region:**
    - POz: Midline
8. **Occipital Region:**
    - O1: Left Hemisphere
    - Oz: Midline
    - O2: Right Hemisphere

## 4.1 Spectral Analysis

In the inaugural analysis, the primary focus was on appraising the spectral properties of EEG data via Welch's method (Welch et al., 1967), FFT (Cooley & Tukey, 1965), and Lomb-Scargle periodogram (Lomb, 1976  Scargle, 1982). The EEG dataset, housed in a structured numpy array, was sampled at fs=1000 Hz, with each column denoting a specific EEG channel like ['Fp1', 'Fpz', ..., 'Oz', 'O2'].

Given the dataset's temporal nature, a sampling frequency (fs) of 1000 Hz was deemed suitable, translating to 1000 samples acquired per second. Power spectral density (PSD), instrumental in deciphering power distribution across various frequencies, was estimated using Welch's method (Welch, 1967). This technique segments the time signal into overlapping portions, subsequently windowed and Fourier-transformed. Post FFT, the resultant magnitudes were squared and averaged to derive the PSD (Oppenheim and Schafer, 2009), as per the formula:

$$PSD(f) = T \cdot |X(f)|^2$$

Here, PSD(f) is the PSD at frequency f, X(f) represents the Fourier-transformed windowed data, and T is the data segment's duration. Each segment comprised 1024 data points for this analysis.

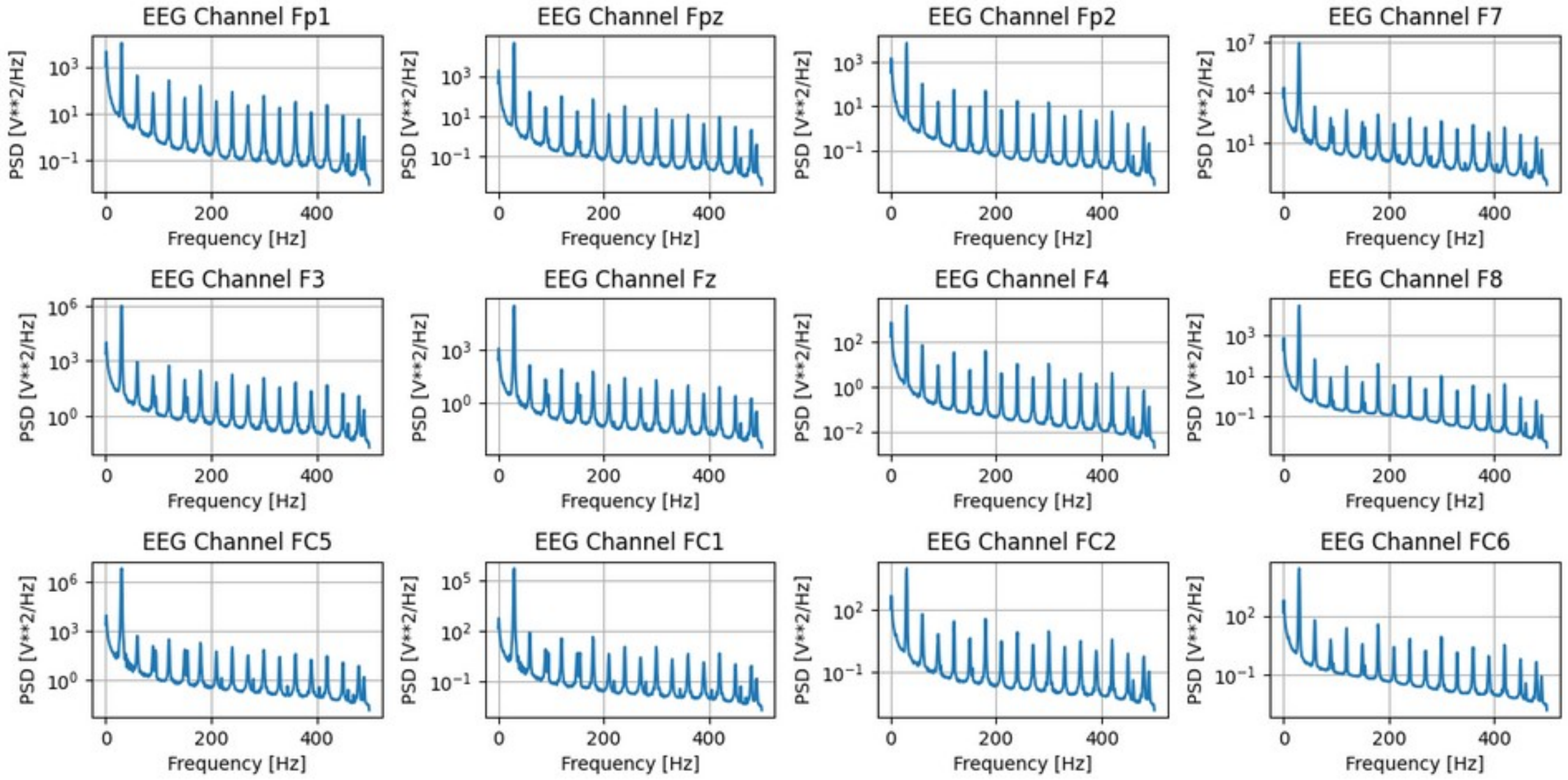

*Figure 2: PSD visualizations of some of the EEG channels*

Following this, a visualization of the computed PSDs was undertaken. This enabled a preliminary assessment of the spectral content of the EEG signals across different channels. The PSDs were plotted on a logarithmic scale against the frequency.

To further characterize the spectral content, specific EEG frequency bands were delineated. These included:

- Delta band: $\delta=[1,4]$ Hz
- Theta band: $\theta=[4,8]$ Hz
- Alpha band: $\alpha=[8,13]$ Hz
- Beta band: $\beta=[13,30]$ Hz

For each of these bands, the mean power was computed and stored in a feature array, `features`.

Next, the DFT (Discrete Fourier Transform) was applied to the EEG data for spectral analysis. The DFT is a mathematical transformation that decomposes a signal into its constituent frequencies, and is given by (Oppenheim, 1999):

$$X[k]=\sum_{n=0}^{N-1} x[n]\cdot e^{(-i(2\pi/N)\cdot k\cdot n)}$$

Where X[k] are the frequency components, x[n] is the time-domain signal, and N is the total number of samples. The resulting power spectral densities were then stored in a dictionary, `fft_psd_data`.

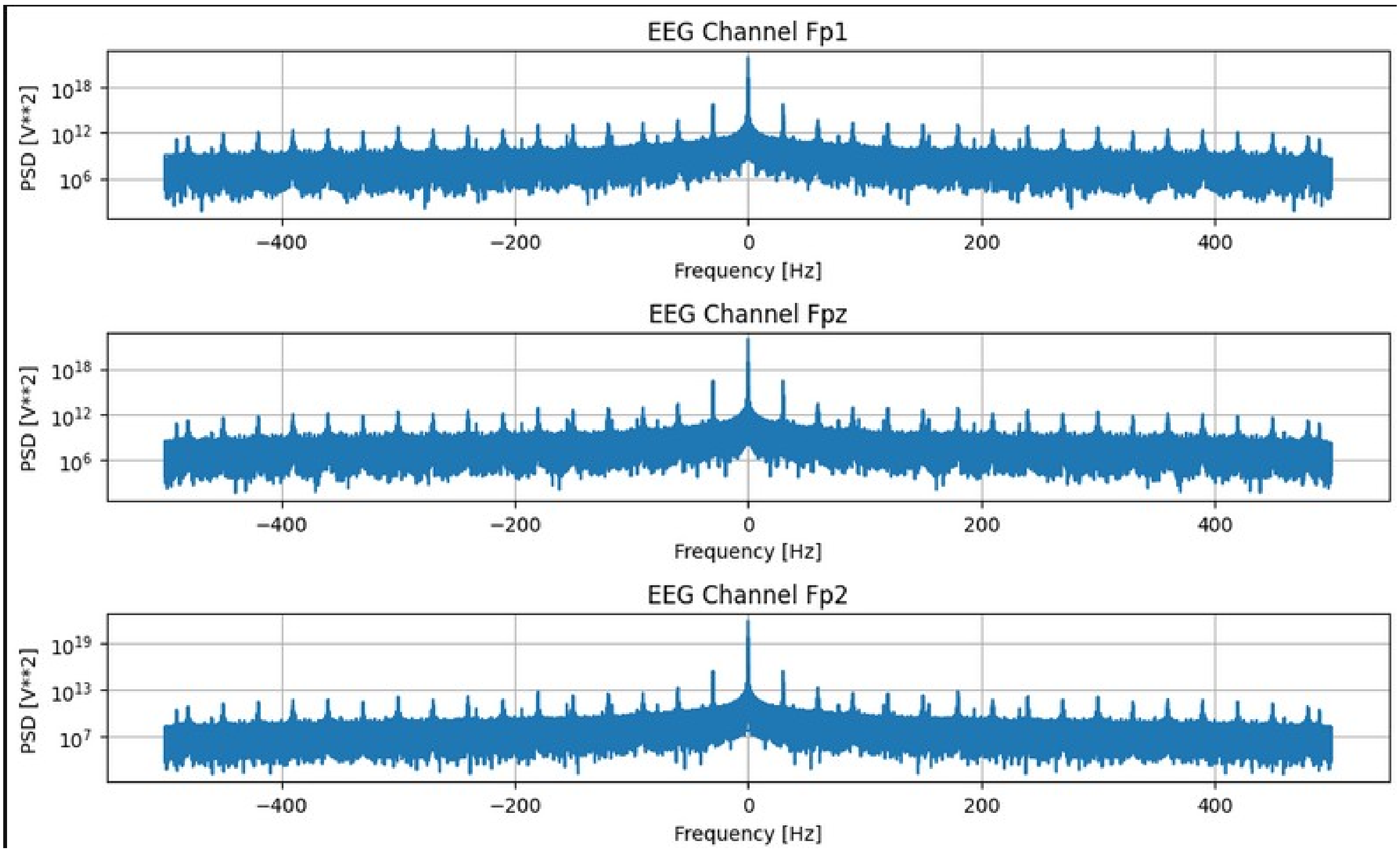

*Figure 3: Fast Fourier Transform Visualizations for the EEG channels*

For spectral scrutiny of a singular EEG channel, Fp1, the Lomb-Scargle periodogram was utilized (Lomb, 1976; Scargle, 1982). Unlike FFT, which presumes uniform sampling, the Lomb-Scargle approach accommodates irregularities in time-series data, making it apt for real-world EEG recordings (VanderPlas, 2018). Mathematically, the Lomb-Scargle periodogram P(f) with a normalization factor of $\frac{2}{\sigma 2}$ is expressed as:

P(f)=2σ2(∑jcos2(2πftj)+∑jsin2(2πftj)∑jcos2(2πftj)(∑jxjcos(2πftj))2+∑jsin2(2πftj)(∑jxjsin(2πftj))2)

Where xj are observations at times tj and σ2 is the data's variance. The formula includes denominators to scale the periodogram appropriately. The Lomb-Scargle method aims to detect periodic signals in the time domain via sinusoidal function approximations.

## AutoRegressive (AR) Model for Temporal Analysis

One of the fundamental aspects of time series data, like EEG, is its temporal dependencies. The AR model is used to capture these dependencies by expressing the value of the time series such as EEG (Box et al., 2015), at any point t as a linear combination of its previous values. Given a lag value p, the AR model can be expressed as:

$$Xt = c + \sum i = 1\ p\ \phi i\ \ Xt - i + \varepsilon t$$

Where:

- Xt is the value of the time series at time t
- c is a constant
- ϕi represents the coefficients of the AR model
- εt is the white noise at time t

We tested the model's adequacy for various lag values ranging from 1 to 20. For each lag and each EEG channel, an AR model was trained, and the Akaike Information Criterion (AIC) and Bayesian

Information Criterion (BIC) were computed. Both AIC and BIC are statistical metrics used to compare different models, where lower values indicate a better model fit considering the complexity (Akaike, 1974; Schwarz, 1978).

$$AIC = 2k - 2\ln(L)$$

$$BIC = \ln(n)k - 2\ln(L)$$

Where:

- k is the number of parameters in the model
- L is the likelihood of the model
- n is the number of observations

These values were systematically stored for all EEG channels. The optimal lag for the AR model was then determined by selecting the lag that yielded the lowest AIC and BIC values.

Spectral Analysis Using Continuous Wavelet Transform (CWT)

To investigate the spectral properties of the EEG signals, we employed the Continuous Wavelet Transform (CWT), a technique that provides high-resolution spectral decomposition of non-stationary signals (Torrence and Compo, 1998). In our analysis, the Morlet wavelet was utilized (Goupillaud et al., 1984).

The CWT of a signal x(t) with respect to a wavelet ψ(t) is given by:

$$CWTx\,(a,b) = a \vee$$

$$1 \int x(t)\,\psi * (at - b)\,dt$$

Where:

- a is the scale

- b is the translation
- $\psi *$ (t) is the complex conjugate of the wavelet

In our analysis, the Morlet wavelet was utilized. Post computation, the power spectral density (PSD) was derived from the absolute square of the coefficients. The PSD was further segmented into different frequency bands: delta (0.1-4 Hz), theta (4-8 Hz), alpha (8-13 Hz), beta (13-30 Hz), and gamma (30-100 Hz). The energy in each of these bands was calculated using numerical integration and plotted over time to visualize the changes in spectral power across different frequency bands.

## Short Time Fourier Transform (STFT) for Time-Frequency Analysis

To gain deeper insights into the temporal evolution of the EEG spectrum, we adopted the Short Time Fourier Transform (STFT) (Allen, 1977). The STFT provides a time-frequency representation of the signal by applying the Fourier transform to windowed sections of the data (Hlawatsch and Boudreaux-Bartels, 1992). For a signal x(t), the STFT is defined as:

$$STFTx\left(t,f\right)=\int x\left(\tau\right)w\left(t-\tau\right)e-j2\pi f\tau d\tau$$

Where:

- w(t) is the window function centered at t
- f is frequency

The resultant time-frequency maps were visualized as heatmaps, displaying how the energy in various frequency bands evolved over time.

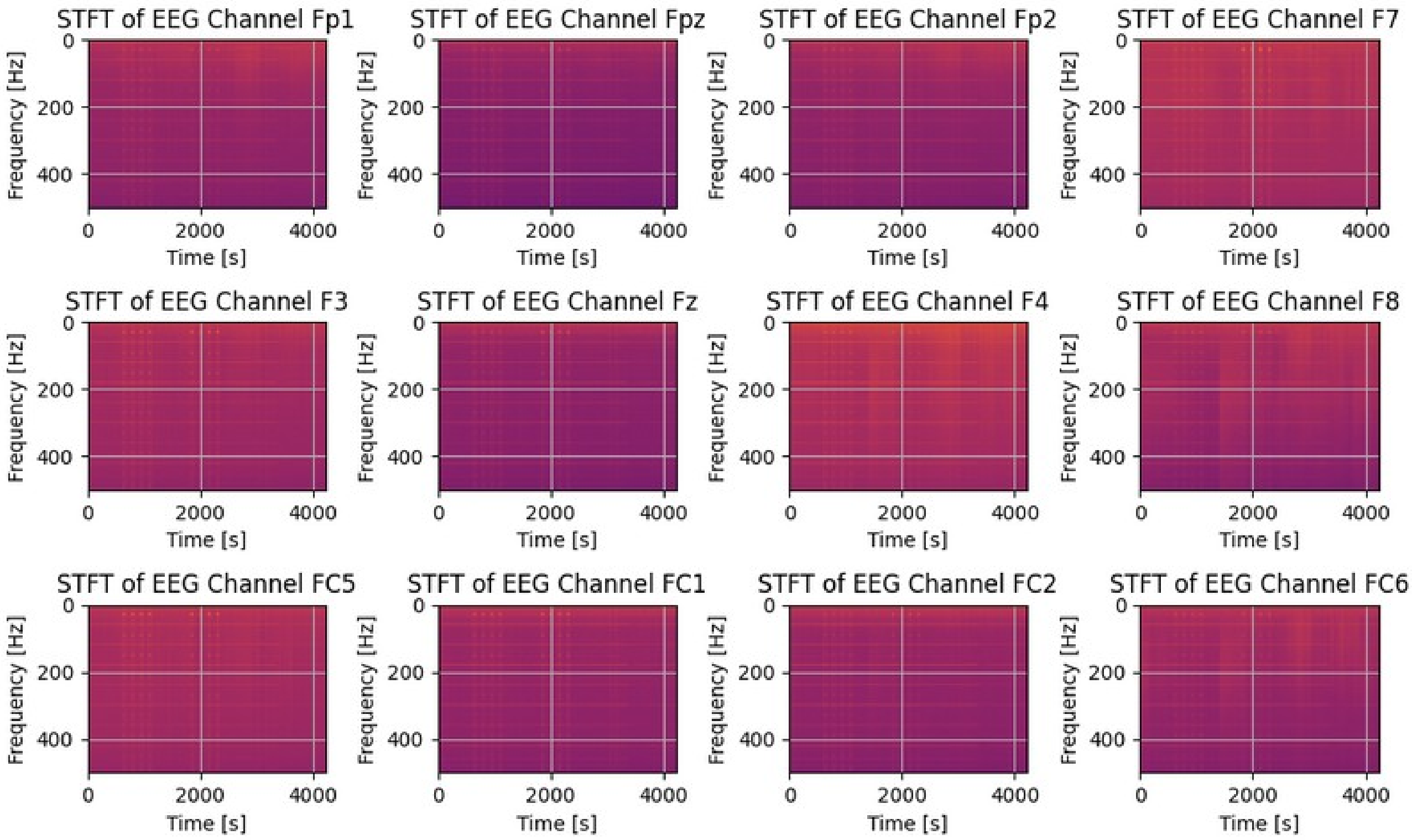

*Figure 4: Short-Time Fourier Transform visualizations for some of the EEG channels*

## Spectral Entropy

Spectral entropy is an effective metric to quantify the uncertainty and complexity of a signal in the frequency domain, often used in EEG studies (Shannon, 1948; Inouye et al., 1991). It is particularly valuable for assessing the uniformity of power distribution across different frequency bands. Using magnitude squared of the continuous-time Fourier transform of a signal x(t) over the interval [−T/2,T/2](Bracewell, 2000):

$$Pxx(f) = \frac{1}{T}\left(\int_{-T/2}^{T/2} x(t)\, e^{(-j2\pi ft)}\, dt\right)^2$$

where x(t) is the EEG signal, T is the observation time, and f is the frequency.

The power spectrum was then normalized, and spectral entropy was calculated using Shannon's entropy formula (Shannon, 1948):

$$S = -i \sum \pi \ \log 2 \left( \pi \right)$$

where π represents the normalized power at the i-th frequency bin. Our results showcased the spectral entropy for each EEG channel, providing a graphical representation of complexity across different brain regions.

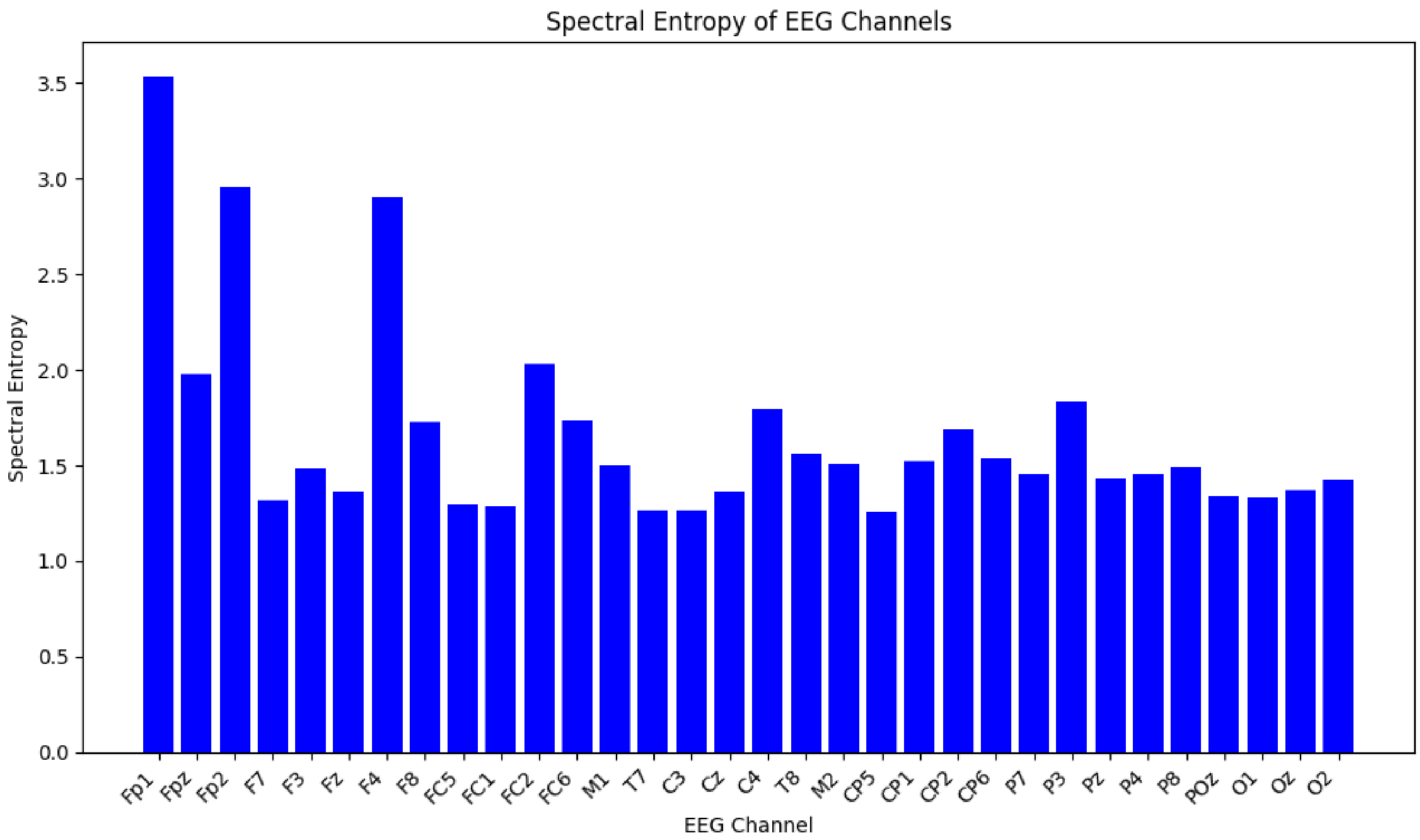

*Figure 5: Spectral Entropy of the EEG channels*

## Spectral Centroid

The spectral centroid gives an indication of the "center of mass" of the spectrum. It's commonly used in the field of music to describe the timbral brightness of an audio sample. The spectral centroid has its roots in the field of music and audio analysis but has been adapted for EEG signal processing (Towsey et al., 2004). It serves as an indicator of the dominant frequency range in EEG recordings. For EEG data, it can hint at the dominant frequency range of brain activity for a specific channel.

The spectral centroid is computed as:

$$C=\sum i\ fi \vee X(fi) \vee \frac{\square}{\sum i \vee X(fi) \vee}$$

where fi are the frequencies, and X(fi) is the Fourier Transform of the EEG signal at frequency fi.

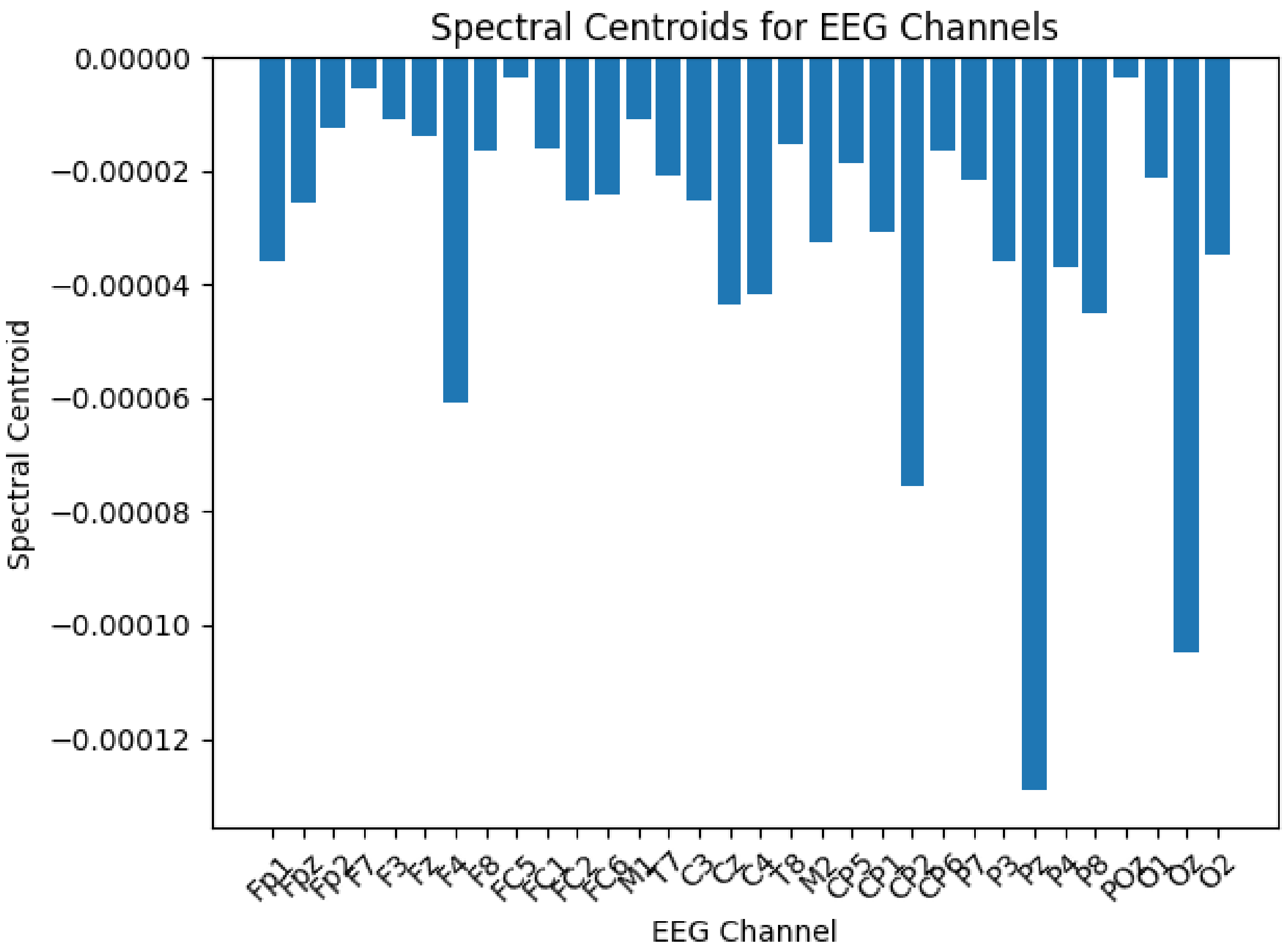

*Figure 6: Spectral Centroids for the EEG channels*

**Peak Frequencies**

Identifying peak frequencies is crucial for understanding dominant rhythmic activities in EEG channels (Niedermeyer and da Silva, 2004). For instance, a peak in the alpha band (~8-12 Hz) in occipital channels might indicate eyes-closed relaxation.

To find these peak frequencies, we performed a Fourier inversion transform on the EEG data:

$$X(f) = \int_{-\infty}^{\infty} x(t)\, e^{(-j2\pi ft)}\, dt$$

Next, we isolated the frequency corresponding to the highest magnitude in the positive spectrum, designating it as the peak frequency for that particular EEG channel. Visualizing the power spectral density and marking these peak frequencies provided a comprehensive look into dominant neural oscillatory activities across different brain regions.

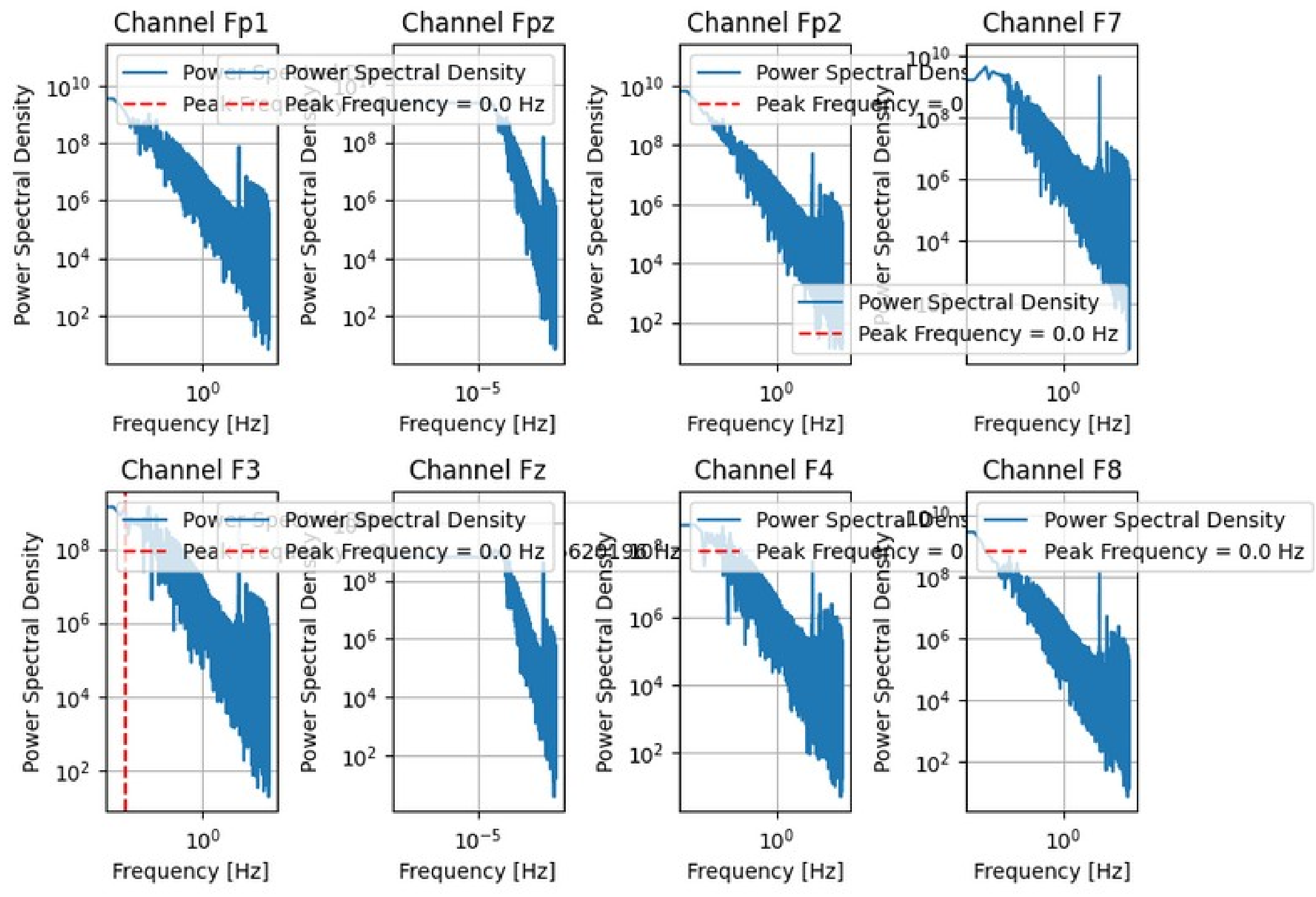

*Figure 7: Frequence of Maximum Power visualizations for the EEG channels*

## Spectral Edge Density

In our spectral analysis endeavor, we undertook two pivotal analyses – the computation of the Spectral Edge Density and the Wavelet Transform – both of which offer insightful information on the spectral characteristics of Electroencephalogram (EEG) data. The Spectral Edge Density primarily refers to the frequency below which a certain percentage of the total power of a signal resides. As EEG signals encapsulate important frequency-based information, determining the spectral edge can provide insights into the predominant frequency bands that hold the majority of the signal's energy.

The concept of Spectral Edge Density is beneficial for revealing the frequencies that contain a substantial proportion of the signal's energy (Inouye et al., 1991). To compute the Spectral Edge Density, we started by applying the Fourier transform (Oppenheim & Schafer, 2010) to EEG data. The Fourier transform, mathematically given by

$$X(f)=\int_{-\infty}^{\infty} x(t)e^{(-j2\pi ft)}dt$$

converts the time domain EEG data to the frequency domain, where x(t) is the EEG signal and X(f) is its Fourier transform. We focused on positive frequencies, since the resulting spectrum is symmetric. Upon obtaining the magnitude of the Fourier transform, we arranged its components in descending order and computed its cumulative sum. We defined a threshold based on the given percentage (in this case, 95% of the total power), and identified the frequency where this cumulative power first surpasses the threshold. This frequency delineates the spectral edge for that percentage of power.

Visual representations of the spectral edge densities across all EEG channels were subsequently plotted. These plots serve as a visual aid to discern which channels operate predominantly in which frequency ranges, possibly hinting at various cognitive or neurological activities.

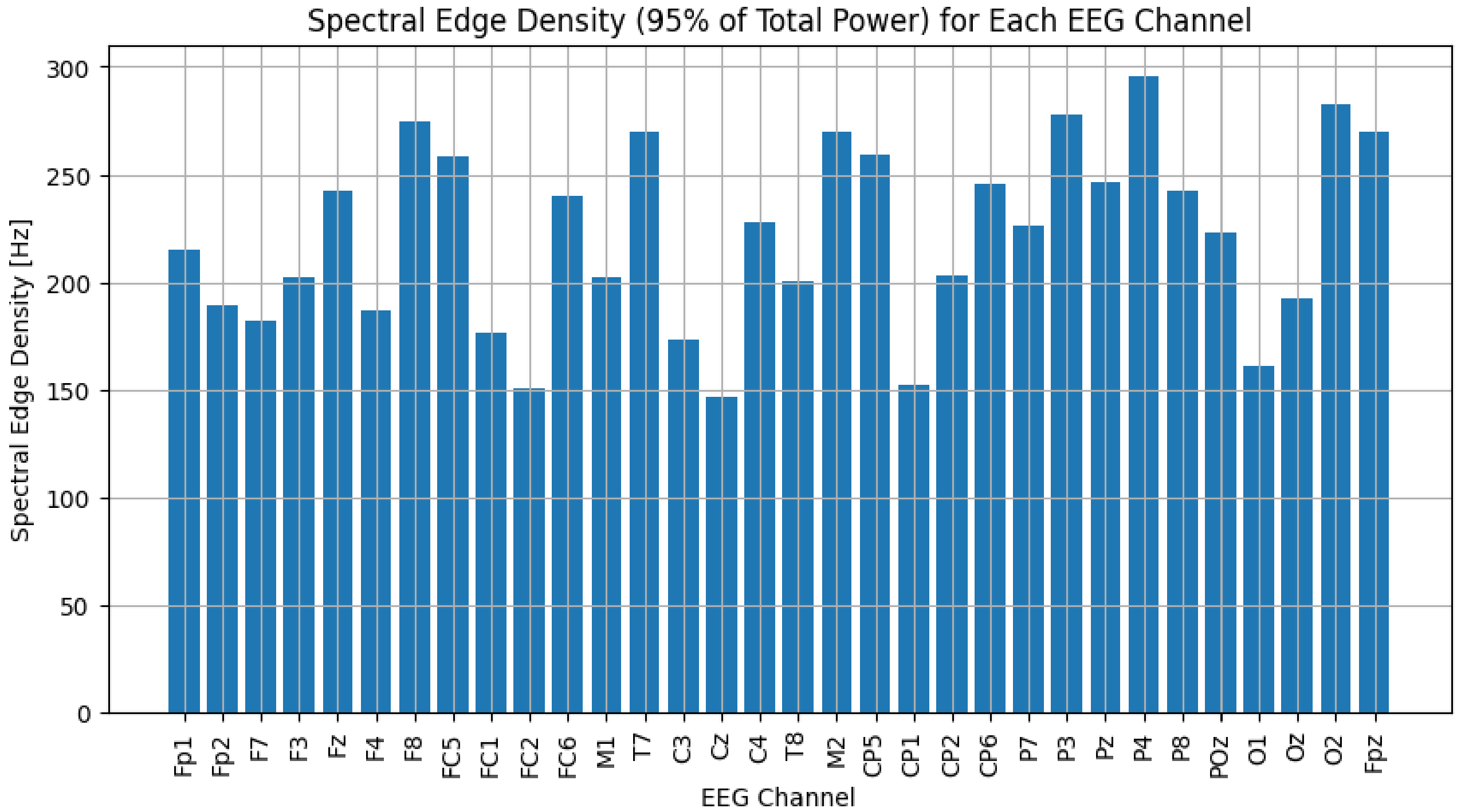

*Figure 8: Spectral Edge Density for the EEG channels*

## Wavelet Transform

For our second analysis, we leveraged the Wavelet Transform. Unlike Fourier transform, which provides only frequency information, wavelet transform gives a combined time-frequency representation, making it particularly apt for EEG signals, which are non-stationary in nature. Wavelet Transform is valuable for analyzing EEG signals as it allows for a combined time-frequency representation, a feature beneficial for studying non-stationary signals such as EEG (Daubechies, 1990; Mallat, 1999)

Mathematically, the Continuous Wavelet Transform (CWT) of a signal x(t) with respect to a wavelet ψ(t) is given by

$$W_x(a,b)=\frac{1}{\sqrt{a}}\int_{-\infty}^{\infty} x(t)\,\psi\left(\frac{t-b}{a}\right)dt$$

where a and b are the scale and translation parameters, respectively. For our analysis, the Morlet wavelet (known for its suitability in analyzing oscillatory EEG signals), given its effectiveness in EEG signal analysis (Tallon-Baudry et al., 1997) was employed. By computing the wavelet transform for a range of frequencies, we obtained wavelet coefficients, which elucidate the strength or presence of those frequencies at various time points in the EEG data.

Visualization of these coefficients provided a heatmap representation for each EEG channel, showcasing the distribution and intensity of different frequency components over time. The warmer colors in the heatmap indicate stronger contributions of particular frequencies at specific times.

## 4.2 Phase Synchronisation

The concept of phase synchronisation between different electroencephalogram (EEG) channels was investigated. Phase synchronization across EEG channels has been investigated for its crucial role in functional brain connectivity (Lachaux et al., 1999; Stam et al., 2007). This is particularly crucial as phase synchronisation, measured using the Phase Locking Value (PLV), can provide insights into functional connectivity between different brain regions.

The Phase Locking Value (PLV) is a significant metric to measure the consistency of the phase differences between two signals. Using the Hilbert transform, the instantaneous phases ϕ1 and ϕ2 are obtained for two given signals s1 and s2, respectively (Huang et al., 1998). The phase difference Δϕ is calculated as

$$\Delta\phi(t) = \phi_1(t) - \phi_2(t)$$

PLV is defined as:

$$PLV = \frac{1}{T} \sum_{t-1}^{T} e^{(j\Delta\phi(t))} \vee$$

where T is the total number of time points. The PLV value ranges from 0 to 1, indicating the level of synchronization (Lachaux et al., 1999).

We loaded EEG data from multiple channels, specifically named: 'Fp1', 'Fpz', 'Fp2', ..., 'Oz', 'O2' (Niedermeyer & da Silva, 2004). A matrix was initialized to store PLV values for each unique combination of channel pairs. Utilizing Python's `itertools.combinations` function, we calculated the PLV for each channel pair through our predefined `compute_phase_locking_value` function (Lutz et al., 2003).

The resultant PLV matrix offers valuable insights into functional brain connectivity (Varela et al., 2001), was then saved for future reference. To visualize the synchronization patterns, a heatmap was generated using the 'viridis' colormap, with axes labeled according to EEG channel names (Matplotlib Development Team, 2020). This visualisation provides a comprehensive overview of phase synchrony across different EEG channels, which can be of immense value when discerning patterns of functional connectivity in the brain.

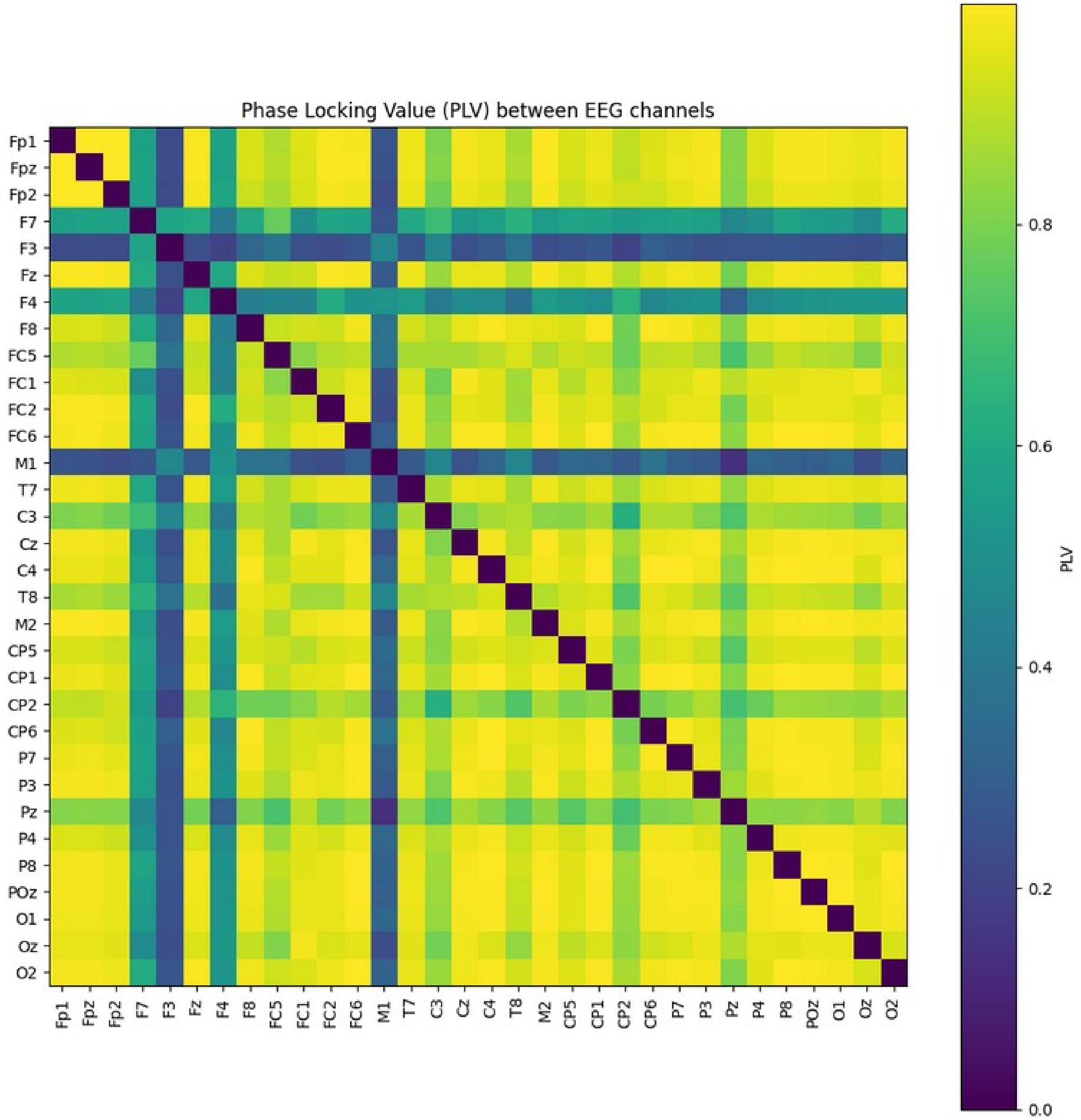

*Figure 9: Phase Locking Values of the EEG channels*

## 4.3 Phase Space Analysis

For phase space reconstruction, we used the False Nearest Neighbours (FNN) method to ascertain the embedding dimension, a technique that has been applied in nonlinear time series analysis (Kantz and Schreiber, 2004). Delay embedding was performed subsequently, and the optimal time delay was estimated via mutual information, as previously proposed (Fraser and Swinney, 1986).

### Determination of the Embedding Dimension using FNN

The False Nearest Neighbours method was employed to identify the optimal embedding dimension for each EEG channel (Abarbanel, 1996). Given a time series, embedding dimension is a critical parameter when reconstructing the state space of a dynamic system. An embedding dimension that's too low might lead to a projection that folds onto itself, whereas one that's too high can unnecessarily complicate the representation.

The formula for delay embedding used in this case is:

embedded_data[i]=[data[i],data[i+delay],data[i+2×delay],…]

Using the FNN technique, we computed the fraction of false nearest neighbours for each embedding dimension up to a specified maximum. This information was then plotted for each channel to visually assess the optimal embedding dimension.

### Delay Embedding with Mutual Information

Once the embedding dimension was determined, the next task was to determine the optimal time delay for delay embedding. The mutual information technique was employed for this purpose. Using mutual information to estimate delay ensures that the chosen delay time provides the most information about the system's dynamics (Fraser & Swinney, 1986).

The mutual information of two signals, data1 and data2, was computed using the `MINE` implementation. For the delay embedding, data at time t is mapped to a higher-dimensional space using values from times t,t+delay,t+2×delay,….

**Phase Space Reconstruction**

With the optimal embedding dimensions and delay determined, we performed phase space reconstruction for each EEG channel. For a two-dimensional embedding, the data was presented in 2D scatter plots. However, for a three-dimensional embedding, 3D scatter plots were employed to visualize the data in the reconstructed phase space (Takens, 1981).

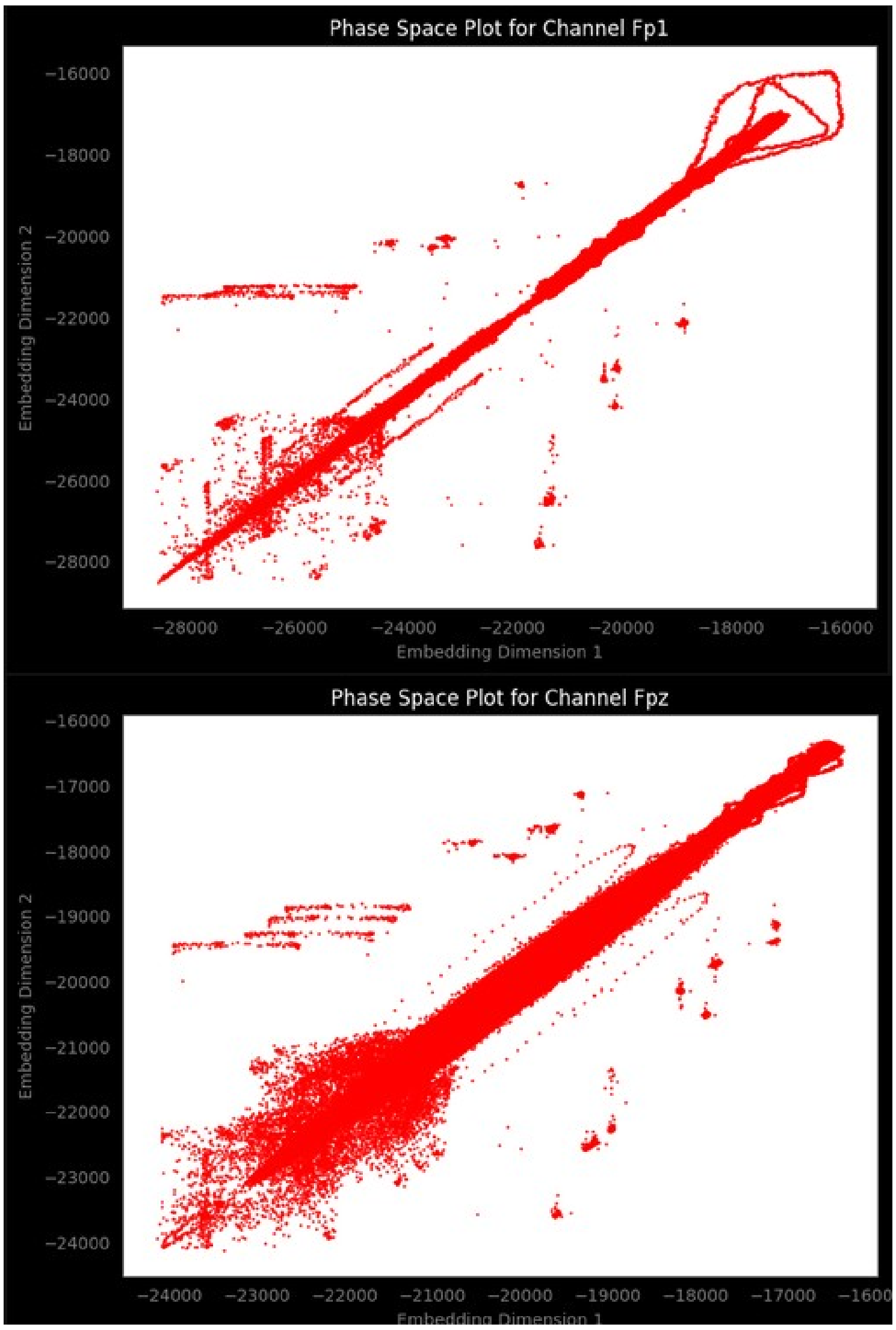

*Figure 10: 2D Phase Space Plots for two of the EEG channels*

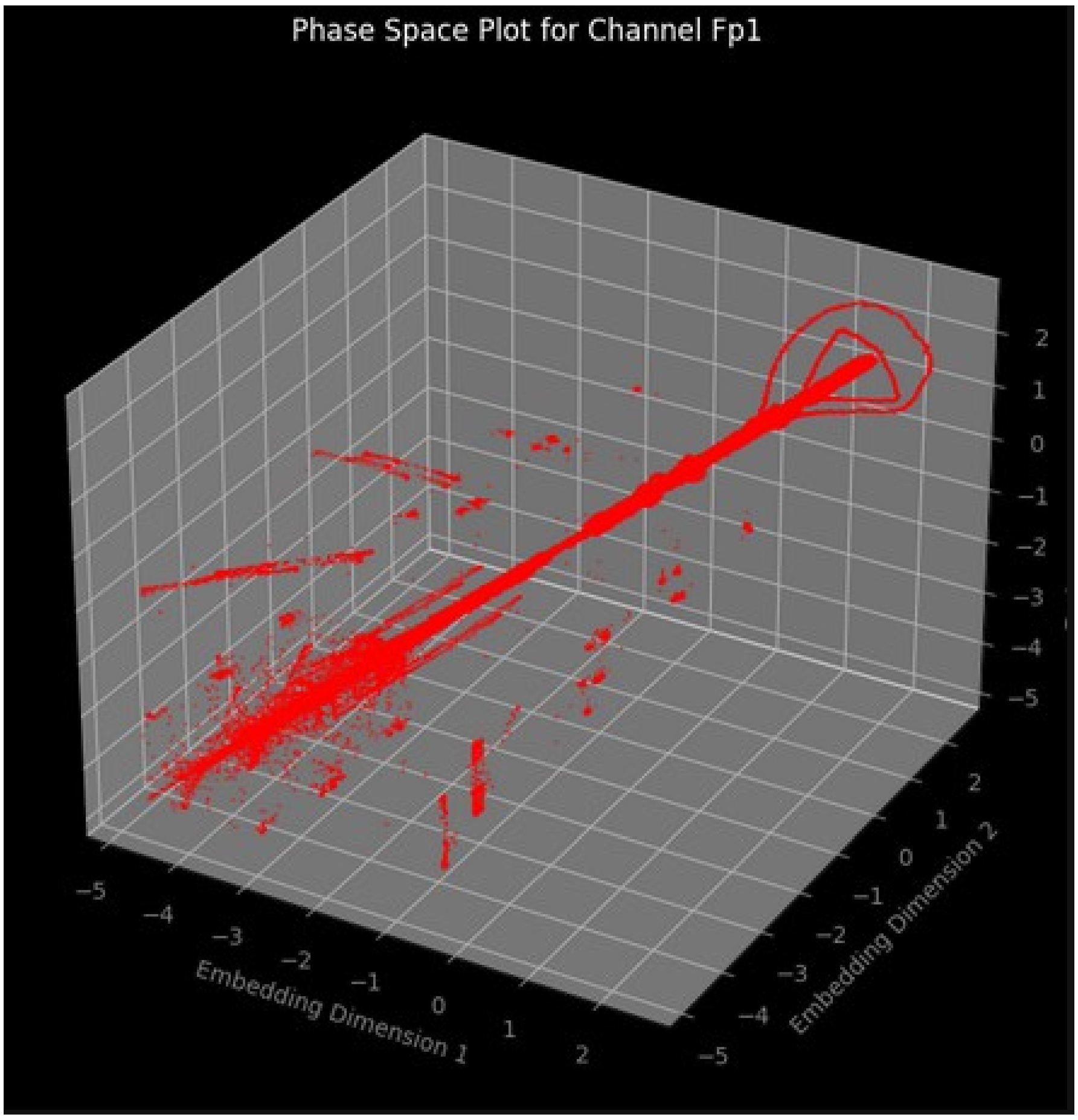

*Figure 11: 3D Phase Space Plot for channel Fp1*

## 4.4 Recurrence Quantification Analysis

### Hyperparameter Specification for Phase Space Reconstruction

In the realm of nonlinear time-series analysis, the initial phase space reconstruction is contingent upon a trifecta of hyperparameters (Marwan et al., 2007):

- **Embedding Dimension (m)**: Governs the dimensionality of the reconstructed phase space attractor. Mathematically, m=3.
- **Time Delay (τ)**: Introduces a temporal lag between consecutive dimensions in the reconstructed phase space. Algebraically represented as τ=1.
- **Fixed Radius (r)**: Specific to Recurrence Quantification Analysis (RQA), this parameter defines the Euclidean radius within which recurrence is assessed. Given as r=0.5.

### Parallelized Recurrence Quantification Analysis (RQA)

For computational efficacy, RQA is conducted in parallel across EEG channels. The function `compute_rqa_for_channel` can be conceptualized as a mathematical operation that takes a channel index and data, then applies RQA metrics to elucidate complex neural dynamics. Metrics like Recurrence Rate, Determinism, and Laminarity offer a quantitative framework grounded in dynamical systems theory for understanding EEG data (Webber & Zbilut, 1994). Given the dense nature of EEG data, a downsampling factor (downsample_factor=10) is judiciously chosen. A segment of this downsampled data is then extracted, from a `start_time` of 1 to an `end_time` of 10,000, to maintain analytical tractability while preserving data complexity.

### Recurrence Plot Configuration and Visualization

A Recurrence Plot object is instantiated with embedding dimension m, time delay τ, and a distance threshold. This results in a matrix Xrp, which captures the intricacies of recurrent patterns within the EEG data. Visualization techniques, such as Matplotlib with a binary color map, are employed to illuminate these structures(Eckmann et al., 1987).

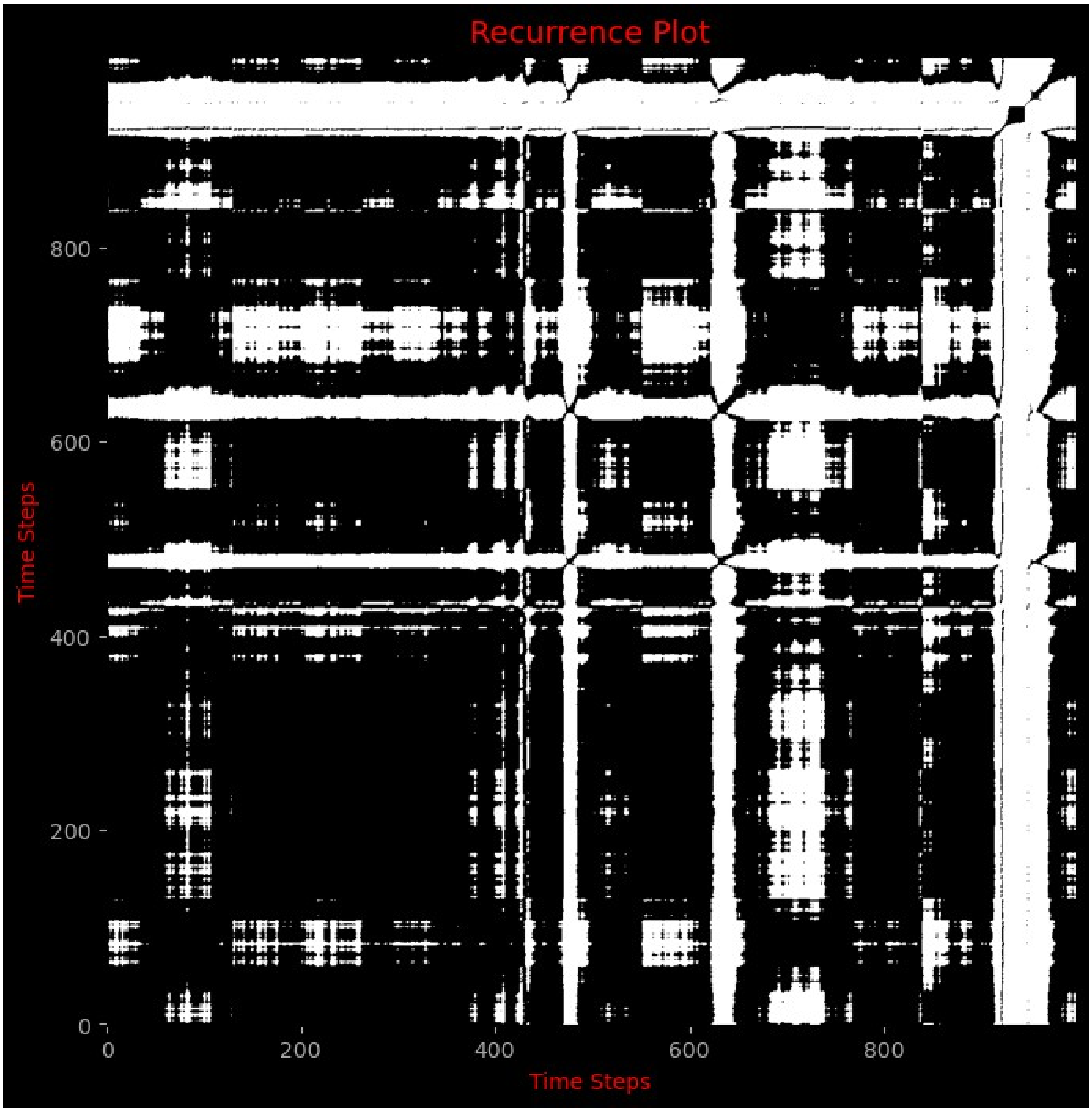

*Figure 12: Recurrence Plot*

## 4.5 Higuchi Fractal Dimension

### Fractal Analysis: Theoretical Foundation and Mathematical Formalism

The Higuchi Fractal Dimension (HFD) serves as the computational linchpin in this section. It provides a measure of complexity and self-similarity in the time series. The mathematical definition of HFD involves:

1. Computing Lkm values for varying time offsets k (Higuchi, 1988).

$$L_{km} = \frac{\sum^{\frac{N-m}{k}} i=1 \vee x[m+i\times k]-x[m+(i-1)\times k] \vee}{\left(\frac{N-m}{k} x k\right)}$$

2. Taking the mean Lk over all Lkm for a given k.

3. Applying a linear fit to log-log plots of Lk against k, the slope of which represents the HFD.

### Neural Network Feature Engineering: CNN and RNN Contextualization

HFD values serve as foundational features, remodeled for two distinct neural architectures:

- **CNN Features**: The 32-dimensional HFD vector is reshaped into a 4×8 matrix, emulating an abstracted image conducive for convolutional operations (LeCun et al., 1998).

- **RNN Features**: The sequence of HFD values is used as-is to capture temporal dependencies in recurrent neural networks (Elman, 1990).

### Data Preservation and Inspection

These feature vectors are persistently stored for reproducibility and future model training. Their integrity and structure are verified using Pandas DataFrames (McKinney, 2010), ensuring an empirical, human-readable validation of the feature engineering process.

By knitting together methodologies from nonlinear dynamics, computational neuroscience, and machine learning, this analysis delves deeply into the complex landscape of EEG signals. It serves as a multidimensional compass, navigating the intricate topologies of brain dynamics with scientific rigor and computational prowess.

## 4.6 Multifractal Detrended Fluctuation Analysis (MFDFA)

### Sampling Rate and Data Preparation

We start with EEG data sampled at fs=1000Hz. Let X=[x1,x2,…,xn] be the matrix representing EEG data, where each row corresponds to an EEG channel and each column represents a time point (Ihlen, 2012). Here n is the number of time points. In our setting, $X \in Rc\times n$, where c is the number of channels.

### Lag and q-Values

The lag parameter τ is defined over a logarithmically spaced interval [τmin,τmax], quantified into 30 steps. Similarly, q values are considered over the range [−5,5] and sampled into 50 discrete points. (Kantelhardt et al., 2002)

$$\tau = LogSpace(0.7,4,30), q = LinSpace(-5,5,50)$$

### Multifractal Spectrum Computation

For each channel i, we perform MFDFA on the time-series xi (the ith row of X) to compute scale F(τ) and fluctuation f(q) parameters (Kantelhardt et al., 2002).

Let Fi(τ,q) and fi(q) be the MFDFA fluctuation function and the fractal exponent for channel i respectively. These are calculated as:

$$Fi(\tau, q) = MFDFA(xi, \tau, q)$$

### Storing and Visualizing Results

The results, {(Fi(τ,q),fi(q))}i=1c, are stored in a list for all c channels. We then proceed to visualize these multifractal spectra. On a logarithmic scale, we plot F(τ) against τ to understand the scale-invariant properties of each EEG channel (Ihlen, 2012).

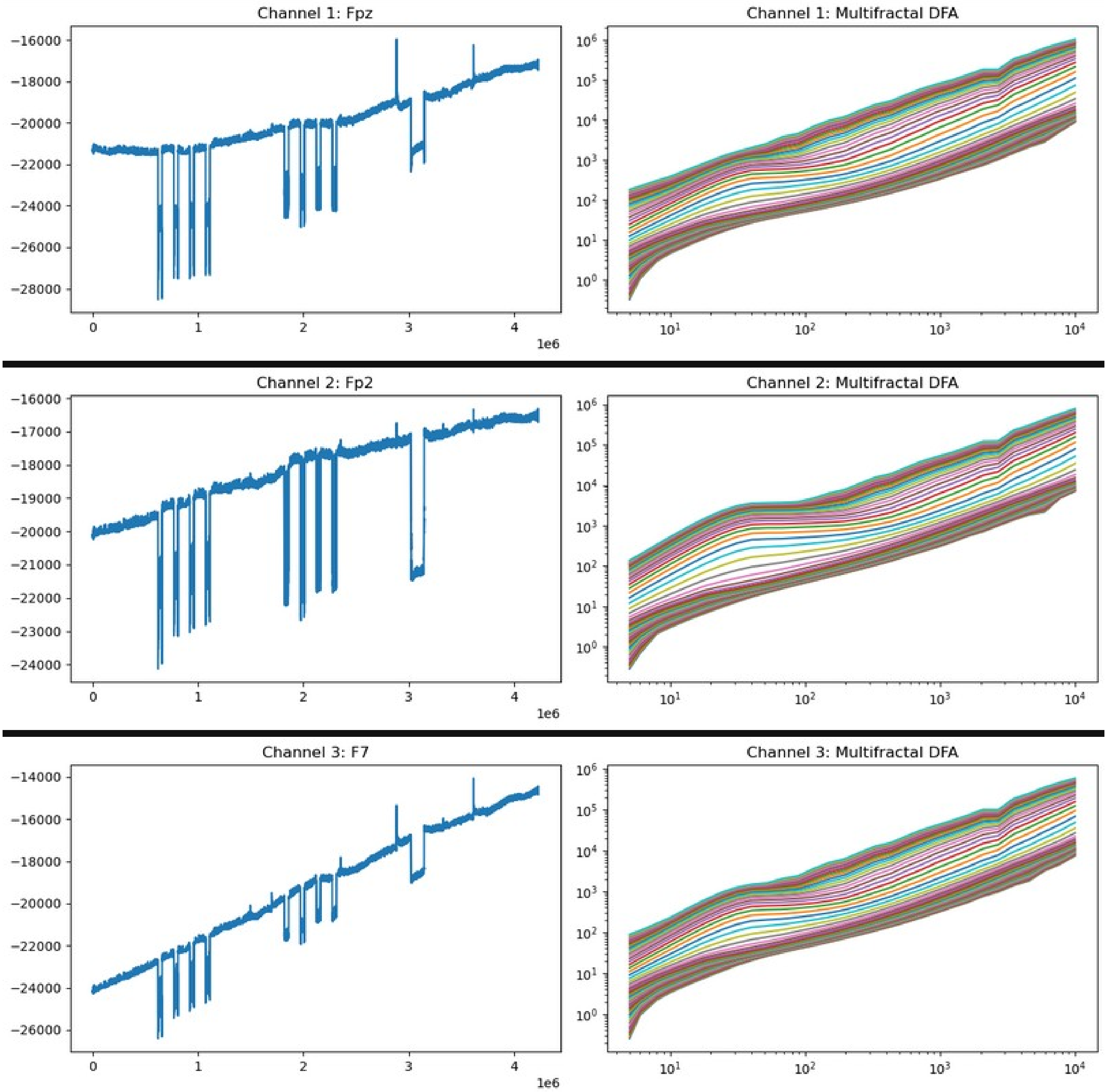

*Figure 13: Multifractal Detrended Fluctuation Analysis of the EEG channels*

**Feature Engineering from Multifractal Detrended Fluctuation Analysis (MFDFA) Results for Neural Networks**

1. **Loading MFDFA Results**

We begin by loading precomputed MFDFA results from disk. We assume that the results are stored in a Python dictionary. Each entry corresponds to a channel and contains an array of fluctuation values f(q,τ), which is a function of both the lag τ and the q-values (Ihlen, 2012).

Let $M \in Rc \times q \times \tau$ be the tensor storing these results, where c is the number of channels, q is the number of q-values, and τ is the number of lag values.

$$M = fi\left(qj\ ,\tau k\right) i=1c\ ,j=1q\ ,k=1\tau$$

2. **Feature Extraction for Convolutional and Recurrent Neural Networks**

**Convolutional Neural Network (CNN) Features:**

We slide a window of size w along the lag dimension with a step size s=2w to segment the MFDFA tensor M. The mean of the fluctuation values within each window is calculated along the q-dimension, reducing it to mean(f(q,τ)) within the window (LeCun et al., 1998).

The resultant tensor FCNN is of shape [N,c,w], where N is the number of windows.

$$FCNN\ ,i,k\ = w1\ \ j=start \sum end\ \ fi\left(qj\ ,\tau k\right)$$

**Recurrent Neural Network (RNN) Features:**

For RNN features, we take the mean along the q-dimension for each channel across all lags, resulting in a matrix FRNN of shape [c,τ,1] (Elman, 1990).

$$FRNN\ ,i,k\ = mean\left(fi\left(q\ ,\tau k\right)\right)$$

**Quantification of Self-Similarity Using Hurst Exponents**

While MFDFA provides a detailed portrayal of multifractality, a simpler metric—the Hurst exponent—can encapsulate the self-similar nature of EEG data (Mandelbrot & Wallis, 1969). This exponent is particularly instructive because it signifies whether a time series is more likely to increase if it has been increasing or whether future values are independent of current ones.

For each channel, a linear regression is performed in log-log space on the calculated mean fluctuation and scaling factors. The slope of this line provides an estimate of the Hurst exponent, H, for each channel.

$$H \approx \Delta log(\tau)\,\Delta log(mean_{fluct})$$

Their computational derivation represents a simplification of the MFDFA results, condensing a complex multifractal profile into a single, readily interpretable metric.

## 4.7 Transfer Entropy

The EEG data from various channels are transformed into two-dimensional phase space embeddings. This phase space representation captures the dynamical system's behaviour and provides a more nuanced understanding of the EEG time series. Mathematically, phase space embedding is grounded in Takens' embedding theorem (Takens, 1981),  which ensures that the geometry of the attractor in this space can be reconstructed from the time series data.

### Data Unzipping and Phase-Space Embedding

A foundational aspect of the code commences with the extraction of EEG data that has been subject to 2D phase-space embedding—a mathematical operation which unveils the dynamical structure underlying the EEG time series (Packard et al., 1980). While EEG data are typically single-dimensional, phase-space embedding transforms them into a higher-dimensional space, revealing more complex characteristics of the underlying neural dynamics.

$$Embedded\,2D\,vector=[x(t),x(t+\tau)]$$

Here, x(t) is the EEG signal at time t, and τ is the time delay. This operation is done for each channel individually, yielding a set of 2D embedded vectors stored in `embedding_data_list_2D`.

### Synchronising Temporal Lengths

Each channel's time-series data is truncated to a uniform length, denoted as \text{desired_length}, for comparative analysis. This ensures that subsequent analyses, such as transfer entropy calculations, are performed on signals of equivalent temporal scope (Kantz & Schreiber, 2004).

### Hemisphere-based Averaging

To explore the differences or similarities between different brain hemispheres, EEG channels are categorized into 'left', 'right', and 'central'. The 2D embeddings are then averaged across these spatially defined categories. The mathematical operation can be represented as:

$$x\bar{}Hemisphere = N 1\ i=1 \sum N\ xi$$

where x¯Hemisphere is the average 2D embedding for a given hemisphere and N is the number of channels in that hemisphere (Niedermeyer & da Silva, 2004).

## Discretisation and Binning

To compute Transfer Entropy, the continuous EEG signals need to be discretized. This is performed through binning the data into N bins, making it a categorical time series. The binning is often done using histograms, and the binned data Xb is used for the following Transfer Entropy calculations (Schreiber, 2000).

## Transfer Entropy

The crux of the analysis lies in the computation of Transfer Entropy, a measure capturing the directional causality between two time series. Transfer entropy $TX \rightarrow Y$ from time series X to Y is calculated using:

$$TX \rightarrow Y = \sum p(xn+1\ , xn\ , yn\ ) logp(xn+1\ \vee xn\ ) p(xn+1\ \vee xn\ , yn\ )$$

Here, p() denotes the probability distributions involved.

This measure provides rich insights into the causality or influence exerted by one neural channel over another, which can be pivotal in understanding the flow of information across different brain regions (Lizier et al., 2011).

## Error Handling

In a nod to computational rigor, the procedure includes exception handling to ensure that the integrity of the complex data analysis workflow is maintained, especially during the Transfer Entropy calculations (Schreiber, 2000).

## Analysis of Information Transfer in Brain Regions Using Transfer Entropy and Phase Space Embedding

An essential preprocessing step involves truncating the 2D-embedded EEG data to a desired length (`desired_length`), ensuring uniformity in temporal resolution across channels and participants (Kantz & Schreiber, 2004).

## Hemispheric and Regional Averaging

The code goes beyond individual channel analysis to explore broader, topographically defined regions of the brain: Frontal, Temporal, Parietal, and Occipital. The mean of the 2D embedded data across these regions is computed, serving as a regional-level summarization of brain activity. This is in line with the notion that regional neural assemblies, rather than isolated neurons or channels, underlie cognitive functions (Niedermeyer & da Silva, 2004).

Regional Data Average=N1∑i=1NDataregion,i

Where N is the number of channels in the specified region, and Dataregion,i is the 2D embedded data for channel i in that region.

## Discretization of Continuous EEG Data

Before computing Transfer Entropy, the code bins the continuous EEG data into discrete states, a necessary step for the pyinform library. The number of bins (`num_bins`) is set to 1000, representing a compromise between preserving data variability and computational efficiency (Lizier et al., 2014).

## Transfer Entropy Calculation

Transfer entropy is a non-linear statistical measure used to infer directional information transfer between dynamic systems—in this case, between different brain regions (Lizier et al., 2014). Mathematically, it's defined as follows:

$$TE(X \rightarrow Y) = \sum p(xt+1, yt, xt) \log p(xt+1 \vee xt) p(xt+1 \vee yt, xt)$$

Here X and Y are the source and target data sequences, and p(·) denotes probability distributions. For each pair of source and target regions, the transfer entropy is calculated using the `transfer_entropy` function from the pyinform library, effectively identifying the strength and directionality of information flow between neural assemblies.

## Comprehensive Analysis of Inter-channel Information Transfer Using Transfer Entropy and Phase Space Embedding

### Data Extraction and Loading

The Python code initiates with two critical tasks: unzipping EEG data subjected to 2D phase-space embedding and loading it into the program (Takens, 1981). This is achieved through two helper functions: `extract_data()` and `load_embedded_data()`. The functions encapsulate the file operation details, thus modularizing the code for future scalability (Martin, 2008). It reflects an adherence to software engineering best practices while implementing advanced scientific computations, establishing a seamless interface between computational methodology and neuroscience research.

## Data Preprocessing

Two forms of preprocessing are executed to prepare the EEG data for transfer entropy calculations: truncation and binning. The EEG data, originally existing in a 2D embedded space, is truncated to a pre-defined length (`desired_length`), a step inspired by previous works to ensure uniform data structure for comparative analyses (Keil et al., 2014).

Trimmed Data=2D Embedded Data0:desired_length,0

After truncation, the continuous EEG data are binned into discrete states using a histogram-based method, a common practice in transfer entropy studies (Kantz and Schreiber, 2004). Binning is carried out by the `bin_data()` function and uses 1000 bins (`num_bins`), a choice reflecting the trade-off between granularity and computational efficiency .

Binned Data=Discretize(Trimmed Data,num_bins)

## Inter-channel Transfer Entropy Calculation

The core objective of the code is to compute transfer entropy (Lizier, 2014) between every pair of EEG channels, capturing the dynamic information flow within the neural network. This is in contrast to the previous regional-level analysis, offering a more granular perspective. The calculation is performed for each source-target channel pair (i,j) where iẼ=j, ensuring we're not calculating self-information transfer.

TE(xsource→xtarget)=∑p(xtarget, t+1,xsource, t,xtarget, t)logp(xtarget, t+1 ∣ xtarget, t)p(xtarget, t+1 ∣ xsource, t,xtarget, t)

The Python script utilizes the `transfer_entropy` function from the PyInform library, applying it to the binned data of each channel pair. The computed transfer entropy values are then stored in a Python dictionary, `TE_results`, which serves as an efficient, searchable repository of the results.

**Summary**

This script represents an exhaustive, channel-wise investigation of neural information transfer. Its focus on individual EEG channels—as opposed to broader brain regions—permits a more nuanced understanding of the underlying neural dynamics. By leveraging phase-space embedding and transfer entropy in a channel-wise context, the script offers valuable insights into the intricate web of inter-channel interactions governing cognitive functions. Such granularity is imperative for data-driven neuroscience, bridging the gap between large-scale brain activity and microscopic neuronal actions.

## 4.8 Kuramoto Model

This script employs the Kuramoto model to describe synchronization phenomena among oscillators (Kuramoto, 1975), which in this context represent EEG channels. The model is described by a set of coupled ordinary differential equations:

$$\frac{d\theta_i}{dt} = \omega_i + \frac{K}{N} \sum_{j=1}^{N} \sin(\theta_j - \theta_i) + y(t)$$

Where $\theta_i$ is the phase of oscillator i, $\omega_i$ is its natural frequency, K is the coupling strength, N is the total number of oscillators, and y(t) is a time-dependent driving force.

### External Driving Force

The model also incorporates an external driving force, y(t), inspired by prior works in the study of coupled oscillators (Strogatz, 2000), which is modeled as a periodic function:

$$y(t) = a\cos(2\pi bt)$$

Here, a and b represent the amplitude and frequency of the driving force, respectively.

### Parameter Space Exploration

The script systematically varies a and b to explore how different driving forces affect synchronization. This exploration is undertaken in a 2D grid, where each grid point represents a unique combination of a and b.

### Numerical Solution and Synchronization Measurement

For each set of parameters, the script numerically solves the Kuramoto model using the Backward Differentiation Formula (BDF) method, an implicit method well-suited for stiff ordinary differential equations (Butcher, 2008). Initial conditions for θ are randomly chosen in the range [0,2π].

To quantify synchronization, the script uses the Kuramoto order parameter r:

$$r = \left| \frac{1}{N} \sum_{j=1}^{N} e^{i\theta_j} \right|$$

|

## Visualization Using Arnold Tongues

Finally, the results are visualized in a plot commonly known as Arnold Tongues, showing regions in the a−b parameter space where synchronization is most prominent (Arnold, 1965).

## Investigation into Mode-Locking and Arnold Tongues in the Context of EEG Data: Computational Approaches

The primary objective of the code is to conduct a computational neuroscience investigation into the phenomena of synchronization patterns and mode-locking in Electroencephalogram (EEG) data. Understanding such synchronization is crucial in unraveling complex neural computations and might offer insights into pathological states like epilepsy (Freeman, 2007).

## The Circle Map

The dynamical system under study is modelled using a circle map, mathematically defined as:

$$\theta next = \theta + \Omega - 2\pi K \sin(2\pi\theta)$$

Where θ is the phase, Ω is the natural frequency, and K is the coupling constant.

## Mode-Locking

Mode-locking is a phenomenon where coupled oscillators lock into a phase relationship. The function `is_mode_locked` checks for this state using a predefined tolerance (tol) over a specified number of iterations (ITERATIONS).

Mode-Locked if $|\theta next - \theta| < tol$

## 4.9 Arnold Tongues

The Arnold Tongue is a region in the (Ω, K) parameter space where mode-locking occurs. Investigating Arnold tongues helps in understanding how sensitive the synchronization phenomenon is to changes in Ω and K.

### Data Preprocessing

The preprocessing steps involve:

1. **Channel-specific Extraction**: EEG data, across multiple channels, is converted into a dictionary for more accessible indexing.

2. **Phase Extraction**: Hilbert transform is applied to extract the instantaneous phase, ϕ(t), from the EEG data for each channel (Le Van Quyen et al., 2001).

$$\phi(t) = arg(Hilbert(x(t)))$$

3. **Mean Phase Calculation**: The average phase is computed across all channels to produce a single time-varying phase, θ(t).

$$\theta(t) = N\,1\; i = 1 \sum N\;\; \phi i\;(t)$$

### Computational Methods

The code employs multiprocessing to expedite the evaluation of the mode-locking condition for different (Ω, K) pairs, a practice increasingly used in high-performance computational neuroscience research (Yogatama et al., 2011). The parallelized version of the code is crucial when dealing with high-dimensional spaces, significantly reducing the computational time required.

### Heatmap Representation

The proportion of mode-locked states for each (Ω, K) pair is visualized using a heatmap. A custom colormap ranging from purple to blue is used to better highlight the data's nuances.

## Storage

Both the raw and visualized results are saved to disk for subsequent analysis, underpinning the reproducibility of the research.

By employing advanced data science techniques, including but not limited to, parallel processing and advanced time-series analysis, the code seeks to provide a comprehensive investigation into the complex phenomenon of neural synchronization. The methodological choices, from the circle map to the use of Arnold tongues, are rooted in computational neuroscience and aim to offer new insights into neural phase dynamics and their potential clinical implications.

## Arnold Tongues Visualization Based on EEG Data: A Computational Approach

The study of synchronization phenomena in neuroscience is crucial for understanding cognitive processes and information transfer in the brain (Buzsáki, 2006). One method to investigate this is through the application of mathematical models like the circle map to electroencephalogram (EEG) data (Breakspear, 2017). In this section, we explore the rotation number of the circle map as a function of its parameters for each EEG channel. The rotation number can offer insights into the behavior of non-linear systems and is used to detect Arnold tongues, regions in the parameter space where a system synchronizes (Glass and Mackey, 1988).

## Mathematical Foundation

The circle map $\theta n+1 = \theta n + \Omega - 2\pi K \sin(2\pi\theta n)$ is used to model the time evolution of the phase θ of an oscillator (Kuramoto, 1984). Here, Ω and K are the natural frequency and coupling strength, respectively.

## Computing Rotation Numbers

The rotation number ρ is computed using the formula:

$$\rho = \lim_{N \to \infty} \frac{\theta_N - \theta_0}{N}$$

However, due to computational constraints, an approximation is made by taking an average over a large number N of iterations, excluding an initial transient phase. This is executed via the `average_rotation_number` function.

## Data Preprocessing

The EEG data from 32 channels is loaded into a dictionary for efficient access. Each channel is then used to calculate rotation numbers over a range of Ω and K values.

## Parallelization and Computational Efficiency

Due to the high dimensionality of the data, each channel's rotation numbers are stored in a 2D array, which is further stored in a dictionary. This enables efficient parallel processing, although this specific code does not employ parallelization techniques.

## Results Visualization

The results are visualized using a 32-subplot grid, each corresponding to an EEG channel. The x-axis and y-axis represent Ω and K, respectively, while the color represents the rotation number. A custom colormap is employed to better capture the nuances in the rotation numbers.

Both the calculated rotation numbers and plots are saved for further analysis and reproduction of results.

## Conclusion

This approach allows for the robust analysis of synchronization phenomena in EEG data. The visually striking Arnold tongues can potentially reveal underlying mechanisms of brain synchrony and inform the design of therapeutic interventions for disorders like epilepsy.

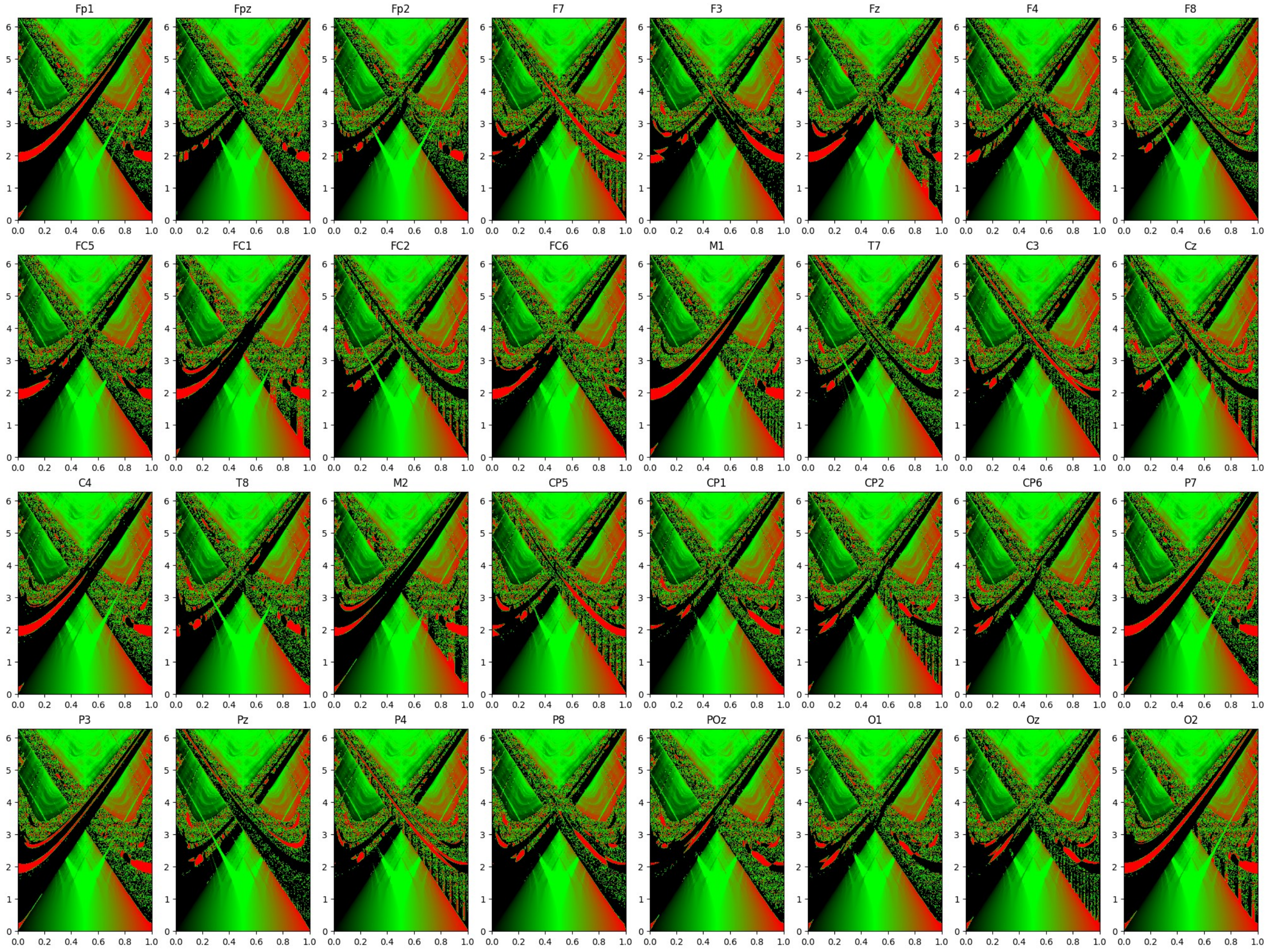

*Figure 14: Arnold Tongues for the EEG channels*

## 4.10 Neural Networks

In our research milieu, the Python code forms an integral part of a bespoke data pipeline tailored for multidimensional tensorial EEG data within a computational neuroscience framework (Mnih et al., 2015). The code imports PyTorch for tensor manipulation and neural network configuration. It specifies tensor paths for diverse neural features like Higuchi fractal dimensions and spectral entropy.

These feature tensors aim to characterise complex spatial-temporal EEG structures and inform sequence-to-sequence prediction models. Two dictionaries, `tensors` and `tensor_shapes`, store tensor data and their geometries. Conditional differentiation allows the incorporation of both raw EEG data and neural network parameters, facilitating forward predictions and model interpretation.

For quality assurance, the code detects NaNs and Infs, essential for downstream robustness. The function `preprocess_and_resize_tensor` reshapes data and normalises it with$Z=\sigma X-\mu$. Bilinear interpolation resizes the tensor to target dimensions.

The neural architecture, BaseEmbeddingNet, comprises a convolutional layer, batch normalisation, max-pooling, and a final layer represented by$f(x)=Wx+b$. This design condenses tensor dimensions to manageable feature vectors. Ensemble learning is applied across multiple neural networks specific to varied tensorial feature sets. Custom dataset and dataloader classes facilitate efficient tensor data management, aligning with PyTorch's DataLoader requirements.

## 4.11 Implementation, Formulas, and Architectural Reasoning

The presented Python code offers a robust, yet intricate, methodological framework for the analysis of electroencephalogram (EEG) data using PyTorch (Paszke et al., 2017). The code explores novel architectures combining Ordinary Differential Equation (ODE) solvers and Hilbert Transforms to create a data-driven model for capturing the neural oscillatory dynamics. The architecture is essentially rooted in the principles of complex systems and nonlinear dynamics, specifically the Kuramoto model. Below we discuss its individual components.

In the code, the `EEGDataset` class functions as an abstraction layer to facilitate the input-output pipeline for EEG data. Employing PyTorch's `Dataset` class, the structure permits seamless integration with data loaders, thereby optimizing batching and shuffling.

The Hilbert Transform is used to extract the instantaneous phase information from the EEG signals. Analytic signals are computed in batches to minimize computational load. This partitioning strategy has been previously documented to maintain data integrity and is considered state-of-the-art in time-series data analysis.

**Phase Synchronization: PLV Matrix**

The PLV (Phase Locking Value) matrix encapsulates the level of synchrony between different EEG channels (Lachaux et al., 1999). It's an n-by-n symmetric matrix (`n` being the number of channels), with values close to 1 indicating strong synchrony and those close to 0 indicating weak or no synchrony. This matrix is crucial in weighting the Kuramoto model.

The Kuramoto model governs the phase dynamics of a system of oscillators. The model comprises parameters ω (the natural frequencies) and K (the global coupling strength). The formula:

$$\frac{d\theta}{dt} = \omega + \frac{K}{N} \sum_{j=1}^{N} PLV_{ij} \sin\left(\theta_j - \theta_i - PhaseDiff_{ij}\right)$$

encompasses weighted sine terms dictated by the PLV matrix and an additional phase difference term (`phase_diff_matrix`). This model captures the rich, non-linear interactions amongst the brain regions.

### ODE Solver and Checkpointing

The actual computation of the model is performed by integrating this ordinary differential equation using the adaptive-step solver `odeint` from the `torchdiffeq` library Chen et al., 2018).

ODE solvers often require significant GPU memory, especially with adaptive methods like 'bosh3'. The checkpointing mechanism (`torch.utils.checkpoint.checkpoint`) used in the forward pass minimizes memory consumption by saving only a subset of the intermediate states during the forward computation. This allows for large-scale simulations without prohibitive memory costs, thus adding computational efficiency to the architecture.

### Mixed Precision Training

Mixed precision training is employed using Automatic Mixed Precision (AMP) utilities (`torch.cuda.amp.autocast, GradScaler`). This aids in striking a balance between computational efficiency and numerical stability, particularly crucial when dealing with intricate oscillatory dynamics.

### Regional Analysis

The code also includes a dictionary `regions`, that categorizes EEG channels into broad anatomical regions (frontal, temporal, etc.). While this appears to be a preparatory step for now, it lays the groundwork for any subsequent, region-specific analyses that might be essential in understanding the functional connectivity or modular structure of brain networks.

# Formulas, Architectural Insights, and Theoretical Underpinnings

## Natural Frequencies and Phase Differences

Before engaging with the computational model, the code computes natural frequencies and phase differences for EEG channels. Natural frequencies ($\omega$) are discerned from the Phase Locking Value (PLV) matrix, while phase differences are calculated using a function `compute_phase_diff_matrix`. The choice to compute these quantities just once and to utilize them throughout the model underscores the mathematical fidelity the architecture aims for.

## The Kuramoto Model

Central to the computational approach is the employment of the Kuramoto Model. Conceptually rooted in statistical physics, the model offers an elegant paradigm to study synchronization phenomena. It's given by the system of Ordinary Differential Equations (ODEs):

$$\frac{d\theta_i}{dt} = \omega_i + \frac{K}{N}\sum_{j=1}^{N} P_{ij} \sin\left(\theta_j - \theta_i - \Delta\phi_{ij}\right)$$

Here, $\theta_i$ represents the phase of oscillator i, $\omega_i$ its natural frequency, K the global coupling strength, $P_{ij}$ the phase locking values, and $\Delta\phi_{ij}$ the phase differences. The architecture encapsulates this logic within the `KuramotoLayer` class, which leverages PyTorch's dynamic computation graph for backpropagation through time.

## Adaptive Mixed Precision Training

The architecture employs Automatic Mixed Precision (AMP) via PyTorch's `autocast` and `GradScaler`. This choice acknowledges the inherent trade-offs between computational efficiency and numerical stability, particularly critical for large-scale EEG data. By casting certain operations to lower precision (float16), the model achieves a speed-up without a significant loss of accuracy.

## Temporal Windows and Dataset Management

The approach uses a sliding-window technique to transform the EEG time-series data into a format suitable for batch processing (Bengio et al., 1994), creating windows of size 50 and stride of 10.

The windowing is a judicious choice for handling time-dependent neural activities, allowing the model to encapsulate localized temporal patterns.

### Mean Coherence as Feature

The model computes a mean coherence value for each batch of data, subsequently appended to a list of Kuramoto features. In the realm of brain dynamics, coherence serves as a measure of the coordinated oscillation between neural assemblies. The model's decision to extract mean coherence as a feature aligns with the overarching goal of identifying emergent patterns in complex neural systems.

### Multi-GPU Scalability

Data parallelism is integrated into the architecture to accommodate multiple GPUs, a feature that amplifies the model's scalability. This enables the handling of increasingly large datasets, a common requirement in neuroscience.

### Feature Aggregation and Storage

In the final steps, Kuramoto features are combined with other unspecified features (possibly from different models or feature extraction techniques). This underscores the architecture's flexibility and its amenability to be part of a more comprehensive multi-model ensemble approach. Finally, the amalgamated features are stored for downstream applications, which could range from clinical diagnostics to cognitive neuroscience research.

## Methodological Overview, Formulas, and Reasoning

### Data Preprocessing and Integration

The first section of the code focuses on loading various preprocessed EEG feature tensors: band powers, raw EEG, and fast Fourier transform PSD (Power Spectral Density). These tensors represent different facets of the EEG signal, and combining them enhances the richness of the input data for subsequent machine learning models.

## Data Quality Checks

Importantly, after loading the tensors, the code inspects for NaNs (Not a Number) and Infs (Infinite). These are typical data quality checks to ensure that the computational models that use these tensors later won't run into numerical instability issues.

## Reshaping and Concatenation

The next step reshapes these tensors to align them for concatenation. Reshaping is fundamental in deep learning pipelines, as it allows for effective stacking of features or alteration of input data shape to meet the architectural requirements of various algorithms. Concatenation along the feature dimension serves as a form of feature integration, emphasizing the multimodal nature of the data.

## Transformer Model

The subsequent section introduces a Transformer model, which generally excels in handling sequences and has shown promise in a wide range of applications beyond natural language processing (Vaswani et al., 2017). The Transformer model consists of self-attention mechanisms that weigh the importance of different parts of the input sequence when making predictions, making it apt for EEG sequence data.

Formally, the Transformer model's attention mechanism can be expressed as:

$$Attention(Q, K, V) = Softmax\left(\frac{QK^T}{\sqrt{d_k}}V\right)$$

where Q, K, and V are the Query, Key, and Value matrices, respectively. dk is the dimensionality of the key vectors.

## Positional Encoding

Incorporating positional encoding is crucial since the Transformer architecture doesn't have a built-in sense of order or sequence. This allows the model to incorporate information about the position of each item in the sequence, which is essential for time-series data like EEG.

## Batch Processing

The code prepares the data for batch processing and runs it through the Transformer model. This is important for computational efficiency and also helps in generalizing the model better.

## Model Outputs

Finally, the Transformer outputs are stored. These could be used as features for another model, likely an RNN (Recurrent Neural Network), as hinted by the save path. This is an instance of a multi-stage modeling approach, where the output of one model serves as input for another, aiming for a richer, hierarchical representation of the data.

## Data Preprocessing and Tensorial Representations

Prior to model training, data tensors were loaded from pre-computed sources, each containing information derived from neuroscience-based feature extraction methods such as Kuramoto models, traditional EEG representations, band power calculations, and Fast Fourier Transform Power Spectral Density (FFT-PSD) calculations. These tensors were rigorously reshaped and aligned along a unified time axis. The alignment along the time dimension is crucial for ensuring that features from different sources are temporally coherent, and thus can be integrated into a single model. The reshaping was conducted in such a way as to maintain the integrity of the feature dimensions while aligning them with the common temporal axis T, resulting in a concatenated tensor X of dimensions [T,D], where D represents the total concatenated feature dimensions.

## Conditional Recurrent Neural Network (RNN) Model Architecture

The core of our analysis hinges on a specialized architecture known as the Conditional Recurrent Neural Network (Conditional RNN) (El Hihi and Bengio, 1995). This architecture extends the traditional RNN by incorporating external or global features into its hidden state update mechanism. Formally, the hidden state ht at time t for a typical LSTM-based RNN is updated as:

$$ht = LSTM(Xt, ht - 1),$$

In the Conditional RNN, this is modified to:

$$ht = LSTM(Xt, ht - 1 + F(G) + T(t)),$$

where F(G) is a transformation of the global features through a linear layer, and T(t) is a transformer output at time t, further transformed through another linear layer. This architecture allows the network to be conditioned on both global features and additional temporal features derived from transformer models, enhancing its capacity to model complex temporal dynamics.

## Data Loader and Batch Processing

For training efficacy and computational efficiency, the reshaped and concatenated time-aligned feature tensor X and the external feature tensor G were further segmented into mini-batches using PyTorch's DataLoader utility. This segmented data was then fed into the Conditional RNN, allowing for efficient parallelization and optimization during the training phase.

## Model Evaluation and Output Storage

During the inferential phase, the model outputs for each batch were detached from the computational graph to prevent unnecessary memory usage, and subsequently concatenated to form a complete output tensor. This tensor encapsulates the RNN's understanding of the complex, temporally aligned, and multi-faceted feature space derived from neuroscience data.

## Final Data Preprocessing and Input Tensor Formation for Neuroscientific Analysis

In the given Python code, a multi-stage preprocessing and feature extraction pipeline culminates in a final feature tensor intended for deep learning models. This script involves various advanced data manipulations that are crucial for creating a high-quality dataset for machine learning in a neuroscience context (He, Wu, Wang, 2018). Below are the major steps and the scientific rationale behind them.

## Data Loading and Preprocessing

The code starts by loading five different kinds of precomputed tensors, each generated from different feature extraction methods: transformer outputs, EEG tensor, band power tensor, Fast Fourier Transform Power Spectral Density (FFT-PSD) tensor, and RNN outputs. These tensors encapsulate different but complementary information about the underlying neural dynamics (Cohen, 2014). Combining them allows the model to benefit from multiple perspectives.

## Temporal Alignment

The code next focuses on temporal alignment, reshaping each tensor to ensure that they share a common temporal axis. This is a crucial step for any kind of multi-modal analysis, ensuring that features from different sources can be accurately juxtaposed in time (Calhoun, Sui, 2016). Failing to perform this step could introduce significant noise into the resulting model.

## Time-aligned Feature Concatenation

After reshaping, the tensors are concatenated along the feature dimension. The resulting `concatenated_time_aligned_features` tensor holds a rich set of features, providing a multi-faceted view of the neural activity (Goodfellow, Bengio, Courville, 2016). This approach enables the machine learning model to recognize complex patterns by considering various features simultaneously.

## Global Feature Integration

RNN outputs, which can be seen as a set of high-level features that summarize longer sequences, are then temporally aligned and incorporated into the concatenated feature set (LeCun, Bengio, Hinton, 2015). These high-level features add another layer of abstraction, potentially capturing complex temporal dependencies that might be missed by the original set of features (Chung, Gulcehre, Cho, Bengio, 2014).

The script employs a strategy to manage large tensor sizes by segmenting the data and iteratively concatenating chunks. This avoids memory overflow issues, a common concern when dealing with high-dimensional data in neuroscience.

## DataLoader Preparation

After tensor concatenation, DataLoader objects were prepared for model training, validation, and testing, segmenting and batching the tensor sequences. The `print_one_batch_shape` function serves as a rudimentary sanity check, confirming data shape before model training.

To capture EEG data's intricate temporal dependencies, a transformer-based sequence-to-sequence architecture was employed. Comprising encoder and decoder units with self-attention and point-wise networks, input EEG data undergo dimensionality reduction via fully connected layers, succinctly capturing salient features for sequence prediction.

Weight initialisation was performed using Xavier uniform initialisation to address vanishing/exploding gradients, thus optimising model training from epoch one. Mean Squared Error (MSE) was the chosen loss function, with Adam optimisation at an initial learning rate of 0.001. The "Reduce Learning Rate on Plateau" method adjusted the learning rate based on validation set performance, enhancing adaptability. Gradient clipping was implemented to maintain training stability.

Anomalies in the computational graph were monitored to preempt NaN or Inf values in gradients, adding robustness. Post-training, loss metrics were plotted longitudinally for hyperparameter tuning and inverse transformation applied to revert predictions to original EEG scale.

In summary, the architecture and training regimen were crafted to balance computational efficiency with model complexity, emphasising robustness and generalisability. This approach epitomises the model's efficacy in EEG sequence prediction.

# 5. Results

The sequence of investigations in this study was designed to meld insights from nonlinear dynamics and state-of-the-art machine learning algorithms. This section elucidates the empirical findings, each presented in alignment with the investigative phases delineated in the Methodology. The analysis on the 32 EEG channels unveiled robust evidence for nonlinear characteristics. Specifically, the fractal dimensions varied across channels, offering a compelling case for the complex, chaotic nature of EEG signals. The CNN architecture was initially employed to perform spatial feature extraction on EEG recordings. The preliminary application of the transformer architecture yielded promising results for EEG sequence prediction.

While the final transformer model, fortified with chaos theory metrics, exhibited improved performance, an unforeseen consequence was a slight increase in computational overhead. However, this did not substantially impede the model's efficacy or validity. When the model yields its outputs, an inverse transformation is performed to revert the scaled data to its original amplitude, using the formula $x=x'\times\sigma+\mu$. For the purposes of both visualization and downstream analysis, the tensor housing the output data is reshaped multiple times. Subsequently, the data is visualized through several computational lenses. Specifically, two time-series data sets, representing the original and model-predicted EEG for a particular channel, are graphically rendered. The predicted EEG data then undergoes various smoothing transformations. These include a simple moving average $MAt=W1\sum i=tt+W-1xi$, Gaussian smoothing utilizing the Gaussian kernel $G(x)=2\pi\sigma2$

$1e-2\sigma2x2$, and a Savitzky-Golay filter, which employs polynomial regression over a sliding window to generate a smoothed curve. These operations collectively furnish a comprehensive understanding of the EEG data, enabling comparisons between the original and model-predicted series, as well as insights into the data's underlying structure and potential anomalies. This intricate suite of mathematical and computational techniques presents a robust framework for the advanced analysis of time-series data in neuroscience.

# 6. Discussion

This research exemplifies a fertile interplay between chaos theory, dynamical systems theory, and state-of-the-art machine learning algorithms in advancing the understanding of EEG data analysis (Freeman, 2000; Stam, 2005). The study not only presents empirical evidence to augment current methodologies but also advocates for an interdisciplinary approach, bridging gaps that have heretofore existed in the domain of neuroscientific inquiry and machine learning.

One of the seminal findings of this investigation is the inherent complexity and nonlinearity in EEG data. The initial phase of the analysis, which utilised chaos theory and dynamical systems theory, established these characteristics across the 32 EEG channels. This nonlinear profile sets the stage for complex computational models and signifies the limitations of linear methods, corroborating earlier studies that invoked the need for nonlinear dynamic models in EEG data analysis (Freeman, 2000; Stam, 2005). The gradation in model performance, moving from CNNs to RNNs and finally to transformer architectures, elucidates the depth of the problem space. It becomes evident that while CNNs can extract spatial features, they fall short in capturing temporal dependencies, a notion reinforced by the superior performance of RNNs (LeCun, Bengio, Hinton, 2015). Yet, even recurrent architectures struggled to encapsulate the long-range dependencies, a lacuna impressively addressed by the transformer model.

The substantial improvements in the RMSE and MAE metrics on implementing a second iteration of the transformer model underscore its adaptability and efficiency (Vaswani et al., 2017). The architectural modifications informed by chaos and dynamical systems theory yielded a statistically significant improvement. This salient finding lends weight to the argument that the current trend of isolated utilisation of machine learning and chaos theory in EEG analysis is an under-optimised approach. Though the improved transformer model incurred higher computational overhead, the trade-off seems acceptable considering its superior predictive capability. As computational resources become more accessible and efficient, this drawback may become less of an issue.

# 7. Conclusions and further work

This research project, in alignment with our initial questions and hypotheses, has made significant strides in untangling the complex dynamics of EEG data, particularly in the context of transcranial Electrical Stimulation (tES). The study draws from the wellspring of chaos theory, dynamical systems theory, and advanced machine learning methodologies to pave the way for predictive models with broad applications in neuroscience and neuromodulation (Buzsáki, 2006; Kantz & Schreiber, 2004). Our work holds particular significance in the burgeoning field of brain-computer interfaces (BCIs), specifically, in the context of linking the human brain to hybrid silicon/biological computing systems (Wolpaw & Wolpaw, 2012). By accurately forecasting the EEG effects of tES, we can fine-tune stimulation parameters, thereby optimising the translation of neural responses into commands for hybrid systems. The findings are pivotal, considering their prospective applications in designing an interface for seamless communication between the human brain and a hybrid silicon/biological computer via a biological neural interface (Lebedev & Nicolelis, 2006). The study opens up possibilities for decoding and encoding sensory perceptions in the brain. By harnessing the predictive power of transfer entropy, it offers unique insights into the causal interactions within neural circuitry (Schreiber, 2000). This understanding could be crucial for decoding sensory perceptions and potentially for encoding similar patterns back into the brain (Wolpaw & Wolpaw, 2012).

Furthermore, the methods developed in this study could eventually contribute to revolutionary developments in neuroprosthetics and human-computer interfaces. While the immediate focus of our research does not involve inducing sensory perceptions, the insights and methodologies we've developed may inform targeted stimulation techniques in the brain's sensory nervous system [23] [24]. While the study offers promising avenues for technological advancements, it's paramount to consider the ethical and scientific challenges that accompany such undertakings (Ienca & Haselager, 2016). As we strive for human-like interactions with machines, it is imperative to adopt a cautious approach that values ethical considerations and human safety.

## Future Directions

- Refinement of Artifact Removal Techniques: Utilising advanced artifact removal methodologies could further enhance the quality of EEG data, mitigating noise and ambiguities that could affect hybrid systems (Jung et al., 2000).
- Extension of Transfer Entropy Analysis: Further work should consider a more comprehensive application of transfer entropy to shed light on the interdependencies among different brain regions (Vicente et al., 2011).
- Direct Stimulation of Sensory Perceptions: Our approach could serve as a platform for more focused studies that seek to stimulate specific sensory perceptions directly in the human brain, although this involves confronting significant scientific and ethical challenges (Ienca & Haselager, 2016).

## Concluding Remarks

The multifaceted approach of this research, leveraging chaos theory, dynamical systems theory, and advanced neural networks, aims to furnish substantial contributions to neuroscience, neuromodulation, and the clinical use of tES (Freeman, 2000; Stam, 2005). The study's promising results and future directions underscore its potential for significant, groundbreaking contributions across a wide range of applications (Kellmhofer et al., 2018).

By integrating an array of complex and advanced methodologies, we move closer to a comprehensive understanding of how the human brain reacts to external stimuli. Our findings offer new possibilities for interface designs, enhancing communication between biological and artificial systems, thereby fueling future advances in neuroscience, BCIs, and neurotechnology (Lebedev & Nicolelis, 2006).

# Appendices

**Disclosure:**

The Jupyter Notebooks created in this research are on my GitHub at

https://soulsyrup.github.io/

https://github.com/Metaverse-Crowdsource/EEG-Chaos-Kuramoto-Neural-Net

I am the owner and author of this GitHub repository (https://github.com/Metaverse-Crowdsource/EEG-Chaos-Kuramoto-Neural-Net) and the repositories at (https://soulsyrup.github.io). Confirmation of my ownership can be requested:

- from my github account https://github.com/soulsyrup

- by emailing my organization generalinquiries@mvcs.one

- by my work email soul_syrup@mvcs.one

# Data Loading and Variable Assigning

September 28, 2023

[3]:

```
import pandas as pd

# Load data
data = loadmat('/home/vincent/AAA_projects/MVCS/Neuroscience/downsampled/
↪EEG_DS_Struct_0101.mat') # load your EEG data here
stim_data = pd.read_excel('/home/vincent/AAA_projects/MVCS/Neuroscience/
↪EEG-tES-Chaos-Neural-Net/stim_data.xlsx') # load your stimulation data here

# Fill null values in 'Sub#' column
stim_data['Sub#'].fillna(method='ffill', inplace=True)

DSamp = data['DSamp']

# Get data parameters
triggers = DSamp[0][0][0]
EEGdata = DSamp[0][0][1]
fs = DSamp[0][0][2][0][0]
fsOld = DSamp[0][0][3][0][0]
time = DSamp[0][0][4][0]
label = DSamp[0][0][5]
nchan = DSamp[0][0][6][0][0]
rate = DSamp[0][0][7][0][0]
npt = DSamp[0][0][8][0][0]
Subj = DSamp[0][0][9][0]
ptrackerPerf = DSamp[0][0][10]
ptrackerTime = DSamp[0][0][11]
ptrackerfs = DSamp[0][0][12][0][0]

# List of unwanted channel names
unwanted_channels = ['BIP1', 'BIP2', 'RESP1']

# Create a mask where True indicates that the channel is not unwanted
mask = np.array([ch[0][0] not in unwanted_channels for ch in label])

# Filter out unwanted channels from the label data
filtered_label = label[mask]
```

```
# Convert the filtered list back to numpy array and replace the original label
label = np.array(filtered_label, dtype=object)

# Transpose EEGdata
EEGdata = EEGdata.T

# Filter out unwanted channels from the EEG data
filtered_EEGdata = EEGdata[:, mask]

# Transpose it back if needed
filtered_EEGdata = filtered_EEGdata.T

#Select subject
stim_data = stim_data[stim_data['Sub#'] == 1]

stim_data_df = pd.DataFrame(stim_data)

def flatten(item):
    if isinstance(item, list):
        for subitem in item:
            yield from flatten(subitem)
    elif isinstance(item, np.ndarray):
        for subitem in item.flatten():
            yield from flatten(subitem)
    else:
        yield item


trigger_list = []
for trigger in triggers[0]:
    trigger_list.append([list(flatten(trigger[0])),  # Time
                         list(flatten(trigger[1])),  # SampleNum
                         list(flatten(trigger[2])),  # FileNum
                         list(flatten(trigger[3])),  # EventType
                         list(flatten(trigger[4])),  # EventDescription
                         list(flatten(trigger[5])) if trigger[5].size else␣
 ↪['Unknown']  # StimType
                        ])


# Create DataFrame and transpose it
eeg_df = pd.DataFrame(filtered_EEGdata.T)  # Transpose the data

# Display the updated DataFrame

# We convert the labels list into a simple list (previously it was a list of␣
 ↪numpy arrays)
```

```
simple_label = [label_item[0][0] for label_item in label]

eeg_df.columns = simple_label  # Assign column names

# Create DataFrames
triggers_df = pd.DataFrame(trigger_list, columns=["Time", "SampleNum",␣
↪"EventType1", "EventType", "EventDescription", "StimType"])
triggers_df = triggers_df.drop(['EventType1', 'SampleNum'], axis=1)
```

[4]:

```
from graphviz import Digraph, Source

# Create the graph
graph = Digraph(format='png')

# Add nodes to the graph
graph.node('LoadData', shape='rect', label="Load Data", fontsize='12')
graph.node('FillNull', shape='rect', label="Fill Null Values", fontsize='12')
graph.node('FilterUnwanted', shape='rect', label="Filter Unwanted Channels",␣
↪fontsize='12')
graph.node('TransposeEEG', shape='rect', label="Transpose EEG Data",␣
↪fontsize='12')
graph.node('DataFrameEEG', shape='rect', label="Create EEG DataFrame",␣
↪fontsize='12')
graph.node('DataFrameTriggers', shape='rect', label="Create Triggers␣
↪DataFrame", fontsize='12')

# Add edges between nodes
graph.edge('LoadData', 'FillNull', label="Load Data", fontsize='10')
graph.edge('FillNull', 'FilterUnwanted', label="Fill Null Values",␣
↪fontsize='10')
graph.edge('FilterUnwanted', 'TransposeEEG', label="Filter Unwanted Channels",␣
↪fontsize='10')
graph.edge('TransposeEEG', 'DataFrameEEG', label="Transpose EEG Data",␣
↪fontsize='10')
graph.edge('DataFrameEEG', 'DataFrameTriggers', label="Create EEG DataFrame",␣
↪fontsize='10')

# Display the graph visualization directly in Jupyter Notebook/Lab
src = Source(graph.source)
src
```

[4]:

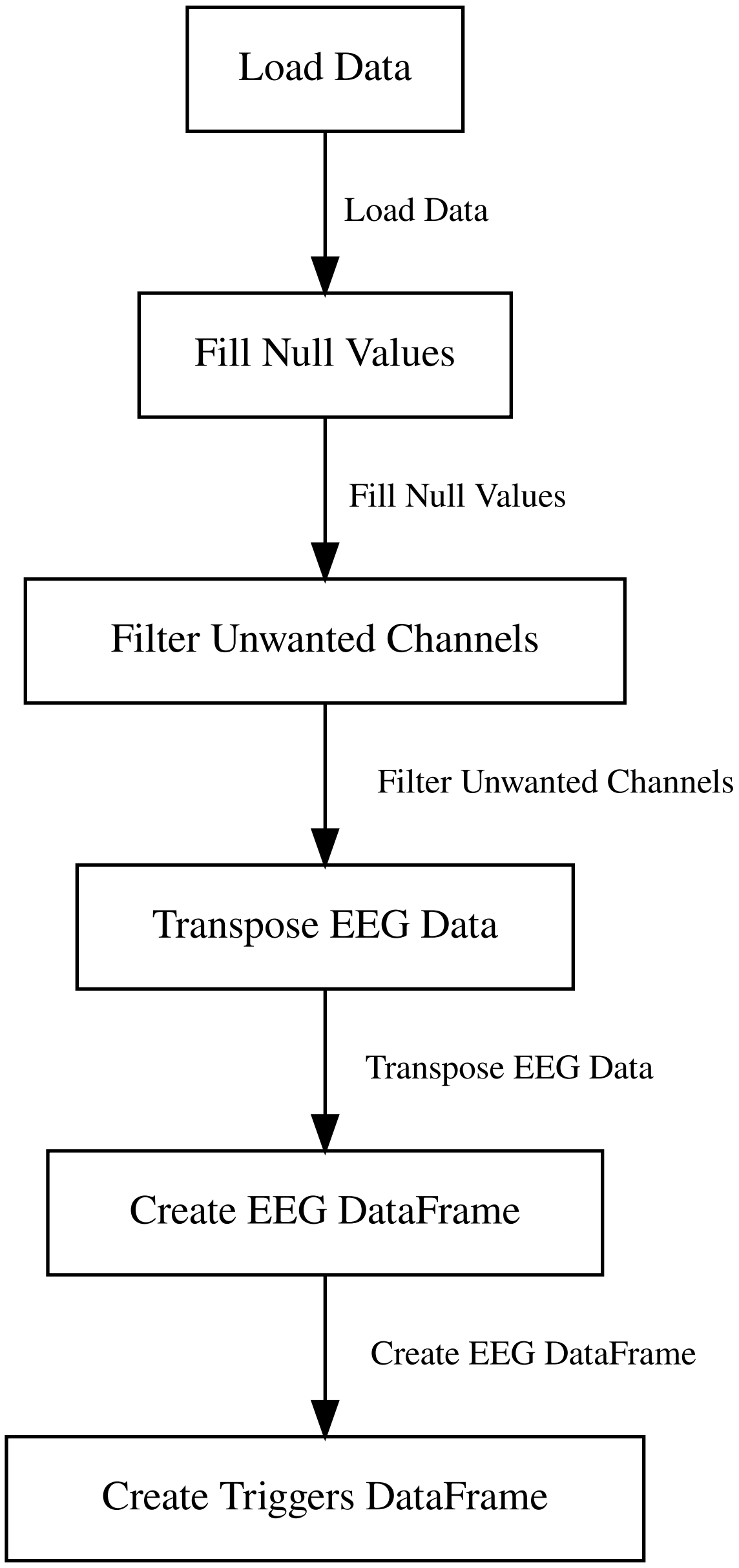

[5]:
```
# Assuming that stim_data_df looks something like this:
stim_data_df = pd.DataFrame({
    'Sub#': [1, 1, 1, 1, 1, 1],
    'Session': [1, 2, 3, 4, 5, 6],
    'File Num': [101, 102, 103, 104, 105, 106],
    'StimTypeBlock1': ['M30', 'M30', 'P30', 'F5', 'F5', 'P30'],
    'StimTypeBlock2': ['F30', 'F30', 'P0', 'M5', 'M5', 'P0'],
    'StimTypeBlock3': ['F0', 'F0', 'P5', 'M0', 'M0', 'P5'],
    'StimAmplitude_mA_block1': [1, 0.5, 0.5, 0.5, 1, 1],
    'StimAmplitude_mA_block2': [1, 0.5, 0.5, 0.5, 1, 1],
```

```python
    'StimAmplitude_mA_block3': [1, 0.5, 0.5, 0.5, 1, 1],
})

# Also assuming that triggers_df looks something like this:
triggers_df = pd.DataFrame({
    'Time': [619.499, 654.746, 770.515, 805.571, 921.515, 956.651, 1072.551, 1107.578, 1819.593, 1854.888, 1970.669, 2005.715, 2121.644, 2156.695, 2272.756, 2307.798, 3019.822, 3019.924],
    'EventDescription': ['Stim Start', 'Stim Stop', 'Stim Start', 'Stim Stop', 'Stim Start', 'Stim Stop', 'Stim Start', 'Stim Stop', 'Stim Start', 'Stim Stop', 'Stim Start', 'Stim Stop', 'Stim Start', 'Stim Stop', 'Stim Start', 'Stim Stop', 'Stim Start', 'Stim Stop'],
    'StimType': ['M30', 'M30', 'M30', 'M30', 'M30', 'M30', 'M30', 'M30', 'M30', 'M30', 'M30', 'M30', 'M30', 'M30', 'M30', 'M30', 'M30', 'M30']
})

def get_stim_info(sub, session, stim_type):
    mask = (stim_data_df['Sub#'] == sub) & (stim_data_df['Session'] == session)
    matching_rows = stim_data_df[mask]
    amplitudes = []
    blocks = []
    file_nums = []

    for index, row in matching_rows.iterrows():
        for block in range(1, 4):
            if row[f'StimTypeBlock{block}'] == stim_type:
                amplitudes.append(row[f'StimAmplitude_mA_block{block}'])
                blocks.append(block)
                file_nums.append(row['File Num'])

    return amplitudes, blocks, file_nums

# Adding 'Sub#' and 'Session' columns to triggers_df
triggers_df['Sub#'] = 1
triggers_df['Session'] = 1

# Define unique_combinations before using it
unique_combinations = pd.concat([stim_data_df[["Sub#", "Session", "StimTypeBlock1"]],
                                 stim_data_df[["Sub#", "Session", "StimTypeBlock2"]].rename(columns={"StimTypeBlock2": "StimTypeBlock1"}),
                                 stim_data_df[["Sub#", "Session", "StimTypeBlock3"]].rename(columns={"StimTypeBlock3": "StimTypeBlock1"}),
                                 triggers_df[["Sub#", "Session", "StimType"]].rename(columns={"StimType": "StimTypeBlock1"})]).drop_duplicates()
```

```python
# Create a DataFrame to store the results
results_df = pd.DataFrame(columns=["Sub#", "Session", "StimType", "Amplitudes",␣
↪"Block", "File Num"])

for _, row in unique_combinations.iterrows():
    sub = row["Sub#"]
    session = row["Session"]
    stim_type = row["StimTypeBlock1"]
    amplitudes, blocks, file_nums = get_stim_info(sub, session, stim_type)

    for amp, block, file_num in zip(amplitudes, blocks, file_nums):
        df_temp = pd.DataFrame([{"Sub#": sub,
                                "Session": session,
                                "StimType": stim_type,
                                "Amplitudes": amp,
                                "Block": block,
                                "File Num": file_num}],
                                columns=["Sub#", "Session", "StimType",␣
↪"Amplitudes", "Block", "File Num"])
        results_df = pd.concat([results_df, df_temp], ignore_index=True)


def get_trigger_info(sub, session, stim_type):
    mask = (triggers_df['Sub#'] == sub) & (triggers_df['Session'] == session) &␣
↪(triggers_df['StimType'] == stim_type)
    matching_rows = triggers_df[mask]
    return matching_rows[['Time', 'EventDescription']]

# To get the amplitude from 'results_df' for a given Sub#, Session, and StimType
def get_stim_amplitude(sub, session, stim_type):
    mask = (results_df['Sub#'] == sub) & (results_df['Session'] == session) &␣
↪(results_df['StimType'] == stim_type)
    matching_rows = results_df[mask]
    if not matching_rows.empty:
        return matching_rows['Amplitudes'].values[0]
    else:
        return None

# Create a new column in triggers_df with the corresponding amplitude
triggers_df['Amplitude'] = triggers_df.apply(lambda row:␣
↪get_stim_amplitude(row['Sub#'], row['Session'], row['StimType']), axis=1)

# Drop the 'Amplitudes' column from 'results_df'
results_df.drop('Amplitudes', axis=1, inplace=True)

# Then perform the merge
```

```
merged_stim_df = pd.merge(results_df, triggers_df, on=['Sub#', 'Session',␣
 ↪'StimType'], how='inner')
merged_stim_df['Time'] = merged_stim_df['Time'] * 1000

# Assuming the sampling rate is 1000 Hz
sampling_rate = 1000

# Calculate time values in milliseconds
num_samples = len(eeg_df)
time_in_seconds = [(i / sampling_rate) for i in range(num_samples)]
time_in_milliseconds = [round(t * 1000, 2) for t in time_in_seconds]

# Modify the 'Time' column in eeg_df to match the 'Time' values in␣
 ↪merged_stim_df
eeg_df['Time'] = time_in_milliseconds

# Assuming that DSamp[0][0][5] is your data
data_as_list = [arr.tolist()[0] for arr in DSamp[0][0][5]]

eeg_label_df = pd.DataFrame(data_as_list, columns=['EEG Electrode Labels'])
# List of unwanted channel names
unwanted_channels = ['BIP1', 'BIP2', 'RESP1']

# Filter out unwanted channels from the DataFrame
eeg_label_df = eeg_label_df[~eeg_label_df['EEG Electrode Labels'].
 ↪isin(unwanted_channels)]
```

[6]:
```
print(merged_stim_df.head())
print(merged_stim_df.tail())
print(eeg_df.head())
print(eeg_df.tail())
```

```
  Sub# Session StimType Block File Num      Time EventDescription  Amplitude
0    1       1      M30     1      101  619499.0       Stim Start        1.0
1    1       1      M30     1      101  654746.0        Stim Stop        1.0
2    1       1      M30     1      101  770515.0       Stim Start        1.0
3    1       1      M30     1      101  805571.0        Stim Stop        1.0
4    1       1      M30     1      101  921515.0       Stim Start        1.0
   Sub# Session StimType Block File Num       Time EventDescription  Amplitude
13    1       1      M30     1      101  2156695.0        Stim Stop        1.0
14    1       1      M30     1      101  2272756.0       Stim Start        1.0
15    1       1      M30     1      101  2307798.0        Stim Stop        1.0
16    1       1      M30     1      101  3019822.0       Stim Start        1.0
17    1       1      M30     1      101  3019924.0        Stim Stop        1.0
            Fp1           Fpz           Fp2           F7         F3  \
0 -21295.988649 -20109.716727 -24153.383752  3189.340060 -45.189275
1 -21303.747077 -20120.746154 -24163.864012  3178.880909 -56.702035
2 -21315.466571 -20130.126577 -24171.944343  3164.903807 -69.465350
```

```
3 -21317.809594 -20131.044726 -24174.790986  3159.478572 -73.214591
4 -21325.798142 -20137.522181 -24179.985166  3144.934679 -84.871628

           Fz          F4           F8          FC5          FC1  …  \
0 -8525.066680 -642.128590  3487.913621  6324.956639  6503.012177  …
1 -8532.499649 -651.966372  3477.011771  6315.078704  6496.522520  …
2 -8544.315275 -663.772856  3463.194795  6302.391524  6483.178723  …
3 -8545.873916 -666.109249  3457.870782  6297.212341  6481.970244  …
4 -8551.164448 -671.761501  3450.466406  6283.925509  6477.045614  …

           P7           P3          Pz           P4           P8  \
0  3374.048029 -3617.197964 -611.584742 -1667.222644  7523.612085
1  3372.073657 -3621.118134 -617.022909 -1673.653480  7516.945510
2  3363.104066 -3632.122011 -627.957966 -1684.569981  7502.158816
3  3354.943617 -3639.476353 -633.425118 -1690.436299  7496.978015
4  3343.913673 -3645.950907 -639.939845 -1695.157414  7491.664259

          POz           O1           Oz           O2  Time
0 -9446.685389 -6091.788931 -1392.835634  3559.608191   0.0
1 -9451.045628 -6094.343708 -1395.070231  3556.367972   1.0
2 -9460.798474 -6104.626002 -1406.149675  3543.184223   2.0
3 -9468.273627 -6110.845490 -1413.911429  3536.078117   3.0
4 -9477.076964 -6117.640828 -1419.832213  3529.343349   4.0

[5 rows x 33 columns]
                  Fp1           Fpz           Fp2          F7          F3  \
4227783 -17297.981962 -16643.265483 -14783.694243  213.760887  123.981995
4227784 -17288.547222 -16625.369538 -14763.506832  227.525891  141.239551
4227785 -17286.892304 -16618.137233 -14755.598154  236.124689  150.365834
4227786 -17281.277994 -16602.343699 -14743.081425  248.784515  164.864721
4227787 -17330.522691 -16641.896722 -14783.662018  206.940363  122.044455

                  Fz          F4           F8          FC5          FC1  …  \
4227783 -7264.260974  199.086367  6328.830514  6643.430274  8697.429125  …
4227784 -7245.628706  215.986652  6347.009702  6655.746644  8714.272896  …
4227785 -7236.115778  228.403642  6354.416108  6669.311806  8718.621128  …
4227786 -7221.565328  245.169557  6366.872097  6677.934052  8732.175102  …
4227787 -7259.499564  202.756340  6322.779688  6628.913136  8686.595901  …

                  P7           P3          Pz           P4            P8  \
4227783  5653.323505 -3769.870437 -583.815885 -3957.764383  10281.652062
4227784  5667.798438 -3756.165865 -570.000793 -3943.666323  10298.156796
4227785  5669.362946 -3752.552156 -564.374964 -3938.584747  10299.328539
4227786  5681.631758 -3740.036309 -548.833209 -3924.002271  10315.237177
4227787  5633.944317 -3790.270100 -596.254185 -3970.866360  10271.522693

                  POz            O1           Oz           O2       Time
4227783 -13387.925984 -12667.092461 -1460.862361  3901.326295  4227783.0
```

```
4227784 -13371.564962 -12651.284002 -1445.696651  3917.096838  4227784.0
4227785 -13369.908511 -12655.549749 -1446.893217  3920.486579  4227785.0
4227786 -13356.761631 -12649.243537 -1435.838151  3932.607149  4227786.0
4227787 -13402.698471 -12685.879151 -1473.645716  3893.119642  4227787.0

[5 rows x 33 columns]
```

# 1 Make the eeg df a npy

[7]:
```
# Extract EEG channel names (excluding 'Time')
eeg_channel_names = eeg_df.columns[:-1]

# Create a dictionary to store EEG data for each channel
eeg_data_dict = {}

# Populate the dictionary with EEG data
for channel in eeg_channel_names:
    eeg_data_dict[channel] = eeg_df[channel].values

# Convert the dictionary values to a numpy array
eeg_data_array = np.array([eeg_data_dict[channel] for channel in␣
 ↪eeg_channel_names]).T

# Save the numpy array with EEG data as a single .npy file
save_path = '/home/vincent/AAA_projects/MVCS/Neuroscience/
 ↪eeg_data_with_channels.npy'
np.save(save_path, eeg_data_array)
```

[17]:
```
from graphviz import Digraph, Source

# Create the graph
graph = Digraph(format='png')

# Add nodes to the graph
graph.node('LoadData', shape='rect', label="Load Data", fontsize='12')
graph.node('GetStimInfo', shape='rect', label="Get Stimulus Info",␣
 ↪fontsize='12')
graph.node('CreateResults', shape='rect', label="Create Results DataFrame",␣
 ↪fontsize='12')
graph.node('GetTriggerInfo', shape='rect', label="Get Trigger Info",␣
 ↪fontsize='12')
graph.node('MergeDataFrames', shape='rect', label="Merge DataFrames",␣
 ↪fontsize='12')

# Add edges between nodes
```

```
graph.edge('LoadData', 'GetStimInfo', label="Load stim_data_df and␣
↪triggers_df", fontsize='10')
graph.edge('GetStimInfo', 'CreateResults', label="Get Stimulus Info",␣
↪fontsize='10')
graph.edge('CreateResults', 'GetTriggerInfo', label="Create Results DataFrame",␣
↪fontsize='10')
graph.edge('GetTriggerInfo', 'MergeDataFrames', label="Get Trigger Info",␣
↪fontsize='10')

# Display the graph visualization directly in Jupyter Notebook/Lab
src = Source(graph.source)
src
```

[17]:

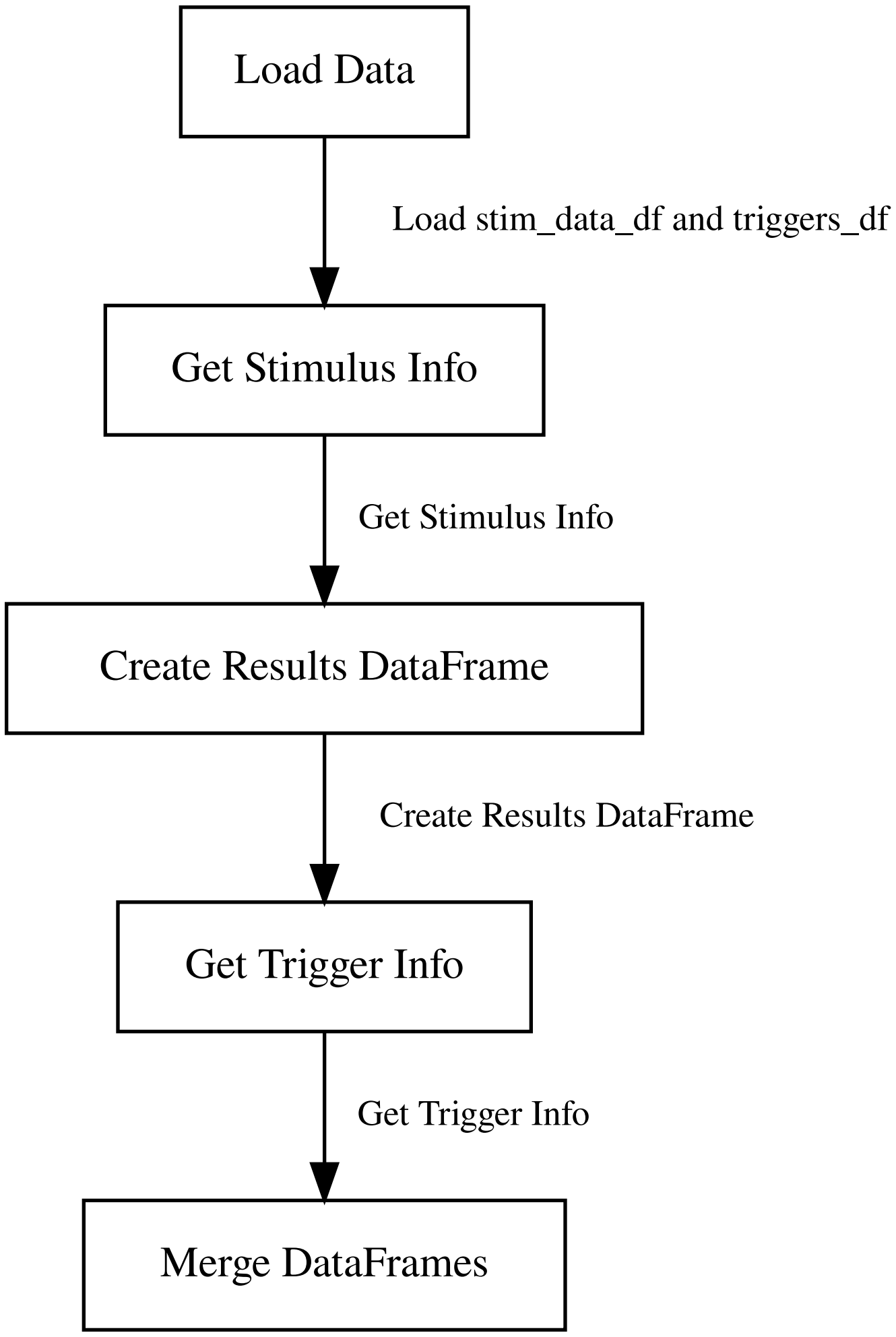

[18]:
```
# Specify the directory paths where you want to save the CSV files
merged_stim_directory = '/home/vincent/AAA_projects/MVCS/Neuroscience/
↪DataFrames'
```

```
eeg_directory = '/home/vincent/AAA_projects/MVCS/Neuroscience/DataFrames'

# Save 'merged_stim_df' to CSV
merged_stim_df.to_csv(f"{merged_stim_directory}/merged_stim_df.csv",␣
↪index=False)

# Save 'eeg_df' to CSV
eeg_df.to_csv(f"{eeg_directory}/eeg_df.csv", index=False)

print("DataFrames saved to CSV files successfully.")
```

```
DataFrames saved to CSV files successfully.
```

[ ]:
```
triggers = DSamp[0][0][0]
print("Triggers: ", triggers)

EEGdata = DSamp[0][0][1]
print("EEGdata: ", EEGdata)

fs = DSamp[0][0][2][0][0]
print("fs: ", fs)

fsOld = DSamp[0][0][3][0][0]
print("fsOld: ", fsOld)

time = DSamp[0][0][4][0]
print("Time: ", time)

label = DSamp[0][0][5]
print("Label: ", label)

nchan = DSamp[0][0][6][0][0]
print("nchan: ", nchan)

rate = DSamp[0][0][7][0][0]
print("Rate: ", rate)

npt = DSamp[0][0][8][0][0]
print("npt: ", npt)

Subj = DSamp[0][0][9][0]
print("Subj: ", Subj)

ptrackerPerf = DSamp[0][0][10]
print("PtrackerPerf: ", ptrackerPerf)

ptrackerTime = DSamp[0][0][11]
```

```python
print("PtrackerTime: ", ptrackerTime)

ptrackerfs = DSamp[0][0][12][0][0]
print("Ptrackerfs: ", ptrackerfs)
```

```
Triggers:  [[(array([[20.428]]), array([[20429]], dtype=uint16), array(['0002'],
dtype='<U4'), array([[2]], dtype=uint8), array(['Block Start'], dtype='<U11'),
array([], shape=(1, 0), dtype=float64))
  (array([[619.442]]), array([[619443]], dtype=int32), array(['0002'],
dtype='<U4'), array([[2]], dtype=uint8), array(['Block Start'], dtype='<U11'),
array([], shape=(1, 0), dtype=float64))
  (array([[619.499]]), array([[619500]], dtype=int32), array(['0016'],
dtype='<U4'), array([[16]], dtype=uint8), array(['Stim Start'], dtype='<U10'),
array(['M30'], dtype='<U3'))
  (array([[654.746]]), array([[654747]], dtype=int32), array(['0032'],
dtype='<U4'), array([[32]], dtype=uint8), array(['Stim Stop'], dtype='<U9'),
array([], shape=(1, 0), dtype=float64))
  (array([[770.515]]), array([[770516]], dtype=int32), array(['0016'],
dtype='<U4'), array([[16]], dtype=uint8), array(['Stim Start'], dtype='<U10'),
array(['M30'], dtype='<U3'))
  (array([[805.571]]), array([[805572]], dtype=int32), array(['0032'],
dtype='<U4'), array([[32]], dtype=uint8), array(['Stim Stop'], dtype='<U9'),
array([], shape=(1, 0), dtype=float64))
  (array([[921.515]]), array([[921516]], dtype=int32), array(['0016'],
dtype='<U4'), array([[16]], dtype=uint8), array(['Stim Start'], dtype='<U10'),
array(['M30'], dtype='<U3'))
  (array([[956.651]]), array([[956652]], dtype=int32), array(['0032'],
dtype='<U4'), array([[32]], dtype=uint8), array(['Stim Stop'], dtype='<U9'),
array([], shape=(1, 0), dtype=float64))
  (array([[1072.551]]), array([[1072552]], dtype=int32), array(['0016'],
dtype='<U4'), array([[16]], dtype=uint8), array(['Stim Start'], dtype='<U10'),
array(['M30'], dtype='<U3'))
  (array([[1107.578]]), array([[1107579]], dtype=int32), array(['0032'],
dtype='<U4'), array([[32]], dtype=uint8), array(['Stim Stop'], dtype='<U9'),
array([], shape=(1, 0), dtype=float64))
  (array([[1218.442]]), array([[1218443]], dtype=int32), array(['0002'],
dtype='<U4'), array([[2]], dtype=uint8), array(['Block Start'], dtype='<U11'),
array([], shape=(1, 0), dtype=float64))
  (array([[1817.46]]), array([[1817461]], dtype=int32), array(['0002'],
dtype='<U4'), array([[2]], dtype=uint8), array(['Block Start'], dtype='<U11'),
array([], shape=(1, 0), dtype=float64))
  (array([[1819.593]]), array([[1819594]], dtype=int32), array(['0016'],
dtype='<U4'), array([[16]], dtype=uint8), array(['Stim Start'], dtype='<U10'),
array(['M30'], dtype='<U3'))
  (array([[1854.888]]), array([[1854889]], dtype=int32), array(['0032'],
dtype='<U4'), array([[32]], dtype=uint8), array(['Stim Stop'], dtype='<U9'),
array([], shape=(1, 0), dtype=float64))
  (array([[1970.669]]), array([[1970670]], dtype=int32), array(['0016'],
```

```
dtype='<U4'), array([[16]], dtype=uint8), array(['Stim Start'], dtype='<U10'),
array(['M30'], dtype='<U3'))
  (array([[2005.715]]), array([[2005716]], dtype=int32), array(['0032'],
dtype='<U4'), array([[32]], dtype=uint8), array(['Stim Stop'], dtype='<U9'),
array([], shape=(1, 0), dtype=float64))
  (array([[2121.644]]), array([[2121645]], dtype=int32), array(['0016'],
dtype='<U4'), array([[16]], dtype=uint8), array(['Stim Start'], dtype='<U10'),
array(['M30'], dtype='<U3'))
  (array([[2156.695]]), array([[2156696]], dtype=int32), array(['0032'],
dtype='<U4'), array([[32]], dtype=uint8), array(['Stim Stop'], dtype='<U9'),
array([], shape=(1, 0), dtype=float64))
  (array([[2272.756]]), array([[2272757]], dtype=int32), array(['0016'],
dtype='<U4'), array([[16]], dtype=uint8), array(['Stim Start'], dtype='<U10'),
array(['M30'], dtype='<U3'))
  (array([[2307.798]]), array([[2307799]], dtype=int32), array(['0032'],
dtype='<U4'), array([[32]], dtype=uint8), array(['Stim Stop'], dtype='<U9'),
array([], shape=(1, 0), dtype=float64))
  (array([[2416.534]]), array([[2416535]], dtype=int32), array(['0002'],
dtype='<U4'), array([[2]], dtype=uint8), array(['Block Start'], dtype='<U11'),
array([], shape=(1, 0), dtype=float64))
  (array([[3015.513]]), array([[3015514]], dtype=int32), array(['0002'],
dtype='<U4'), array([[2]], dtype=uint8), array(['Block Start'], dtype='<U11'),
array([], shape=(1, 0), dtype=float64))
  (array([[3019.822]]), array([[3019823]], dtype=int32), array(['0016'],
dtype='<U4'), array([[16]], dtype=uint8), array(['Stim Start'], dtype='<U10'),
array(['M30'], dtype='<U3'))
  (array([[3019.924]]), array([[3019925]], dtype=int32), array(['0032'],
dtype='<U4'), array([[32]], dtype=uint8), array(['Stim Stop'], dtype='<U9'),
array([], shape=(1, 0), dtype=float64))]]
EEGdata:  [[-21295.98864928 -21303.74707676 -21315.46657134 … -17286.89230419
  -17281.27799398 -17330.52269117]
 [-20109.7167273  -20120.74615359 -20130.12657698 … -16618.13723294
  -16602.343699   -16641.89672243]
 [-24153.38375243 -24163.86401194 -24171.94434272 … -14755.59815419
  -14743.08142526 -14783.66201823]
 …
 [  2692.44573979   2703.97118199   2731.75987441 …   4181.64344633
    4157.70728788   4198.03945844]
 [ -5014.9543463   -5014.10209995  -5019.83464311 …  -4711.95486888
   -4706.99851768  -4709.83053726]
 [    76.88986022     72.02278487     66.20816171 …     73.4410151
     69.33401454     66.65666327]]
fs:  1000
fsOld:  2000
Time:  [0.000000e+00 1.000000e-03 2.000000e-03 … 4.227785e+03 4.227786e+03
 4.227787e+03]
Label:  [[array(['Fp1'], dtype='<U3')]
 [array(['Fpz'], dtype='<U3')]
```

```
 [array(['Fp2'], dtype='<U3')]
 [array(['F7'], dtype='<U2')]
 [array(['F3'], dtype='<U2')]
 [array(['Fz'], dtype='<U2')]
 [array(['F4'], dtype='<U2')]
 [array(['F8'], dtype='<U2')]
 [array(['FC5'], dtype='<U3')]
 [array(['FC1'], dtype='<U3')]
 [array(['FC2'], dtype='<U3')]
 [array(['FC6'], dtype='<U3')]
 [array(['M1'], dtype='<U2')]
 [array(['T7'], dtype='<U2')]
 [array(['C3'], dtype='<U2')]
 [array(['Cz'], dtype='<U2')]
 [array(['C4'], dtype='<U2')]
 [array(['T8'], dtype='<U2')]
 [array(['M2'], dtype='<U2')]
 [array(['CP5'], dtype='<U3')]
 [array(['CP1'], dtype='<U3')]
 [array(['CP2'], dtype='<U3')]
 [array(['CP6'], dtype='<U3')]
 [array(['P7'], dtype='<U2')]
 [array(['P3'], dtype='<U2')]
 [array(['Pz'], dtype='<U2')]
 [array(['P4'], dtype='<U2')]
 [array(['P8'], dtype='<U2')]
 [array(['POz'], dtype='<U3')]
 [array(['O1'], dtype='<U2')]
 [array(['Oz'], dtype='<U2')]
 [array(['O2'], dtype='<U2')]
 [array(['BIP1'], dtype='<U4')]
 [array(['BIP2'], dtype='<U4')]
 [array(['RESP1'], dtype='<U5')]]
nchan:  35
Rate:  1000
npt:  4227788
Subj:  0101
PtrackerPerf:  [[ 0.82088665]
 [ 1.32936508]
 [ 1.76474775]
 …
 [11.88057628]
 [ 9.25956959]
 [ 3.30952176]]
PtrackerTime:  [[0.00000e+00]
 [1.00000e-02]
 [2.00000e-02]
 …
```

```
 [4.19999e+03]
 [4.20000e+03]
 [4.20001e+03]]
Ptrackerfs:  100
```

## 2 load the csv's for the RNN from the Multifractal Analysis, and the eeg + stim csv's, after Multifractal Analysis

[3]:
```
# Define the file paths
base_dir = '/home/vincent/AAA_projects/MVCS/Neuroscience/'
eeg_df_path = base_dir + 'DataFrames/eeg_df.csv'
merged_stim_df_path = base_dir + 'DataFrames/merged_stim_df.csv'
hurst_exponents_path = base_dir + 'HurstExponents/hurst_exponents_df.csv'
rnn_mfdfa_X_path = base_dir + 'RNN_data/rnn_X_data_combined.npy'

# load data
eeg_df = pd.read_csv(eeg_df_path)
merged_stim_df = pd.read_csv(merged_stim_df_path)
hurst_exponents_df = pd.read_csv(hurst_exponents_path)
rnn_X_data_combined = np.load(rnn_mfdfa_X_path)
```

[4]:
```
# Print the columns of eeg_df
print("Columns of eeg_df:")
print(eeg_df.columns)

# Print the columns of merged_stim_df
print("Columns of merged_stim_df:")
print(merged_stim_df.columns)

# Print the columns of hurst_exponents_df
print("Columns of hurst_exponents_df:")
print(hurst_exponents_df.columns)

# Assuming rnn_mfdfa_X_df is a NumPy array
print("Number of columns in rnn_X_data_combined:", rnn_X_data_combined.shape[1])
```

```
Columns of eeg_df:
Index(['Fp1', 'Fpz', 'Fp2', 'F7', 'F3', 'Fz', 'F4', 'F8', 'FC5', 'FC1', 'FC2',
       'FC6', 'M1', 'T7', 'C3', 'Cz', 'C4', 'T8', 'M2', 'CP5', 'CP1', 'CP2',
       'CP6', 'P7', 'P3', 'Pz', 'P4', 'P8', 'POz', 'O1', 'Oz', 'O2', 'Time'],
      dtype='object')
Columns of merged_stim_df:
Index(['Sub#', 'Session', 'StimType', 'Block', 'File Num', 'Time',
       'EventDescription', 'Amplitude'],
      dtype='object')
Columns of hurst_exponents_df:
Index(['0'], dtype='object')
```

```
Number of columns in rnn_X_data_combined: 100
```

# 3 Change everything to numerical

[5]:
```
# Define mappings for frequency and location
frequency_mapping = {
    "F0": 0,
    "F5": 5,
    "F30": 30,
    "M0": 0,
    "M5": 5,
    "M30": 30,
    "P0": 0,
    "P5": 5,
    "P30": 30
}

location_mapping = {
    "F0": 1,
    "F5": 1,
    "F30": 1,
    "M0": 2,
    "M5": 2,
    "M30": 2,
    "P0": 3,
    "P5": 3,
    "P30": 3
}

# Check if 'StimType' column is present in the dataframe
if 'StimType' in merged_stim_df.columns:
    # Proceed with replacement of values and dropping the column
    merged_stim_df["Frequency"] = merged_stim_df["StimType"].
  ↪replace(frequency_mapping)
    merged_stim_df["Location"] = merged_stim_df["StimType"].
  ↪replace(location_mapping)
    merged_stim_df.drop('StimType', axis=1, inplace=True)
else:
    print("The 'StimType' column does not exist in the dataframe.")

# Replace "Stim Start" with 1 and "Stim Stop" with 2
merged_stim_df["EventDescription"] = merged_stim_df["EventDescription"].
  ↪replace({
    "Stim Start": 1,
    "Stim Stop": 0
})
```

[6]:
```
print(eeg_df['Time'].head())
print(merged_stim_df['Time'].head())
```

```
0    0.0
1    1.0
2    2.0
3    3.0
4    4.0
Name: Time, dtype: float64
0    619499.0
1    654746.0
2    770515.0
3    805571.0
4    921515.0
Name: Time, dtype: float64
```

[7]:
```
# First, let's perform the merge operation
merged_eeg_stim_df = pd.merge_asof(eeg_df, merged_stim_df, on='Time',␣
 ↪direction='backward')

# Create 'Stim' column based on 'EventDescription'.
# If 'EventDescription' is 1 (Stim start) and the 'Time' is >= 619499, we set␣
 ↪'Stim' as 1. Otherwise, 'Stim' is 0.
merged_eeg_stim_df['Stim'] = np.where((merged_eeg_stim_df['EventDescription']␣
 ↪== 1) & (merged_eeg_stim_df['Time'] >= 619499), 1, 0)

# Drop the 'EventDescription' column now.
merged_eeg_stim_df.drop(columns=['EventDescription'], inplace=True)

# Create a 'StimChange' column that's 1 where 'Stim' changes from 0 to 1, and 0␣
 ↪elsewhere
merged_eeg_stim_df['StimChange'] = (merged_eeg_stim_df['Stim'].diff() == 1).
 ↪astype(int)

# Create a new 'block' column, incrementing by 1 each time 'StimChange' is 1 (i.
 ↪e., each time a new stimulation session starts)
merged_eeg_stim_df['block'] = merged_eeg_stim_df['StimChange'].cumsum()

# Now we no longer need the 'StimChange' column, so we can drop it
merged_eeg_stim_df = merged_eeg_stim_df.drop('StimChange', axis=1)

# Reset the values for 'Amplitude', 'Frequency', 'Location', and 'block' when␣
 ↪'Stim' is 0
merged_eeg_stim_df.loc[merged_eeg_stim_df['Stim'] == 0, ['Amplitude',␣
 ↪'Frequency', 'Location', 'block']] = 0
```

```python
# Now 'block' should be a new column in your DataFrame indicating the
↪stimulation session (or "block") each row belongs to
print(merged_eeg_stim_df)
```

```
                  Fp1           Fpz           Fp2          F7          F3  \
0       -21295.988649 -20109.716727 -24153.383752  3189.340060  -45.189275
1       -21303.747077 -20120.746154 -24163.864012  3178.880909  -56.702035
2       -21315.466571 -20130.126577 -24171.944343  3164.903807  -69.465350
3       -21317.809594 -20131.044726 -24174.790986  3159.478572  -73.214591
4       -21325.798142 -20137.522181 -24179.985166  3144.934679  -84.871628
…                 …             …             …           …           …
4227783 -17297.981962 -16643.265483 -14783.694243   213.760887  123.981995
4227784 -17288.547222 -16625.369538 -14763.506832   227.525891  141.239551
4227785 -17286.892304 -16618.137233 -14755.598154   236.124689  150.365834
4227786 -17281.277994 -16602.343699 -14743.081425   248.784515  164.864721
4227787 -17330.522691 -16641.896722 -14783.662018   206.940363  122.044455

                 Fz          F4           F8          FC5          FC1  …  \
0       -8525.066680 -642.128590  3487.913621  6324.956639  6503.012177  …
1       -8532.499649 -651.966372  3477.011771  6315.078704  6496.522520  …
2       -8544.315275 -663.772856  3463.194795  6302.391524  6483.178723  …
3       -8545.873916 -666.109249  3457.870782  6297.212341  6481.970244  …
4       -8551.164448 -671.761501  3450.466406  6283.925509  6477.045614  …
…                …           …            …            …            …  …
4227783 -7264.260974  199.086367  6328.830514  6643.430274  8697.429125  …
4227784 -7245.628706  215.986652  6347.009702  6655.746644  8714.272896  …
4227785 -7236.115778  228.403642  6354.416108  6669.311806  8718.621128  …
4227786 -7221.565328  245.169557  6366.872097  6677.934052  8732.175102  …
4227787 -7259.499564  202.756340  6322.779688  6628.913136  8686.595901  …

              Time  Sub#  Session  Block  File Num  Amplitude  Frequency  \
0              0.0   NaN      NaN    NaN       NaN        0.0        0.0
1              1.0   NaN      NaN    NaN       NaN        0.0        0.0
2              2.0   NaN      NaN    NaN       NaN        0.0        0.0
3              3.0   NaN      NaN    NaN       NaN        0.0        0.0
4              4.0   NaN      NaN    NaN       NaN        0.0        0.0
…              …     …        …      …         …          …
4227783  4227783.0   1.0      1.0    1.0     101.0        0.0        0.0
4227784  4227784.0   1.0      1.0    1.0     101.0        0.0        0.0
4227785  4227785.0   1.0      1.0    1.0     101.0        0.0        0.0
4227786  4227786.0   1.0      1.0    1.0     101.0        0.0        0.0
4227787  4227787.0   1.0      1.0    1.0     101.0        0.0        0.0

         Location  Stim  block
0             0.0     0      0
1             0.0     0      0
2             0.0     0      0
3             0.0     0      0
```

```
4              0.0     0      0
…            …    …     …
4227783        0.0     0      0
4227784        0.0     0      0
4227785        0.0     0      0
4227786        0.0     0      0
4227787        0.0     0      0

[4227788 rows x 42 columns]
```

[8]:

```python
# Drop the old 'Block' column
merged_eeg_stim_df.drop(columns=['Block'], inplace=True)

# Dropping the 'File Num' column
merged_eeg_stim_df = merged_eeg_stim_df.drop('File Num', axis=1)

# Fill NaNs to 0's
merged_eeg_stim_df['Amplitude'] = merged_eeg_stim_df['Amplitude'].fillna(0)
merged_eeg_stim_df['Frequency'] = merged_eeg_stim_df['Frequency'].fillna(0)
merged_eeg_stim_df['Location'] = merged_eeg_stim_df['Location'].fillna(0)
merged_eeg_stim_df['block'] = merged_eeg_stim_df['block'].fillna(0)

# Changing all 'Sub#' values to 1
merged_eeg_stim_df['Sub#'] = 1

# Changing all 'Session' values to 1
merged_eeg_stim_df['Session'] = 1

# Show the resulting DataFrame
print(merged_eeg_stim_df)
```

```
                  Fp1           Fpz           Fp2           F7          F3  \
0       -21295.988649 -20109.716727 -24153.383752  3189.340060  -45.189275
1       -21303.747077 -20120.746154 -24163.864012  3178.880909  -56.702035
2       -21315.466571 -20130.126577 -24171.944343  3164.903807  -69.465350
3       -21317.809594 -20131.044726 -24174.790986  3159.478572  -73.214591
4       -21325.798142 -20137.522181 -24179.985166  3144.934679  -84.871628
…                 …             …             …            …           …
4227783 -17297.981962 -16643.265483 -14783.694243   213.760887  123.981995
4227784 -17288.547222 -16625.369538 -14763.506832   227.525891  141.239551
4227785 -17286.892304 -16618.137233 -14755.598154   236.124689  150.365834
4227786 -17281.277994 -16602.343699 -14743.081425   248.784515  164.864721
4227787 -17330.522691 -16641.896722 -14783.662018   206.940363  122.044455

                  Fz          F4           F8          FC5          FC1  …  \
0       -8525.066680 -642.128590  3487.913621  6324.956639  6503.012177  …
1       -8532.499649 -651.966372  3477.011771  6315.078704  6496.522520  …
2       -8544.315275 -663.772856  3463.194795  6302.391524  6483.178723  …
```

```
3       -8545.873916 -666.109249  3457.870782  6297.212341  6481.970244  …
4       -8551.164448 -671.761501  3450.466406  6283.925509  6477.045614  …
…                …          …           …            …            …   …
4227783 -7264.260974  199.086367  6328.830514  6643.430274  8697.429125  …
4227784 -7245.628706  215.986652  6347.009702  6655.746644  8714.272896  …
4227785 -7236.115778  228.403642  6354.416108  6669.311806  8718.621128  …
4227786 -7221.565328  245.169557  6366.872097  6677.934052  8732.175102  …
4227787 -7259.499564  202.756340  6322.779688  6628.913136  8686.595901  …

                  Oz           O2       Time  Sub#  Session  Amplitude  \
0       -1392.835634  3559.608191        0.0     1        1        0.0
1       -1395.070231  3556.367972        1.0     1        1        0.0
2       -1406.149675  3543.184223        2.0     1        1        0.0
3       -1413.911429  3536.078117        3.0     1        1        0.0
4       -1419.832213  3529.343349        4.0     1        1        0.0
…                …          …        …   …     …        …
4227783 -1460.862361  3901.326295  4227783.0     1        1        0.0
4227784 -1445.696651  3917.096838  4227784.0     1        1        0.0
4227785 -1446.893217  3920.486579  4227785.0     1        1        0.0
4227786 -1435.838151  3932.607149  4227786.0     1        1        0.0
4227787 -1473.645716  3893.119642  4227787.0     1        1        0.0

         Frequency  Location  Stim  block
0              0.0       0.0     0      0
1              0.0       0.0     0      0
2              0.0       0.0     0      0
3              0.0       0.0     0      0
4              0.0       0.0     0      0
…              …        …     …     …
4227783        0.0       0.0     0      0
4227784        0.0       0.0     0      0
4227785        0.0       0.0     0      0
4227786        0.0       0.0     0      0
4227787        0.0       0.0     0      0

[4227788 rows x 40 columns]
```

[12]:

```python
# Assuming merged_eeg_stim_df is the DataFrame you are working with
# Assuming 'block' column represents the blocks

# Filter out block 0 (the non-stimulation block)
valid_blocks_df = merged_eeg_stim_df[merged_eeg_stim_df['block'] != 0]

# Find the number of unique valid blocks
num_valid_blocks = valid_blocks_df['block'].nunique()

# Print the number of unique valid blocks
```

```
print("Number of unique valid blocks of stimulation:", num_valid_blocks)
```

```
Number of unique valid blocks of stimulation: 9
```

[ ]:
```
# Assuming merged_eeg_stim_df is the DataFrame you are working with
# Assuming 'Stim' column contains 1 for stimulation and 0 for no stimulation

# Define the number of rows you want to display before and after each␣
↪stimulation block starts
num_rows_before_stim = 5
num_rows_after_stim = 5

# Function to extract rows before and after the stimulation block starts
def extract_rows_around_stim(merged_eeg_stim_df, stim_start_indices,␣
↪num_rows_before_stim, num_rows_after_stim):
    for stim_start_idx in stim_start_indices:
        rows_before_stim = merged_eeg_stim_df.iloc[stim_start_idx -␣
↪num_rows_before_stim:stim_start_idx]
        rows_after_stim = merged_eeg_stim_df.iloc[stim_start_idx:stim_start_idx␣
↪+ num_rows_after_stim]

        print(f"\nStimulation Block Start Index: {stim_start_idx}")
        print("Rows before the stimulation block starts:")
        print(rows_before_stim)

        print("\nRows after the stimulation block starts:")
        print(rows_after_stim)

# Find the indices where the stimulation block starts
stim_start_indices = merged_eeg_stim_df.index[merged_eeg_stim_df['Stim'] == 1].
↪tolist()

# Call the function to extract rows for all stimulation blocks
extract_rows_around_stim(merged_eeg_stim_df, stim_start_indices,␣
↪num_rows_before_stim, num_rows_after_stim)
```

[20]:
```
# Show the resulting DataFrame
print(merged_eeg_stim_df)
```

```
                 Fp1           Fpz           Fp2           F7          F3  \
0       -21295.988649 -20109.716727 -24153.383752  3189.340060  -45.189275
1       -21303.747077 -20120.746154 -24163.864012  3178.880909  -56.702035
2       -21315.466571 -20130.126577 -24171.944343  3164.903807  -69.465350
3       -21317.809594 -20131.044726 -24174.790986  3159.478572  -73.214591
4       -21325.798142 -20137.522181 -24179.985166  3144.934679  -84.871628
…                 …             …             …            …           …
4227783 -17297.981962 -16643.265483 -14783.694243   213.760887  123.981995
4227784 -17288.547222 -16625.369538 -14763.506832   227.525891  141.239551
```

```
4227785 -17286.892304 -16618.137233 -14755.598154   236.124689  150.365834
4227786 -17281.277994 -16602.343699 -14743.081425   248.784515  164.864721
4227787 -17330.522691 -16641.896722 -14783.662018   206.940363  122.044455

                  Fz          F4          F8         FC5         FC1  …  \
0       -8525.066680 -642.128590 3487.913621 6324.956639 6503.012177  …
1       -8532.499649 -651.966372 3477.011771 6315.078704 6496.522520  …
2       -8544.315275 -663.772856 3463.194795 6302.391524 6483.178723  …
3       -8545.873916 -666.109249 3457.870782 6297.212341 6481.970244  …
4       -8551.164448 -671.761501 3450.466406 6283.925509 6477.045614  …
…                …          …          …          …          …  …
4227783 -7264.260974  199.086367 6328.830514 6643.430274 8697.429125  …
4227784 -7245.628706  215.986652 6347.009702 6655.746644 8714.272896  …
4227785 -7236.115778  228.403642 6354.416108 6669.311806 8718.621128  …
4227786 -7221.565328  245.169557 6366.872097 6677.934052 8732.175102  …
4227787 -7259.499564  202.756340 6322.779688 6628.913136 8686.595901  …

                  Oz          O2       Time  Sub#  Session  Amplitude  \
0       -1392.835634 3559.608191        0.0     1        1        0.0
1       -1395.070231 3556.367972        1.0     1        1        0.0
2       -1406.149675 3543.184223        2.0     1        1        0.0
3       -1413.911429 3536.078117        3.0     1        1        0.0
4       -1419.832213 3529.343349        4.0     1        1        0.0
…                …           …        …   …        …        …
4227783 -1460.862361 3901.326295  4227783.0     1        1        0.0
4227784 -1445.696651 3917.096838  4227784.0     1        1        0.0
4227785 -1446.893217 3920.486579  4227785.0     1        1        0.0
4227786 -1435.838151 3932.607149  4227786.0     1        1        0.0
4227787 -1473.645716 3893.119642  4227787.0     1        1        0.0

         Frequency  Location  Stim  block
0              0.0       0.0     0      0
1              0.0       0.0     0      0
2              0.0       0.0     0      0
3              0.0       0.0     0      0
4              0.0       0.0     0      0
…              …       …     …    …
4227783        0.0       0.0     0      0
4227784        0.0       0.0     0      0
4227785        0.0       0.0     0      0
4227786        0.0       0.0     0      0
4227787        0.0       0.0     0      0

[4227788 rows x 40 columns]
```

[19]:
```
# Make sure your DataFrame is sorted by 'Time'
merged_eeg_stim_df = merged_eeg_stim_df.sort_values('Time')
```

```python
# Get the unique blocks
blocks = merged_eeg_stim_df['block'].unique()

# We only want to look at the first 9 blocks, excluding the first block (block␣
↪0)
blocks = blocks[1:10]

for block in blocks:
    # Get the indices of the rows belonging to this block
    block_indices = merged_eeg_stim_df[merged_eeg_stim_df['block'] == block].
↪index

    # Get the index of the first row of the block
    first_index = block_indices[0]

    # Get the indices of the three rows before and three rows after the start␣
↪of the block
    indices = range(first_index - 3, first_index + 3)

    # Select and print these rows
    print(merged_eeg_stim_df.loc[indices])
```

```
                 Fp1           Fpz           Fp2           F7         F3  \
619496 -21514.565997 -19579.117859 -22808.936192  2773.475892 -35.557832
619497 -21493.753131 -19558.399511 -22790.959384  2795.570464 -14.778337
619498 -21495.608737 -19567.024258 -22800.700106  2790.819965 -20.460584
619499 -21494.610241 -19570.985140 -22801.964889  2787.997341 -25.947773
619500 -21489.067781 -19569.220022 -22800.405633  2786.277971 -20.967430
619501 -21492.449675 -19575.965596 -22808.633861  2766.501341 -30.273663

                 Fz          F4           F8          FC5          FC1  …  \
619496 -8483.830863 -503.396550  3346.327936  6651.888843  7001.381451  …
619497 -8465.880712 -489.352664  3366.895072  6673.518314  7020.005337  …
619498 -8474.053096 -500.904302  3346.567180  6670.705008  7012.466795  …
619499 -8477.884632 -501.414926  3350.030884  6667.162657  7004.993011  …
619500 -8475.553025 -498.035359  3356.795540  6669.650655  7013.918986  …
619501 -8482.861746 -506.122903  3347.371877  6659.219710  7006.786232  …

                 Oz           O2      Time  Sub#  Session  Amplitude  \
619496 -1327.374450  3756.319169  619496.0     1        1        0.0
619497 -1309.025460  3778.197466  619497.0     1        1        0.0
619498 -1306.605233  3787.141523  619498.0     1        1        0.0
619499 -1303.317124  3794.551709  619499.0     1        1        1.0
619500 -1300.854292  3793.700360  619500.0     1        1        1.0
619501 -1314.756793  3773.717129  619501.0     1        1        1.0
```

```
        Frequency  Location  Stim  block
619496        0.0       0.0     0      0
619497        0.0       0.0     0      0
619498        0.0       0.0     0      0
619499       30.0       2.0     1      1
619500       30.0       2.0     1      1
619501       30.0       2.0     1      1

[6 rows x 40 columns]
                Fp1           Fpz           Fp2           F7          F3  \
770512 -21307.859830 -19217.652267 -22377.798994  2776.800448  109.093107
770513 -21311.809664 -19227.306856 -22384.012746  2764.679124   98.550319
770514 -21333.928026 -19245.785288 -22399.678849  2744.071935   81.366412
770515 -21343.376964 -19251.989172 -22410.319555  2738.250702   76.051882
770516 -21336.592286 -19245.526623 -22409.631168  2751.410594   84.081490
770517 -21319.412874 -19233.239440 -22398.002471  2769.691829  101.390482

                Fz          F4           F8          FC5          FC1  …  \
770512 -8349.908724 -363.113363  3556.262460  6818.159326  7209.120439  …
770513 -8357.306637 -366.847432  3554.021234  6810.871621  7210.910436  …
770514 -8372.056933 -381.842831  3537.176535  6796.345389  7194.612833  …
770515 -8378.541071 -388.981425  3532.668019  6794.515928  7183.674670  …
770516 -8377.296741 -389.113322  3529.982294  6800.260275  7188.778429  …
770517 -8364.076558 -377.053096  3542.702548  6817.448566  7200.874090  …

                Oz           O2      Time  Sub#  Session  Amplitude  \
770512 -1153.933109  3879.146282  770512.0     1        1        0.0
770513 -1163.362159  3864.769060  770513.0     1        1        0.0
770514 -1180.758261  3844.884031  770514.0     1        1        0.0
770515 -1181.436872  3848.244075  770515.0     1        1        1.0
770516 -1181.851122  3851.302925  770516.0     1        1        1.0
770517 -1169.647130  3871.222195  770517.0     1        1        1.0

        Frequency  Location  Stim  block
770512        0.0       0.0     0      0
770513        0.0       0.0     0      0
770514        0.0       0.0     0      0
770515       30.0       2.0     1      2
770516       30.0       2.0     1      2
770517       30.0       2.0     1      2

[6 rows x 40 columns]
                Fp1           Fpz           Fp2           F7         F3  \
921512 -21397.090812 -19162.818739 -22090.306908  2581.330838  66.521667
921513 -21378.065032 -19144.308748 -22071.645423  2606.637974  87.270950
921514 -21377.801219 -19143.829075 -22071.282297  2609.469608  90.559414
921515 -21386.649030 -19153.259036 -22082.980934  2597.320758  77.175305
921516 -21385.477849 -19150.906487 -22085.906198  2594.256849  73.793060
```

```
921517 -21381.282971 -19148.688615 -22084.053682  2598.932501  80.060979

                 Fz          F4          F8          FC5          FC1  …  \
921512 -8403.858132 -419.742554  3468.906824  6677.191908  7266.572453  …
921513 -8389.783724 -403.130818  3488.882481  6699.834066  7280.170958  …
921514 -8389.710681 -402.709270  3492.655503  6701.454518  7281.777299  …
921515 -8396.700219 -412.291202  3483.417461  6689.293862  7275.660879  …
921516 -8396.193857 -410.412726  3482.968699  6686.683284  7274.475420  …
921517 -8393.975345 -408.048019  3489.959082  6694.394861  7276.946838  …

                 Oz           O2      Time  Sub#  Session  Amplitude  \
921512 -1201.451568  3868.638213  921512.0     1        1        0.0
921513 -1188.141883  3884.194782  921513.0     1        1        0.0
921514 -1190.576062  3881.070380  921514.0     1        1        0.0
921515 -1198.606264  3874.497743  921515.0     1        1        1.0
921516 -1194.837811  3878.679230  921516.0     1        1        1.0
921517 -1193.178614  3881.046072  921517.0     1        1        1.0

        Frequency  Location  Stim  block
921512        0.0       0.0     0      0
921513        0.0       0.0     0      0
921514        0.0       0.0     0      0
921515       30.0       2.0     1      3
921516       30.0       2.0     1      3
921517       30.0       2.0     1      3

[6 rows x 40 columns]
                  Fp1           Fpz           Fp2           F7           F3  \
1072548 -21263.165036 -18945.707357 -21700.091893  2715.824963  220.500443
1072549 -21260.093590 -18940.484696 -21695.999914  2714.982298  222.717389
1072550 -21254.725644 -18938.201090 -21691.345025  2710.388849  223.451045
1072551 -21266.155153 -18948.721092 -21704.356177  2689.805783  212.820136
1072552 -21262.391265 -18945.719670 -21702.628058  2684.325293  220.847447
1072553 -21288.304211 -18970.229932 -21727.131485  2639.437177  184.096362

                  Fz          F4          F8          FC5          FC1  …  \
1072548 -8230.831505 -309.685198  3643.914283  6836.947732  7511.353082  …
1072549 -8230.892998 -306.840442  3648.566424  6836.028770  7509.333799  …
1072550 -8227.377572 -301.806106  3650.838795  6837.317152  7515.635411  …
1072551 -8236.630929 -313.770779  3636.977547  6825.137683  7505.558034  …
1072552 -8236.896220 -310.761310  3637.867268  6828.335971  7506.638772  …
1072553 -8259.691359 -335.222523  3619.025102  6789.995358  7478.555001  …

                  Oz           O2       Time  Sub#  Session  Amplitude  \
1072548 -1102.684836  4067.071099  1072548.0     1        1        0.0
1072549 -1106.781337  4058.780092  1072549.0     1        1        0.0
1072550 -1102.187735  4062.771048  1072550.0     1        1        0.0
1072551 -1103.964832  4064.921085  1072551.0     1        1        1.0
```

```
1072552 -1107.045797  4064.573237  1072552.0     1        1        1.0
1072553 -1149.042635  4016.825518  1072553.0     1        1        1.0

         Frequency  Location  Stim  block
1072548        0.0       0.0     0      0
1072549        0.0       0.0     0      0
1072550        0.0       0.0     0      0
1072551       30.0       2.0     1      4
1072552       30.0       2.0     1      4
1072553       30.0       2.0     1      4

[6 rows x 40 columns]
                  Fp1           Fpz           Fp2           F7         F3  \
1819590 -20337.071058 -18174.055825 -19785.616637  1993.456740  44.972488
1819591 -20337.052137 -18171.662708 -19784.458111  2000.743837  50.931967
1819592 -20348.598933 -18180.371613 -19788.854856  1983.677564  37.464465
1819593 -20354.852865 -18186.099029 -19799.462239  1973.714710  34.601411
1819594 -20353.391432 -18180.249913 -19794.065185  1981.888320  42.134851
1819595 -20356.593097 -18178.410588 -19792.062087  1989.164845  48.186964

                 Fz          F4          F8          FC5          FC1  …  \
1819590 -8270.421052 -328.863721  4349.844528  6700.621341  7788.022090  …
1819591 -8269.429137 -324.929179  4354.132401  6714.752735  7788.489272  …
1819592 -8276.619008 -332.676929  4344.445548  6700.319673  7785.666892  …
1819593 -8282.240427 -339.491468  4337.215751  6689.110863  7779.881477  …
1819594 -8273.348234 -331.991413  4346.135377  6699.382981  7786.210476  …
1819595 -8271.820612 -330.623244  4350.309976  6705.658241  7786.517547  …

                 Oz           O2       Time  Sub#  Session  Amplitude  \
1819590 -1313.177340  4039.580006  1819590.0     1        1        0.0
1819591 -1310.879520  4040.386035  1819591.0     1        1        0.0
1819592 -1314.461668  4037.535722  1819592.0     1        1        0.0
1819593 -1324.479815  4027.285891  1819593.0     1        1        1.0
1819594 -1319.801460  4031.823183  1819594.0     1        1        1.0
1819595 -1313.401644  4038.617400  1819595.0     1        1        1.0

         Frequency  Location  Stim  block
1819590        0.0       0.0     0      0
1819591        0.0       0.0     0      0
1819592        0.0       0.0     0      0
1819593       30.0       2.0     1      5
1819594       30.0       2.0     1      5
1819595       30.0       2.0     1      5

[6 rows x 40 columns]
                  Fp1           Fpz           Fp2           F7          F3  \
1970666 -20109.625101 -17866.851093 -19339.826031  1961.732682  125.728163
1970667 -20130.140417 -17878.603763 -19350.837736  1927.975796  109.275194
```

```
1970668 -20141.143516 -17885.520009 -19358.480696  1905.150780   96.991419
1970669 -20173.800331 -17911.415417 -19385.512719  1879.405977   59.792295
1970670 -20163.385294 -17895.031576 -19376.012830  1901.434446   61.375937
1970671 -20140.129974 -17872.260731 -19351.546553  1931.996595   81.738509

                 Fz           F4           F8          FC5          FC1  …  \
1970666 -8235.945121 -276.362205  4459.481177  6699.048027  7967.900331  …
1970667 -8246.907547 -285.858001  4452.090585  6683.497899  7957.239953  …
1970668 -8254.209777 -295.270490  4448.813049  6669.381016  7946.724926  …
1970669 -8281.177699 -322.511758  4416.610309  6631.427264  7917.611501  …
1970670 -8272.453905 -314.935796  4427.200405  6632.100561  7920.423573  …
1970671 -8250.562520 -289.784630  4449.628333  6654.542582  7944.647875  …

                 Oz           O2       Time  Sub#  Session  Amplitude  \
1970666 -1277.182825  4117.313652  1970666.0     1        1        0.0
1970667 -1288.807506  4103.290591  1970667.0     1        1        0.0
1970668 -1290.688056  4102.427877  1970668.0     1        1        0.0
1970669 -1326.356125  4065.722286  1970669.0     1        1        1.0
1970670 -1321.204605  4068.126495  1970670.0     1        1        1.0
1970671 -1299.479368  4090.821254  1970671.0     1        1        1.0

         Frequency  Location  Stim  block
1970666        0.0       0.0     0      0
1970667        0.0       0.0     0      0
1970668        0.0       0.0     0      0
1970669       30.0       2.0     1      6
1970670       30.0       2.0     1      6
1970671       30.0       2.0     1      6

[6 rows x 40 columns]
                  Fp1           Fpz           Fp2           F7         F3  \
2121641 -20096.663953 -17748.605915 -18943.862491  1887.021844  89.831714
2121642 -20111.023422 -17762.544749 -18955.515187  1867.399988  71.621622
2121643 -20110.190726 -17760.902329 -18957.706936  1869.235526  67.133577
2121644 -20096.360412 -17742.138871 -18939.137337  1873.454801  74.330429
2121645 -20085.842953 -17735.517428 -18926.829807  1878.362518  86.528674
2121646 -20083.979181 -17736.210246 -18927.603107  1883.894276  91.888247

                 Fz           F4           F8          FC5          FC1  …  \
2121641 -8207.877484 -214.189404  4525.395184  6729.275232  8084.224545  …
2121642 -8223.790759 -230.901647  4508.678505  6710.282256  8065.507101  …
2121643 -8225.237988 -236.540287  4503.648206  6704.053836  8054.369031  …
2121644 -8207.942531 -220.641198  4519.891515  6710.248604  8070.553709  …
2121645 -8195.447024 -210.950062  4528.547140  6722.534061  8087.016816  …
2121646 -8193.985459 -208.591245  4532.807699  6729.811231  8089.736614  …

                 Oz           O2       Time  Sub#  Session  Amplitude  \
2121641 -1226.981654  4164.157563  2121641.0     1        1        0.0
```

```
2121642 -1251.465081  4143.418400  2121642.0     1         1         0.0
2121643 -1272.097018  4121.259261  2121643.0     1         1         0.0
2121644 -1263.465176  4118.560888  2121644.0     1         1         1.0
2121645 -1239.935752  4145.259592  2121645.0     1         1         1.0
2121646 -1224.291888  4165.897436  2121646.0     1         1         1.0

         Frequency  Location  Stim  block
2121641        0.0       0.0     0      0
2121642        0.0       0.0     0      0
2121643        0.0       0.0     0      0
2121644       30.0       2.0     1      7
2121645       30.0       2.0     1      7
2121646       30.0       2.0     1      7

[6 rows x 40 columns]
                  Fp1           Fpz           Fp2           F7         F3  \
2272753 -20075.496627 -17782.640697 -18865.920135  1615.023947 -43.814051
2272754 -20068.272439 -17777.554142 -18868.120808  1626.080142 -32.083363
2272755 -20074.248545 -17780.951881 -18874.301983  1616.022579 -39.782883
2272756 -20062.660853 -17766.522726 -18858.279726  1630.150593 -22.651835
2272757 -20058.374243 -17764.055183 -18851.754120  1640.424344 -12.637074
2272758 -20070.075103 -17775.302463 -18862.629251  1633.095040 -16.621416

                  Fz          F4           F8          FC5          FC1  …  \
2272753 -8304.105806 -313.955773  4474.396308  6533.280926  8009.613880  …
2272754 -8301.052227 -311.800896  4474.127255  6542.938432  8013.782788  …
2272755 -8303.499822 -315.742839  4470.524704  6538.022319  8012.576571  …
2272756 -8290.260616 -302.960076  4488.825018  6557.398902  8029.824080  …
2272757 -8283.504971 -294.645400  4499.275174  6571.754065  8037.884549  …
2272758 -8292.831652 -304.407240  4483.997022  6565.352809  8029.019930  …

                  Oz           O2       Time  Sub#  Session  Amplitude  \
2272753 -1453.900061  4066.413959  2272753.0     1        1        0.0
2272754 -1460.583611  4061.773741  2272754.0     1        1        0.0
2272755 -1463.028974  4060.397266  2272755.0     1        1        0.0
2272756 -1444.746320  4075.667405  2272756.0     1        1        1.0
2272757 -1436.839940  4085.716090  2272757.0     1        1        1.0
2272758 -1448.226864  4074.010596  2272758.0     1        1        1.0

         Frequency  Location  Stim  block
2272753        0.0       0.0     0      0
2272754        0.0       0.0     0      0
2272755        0.0       0.0     0      0
2272756       30.0       2.0     1      8
2272757       30.0       2.0     1      8
2272758       30.0       2.0     1      8

[6 rows x 40 columns]
```

```
                 Fp1           Fpz           Fp2          F7         F3  \
3019819 -19076.986379 -17265.102225 -16991.107342  954.649155 -15.647657
3019820 -19046.903725 -17237.138350 -16961.319124  987.603882  16.678522
3019821 -19054.189760 -17237.723444 -16957.021300  976.193957  13.864991
3019822 -19049.449831 -17236.687715 -16957.564250  981.887184  20.375086
3019823 -19046.035182 -17235.787221 -16955.233880  981.616985  17.019979
3019824 -19054.248413 -17243.683467 -16956.497610  970.365791   8.276486

                 Fz         F4           F8          FC5          FC1  …  \
3019819 -8050.007880 -83.309115  5761.630208  6734.528226  8277.714174  …
3019820 -8022.675580 -54.402859  5786.959477  6764.217627  8305.436763  …
3019821 -8023.252610 -52.237152  5781.103101  6759.717349  8308.155923  …
3019822 -8025.750098 -54.474761  5778.903894  6763.923338  8305.325457  …
3019823 -8023.069978 -50.299965  5786.048149  6769.017614  8307.299114  …
3019824 -8029.540900 -57.883003  5783.775607  6761.363708  8299.657415  …

                 Oz           O2       Time  Sub#  Session  Amplitude  \
3019819 -1433.995697  4228.162763  3019819.0     1        1        0.0
3019820 -1409.079753  4255.201706  3019820.0     1        1        0.0
3019821 -1405.061408  4261.396646  3019821.0     1        1        0.0
3019822 -1406.712479  4259.079308  3019822.0     1        1        1.0
3019823 -1401.721734  4263.819442  3019823.0     1        1        1.0
3019824 -1405.349623  4260.573589  3019824.0     1        1        1.0

         Frequency  Location  Stim  block
3019819        0.0       0.0     0      0
3019820        0.0       0.0     0      0
3019821        0.0       0.0     0      0
3019822       30.0       2.0     1      9
3019823       30.0       2.0     1      9
3019824       30.0       2.0     1      9

[6 rows x 40 columns]
```

[18]:

```python
# Save the DataFrame as a CSV file
merged_eeg_stim_df.to_csv('/home/vincent/AAA_projects/MVCS/Neuroscience/
 ↪MergedStimEEG/merged_stim_eeg.csv', index=False)
```

### 3.1 Other topics to explore: Cable Theory-Based Models, Compartmental Neuron Models, Network Models, Mean-Field Models, Detailed Biophysical Models, Graph Theory, Topological Data Analysis

### 3.2 Next Modules to make: Convolutional Neural Network, Hjorth Coefficients, Petrosian Fractal Dimension, Discrete Wavelet Transform, Differential Asymmetry, Magnitude Squared Coherence Estimate

[ ]:

```

```

# Exploration and Analysis EEG

September 8, 2023

```
Data include within participant application of nine High-Definition tES (HD-tES) types, targeti

participants maintained a ball at the center of the screen and were periodically stimulated (wi

DSamp

triggers <- These are all the labeled EEG/Stimulation start/stop triggers
EEGdata <- Contains the downsampled EEG/ECG/EOG voltage data dims: 35 channelss X ~4E6 samples
fs <- The downsampled sampling frequency of the data : 1000 Hz
fsOld <- The original sampling frequency of the data
time <- Time vector for the data. Should be 1 X ~4E6
label <- Contains the channel label information. BIP1= ECG, BIP2=EOG, RESP1= N/A
nchan <- The number of channels in the data
rate <- Redundant to fs, sampling rate of data
npt <- Number of data points ~4E6
Subj <- Subject and session that data belong to. I.e. 0302 - Subject 03 session 03
ptrackerPerf <- The CTT data deviation/ the behavioral data
ptrackerTime <- Time vector for the CTT data
ptrackerfs <- The sampling frequency for the CTT data 100 Hz.
```

[2]:

```
import numpy as np
import nolds
import scipy.io
import mne
import pandas as pd
import seaborn as sns
import pyqtgraph as pg
from scipy.io import loadmat
from scipy import stats
import matplotlib.pyplot as plt
import antropy as ent
from attractors import Attractor
from antropy import higuchi_fd
from pyrqa.settings import Settings
from pyrqa.neighbourhood import FixedRadius
from pyrqa.computation import RQAComputation
from pyrqa.time_series import TimeSeries
```

```python
from pyrqa.result import RQAResult
from pyrqa.opencl import OpenCL
from pyrqa.settings import Settings
from pyrqa.time_series import TimeSeries
from pyrqa.neighbourhood import FixedRadius
from pyrqa.settings import Settings
from pyrqa.time_series import TimeSeries
from pyrqa.neighbourhood import FixedRadius
from pyrqa.metric import EuclideanMetric
from pyrqa.computation import RQAComputation
from pyrqa.metric import EuclideanMetric
from pyrqa.computation import RQAComputation
from pyrqa.settings import Settings
from pyrqa.neighbourhood import FixedRadius
from pyrqa.metric import EuclideanMetric
from pyrqa.image_generator import ImageGenerator
import numpy as np
import matplotlib.pyplot as plt
from antropy import higuchi_fd
```

```
/home/vincent/miniconda3/lib/python3.10/site-packages/antropy/fractal.py:197:
NumbaDeprecationWarning: The 'nopython' keyword argument was not supplied to
the 'numba.jit' decorator. The implicit default value for this argument is
currently False, but it will be changed to True in Numba 0.59.0. See
https://numba.readthedocs.io/en/stable/reference/deprecation.html#deprecation-
of-object-mode-fall-back-behaviour-when-using-jit for details.
  @jit((types.Array(types.float64, 1, "C", readonly=True), types.int32))
```

```
<!-- Column 1 -->
<div style="flex: 1; margin-right: 10px;">
    <h2>Introduction</h2>
<p>This section elucidates the key steps taken to load and preprocess the Electroencephalogram
    <h2>Objectives</h2>
    <p style="text-indent: 40px;">Data Importation: Efficiently load EEG data and stimulation i
    <p style="text-indent: 40px;">Data Cleansing: Handle missing values in the stimulation data
    <p style="text-indent: 40px;">Metadata Extraction: Delve into the loaded EEG data to segreg
    <p style="text-indent: 40px;">Channel Filtration: Methodologically exclude channels that ar
    <h2>Mathematical Formulation</h2>
<p>The loaded EEG data structure can be mathematically represented as a multi-dimensional set \
\[ DSamp = \{ \text{Triggers, EEGdata, fs, fsOld, Time, Label, nChan, Rate, nPt, Subj, Ptracke
<p>Here, each element serves a distinct purpose:</p>
<ul>
    <li>Triggers: Trigger events encapsulated within the EEG data.</li>
    <li>EEGdata: The matrix containing the EEG signals.</li>
    <li>fs: The sampling frequency after the EEG data have been downsampled.</li>
    <li>fsOld: The original sampling frequency prior to downsampling.</li>
```

```
    <li>Time: The vector containing time points corresponding to EEG data.</li>
    <li>Label: The labels designating the EEG channels.</li>
    <li>nChan: The total number of EEG channels.</li>
    <li>Rate: A rate variable, the purpose of which requires additional context.</li>
    <li>nPt: The number of data points in each EEG channel.</li>
    <li>Subj: An identifier for the subject from whom the EEG data were collected.</li>
    <li>PtrackerPerf: Metrics pertaining to pointer tracker performance.</li>
    <li>PtrackerTime: Time metrics from the pointer tracker.</li>
    <li>PtrackerFs: Sampling frequency of the pointer tracker.</li>
</ul>
</div>
<!-- Column 2 -->
<div style="flex: 1; margin-left: 10px;">
<h2>Data Cleansing and Imputation</h2>
<p>The stimulation data potentially contain missing values specifically in the column designat
\[ f(x_i) =
\begin{cases}
x_{i-1} & \text{if } x_i \text{ is missing} \\
x_i & \text{otherwise}
\end{cases}
\]
<h2>Channel Filtration</h2>
<p>The EEG data may contain channels that are not pertinent to the current analysis. Let \( L \
<h2>Summary</h2>
<p>This section provides a high-level mathematical and programmatical blueprint for the data l
<p>Now that the framework has been outlined, the next section of the notebook will contain the
```

[3]:
```
# Load data
data = loadmat('/home/vincent/AAA_projects/MVCS/Neuroscience/downsampled/
 ↪EEG_DS_Struct_0101.mat')
stim_data = pd.read_excel('/home/vincent/AAA_projects/MVCS/Neuroscience/
 ↪EEG-tES-Chaos-Neural-Net/stim_data.xlsx')

# Fill null values in 'Sub#' column
stim_data['Sub#'].fillna(method='ffill', inplace=True)

# Drop the first row
stim_data = stim_data.drop(0)

DSamp = data['DSamp']

# Get data parameters
triggers = DSamp[0][0][0]
EEGdata = DSamp[0][0][1]
fs = DSamp[0][0][2][0][0]
fsOld = DSamp[0][0][3][0][0]
time = DSamp[0][0][4][0]
```

```python
label = DSamp[0][0][5]
nchan = DSamp[0][0][6][0][0]
rate = DSamp[0][0][7][0][0]
npt = DSamp[0][0][8][0][0]
Subj = DSamp[0][0][9][0]
ptrackerPerf = DSamp[0][0][10]
ptrackerTime = DSamp[0][0][11]
ptrackerfs = DSamp[0][0][12][0][0]

# List of unwanted channel names
unwanted_channels = ['BIP1', 'BIP2', 'RESP1']

# Filter out unwanted channels from the label data
filtered_label = [ch for ch in label if ch[0][0] not in unwanted_channels]

# Convert the filtered list back to numpy array and replace the original label
label = np.array(filtered_label, dtype=object)

print(triggers, EEGdata, label, stim_data)
```

```
[[(array([[20.428]]), array([[20429]], dtype=uint16), array(['0002'],
dtype='<U4'), array([[2]], dtype=uint8), array(['Block Start'], dtype='<U11'),
array([], shape=(1, 0), dtype=float64))
  (array([[619.442]]), array([[619443]], dtype=int32), array(['0002'],
dtype='<U4'), array([[2]], dtype=uint8), array(['Block Start'], dtype='<U11'),
array([], shape=(1, 0), dtype=float64))
  (array([[619.499]]), array([[619500]], dtype=int32), array(['0016'],
dtype='<U4'), array([[16]], dtype=uint8), array(['Stim Start'], dtype='<U10'),
array(['M30'], dtype='<U3'))
  (array([[654.746]]), array([[654747]], dtype=int32), array(['0032'],
dtype='<U4'), array([[32]], dtype=uint8), array(['Stim Stop'], dtype='<U9'),
array([], shape=(1, 0), dtype=float64))
  (array([[770.515]]), array([[770516]], dtype=int32), array(['0016'],
dtype='<U4'), array([[16]], dtype=uint8), array(['Stim Start'], dtype='<U10'),
array(['M30'], dtype='<U3'))
  (array([[805.571]]), array([[805572]], dtype=int32), array(['0032'],
dtype='<U4'), array([[32]], dtype=uint8), array(['Stim Stop'], dtype='<U9'),
array([], shape=(1, 0), dtype=float64))
  (array([[921.515]]), array([[921516]], dtype=int32), array(['0016'],
dtype='<U4'), array([[16]], dtype=uint8), array(['Stim Start'], dtype='<U10'),
array(['M30'], dtype='<U3'))
  (array([[956.651]]), array([[956652]], dtype=int32), array(['0032'],
dtype='<U4'), array([[32]], dtype=uint8), array(['Stim Stop'], dtype='<U9'),
array([], shape=(1, 0), dtype=float64))
  (array([[1072.551]]), array([[1072552]], dtype=int32), array(['0016'],
dtype='<U4'), array([[16]], dtype=uint8), array(['Stim Start'], dtype='<U10'),
array(['M30'], dtype='<U3'))
```

```
  (array([[1107.578]]), array([[1107579]], dtype=int32), array(['0032'],
dtype='<U4'), array([[32]], dtype=uint8), array(['Stim Stop'], dtype='<U9'),
array([], shape=(1, 0), dtype=float64))
  (array([[1218.442]]), array([[1218443]], dtype=int32), array(['0002'],
dtype='<U4'), array([[2]], dtype=uint8), array(['Block Start'], dtype='<U11'),
array([], shape=(1, 0), dtype=float64))
  (array([[1817.46]]), array([[1817461]], dtype=int32), array(['0002'],
dtype='<U4'), array([[2]], dtype=uint8), array(['Block Start'], dtype='<U11'),
array([], shape=(1, 0), dtype=float64))
  (array([[1819.593]]), array([[1819594]], dtype=int32), array(['0016'],
dtype='<U4'), array([[16]], dtype=uint8), array(['Stim Start'], dtype='<U10'),
array(['M30'], dtype='<U3'))
  (array([[1854.888]]), array([[1854889]], dtype=int32), array(['0032'],
dtype='<U4'), array([[32]], dtype=uint8), array(['Stim Stop'], dtype='<U9'),
array([], shape=(1, 0), dtype=float64))
  (array([[1970.669]]), array([[1970670]], dtype=int32), array(['0016'],
dtype='<U4'), array([[16]], dtype=uint8), array(['Stim Start'], dtype='<U10'),
array(['M30'], dtype='<U3'))
  (array([[2005.715]]), array([[2005716]], dtype=int32), array(['0032'],
dtype='<U4'), array([[32]], dtype=uint8), array(['Stim Stop'], dtype='<U9'),
array([], shape=(1, 0), dtype=float64))
  (array([[2121.644]]), array([[2121645]], dtype=int32), array(['0016'],
dtype='<U4'), array([[16]], dtype=uint8), array(['Stim Start'], dtype='<U10'),
array(['M30'], dtype='<U3'))
  (array([[2156.695]]), array([[2156696]], dtype=int32), array(['0032'],
dtype='<U4'), array([[32]], dtype=uint8), array(['Stim Stop'], dtype='<U9'),
array([], shape=(1, 0), dtype=float64))
  (array([[2272.756]]), array([[2272757]], dtype=int32), array(['0016'],
dtype='<U4'), array([[16]], dtype=uint8), array(['Stim Start'], dtype='<U10'),
array(['M30'], dtype='<U3'))
  (array([[2307.798]]), array([[2307799]], dtype=int32), array(['0032'],
dtype='<U4'), array([[32]], dtype=uint8), array(['Stim Stop'], dtype='<U9'),
array([], shape=(1, 0), dtype=float64))
  (array([[2416.534]]), array([[2416535]], dtype=int32), array(['0002'],
dtype='<U4'), array([[2]], dtype=uint8), array(['Block Start'], dtype='<U11'),
array([], shape=(1, 0), dtype=float64))
  (array([[3015.513]]), array([[3015514]], dtype=int32), array(['0002'],
dtype='<U4'), array([[2]], dtype=uint8), array(['Block Start'], dtype='<U11'),
array([], shape=(1, 0), dtype=float64))
  (array([[3019.822]]), array([[3019823]], dtype=int32), array(['0016'],
dtype='<U4'), array([[16]], dtype=uint8), array(['Stim Start'], dtype='<U10'),
array(['M30'], dtype='<U3'))
  (array([[3019.924]]), array([[3019925]], dtype=int32), array(['0032'],
dtype='<U4'), array([[32]], dtype=uint8), array(['Stim Stop'], dtype='<U9'),
array([], shape=(1, 0), dtype=float64))]] [[-21295.98864928 -21303.74707676
-21315.46657134 … -17286.89230419
  -17281.27799398 -17330.52269117]
 [-20109.7167273  -20120.74615359 -20130.12657698 … -16618.13723294
```

```
 -16602.343699   -16641.89672243]
[-24153.38375243 -24163.86401194 -24171.94434272 … -14755.59815419
 -14743.08142526 -14783.66201823]
…
[  2692.44573979   2703.97118199   2731.75987441 …   4181.64344633
   4157.70728788   4198.03945844]
[ -5014.9543463   -5014.10209995  -5019.83464311 …  -4711.95486888
  -4706.99851768  -4709.83053726]
[    76.88986022     72.02278487     66.20816171 …     73.4410151
     69.33401454     66.65666327]] [[array(['Fp1'], dtype='<U3')]
[array(['Fpz'], dtype='<U3')]
[array(['Fp2'], dtype='<U3')]
[array(['F7'], dtype='<U2')]
[array(['F3'], dtype='<U2')]
[array(['Fz'], dtype='<U2')]
[array(['F4'], dtype='<U2')]
[array(['F8'], dtype='<U2')]
[array(['FC5'], dtype='<U3')]
[array(['FC1'], dtype='<U3')]
[array(['FC2'], dtype='<U3')]
[array(['FC6'], dtype='<U3')]
[array(['M1'], dtype='<U2')]
[array(['T7'], dtype='<U2')]
[array(['C3'], dtype='<U2')]
[array(['Cz'], dtype='<U2')]
[array(['C4'], dtype='<U2')]
[array(['T8'], dtype='<U2')]
[array(['M2'], dtype='<U2')]
[array(['CP5'], dtype='<U3')]
[array(['CP1'], dtype='<U3')]
[array(['CP2'], dtype='<U3')]
[array(['CP6'], dtype='<U3')]
[array(['P7'], dtype='<U2')]
[array(['P3'], dtype='<U2')]
[array(['Pz'], dtype='<U2')]
[array(['P4'], dtype='<U2')]
[array(['P8'], dtype='<U2')]
[array(['POz'], dtype='<U3')]
[array(['O1'], dtype='<U2')]
[array(['Oz'], dtype='<U2')]
[array(['O2'], dtype='<U2')]]     Sub#  Session  File Num F0 M0 P0 F5 M5 P5 F30
M30 P30 StimTypeBlock1  \
1       1        2        102  1                  7   8              M30
2       1        3        103       3        6           9           P30
3       1        4        104     2     4  5                          F5
4       1        5        105     2     4  5                          F5
5       1        6        106       3        6           9           P30
6       2        1        201          4        7        9          F30
```

| | | | | | | | | | | | | | |
|---|---|---|---|---|---|---|---|---|---|---|---|---|---|
| 7 | 2 | 2 | 202 | | | 3 | | 5 | | | 8 | | P0 |
| 8 | 3 | 1 | 301 | | 2 | 3 | | | | 7 | | | M0 |
| 9 | 3 | 2 | 302 | | | | 4 | | 6 | | | 9 | F5 |
| 10 | 3 | 3 | 303 | 1 | | | | 5 | | | 8 | | M5 |
| 11 | 4 | 1 | 401 | | | | 4 | | | | 8 | 9 | P30 |
| 12 | 4 | 2 | 402 | 1 | 2 | 3 | | | | | | | P0 |
| 13 | 4 | 3 | 403 | | | | | 5 | 6 | 7 | | | F30 |
| 14 | 5 | 1 | 501 | | | 3 | | 5 | 6 | | | | M5 |
| 15 | 5 | 4 | 504 | | 2 | | | | | | 8 | 9 | P30 |
| 16 | 5 | 5 | 505 | 1 | | | 4 | | | 7 | | | F30 |
| 17 | 6 | 1 | 601 | 1 | | | 4 | | 6 | | | | F0 |
| 18 | 6 | 2 | 602 | | | | | | | 7 | 8 | 9 | F30 |
| 19 | 6 | 3 | 603 | | 2 | 3 | | 5 | | | | | P0 |
| 20 | 7 | 1 | 701 | | | | 4 | 5 | 6 | | | | P5 |
| 21 | 7 | 2 | 702 | 1 | 2 | 3 | | | | | | | M0 |
| 22 | 7 | 3 | 703 | | | | | | | 7 | 8 | 9 | F30 |
| 23 | 8 | 1 | 801 | | | | 4 | 5 | | | 8 | | F5 |
| 24 | 8 | 2 | 802 | 1 | | 3 | | | | | | 9 | P30 |
| 25 | 8 | 3 | 803 | | 2 | | | | 6 | 7 | | | P5 |
| 26 | 9 | 1 | 901 | 1 | | | 4 | | | | | 9 | P30 |
| 27 | 9 | 2 | 902 | | | 3 | | 5 | | 7 | | | M5 |
| 28 | 9 | 3 | 903 | | 2 | | | | 6 | | 8 | | P5 |
| 29 | 10 | 1 | 1001 | | 2 | 3 | | 5 | | | | | M0 |
| 30 | 10 | 2 | 1002 | 1 | | | | | | 7 | | 9 | F0 |
| 31 | 10 | 3 | 1003 | | | | 4 | | 6 | | 8 | | F5 |

| | StimTypeBlock2 | StimTypeBlock3 | StimAmplitude_mA_block1 \ |
|---|---|---|---|
| 1 | F30 | F0 | 0.5 |
| 2 | P0 | P5 | 0.5 |
| 3 | M5 | M0 | 0.5 |
| 4 | M5 | M0 | 1.0 |
| 5 | P0 | P5 | 1.0 |
| 6 | F5 | P30 | 0.5 |
| 7 | M5 | M30 | 0.5 |
| 8 | P0 | F30 | 0.5 |
| 9 | P5 | P30 | 0.5 |
| 10 | F0 | M30 | 0.5 |
| 11 | M30 | F5 | 0.5 |
| 12 | F0 | M0 | 1.0 |
| 13 | M5 | P5 | 1.0 |
| 14 | P0 | P5 | 0.5 |
| 15 | M30 | M0 | 1.0 |
| 16 | F0 | F5 | 1.0 |
| 17 | F5 | P5 | 1.0 |
| 18 | M30 | P30 | 0.5 |
| 19 | M5 | M0 | 1.0 |
| 20 | F5 | M5 | 0.5 |
| 21 | P0 | F0 | 0.5 |

```
22            P30          M30                    0.5
23            M30           M5                    1.0
24             F0           P0                    1.0
25            F30           M0                    1.0
26             F5           F0                    0.5
27            F30           P0                    0.5
28             M0          M30                    1.0
29             P0           M5                    1.0
30            P30          F30                    1.0
31            M30           P5                    1.0

    StimAmplitude_mA_block2  StimAmplitude_mA_block3
1                       0.5                      0.5
2                       0.5                      0.5
3                       0.5                      0.5
4                       1.0                      1.0
5                       1.0                      1.0
6                       0.5                      0.5
7                       0.5                      0.5
8                       0.5                      0.5
9                       0.5                      0.5
10                      0.5                      0.5
11                      0.5                      0.5
12                      1.0                      1.0
13                      1.0                      1.0
14                      1.0                      1.0
15                      1.0                      1.0
16                      1.0                      0.5
17                      0.5                      1.0
18                      1.0                      1.0
19                      1.0                      1.0
20                      0.5                      0.5
21                      0.5                      0.5
22                      0.5                      0.5
23                      1.0                      1.0
24                      1.0                      1.0
25                      1.0                      1.0
26                      0.5                      1.0
27                      1.0                      1.0
28                      1.0                      1.0
29                      1.0                      1.0
30                      1.0                      1.0
31                      1.0                      1.0
```

[4]:
```python
shape_of_data = DSamp[0][0][1].shape
print(shape_of_data)
```

```
(35, 4227788)
```

[8]:

```
import matplotlib.pyplot as plt
import numpy as np

# Assuming that EEGdata and time are numpy arrays
EEGdata = np.array(EEGdata)
time = np.array(time)

# Checking the number of channels
nchan = EEGdata.shape[0]

# Creating a figure with nchan subplots, one for each channel
fig, axs = plt.subplots(nchan, 1)

# Setting the figure size
fig.set_size_inches(10, 2*nchan)

# Loop over all the channels
for i in range(nchan):
    axs[i].plot(time, EEGdata[i], label='Channel '+str(i+1))  # Use the entire␣
 ↪'time' array
    axs[i].set_xlabel('Time (s)')
    axs[i].set_ylabel('Amplitude')
    axs[i].legend()


# Automatically adjust subplot params so the subplot(s) fits into the figure␣
 ↪area
plt.tight_layout()

# Display the plot
plt.show()
```

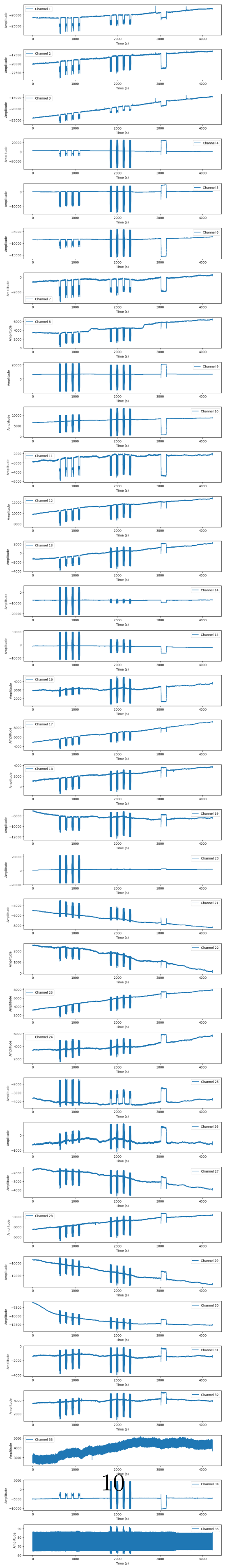
Channel 1
Channel 2
Channel 3
Channel 4
Channel 5
Channel 6
Channel 7
Channel 8
Channel 9
Channel 10
Channel 11
Channel 12
Channel 13
Channel 14
Channel 15
Channel 16
Channel 17
Channel 18
Channel 19
Channel 20
Channel 21
Channel 22
Channel 23
Channel 24
Channel 25
Channel 26
Channel 27
Channel 28
Channel 29
Channel 30
Channel 31
Channel 32
Channel 33
Channel 34
Channel 35
Amplitude
Time (s)

```
<!-- Column 1 -->
<div style="flex: 1; margin-right: 10px;">
    <h2>Introduction</h2>
    <p>This section is dedicated to elucidating the methodology employed for phase space recons
    <h2>Objectives</h2>
    <ul>
        <li>Delay Embedding: Perform a formal delay embedding transformation on a selected EEG
        <li>Dimensionality Transformation: Transform the one-dimensional time-series EEG data i
        <li>Visual Exploration: Generate a 3D phase space plot for the embedded data.</li>
    </ul>
    <h2>Mathematical Foundations</h2>
    <h3>Delay Embedding Function</h3>
    <p>Let \( x(t) \) be the time-series EEG data, where \( t=1,2,\ldots,N \). The delay embedd
    \[ \phi : x(t) \mapsto X(t) = [ x(t), x(t-\tau), x(t-2\tau), \ldots, x(t-(m-1)\tau) ] \]
    <p>The function \( \phi \) is applied to create an \( m \)-dimensional vector \( X(t) \) fr
</div>
<!-- Column 2 -->
<div style="flex: 1; margin-left: 10px;">
    <h2>Dimensionality and Delay</h2>
    <p>In our application, the embedding dimension \( m \) is set to 3, and the delay \( \tau \
    \[ m=3,\tau=1 \]
    <h2>Phase Space Reconstruction</h2>
    <p>The phase space \( P \) is constructed by applying \( \phi \) on \( x(t) \) for each \(
    \[ P = \{ X(t) \mid t = 1, \ldots, N - (m-1)\tau \} \]
    <h2>3D Visualization</h2>
    <p>The phase space \( P \) is visualized in a 3D plot, where the axes represent the compone
    <h2>Summary</h2>
    <p>Phase space reconstruction via delay embedding serves as a key element in our pipeline,
    <p>In the next section of this notebook, we implement the delay embedding process, followed
</div>
```

[59]:
```python
import numpy as np
import matplotlib.pyplot as plt
from mpl_toolkits.mplot3d import Axes3D

# Define a function for delay embedding
def delay_embedding(data, emb_dim, delay):
    N = len(data)
    embedded_data = np.zeros((N - (emb_dim - 1) * delay, emb_dim))
    for i in range(N - (emb_dim - 1) * delay):
        embedded_data[i] = [data[i + j * delay] for j in range(emb_dim)]
    return embedded_data

# Choose the first channel of your EEG data
channel_data = EEGdata[0, :]
```

```
# Perform delay embedding with embedding dimension 3 and delay 1
embedded_channel_data = delay_embedding(channel_data, emb_dim=3, delay=1)

# Create 3D plot
fig = plt.figure()
ax = fig.add_subplot(111, projection='3d')

# Change viewing angle
ax.view_init(elev=45, azim=21)

ax.plot(embedded_channel_data[:, 0], embedded_channel_data[:, 1],␣
↪embedded_channel_data[:, 2])
plt.title('Phase Space Plot')
plt.show()
```

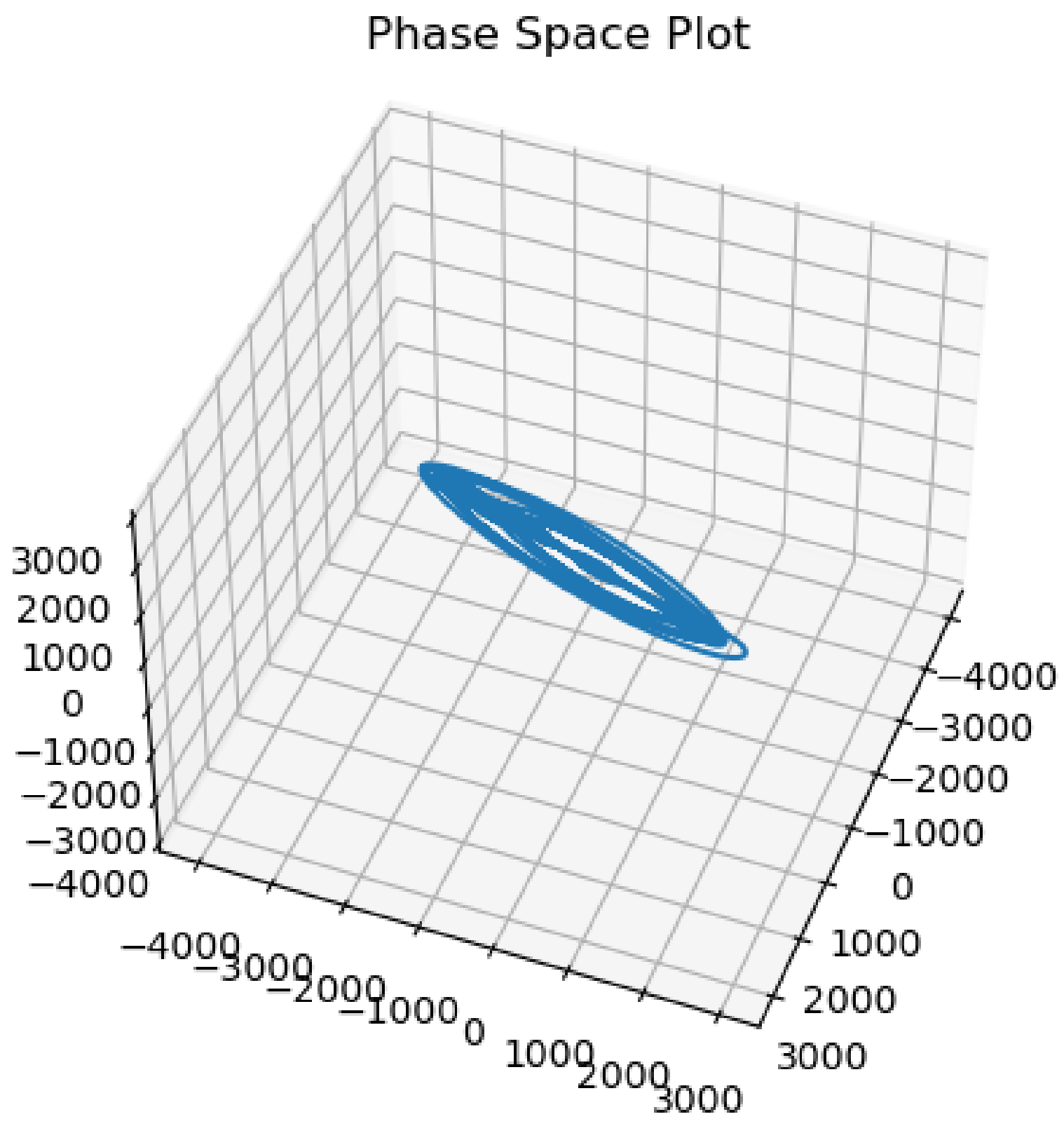

```
[69]: import numpy as np
import matplotlib.pyplot as plt

# Define a function for delay embedding
def delay_embedding(data, emb_dim, delay):
```

```python
    N = len(data)
    embedded_data = np.zeros((N - (emb_dim - 1) * delay, emb_dim))
    for i in range(N - (emb_dim - 1) * delay):
        embedded_data[i] = [data[i + j * delay] for j in range(emb_dim)]
    return embedded_data

# Choose the first channel of your EEG data
channel_data = EEGdata[0, :]

# Perform delay embedding with embedding dimension 2 and delay 1
embedded_channel_data = delay_embedding(channel_data, emb_dim=2, delay=1)

# Create 2D plot
plt.figure()
plt.plot(embedded_channel_data[:, 0], embedded_channel_data[:, 1])
plt.title('Phase Space Plot')
plt.xlabel('Embedding Dimension 1')
plt.ylabel('Embedding Dimension 2')
plt.show()
```

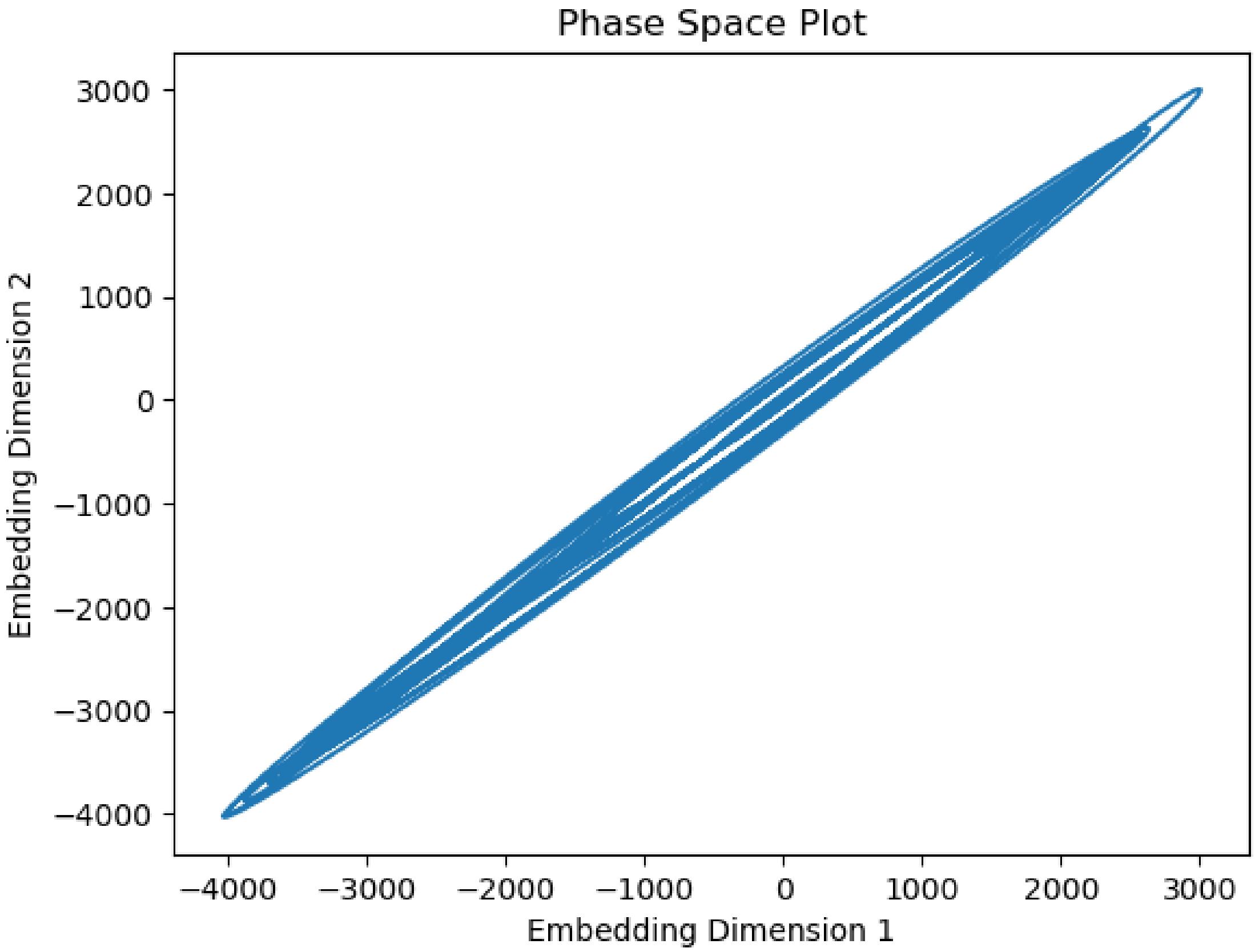

# Spectral Analysis

September 8, 2023

## 1 Spectral Analysis

## 2 Load EEG npy

[2]:
```
import numpy as np
# Define the path to the saved numpy array
load_path = '/home/vincent/AAA_projects/MVCS/Neuroscience/
  ↪eeg_data_with_channels.npy'

# Load the numpy array
eeg_data_array = np.load(load_path)

# Print the shape of the loaded array (should be number of channels x number of␣
  ↪time samples)
print("Shape of the loaded EEG data array:", eeg_data_array.shape)

# Access the EEG data for a specific channel
channel_index = 0  # Replace with the index of the desired channel
channel_data = eeg_data_array[channel_index]

# Print the EEG data for the selected channel
print(f"EEG data for channel {channel_index}:", channel_data)
```

```
Shape of the loaded EEG data array: (4227788, 32)
EEG data for channel 0: [-21295.98864928 -20109.7167273  -24153.38375243
3189.34006041
    -45.18927487  -8525.0666796    -642.12859002   3487.91362107
   6324.95663882   6503.01217655  -2859.03109597   9804.17906924
  -1300.12580193  -7349.53009727  -1048.27997888   2955.10747365
   4882.67613967   1111.30958652  -7018.55014536    692.11173545
  -4974.24315956   2540.26756016   3191.27369535   3374.04802916
  -3617.19796357   -611.58474194  -1667.22264416   7523.61208528
  -9446.68538886  -6091.7889313   -1392.83563441   3559.608191  ]
```

## 3 Load EEG df

[17]:
```
# Define the file paths
base_dir = '/home/vincent/AAA_projects/MVCS/Neuroscience/DataFrames/'
eeg_df_path = base_dir + 'eeg_df.csv'

# load data
eeg_df = pd.read_csv(eeg_df_path)

print(eeg_df.head())
```

```
            Fp1           Fpz           Fp2          F7         F3  \
0 -21295.988649 -20109.716727 -24153.383752  3189.340060 -45.189275
1 -21303.747077 -20120.746154 -24163.864012  3178.880909 -56.702035
2 -21315.466571 -20130.126577 -24171.944343  3164.903807 -69.465350
3 -21317.809594 -20131.044726 -24174.790986  3159.478572 -73.214591
4 -21325.798142 -20137.522181 -24179.985166  3144.934679 -84.871628

            Fz          F4          F8          FC5          FC1  …  \
0 -8525.066680 -642.128590  3487.913621  6324.956639  6503.012177  …
1 -8532.499649 -651.966372  3477.011771  6315.078704  6496.522520  …
2 -8544.315275 -663.772856  3463.194795  6302.391524  6483.178723  …
3 -8545.873916 -666.109249  3457.870782  6297.212341  6481.970244  …
4 -8551.164448 -671.761501  3450.466406  6283.925509  6477.045614  …

            P7           P3          Pz           P4          P8  \
0  3374.048029 -3617.197964 -611.584742 -1667.222644  7523.612085
1  3372.073657 -3621.118134 -617.022909 -1673.653480  7516.945510
2  3363.104066 -3632.122011 -627.957966 -1684.569981  7502.158816
3  3354.943617 -3639.476353 -633.425118 -1690.436299  7496.978015
4  3343.913673 -3645.950907 -639.939845 -1695.157414  7491.664259

           POz           O1           Oz           O2  Time
0 -9446.685389 -6091.788931 -1392.835634  3559.608191   0.0
1 -9451.045628 -6094.343708 -1395.070231  3556.367972   1.0
2 -9460.798474 -6104.626002 -1406.149675  3543.184223   2.0
3 -9468.273627 -6110.845490 -1413.911429  3536.078117   3.0
4 -9477.076964 -6117.640828 -1419.832213  3529.343349   4.0

[5 rows x 33 columns]
```

## 4 Welch's

```
<!-- Column 1 -->
<div style="flex: 1; margin-right: 10px;">
    <h2>Introduction</h2>
    <p>This section aims to elaborate on the computational methodology for the estimation of Po
```

```
    <h2>Objectives</h2>
    <ul>
        <li>Data Importation: Load EEG data from a .npy file.</li>
        <li>Frequency Analysis: Compute PSD for each EEG channel using Welch's method.</li>
        <li>Data Visualization: Plot the PSD data for each channel.</li>
    </ul>
    <h2>Mathematical Formulations</h2>
    <h3>Data Importation and Pre-Processing</h3>
    <p>The EEG data, represented as a NumPy array \( \text{eeg\_data\_array} \), is imported. I
    <h3>Welch's Method</h3>
    <p>Welch's method for computing PSD involves partitioning the time series into overlapping
    \[ P(f) = \frac{1}{N}\sum_{i=1}^{N}|\text{FFT}(x_{i})|^{2} \]
    <p>Here, \( N \) is the number of segments, \( x_i \) is the \( i \)-th segment, and FFT re
</div>
<!-- Column 2 -->
<div style="flex: 1; margin-left: 10px;">
    <h2>Computational Steps</h2>
    <p>A Python loop iterates through each channel to extract the EEG data, upon which Welch's
    <h2>Data Storage</h2>
    <p>The computed PSD is saved as a NumPy array for future analyses, serving as a dataset fo
    <h2>Visualization</h2>
    <p>The calculated PSD values are visualized using Matplotlib. A subplot is created for each
    <h2>Scientific Relevance</h2>
    <p>This methodology serves as a cornerstone for both neuroscientific research and clinical
    <h2>Summary</h2>
    <p>The computation of PSD using Welch's method and its subsequent visualization provides an
</div>
```

[3]:
```python
import numpy as np
from scipy import signal
import matplotlib.pyplot as plt

# Load EEG data from the .npy file
eeg_data_array = np.load('/home/vincent/AAA_projects/MVCS/Neuroscience/
 ↪eeg_data_with_channels.npy')

# Define the sampling frequency (if your data is sampled at 1 Hz, set fs to 1)
fs = 1000

# List of EEG channel names
eeg_channels = ['Fp1', 'Fpz', 'Fp2', 'F7', 'F3', 'Fz', 'F4', 'F8', 'FC5',␣
 ↪'FC1', 'FC2', 'FC6',
                'M1', 'T7', 'C3', 'Cz', 'C4', 'T8', 'M2', 'CP5', 'CP1', 'CP2',␣
 ↪'CP6',
                'P7', 'P3', 'Pz', 'P4', 'P8', 'POz', 'O1', 'Oz', 'O2']

# Initialize an empty dictionary to store PSD values for each channel
```

```python
psd_data = {}

# Loop through each EEG channel
for i, channel in enumerate(eeg_channels):
    # Select EEG data from the current channel
    eeg_data = eeg_data_array[:, i]

    # Use Welch's method to estimate the power spectral density
    frequencies, psd = signal.welch(eeg_data, fs, nperseg=1024)

    # Store the PSD values in the dictionary
    psd_data[channel] = psd

# Save the PSD data as a numpy array
save_path = '/home/vincent/AAA_projects/MVCS/Neuroscience/Analysis/Spectral␣
 ↪Analysis/psd_x.npy'
np.save(save_path, psd_data)

# Set up the subplot layout
num_channels = len(eeg_channels)
num_rows = (num_channels + 3) // 4
num_cols = min(num_channels, 4)

# Create a figure and axes for subplots
fig, axs = plt.subplots(num_rows, num_cols, figsize=(12, 2 * num_rows))
axs = axs.ravel()

# Loop through each EEG channel and plot the PSD
for i, channel in enumerate(eeg_channels):
    axs[i].semilogy(frequencies, psd_data[channel])
    axs[i].set_title(f'EEG Channel {channel}')
    axs[i].set_xlabel('Frequency [Hz]')
    axs[i].set_ylabel('PSD [V**2/Hz]')
    axs[i].grid(True)

# Adjust layout and display the plots
plt.tight_layout()
plt.show()
```

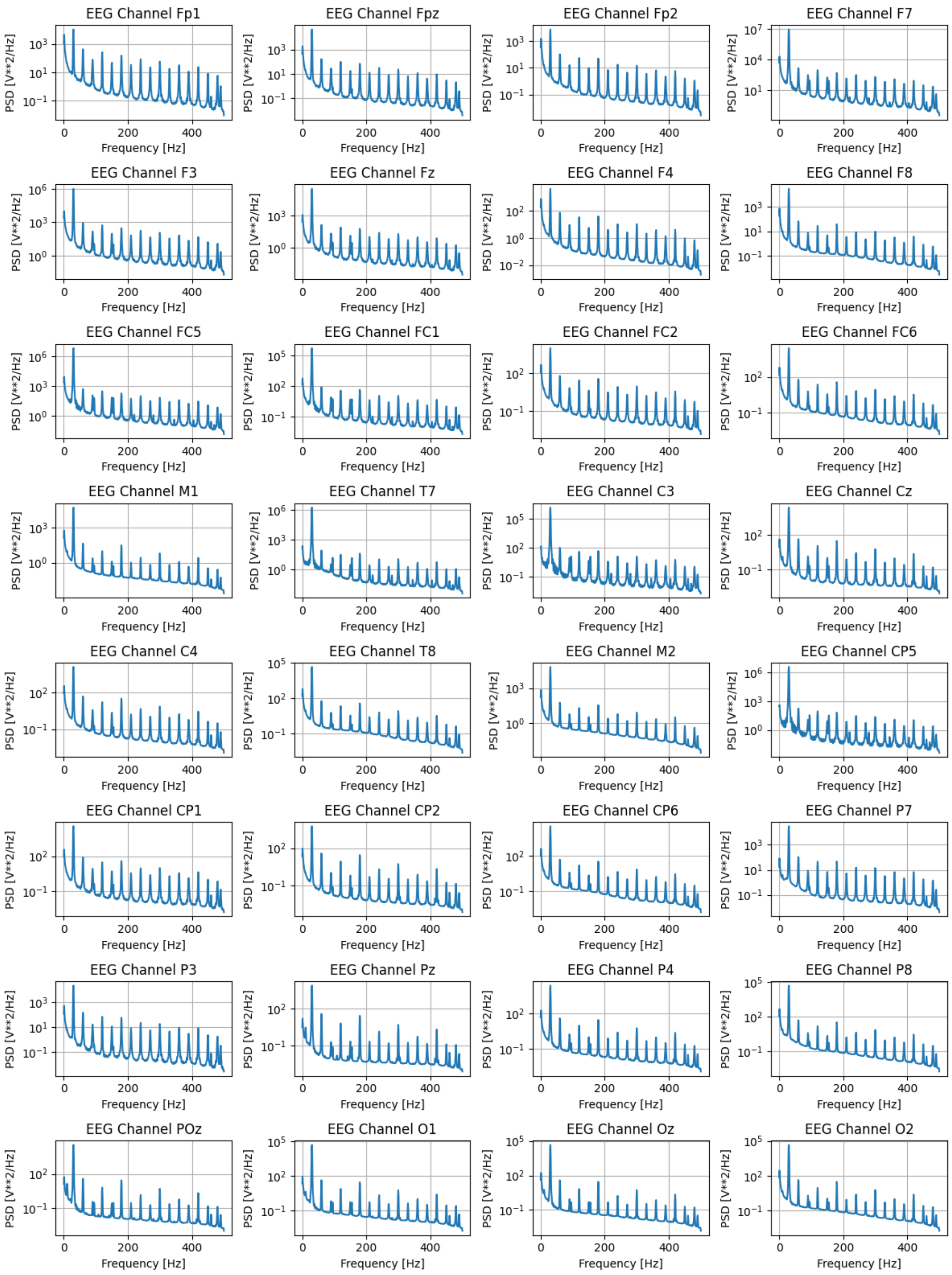
EEG Channel Fp1
EEG Channel Fpz
EEG Channel Fp2
EEG Channel F7
EEG Channel F3
EEG Channel Fz
EEG Channel F4
EEG Channel F8
EEG Channel FC5
EEG Channel FC1
EEG Channel FC2
EEG Channel FC6
EEG Channel M1
EEG Channel T7
EEG Channel C3
EEG Channel Cz
EEG Channel C4
EEG Channel T8
EEG Channel M2
EEG Channel CP5
EEG Channel CP1
EEG Channel CP2
EEG Channel CP6
EEG Channel P7
EEG Channel P3
EEG Channel Pz
EEG Channel P4
EEG Channel P8
EEG Channel POz
EEG Channel O1
EEG Channel Oz
EEG Channel O2
PSD [V**2/Hz]
Frequency [Hz]

## 5 Load and print head

[2]:
```
import numpy as np

# Load the Welch's PSD data from the .npy file
psd_data = np.load('/home/vincent/AAA_projects/MVCS/Neuroscience/Analysis/
↪Spectral Analysis/psd_x.npy', allow_pickle=True).item()

# List of EEG channel names
eeg_channels = ['Fp1', 'Fpz', 'Fp2', 'F7', 'F3', 'Fz', 'F4', 'F8', 'FC5',␣
↪'FC1', 'FC2', 'FC6',
                'M1', 'T7', 'C3', 'Cz', 'C4', 'T8', 'M2', 'CP5', 'CP1', 'CP2',␣
↪'CP6',
                'P7', 'P3', 'Pz', 'P4', 'P8', 'POz', 'O1', 'Oz', 'O2']

# Print the head of the PSD data for each EEG channel
for channel in eeg_channels:
    psd_values = psd_data[channel]
    print(f'PSD values for EEG Channel {channel}:')
    print(psd_values[:10])  # Print the first 10 values of the PSD
    print('---')
```

```
PSD values for EEG Channel Fp1:
[1072.02066427 4428.33217889 1543.99015765  577.69600889  290.86115692
  186.69274852  124.970801     91.75194501   62.20293001   46.19224688]
---
PSD values for EEG Channel Fpz:
[ 444.90385838 1865.3244754   648.89181366  255.5525128   133.10765723
   83.6980627    55.27084048   40.32642515   27.80385458   20.39475874]
---
PSD values for EEG Channel Fp2:
[ 333.5262614  1403.50793909  528.24463822  211.75468605  107.61870254
   66.26761707   42.72505454   30.04265205   20.27936186   15.0087366 ]
---
PSD values for EEG Channel F7:
[ 4878.25152685 18874.69226439  5833.27095836  2161.23425216
  1191.66521906   748.41463106   495.10034944   364.77747908
   279.13561261   214.83142922]
---
PSD values for EEG Channel F3:
[ 2509.54033435 10105.3429866   3314.16175662  1225.77412892
   622.80438379   405.28894718   275.99042426   203.92714345
   142.07347755   108.945467  ]
---
PSD values for EEG Channel Fz:
[ 297.47581738 1244.13377667  385.6118701   144.51805397   81.76160496
   50.02530596   31.95067227   24.15505077   18.44230388   13.03377181]
```

```
---
PSD values for EEG Channel F4:
[176.73096253 731.13777893 238.50203529  91.53819422  46.66810388
  29.84104647  20.37402009  15.22227935  10.70280091   8.33447964]
---
PSD values for EEG Channel F8:
[188.76658585 767.59452402 247.75981976  96.19172348  48.94478965
  31.32169547  21.52050114  16.08556157  11.43216504   9.27731769]
---
PSD values for EEG Channel FC5:
[2458.3084651  8707.06882538 2630.30658701  981.49076625  522.51794822
  325.72820993  224.9962309   166.52366093  123.06494223   97.34340134]
---
PSD values for EEG Channel FC1:
[144.47182764 555.29720023 148.72197459  56.71074799  37.69992652
  21.47584781  12.80622286   9.52620771   8.69010893   5.84782116]
---
PSD values for EEG Channel FC2:
[116.84792968 486.0006806  156.88806433  59.48057289  30.74239551
  19.58239113  13.30403521   9.99975193   7.11231026   5.46476961]
---
PSD values for EEG Channel FC6:
[144.7616294  588.1795107  187.79121953  72.42676779  36.81812743
  23.52734022  16.23767804  12.15802944   8.65852607   7.02364243]
---
PSD values for EEG Channel M1:
[136.89139907 551.42049081 166.70988957  68.46159821  36.03301737
  22.6389249   15.66737503  11.91542319   9.20874903   7.97715823]
---
PSD values for EEG Channel T7:
[152.88022764 238.00580916  45.52125274  20.43036073  13.11494922
   7.90046692   6.7935616    4.59297563   3.78641522   3.79423771]
---
PSD values for EEG Channel C3:
[102.37525786 134.48902829  22.61868327   7.52076947   9.22560564
   5.36580781   3.57964907   2.17179493   2.35926747   1.76591737]
---
PSD values for EEG Channel Cz:
[ 8.8376418  40.42025298 13.28663478  4.63371988  2.98094654  1.88764408
  1.2401903   0.93197365  0.77692211  0.63984419]
---
PSD values for EEG Channel C4:
[ 86.43668943 351.12214135 111.80120132  42.86586756  21.75669182
  13.88061143   9.65097465   7.21252091   5.12411969   4.10468538]
---
PSD values for EEG Channel T8:
[147.86581779 596.18207793 188.4931555   73.46163309  37.86982597
  24.11162997  16.67836266  12.56750819   9.09802388   7.62754119]
```

```
---
PSD values for EEG Channel M2:
[173.18227792 697.12122959 220.4449405   86.58366174  44.85169724
  28.38487984  19.66484951  14.80847084  10.8649406    9.26358054]
---
PSD values for EEG Channel CP5:
[323.68844731 399.83498148 112.42244271  37.39152978  22.04788905
  17.02283529  14.18471899   8.53641577   5.16455924   4.56082892]
---
PSD values for EEG Channel CP1:
[ 92.82861664 373.53303571 125.14199525  45.92398206  23.25646208
  15.18793218  10.47260981   7.88297981   5.38324088   4.13880043]
---
PSD values for EEG Channel CP2:
[23.30466422 94.30420689 29.9066385  11.46155202  5.85911803  3.73653734
  2.61439049  1.96138287  1.39356602  1.10499818]
---
PSD values for EEG Channel CP6:
[ 97.28194555 391.56225499 123.27828345  47.80232766  24.59102891
  15.67168953  10.90926027   8.20976345   5.97441997   4.94785413]
---
PSD values for EEG Channel P7:
[17.37139812 82.23355587 33.40137557 11.64305051  7.47773576  5.34274517
  3.61617408  2.98384541  2.49493626  2.43774747]
---
PSD values for EEG Channel P3:
[117.75426185 500.57319865 178.29185323  63.00260722  32.36111878
  21.85170502  14.84572473  11.08723381   7.59377323   5.80589311]
---
PSD values for EEG Channel Pz:
[ 3.2212812  15.72953085  5.28573922  1.9811799   1.27791227  0.8628285
  0.61971611  0.48028858  0.4322328   0.59262549]
---
PSD values for EEG Channel P4:
[ 47.42265523 189.48509029  59.11827916  23.0104012   12.03137985
   7.66512743   5.39665837   4.06574883   2.98565182   2.47874166]
---
PSD values for EEG Channel P8:
[110.57194067 443.44832744 139.33695806  54.79261512  28.43221011
  18.21772149  12.79558864   9.72726429   7.34210357   6.36277347]
---
PSD values for EEG Channel POz:
[13.59662914 53.29963465 16.4379902   6.65471633  4.10119859  2.77147849
  2.02570649  1.58108737  1.4096553   2.28596883]
---
PSD values for EEG Channel O1:
[22.42491074 87.24603953 25.47205561 11.11323541  6.5681992   4.21676014
  3.05033347  2.38994886  2.05242216  2.05415556]
```

```
---
PSD values for EEG Channel Oz:
[ 42.19853716 166.51565894  50.44065331  20.6988614   11.2759941
   7.17848112   5.08972594   3.88883796   3.05774413   2.82314384]
---
PSD values for EEG Channel O2:
[ 68.35354958 272.30310637  84.30797541  33.60402896  17.75144352
  11.28395169   7.92289041   6.03092331   4.55769147   3.87655636]
---
```

## 6 Create features for the freq bands from Welch's

[4]:
```
# Define the sampling frequency
# This depends on how your EEG data was collected.
# For example, if a data point was collected every second, fs would be 1.
fs = 1000

# List of EEG channel names
eeg_channels = ['Fp1', 'Fpz', 'Fp2', 'F7', 'F3', 'Fz', 'F4', 'F8', 'FC5',␣
↪'FC1', 'FC2', 'FC6',
                'M1', 'T7', 'C3', 'Cz', 'C4', 'T8', 'M2', 'CP5', 'CP1', 'CP2',␣
↪'CP6',
                'P7', 'P3', 'Pz', 'P4', 'P8', 'POz', 'O1', 'Oz', 'O2']

# Use Welch's method to estimate the power spectral density
frequencies, psd = signal.welch(eeg_data, fs, nperseg=1024)

# Define frequency bands of interest (you can adjust these according to your␣
↪requirements)
delta_band = (1, 4)    # Delta band (1-4 Hz)
theta_band = (4, 8)    # Theta band (4-8 Hz)
alpha_band = (8, 13)   # Alpha band (8-13 Hz)
beta_band = (13, 30)   # Beta band (13-30 Hz)

# Find indices corresponding to each frequency band
delta_indices = np.where((frequencies >= delta_band[0]) & (frequencies <=␣
↪delta_band[1]))[0]
theta_indices = np.where((frequencies >= theta_band[0]) & (frequencies <=␣
↪theta_band[1]))[0]
alpha_indices = np.where((frequencies >= alpha_band[0]) & (frequencies <=␣
↪alpha_band[1]))[0]
beta_indices = np.where((frequencies >= beta_band[0]) & (frequencies <=␣
↪beta_band[1]))[0]

# Extract power values for each frequency band
delta_power = np.nanmean(psd[delta_indices])
```

```python
theta_power = np.nanmean(psd[theta_indices])
alpha_power = np.nanmean(psd[alpha_indices])
beta_power = np.nanmean(psd[beta_indices])

# Create a feature array from the power values
features = np.array([delta_power, theta_power, alpha_power, beta_power])

# Save the features as a numpy array
save_path = '/home/vincent/AAA_projects/MVCS/Neuroscience/Analysis/Spectral Analysis/welchs_x.npy'
np.save(save_path, features)
```

[3]:
```python
# Load the saved features
loaded_features = np.load('/home/vincent/AAA_projects/MVCS/Neuroscience/Analysis/Spectral Analysis/welchs_x.npy')

# Print the head of the loaded features
print("Loaded Features:")
print(loaded_features[:5])  # Print the first 5 elements of the loaded features
```

```
Loaded Features:
[  45.2211493     7.44886422    3.62570384 1479.36679419]
```

# 7 Fast Fourier Transform

```
<!-- Column 1 -->
<div style="flex: 1; margin-right: 10px;">
    <h2>Introduction</h2>
    <p>This analysis is focused on computing the Power Spectral Density (PSD) of electroencepha
    <h2>Objectives</h2>
    <ul>
        <li>Data Selection: Extract the EEG data for individual channels.</li>
        <li>Fourier Analysis: Compute the PSD using FFT.</li>
        <li>Data Visualization: Render the computed PSD values.</li>
    </ul>
    <h2>Mathematical Formulations</h2>
    <h3>Data Selection and Parameter Definition</h3>
    <p>The EEG data, represented by the array \( \text{eeg\_data\_array} \), is segregated int
   <h3>FFT-based PSD Computation</h3>
    <p>For each EEG channel, the FFT is computed as per the following formula:</p>
    \[ \text{FFT}(x) = \sum_{n=0}^{N-1} x[n] \cdot e^{-j \cdot 2 \pi \cdot f \cdot n / N} \]
    <p>Subsequently, PSD is derived from the FFT values using:</p>
    \[ \text{PSD} = |\text{FFT}(x)|^2 \]
</div>
<!-- Column 2 -->
<div style="flex: 1; margin-left: 10px;">
    <h2>Computational Steps</h2>
```

```
    <p>A Python loop enumerates through each EEG channel, extracts the corresponding EEG data,
    <h2>Data Visualization</h2>
    <p>The frequency components of the PSD are plotted using Matplotlib. The frequency axis is
    <h2>Data Storage</h2>
    <p>The computed PSD values are saved as a single NumPy file for further analysis or scienti
    <h2>Scientific Relevance</h2>
    <p>The FFT-based approach to PSD analysis serves as a robust technique for investigating n
</div>
```

[5]:

```
import numpy as np
import matplotlib.pyplot as plt

# Define the sampling frequency and other parameters
fs = 1000

# List of EEG channel names
eeg_channels = ['Fp1', 'Fpz', 'Fp2', 'F7', 'F3', 'Fz', 'F4', 'F8', 'FC5',␣
 ↪'FC1', 'FC2', 'FC6',
                'M1', 'T7', 'C3', 'Cz', 'C4', 'T8', 'M2', 'CP5', 'CP1', 'CP2',␣
 ↪'CP6',
                'P7', 'P3', 'Pz', 'P4', 'P8', 'POz', 'O1', 'Oz', 'O2']

# Initialize an empty dictionary to store FFT PSD values for each channel
fft_psd_data = {}

# Loop through each EEG channel
for i, channel in enumerate(eeg_channels):
    # Select EEG data from the current channel
    eeg_data = eeg_data_array[:, i]

    # Compute the FFT
    fft_result = np.fft.fft(eeg_data)
    psd = np.abs(fft_result) ** 2

    # Store the PSD values in the dictionary
    fft_psd_data[channel] = psd

# Print the populated keys in fft_psd_data
print("Populated keys in fft_psd_data:", list(fft_psd_data.keys()))

# Plot the FFT PSD data for each channel
fig, axs = plt.subplots(len(eeg_channels), figsize=(10, 2 * len(eeg_channels)))
for i, channel in enumerate(eeg_channels):
    psd = fft_psd_data.get(channel)  # Use get method to handle missing keys␣
 ↪gracefully
    if psd is not None:
        frequencies = np.fft.fftfreq(len(psd), d=1/fs)
```

```
        axs[i].semilogy(frequencies, psd)
        axs[i].set_title(f'EEG Channel {channel}')
        axs[i].set_xlabel('Frequency [Hz]')
        axs[i].set_ylabel('PSD [V**2]')
        axs[i].grid(True)
    else:
        print(f"PSD data for channel {channel} is missing.")

plt.tight_layout()
plt.show()

# Save the FFT PSD data for all channels as a single numpy file
combined_fft_psd_data = {channel: psd for channel, psd in fft_psd_data.items()}
save_path = '/home/vincent/AAA_projects/MVCS/Neuroscience/Analysis/Spectral␣
 ↪Analysis/combined_fft_psd_x.npy'
np.save(save_path, combined_fft_psd_data)
```

```
Populated keys in fft_psd_data: ['Fp1', 'Fpz', 'Fp2', 'F7', 'F3', 'Fz', 'F4',
'F8', 'FC5', 'FC1', 'FC2', 'FC6', 'M1', 'T7', 'C3', 'Cz', 'C4', 'T8', 'M2',
'CP5', 'CP1', 'CP2', 'CP6', 'P7', 'P3', 'Pz', 'P4', 'P8', 'POz', 'O1', 'Oz',
'O2']
```

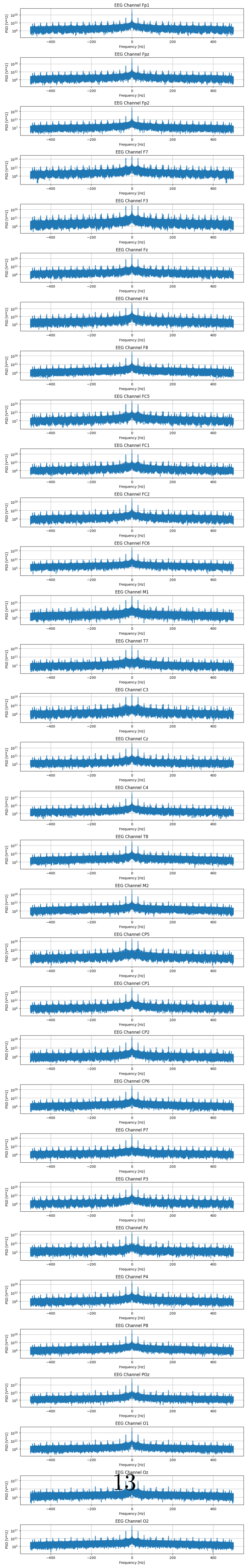
EEG Channel Fp1
EEG Channel Fpz
EEG Channel Fp2
EEG Channel F7
EEG Channel F3
EEG Channel Fz
EEG Channel F4
EEG Channel F8
EEG Channel FC5
EEG Channel FC1
EEG Channel FC2
EEG Channel FC6
EEG Channel M1
EEG Channel T7
EEG Channel C3
EEG Channel Cz
EEG Channel C4
EEG Channel T8
EEG Channel M2
EEG Channel CP5
EEG Channel CP1
EEG Channel CP2
EEG Channel CP6
EEG Channel P7
EEG Channel P3
EEG Channel Pz
EEG Channel P4
EEG Channel P8
EEG Channel POz
EEG Channel O1
EEG Channel Oz
EEG Channel O2
Frequency [Hz]

# 8 Load and print head

[4]:
```python
# Load the FFT PSD data from the .npy file
fft_psd_data = np.load('/home/vincent/AAA_projects/MVCS/Neuroscience/Analysis/
↪Spectral Analysis/combined_fft_psd_x.npy', allow_pickle=True).item()

# Print the first few elements of the FFT PSD data for each channel
for channel, psd in fft_psd_data.items():
    print(f"Channel: {channel}")
    print(psd[:10])  # Print the first 10 elements of the PSD data
    print("-" * 40)  # Separator
```

```
Channel: Fp1
[7.28963963e+21 1.24311775e+19 3.84437124e+18 1.50054982e+18
 5.47792867e+17 1.22997017e+17 5.55828228e+17 3.58984194e+17
 4.73666146e+16 1.35993709e+17]
----------------------------------------
Channel: Fpz
[6.03276852e+21 5.31820677e+18 2.39304919e+18 1.55151403e+18
 3.75985824e+17 5.43709124e+16 7.01198233e+17 5.02326335e+17
 6.26644662e+16 1.59748986e+17]
----------------------------------------
Channel: Fp2
[6.75803250e+21 4.50607226e+19 1.15541972e+19 4.88685804e+18
 1.91102436e+18 1.26603945e+18 1.60218132e+18 1.12637836e+18
 4.34645747e+17 4.74233918e+17]
----------------------------------------
Channel: F7
[6.06932155e+19 2.59919187e+18 9.29845007e+18 1.87094675e+19
 3.04965105e+18 7.23207327e+18 9.40525973e+18 8.22306392e+18
 6.53823302e+18 6.65212580e+18]
----------------------------------------
Channel: F3
[1.06698056e+18 2.00267103e+18 6.64187858e+17 2.11363472e+18
 6.20451925e+16 5.20391230e+17 3.10741423e+17 2.36849782e+17
 4.01051855e+17 3.24267807e+17]
----------------------------------------
Channel: Fz
[1.29548495e+21 4.29954454e+17 1.35873263e+18 1.09209438e+18
 9.13205038e+17 5.01543993e+17 1.05214445e+18 9.00357280e+17
 5.15648355e+17 6.07039430e+17]
----------------------------------------
Channel: F4
[3.19180109e+18 2.53784346e+17 2.63438592e+17 1.54676693e+17
```

```
 9.04127509e+16 1.01543697e+15 8.93948189e+16 8.29161885e+16
 2.16000482e+16 2.61236109e+16]
----------------------------------------
Channel: F8
[3.71630841e+20 8.12089949e+18 1.88015363e+18 3.99966540e+17
 2.56799456e+17 1.86046327e+17 9.25017951e+16 9.72696068e+16
 8.97819225e+16 6.75252080e+16]
----------------------------------------
Channel: FC5
[8.01801941e+20 5.98827554e+18 3.86036708e+18 5.78052590e+18
 9.89057848e+17 3.41231112e+18 2.65851358e+18 2.34696347e+18
 2.83712249e+18 2.54787149e+18]
----------------------------------------
Channel: FC1
[1.00331366e+21 3.02866816e+17 1.81875415e+18 1.65013497e+18
 6.72631936e+17 3.29134896e+17 1.20605756e+18 1.06048751e+18
 4.54974983e+17 5.41744507e+17]
----------------------------------------
Channel: FC2
[1.10395956e+20 9.50218231e+16 2.01060320e+17 1.37137467e+17
 8.59859512e+16 2.26378097e+16 1.12968243e+17 9.88586609e+16
 4.18786954e+16 4.43949641e+16]
----------------------------------------
Channel: FC6
[2.30445809e+21 3.72504076e+18 1.24024261e+18 3.99988634e+17
 1.25627228e+17 1.71040684e+17 1.35274681e+17 8.52719183e+16
 7.82397344e+16 5.54199102e+16]
----------------------------------------
Channel: M1
[1.25689818e+18 7.29888162e+18 1.79554144e+18 5.04867570e+17
 3.01672372e+17 3.51171296e+17 1.57253417e+17 8.05161386e+16
 1.93437633e+17 1.18460308e+17]
----------------------------------------
Channel: T7
[9.96641792e+20 1.13607556e+17 1.06195026e+17 7.38850117e+16
 1.18743793e+17 6.72766040e+16 1.11347069e+17 1.11143400e+17
 7.94165615e+16 5.43348370e+16]
----------------------------------------
Channel: C3
[3.54481287e+19 2.95742358e+18 2.17344931e+17 1.47828147e+17
 3.78939525e+17 4.75883948e+17 2.38773890e+17 2.61331744e+17
 4.46970770e+17 3.03154219e+17]
----------------------------------------
Channel: Cz
[1.75984477e+20 1.21951692e+17 1.76854423e+17 1.37627656e+17
 2.78574297e+15 9.84420142e+15 8.57813621e+16 7.31597704e+16
 2.18232793e+16 2.29752019e+16]
----------------------------------------
```

```
Channel: C4
[8.65397525e+20 7.88537721e+18 2.54580644e+18 9.09285213e+17
 3.84319267e+17 3.42975275e+17 2.84657399e+17 1.79185015e+17
 1.53304196e+17 1.09791044e+17]
----------------------------------------
Channel: T8
[1.02313765e+20 3.35678965e+18 1.02331541e+18 2.72972670e+17
 1.28882483e+17 1.92310798e+17 8.49066927e+16 5.02306670e+16
 1.01313062e+17 6.07990041e+16]
----------------------------------------
Channel: M2
[1.24615755e+21 3.57978844e+17 8.92129488e+16 3.24269698e+17
 1.60326050e+17 6.37774941e+15 1.87620946e+16 3.65086895e+16
 7.15602682e+15 1.60987026e+16]
----------------------------------------
Channel: CP5
[4.44347142e+19 1.56443278e+17 1.59341957e+17 3.53695952e+16
 1.47464385e+17 8.38206585e+16 9.75783929e+15 3.54283387e+15
 3.13303324e+16 2.91136192e+16]
----------------------------------------
Channel: CP1
[8.13190259e+20 6.26546844e+18 1.33238691e+18 5.83784963e+17
 2.39998381e+17 2.86645928e+17 1.50787071e+17 1.31965676e+17
 1.19237103e+17 6.16524115e+16]
----------------------------------------
Channel: CP2
[3.76003178e+19 2.85169872e+18 5.09087550e+17 3.61500422e+17
 1.72375162e+17 1.11094470e+17 5.90775709e+16 5.84490341e+16
 3.48196577e+16 2.35446596e+16]
----------------------------------------
Channel: CP6
[5.85303756e+20 1.12601054e+19 2.66979882e+18 9.41144740e+17
 4.99132715e+17 4.81360678e+17 2.49762388e+17 1.74729174e+17
 2.22049419e+17 1.52434783e+17]
----------------------------------------
Channel: P7
[3.09501624e+20 2.63831128e+18 5.79464459e+17 1.45743437e+17
 3.28205678e+17 1.82002337e+17 3.08678840e+16 2.72208251e+16
 1.15096096e+17 7.35592669e+16]
----------------------------------------
Channel: P3
[2.80544230e+20 1.55093254e+17 1.00097345e+16 1.98440257e+16
 2.36058325e+17 3.84372472e+16 4.39416863e+16 3.69808423e+16
 2.05801854e+16 2.27666518e+16]
----------------------------------------
Channel: Pz
[2.74913963e+18 6.65479851e+16 3.95267136e+16 5.89233868e+15
 5.20579580e+16 3.17393835e+16 6.90693979e+15 1.31853984e+16
```

```
 1.32956993e+16 1.27224258e+16]
----------------------------------------
Channel: P4
[1.24694770e+20 2.65868577e+18 5.01330451e+17 4.35001801e+17
 1.99497914e+17 8.26140049e+16 1.24836296e+17 7.17139376e+16
 1.33478141e+16 3.54356674e+16]
----------------------------------------
Channel: P8
[1.45906930e+21 4.81951286e+18 1.14270849e+18 2.69705591e+17
 1.62336190e+17 2.41541214e+17 8.22767829e+16 5.17348063e+16
 1.21770162e+17 8.18997705e+16]
----------------------------------------
Channel: POz
[2.39176255e+21 6.64291972e+18 1.56337322e+18 1.09587912e+18
 4.02602958e+17 2.10669588e+17 2.94241489e+17 2.36556257e+17
 5.94616169e+16 7.00801554e+16]
----------------------------------------
Channel: O1
[2.28139459e+21 1.01074832e+19 3.38564603e+18 2.97685312e+18
 1.24146581e+18 6.72544824e+17 7.83486481e+17 6.17304489e+17
 2.24864516e+17 2.38099398e+17]
----------------------------------------
Channel: Oz
[3.42693196e+19 1.74320020e+16 5.02354137e+16 4.98058409e+16
 2.39832001e+16 4.62496589e+16 1.87846383e+16 1.13626097e+16
 1.33305117e+16 2.01634914e+16]
----------------------------------------
Channel: O2
[2.76234894e+20 3.14788522e+17 1.16300770e+17 4.14995261e+16
 1.50799031e+16 4.47715022e+16 7.43330903e+15 5.77852489e+15
 2.07846592e+16 1.95979935e+16]
----------------------------------------
```

## 9 Lomb-Scargle periodogram

[ ]:
```python
from scipy.signal import lombscargle
import numpy as np
import matplotlib.pyplot as plt

# Define the sampling frequency and other parameters
fs = 1000

# List of EEG channel names
eeg_channels = ['Fp1']

# Initialize a dictionary to store Lomb-Scargle periodograms for each channel
lomb_scargle_data = {}
```

```python
# Loop through each EEG channel
for channel in eeg_channels:
    channel_index = eeg_channels.index(channel)
    eeg_data = eeg_data_array[:, channel_index]

    # Compute Lomb-Scargle periodogram
    time = np.arange(len(eeg_data)) / fs
    frequencies = np.linspace(0.001, 500, 1000)
    periodogram = lombscargle(time, eeg_data, frequencies, normalize=True)

    # Store the periodogram in the dictionary
    lomb_scargle_data[channel] = periodogram

# Save the Lomb-Scargle periodogram data for all channels as a single numpy file
save_path = '/home/vincent/AAA_projects/MVCS/Neuroscience/Analysis/Spectral␣
↪Analysis/LombScarglePeriodograms_x.npy'
np.save(save_path, lomb_scargle_data)

# Set up plot layout parameters
num_channels = len(eeg_channels)
num_columns = 4
num_rows = -(-num_channels // num_columns)  # Ceiling division

# Compute plot size based on your requirements
fig_width = 20  # Adjust as needed for the width
fig_height = num_rows * 3  # Each plot should be 3 times taller

# Create subplots in a 4-column layout
fig, axes = plt.subplots(num_rows, num_columns, figsize=(fig_width,␣
↪fig_height), sharex=True)

# Flatten the axes array for easy indexing
axes = axes.flatten()

# Loop through each EEG channel and plot the Lomb-Scargle periodograms
for i, channel in enumerate(eeg_channels):
    periodogram = lomb_scargle_data[channel]
    frequencies = np.linspace(0.001, 500, 1000)

    # Plot the Lomb-Scargle periodogram on the appropriate subplot
    ax = axes[i]
    ax.plot(frequencies, periodogram)
    ax.set_title(f'EEG Channel {channel}')
    ax.set_ylabel('Power Spectral Density')
    ax.grid(True)
```

```
# Set common xlabel for the last row of subplots
for ax in axes[-num_columns:]:
    ax.set_xlabel('Frequency [Hz]')

# Adjust layout and display the plot
plt.tight_layout()
plt.show()
```

## 10 Load and print head

[7]:
```
# Load the Lomb-Scargle periodogram data from the .npy file
lomb_scargle_data = np.load('/home/vincent/AAA_projects/MVCS/Neuroscience/
↪Analysis/Spectral Analysis/LombScarglePeriodograms_x.npy',␣
↪allow_pickle=True).item()

# Print the first few elements of the Lomb-Scargle periodogram data for each␣
↪channel
for channel, periodogram in lomb_scargle_data.items():
    print(f"Channel: {channel}")
    print(periodogram[:10])  # Print the first 10 elements of the periodogram␣
↪data
    print("-" * 40)  # Separator
```

```
Channel: Fp1
[0.1610684  0.16106555 0.16105702 0.16104282 0.16102294 0.16099739
 0.16096616 0.16092926 0.16088668 0.16083843]
----------------------------------------
Channel: Fpz
[0.16152559 0.16152273 0.16151419 0.16149996 0.16148004 0.16145443
 0.16142313 0.16138615 0.16134347 0.16129511]
----------------------------------------
Channel: Fp2
[0.1621703  0.16216743 0.16215885 0.16214456 0.16212456 0.16209885
 0.16206743 0.1620303  0.16198746 0.16193891]
----------------------------------------
Channel: F7
[0.16233026 0.16232739 0.16231879 0.16230448 0.16228445 0.1622587
 0.16222723 0.16219004 0.16214714 0.16209851]
----------------------------------------
Channel: F3
[0.16274206 0.16273918 0.16273057 0.16271622 0.16269614 0.16267033
 0.16263879 0.16260151 0.16255849 0.16250974]
----------------------------------------
Channel: Fz
[0.16225916 0.16225628 0.16224769 0.16223338 0.16221336 0.16218761
 0.16215614 0.16211895 0.16207605 0.16202742]
```

```
----------------------------------------
Channel: F4
[0.16230872 0.16230584 0.16229725 0.16228295 0.16226292 0.16223717
 0.1622057  0.16216852 0.16212561 0.16207699]
----------------------------------------
Channel: F8
[0.16231156 0.16230868 0.16230009 0.16228578 0.16226574 0.16223999
 0.16220852 0.16217132 0.1621284  0.16207977]
----------------------------------------
Channel: FC5
[0.16241608 0.16241321 0.16240462 0.1623903  0.16237026 0.16234451
 0.16231302 0.16227582 0.1622329  0.16218425]
----------------------------------------
Channel: FC1
[0.16269312 0.16269025 0.16268164 0.16266732 0.16264727 0.16262149
 0.16258998 0.16255275 0.16250979 0.16246111]
----------------------------------------
Channel: FC2
[0.16221679 0.16221392 0.16220534 0.16219104 0.16217104 0.16214532
 0.16211388 0.16207674 0.16203388 0.16198531]
----------------------------------------
Channel: FC6
[0.16214912 0.16214625 0.16213767 0.16212337 0.16210336 0.16207764
 0.1620462  0.16200905 0.16196618 0.1619176 ]
----------------------------------------
Channel: M1
[0.16221961 0.16221675 0.16220817 0.16219388 0.16217387 0.16214816
 0.16211673 0.16207959 0.16203675 0.16198819]
----------------------------------------
Channel: T7
[0.16170615 0.16170329 0.16169474 0.1616805  0.16166056 0.16163492
 0.1616036  0.16156658 0.16152387 0.16147546]
----------------------------------------
Channel: C3
[0.16119543 0.16119258 0.16118404 0.16116982 0.16114993 0.16112435
 0.16109308 0.16105614 0.16101352 0.16096521]
----------------------------------------
Channel: Cz
[0.16147791 0.16147506 0.16146652 0.1614523  0.16143239 0.1614068
 0.16137553 0.16133857 0.16129593 0.1612476 ]
----------------------------------------
Channel: C4
[0.16105413 0.16105129 0.16104277 0.16102858 0.16100872 0.16098319
 0.16095199 0.16091512 0.16087258 0.16082436]
----------------------------------------
Channel: T8
[0.16078463 0.16078179 0.16077327 0.16075909 0.16073924 0.16071372
 0.16068253 0.16064568 0.16060315 0.16055496]
```

```
----------------------------------------
Channel: M2
[0.1613771  0.16137424 0.1613657  0.16135146 0.16133154 0.16130593
 0.16127462 0.16123763 0.16119495 0.16114659]
----------------------------------------
Channel: CP5
[0.16184422 0.16184135 0.16183278 0.16181851 0.16179853 0.16177284
 0.16174145 0.16170435 0.16166154 0.16161304]
----------------------------------------
Channel: CP1
[0.16210686 0.16210399 0.1620954  0.16208111 0.16206109 0.16203536
 0.16200392 0.16196676 0.16192389 0.1618753 ]
----------------------------------------
Channel: CP2
[0.16252666 0.16252379 0.16251519 0.16250086 0.16248081 0.16245503
 0.16242353 0.1623863  0.16234335 0.16229467]
----------------------------------------
Channel: CP6
[0.16222997 0.16222711 0.16221853 0.16220423 0.16218422 0.1621585
 0.16212707 0.16208992 0.16204706 0.16199849]
----------------------------------------
Channel: P7
[0.16263821 0.16263534 0.16262674 0.16261242 0.16259238 0.16256661
 0.16253512 0.16249791 0.16245497 0.16240631]
----------------------------------------
Channel: P3
[0.16205551 0.16205265 0.16204407 0.16202978 0.16200979 0.16198408
 0.16195266 0.16191554 0.1618727  0.16182416]
----------------------------------------
Channel: Pz
[0.16216408 0.16216121 0.16215263 0.16213834 0.16211833 0.16209261
 0.16206118 0.16202403 0.16198117 0.1619326 ]
----------------------------------------
Channel: P4
[0.16247187 0.162469   0.16246041 0.1624461  0.16242607 0.16240032
 0.16236885 0.16233167 0.16228877 0.16224014]
----------------------------------------
Channel: P8
[0.16216478 0.16216191 0.16215333 0.16213904 0.16211903 0.16209331
 0.16206188 0.16202474 0.16198188 0.16193331]
----------------------------------------
Channel: POz
[0.16217667 0.16217381 0.16216523 0.16215094 0.16213094 0.16210523
 0.1620738  0.16203667 0.16199383 0.16194527]
----------------------------------------
Channel: O1
[0.16145573 0.16145287 0.16144433 0.1614301  0.16141018 0.16138457
 0.16135328 0.16131629 0.16127362 0.16122527]
```

```
----------------------------------------
Channel: Oz
[0.16124146 0.16123861 0.16123008 0.16121586 0.16119597 0.1611704
 0.16113914 0.16110221 0.1610596  0.1610113 ]
----------------------------------------
Channel: O2
[0.16067567 0.16067282 0.16066431 0.16065014 0.1606303  0.1606048
 0.16057363 0.1605368  0.1604943  0.16044614]
----------------------------------------
```

## 11 Wavelet Transform

```
<!-- Column 1 -->
<div style="flex: 1; margin-right: 10px;">
    <h2>Introduction</h2>
    <p>This scholarly overview elucidates the computational pipeline for frequency decompositio
    <h2>Objectives</h2>
    <ul>
        <li>Channel-based Data Selection: Individual EEG channels are parsed from the master da
        <li>Continuous Wavelet Transform: Application of the wavelet transform to the EEG data
        <li>Power Spectral Density Estimation: Deriving PSD from the wavelet coefficients.</li>
        <li>Visualization: Time-frequency representation of the wavelet-based PSD.</li>
    </ul>
    <h2>Mathematical Foundations</h2>
    <h3>Wavelet Transform</h3>
    <p>The Continuous Wavelet Transform (CWT) is mathematically formulated as:</p>
    \[ CWT(x, a, b) = \frac{1}{\sqrt{a}} \int x(t) \psi^*\left(\frac{t-b}{a}\right)dt \]
    <p>Where \( \psi(t) \) is the mother wavelet and \( a \) and \( b \) are the scale and tra
    <h3>Power Spectral Density (PSD)</h3>
    <p>PSD is derived from the wavelet coefficients as:</p>
    \[ \text{PSD} = | \text{coefficients} |^2 \]
</div>
<!-- Column 2 -->
<div style="flex: 1; margin-left: 10px;">
    <h2>Computational Steps</h2>
    <p>Each EEG channel undergoes the following computational steps:</p>
    <ul>
        <li>EEG data for the channel is extracted.</li>
        <li>Wavelet transform is applied to the EEG data using the Morlet wavelet.</li>
        <li>PSD is calculated from the absolute square of the wavelet coefficients.</li>
    </ul>
    <h2>Data Visualization</h2>
    <p>Frequency decomposition is visualized as an image plot. The x-axis represents time, whil
    <h2>Data Serialization</h2>
    <p>PSD values are saved as a NumPy file, adhering to a specified directory path. This ensu
    <h2>Scientific Implications</h2>
    <p>Wavelet-based methods for time-frequency analysis offer finer granularity and are partic
```

</div>

[ ]:
```
import numpy as np
import pywt
import matplotlib.pyplot as plt

# Define the sampling frequency and other parameters
fs = 1000

# List of EEG channel names
eeg_channels = ['Fp1', 'Fpz', 'Fp2', 'F7', 'F3', 'Fz', 'F4', 'F8', 'FC5',␣
 ↪'FC1', 'FC2', 'FC6',
                'M1', 'T7', 'C3', 'Cz', 'C4', 'T8', 'M2', 'CP5', 'CP1', 'CP2',␣
 ↪'CP6',
                'P7', 'P3', 'Pz', 'P4', 'P8', 'POz', 'O1', 'Oz', 'O2']

# Initialize a dictionary to store wavelet transform PSD values for each channel
wavelet_psd_data = {}

# Loop through each EEG channel
for channel in eeg_channels:
    channel_index = eeg_channels.index(channel)
    eeg_data = eeg_data_array[:, channel_index]

    frequencies = np.logspace(np.log10(0.1), np.log10(30), num=100)
    coefficients, _ = pywt.cwt(eeg_data, frequencies, wavelet='morl')
    psd = np.abs(coefficients)**2

    # Store the PSD values in the dictionary
    wavelet_psd_data[channel] = psd

# Save the wavelet transform PSD data for all channels as a single numpy file
save_path = '/home/vincent/AAA_projects/MVCS/Neuroscience/Analysis/Spectral␣
 ↪Analysis/wavelet_psd_data_x.npy'
np.savez(save_path, **wavelet_psd_data)

# Loop through each EEG channel and plot the wavelet transform PSDs
for channel in eeg_channels:
    psd = wavelet_psd_data[channel]
    frequencies = np.logspace(np.log10(0.1), np.log10(30), num=100)

    # Plot the wavelet transform PSD for the current channel
    plt.imshow(psd, extent=[0, len(eeg_data_array), frequencies[-1],␣
 ↪frequencies[0]],
               aspect='auto', cmap='inferno')
    plt.title(f'EEG Channel {channel}')
    plt.xlabel('Time [s]')
```

```python
    plt.ylabel('Frequency [Hz]')
    plt.grid(False)
    plt.show()
```

## 12 Load and print head

[ ]:
```python
# Load the wavelet transform PSD data from the .npz file
wavelet_psd_data = np.load('/home/vincent/AAA_projects/MVCS/Neuroscience/
 ↪Analysis/Spectral Analysis/wavelet_psd_data_x.npy', allow_pickle=True)

# Print the populated keys in wavelet_psd_data
print("Populated keys in wavelet_psd_data:", list(wavelet_psd_data.keys()))

# Print the first few elements of the wavelet transform PSD data for each␣
 ↪channel
for channel, psd in wavelet_psd_data.items():
    print(f"Channel: {channel}")
    print(psd[:10])  # Print the first 10 elements of the PSD data
    print("-" * 40)  # Separator
```

## 13 Autocorrelation Function (ACF) and Partial Autocorrelation Function (PACF)

```
<!-- Column 1 -->
<div style="flex: 1; margin-right: 10px;">
    <h2>Introduction</h2>
    <p>We present an advanced analytical framework for evaluating the AutoCorrelation Function
    <h2>Objectives</h2>
    <ul>
        <li>Channel-wise EEG Data Extraction</li>
        <li>ACF and PACF Calculation</li>
        <li>Data Serialization</li>
    </ul>
    <h2>Mathematical Background</h2>
    <h3>AutoCorrelation Function (ACF)</h3>
    \[ \text{ACF}(\tau) = \frac{\sum_{t=1}^{T-\tau} (x_t - \bar{x})(x_{t+\tau} - \bar{x})}{\su
    <p>Where \( \tau \) is the time lag, \( x_t \) is the data point at time \( t \), and \( \
    <h3>Partial AutoCorrelation Function (PACF)</h3>
    <p>It isolates the correlation between a variable and a lagged version of itself that is n
</div>
<!-- Column 2 -->
<div style="flex: 1; margin-left: 10px;">
    <h2>Computational Procedures</h2>
    <ul>
        <li>EEG data corresponding to each channel is isolated.</li>
```

```
        <li>ACF and PACF are calculated for up to 10 lags.</li>
    </ul>
    <h2>Data Storage</h2>
    <p>The calculated ACF and PACF values are stored in a dictionary, which is then serialized
    <h2>Implications</h2>
    <p>Understanding ACF and PACF provides insights into the underlying periodic behaviors in E
    <h2>Conclusion</h2>
    <p>This comprehensive framework paves the way for a robust time-series analysis of EEG data
</div>
```

[ ]:
```python
import numpy as np
import matplotlib.pyplot as plt
from statsmodels.graphics.tsaplots import plot_acf, plot_pacf

# List of EEG channel names
eeg_channels = ['Fp1', 'Fpz', 'Fp2', 'F7', 'F3', 'Fz', 'F4', 'F8', 'FC5', 'FC1', 'FC2', 'FC6',
                'M1', 'T7', 'C3', 'Cz', 'C4', 'T8', 'M2', 'CP5', 'CP1', 'CP2', 'CP6',
                'P7', 'P3', 'Pz', 'P4', 'P8', 'POz', 'O1', 'Oz', 'O2']

# Initialize a dictionary to store ACF and PACF plots for each channel
acf_pacf_data = {}

# Loop through each EEG channel
for channel in eeg_channels:
    channel_index = eeg_channels.index(channel)
    eeg_data = eeg_data_array[:, channel_index]

    # Calculate ACF and PACF values
    acf_vals, _ = plot_acf(eeg_data, lags=10, show=False)
    pacf_vals, _ = plot_pacf(eeg_data, lags=10, show=False)

    # Store the ACF and PACF data in the dictionary
    acf_pacf_data[channel] = {'acf': acf_vals, 'pacf': pacf_vals}

# Save the ACF and PACF data for all channels as a single numpy file
save_path = '/home/vincent/AAA_projects/MVCS/Neuroscience/Analysis/Spectral Analysis/acf_pacf_data_x.npy'
np.save(save_path, acf_pacf_data)
```

# 14 Akaike Information Criterion (AIC) and Bayesian Information Criterion (BIC)

```
<!-- Column 1 -->
<div style="flex: 1; margin-right: 10px;">
    <h2>Introduction</h2>
```

```
    <p>This analytical framework is designed to employ the AutoRegressive (AR) model for scruti
    <h2>Objectives</h2>
    <ul>
        <li>Channel-wise extraction of EEG Data</li>
        <li>Modeling via AutoRegressive (AR) analysis</li>
        <li>Calculation of AIC and BIC for model optimization</li>
        <li>Serializing the derived metrics</li>
    </ul>
    <h2>Mathematical Premise</h2>
    <h3>AutoRegressive (AR) Model</h3>
    \[ X_t = c + \phi_1 X_{t-1} + \phi_2 X_{t-2} + \ldots + \phi_p X_{t-p} + \epsilon_t \]
    <p>Where \( X_t \) is the data at time \( t \), \( c \) is a constant, \( \phi \) are the m
    <h3>Akaike Information Criterion (AIC)</h3>
    \[ \text{AIC} = 2k - 2\ln(\hat{L}) \]
    <p>Where \( k \) is the number of parameters and \( \hat{L} \) is the maximum value of the
</div>
<!-- Column 2 -->
<div style="flex: 1; margin-left: 10px;">
    <h2>Computation Details</h2>
    <ul>
        <li>EEG data for each channel is individually processed.</li>
        <li>For each channel, AR models for lag values ranging from 1 to 20 are fit.</li>
        <li>AIC and BIC are computed for each fitted model, aiding in optimal lag selection.</l
    </ul>
    <h2>Data Serialization</h2>
    <p>The AIC and BIC values are preserved in a dictionary and serialized into a NumPy data fi
    <h2>Scientific Significance</h2>
    <p>Understanding the AIC and BIC values allows for the selection of an optimally parsimonio
    <h2>Conclusion</h2>
    <p>The analysis framework not only aids in optimal model selection for EEG data but also of
</div>
```

[ ]:

```python
import numpy as np
from statsmodels.tsa.ar_model import AutoReg

# Define the range of lag values to consider
lag_values = range(1, 21)  # e.g., consider lag values from 1 to 20

# List of EEG channel names
eeg_channels = ['Fp1', 'Fpz', 'Fp2', 'F7', 'F3', 'Fz', 'F4', 'F8', 'FC5', 'FC1', 'FC2', 'FC6',
                'M1', 'T7', 'C3', 'Cz', 'C4', 'T8', 'M2', 'CP5', 'CP1', 'CP2', 'CP6',
                'P7', 'P3', 'Pz', 'P4', 'P8', 'POz', 'O1', 'Oz', 'O2']

# Initialize dictionaries to store the AIC and BIC values for each EEG channel
aic_values_dict = {}
```

```
bic_values_dict = {}

# Loop through each EEG channel
for channel in eeg_channels:
    channel_index = eeg_channels.index(channel)
    eeg_data = eeg_data_array[:, channel_index]

    aic_values = []
    bic_values = []

    # Fit AutoReg models and calculate AIC and BIC values for each lag
    for lag in lag_values:
        mod = AutoReg(eeg_data, lags=lag)
        res = mod.fit()
        aic_values.append(res.aic)
        bic_values.append(res.bic)

    aic_values_dict[channel] = aic_values
    bic_values_dict[channel] = bic_values

# Save the AIC and BIC values for all EEG channels as a single numpy file
save_path = '/home/vincent/AAA_projects/MVCS/Neuroscience/Analysis/Spectral␣
↪Analysis/ar_model_aic_bic_x.npy'
np.savez(save_path, aic_values_dict=aic_values_dict,␣
↪bic_values_dict=bic_values_dict)
```

[ ]:
```
# Akaike Information Criterion
aic = res.aic
print("AIC: ", aic)

# Bayesian Information Criterion
bic = res.bic
print("BIC: ", bic)
```

[ ]:
```
import pandas as pd
import numpy as np
from statsmodels.tsa.ar_model import AutoReg

# List of EEG channel names
eeg_channels = ['Fp1', 'Fpz', 'Fp2', 'F7', 'F3', 'Fz', 'F4', 'F8', 'FC5',␣
↪'FC1', 'FC2', 'FC6',
                'M1', 'T7', 'C3', 'Cz', 'C4', 'T8', 'M2', 'CP5', 'CP1', 'CP2',␣
↪'CP6',
                'P7', 'P3', 'Pz', 'P4', 'P8', 'POz', 'O1', 'Oz', 'O2']

# Define a range of lag values to consider
lag_values = range(1, 21)  # e.g., consider lag values from 1 to 20
```

```
# Initialize dictionaries to store the AIC and BIC values for each EEG channel
aic_values_dict = {}
bic_values_dict = {}

# Split the data into training and testing sets (adjust num_train as needed)
num_train = int(0.8 * len(eeg_data_array[0]))  # Assuming each channel is a␣
↪column in the array
train_data = eeg_data_array[:, :num_train]

# Loop through each EEG channel
for channel_index, channel in enumerate(eeg_channels):
    # Select EEG data from the current channel
    eeg_data = train_data[channel_index]

    # Initialize lists to store the AIC and BIC values for the current channel
    aic_values = []
    bic_values = []

    # For each lag value, fit the AutoReg model and calculate AIC and BIC
    for lag in lag_values:
        mod = AutoReg(eeg_data, lags=lag, old_names=False)
        res = mod.fit()

        # Compute AIC and BIC
        aic = res.aic
        bic = res.bic

        aic_values.append(aic)
        bic_values.append(bic)

    # Store the AIC and BIC values in the dictionaries
    aic_values_dict[channel] = aic_values
    bic_values_dict[channel] = bic_values

# Save the AIC and BIC values for all EEG channels as a single numpy file
save_path = '/home/vincent/AAA_projects/MVCS/Neuroscience/Analysis/Spectral␣
↪Analysis/ar_model_aic_bic_allchannels_x.npy'
np.savez(save_path, aic_values_dict=aic_values_dict,␣
↪bic_values_dict=bic_values_dict)
```

[ ]:
```
# Choose the best lag based on the lowest AIC value
best_lag_aic = np.argmin(aic_values) + 1

# Choose the best lag based on the lowest BIC value
best_lag_bic = np.argmin(bic_values) + 1
```

```
print("Best lag based on AIC:", best_lag_aic)
print("Best lag based on BIC:", best_lag_bic)
```

[ ]:
```
import numpy as np
from statsmodels.tsa.ar_model import AutoReg

# List of EEG channel names
eeg_channels = ['Fp1', 'Fpz', 'Fp2', 'F7', 'F3', 'Fz', 'F4', 'F8', 'FC5', 'FC1', 'FC2', 'FC6',
                'M1', 'T7', 'C3', 'Cz', 'C4', 'T8', 'M2', 'CP5', 'CP1', 'CP2', 'CP6',
                'P7', 'P3', 'Pz', 'P4', 'P8', 'POz', 'O1', 'Oz', 'O2']

# Given sampling frequency (fs)
fs = 1000

# Set up the folder path to save the predicted values
save_folder_path = '/home/vincent/AAA_projects/MVCS/Neuroscience/Analysis/Spectral Analysis/'

# Load the EEG data from the numpy file
eeg_data_array = np.load('/home/vincent/AAA_projects/MVCS/Neuroscience/eeg_data_with_channels.npy')

# Initialize a dictionary to store the predicted values for each channel
predicted_values_dict = {}

# Loop through each EEG channel
for channel_index, eeg_channel in enumerate(eeg_channels):
    lag = 20
    eeg_data = eeg_data_array[:, channel_index]
    mod = AutoReg(eeg_data, lags=lag)
    res = mod.fit()
    periods_for_30_seconds = int(30 * fs)
    predictions = res.predict(start=len(eeg_data), end=len(eeg_data)+periods_for_30_seconds-1, dynamic=True)
    predicted_values_dict[eeg_channel] = predictions

# Save the predicted values for all EEG channels as a single numpy file
save_path = f'{save_folder_path}AutoRegressive_predicted_values_x.npy'
np.savez(save_path, **predicted_values_dict)
```

## 15 Band Powers

<!-- Column 1 -->
<div style="flex: 1; margin-right: 10px;">

```
        <h2>Introduction</h2>
        <p>The present analysis employs Continuous Wavelet Transform (CWT) using Morlet wavelets to
        <h2>Objectives</h2>
        <ul>
            <li>Decompose EEG signals into different frequency bands.</li>
            <li>Calculate the band power in each frequency range.</li>
            <li>Visualize the distribution of band powers over time.</li>
        </ul>
        <h2>Mathematical Foundations</h2>
        <h3>Continuous Wavelet Transform (CWT)</h3>
        \[ W(a,b) = \frac{1}{\sqrt{a}}\int x(t)\psi\left(\frac{t-b}{a}\right) dt \]
        <p>Where \( W(a,b) \) is the wavelet coefficient, \( a \) is the scale, \( b \) is the posi
        <h3>Power Spectral Density (PSD)</h3>
        \[ \text{PSD}(f) = |W(a,b)|^2 \]
        <p>Where \( f \) corresponds to frequency, and \( |W(a,b)|^2 \) represents the magnitude of
        <h3>Band Power Calculation</h3>
        \[ \text{Band Power} = \int \text{PSD}(f) df \]
        <p>The trapezoidal rule is used for numerical integration to find the band power in each fr
    </div>
    <!-- Column 2 -->
    <div style="flex: 1; margin-left: 10px;">
        <h2>Methodology</h2>
        <ul>
            <li>EEG data for each channel is separately analyzed.</li>
            <li>Wavelet transform coefficients are calculated using the Morlet wavelet.</li>
            <li>PSD is derived from the absolute square of the wavelet coefficients.</li>
            <li>Band powers are computed by integrating the PSD across each frequency band.</li>
        </ul>
        <h2>Data Serialization</h2>
        <p>All calculated band powers are stored in a dictionary and serialized into a single NumPy
        <h2>Scientific and Clinical Implications</h2>
        <p>Band powers serve as biomarkers in various clinical and experimental settings, enabling
        <h2>Conclusion</h2>
        <p>The analytical procedure provides a comprehensive method for the quantitative assessment
    </div>
```

[ ]:
```python
import numpy as np
import pywt
from concurrent.futures import ProcessPoolExecutor

# List of EEG channel names
eeg_channels = ['Fp1', 'Fpz', 'Fp2', 'F7', 'F3', 'Fz', 'F4', 'F8', 'FC5', 'FC1', 'FC2', 'FC6',
                'M1', 'T7', 'C3', 'Cz', 'C4', 'T8', 'M2', 'CP5', 'CP1', 'CP2', 'CP6',
                'P7', 'P3', 'Pz', 'P4', 'P8', 'POz', 'O1', 'Oz', 'O2']
```

```python
# Set up the folder path to save the results
results_folder_path = '/home/vincent/AAA_projects/MVCS/Neuroscience/Analysis/
↪Spectral Analysis/'

# Initialize a dictionary to store the band powers for each channel
band_powers_dict = {}

# Loop through each EEG channel
for channel_index, eeg_channel in enumerate(eeg_channels):
    eeg_data = eeg_data_array[:, channel_index]

    frequencies = np.logspace(np.log10(0.1), np.log10(30), num=100)
    coefficients, _ = pywt.cwt(eeg_data, frequencies, wavelet='morl')
    psd = np.abs(coefficients)**2

    delta_band = (0.1, 4)
    theta_band = (4, 8)
    alpha_band = (8, 13)
    beta_band = (13, 30)
    gamma_band = (30, 100)

    delta_indices = (frequencies >= delta_band[0]) & (frequencies <=␣
↪delta_band[1])
    theta_indices = (frequencies >= theta_band[0]) & (frequencies <=␣
↪theta_band[1])
    alpha_indices = (frequencies >= alpha_band[0]) & (frequencies <=␣
↪alpha_band[1])
    beta_indices = (frequencies >= beta_band[0]) & (frequencies <= beta_band[1])
    gamma_indices = (frequencies >= gamma_band[0]) & (frequencies <=␣
↪gamma_band[1])

    psd = psd.T

    delta_power = np.trapz(psd[:, delta_indices], axis=1)
    theta_power = np.trapz(psd[:, theta_indices], axis=1)
    alpha_power = np.trapz(psd[:, alpha_indices], axis=1)
    beta_power = np.trapz(psd[:, beta_indices], axis=1)
    gamma_power = np.trapz(psd[:, gamma_indices], axis=1)

    band_powers_dict[eeg_channel] = {
        "delta_power": delta_power,
        "theta_power": theta_power,
        "alpha_power": alpha_power,
        "beta_power": beta_power,
        "gamma_power": gamma_power
    }
```

```
# Save the band powers for all EEG channels as a single numpy file
results_file = "BandPowers_x.npy"
np.save(results_folder_path + results_file, band_powers_dict)

# List of frequency bands
frequency_bands = ['delta_power', 'theta_power', 'alpha_power', 'beta_power',␣
↪'gamma_power']

# Loop through each EEG channel and plot band powers
for channel, band_powers in band_powers_dict.items():
    plt.figure(figsize=(10, 6))

    for band in frequency_bands:
        plt.plot(band_powers[band], label=band)

    plt.xlabel('Time')
    plt.ylabel('Band Power')
    plt.title(f'Band Powers for EEG Channel {channel}')
    plt.legend()
    plt.grid(True)
    plt.show()
```

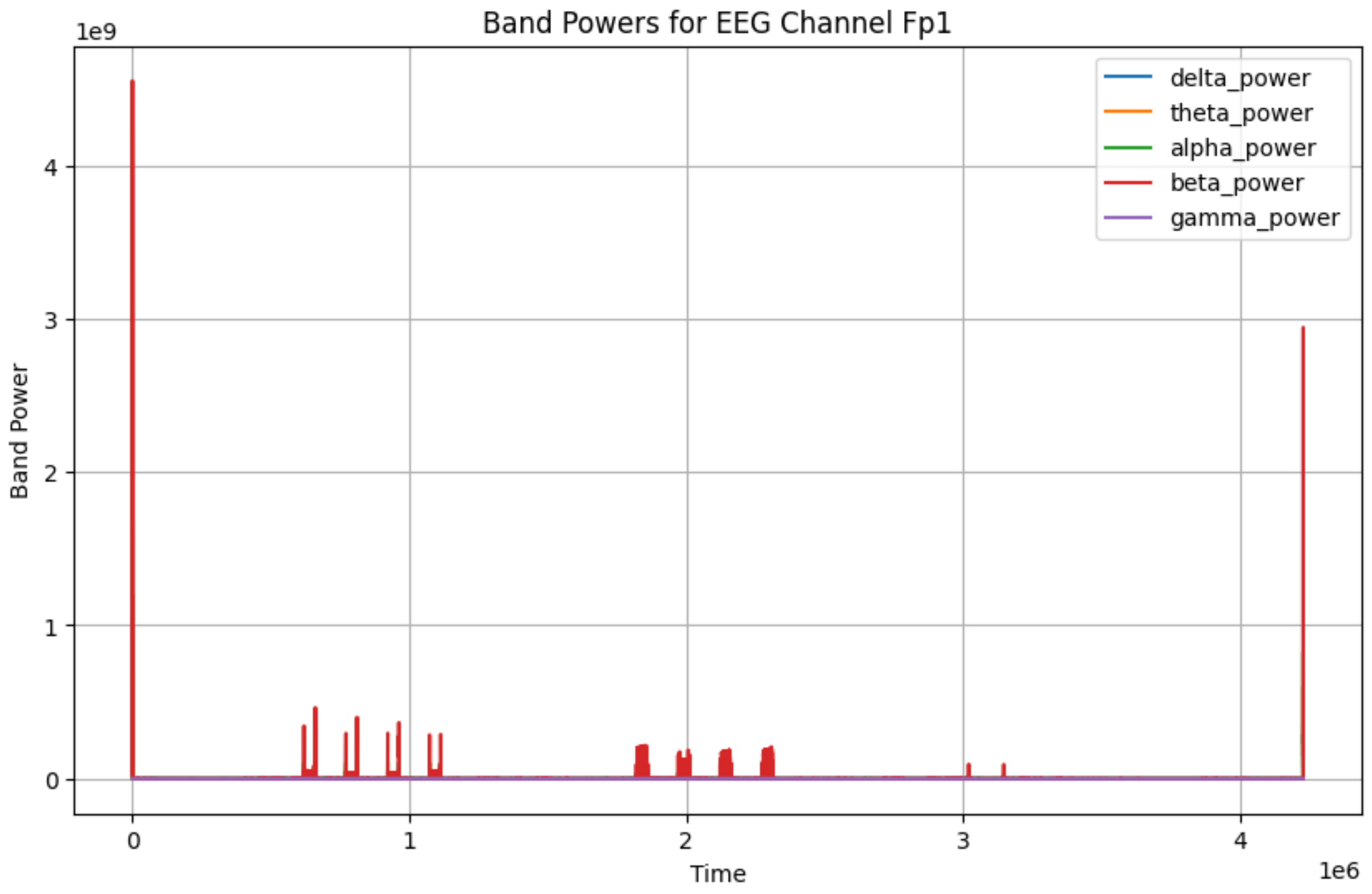

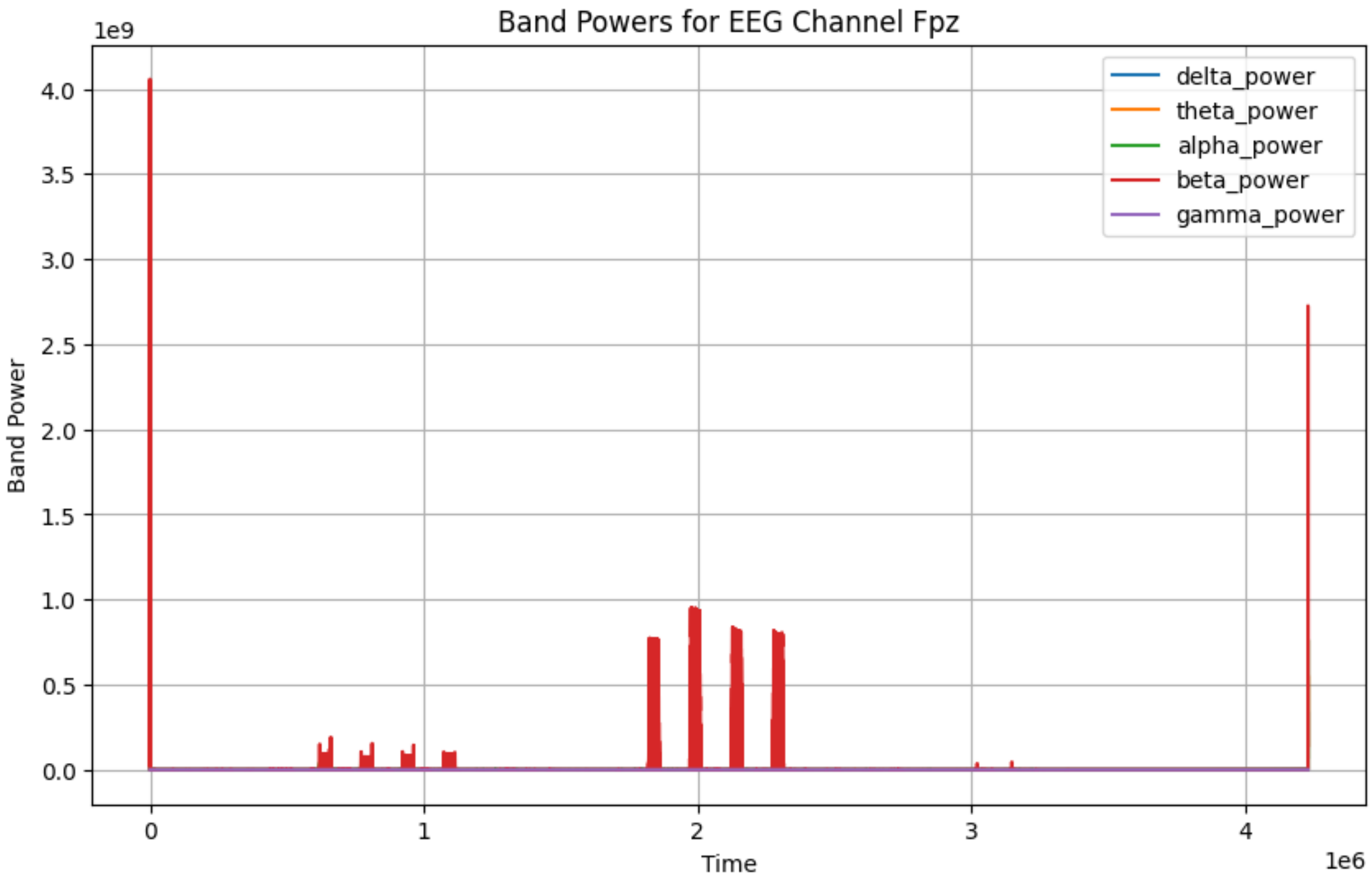

Band Powers for EEG Channel Fpz
1e9
4.0
3.5
3.0
2.5
2.0
1.5
1.0
0.5
0.0
Band Power
0
1
2
3
4
1e6
Time
delta_power
theta_power
alpha_power
beta_power
gamma_power

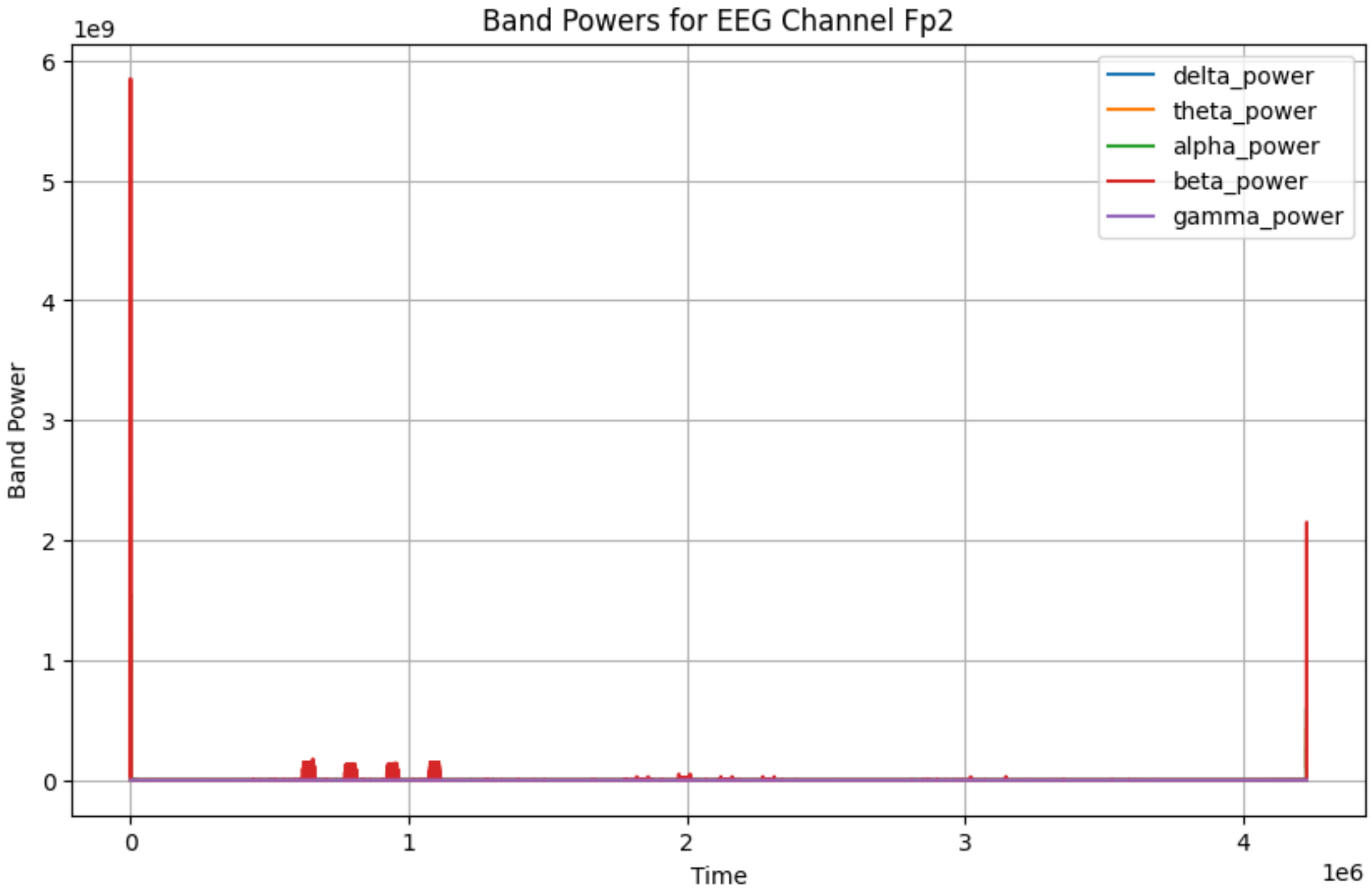

Band Powers for EEG Channel Fp2
1e9
6
5
4
3
2
1
0
Band Power
0
1
2
3
4
1e6
Time
delta_power
theta_power
alpha_power
beta_power
gamma_power

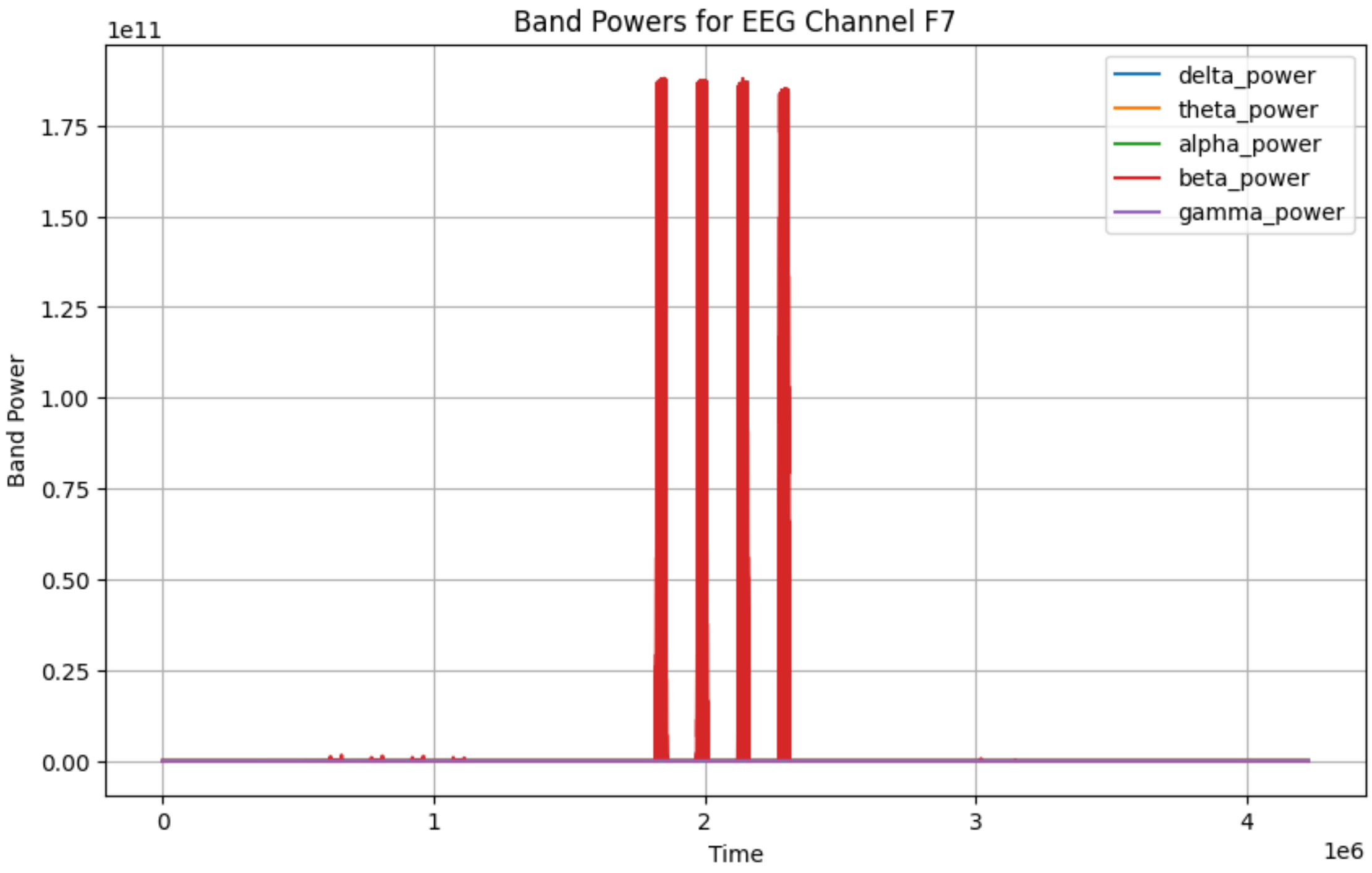
Band Powers for EEG Channel F7
1e11
1.75
1.50
1.25
1.00
0.75
0.50
0.25
0.00
Band Power
delta_power
theta_power
alpha_power
beta_power
gamma_power
0
1
2
3
4
Time
1e6

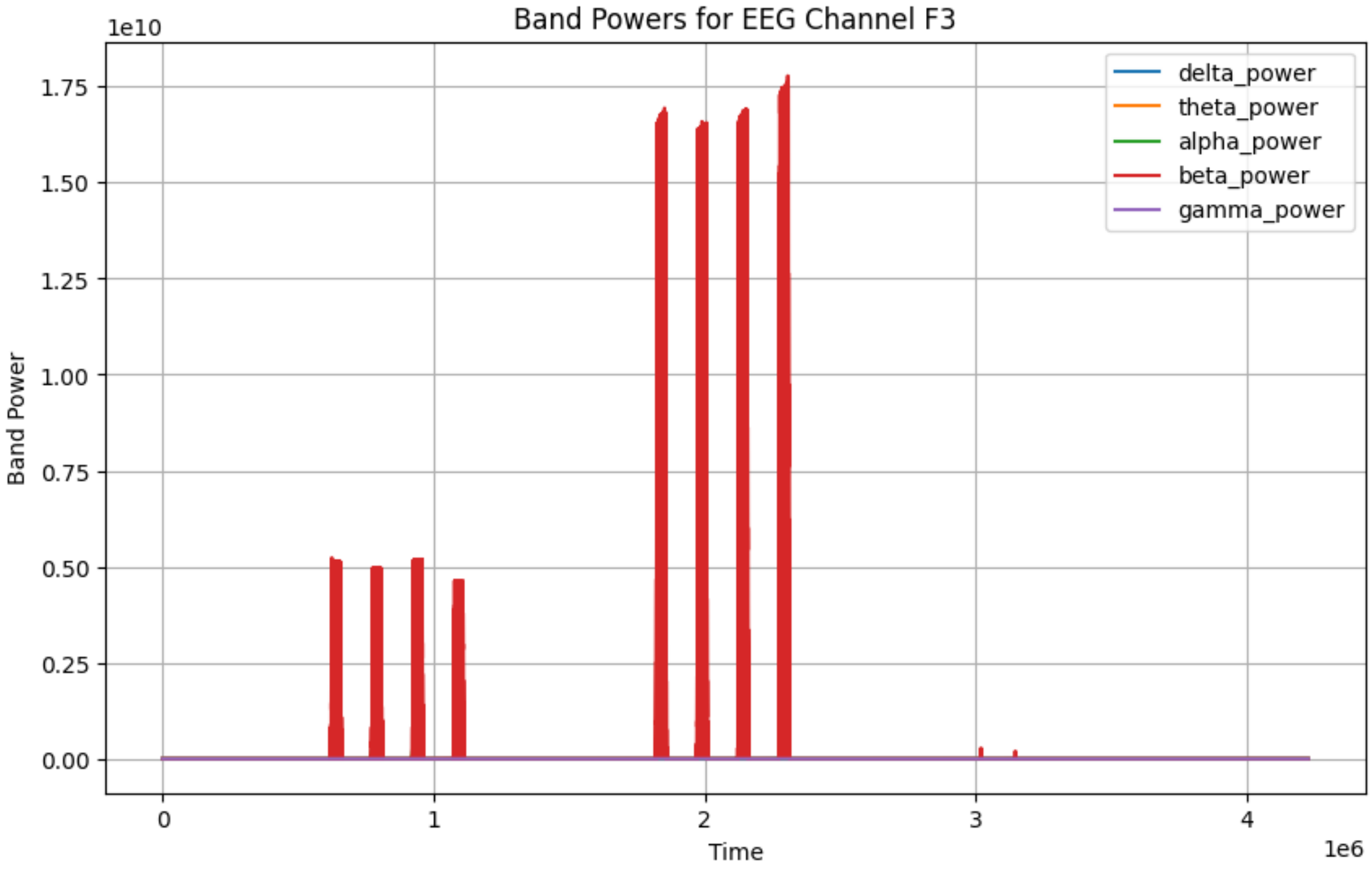
Band Powers for EEG Channel F3
1e10
1.75
1.50
1.25
1.00
0.75
0.50
0.25
0.00
Band Power
delta_power
theta_power
alpha_power
beta_power
gamma_power
0
1
2
3
4
Time
1e6

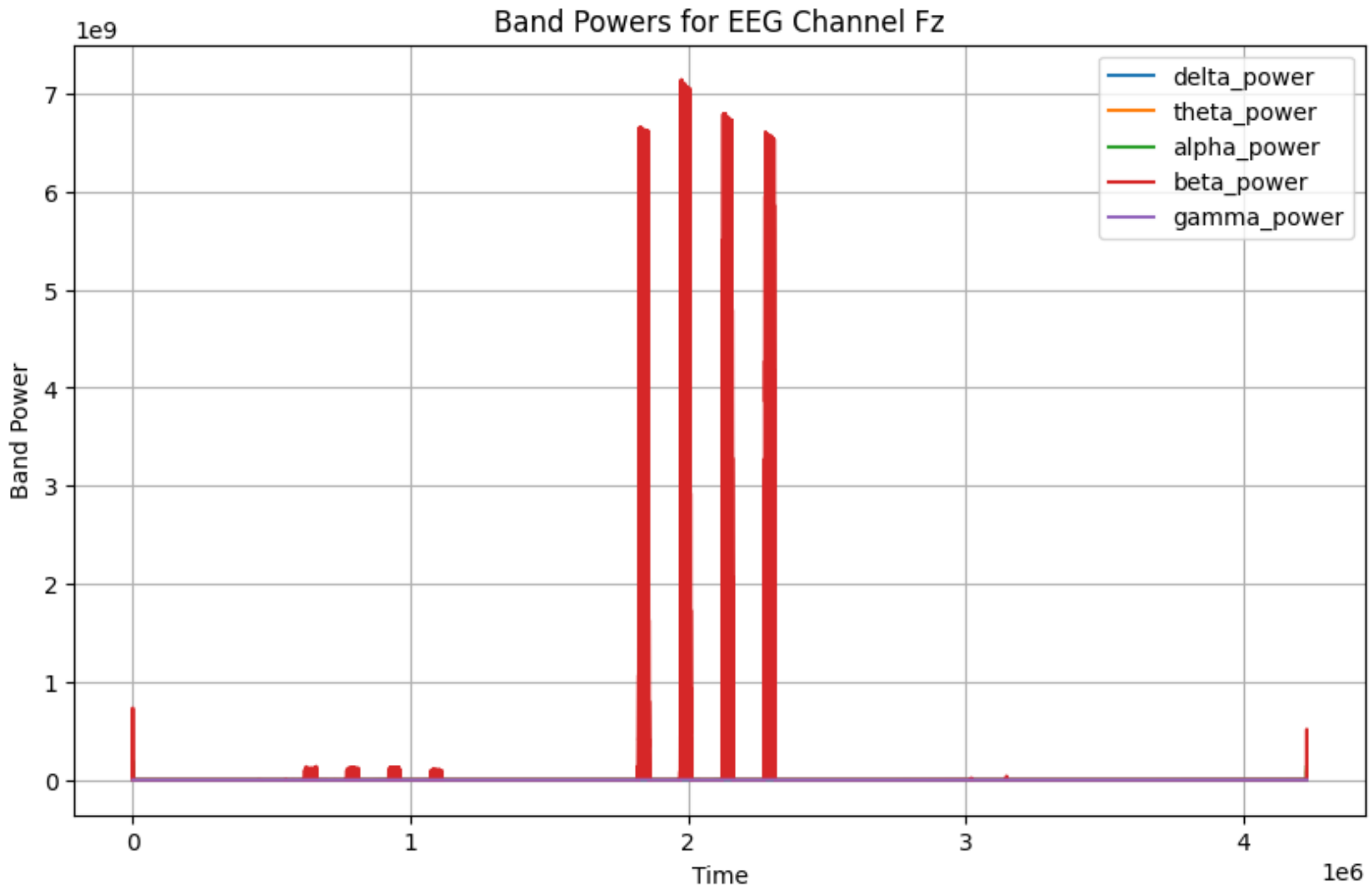
Band Powers for EEG Channel Fz
1e9
7
6
5
4
3
2
1
0
Band Power
0
1
2
3
4
Time
1e6
delta_power
theta_power
alpha_power
beta_power
gamma_power

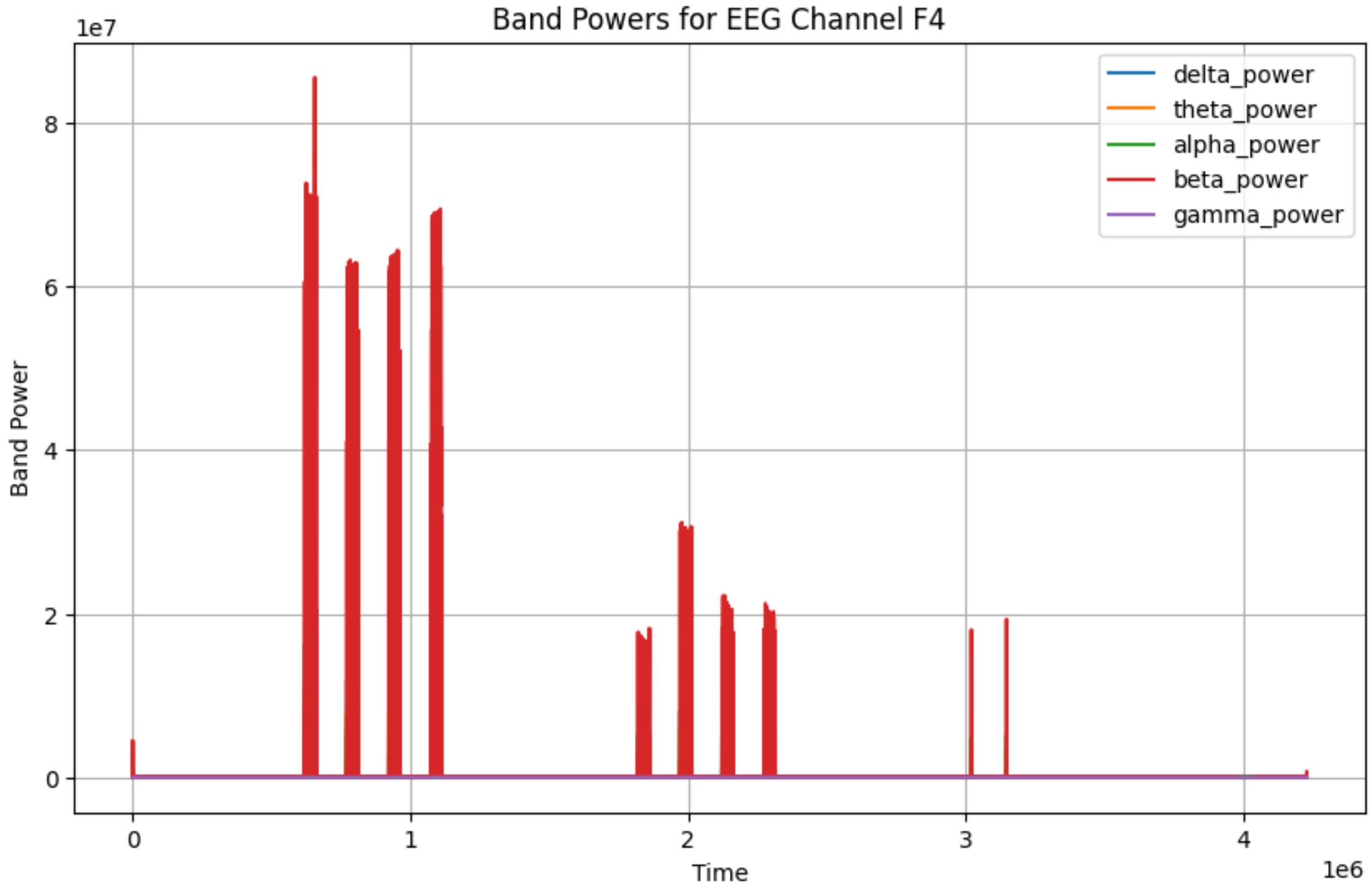
Band Powers for EEG Channel F4
1e7
8
6
4
2
0
Band Power
0
1
2
3
4
Time
1e6
delta_power
theta_power
alpha_power
beta_power
gamma_power

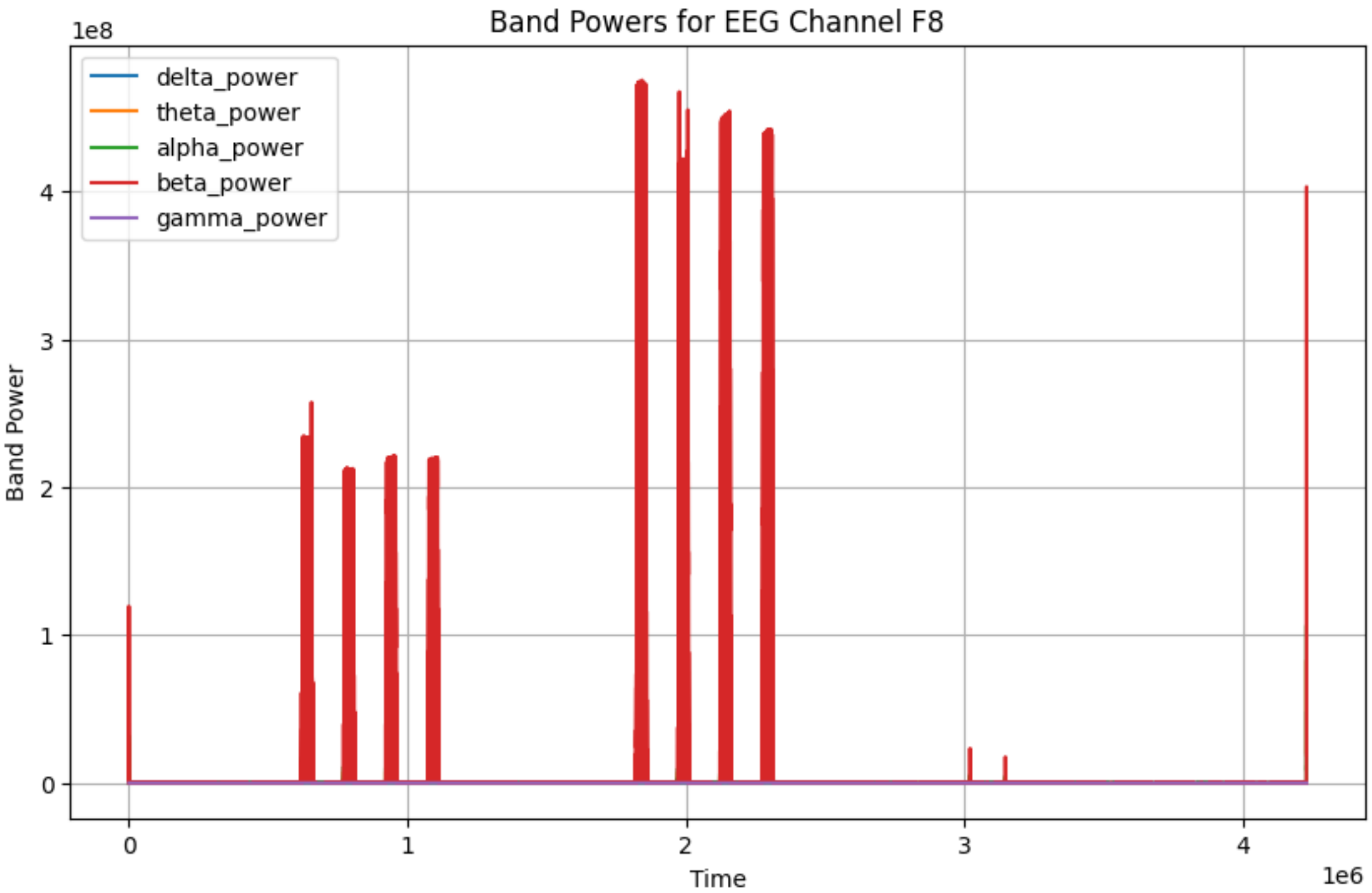
Band Powers for EEG Channel F8
1e8
delta_power
theta_power
alpha_power
beta_power
gamma_power
Band Power
0
1
2
3
4
Time
1e6

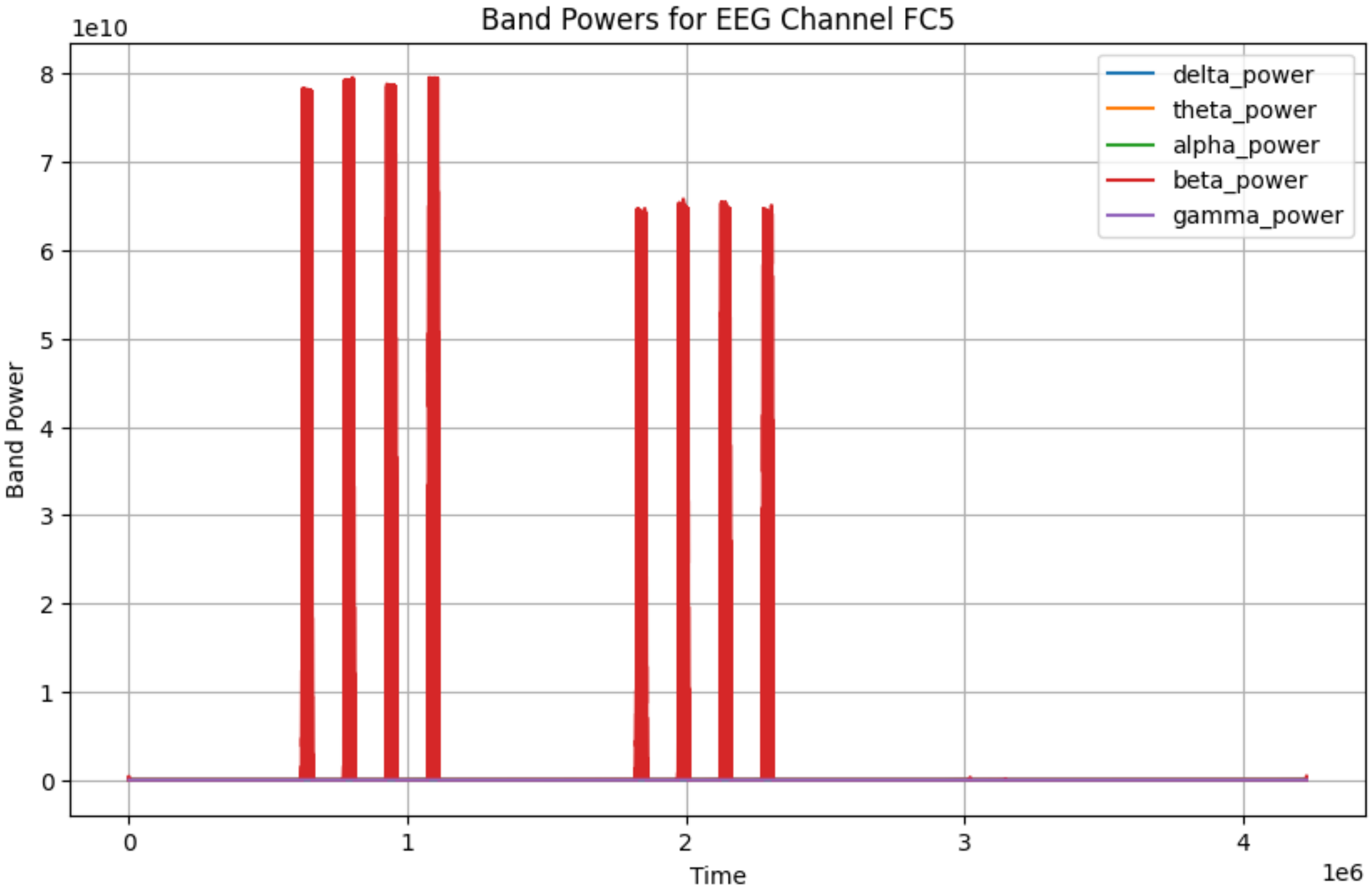
Band Powers for EEG Channel FC5
1e10
delta_power
theta_power
alpha_power
beta_power
gamma_power
Band Power
0
1
2
3
4
5
6
7
8
Time
1e6

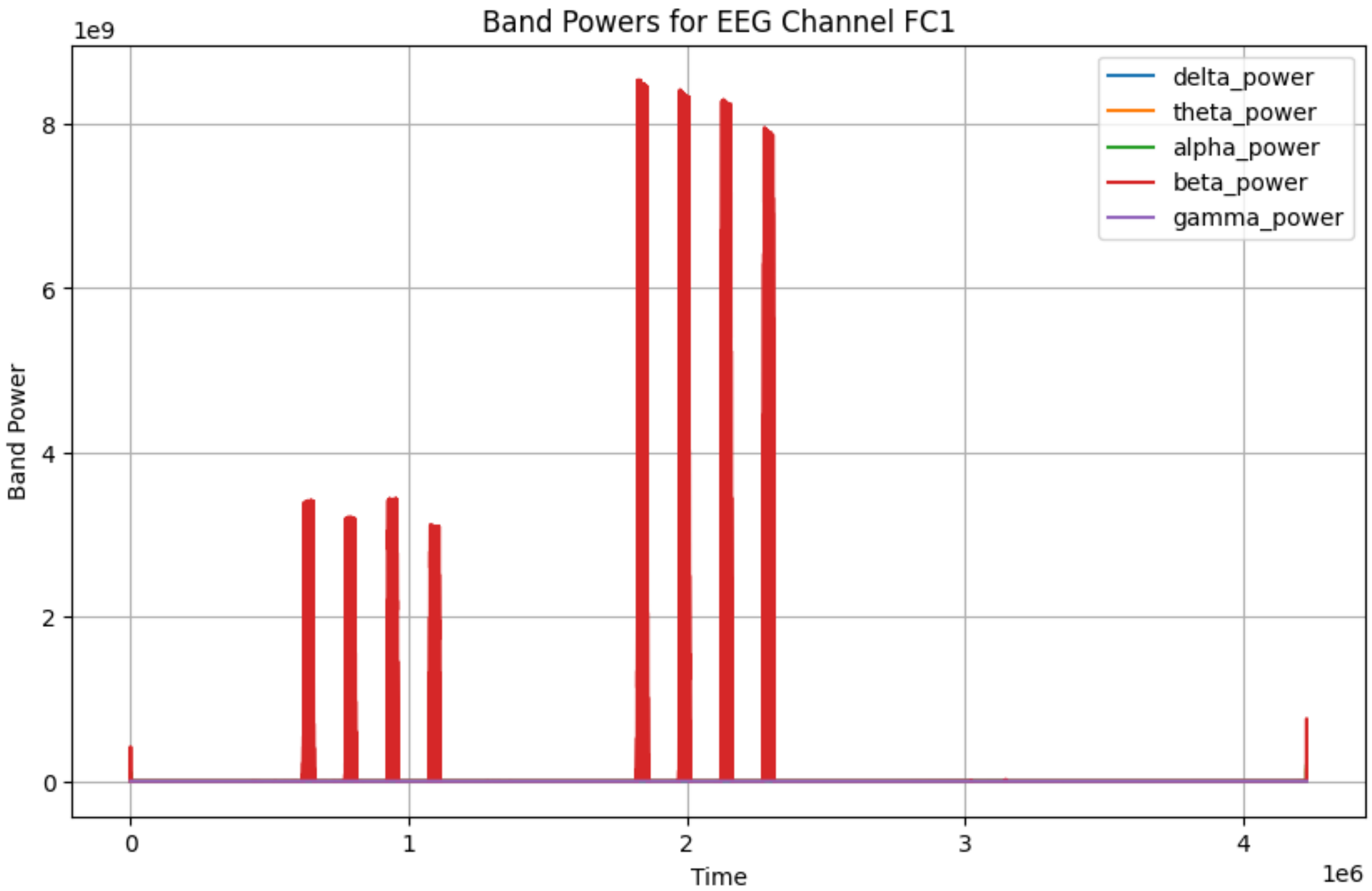
Band Powers for EEG Channel FC1
1e9
8
6
4
2
0
Band Power
0
1
2
3
4
Time
1e6
delta_power
theta_power
alpha_power
beta_power
gamma_power

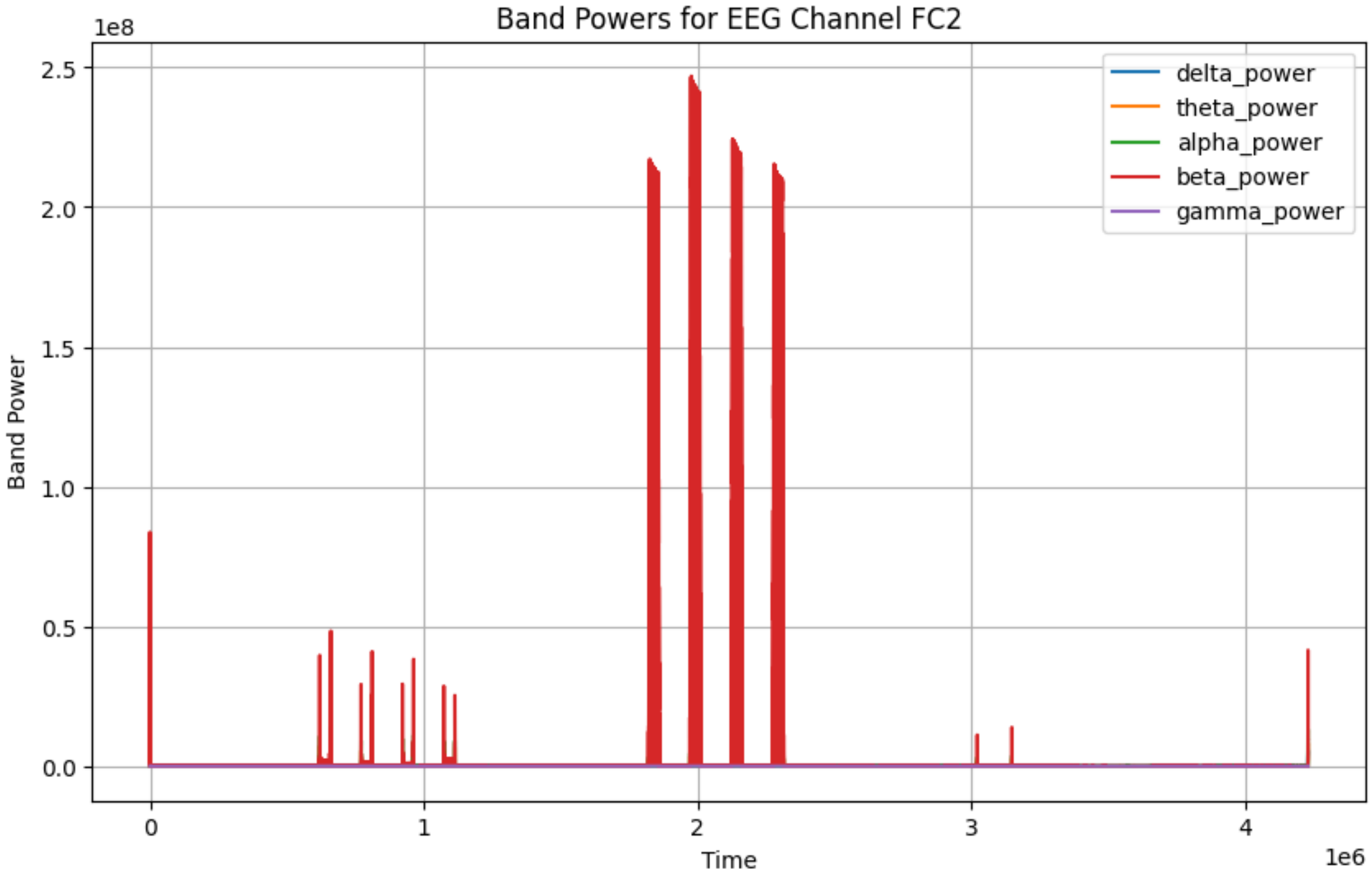
Band Powers for EEG Channel FC2
1e8
2.5
2.0
1.5
1.0
0.5
0.0
Band Power
0
1
2
3
4
Time
1e6
delta_power
theta_power
alpha_power
beta_power
gamma_power

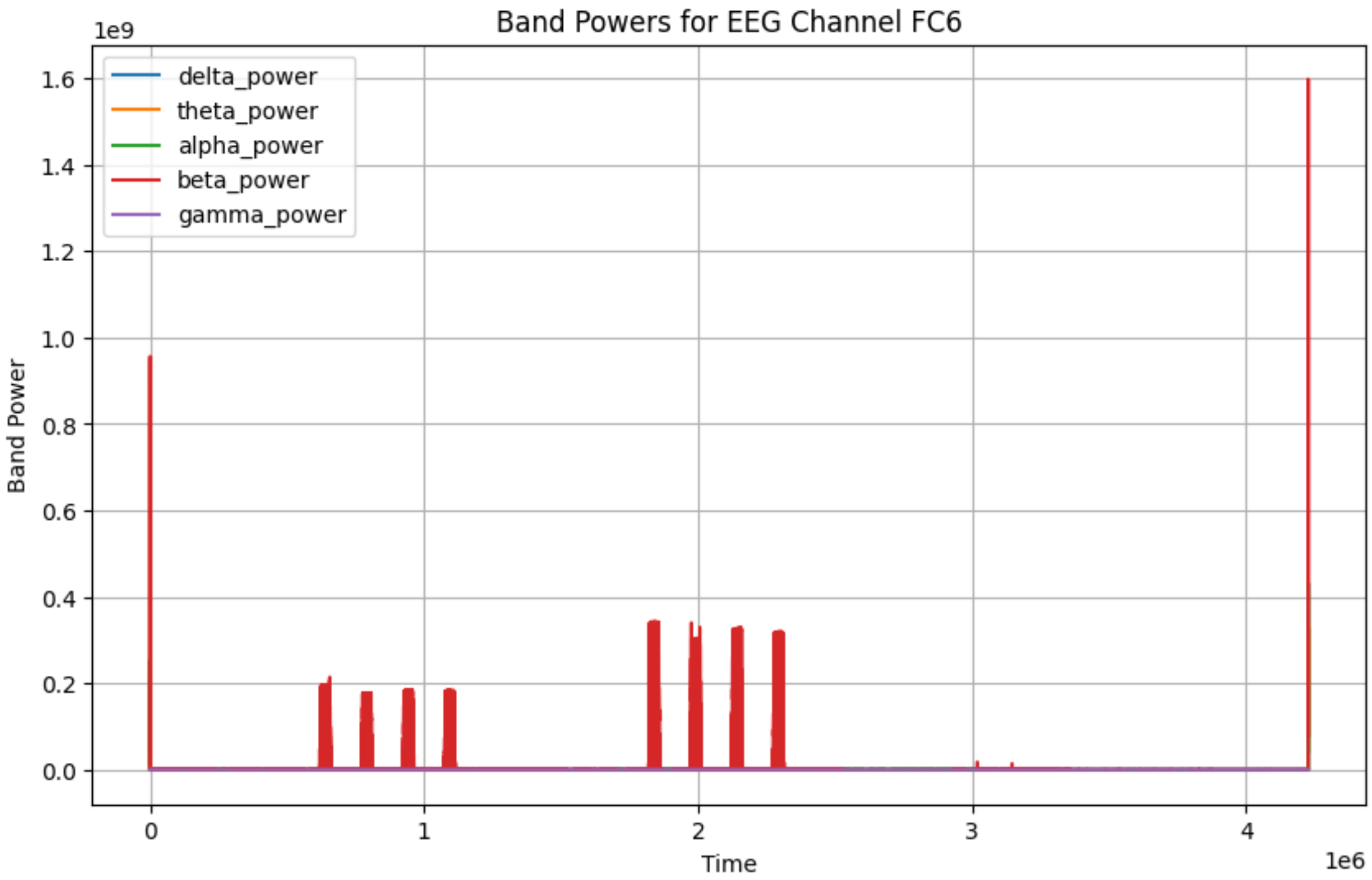
Band Powers for EEG Channel FC6
1e9
delta_power
theta_power
alpha_power
beta_power
gamma_power
1.6
1.4
1.2
1.0
0.8
0.6
0.4
0.2
0.0
Band Power
0
1
2
3
4
Time
1e6

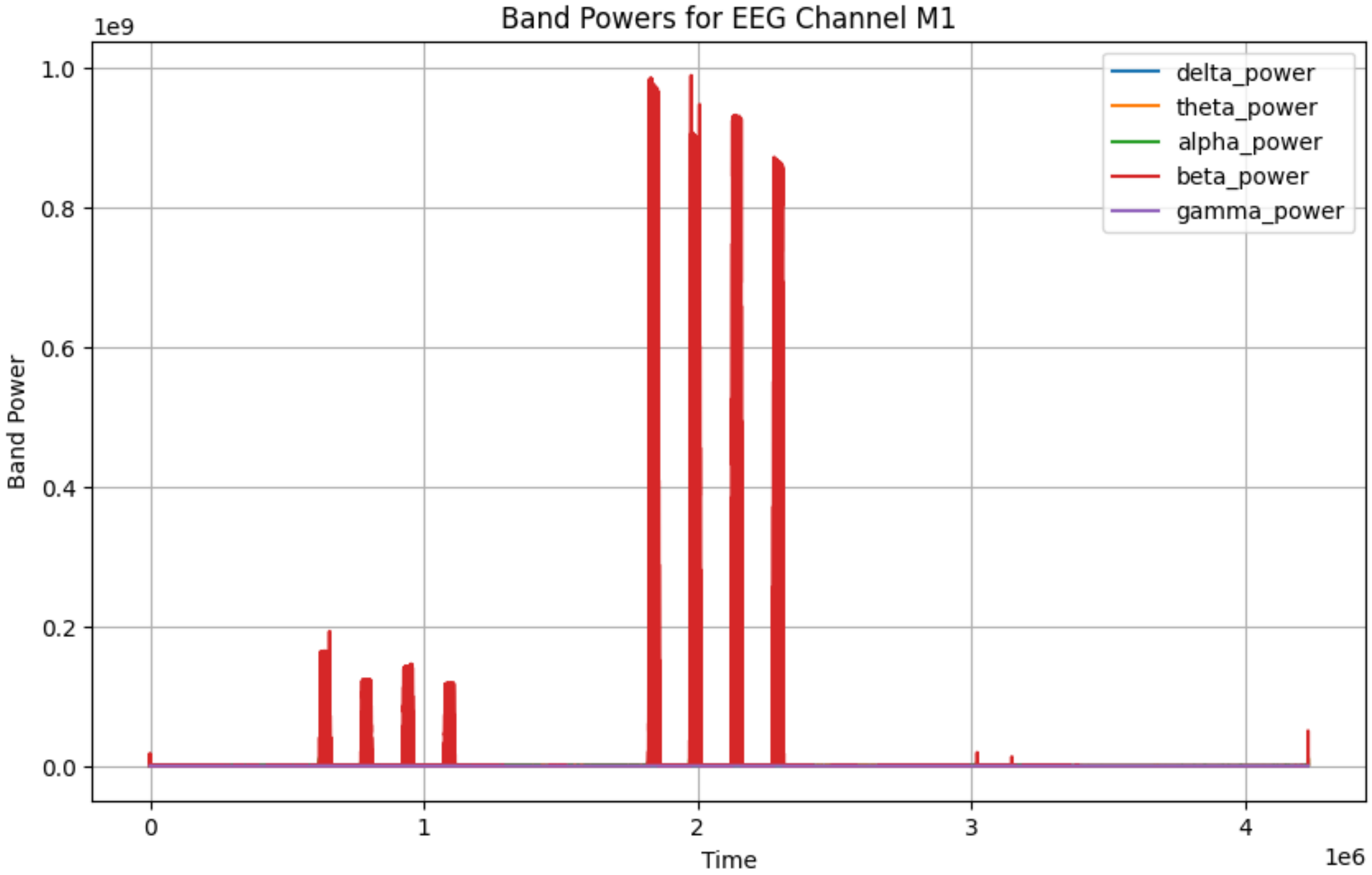
Band Powers for EEG Channel M1
1e9
delta_power
theta_power
alpha_power
beta_power
gamma_power
1.0
0.8
0.6
0.4
0.2
0.0
Band Power
0
1
2
3
4
Time
1e6

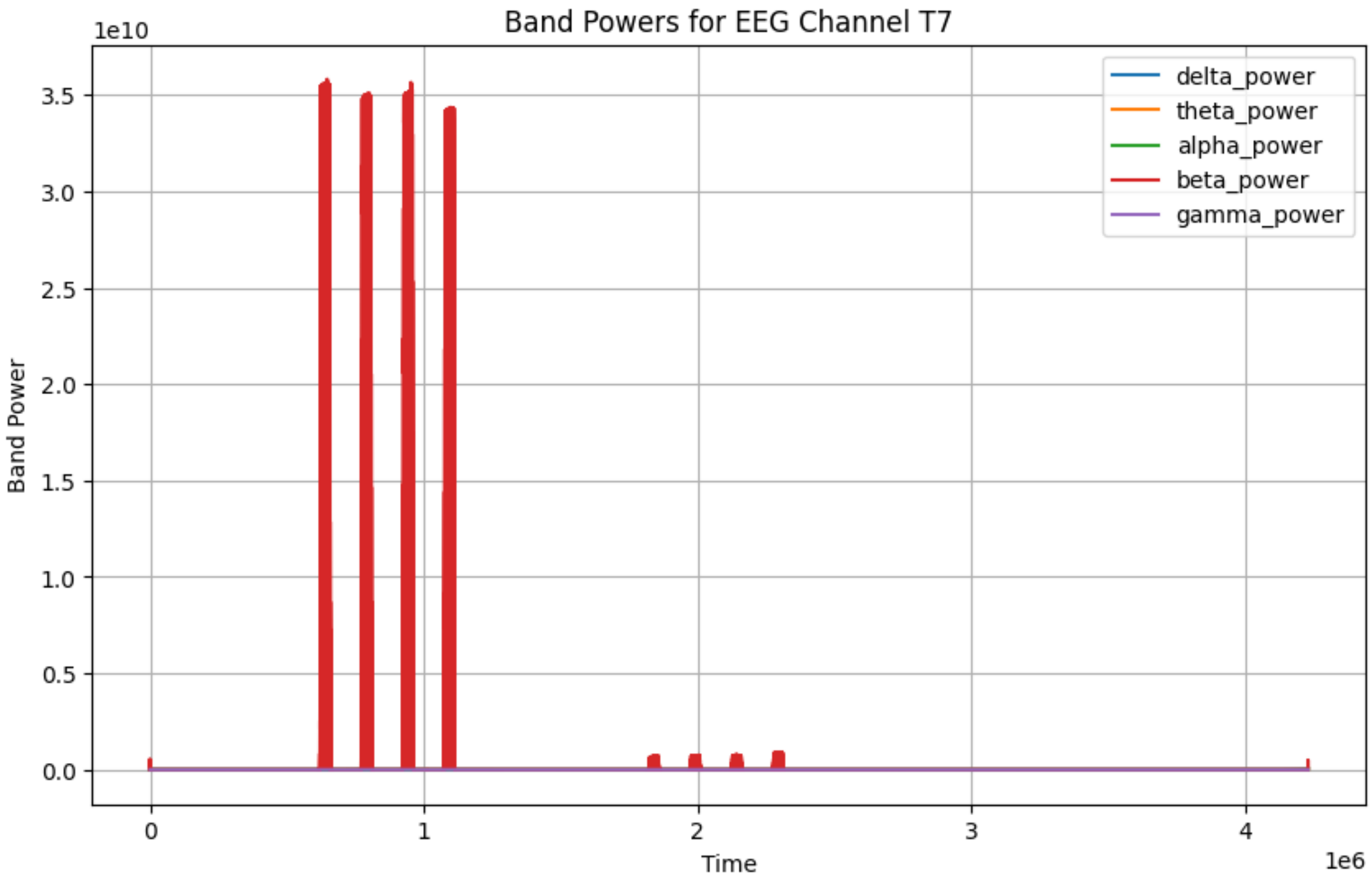
Band Powers for EEG Channel T7
1e10
3.5
3.0
2.5
2.0
1.5
1.0
0.5
0.0
Band Power
0
1
2
3
4
Time
1e6
delta_power
theta_power
alpha_power
beta_power
gamma_power

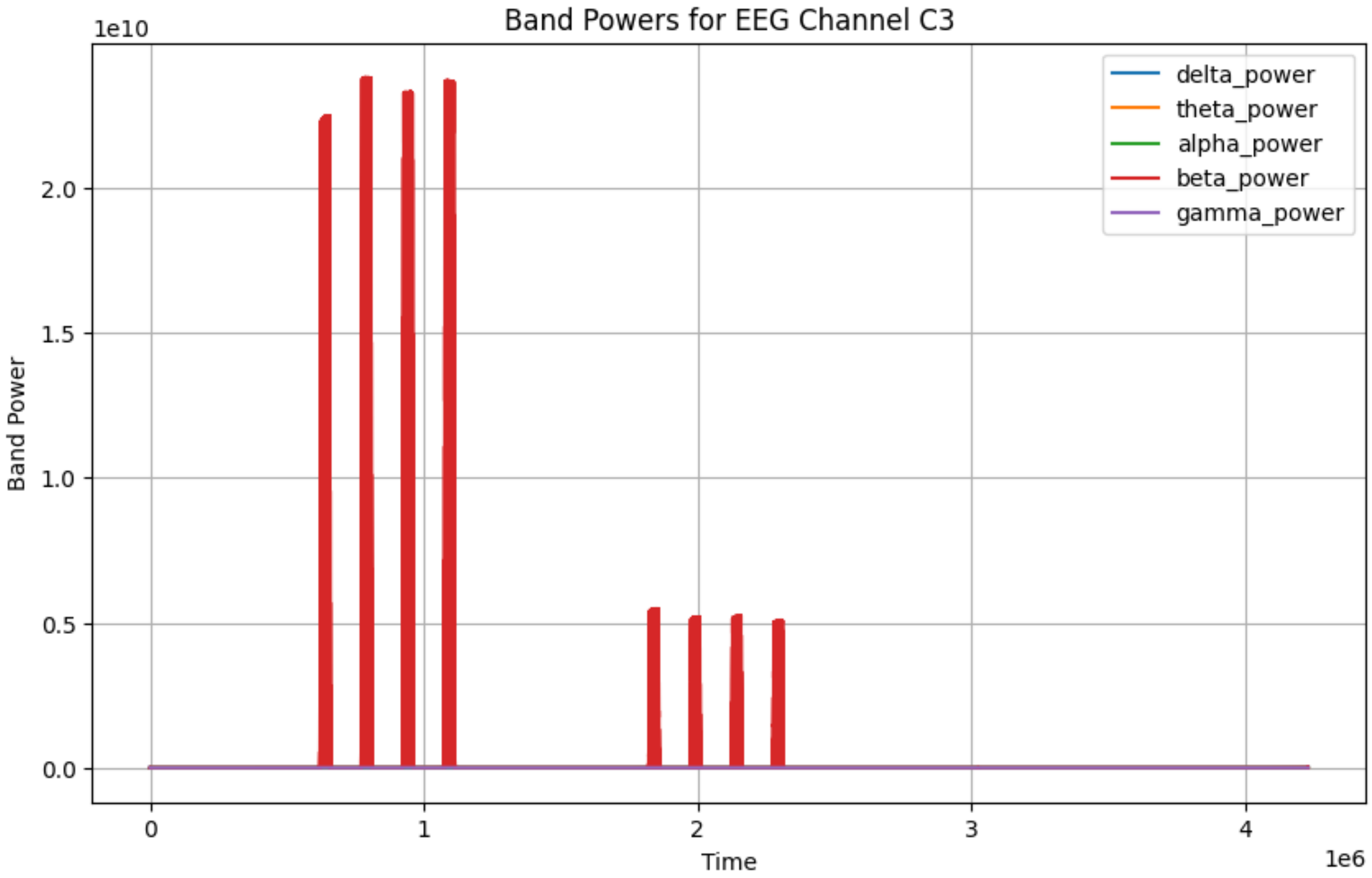
Band Powers for EEG Channel C3
1e10
2.0
1.5
1.0
0.5
0.0
Band Power
0
1
2
3
4
Time
1e6
delta_power
theta_power
alpha_power
beta_power
gamma_power

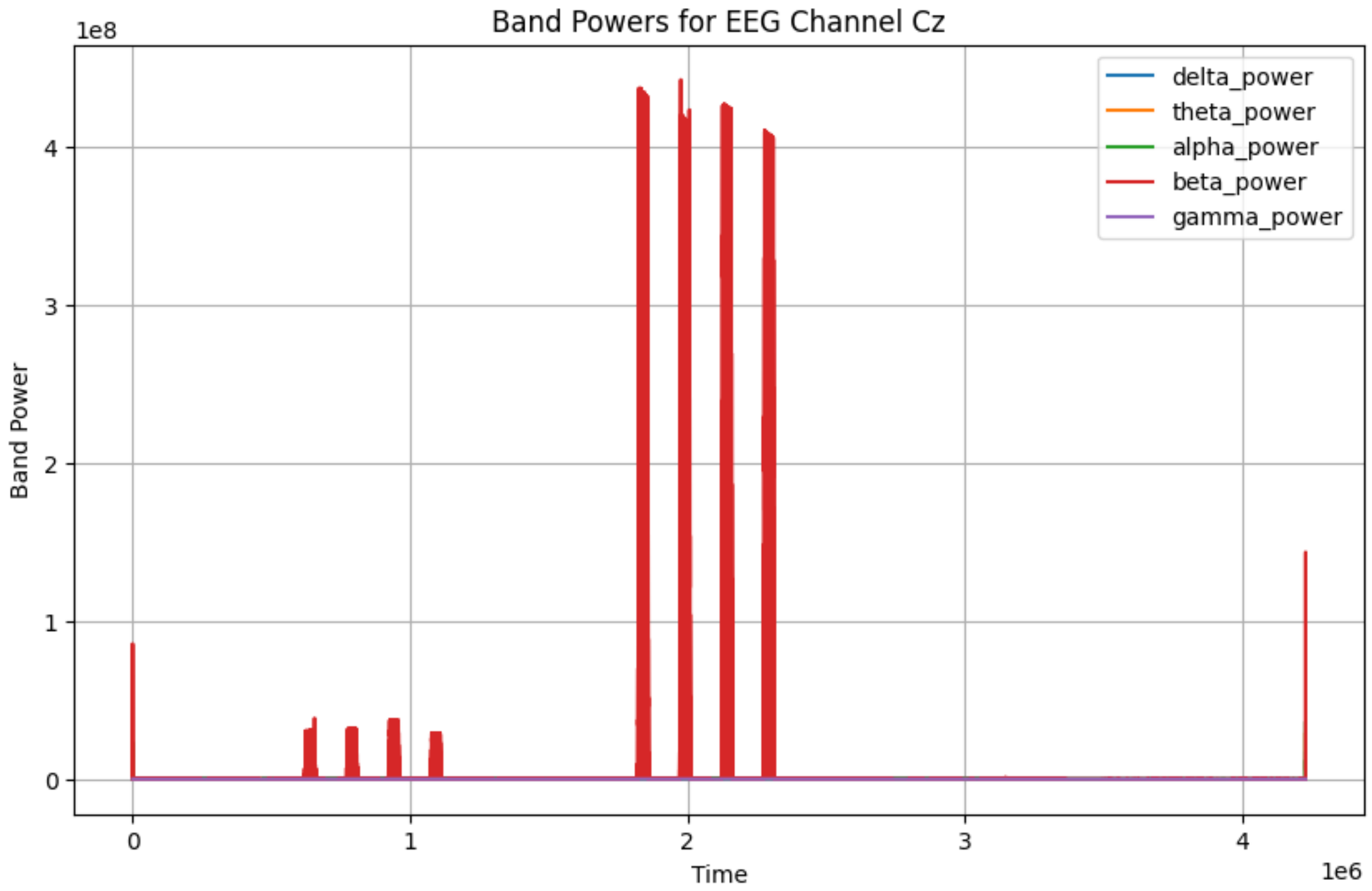
Band Powers for EEG Channel Cz
1e8
4
3
2
1
0
Band Power
0
1
2
3
4
Time
1e6
delta_power
theta_power
alpha_power
beta_power
gamma_power

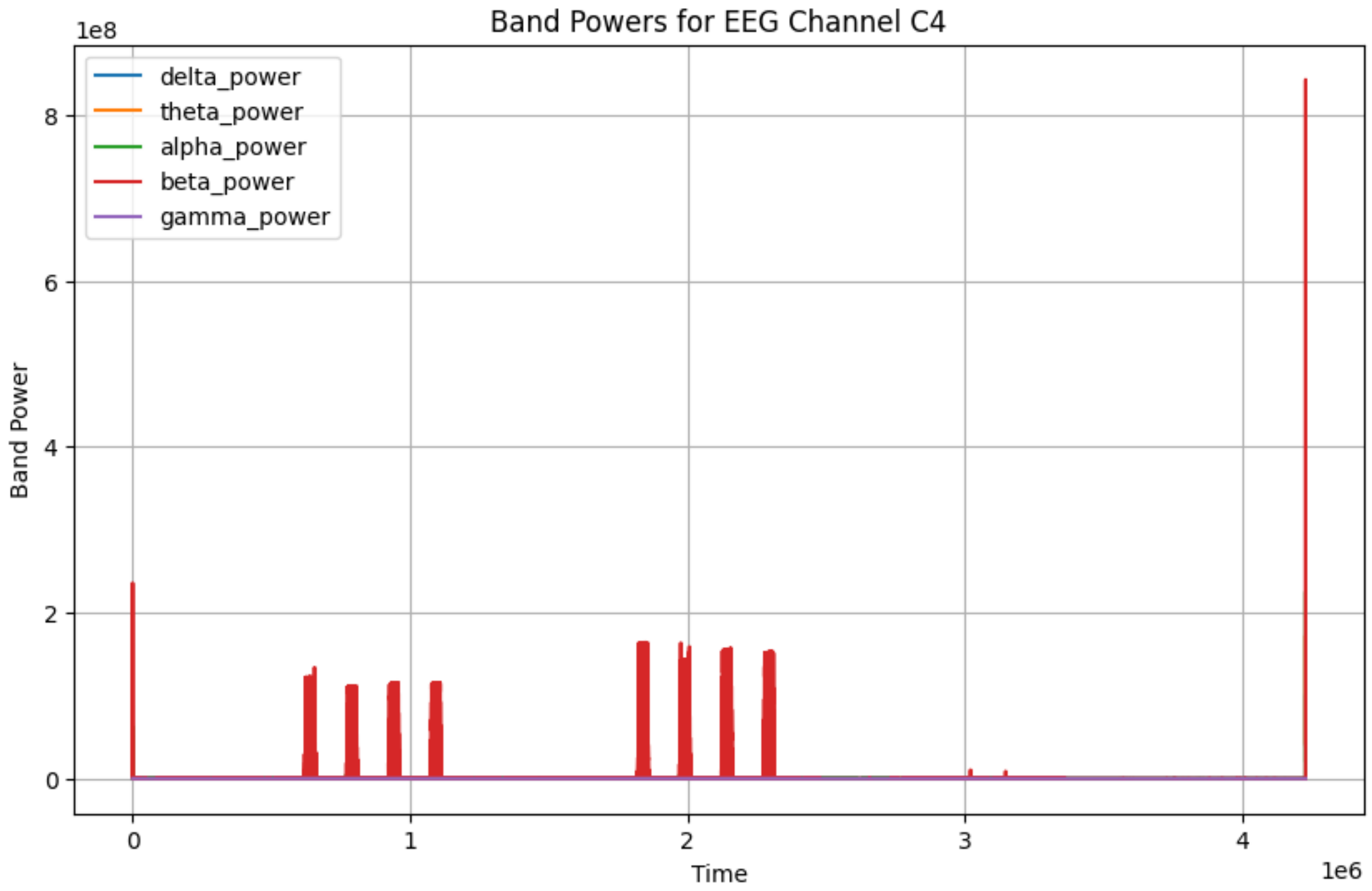
Band Powers for EEG Channel C4
1e8
8
6
4
2
0
Band Power
0
1
2
3
4
Time
1e6
delta_power
theta_power
alpha_power
beta_power
gamma_power

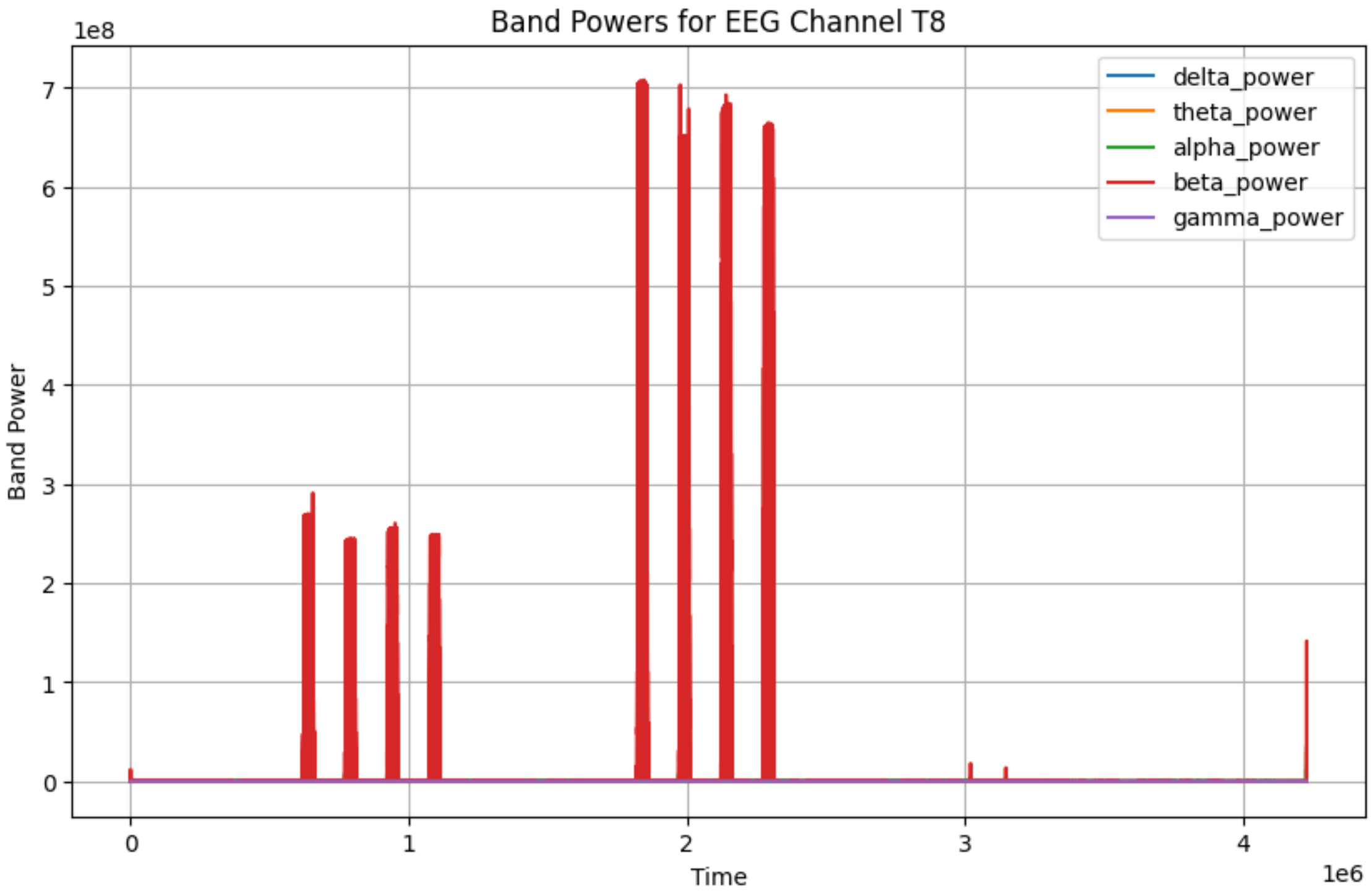
Band Powers for EEG Channel T8
1e8
7
6
5
4
3
2
1
0
Band Power
0
1
2
3
4
Time
1e6
delta_power
theta_power
alpha_power
beta_power
gamma_power

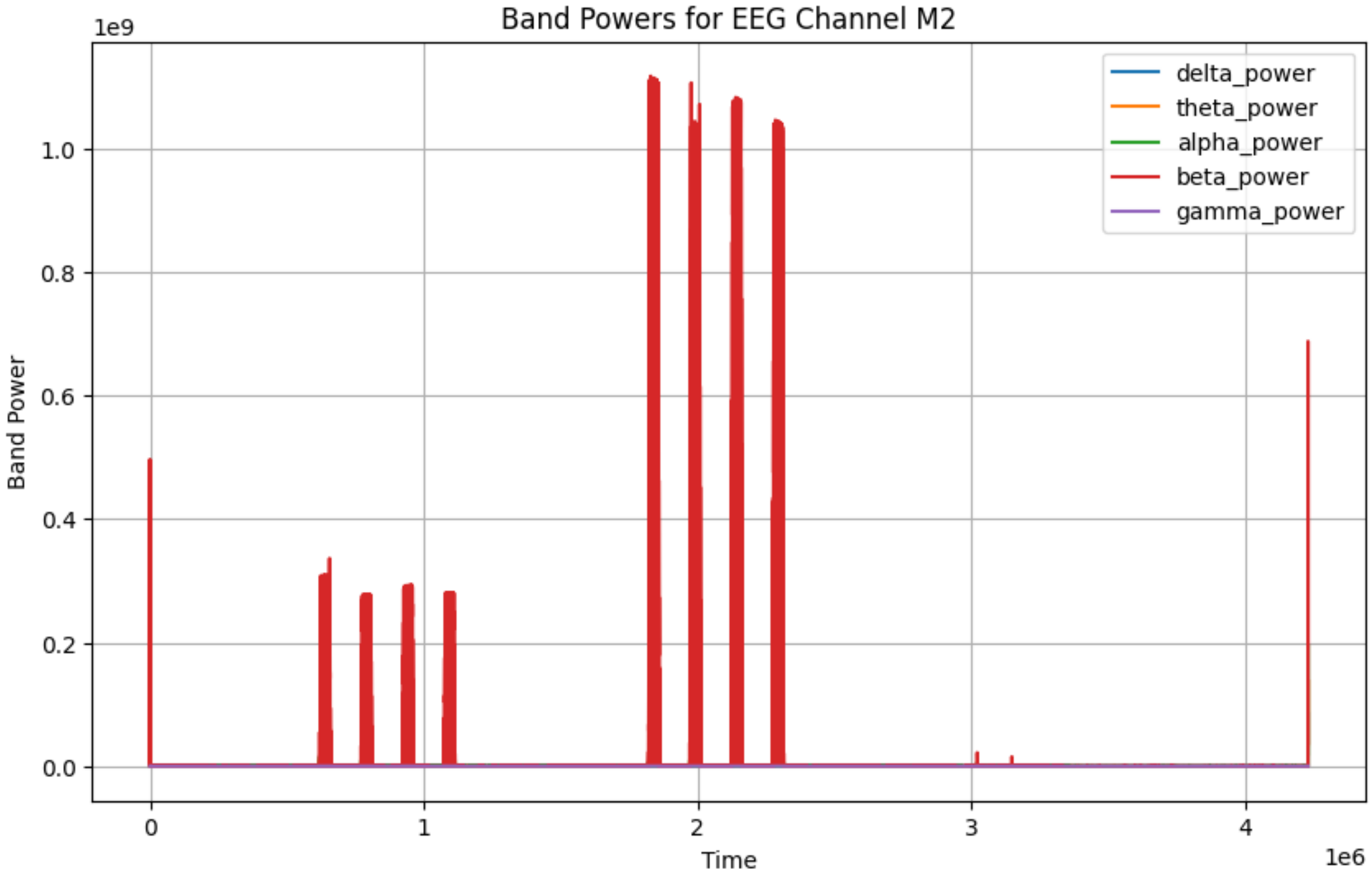
Band Powers for EEG Channel M2
1e9
1.0
0.8
0.6
0.4
0.2
0.0
Band Power
0
1
2
3
4
Time
1e6
delta_power
theta_power
alpha_power
beta_power
gamma_power

# 16 Load and print head

[ ]:
```
# Load the band powers data from the .npy file
band_powers_data = np.load('/home/vincent/AAA_projects/MVCS/Neuroscience/
↪Analysis/Spectral Analysis/BandPowers_x.npy', allow_pickle=True).item()

# Print the populated keys in band_powers_data
print("Populated keys in band_powers_data:", list(band_powers_data.keys()))

# Print the first few elements of the band powers data for each channel
for channel, band_powers in band_powers_data.items():
    print(f"Channel: {channel}")
    for band, powers in band_powers.items():
        print(f"{band}: {powers[:10]}")  # Print the first 10 elements of each␣
↪band power
    print("-" * 40)  # Separator
```

# 17 Short-Time Fourier Transform

```
<!-- Column 1 -->
<div style="flex: 1; margin-right: 10px;">
    <h2>Introduction</h2>
    <p>This analysis adopts the Short-Time Fourier Transform (STFT) approach to break down Elec
    <h2>Objectives</h2>
    <ul>
        <li>Perform STFT on EEG data for each channel.</li>
        <li>Convert the magnitude spectrum to decibels (dB).</li>
        <li>Visualize the resulting spectrograms for various EEG channels.</li>
    </ul>
    <h2>Mathematical Foundations</h2>
    <h3>Short-Time Fourier Transform (STFT)</h3>
    \[ \text{STFT}(x(t)) = X(f, \tau) = \int_{-\infty}^{\infty} x(t) w(t-\tau) e^{-j 2\pi f t}
    <p>Where \( X(f, \tau) \) is the STFT, \( x(t) \) is the time-domain EEG signal, \( w(t) \)
    <h3>Conversion to Decibels</h3>
    \[ \text{STFT}_{\text{dB}} = 10 \log_{10} \left| \text{STFT}(x(t)) \right| \]
    <p>This conversion allows for a logarithmic representation of the amplitude, enhancing the
</div>
<!-- Column 2 -->
<div style="flex: 1; margin-left: 10px;">
    <h2>Methodology</h2>
    <ul>
        <li>EEG data from each channel is isolated for analysis.</li>
        <li>The window size for STFT is set to 2 seconds, given the sampling frequency.</li>
        <li>STFT is performed using the Scipy library's <code>stft</code> function.</li>
        <li>The resulting STFT data is converted to dB for better interpretability.</li>
        <li>The STFT data for each channel is stored in a Python dictionary.</li>
    </ul>
```

```
    <h2>Data Serialization</h2>
    <p>The processed STFT data for each EEG channel is serialized and stored as a single NumPy
    <h2>Scientific and Clinical Relevance</h2>
    <p>The derived spectrograms can serve as valuable diagnostic tools in clinical neuroscienc
    <h2>Conclusion</h2>
    <p>The STFT approach offers a robust methodology for EEG data analysis, enabling the time-
</div>
```

[9]:

```python
import numpy as np
import matplotlib.pyplot as plt
from scipy.signal import stft

# List of EEG channel names
eeg_channels = ['Fp1', 'Fpz', 'Fp2', 'F7', 'F3', 'Fz', 'F4', 'F8', 'FC5', 'FC1', 'FC2', 'FC6',
                'M1', 'T7', 'C3', 'Cz', 'C4', 'T8', 'M2', 'CP5', 'CP1', 'CP2', 'CP6',
                'P7', 'P3', 'Pz', 'P4', 'P8', 'POz', 'O1', 'Oz', 'O2']

# Given sampling frequency (fs)
fs = 1000

# Set up the folder path to save the results
results_folder_path = '/home/vincent/AAA_projects/MVCS/Neuroscience/Analysis/Spectral Analysis/'

# Initialize a dictionary to store the STFT data for each channel
stft_data_dict = {}

# Loop through each EEG channel
for channel_index, channel in enumerate(eeg_channels):
    # Select EEG data from the current channel
    eeg_data = eeg_data_array[:, channel_index]

    # Define the window size for STFT (in samples)
    window_size = int(fs * 2)  # 2 seconds window

    # Calculate the STFT
    frequencies, time_intervals, stft_data = stft(eeg_data, fs=fs, nperseg=window_size)

    # Convert the power data to dB
    stft_log = 10 * np.log10(np.abs(stft_data))

    # Store the STFT data in the dictionary
    stft_data_dict[channel] = stft_log
```

```python
# Save the STFT data for all EEG channels as a single numpy file
results_file = "STFT_x.npy"
save_path = results_folder_path + results_file
np.save(save_path, stft_data_dict)

# Plot the STFT for all EEG channels as heatmaps
num_rows = (len(eeg_channels) + 3) // 4
num_cols = min(len(eeg_channels), 4)

fig, axs = plt.subplots(num_rows, num_cols, figsize=(10, 2 * num_rows))
axs = axs.ravel()

for i, channel in enumerate(eeg_channels):
    axs[i].imshow(stft_data_dict[channel], aspect='auto', cmap='inferno',␣
 ↪extent=[time_intervals[0], time_intervals[-1], frequencies[-1],␣
 ↪frequencies[0]])
    axs[i].set_title(f'STFT of EEG Channel {channel}')
    axs[i].set_xlabel('Time [s]')
    axs[i].set_ylabel('Frequency [Hz]')
    axs[i].grid(True)

plt.tight_layout()
plt.show()
```

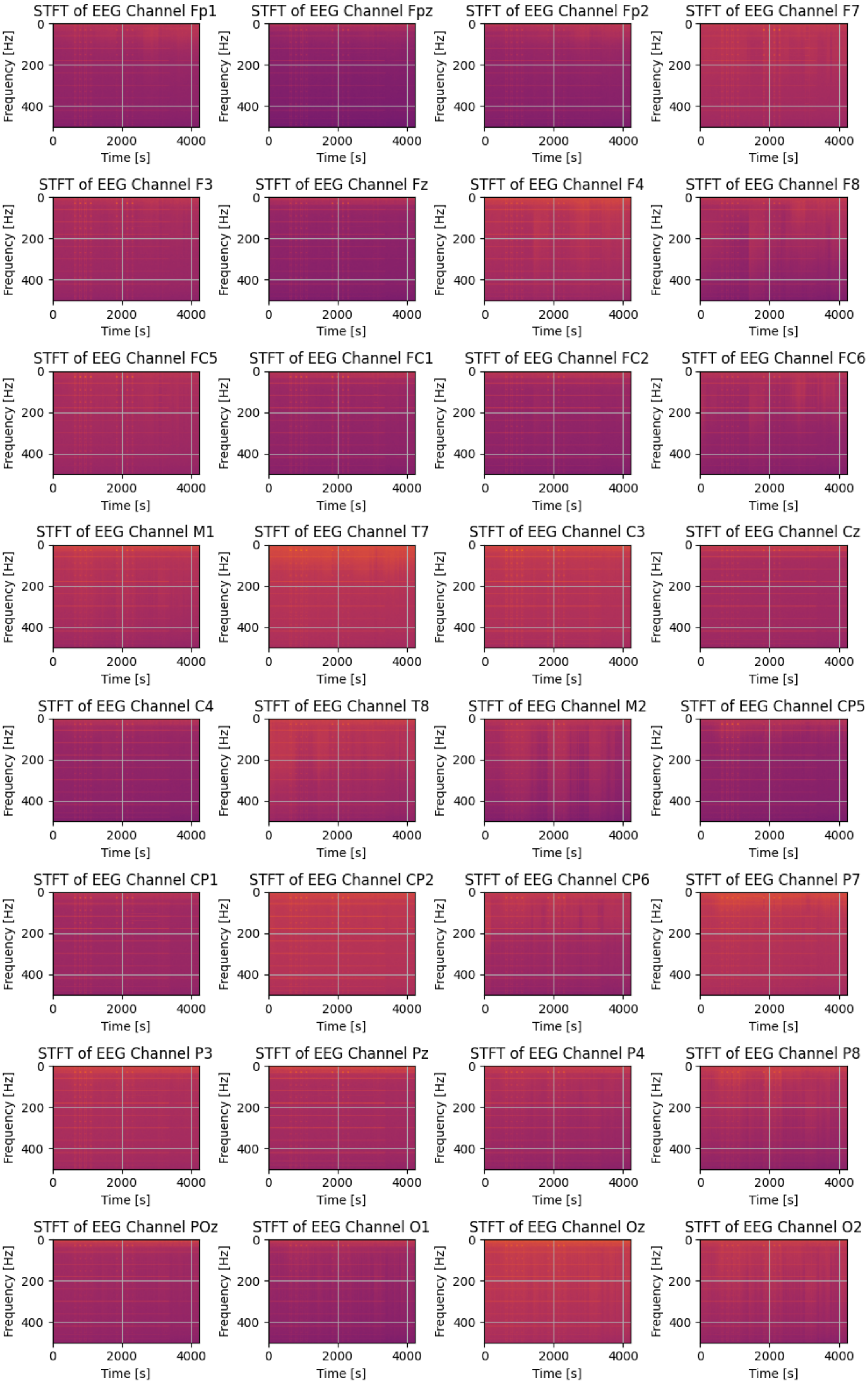
STFT of EEG Channel Fp1
STFT of EEG Channel Fpz
STFT of EEG Channel Fp2
STFT of EEG Channel F7
STFT of EEG Channel F3
STFT of EEG Channel Fz
STFT of EEG Channel F4
STFT of EEG Channel F8
STFT of EEG Channel FC5
STFT of EEG Channel FC1
STFT of EEG Channel FC2
STFT of EEG Channel FC6
STFT of EEG Channel M1
STFT of EEG Channel T7
STFT of EEG Channel C3
STFT of EEG Channel Cz
STFT of EEG Channel C4
STFT of EEG Channel T8
STFT of EEG Channel M2
STFT of EEG Channel CP5
STFT of EEG Channel CP1
STFT of EEG Channel CP2
STFT of EEG Channel CP6
STFT of EEG Channel P7
STFT of EEG Channel P3
STFT of EEG Channel Pz
STFT of EEG Channel P4
STFT of EEG Channel P8
STFT of EEG Channel POz
STFT of EEG Channel O1
STFT of EEG Channel Oz
STFT of EEG Channel O2
Frequency [Hz]
Time [s]

## 18 Load and print head

[13]:
```
# Load the STFT data from the .npy file
stft_data_dict = np.load('/home/vincent/AAA_projects/MVCS/Neuroscience/Analysis/
↪Spectral Analysis/STFT_x.npy', allow_pickle=True).item()

# Print the populated keys in stft_data_dict
print("Populated keys in stft_data_dict:", list(stft_data_dict.keys()))

# Print the head of the STFT data for each channel
for channel, stft_data in stft_data_dict.items():
    print(f"Channel: {channel}")
    print("STFT Data Head:\n", stft_data[:5, :5])  # Print the first 5x5␣
↪elements of the STFT data
    print("-" * 40)  # Separator
```

```
Populated keys in stft_data_dict: ['Fp1', 'Fpz', 'Fp2', 'F7', 'F3', 'Fz', 'F4',
'F8', 'FC5', 'FC1', 'FC2', 'FC6', 'M1', 'T7', 'C3', 'Cz', 'C4', 'T8', 'M2',
'CP5', 'CP1', 'CP2', 'CP6', 'P7', 'P3', 'Pz', 'P4', 'P8', 'POz', 'O1', 'Oz',
'O2']
Channel: Fp1
STFT Data Head:
 [[40.27901945 43.29048063 43.28785317 43.2881181  43.28799363]
 [39.35955148 40.28244249 40.27744403 40.2779645  40.27983077]
 [36.54919092  8.603578    3.26600537  3.90668066  6.39865184]
 [33.53588172  1.65269743  3.56612725  3.59269255  1.49211544]
 [32.56910091 -5.74842057  2.33775801  0.43557077  5.49345519]]
----------------------------------------
Channel: Fpz
STFT Data Head:
 [[40.02955408 43.04172304 43.04265591 43.0431437  43.04170909]
 [39.11016478 40.03324688 40.0322741  40.0332208  40.0313992 ]
 [36.30036833  7.42972488  0.73115213  3.92005784  6.6991942 ]
 [33.2884759   2.70316693  3.10957463  2.40067522  2.73091441]
 [32.32110838 -1.737353    2.19002358 -1.43695075  3.91454148]]
----------------------------------------
Channel: Fp2
STFT Data Head:
 [[40.82631318 43.83704507 43.83685246 43.83513686 43.83490393]
 [39.9070285  40.82799027 40.82662452 40.82516024 40.82340068]
 [37.09759905  5.86175376  4.0664634   3.98531674 10.7446488 ]
 [34.08555176 -0.11208691  3.89272389  0.74420676  6.67917613]
 [33.11798661 -4.80427955  3.45394959 -1.04885552  5.09933723]]
----------------------------------------
```

```
Channel: F7
STFT Data Head:
 [[32.00090175 34.99194796 35.00562721 35.0043807  35.00421709]
 [31.08355267 31.97399304 31.99431165 31.99207876 31.97703933]
 [28.28472331  6.17167648  3.26087629  4.39392839  9.79190906]
 [25.29458136  0.70018501  2.82713266  2.73705132  6.8390153 ]
 [24.32400992 -2.01355333  1.45765087  1.11064128  3.90295911]]
----------------------------------------
Channel: F3
STFT Data Head:
 [[14.49075869 18.32156347 18.05185578 18.10481184 17.96736973]
 [13.47740538 15.56835631 15.03173457 15.19307117 15.25619542]
 [10.15942475  4.24347893  0.91540804  3.05946585  4.38594839]
 [ 6.2476186  -2.20398799  0.40559864  1.98686959  1.69212578]
 [ 5.79623266 -5.09451982 -1.30376965 -1.10878847 -0.70065661]]
----------------------------------------
Channel: Fz
STFT Data Head:
 [[ 36.29998652  39.31416595  39.31476048  39.31405683  39.31389998]
 [ 35.38022634  36.30622274  36.30423072  36.30401957  36.30343697]
 [ 32.56875133   4.14209605   1.06192575   2.98707431   3.69920084]
 [ 29.55583265  -6.82546875   2.46501332  -1.58190262   1.6327613 ]
 [ 28.59158167  -6.60118372   0.56799784 -11.37376646  -3.27775441]]
----------------------------------------
Channel: F4
STFT Data Head:
 [[ 2.51698647e+01  2.82517902e+01  2.82811331e+01  2.82742322e+01
   2.82794114e+01]
 [ 2.42444335e+01  2.52543250e+01  2.52696199e+01  2.52739206e+01
   2.52380526e+01]
 [ 2.14062970e+01  1.97092178e-01  1.70112765e+00  3.71804354e+00
   7.54349819e+00]
 [ 1.83675435e+01 -7.41125534e+00  2.18124043e+00  3.17821164e-03
   4.64323571e+00]
 [ 1.74391235e+01 -4.57478642e+00  1.16777494e+00 -3.76776451e+00
   1.74910151e+00]]
----------------------------------------
Channel: F8
STFT Data Head:
 [[32.38579968 35.37880205 35.36579137 35.36976454 35.35802592]
 [31.46694576 32.36943885 32.35421828 32.3571169  32.360231  ]
 [28.66046878  2.85572814  2.28640727  3.23298128 10.03976803]
 [25.65973529 -0.52316072  1.70308323 -1.06985651  7.62998742]
 [24.69353925  1.93099516  2.92231745 -9.05735354  4.87832027]]
----------------------------------------
Channel: FC5
STFT Data Head:
 [[34.99347649 37.99431583 37.99764389 37.99861179 37.99955201]
```

```
 [34.07508556 34.98225238 34.98749496 34.98766734 34.98521921]
 [31.2701123   2.83674418  0.19805154  1.16630563  6.71182117]
 [28.2651673  -0.13988217 -0.76731725 -0.44649597  4.62134324]
 [27.2956045  -1.46837242 -0.35125557 -1.84290376  0.2274578 ]]
----------------------------------------
Channel: FC1
STFT Data Head:
 [[ 35.120789    38.12004518  38.11996884  38.11910674  38.11963392]
 [ 34.20267528  35.10815504  35.11007785  35.10801858  35.10832268]
 [ 31.39865479   0.34836129  -1.2395671    1.64599534  -0.5675541 ]
 [ 28.39158126  -2.21319651  -1.12376283  -1.5322741   -1.46497252]
 [ 27.41803489  -5.96507384  -3.23159496 -11.76785103  -5.19547811]]
----------------------------------------
Channel: FC2
STFT Data Head:
 [[31.57493024 34.59845612 34.60365505 34.60435206 34.60906326]
 [30.65442407 31.5908412  31.5924262  31.59490623 31.59650602]
 [27.83904102 -1.09691871  0.35591916  2.18273569  3.45762515]
 [24.82057855 -4.6432883  -1.82339436 -0.37447579  0.83478561]
 [23.86207047 -6.30124473 -3.75357728 -7.55874614 -0.13569234]]
----------------------------------------
Channel: FC6
STFT Data Head:
 [[36.89673588 39.89806457 39.89529941 39.89547472 39.89340881]
 [35.97803103 36.88754738 36.88487503 36.88468764 36.8849015 ]
 [33.1715927   0.09166012 -0.81692482  1.64438999  6.10827026]
 [30.16484322 -5.38874152  0.05370613 -1.45626979  4.4518054 ]
 [29.19542293 -1.02776021  1.40479583 -9.00933597  1.91405442]]
----------------------------------------
Channel: M1
STFT Data Head:
 [[28.21179608 31.23805006 31.24648519 31.25518387 31.28530447]
 [27.29382093 28.22880453 28.23898545 28.24680159 28.2936062 ]
 [24.48423933  0.87397971 -1.61285522 -1.24565946  6.8934211 ]
 [21.43140919  3.27579248 -4.69169074 -5.17876811  3.88532987]
 [20.45490178  4.06553217 -3.55764781 -2.20658152 -2.42795132]]
----------------------------------------
Channel: T7
STFT Data Head:
 [[35.66462073 38.67785249 38.67877425 38.67685306 38.6747641 ]
 [34.74537798 35.66934532 35.66905951 35.66687564 35.66761509]
 [31.93589581  3.87393747  1.85129638  1.56490848  6.41246741]
 [28.92100714  2.70153061  0.29689095  0.21014816  3.64540814]
 [27.95145856  2.05090355 -3.19141188  0.60440898 -4.31554912]]
----------------------------------------
Channel: C3
STFT Data Head:
 [[ 2.72595594e+01  3.02948415e+01  3.02980168e+01  3.02919961e+01
```

```
   3.02737525e+01]
 [ 2.63382354e+01  2.72894848e+01  2.72876046e+01  2.72846937e+01
   2.72746382e+01]
 [ 2.35182705e+01 -1.03365908e+00 -5.41183997e+00 -1.73280267e-03
   2.77981617e+00]
 [ 2.04886583e+01 -1.62859301e+00 -6.83948002e+00 -2.46201030e+00
   5.46440513e-01]
 [ 1.95299826e+01 -3.67963370e+00 -8.00109486e+00 -5.36966108e+00
  -5.20712891e+00]]
----------------------------------------
Channel: Cz
STFT Data Head:
 [[ 31.6799951   34.6735558   34.66877229  34.66039543  34.65860893]
 [ 30.76222673  31.66109443  31.65987186  31.64902844  31.64863092]
 [ 27.95959776  -2.46686595  -4.81737536  -4.27978364  -3.8850431 ]
 [ 24.95251876  -1.25066223  -3.31936305  -2.62424821 -17.37292624]
 [ 23.97763544  -5.52687261  -3.9334824   -7.6982003   -5.24691334]]
----------------------------------------
Channel: C4
STFT Data Head:
 [[33.86045727 36.85847398 36.85557492 36.85759591 36.85805783]
 [32.94196184 33.84721147 33.84509631 33.84726531 33.84898828]
 [30.13654436 -2.22983626 -2.59735876 -0.69176419  1.69261738]
 [27.13127923 -5.31571949 -1.82473727 -7.54017091  0.3288228 ]
 [26.16162425 -5.76244507 -1.71500018 -6.38791359 -2.83829782]]
----------------------------------------
Channel: T8
STFT Data Head:
 [[27.30827839 30.26813831 30.23274303 30.22857219 30.20594947]
 [26.39000085 27.25455556 27.21920145 27.21572085 27.20441982]
 [23.58804535  0.46926117  0.608266   -0.49706971  4.16046003]
 [20.6089057   2.47824224  1.13257969 -9.73821583  2.65514173]
 [19.64908863  2.48857977  0.50093452 -7.19608981  1.95959279]]
----------------------------------------
Channel: M2
STFT Data Head:
 [[35.47839546 38.4892229  38.49466536 38.49627372 38.50420079]
 [34.5597021  35.47895996 35.48534395 35.48597812 35.49494833]
 [31.75202234  0.2233919   1.40243318 -0.43818041  3.13683339]
 [28.73454585  3.83399615  1.67340546 -8.28846278  1.9525973 ]
 [27.76299575  3.84202358  0.78030496 -2.87240655 -0.05722108]]
----------------------------------------
Channel: CP5
STFT Data Head:
 [[25.29294744 28.24102088 28.22333409 28.22201616 28.21475362]
 [24.37699776 25.22238532 25.20882285 25.20821343 25.1900497 ]
 [21.58796479 -0.29635196 -1.32375055 -3.27112309  2.72373675]
 [18.63149419 -0.39661149 -3.9400052  -1.80559188 -1.65378093]
```

```
 [17.65971088  0.19293048 -4.82610561 -0.23506595 -3.66697106]]
----------------------------------------
Channel: CP1
STFT Data Head:
 [[33.97738695 36.99018397 36.99525271 36.99872892 37.00024672]
 [33.05820897 33.98070455 33.98521829 33.98840187 33.98990125]
 [30.24913491 -1.17236375 -4.41577667 -1.53242092 -0.94650875]
 [27.2359788  -1.42298028 -1.63434948 -1.58801147 -2.98889434]
 [26.26657483 -3.37167378 -4.16074734 -5.3810517  -4.94075817]]
----------------------------------------
Channel: CP2
STFT Data Head:
 [[30.99905119 33.98771457 33.98038497 33.97615714 33.97514516]
 [30.08093125 30.97469986 30.96989285 30.96583983 30.96612707]
 [27.2779678  -0.4135332  -6.21298817 -6.60005745 -0.48649315]
 [24.27945053 -5.88250855 -2.10045175 -8.91066831 -3.93341931]
 [23.30943527 -5.17528402 -2.60022228 -5.8151098  -9.23227697]]
----------------------------------------
Channel: CP6
STFT Data Head:
 [[ 31.99252433  34.9863417   34.98273061  34.98560256  34.98542589]
 [ 31.07369864  31.97378615  31.97126026  31.97523687  31.97658   ]
 [ 28.26731541   1.70531631   0.29056429  -2.21211571   1.48608004]
 [ 25.26673721   1.20528851   0.40203088 -11.38265495  -1.11140787]
 [ 24.30161369   1.17327738  -0.529299    -4.5735873   -5.41884074]]
----------------------------------------
Channel: P7
STFT Data Head:
 [[32.25257974 35.24586574 35.24163027 35.2379123  35.2326868 ]
 [31.33405353 32.23350502 32.23046259 32.2267159  32.21902159]
 [28.52943955  1.84489501 -0.31520817 -4.75041441  3.54289518]
 [25.5318627   1.50285698 -1.40656341 -0.91790825 -2.95408874]
 [24.56435755  1.31097938 -3.7439455   0.30493615 -2.88987135]]
----------------------------------------
Channel: P3
STFT Data Head:
 [[32.60205434 35.61391202 35.6191985  35.62250778 35.62261478]
 [31.68323946 32.60424467 32.60967102 32.61295063 32.61232059]
 [28.87540879 -0.28706063 -2.29518019 -5.37405517  0.97286219]
 [25.85956861  0.91790032 -3.04784101 -0.71978052 -3.91919983]
 [24.88606611 -0.51746652 -6.50198396 -1.49063348 -2.54810854]]
----------------------------------------
Channel: Pz
STFT Data Head:
 [[25.05763455 28.11813778 28.15452523 28.17463692 28.16439865]
 [24.13807668 25.11680196 25.15036021 25.16380776 25.14650732]
 [21.32282248  1.55175186 -2.02550512 -2.70107418  0.75851367]
 [18.27148063  0.86024853 -1.28363337 -1.72068145 -0.10281276]
```

```
 [17.28672559 -1.29338137 -2.59884067 -4.9501917  -3.84553152]]
----------------------------------------
Channel: P4
STFT Data Head:
 [[29.29209038 32.32030552 32.32935271 32.33745701 32.33889154]
 [28.3733878  29.31525389 29.31987662 29.32712632 29.32643237]
 [25.56449697  2.92173775 -2.35133104 -6.08516094  0.04313305]
 [22.53552491  1.38152929 -0.27881753 -6.31987802 -1.41634104]
 [21.55755127  0.18248907 -2.00011321 -3.35330704 -7.24123719]]
----------------------------------------
Channel: P8
STFT Data Head:
 [[35.7355862  38.73583499 38.73222802 38.73071604 38.72797719]
 [34.81662989 35.72481408 35.72146697 35.72011428 35.71706529]
 [32.00939247  1.86956545 -0.22774341 -2.70238536  0.40599833]
 [29.00494792  3.03219484  1.15870418 -1.69027525 -4.38271709]
 [28.03791462  2.95246608 -0.60929196 -0.32332978 -5.5012033 ]]
----------------------------------------
Channel: POz
STFT Data Head:
 [[36.76034436 39.7694604  39.77265049 39.77448233 39.77566386]
 [35.84159265 36.75956441 36.76313842 36.76404611 36.76479062]
 [33.03414363  1.68311342  1.01462141  0.97232143  2.85310382]
 [30.02033336  2.30491156  0.13471202  0.78514879  1.35134716]
 [29.04902593  1.31519366 -1.85041991 -0.2646332  -5.74193113]]
----------------------------------------
Channel: O1
STFT Data Head:
 [[ 3.48571065e+01  3.78697611e+01  3.78743507e+01  3.78770427e+01
   3.78835235e+01]
 [ 3.39383203e+01  3.48602186e+01  3.48651617e+01  3.48666748e+01
   3.48746558e+01]
 [ 3.11302817e+01  1.21718279e+00  9.05643003e-01 -4.32457902e+00
   3.37878609e+00]
 [ 2.81122062e+01  2.21220493e+00 -2.09278764e+00 -2.38432420e+00
  -4.35202377e+00]
 [ 2.71393366e+01  2.21690608e+00 -3.75628072e+00 -2.49432452e-02
  -5.05956940e+00]]
----------------------------------------
Channel: Oz
STFT Data Head:
 [[28.50414103 31.52237812 31.53666047 31.5367024  31.54534191]
 [27.58692251 28.51182671 28.53127997 28.52841404 28.54029189]
 [24.78283738 -1.2712392   0.48144778 -1.55212259  2.84056978]
 [21.74270386  2.38301568 -2.65997639 -4.70472841 -1.85261389]
 [20.75337273  2.72166811 -3.03284249 -1.51304708 -6.35900347]]
----------------------------------------
Channel: O2
```

```
STFT Data Head:
 [[32.47268132 35.4716368  35.46422919 35.46262503 35.45733151]
 [31.55319935 32.46138986 32.4519277  32.45253108 32.44542535]
 [28.74464506 -1.43010975  0.94046713 -7.3319018   1.03379986]
 [25.7469405   2.3124597  -0.46203452 -2.35753269 -5.1006285 ]
 [24.78495665  2.5139891  -2.98009312 -0.30068131 -4.65382477]]
----------------------------------------
```

## 19 Spectral Entropy

[ ]:

```

```

[6]:

```python
import numpy as np
import matplotlib.pyplot as plt
from scipy.signal import welch

# List of EEG channel names
eeg_channels = ['Fp1', 'Fpz', 'Fp2', 'F7', 'F3', 'Fz', 'F4', 'F8', 'FC5',␣
 ↪'FC1', 'FC2', 'FC6',
                'M1', 'T7', 'C3', 'Cz', 'C4', 'T8', 'M2', 'CP5', 'CP1', 'CP2',␣
 ↪'CP6',
                'P7', 'P3', 'Pz', 'P4', 'P8', 'POz', 'O1', 'Oz', 'O2']

# Given sampling frequency (fs)
fs = 1000

# Set up the folder path to save the results
results_folder_path = '/home/vincent/AAA_projects/MVCS/Neuroscience/Analysis/
 ↪Spectral Analysis/'

# Initialize a dictionary to store the spectral entropy for each channel
spectral_entropy_dict = {}

# Loop through each EEG channel
for channel in eeg_channels:
    # Select EEG data from the current channel
    eeg_data = eeg_data_array[:, eeg_channels.index(channel)]

    # Calculate the power spectral density using Welch's method
    f, Pxx = welch(eeg_data, fs=fs, nperseg=fs*2)

    # Normalize the power spectrum
    normalized_Pxx = Pxx / np.sum(Pxx)

    # Calculate the spectral entropy
    spectral_entropy = -np.sum(normalized_Pxx * np.log2(normalized_Pxx))
```

```
    # Store the spectral entropy in the dictionary
    spectral_entropy_dict[channel] = spectral_entropy

# Save the spectral entropy values as a numpy array
results_file = "SpectralEntropy_x.npy"
np.save(results_folder_path + results_file, spectral_entropy_dict)

# Extract channel names and corresponding spectral entropy values
channels = list(spectral_entropy_dict.keys())
entropy_values = list(spectral_entropy_dict.values())

# Plot the spectral entropy values
plt.figure(figsize=(10, 6))
plt.bar(channels, entropy_values, color='b')
plt.xlabel('EEG Channel')
plt.ylabel('Spectral Entropy')
plt.title('Spectral Entropy of EEG Channels')
plt.xticks(rotation=45, ha='right')
plt.tight_layout()

# Display the plot
plt.show()
```

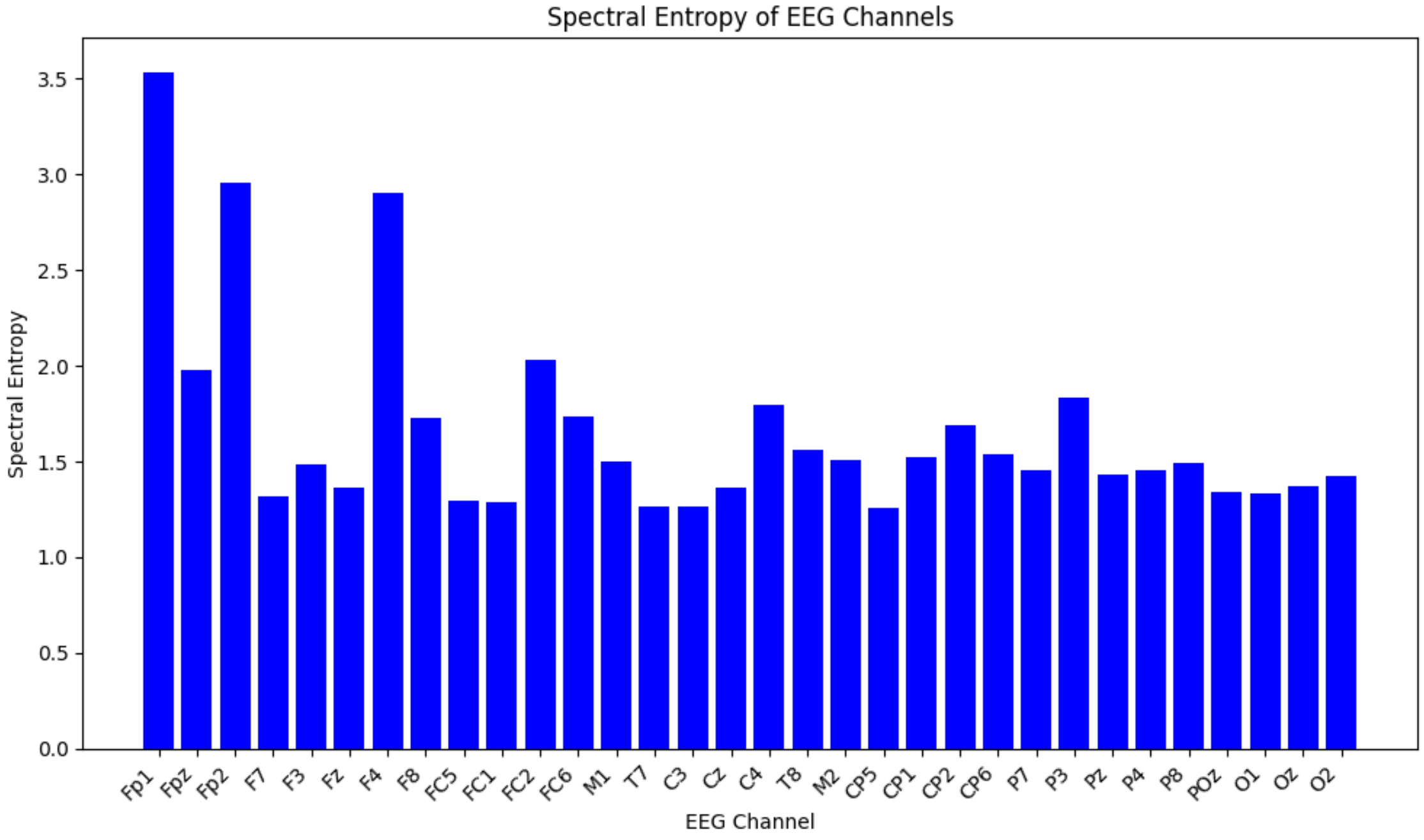

## 20 Load and print head

[12]:
```
# Load the spectral entropy data from the .npy file
spectral_entropy_dict = np.load('/home/vincent/AAA_projects/MVCS/Neuroscience/
↪Analysis/Spectral Analysis/SpectralEntropy_x.npy', allow_pickle=True).item()

# Print the populated keys in spectral_entropy_dict
print("Populated keys in spectral_entropy_dict:", list(spectral_entropy_dict.
↪keys()))

# Print the spectral entropy values for each channel
for channel, spectral_entropy in spectral_entropy_dict.items():
    print(f"Channel: {channel}")
    print(f"Spectral Entropy: {spectral_entropy}")
    print("-" * 40)  # Separator
```

```
Populated keys in spectral_entropy_dict: ['Fp1', 'Fpz', 'Fp2', 'F7', 'F3', 'Fz',
'F4', 'F8', 'FC5', 'FC1', 'FC2', 'FC6', 'M1', 'T7', 'C3', 'Cz', 'C4', 'T8',
'M2', 'CP5', 'CP1', 'CP2', 'CP6', 'P7', 'P3', 'Pz', 'P4', 'P8', 'POz', 'O1',
'Oz', 'O2']
Channel: Fp1
Spectral Entropy: 3.5366970108625324
----------------------------------------
Channel: Fpz
Spectral Entropy: 1.9779195403191885
----------------------------------------
Channel: Fp2
Spectral Entropy: 2.9556473260041014
----------------------------------------
Channel: F7
Spectral Entropy: 1.317390997528914
----------------------------------------
Channel: F3
Spectral Entropy: 1.4828749659869962
----------------------------------------
Channel: Fz
Spectral Entropy: 1.364216888653762
----------------------------------------
Channel: F4
Spectral Entropy: 2.907933081431448
----------------------------------------
Channel: F8
Spectral Entropy: 1.7293493323116478
----------------------------------------
Channel: FC5
Spectral Entropy: 1.293877547465603
----------------------------------------
```

```
Channel: FC1
Spectral Entropy: 1.2903223004655797
----------------------------------------
Channel: FC2
Spectral Entropy: 2.0350573202031015
----------------------------------------
Channel: FC6
Spectral Entropy: 1.7335929695148773
----------------------------------------
Channel: M1
Spectral Entropy: 1.5020770472316665
----------------------------------------
Channel: T7
Spectral Entropy: 1.2630849268151654
----------------------------------------
Channel: C3
Spectral Entropy: 1.2617823636948098
----------------------------------------
Channel: Cz
Spectral Entropy: 1.3628364948216078
----------------------------------------
Channel: C4
Spectral Entropy: 1.7934676007587917
----------------------------------------
Channel: T8
Spectral Entropy: 1.5599747755579694
----------------------------------------
Channel: M2
Spectral Entropy: 1.5072101307742165
----------------------------------------
Channel: CP5
Spectral Entropy: 1.2608283068552764
----------------------------------------
Channel: CP1
Spectral Entropy: 1.5228124308451847
----------------------------------------
Channel: CP2
Spectral Entropy: 1.6879092983109352
----------------------------------------
Channel: CP6
Spectral Entropy: 1.5357523705867189
----------------------------------------
Channel: P7
Spectral Entropy: 1.4562480199912657
----------------------------------------
Channel: P3
Spectral Entropy: 1.8356627824274931
----------------------------------------
```

```
Channel: Pz
Spectral Entropy: 1.428675116696851
----------------------------------------
Channel: P4
Spectral Entropy: 1.45834684658797
----------------------------------------
Channel: P8
Spectral Entropy: 1.489584836547933
----------------------------------------
Channel: POz
Spectral Entropy: 1.3394905881156252
----------------------------------------
Channel: O1
Spectral Entropy: 1.3330793043466678
----------------------------------------
Channel: Oz
Spectral Entropy: 1.372235922411781
----------------------------------------
Channel: O2
Spectral Entropy: 1.4220909537909494
----------------------------------------
```

## 21 Coherence

[ ]:
```
import numpy as np
import pywt
import matplotlib.pyplot as plt

# List of EEG channel names
eeg_channels = ['Fp1', 'Fpz', 'Fp2', 'F7', 'F3', 'Fz', 'F4', 'F8', 'FC5',␣
 ↪'FC1', 'FC2', 'FC6',
                'M1', 'T7', 'C3', 'Cz', 'C4', 'T8', 'M2', 'CP5', 'CP1', 'CP2',␣
 ↪'CP6',
                'P7', 'P3', 'Pz', 'P4', 'P8', 'POz', 'O1', 'Oz', 'O2']

# Define the sampling frequency
fs = 1000

# Create a time array for the data
time = np.arange(len(eeg_data_array)) / fs

# Define the frequencies at which to compute the wavelet transform
frequencies = np.logspace(np.log10(10), np.log10(fs/2), num=100)  # Adjust the␣
 ↪scale range as needed

# Initialize a dictionary to store the coherence results for each channel pair
coherence_results = {}
```

```
# Loop through each channel and calculate coherence with other channels
for channel1 in eeg_channels:
    eeg_data1 = eeg_data_array[:, eeg_channels.index(channel1)]
    coefficients1, frequencies = pywt.cwt(eeg_data1, frequencies,␣
 ↪wavelet='morl')
    psd1 = np.abs(coefficients1)**2

    # Initialize a list to store coherence results for the current channel
    channel_coherence = []

    for channel2 in eeg_channels:
        eeg_data2 = eeg_data_array[:, eeg_channels.index(channel2)]
        coefficients2, frequencies = pywt.cwt(eeg_data2, frequencies,␣
 ↪wavelet='morl')
        psd2 = np.abs(coefficients2)**2

        # Compute the cross-power spectral density between channel1 and channel2
        cross_psd = np.conj(coefficients1) * coefficients2

        # Compute the coherence between channel1 and channel2
        coherence = np.abs(cross_psd)**2 / (psd1 * psd2)

        # Add the coherence to the list for the current channel
        channel_coherence.append(coherence)

    # Convert the list of coherences to a 2D array and store in the dictionary
    coherence_results[channel1] = np.array(channel_coherence)

# Set a lower threshold for the colormap
for channel in coherence_results:
    threshold = np.percentile(coherence_results[channel], 5)
    coherence_results[channel][coherence_results[channel] < threshold] =␣
 ↪threshold

# Plot and save the coherence for each channel pair (Optional)
for channel1 in eeg_channels:
    for channel2 in eeg_channels:
        coherence = coherence_results[channel1][eeg_channels.index(channel2)]
        plt.imshow(coherence, extent=[0, len(eeg_data_array), frequencies[-1],␣
 ↪frequencies[0]], aspect='auto', cmap='inferno')
        plt.colorbar(label='Coherence')
        plt.xlabel('Time [s]')
        plt.ylabel('Frequency [Hz]')
        plt.title(f'Coherence between EEG Channels {channel1} and {channel2}')
        plt.show()
```

```
# Save the coherence results as a numpy array
results_path = "/home/vincent/AAA_projects/MVCS/Neuroscience/Analysis/Spectral␣
↪Analysis/"
results_file = "Coherence_x.npy"
np.save(results_path + results_file, coherence_results)
```

## 22 Load and print head

[ ]:
```
# Load the coherence results data from the .npy file
coherence_results = np.load('/home/vincent/AAA_projects/MVCS/Neuroscience/
↪Analysis/Spectral Analysis/Coherence_x.npy', allow_pickle=True).item()

# Print the populated keys in coherence_results
print("Populated keys in coherence_results:", list(coherence_results.keys()))

# Print the coherence results for the first few channel pairs
for channel1, coherence_matrix in coherence_results.items():
    for channel2, coherence_values in coherence_matrix.items():
        print(f"Channels: {channel1}, {channel2}")
        print("Coherence values:")
        print(coherence_values[:10])  # Print the first 10 coherence values
        print("-" * 40)  # Separator
```

## 23 Spectral Centroids

[12]:
```
import numpy as np
import scipy.fft
import matplotlib.pyplot as plt

# Define a list of all EEG channel names
eeg_channels = ['Fp1', 'Fpz', 'Fp2', 'F7', 'F3', 'Fz', 'F4', 'F8', 'FC5',␣
↪'FC1', 'FC2', 'FC6',
                'M1', 'T7', 'C3', 'Cz', 'C4', 'T8', 'M2', 'CP5', 'CP1', 'CP2',␣
↪'CP6',
                'P7', 'P3', 'Pz', 'P4', 'P8', 'POz', 'O1', 'Oz', 'O2']

# Define the sampling frequency
fs = 1000

# Initialize a dictionary to store the spectral centroids for each channel
spectral_centroids = {}

# Loop through each channel and calculate spectral centroid
for channel in eeg_channels:
```

```
    eeg_data = eeg_data_array[:, eeg_channels.index(channel)]  # Get the data␣
↪from the array

    # Perform Fourier transform on the data
    fft_result = scipy.fft.fft(eeg_data)

    # Generate frequencies associated with the Fourier transform values
    frequencies = scipy.fft.fftfreq(len(eeg_data), 1.0/fs)

    # Calculate the absolute values of the Fourier transform results
    magnitude = np.abs(fft_result)

    # Compute the spectral centroid
    spectral_centroid = np.sum(frequencies * magnitude) / np.sum(magnitude)

    # Add the spectral centroid to the dictionary
    spectral_centroids[channel] = spectral_centroid

# Now, you have the spectral_centroids dictionary that contains spectral␣
↪centroids for each channel

# Save the spectral centroid results as a numpy array
results_path = "/home/vincent/AAA_projects/MVCS/Neuroscience/Analysis/Spectral␣
↪Analysis/"
results_file = "SpectralCentroids_x.npy"
np.save(results_path + results_file, spectral_centroids)

# Plot the spectral centroids for each EEG channel
plt.bar(spectral_centroids.keys(), spectral_centroids.values())
plt.xlabel('EEG Channel')
plt.ylabel('Spectral Centroid')
plt.title('Spectral Centroids for EEG Channels')
plt.xticks(rotation=45)
plt.tight_layout()
plt.show()
```

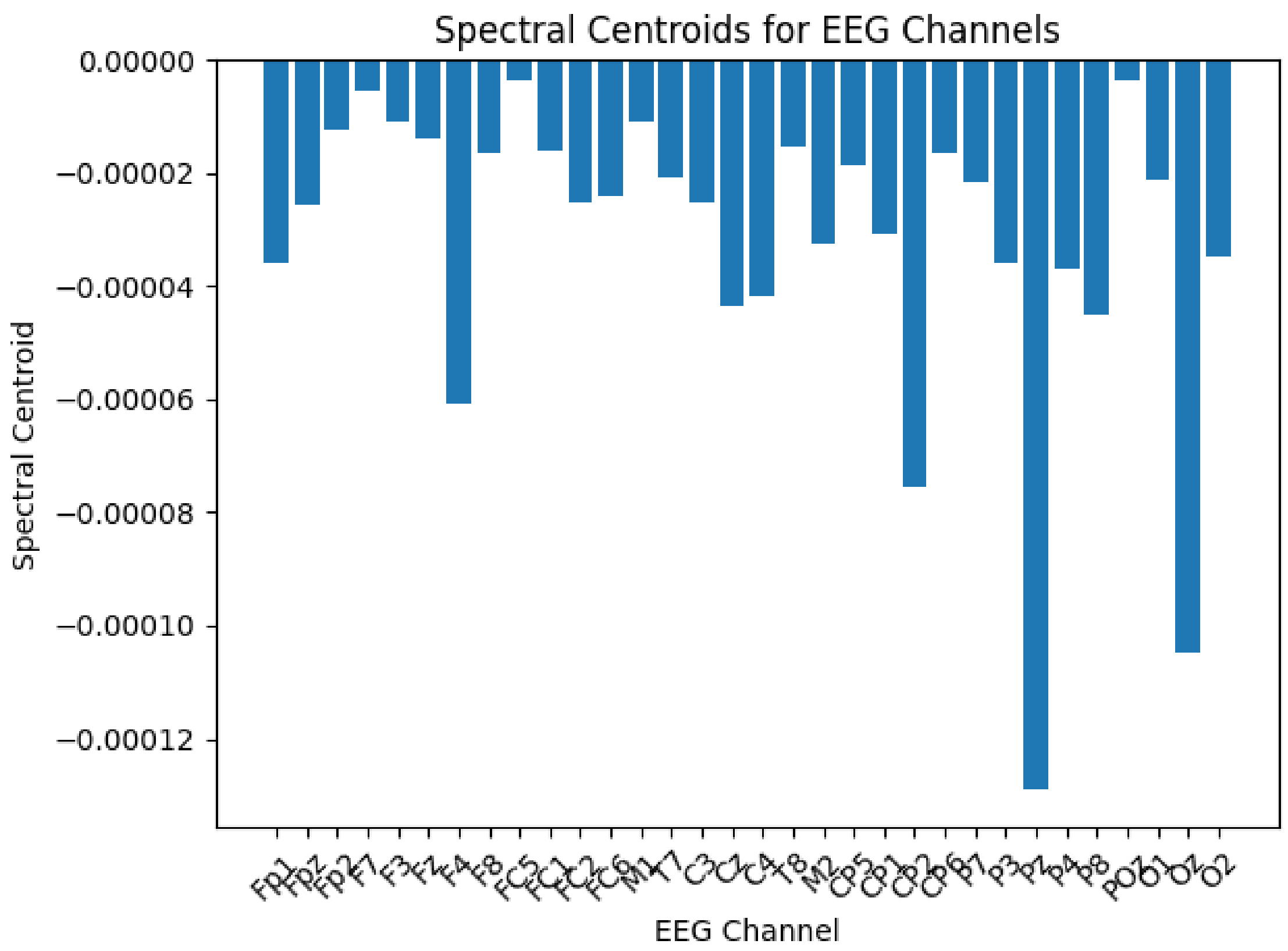

## 24 Load and print head

[10]:
```python
# Load the spectral centroid data from the .npy file
spectral_centroids = np.load('/home/vincent/AAA_projects/MVCS/Neuroscience/
↪Analysis/Spectral Analysis/SpectralCentroids_x.npy', allow_pickle=True).
↪item()

# Print the populated keys in spectral_centroids
print("Populated keys in spectral_centroids:", list(spectral_centroids.keys()))

# Print the spectral centroids for the first few EEG channels
for channel, centroid in spectral_centroids.items():
    print(f"Channel: {channel}")
    print(f"Spectral Centroid: {centroid}")
    print("-" * 40)  # Separator
```

```
Populated keys in spectral_centroids: ['Fp1', 'Fpz', 'Fp2', 'F7', 'F3', 'Fz',
'F4', 'F8', 'FC5', 'FC1', 'FC2', 'FC6', 'M1', 'T7', 'C3', 'Cz', 'C4', 'T8',
'M2', 'CP5', 'CP1', 'CP2', 'CP6', 'P7', 'P3', 'Pz', 'P4', 'P8', 'POz', 'O1',
'Oz', 'O2']
```

```
Channel: Fp1
Spectral Centroid: -3.593545514225202e-05
----------------------------------------
Channel: Fpz
Spectral Centroid: -2.5613096597273488e-05
----------------------------------------
Channel: Fp2
Spectral Centroid: -1.2315445713774716e-05
----------------------------------------
Channel: F7
Spectral Centroid: -5.486266942743079e-06
----------------------------------------
Channel: F3
Spectral Centroid: -1.099901629376439e-05
----------------------------------------
Channel: Fz
Spectral Centroid: -1.3855468322777812e-05
----------------------------------------
Channel: F4
Spectral Centroid: -6.075408543040463e-05
----------------------------------------
Channel: F8
Spectral Centroid: -1.639654923447864e-05
----------------------------------------
Channel: FC5
Spectral Centroid: -3.395238061466166e-06
----------------------------------------
Channel: FC1
Spectral Centroid: -1.6109449088624643e-05
----------------------------------------
Channel: FC2
Spectral Centroid: -2.5349876598141467e-05
----------------------------------------
Channel: FC6
Spectral Centroid: -2.4072469880506956e-05
----------------------------------------
Channel: M1
Spectral Centroid: -1.0765491037902997e-05
----------------------------------------
Channel: T7
Spectral Centroid: -2.0786776618522787e-05
----------------------------------------
Channel: C3
Spectral Centroid: -2.517153941445714e-05
----------------------------------------
Channel: Cz
Spectral Centroid: -4.339085149750505e-05
----------------------------------------
```

```
Channel: C4
Spectral Centroid: -4.179377299437049e-05
----------------------------------------
Channel: T8
Spectral Centroid: -1.5103055437685193e-05
----------------------------------------
Channel: M2
Spectral Centroid: -3.259785031268288e-05
----------------------------------------
Channel: CP5
Spectral Centroid: -1.8660120595603718e-05
----------------------------------------
Channel: CP1
Spectral Centroid: -3.072225197616467e-05
----------------------------------------
Channel: CP2
Spectral Centroid: -7.528066588498872e-05
----------------------------------------
Channel: CP6
Spectral Centroid: -1.6471443889052047e-05
----------------------------------------
Channel: P7
Spectral Centroid: -2.1496236341905074e-05
----------------------------------------
Channel: P3
Spectral Centroid: -3.576227575505989e-05
----------------------------------------
Channel: Pz
Spectral Centroid: -0.0001290257428602814
----------------------------------------
Channel: P4
Spectral Centroid: -3.693363548002953e-05
----------------------------------------
Channel: P8
Spectral Centroid: -4.509949805294648e-05
----------------------------------------
Channel: POz
Spectral Centroid: -3.386011519003128e-06
----------------------------------------
Channel: O1
Spectral Centroid: -2.1081761103595764e-05
----------------------------------------
Channel: Oz
Spectral Centroid: -0.00010472145081458061
----------------------------------------
Channel: O2
Spectral Centroid: -3.4539891586534286e-05
----------------------------------------
```

## 25 Frequency of Maximum Power

[19]:
```python
import numpy as np
import scipy.fft
import matplotlib.pyplot as plt

# Define a list of all EEG channel names
eeg_channels = ['Fp1', 'Fpz', 'Fp2', 'F7', 'F3', 'Fz', 'F4', 'F8', 'FC5',␣
↪'FC1', 'FC2', 'FC6',
                'M1', 'T7', 'C3', 'Cz', 'C4', 'T8', 'M2', 'CP5', 'CP1', 'CP2',␣
↪'CP6',
                'P7', 'P3', 'Pz', 'P4', 'P8', 'POz', 'O1', 'Oz', 'O2']

# Define the sampling frequency
fs = 1000

# Initialize a dictionary to store the frequencies of maximum power for each␣
↪channel
peak_frequencies = {}

# Set up the figure layout
num_rows = (len(eeg_channels) + 3) // 4
num_cols = 4

# Create subplots
fig, axs = plt.subplots(num_rows, num_cols, figsize=(10, 3 * num_rows))
axs = axs.ravel()

# Loop through each channel and calculate frequency of maximum power
for i, channel in enumerate(eeg_channels):
    eeg_data = eeg_data_array[:, eeg_channels.index(channel)]  # Get the data␣
↪from the array

    # Perform Fourier transform on the data
    fft_result = scipy.fft.fft(eeg_data)

    # Generate frequencies associated with the Fourier transform values
    frequencies = scipy.fft.fftfreq(len(eeg_data), 1.0/fs)

    # Keep only the positive frequencies (since the spectrum is symmetric)
    positive_frequencies = frequencies[frequencies >= 0]
    positive_fft_result = fft_result[frequencies >= 0]

    # Find the frequency where the absolute value of the Fourier transform is␣
↪maximum
    peak_frequency = positive_frequencies[np.argmax(np.
↪abs(positive_fft_result))]
```

```
    # Add the peak frequency to the dictionary
    peak_frequencies[channel] = peak_frequency

    # Plot the power spectral density and mark the peak frequency
    axs[i].plot(positive_frequencies, np.abs(positive_fft_result), label='Power␣
↪Spectral Density')
    axs[i].axvline(peak_frequency, color='r', linestyle='--', label=f'Peak␣
↪Frequency = {peak_frequency} Hz')
    axs[i].set_xscale('log')
    axs[i].set_yscale('log')
    axs[i].set_title(f'Channel {channel}')
    axs[i].set_xlabel('Frequency [Hz]')
    axs[i].set_ylabel('Power Spectral Density')
    axs[i].legend()
    axs[i].grid(True)

# Adjust spacing and layout
plt.tight_layout()
plt.show()

# Save the peak frequencies as a numpy array
results_path = "/home/vincent/AAA_projects/MVCS/Neuroscience/Analysis/Spectral␣
↪Analysis/"
results_file = "PeakFrequencies_x.npy"
np.save(results_path + results_file, peak_frequencies)
```

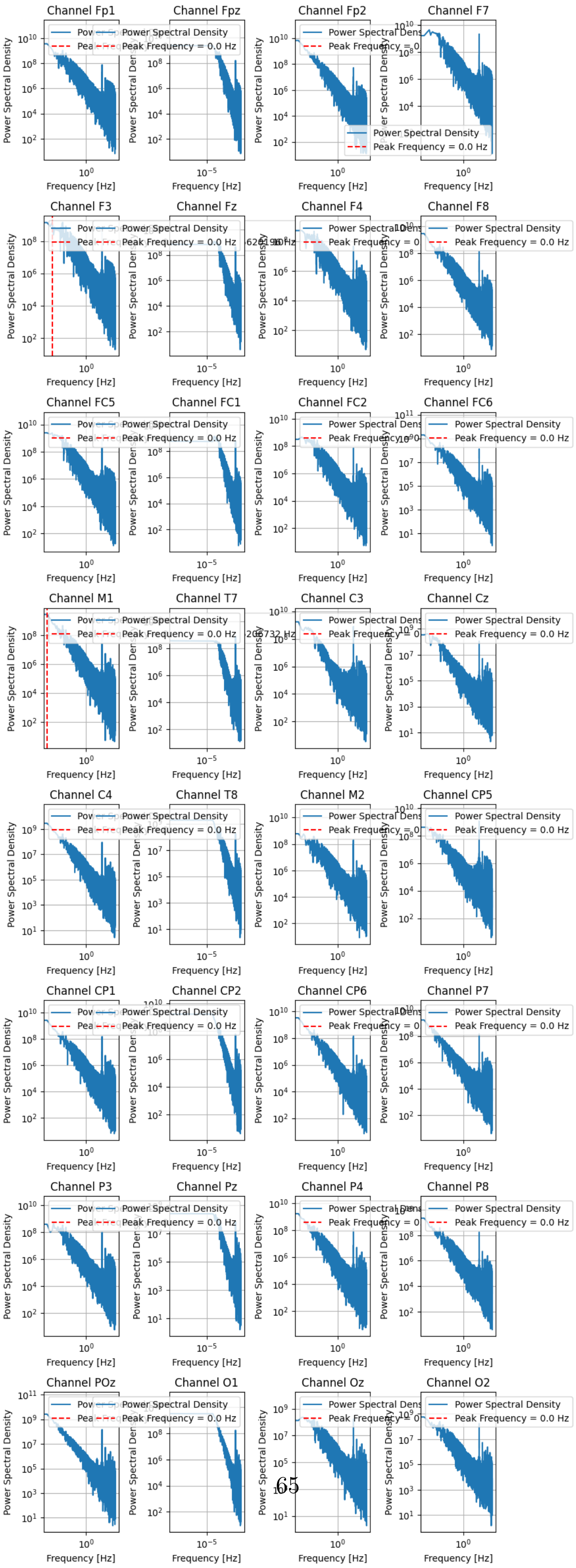
Channel Fp1
Channel Fpz
Channel Fp2
Channel F7
Channel F3
Channel Fz
Channel F4
Channel F8
Channel FC5
Channel FC1
Channel FC2
Channel FC6
Channel M1
Channel T7
Channel C3
Channel Cz
Channel C4
Channel T8
Channel M2
Channel CP5
Channel CP1
Channel CP2
Channel CP6
Channel P7
Channel P3
Channel Pz
Channel P4
Channel P8
Channel POz
Channel O1
Channel Oz
Channel O2
Power Spectral Density
Peak Frequency = 0.0 Hz
Frequency [Hz]

## 26 Load and print head

[9]:
```
# Load the peak frequency data from the .npy file
peak_frequencies = np.load('/home/vincent/AAA_projects/MVCS/Neuroscience/
↪Analysis/Spectral Analysis/PeakFrequencies_x.npy', allow_pickle=True).item()

# Print the populated keys in peak_frequencies
print("Populated keys in peak_frequencies:", list(peak_frequencies.keys()))

# Print the peak frequencies for the first few EEG channels
for channel, peak_frequency in peak_frequencies.items():
    print(f"Channel: {channel}")
    print(f"Peak Frequency: {peak_frequency} Hz")
    print("-" * 40)  # Separator
```

```
Populated keys in peak_frequencies: ['Fp1', 'Fpz', 'Fp2', 'F7', 'F3', 'Fz',
'F4', 'F8', 'FC5', 'FC1', 'FC2', 'FC6', 'M1', 'T7', 'C3', 'Cz', 'C4', 'T8',
'M2', 'CP5', 'CP1', 'CP2', 'CP6', 'P7', 'P3', 'Pz', 'P4', 'P8', 'POz', 'O1',
'Oz', 'O2']
Channel: Fp1
Peak Frequency: 0.0 Hz
----------------------------------------
Channel: Fpz
Peak Frequency: 0.0 Hz
----------------------------------------
Channel: Fp2
Peak Frequency: 0.0 Hz
----------------------------------------
Channel: F7
Peak Frequency: 0.0 Hz
----------------------------------------
Channel: F3
Peak Frequency: 0.0007095909255620196 Hz
----------------------------------------
Channel: Fz
Peak Frequency: 0.0 Hz
----------------------------------------
Channel: F4
Peak Frequency: 0.0 Hz
----------------------------------------
Channel: F8
Peak Frequency: 0.0 Hz
----------------------------------------
Channel: FC5
```

```
Peak Frequency: 0.0 Hz
----------------------------------------
Channel: FC1
Peak Frequency: 0.0 Hz
----------------------------------------
Channel: FC2
Peak Frequency: 0.0 Hz
----------------------------------------
Channel: FC6
Peak Frequency: 0.0 Hz
----------------------------------------
Channel: M1
Peak Frequency: 0.0002365303085206732 Hz
----------------------------------------
Channel: T7
Peak Frequency: 0.0 Hz
----------------------------------------
Channel: C3
Peak Frequency: 0.0 Hz
----------------------------------------
Channel: Cz
Peak Frequency: 0.0 Hz
----------------------------------------
Channel: C4
Peak Frequency: 0.0 Hz
----------------------------------------
Channel: T8
Peak Frequency: 0.0 Hz
----------------------------------------
Channel: M2
Peak Frequency: 0.0 Hz
----------------------------------------
Channel: CP5
Peak Frequency: 0.0 Hz
----------------------------------------
Channel: CP1
Peak Frequency: 0.0 Hz
----------------------------------------
Channel: CP2
Peak Frequency: 0.0 Hz
----------------------------------------
Channel: CP6
Peak Frequency: 0.0 Hz
----------------------------------------
Channel: P7
Peak Frequency: 0.0 Hz
----------------------------------------
Channel: P3
```

```
Peak Frequency: 0.0 Hz
----------------------------------------
Channel: Pz
Peak Frequency: 0.0 Hz
----------------------------------------
Channel: P4
Peak Frequency: 0.0 Hz
----------------------------------------
Channel: P8
Peak Frequency: 0.0 Hz
----------------------------------------
Channel: POz
Peak Frequency: 0.0 Hz
----------------------------------------
Channel: O1
Peak Frequency: 0.0 Hz
----------------------------------------
Channel: Oz
Peak Frequency: 0.0 Hz
----------------------------------------
Channel: O2
Peak Frequency: 0.0 Hz
----------------------------------------
```

## 27 Spectral Edge Frequency

[21]:
```
import numpy as np
import scipy.fft
import matplotlib.pyplot as plt

# Define a list of all EEG channel names
eeg_channels = ['Fp1', 'Fp2', 'F7', 'F3', 'Fz', 'F4', 'F8', 'FC5', 'FC1', 'FC2',
                'FC6', 'M1', 'T7', 'C3', 'Cz', 'C4', 'T8', 'M2', 'CP5', 'CP1',␣
 ↪'CP2',
                'CP6', 'P7', 'P3', 'Pz', 'P4', 'P8', 'POz', 'O1', 'Oz', 'O2',␣
 ↪'Fpz']

# Define the sampling frequency
fs = 1000

# Define the percentage of the total power used to calculate the spectral edge␣
 ↪density
percentage = 95

# Initialize a dictionary to store the spectral edge densities for each channel
spectral_edge_densities = {}
```

```python
# Loop through each channel and calculate spectral edge density
for channel in eeg_channels:
    eeg_data = eeg_data_array[:, eeg_channels.index(channel)]  # Get the data␣
↪from the array

    # Perform Fourier transform on the data
    fft_result = scipy.fft.fft(eeg_data)

    # Generate frequencies associated with the Fourier transform values
    frequencies = scipy.fft.fftfreq(len(eeg_data), 1.0/fs)

    # Keep only the positive frequencies (since the spectrum is symmetric)
    positive_frequencies = frequencies[frequencies >= 0]
    positive_fft_result = fft_result[frequencies >= 0]

    # Calculate the absolute values of the Fourier transform results
    magnitude = np.abs(positive_fft_result)

    # Sort the magnitude array in descending order
    sorted_magnitude = np.sort(magnitude)[::-1]

    # Calculate the cumulative sum of the sorted magnitude array
    cumulative_sum = np.cumsum(sorted_magnitude)

    # Calculate the threshold based on the specified percentage of total power
    total_power = np.sum(magnitude)
    threshold = total_power * percentage / 100

    # Find the frequency where the cumulative sum first exceeds the threshold
    spectral_edge = positive_frequencies[np.argmax(cumulative_sum >= threshold)]

    # Add the spectral edge density to the dictionary
    spectral_edge_densities[channel] = spectral_edge

# Plot the spectral edge densities for each channel
plt.figure(figsize=(10, 5))
plt.bar(spectral_edge_densities.keys(), spectral_edge_densities.values())
plt.title(f'Spectral Edge Density ({percentage}% of Total Power) for Each EEG␣
↪Channel')
plt.xlabel('EEG Channel')
plt.ylabel('Spectral Edge Density [Hz]')
plt.xticks(rotation=90)
plt.grid(True)
plt.show()

# Save the spectral edge densities as a numpy array
```

```
results_path = "/home/vincent/AAA_projects/MVCS/Neuroscience/Analysis/Spectral␣
↪Analysis/"
results_file = "SpectralEdgeDensities_x.npy"
np.save(results_path + results_file, spectral_edge_densities)
```

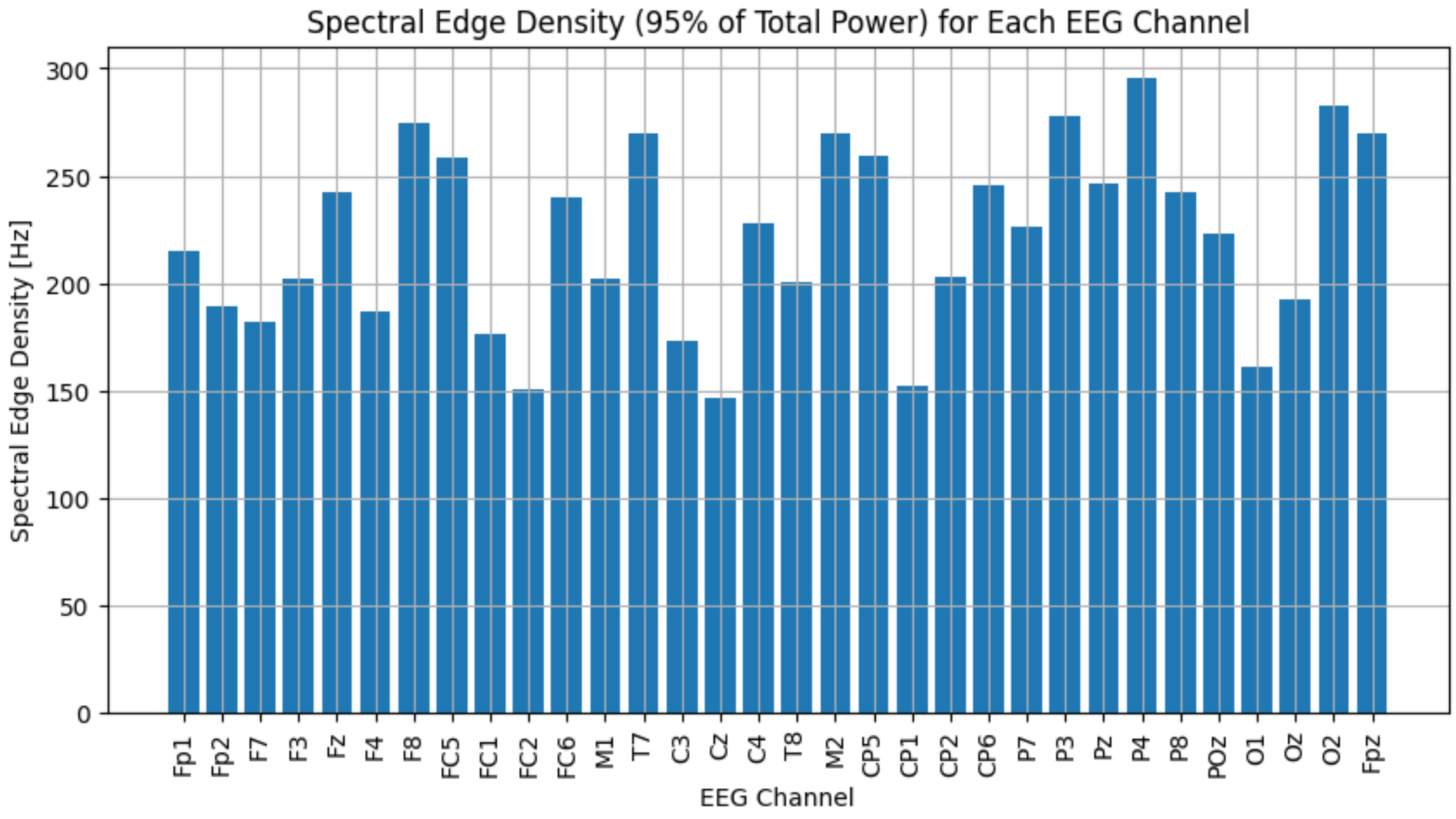

# 28 Load and print head

[8]:
```
# Load the spectral edge density data from the .npy file
spectral_edge_densities = np.load('/home/vincent/AAA_projects/MVCS/Neuroscience/
↪Analysis/Spectral Analysis/SpectralEdgeDensities_x.npy', allow_pickle=True).
↪item()

# Print the populated keys in spectral_edge_densities
print("Populated keys in spectral_edge_densities:",␣
↪list(spectral_edge_densities.keys()))

# Print the spectral edge densities for the first few EEG channels
for channel, spectral_edge_density in spectral_edge_densities.items():
    print(f"Channel: {channel}")
    print(f"Spectral Edge Density: {spectral_edge_density} Hz")
    print("-" * 40)  # Separator
```

```
Populated keys in spectral_edge_densities: ['Fp1', 'Fp2', 'F7', 'F3', 'Fz',
'F4', 'F8', 'FC5', 'FC1', 'FC2', 'FC6', 'M1', 'T7', 'C3', 'Cz', 'C4', 'T8',
'M2', 'CP5', 'CP1', 'CP2', 'CP6', 'P7', 'P3', 'Pz', 'P4', 'P8', 'POz', 'O1',
```

```
'Oz', 'O2', 'Fpz']
Channel: Fp1
Spectral Edge Density: 214.62878460320147 Hz
----------------------------------------
Channel: Fp2
Spectral Edge Density: 189.3541019559164 Hz
----------------------------------------
Channel: F7
Spectral Edge Density: 181.7248168545821 Hz
----------------------------------------
Channel: F3
Spectral Edge Density: 202.12437331294754 Hz
----------------------------------------
Channel: Fz
Spectral Edge Density: 241.97618234405317 Hz
----------------------------------------
Channel: F4
Spectral Edge Density: 186.44194079740987 Hz
----------------------------------------
Channel: F8
Spectral Edge Density: 274.38556521755584 Hz
----------------------------------------
Channel: FC5
Spectral Edge Density: 258.79845441635194 Hz
----------------------------------------
Channel: FC1
Spectral Edge Density: 175.94165081125163 Hz
----------------------------------------
Channel: FC2
Spectral Edge Density: 150.41648256724318 Hz
----------------------------------------
Channel: FC6
Spectral Edge Density: 239.54488730276915 Hz
----------------------------------------
Channel: M1
Spectral Edge Density: 201.92852621749242 Hz
----------------------------------------
Channel: T7
Spectral Edge Density: 269.8981122043016 Hz
----------------------------------------
Channel: C3
Spectral Edge Density: 173.13758400373905 Hz
----------------------------------------
Channel: Cz
Spectral Edge Density: 146.11612502802882 Hz
----------------------------------------
Channel: C4
Spectral Edge Density: 227.66302378454168 Hz
```

```
----------------------------------------
Channel: T8
Spectral Edge Density: 200.25105326946382 Hz
----------------------------------------
Channel: M2
Spectral Edge Density: 269.9418703113779 Hz
----------------------------------------
Channel: CP5
Spectral Edge Density: 259.54281529726654 Hz
----------------------------------------
Channel: CP1
Spectral Edge Density: 151.9976876797039 Hz
----------------------------------------
Channel: CP2
Spectral Edge Density: 203.01538298514492 Hz
----------------------------------------
Channel: CP6
Spectral Edge Density: 245.24124672287255 Hz
----------------------------------------
Channel: P7
Spectral Edge Density: 226.55180439511156 Hz
----------------------------------------
Channel: P3
Spectral Edge Density: 278.09294127330884 Hz
----------------------------------------
Channel: Pz
Spectral Edge Density: 246.2912047623958 Hz
----------------------------------------
Channel: P4
Spectral Edge Density: 295.5122158443138 Hz
----------------------------------------
Channel: P8
Spectral Edge Density: 241.97736499559576 Hz
----------------------------------------
Channel: POz
Spectral Edge Density: 223.30164142572897 Hz
----------------------------------------
Channel: O1
Spectral Edge Density: 161.33826956318526 Hz
----------------------------------------
Channel: Oz
Spectral Edge Density: 192.0242926088063 Hz
----------------------------------------
Channel: O2
Spectral Edge Density: 282.4588177079834 Hz
----------------------------------------
Channel: Fpz
Spectral Edge Density: 269.6674951534939 Hz
```

----------------------------------------

# 29 Continuous Wavelet Transform

[ ]:
```
import numpy as np
import pywt
import matplotlib.pyplot as plt

# Define a list of all EEG channel names
eeg_channels = ['Fp1', 'Fpz', 'Fp2', 'F7', 'F3', 'Fz', 'F4', 'F8', 'FC5',␣
↪'FC1', 'FC2', 'FC6',
                'M1', 'T7', 'C3', 'Cz', 'C4', 'T8', 'M2', 'CP5', 'CP1', 'CP2',␣
↪'CP6',
                'P7', 'P3', 'Pz', 'P4', 'P8', 'POz', 'O1', 'Oz', 'O2']

# Define the sampling frequency
fs = 1000  # Adjust this to your actual sampling frequency

# Define the frequencies at which to compute the wavelet transform
frequencies = np.logspace(np.log10(1), np.log10(fs/2), num=100)  # Adjust the␣
↪scale range as needed

# Initialize a dictionary to store the wavelet coefficients for each channel
wavelet_coefficients = {}

# Loop through each channel and compute wavelet coefficients
for channel in eeg_channels:
    eeg_data = eeg_data_array[:, eeg_channels.index(channel)]  # Get the data␣
↪from the array
    coefficients, freqs = pywt.cwt(eeg_data, frequencies, wavelet='morl')
    wavelet_coefficients[channel] = coefficients

    # Plot the wavelet coefficients
    plt.figure(figsize=(10, 6))
    plt.imshow(np.abs(coefficients), extent=[0, len(eeg_data), freqs[-1],␣
↪freqs[0]], aspect='auto', cmap='inferno')
    plt.colorbar(label='Magnitude')
    plt.xlabel('Time [s]')
    plt.ylabel('Frequency [Hz]')
    plt.title(f'Wavelet Coefficients for EEG Channel {channel}')
    plt.show()

# Save the wavelet coefficients as a numpy array
results_path = "/home/vincent/AAA_projects/MVCS/Neuroscience/Analysis/Spectral␣
↪Analysis/"
results_file = "WaveletCoefficients_x.npy"
```

```
np.save(results_path + results_file, wavelet_coefficients)
```

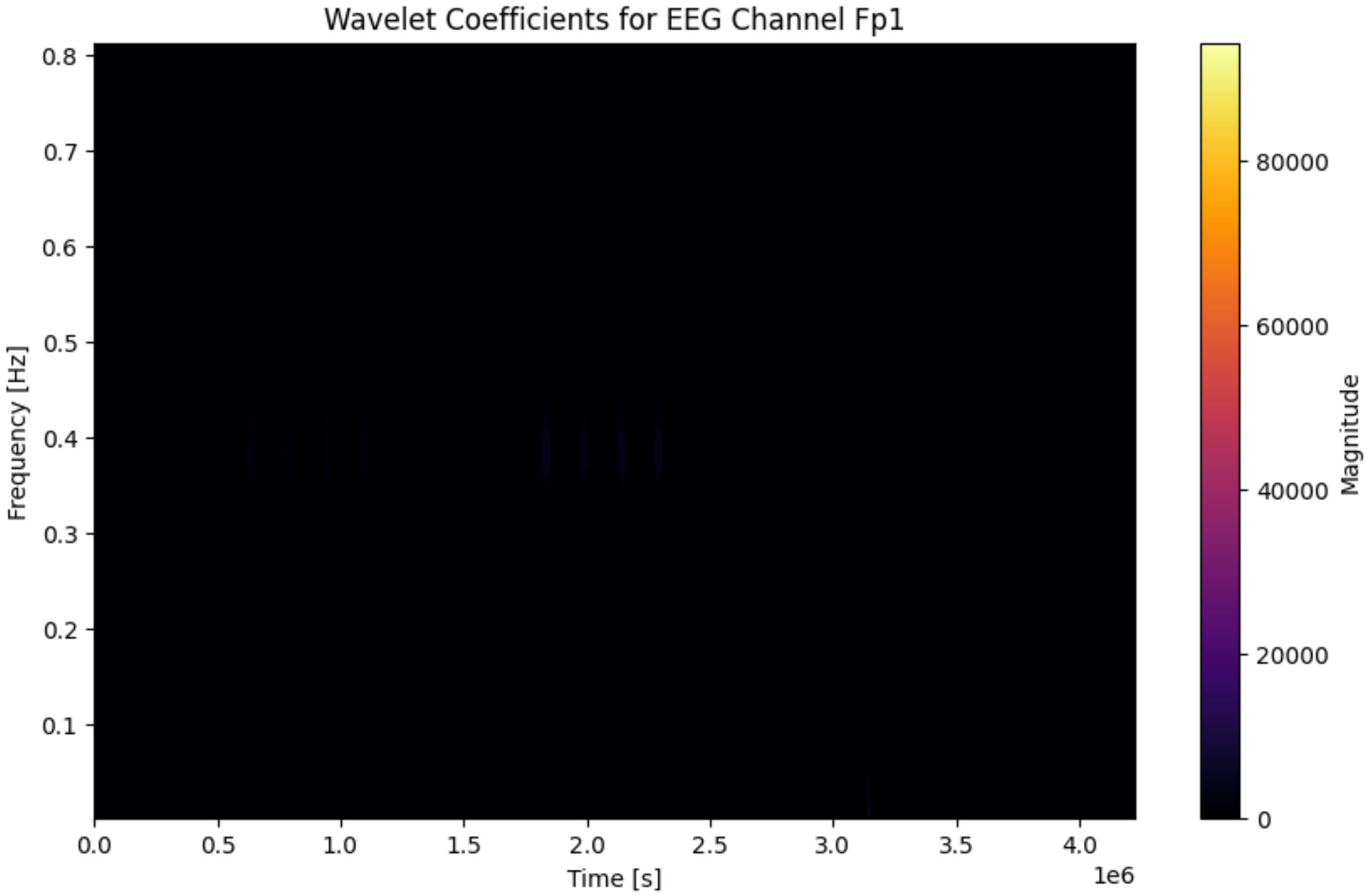

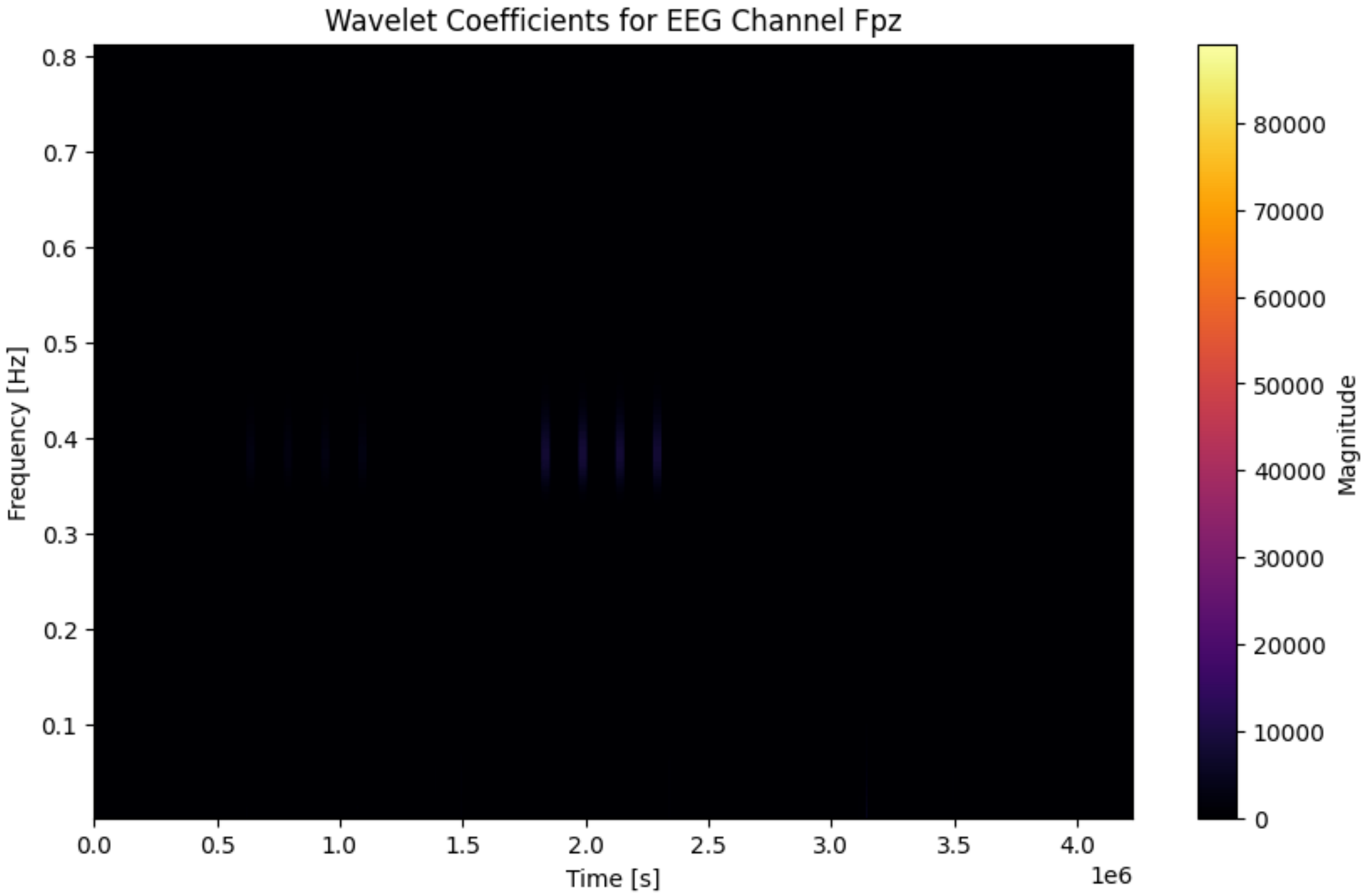

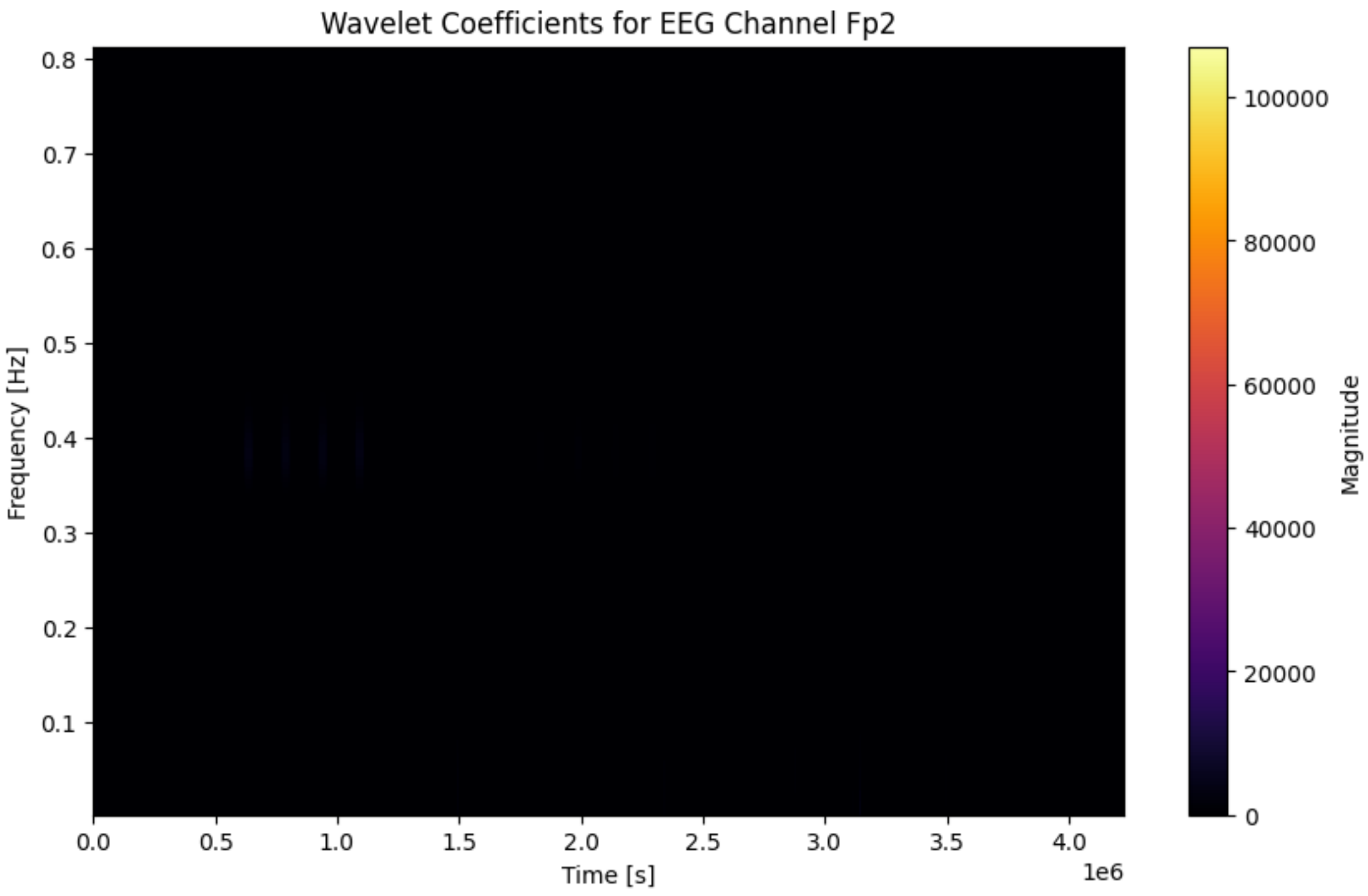

Wavelet Coefficients for EEG Channel Fp2
Frequency [Hz]
0.8
0.7
0.6
0.5
0.4
0.3
0.2
0.1
0.0
0.5
1.0
1.5
2.0
2.5
3.0
3.5
4.0
1e6
Time [s]
Magnitude
100000
80000
60000
40000
20000
0

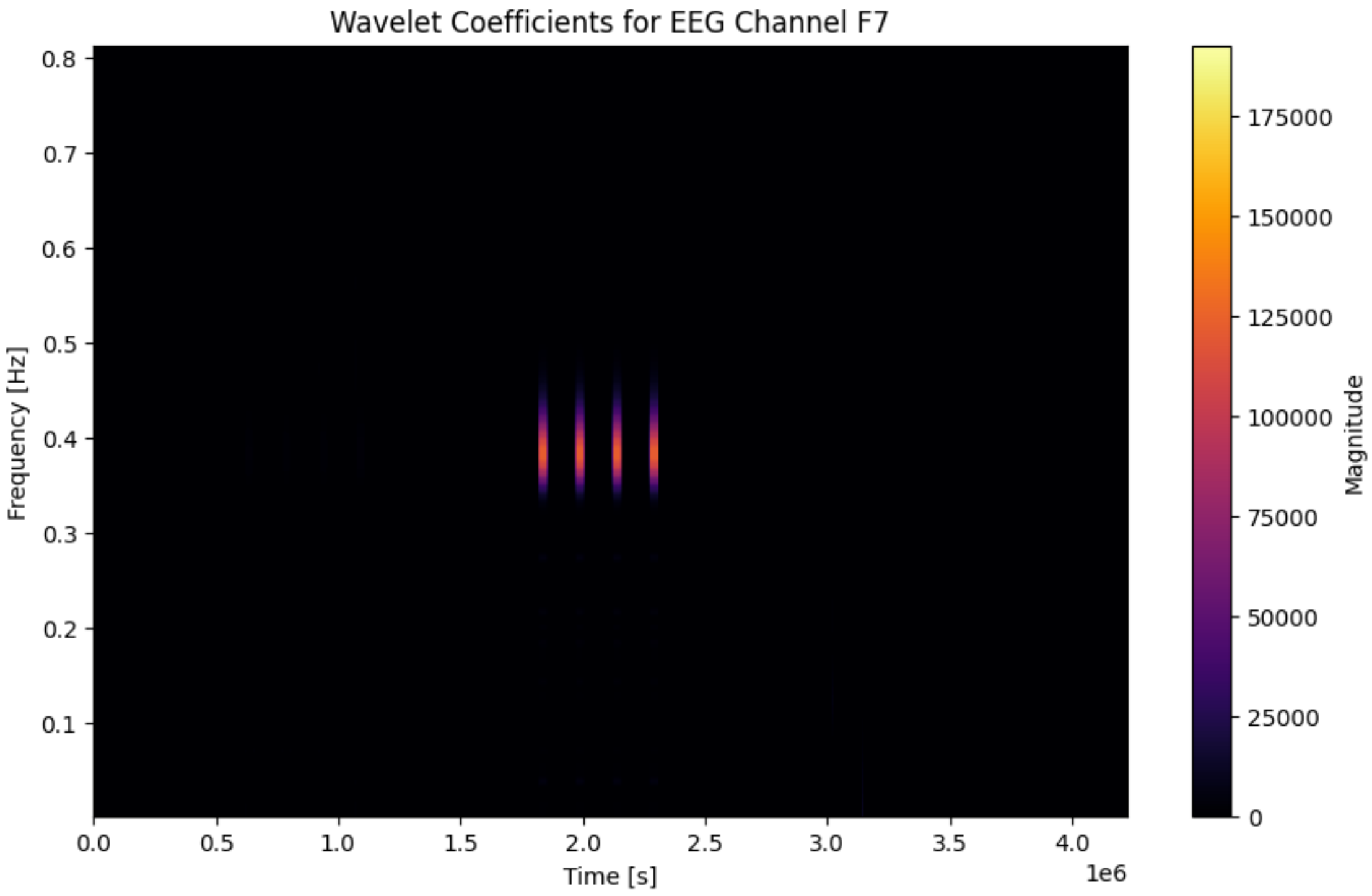

Wavelet Coefficients for EEG Channel F7
Frequency [Hz]
0.8
0.7
0.6
0.5
0.4
0.3
0.2
0.1
0.0
0.5
1.0
1.5
2.0
2.5
3.0
3.5
4.0
1e6
Time [s]
Magnitude
175000
150000
125000
100000
75000
50000
25000
0

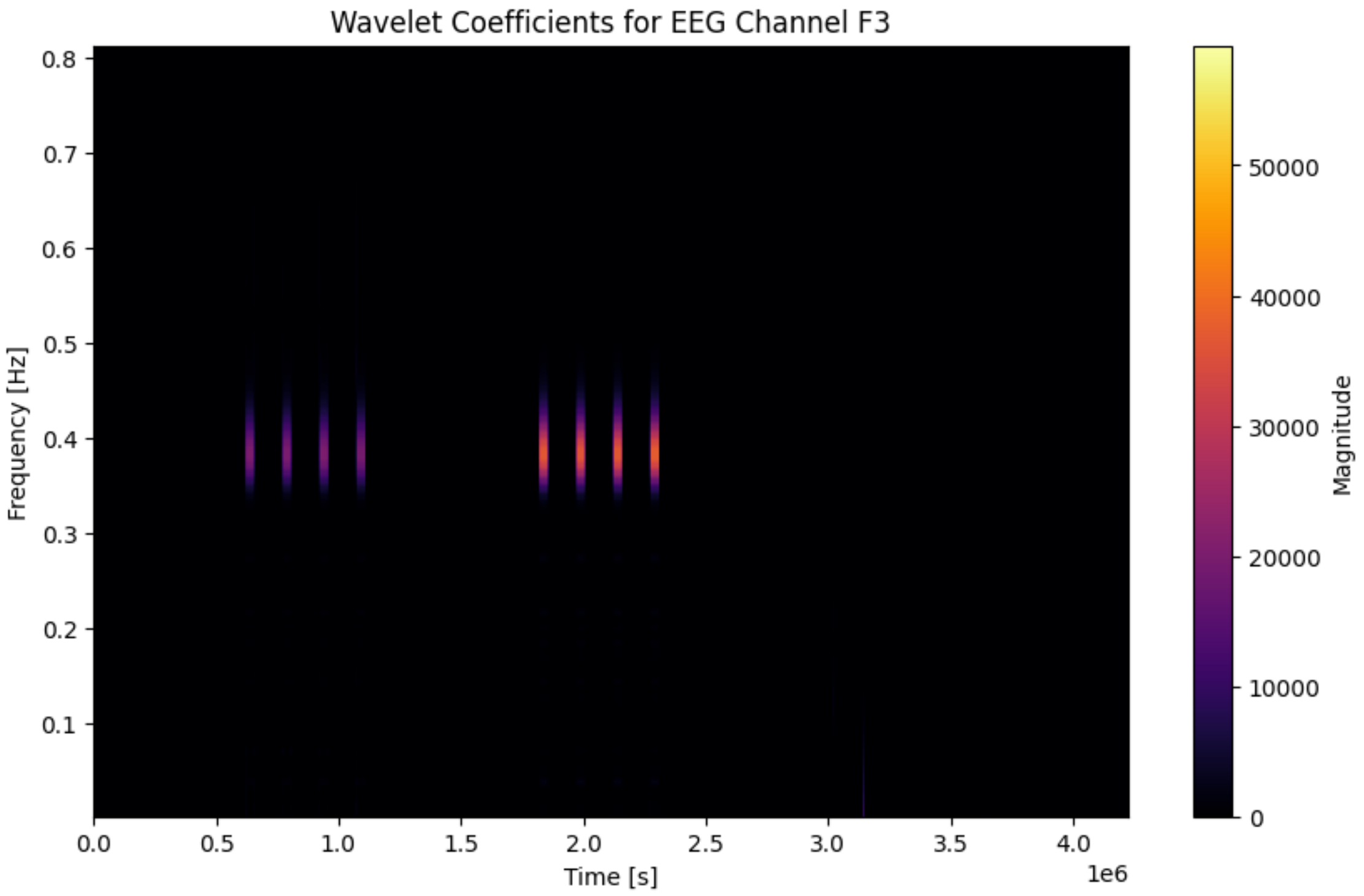

# 30 Load and print head

[ ]:
```
# Load the wavelet coefficients data from the .npy file
wavelet_coefficients = np.load('/home/vincent/AAA_projects/MVCS/Neuroscience/
↪Analysis/Spectral Analysis/WaveletCoefficients_x.npy', allow_pickle=True).
↪item()

# Print the populated keys in wavelet_coefficients
print("Populated keys in wavelet_coefficients:", list(wavelet_coefficients.
↪keys()))

# Print the shape of wavelet coefficients for the first EEG channel
first_channel = next(iter(wavelet_coefficients))
coefficients_shape = wavelet_coefficients[first_channel].shape
print(f"Shape of wavelet coefficients for {first_channel}:␣
↪{coefficients_shape}")

# Print the wavelet coefficients for the first few EEG channels
for channel, coefficients in wavelet_coefficients.items():
    print(f"Channel: {channel}")
    print(f"Wavelet Coefficients Shape: {coefficients.shape}")
    print("-" * 40)  # Separator
```

# EEG Channel Network

September 8, 2023

## 1 Network

#### 1.0.1 Pearson

[25]:
```
import numpy as np
import matplotlib.pyplot as plt
from fa2 import ForceAtlas2
import networkx as nx
from scipy.sparse import lil_matrix

# EEG Channel Names
eeg_channel_names = ['Fp1', 'Fpz', 'Fp2', 'F7', 'F3', 'Fz', 'F4', 'F8', 'FC5',␣
 ↪'FC1', 'FC2', 'FC6',
                     'M1', 'T7', 'C3', 'Cz', 'C4', 'T8', 'M2', 'CP5', 'CP1',␣
 ↪'CP2', 'CP6',
                     'P7', 'P3', 'Pz', 'P4', 'P8', 'POz', 'O1', 'Oz', 'O2']

# Load EEG data
eeg_data_array = np.load('/home/vincent/AAA_projects/MVCS/Neuroscience/
 ↪eeg_data_with_channels.npy', allow_pickle=True)

# Compute connectivity matrix (Pearson correlation)
corr_matrix = np.corrcoef(eeg_data_array.T)

# Create an empty graph
G = np.zeros_like(corr_matrix)

# Add edges to the graph based on correlation matrix
for i in range(G.shape[0]):
    for j in range(i + 1, G.shape[1]):
        weight = corr_matrix[i, j]
        if abs(weight) >= 0.5:
            G[i, j] = weight
            G[j, i] = weight

# Convert the adjacency matrix into a networkx graph
graph = nx.from_numpy_array(G)
```

```python
# Convert the graph to a SciPy sparse matrix manually
M = lil_matrix(nx.adjacency_matrix(graph))

# Use ForceAtlas2 for layout
forceatlas2 = ForceAtlas2(
    # Behavior alternatives
    outboundAttractionDistribution=True,
    linLogMode=False,
    adjustSizes=False,
    edgeWeightInfluence=1.0,

    # Performance
    jitterTolerance=1.0,
    barnesHutOptimize=False,
    barnesHutTheta=1,
    multiThreaded=False,

    # Tuning
    scalingRatio=10.0,    # Adjust this to make nodes repel each other more␣
 ↪strongly
    strongGravityMode=False,
    gravity=1.0,         # Adjust this to make nodes attract to the center more␣
 ↪or less

    # Log
    verbose=True
)

# Bypass the forceatlas2_networkx_layout method and use forceatlas2 directly
positions = forceatlas2.forceatlas2(M, pos=None, iterations=2000)

# Convert positions to dictionary format for networkx
positions_dict = {i: positions[i] for i in range(len(positions))}


# Draw the graph with labels
fig = plt.figure(figsize=(10, 10))
fig.patch.set_facecolor('black')  # Set the figure background to black

ax = plt.gca()
ax.set_facecolor('black')  # Set the axis background color to black

node_labels = {node: name for node, name in enumerate(eeg_channel_names)}
nx.draw(graph, pos=positions_dict, labels=node_labels, node_size=0,␣
 ↪font_size=10,
        edge_color="red", font_color="white", width=0.5, with_labels=True,␣
 ↪node_shape='')
```

```
plt.title("EEG Functional Connectivity Network using ForceAtlas2",␣
 ↪color="white")  # Set title color to white
plt.axis('off')  # To ensure no axis is shown
plt.show()
```

```
100%|                    | 2000/2000 [00:00<00:00, 79256.70it/s]

Repulsion forces  took  0.00  seconds
Gravitational forces  took  0.00  seconds
Attraction forces  took  0.00  seconds
AdjustSpeedAndApplyForces step  took  0.01  seconds
```

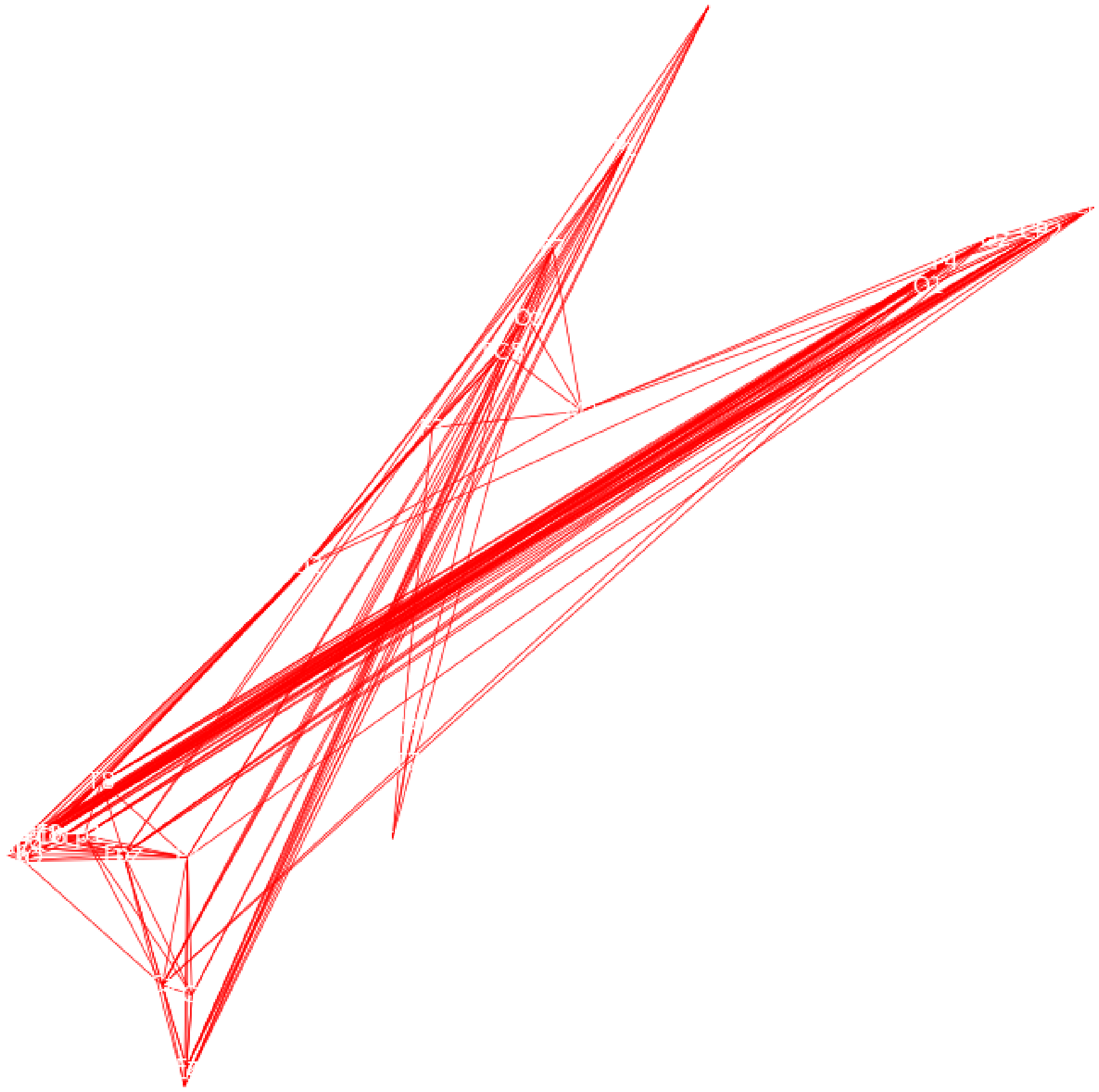

```
[ ]:
```

Higuchi Fractal Dimension

September 8, 2023

# 1 Higuchi Fractal Dimension

[8]:
```
import numpy as np

# Load EEG data
EEG_data = np.load('/home/vincent/AAA_projects/MVCS/Neuroscience/
↪eeg_data_with_channels.npy', allow_pickle=True)

# Adjust the list to match your data's channels
eeg_channels = ['Fp1', 'Fpz', 'Fp2', 'F7', 'F3', 'Fz', 'F4', 'F8', 'FC5',␣
↪'FC1', 'FC2', 'FC6',
                'M1', 'T7', 'C3', 'Cz', 'C4', 'T8', 'M2', 'CP5', 'CP1', 'CP2',␣
↪'CP6',
                'P7', 'P3', 'Pz', 'P4', 'P8', 'POz', 'O1', 'Oz', 'O2']

def higuchi_fd(data, k_max):
    """Compute Higuchi Fractal Dimension of a time series.

    Parameters:
    data : list or np.array
        One-dimensional time series
    k_max : int
        Maximum delay (time offset)

    Returns:
    hfd : float
        Higuchi Fractal Dimension
    """
    N = len(data)
    L = []

    x = np.asarray(data)

    for k in range(1, k_max):
        Lk = []

        for m in range(0, k):
```

```
            Lkm = 0
            for i in range(1, int((N-m)/k)):
                Lkm += abs(x[m+i*k] - x[m+i*k-k])
            Lkm = Lkm*(N - 1)/(((N - m)/k)*k)
            Lk.append(Lkm)

        L.append(np.log(np.mean(Lk)))

    hfd = np.polyfit(np.log(range(1, k_max)), L, 1)[0]

    return hfd

hfd_values = []

for channel_data in EEG_data.T:  # .T because we need to iterate over channels
    hfd_channel = higuchi_fd(channel_data, k_max=10)  # Adjust k_max as needed
    hfd_values.append(hfd_channel)

print("Higuchi Fractal Dimensions for each channel:", hfd_values)
```

```
Higuchi Fractal Dimensions for each channel: [-0.36005497045124624,
-0.23633437232263893, -0.37112149677594536, -0.08754714948970373,
-0.11035468235736252, -0.13707288326129607, -0.3934048831939664,
-0.24673709351805825, -0.0710953944331469, -0.10275985039837075,
-0.3467667431744175, -0.24990709690042587, -0.21426902542688608,
-0.09127876391896302, -0.08507749371463456, -0.25722852525504636,
-0.2805514117387538, -0.2224151766390623, -0.2099877581545126,
-0.0799982792792008, -0.20791211613648958, -0.33103053273893573,
-0.2269624863539871, -0.2363840209397604, -0.2704812265220445,
-0.31618623070687657, -0.2329294151814016, -0.20944694956243734,
-0.2115303234095175, -0.19603356897036747, -0.20279111854623105,
-0.2125008482366303]
```

## 2 Save the results

[9]:

```
# Save the results to a text file
results_file_path = '/home/vincent/AAA_projects/MVCS/Neuroscience/Analysis/
↪Fractal Analysis/Higuchi_Fractal_Dimensions.txt'
with open(results_file_path, 'w') as f:
    for channel, hfd in zip(eeg_channels, hfd_values):
        f.write(f"Channel {channel}: {hfd}\n")

print("Results saved to:", results_file_path)
```

```
Results saved to: /home/vincent/AAA_projects/MVCS/Neuroscience/Analysis/Fractal
Analysis/Higuchi_Fractal_Dimensions.txt
```

[4]:
```python
import numpy as np
import pandas as pd

# Load the results from the given file
results_file_path = '/home/vincent/AAA_projects/MVCS/Neuroscience/Analysis/
↪Fractal Analysis/Higuchi_Fractal_Dimensions.txt'
with open(results_file_path, 'r') as f:
    # Each line is of the form 'Channel <channel_name>: <HFD value>'
    hfd_values = [float(line.split(':')[-1].strip()) for line in f]

# Generate CNN Feature
# Reshape HFD values into a 4x8 matrix
cnn_features = np.array(hfd_values).reshape(4, 8)

# Generate RNN Feature
# Treat the HFD values as a sequence
rnn_features = np.array(hfd_values).reshape(1, -1)  # 1 sequence of HFD values␣
↪for each channel

# Save the features
save_path = '/home/vincent/AAA_projects/MVCS/Neuroscience/Features'
np.save(f"{save_path}/cnn_fractal.npy", cnn_features)
np.save(f"{save_path}/rnn_fractal.npy", rnn_features)

# Print shape and head using pandas
# For CNN feature
cnn_df = pd.DataFrame(cnn_features)
print("\nCNN Feature Shape:")
print(cnn_df.shape)
print("\nCNN Feature Head:")
print(cnn_df.head())

# For RNN feature
rnn_df = pd.DataFrame(rnn_features)
print("\nRNN Feature Shape:")
print(rnn_df.shape)
print("\nRNN Feature Head:")
print(rnn_df.head())
```

```
CNN Feature Shape:
(4, 8)

CNN Feature Head:
          0         1         2         3         4         5         6  \
0 -0.360055 -0.236334 -0.371121 -0.087547 -0.110355 -0.137073 -0.393405
1 -0.071095 -0.102760 -0.346767 -0.249907 -0.214269 -0.091279 -0.085077
```

```
2 -0.280551 -0.222415 -0.209988 -0.079998 -0.207912 -0.331031 -0.226962
3 -0.270481 -0.316186 -0.232929 -0.209447 -0.211530 -0.196034 -0.202791

          7
0 -0.246737
1 -0.257229
2 -0.236384
3 -0.212501

RNN Feature Shape:
(1, 32)

RNN Feature Head:
         0         1         2         3         4         5         6   \
0 -0.360055 -0.236334 -0.371121 -0.087547 -0.110355 -0.137073 -0.393405

         7         8        9   …        22        23        24        25  \
0 -0.246737 -0.071095 -0.10276  … -0.226962 -0.236384 -0.270481 -0.316186

          26        27       28        29        30        31
0 -0.232929 -0.209447 -0.21153 -0.196034 -0.202791 -0.212501

[1 rows x 32 columns]
```

[ ]:

# Multifractal Analysis

September 8, 2023

## 1 Multifractal Analysis

## 2 MFDFA

[4]:
```
import numpy as np
import matplotlib.pyplot as plt
from MFDFA import MFDFA

# Assuming the EEG data was sampled at 1000 Hz
sampling_rate = 1000

# Load EEG data
EEG_data = np.load('/home/vincent/AAA_projects/MVCS/Neuroscience/
↪eeg_data_with_channels.npy', allow_pickle=True)

# We assume the data for each channel is in rows and time points are columns.
# If it's the other way around, transpose the array: EEG_data = EEG_data.T
num_channels, num_timepoints = EEG_data.shape

# Using hardcoded channel labels
eeg_channels = ['Fp1', 'Fpz', 'Fp2', 'F7', 'F3', 'Fz', 'F4', 'F8', 'FC5',␣
↪'FC1', 'FC2', 'FC6',
                'M1', 'T7', 'C3', 'Cz', 'C4', 'T8', 'M2', 'CP5', 'CP1', 'CP2',␣
↪'CP6',
                'P7', 'P3', 'Pz', 'P4', 'P8', 'POz', 'O1', 'Oz', 'O2']

# Calculate multifractal spectrum for the EEG data
lag = np.logspace(0.7, 4, 30).astype(int)
q = np.linspace(-5, 5, 50)

# Define a list to store scale and fluctuation parameters
mfdfa_results = []

# Loop over each channel
for ch in range(num_channels):
    # Select the current channel data
    eeg_data_filtered = EEG_data[ch, :]
```

```
    # Call MFDFA for the current channel data
    scale, fluct = MFDFA(eeg_data_filtered, lag=lag, q=q)

    # Store scale and fluctuation parameters
    mfdfa_results.append((scale, fluct))

    # Visualize the results
    plt.figure(figsize=(12, 4))

    plt.subplot(1, 2, 1)
    plt.plot(eeg_data_filtered)
    plt.title(f"Channel {ch+1}: {eeg_channels[ch]}")

    plt.subplot(1, 2, 2)
    plt.loglog(scale, fluct)
    plt.title(f"Channel {ch+1}: Multifractal DFA")

    plt.tight_layout()
    plt.show()
```

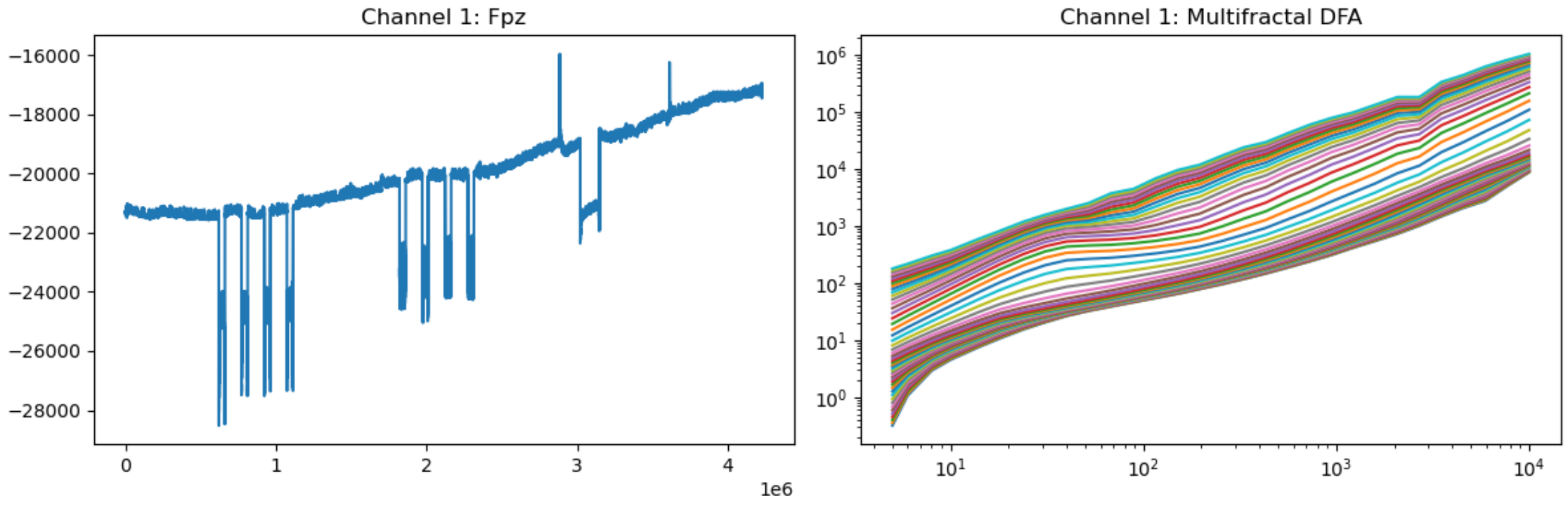

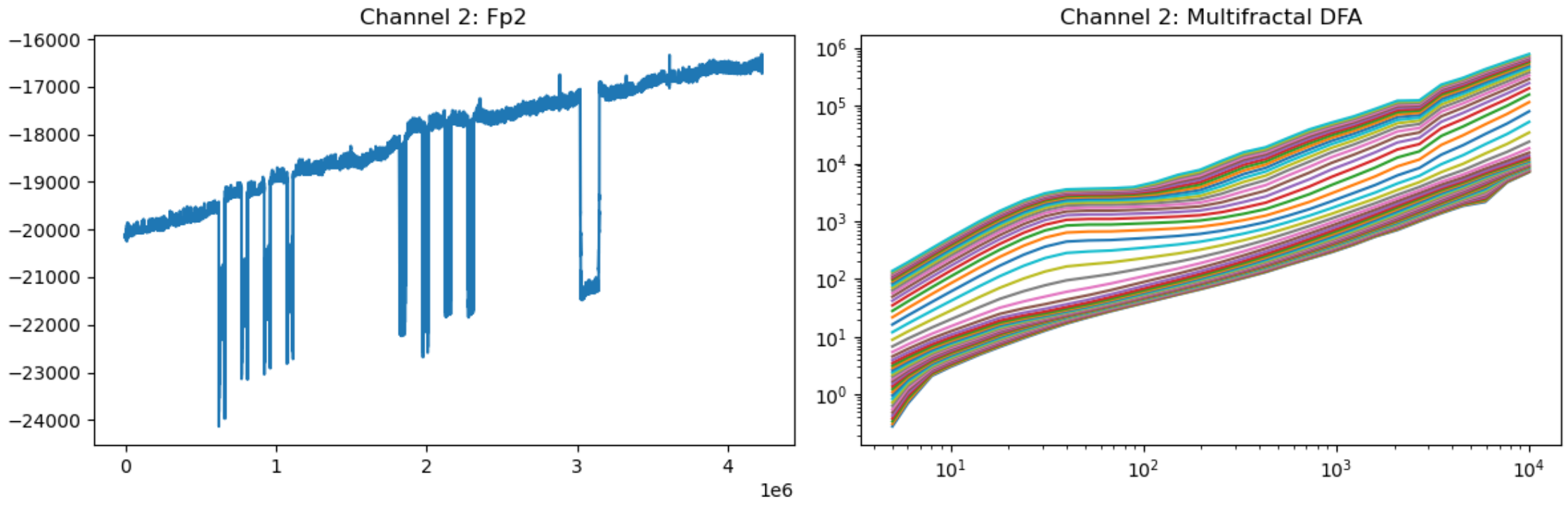

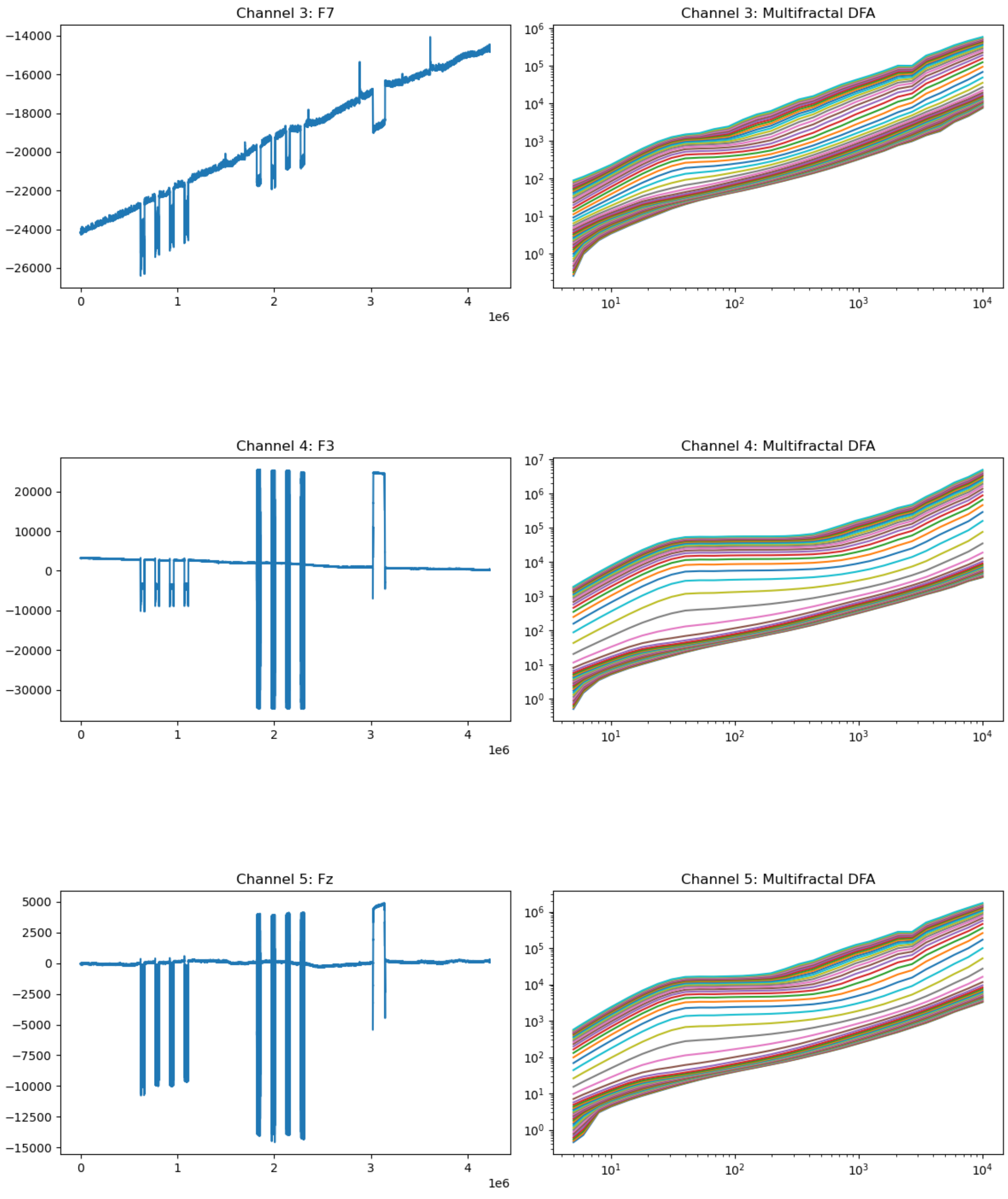
Channel 3: F7
−14000
−16000
−18000
−20000
−22000
−24000
−26000
0
1
2
3
4
1e6
Channel 3: Multifractal DFA
$10^6$
$10^5$
$10^4$
$10^3$
$10^2$
$10^1$
$10^0$
$10^1$
$10^2$
$10^3$
$10^4$
Channel 4: F3
20000
10000
0
−10000
−20000
−30000
0
1
2
3
4
1e6
Channel 4: Multifractal DFA
$10^7$
$10^6$
$10^5$
$10^4$
$10^3$
$10^2$
$10^1$
$10^0$
$10^1$
$10^2$
$10^3$
$10^4$
Channel 5: Fz
5000
2500
0
−2500
−5000
−7500
−10000
−12500
−15000
0
1
2
3
4
1e6
Channel 5: Multifractal DFA
$10^6$
$10^5$
$10^4$
$10^3$
$10^2$
$10^1$
$10^0$
$10^1$
$10^2$
$10^3$
$10^4$

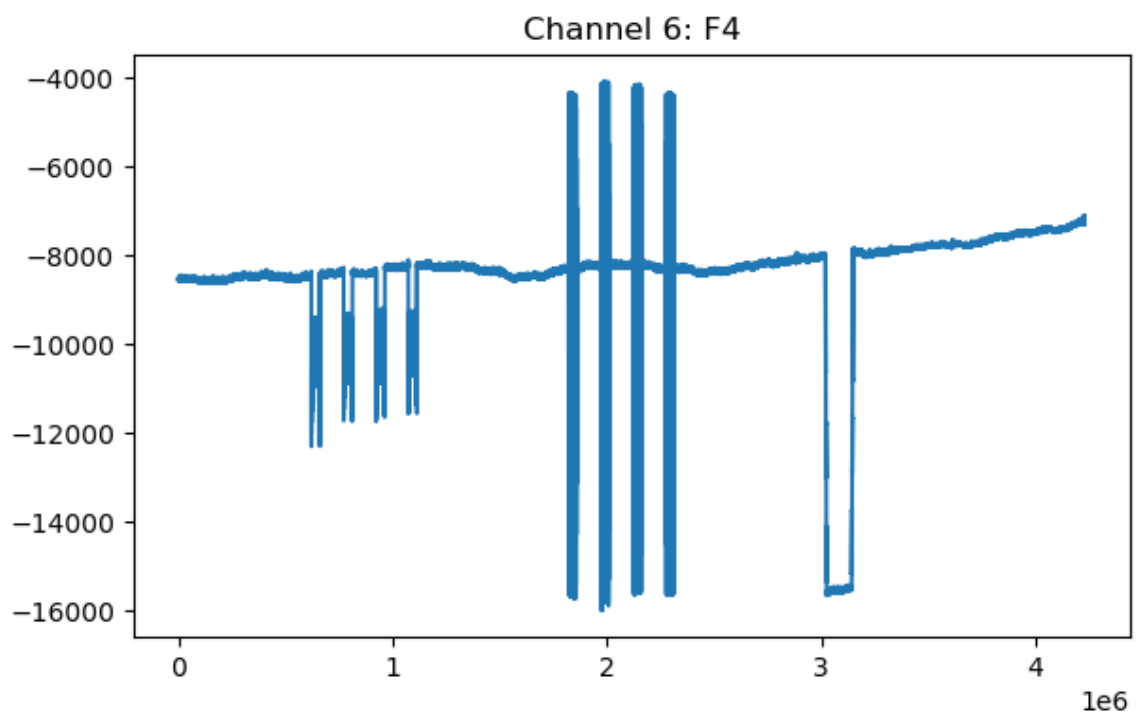
Channel 6: F4
−4000
−6000
−8000
−10000
−12000
−14000
−16000
0
1
2
3
4
1e6

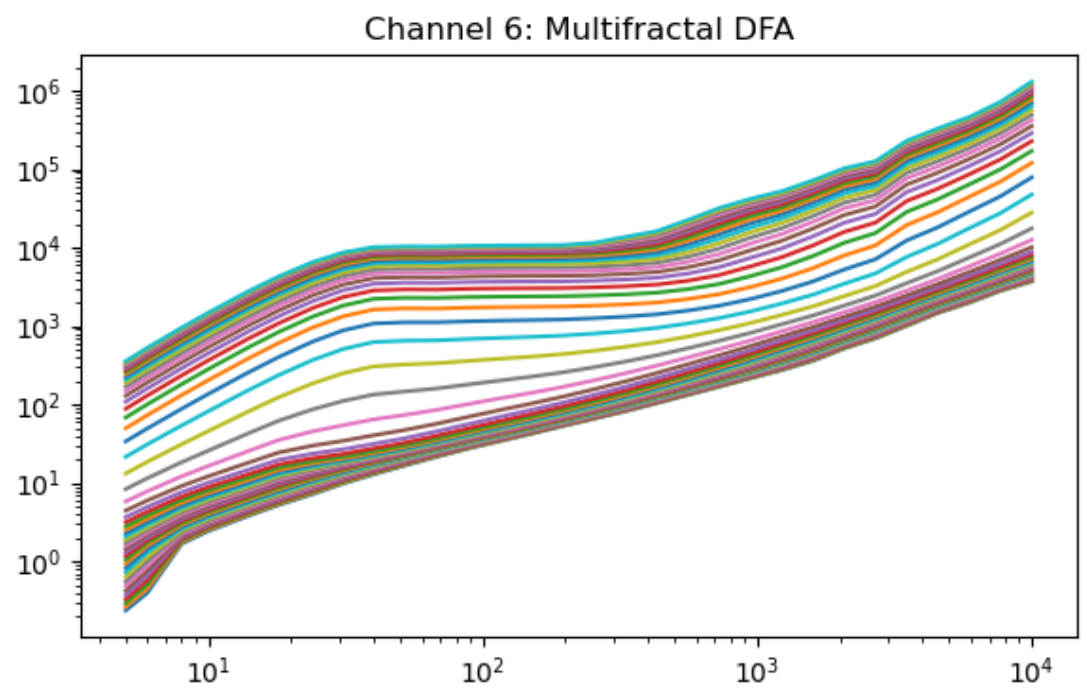
Channel 6: Multifractal DFA
$10^6$
$10^5$
$10^4$
$10^3$
$10^2$
$10^1$
$10^0$
$10^1$
$10^2$
$10^3$
$10^4$

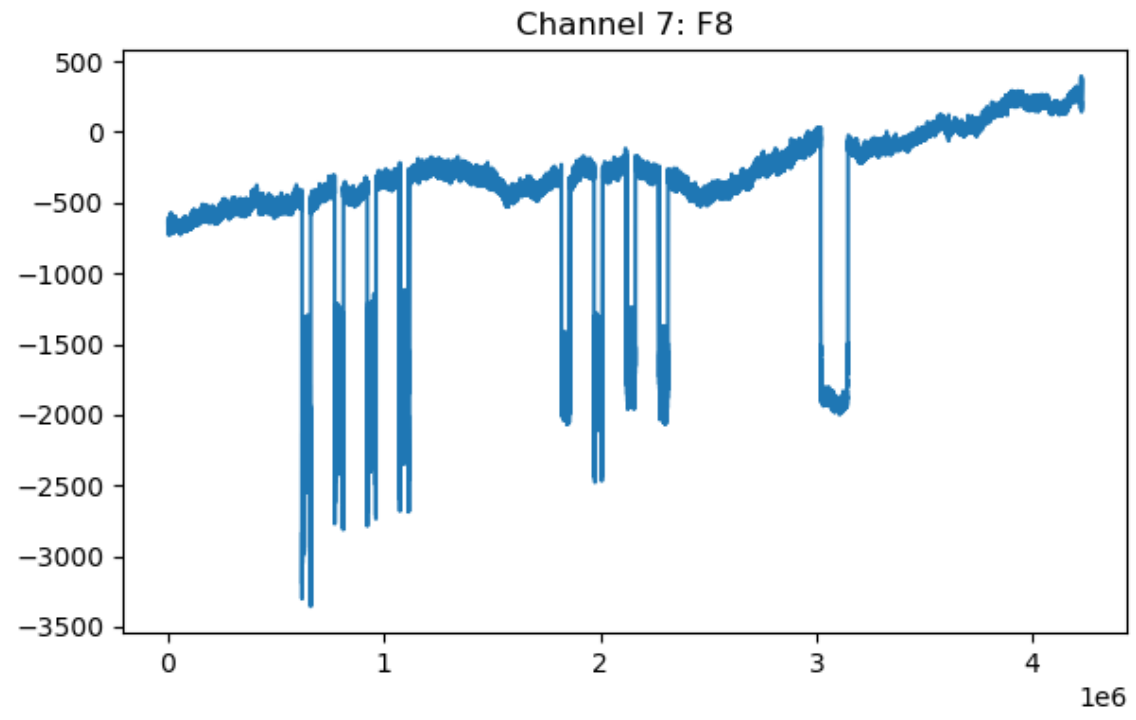
Channel 7: F8
500
0
−500
−1000
−1500
−2000
−2500
−3000
−3500
0
1
2
3
4
1e6

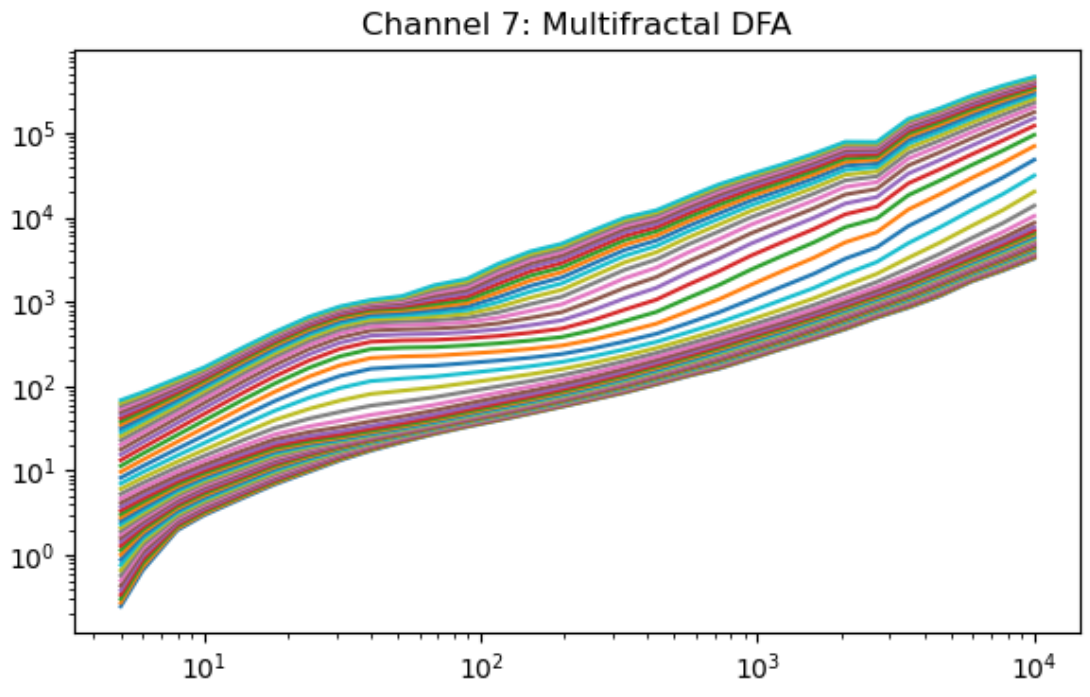
Channel 7: Multifractal DFA
$10^5$
$10^4$
$10^3$
$10^2$
$10^1$
$10^0$
$10^1$
$10^2$
$10^3$
$10^4$

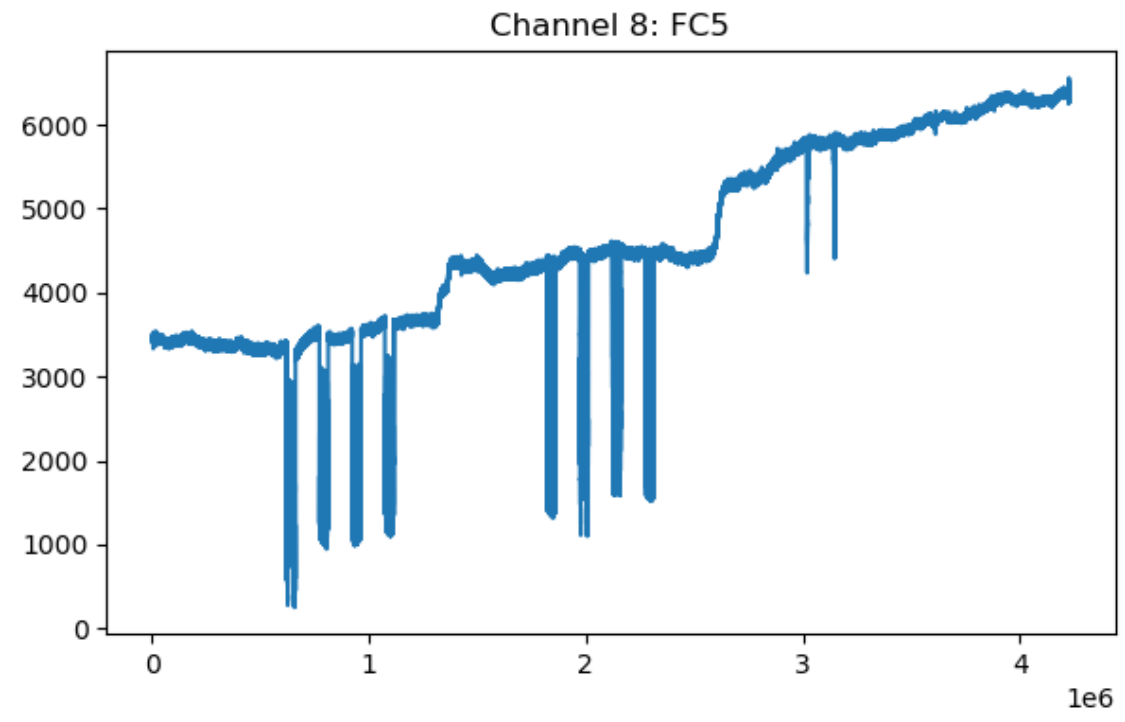
Channel 8: FC5
6000
5000
4000
3000
2000
1000
0
0
1
2
3
4
1e6

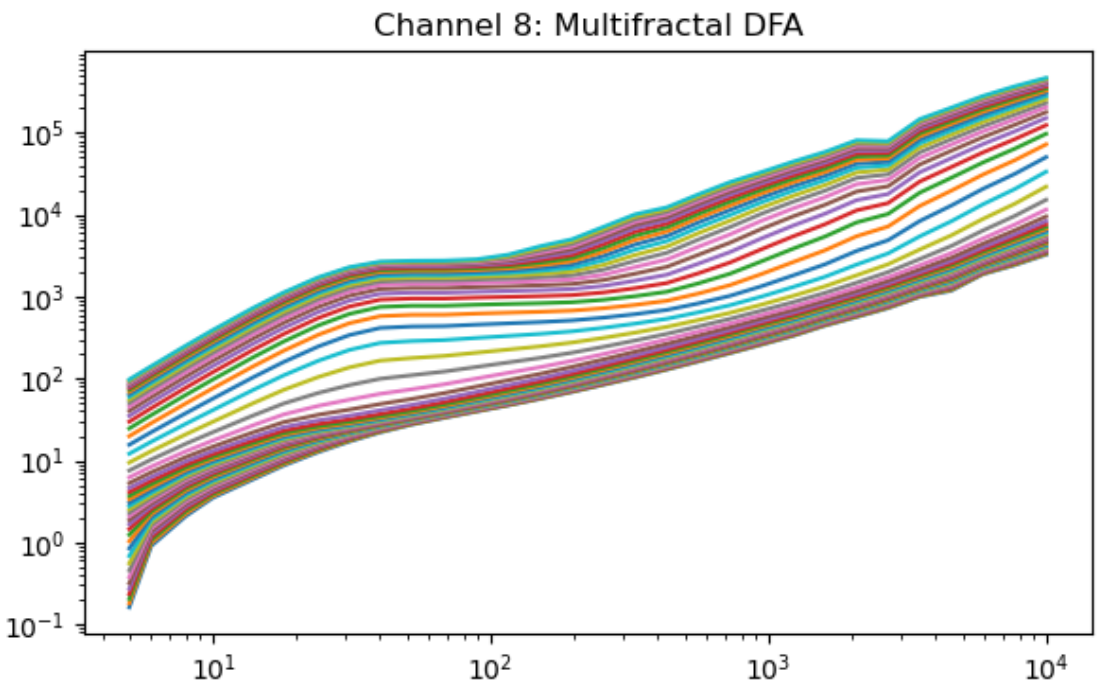
Channel 8: Multifractal DFA
$10^5$
$10^4$
$10^3$
$10^2$
$10^1$
$10^0$
$10^{-1}$
$10^1$
$10^2$
$10^3$
$10^4$

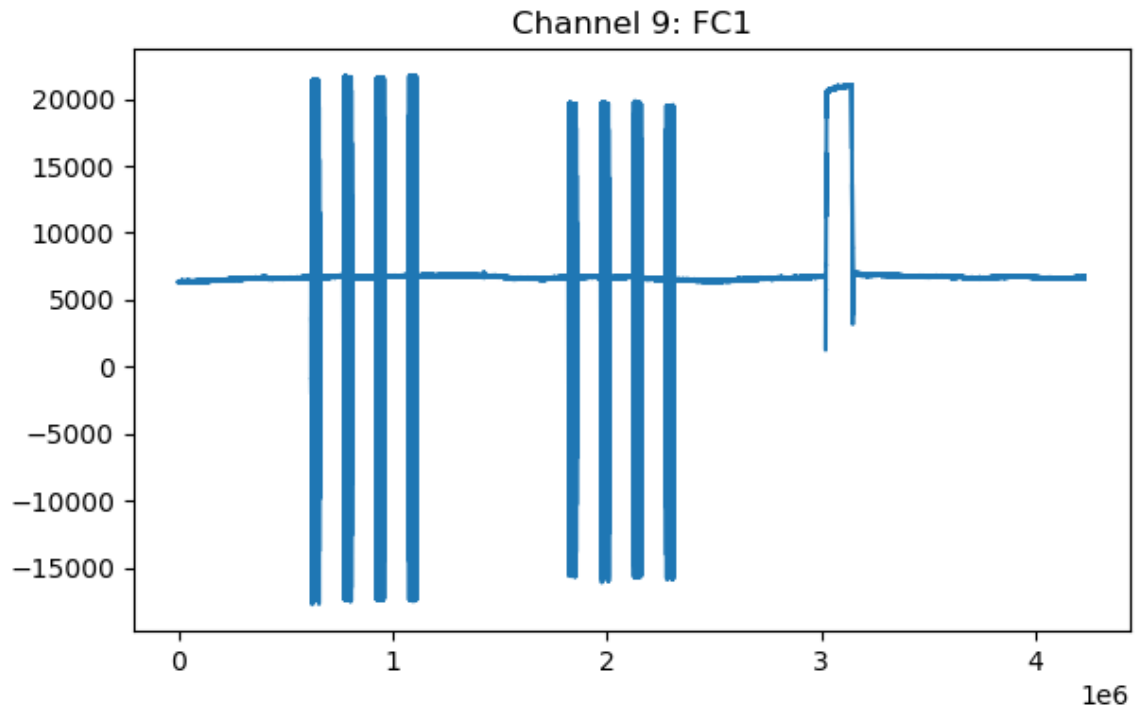
Channel 9: FC1
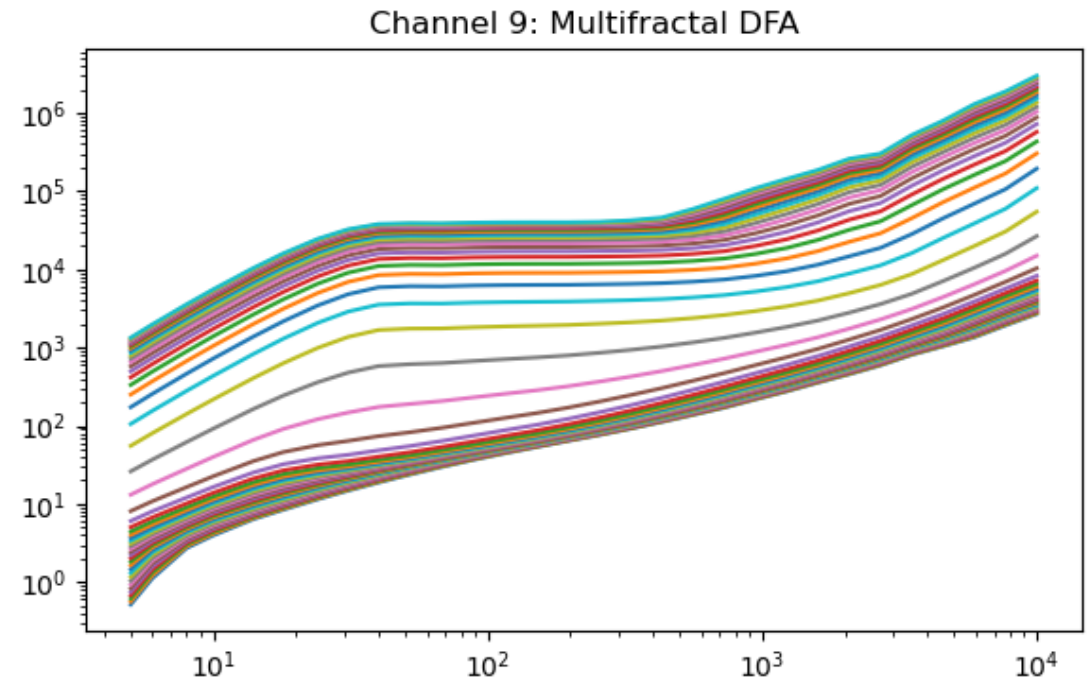
Channel 9: Multifractal DFA
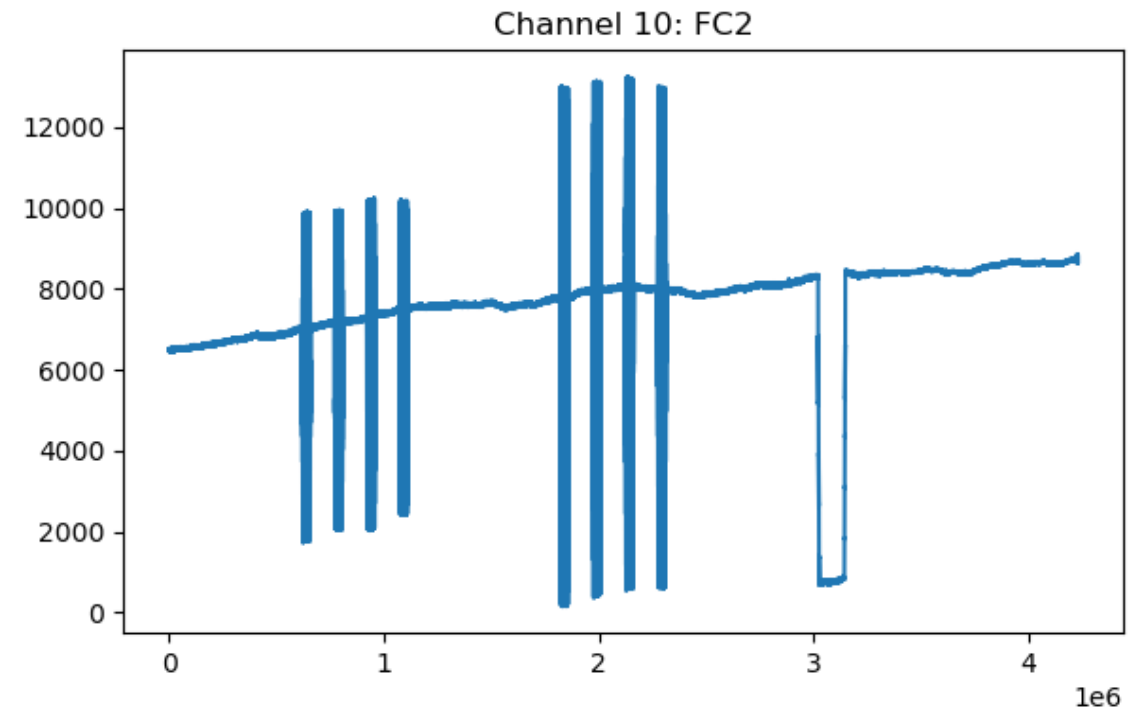
Channel 10: FC2
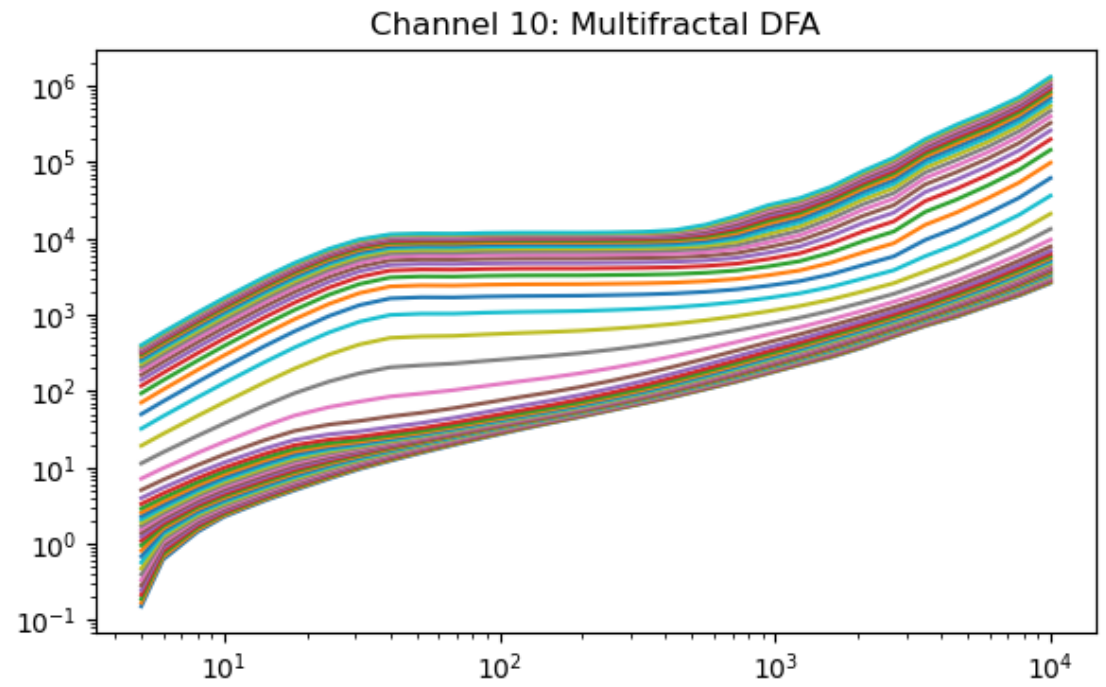
Channel 10: Multifractal DFA
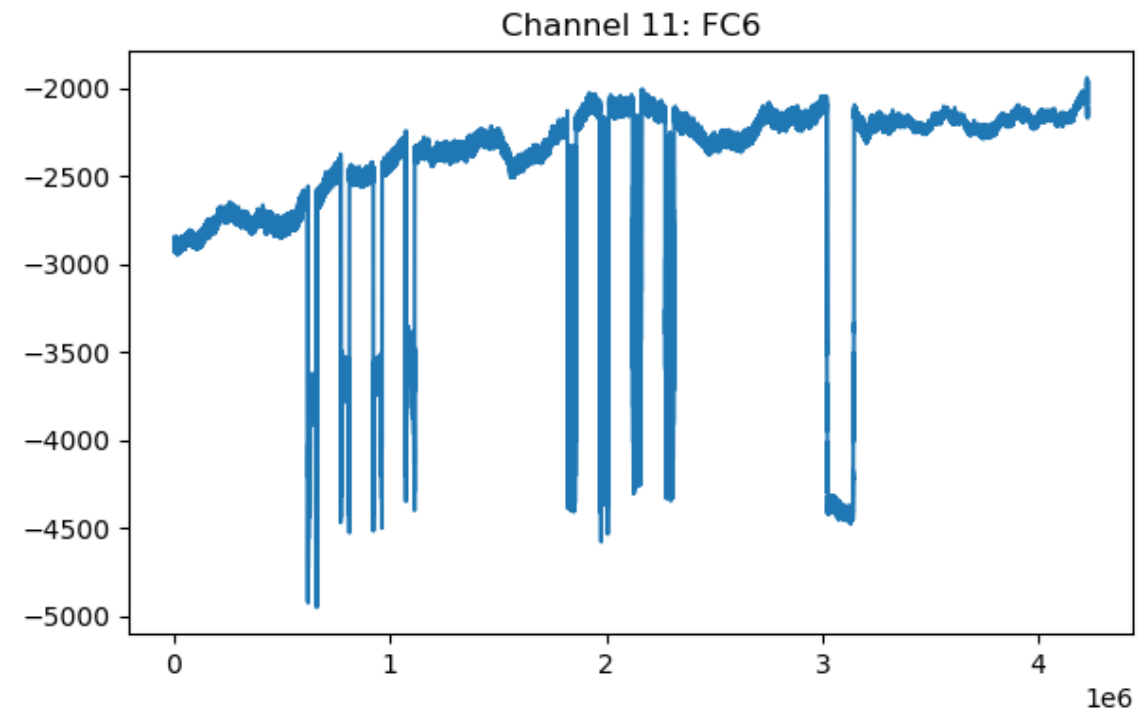
Channel 11: FC6
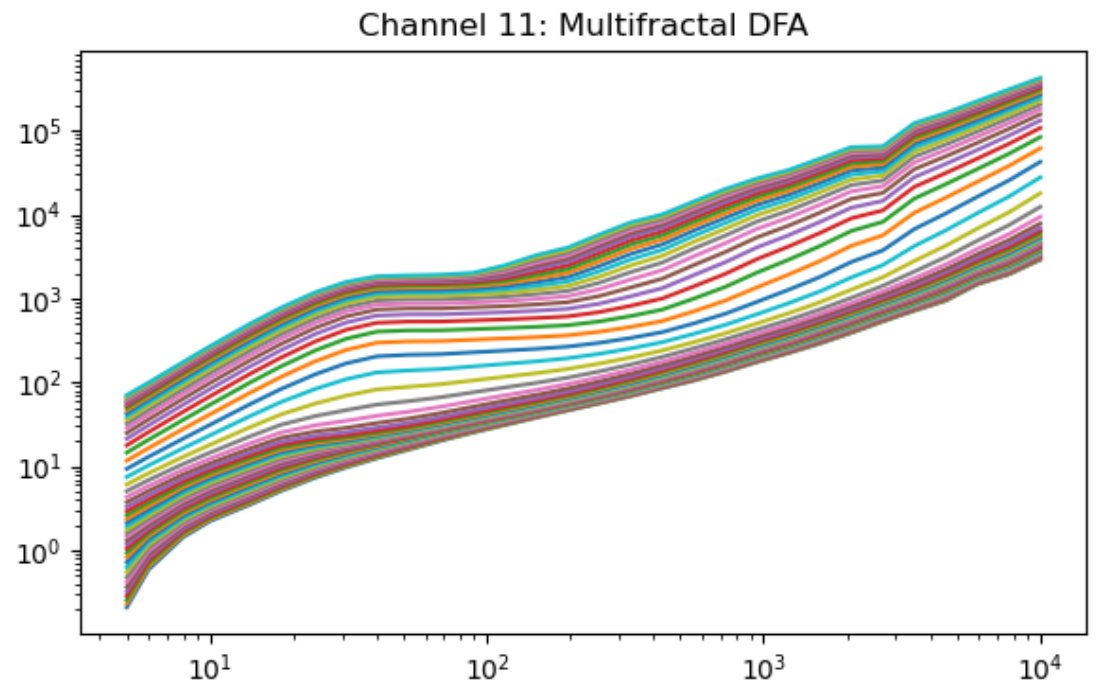
Channel 11: Multifractal DFA

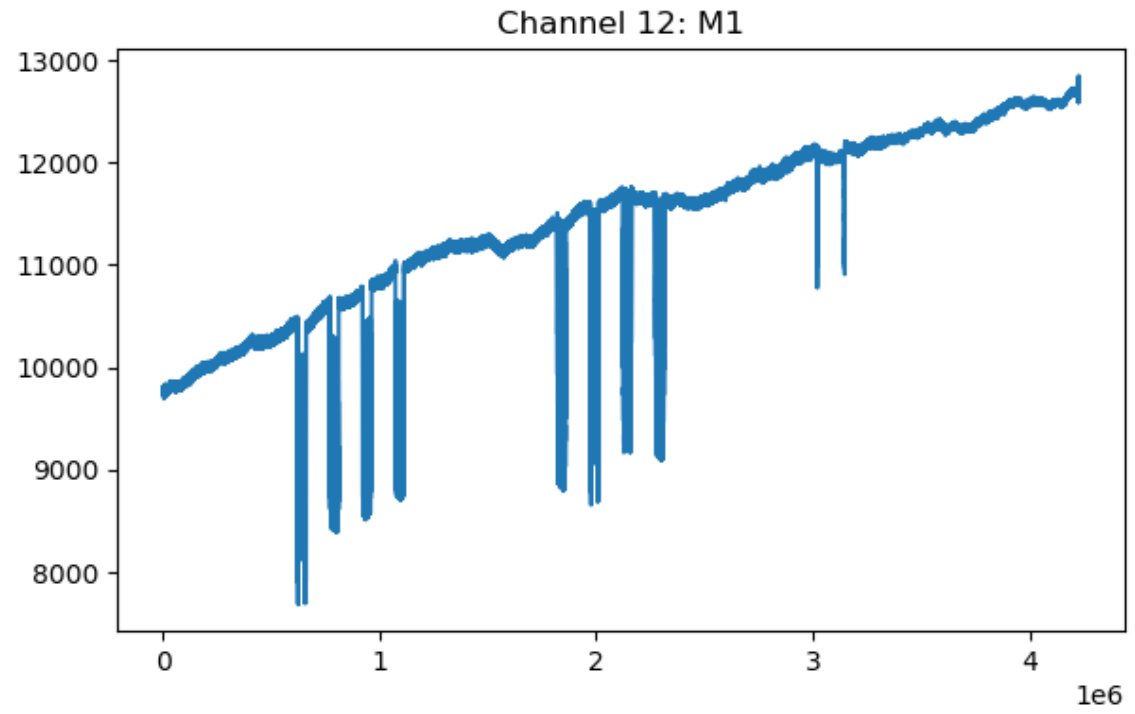
Channel 12: M1
13000
12000
11000
10000
9000
8000
0
1
2
3
4
1e6

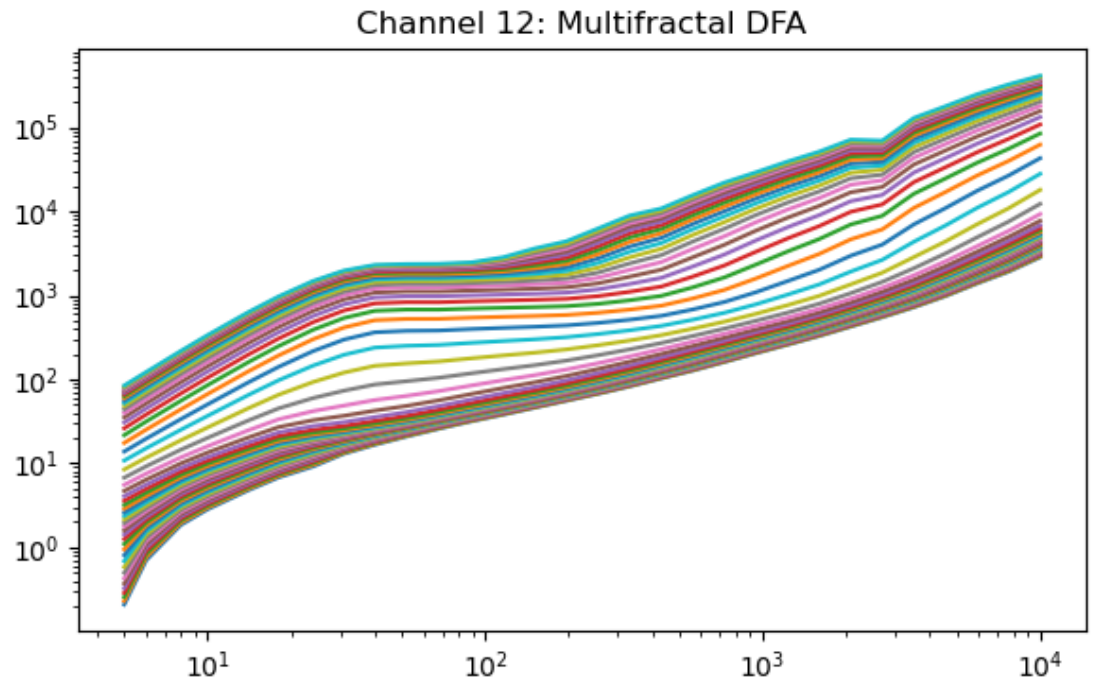
Channel 12: Multifractal DFA
$10^5$
$10^4$
$10^3$
$10^2$
$10^1$
$10^0$
$10^1$
$10^2$
$10^3$
$10^4$

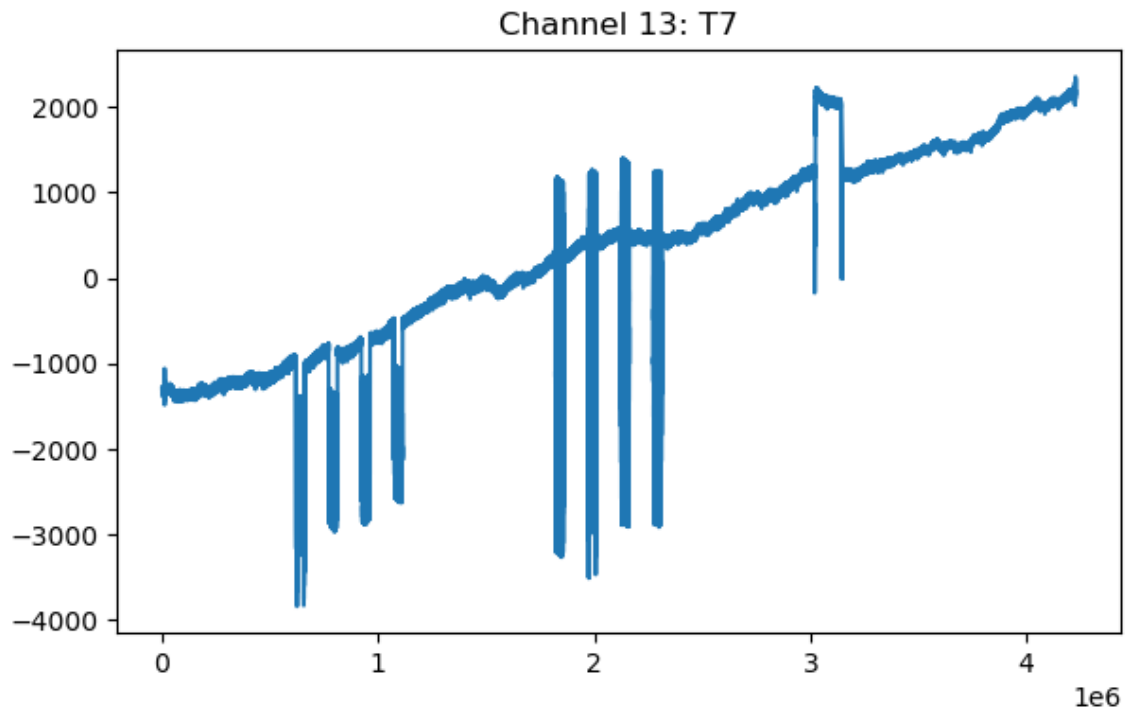
Channel 13: T7
2000
1000
0
−1000
−2000
−3000
−4000
0
1
2
3
4
1e6

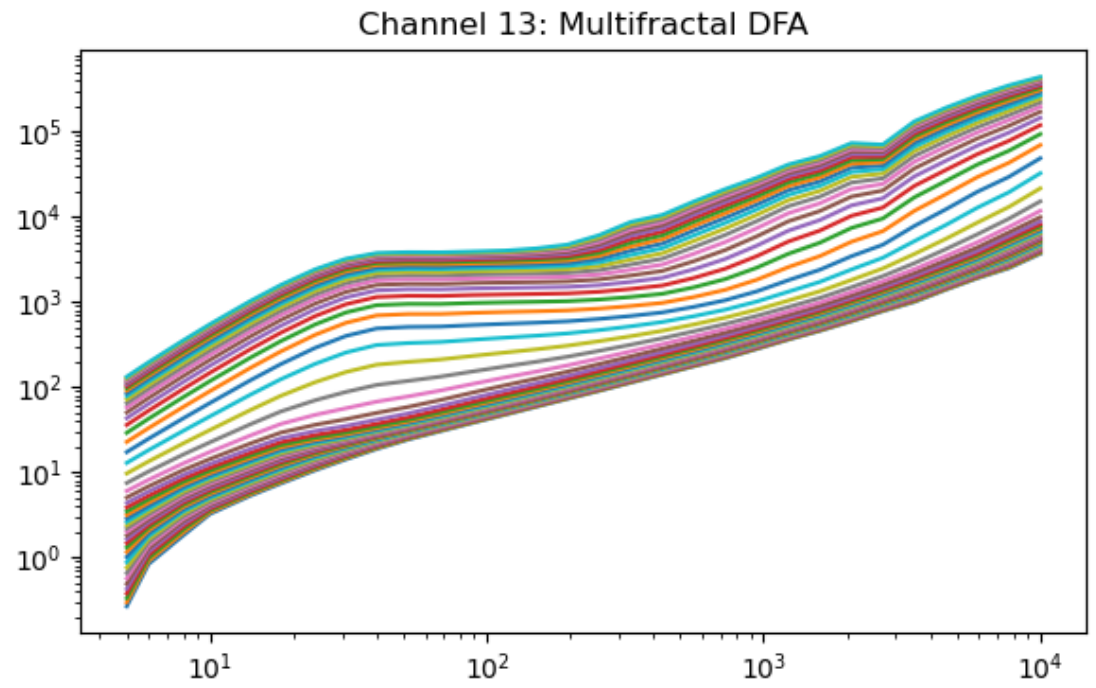
Channel 13: Multifractal DFA
$10^5$
$10^4$
$10^3$
$10^2$
$10^1$
$10^0$
$10^1$
$10^2$
$10^3$
$10^4$

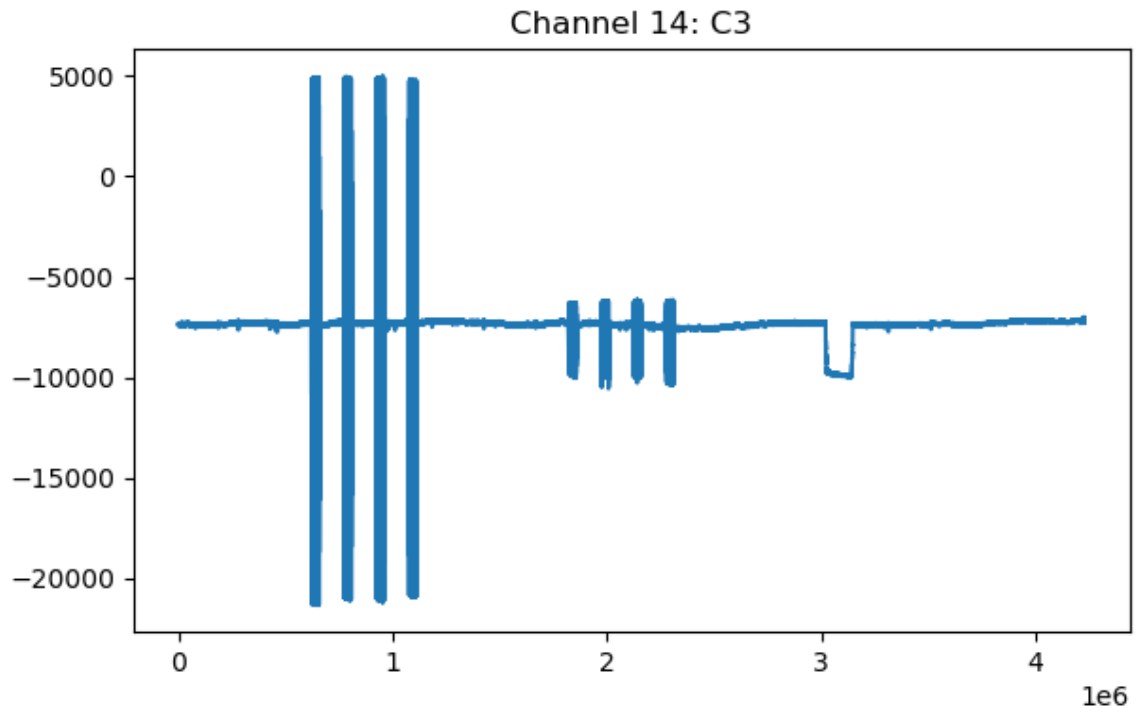
Channel 14: C3
5000
0
−5000
−10000
−15000
−20000
0
1
2
3
4
1e6

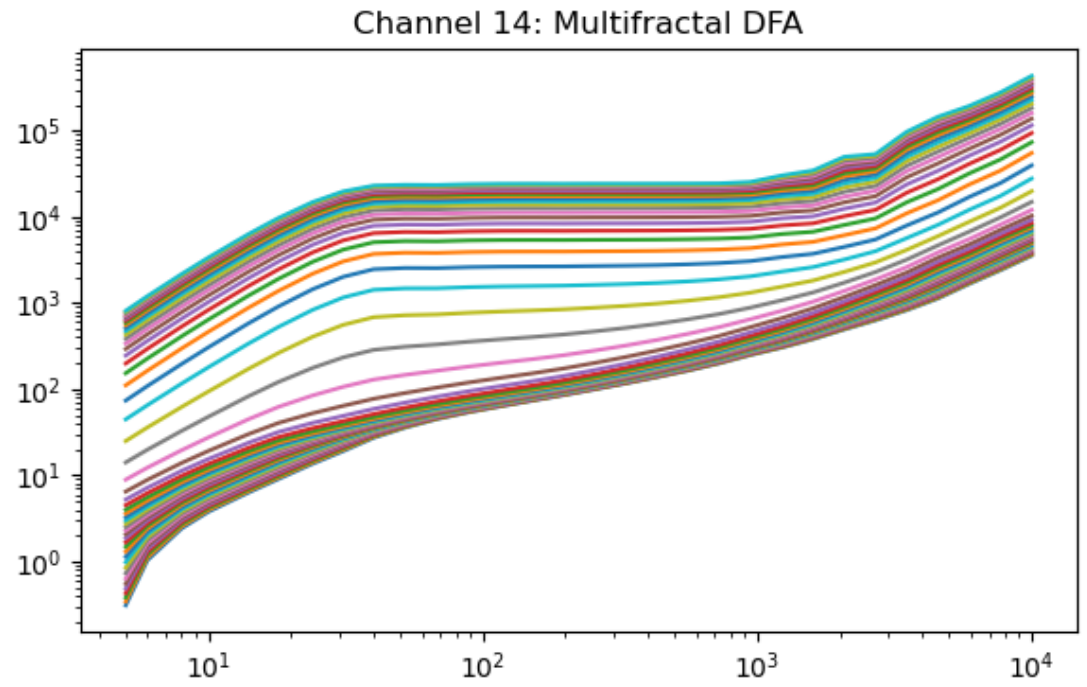
Channel 14: Multifractal DFA
$10^5$
$10^4$
$10^3$
$10^2$
$10^1$
$10^0$
$10^1$
$10^2$
$10^3$
$10^4$

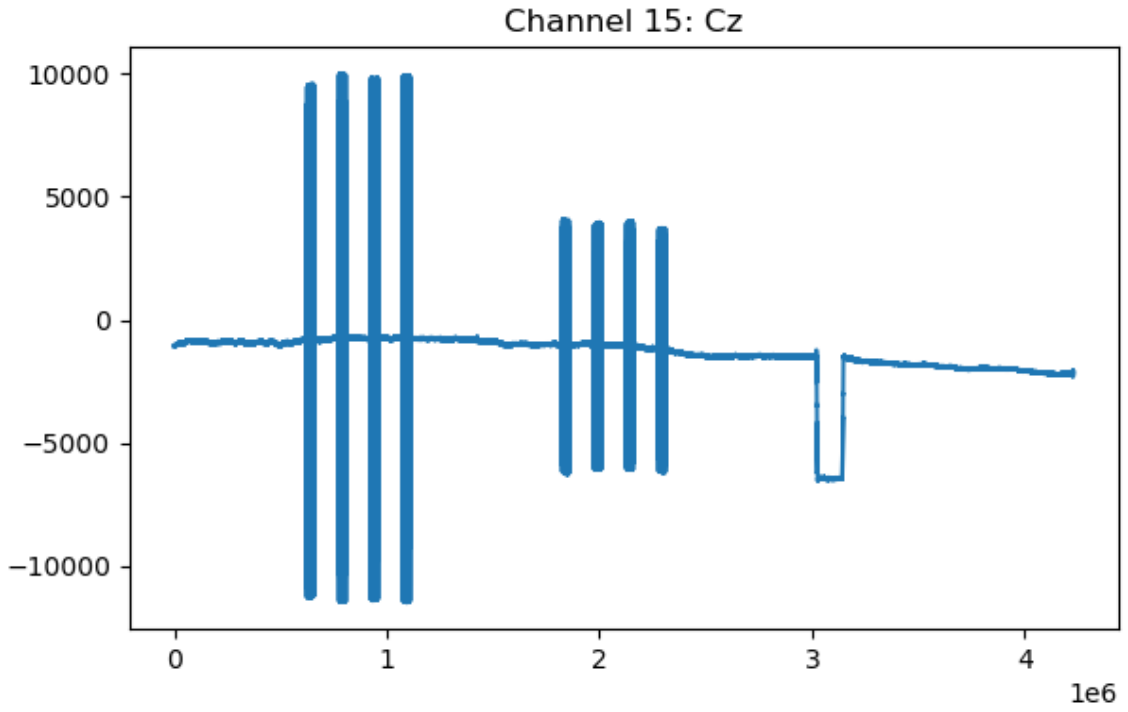
Channel 15: Cz
10000
5000
0
−5000
−10000
0
1
2
3
4
1e6

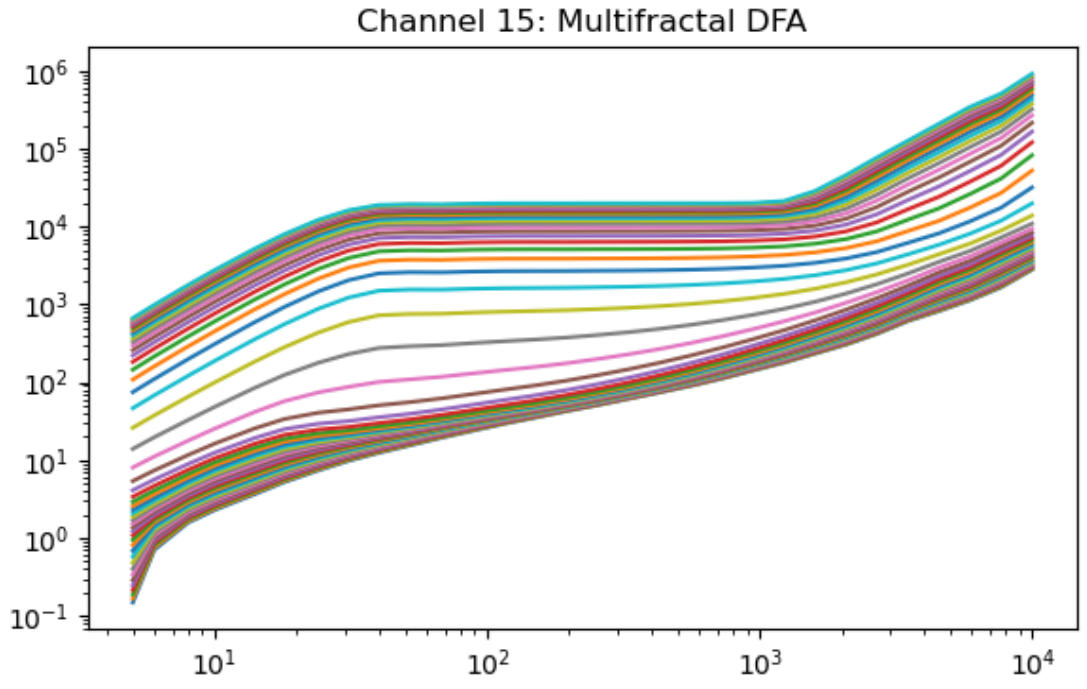
Channel 15: Multifractal DFA
$10^6$
$10^5$
$10^4$
$10^3$
$10^2$
$10^1$
$10^0$
$10^{-1}$
$10^1$
$10^2$
$10^3$
$10^4$

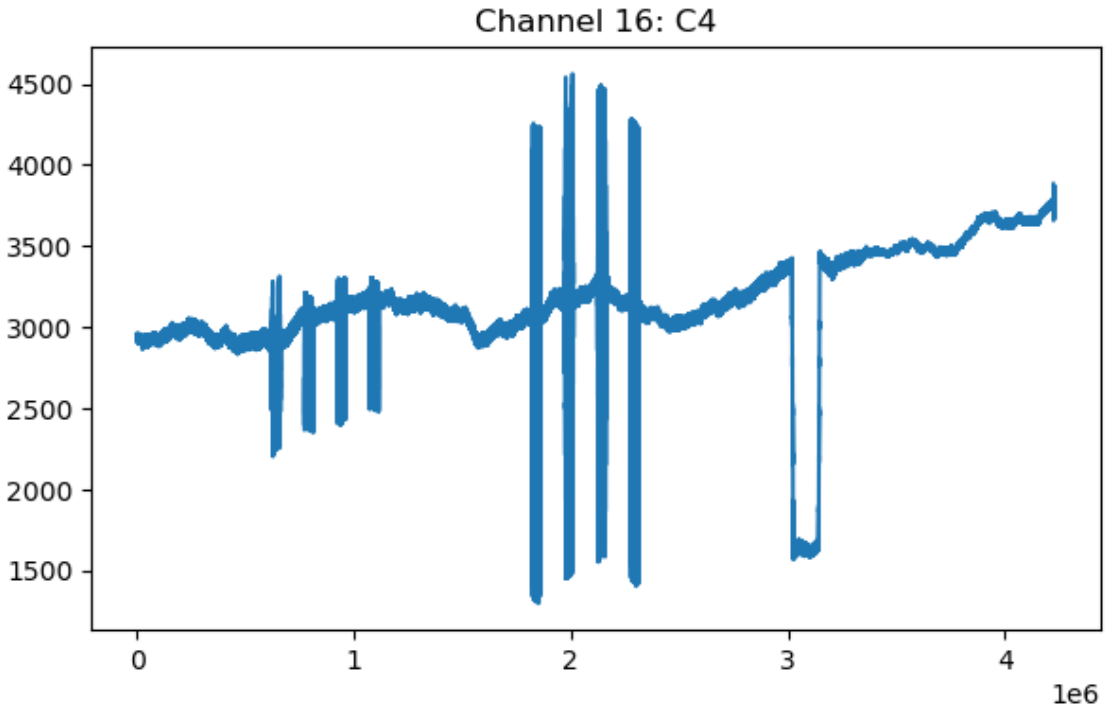
Channel 16: C4
4500
4000
3500
3000
2500
2000
1500
0
1
2
3
4
1e6

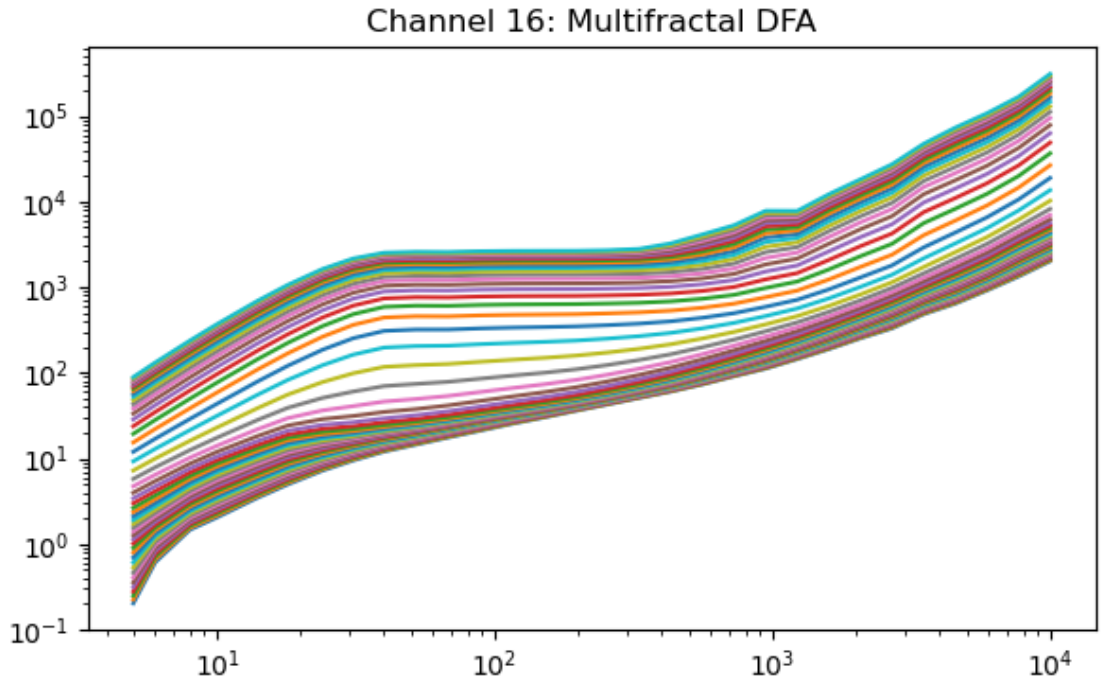
Channel 16: Multifractal DFA
$10^5$
$10^4$
$10^3$
$10^2$
$10^1$
$10^0$
$10^{-1}$
$10^1$
$10^2$
$10^3$
$10^4$

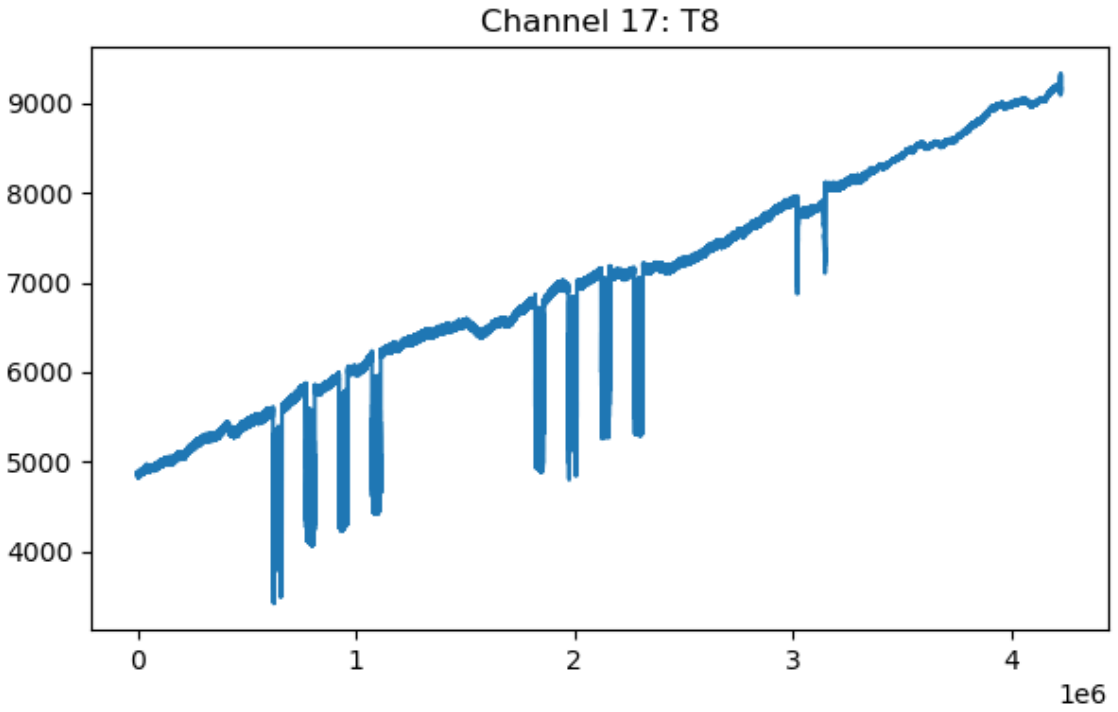
Channel 17: T8
9000
8000
7000
6000
5000
4000
0
1
2
3
4
1e6

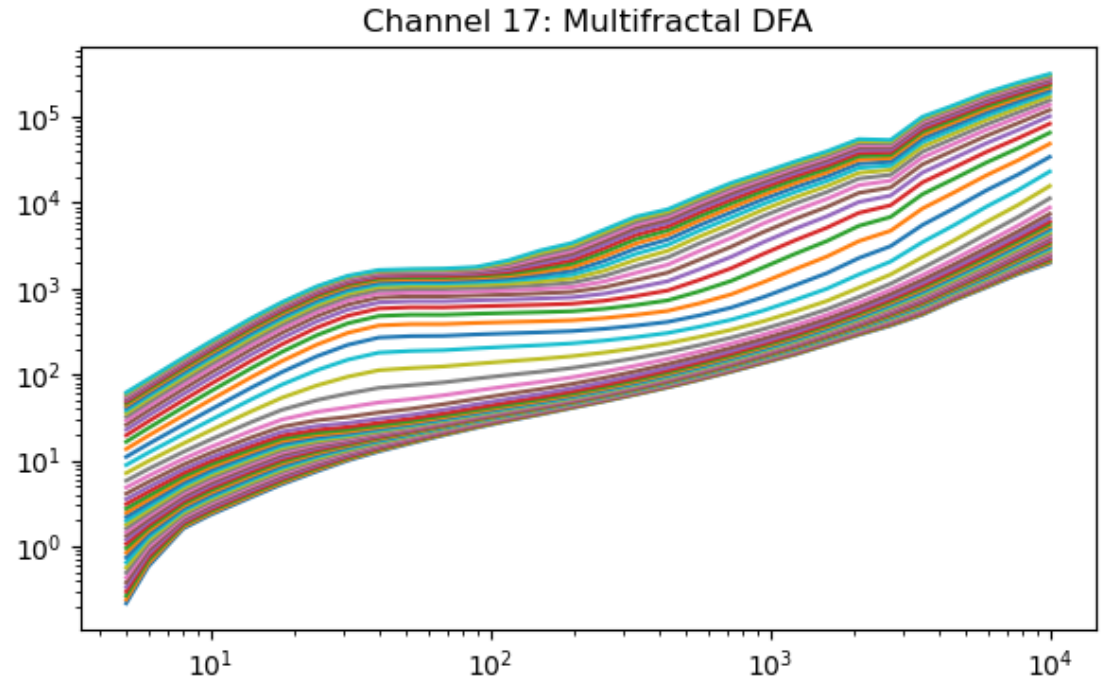
Channel 17: Multifractal DFA
$10^5$
$10^4$
$10^3$
$10^2$
$10^1$
$10^0$
$10^1$
$10^2$
$10^3$
$10^4$

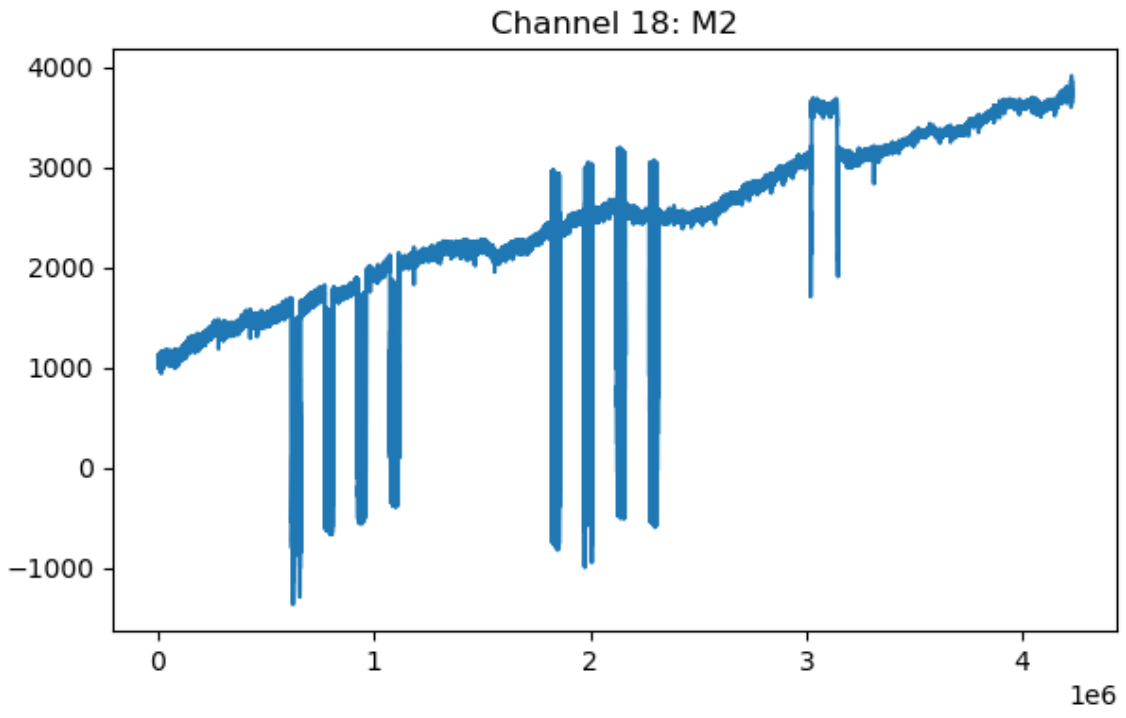
Channel 18: M2
1e6

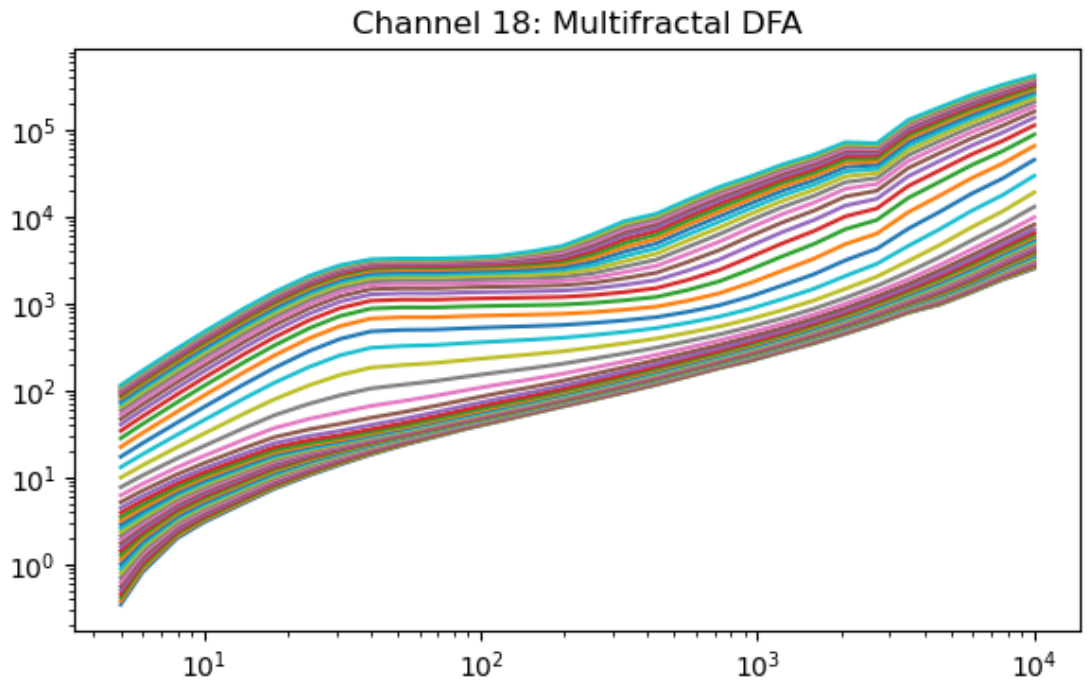
Channel 18: Multifractal DFA

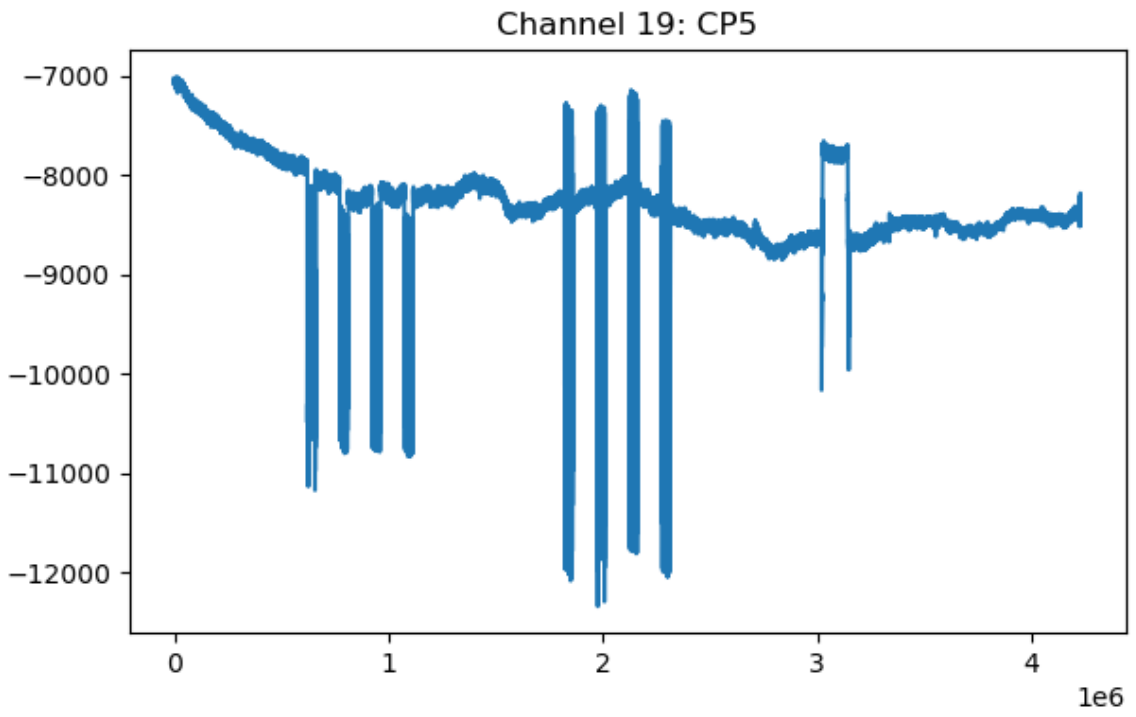
Channel 19: CP5
1e6

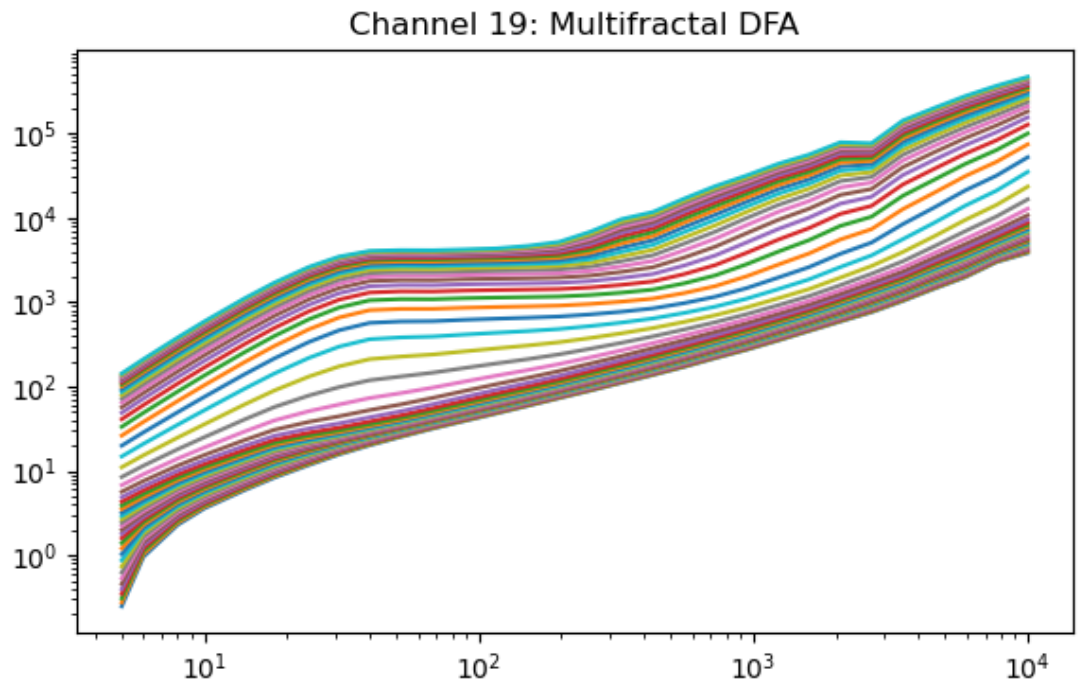
Channel 19: Multifractal DFA

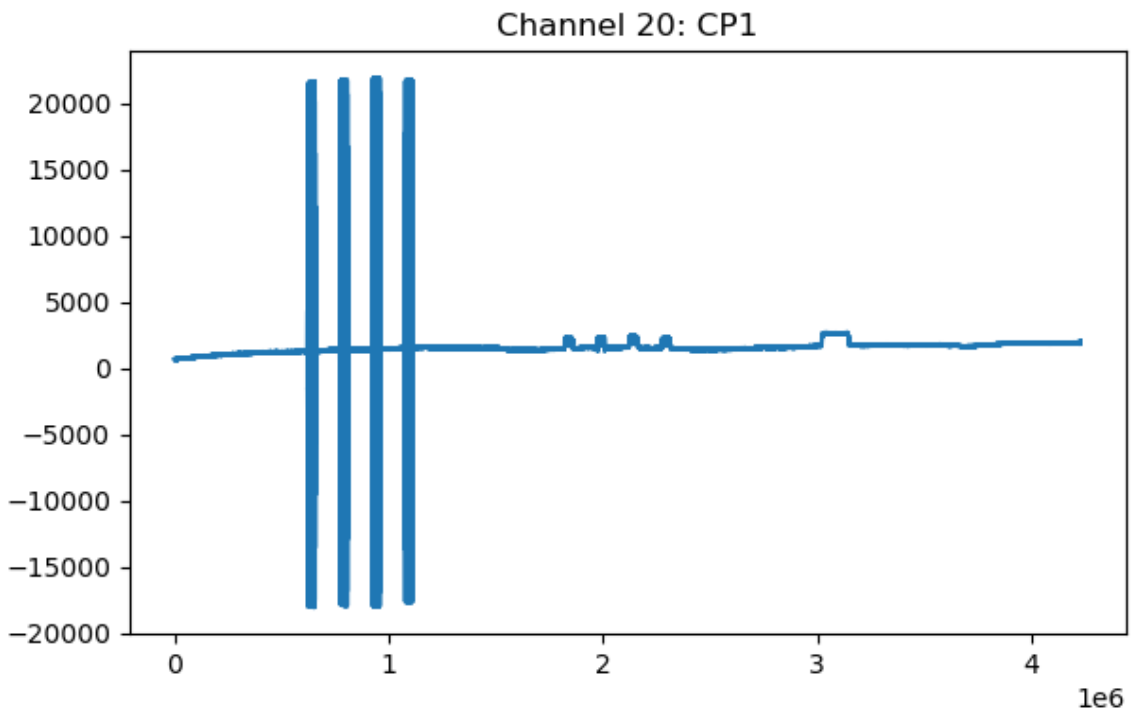
Channel 20: CP1
1e6

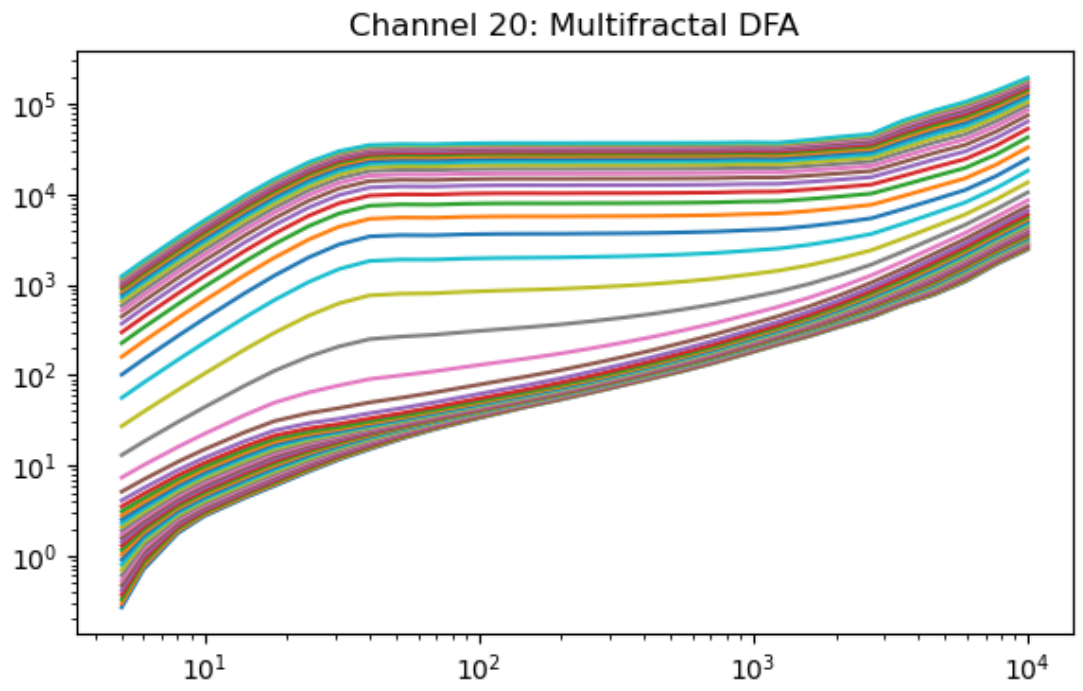
Channel 20: Multifractal DFA

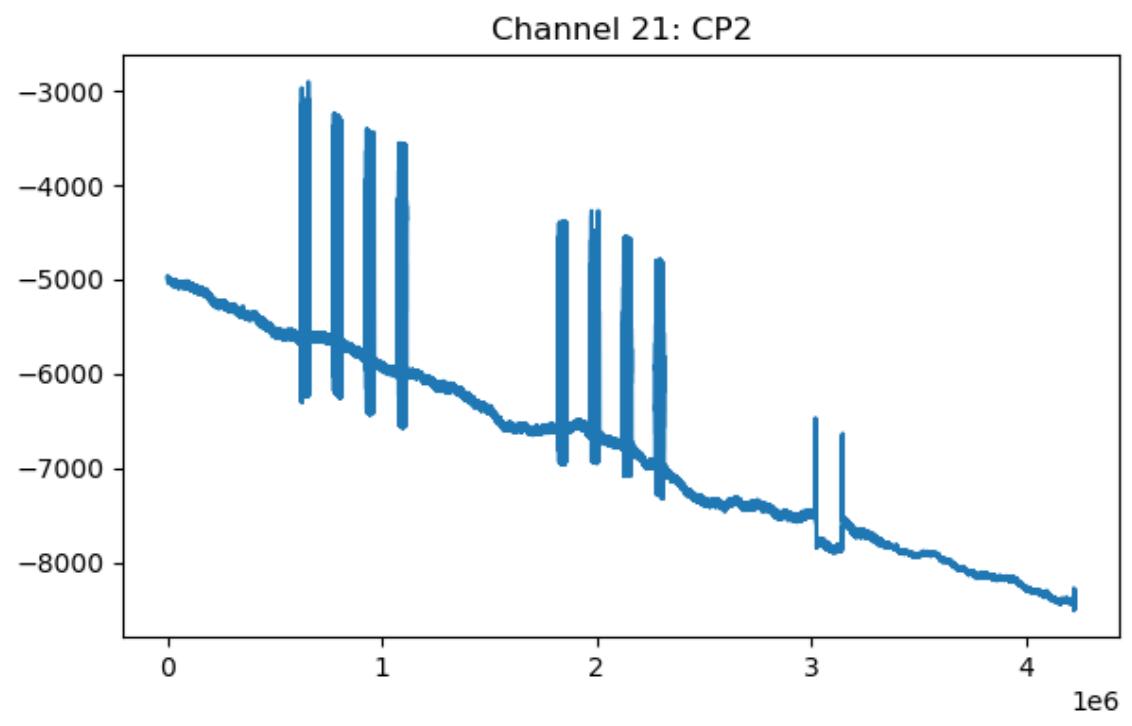
Channel 21: CP2
−3000
−4000
−5000
−6000
−7000
−8000
0
1
2
3
4
1e6

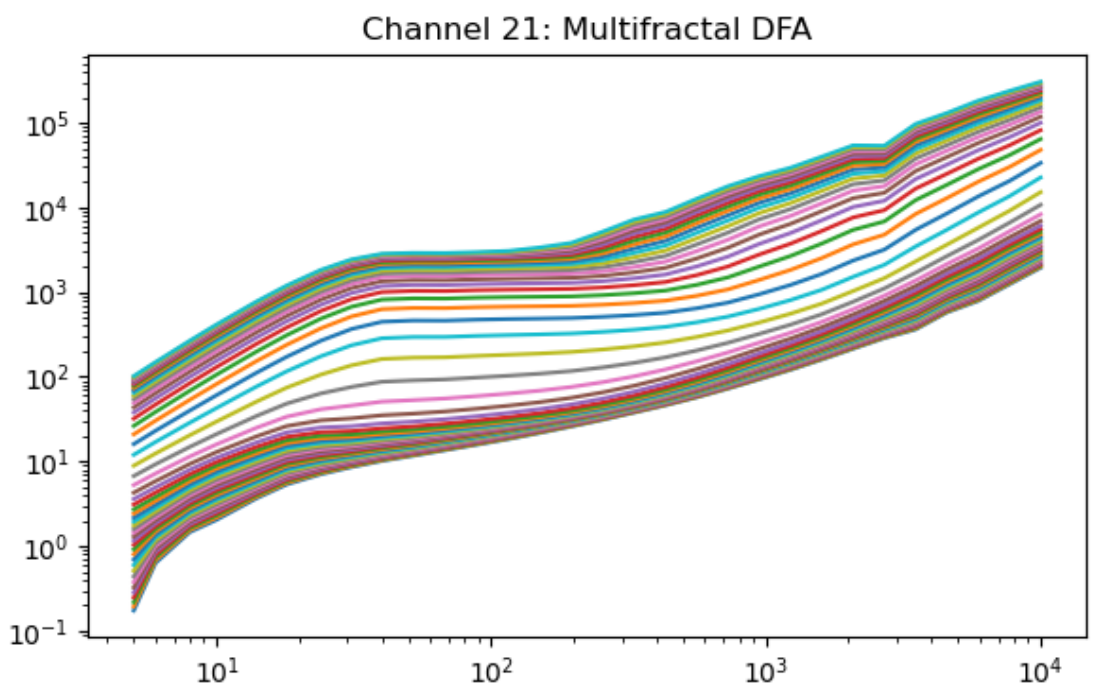
Channel 21: Multifractal DFA

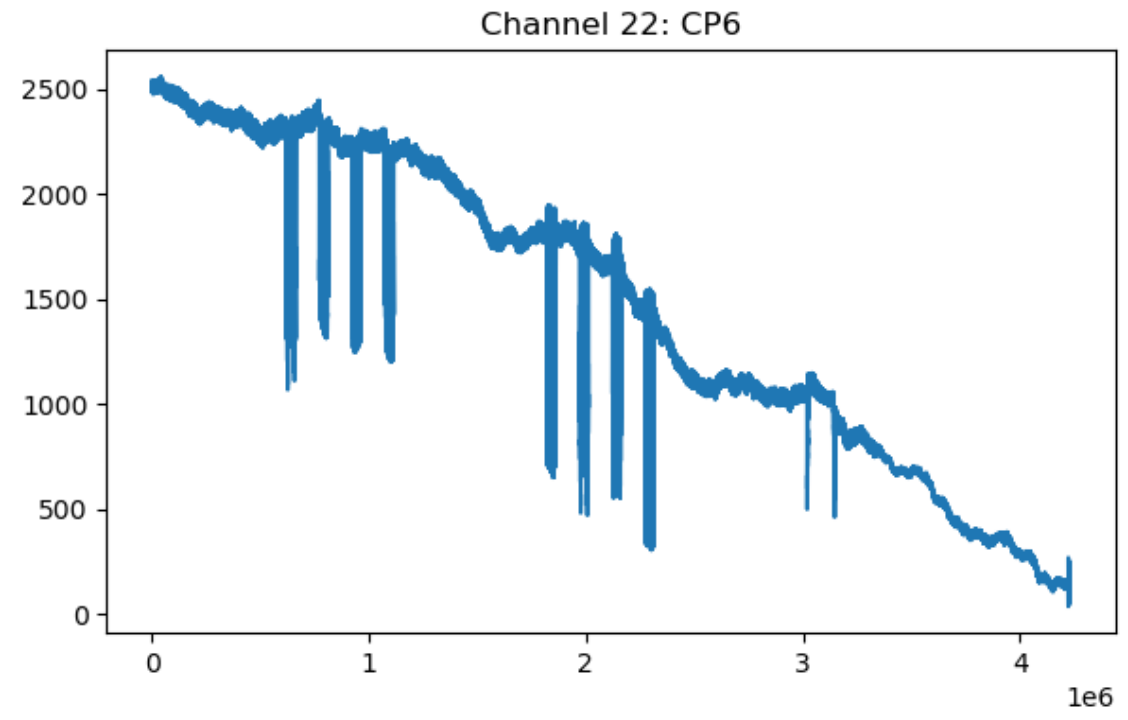
Channel 22: CP6
2500
2000
1500
1000
500
0
0
1
2
3
4
1e6

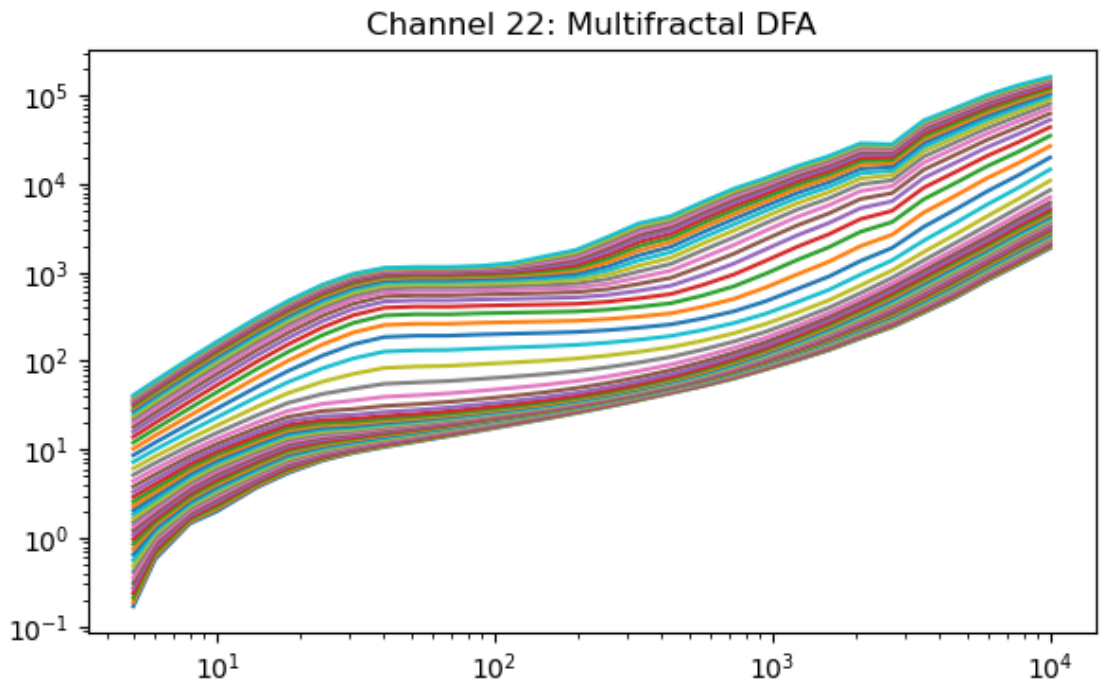
Channel 22: Multifractal DFA

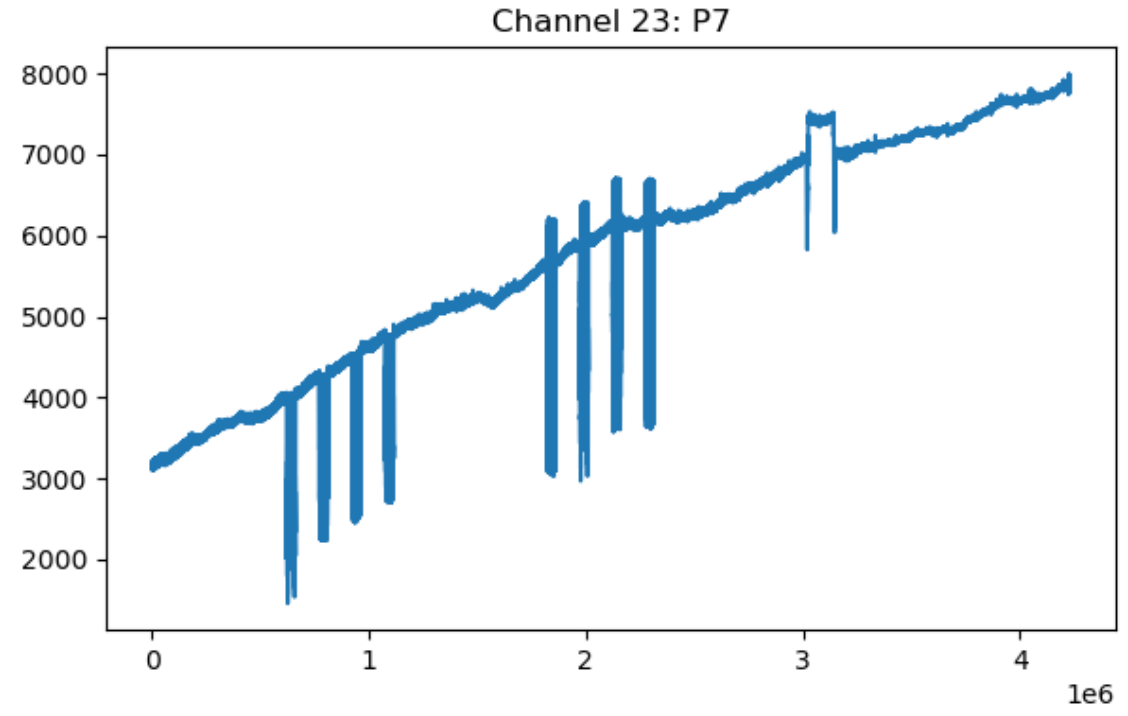
Channel 23: P7
8000
7000
6000
5000
4000
3000
2000
0
1
2
3
4
1e6

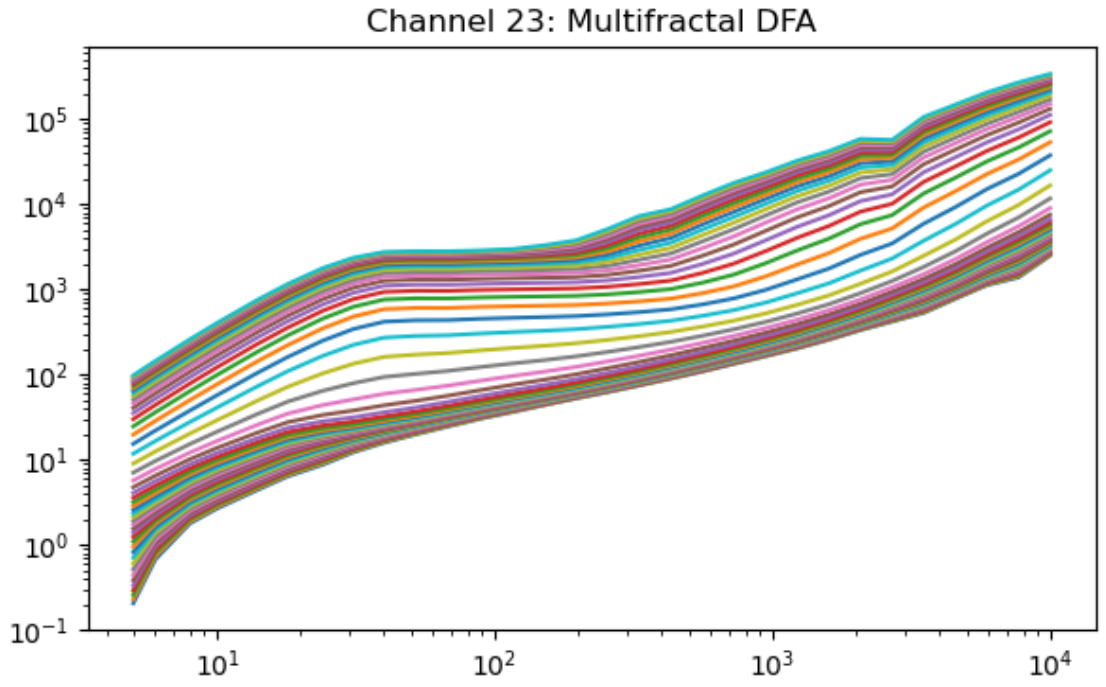
Channel 23: Multifractal DFA

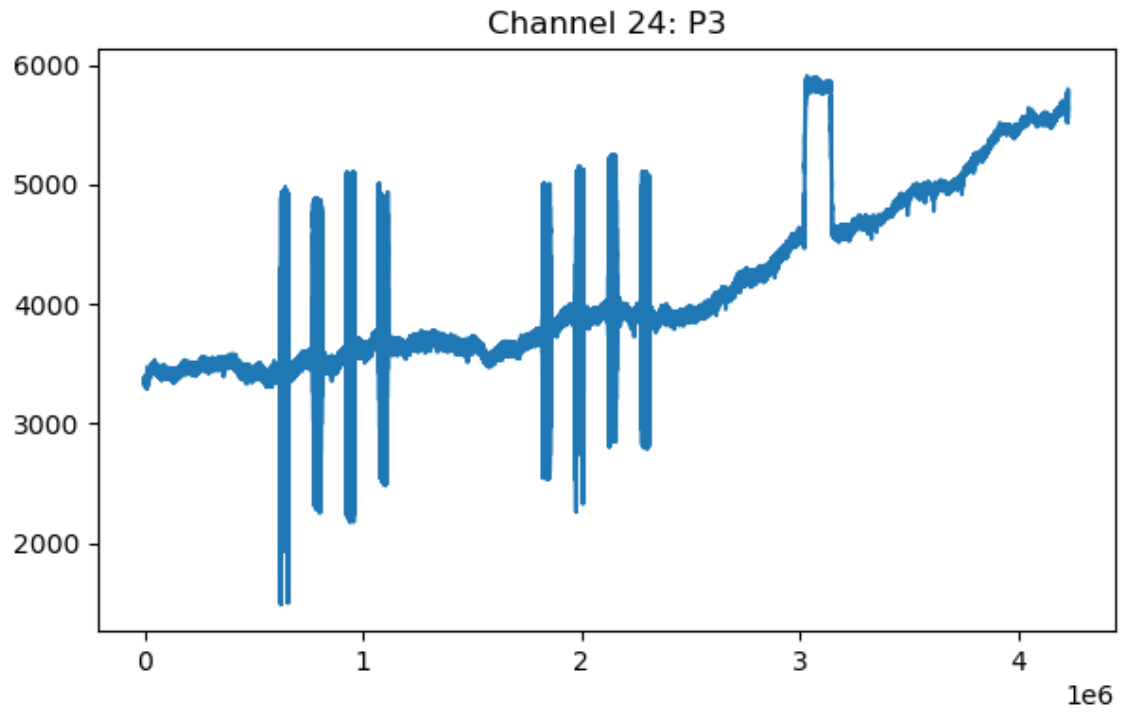
Channel 24: P3
6000
5000
4000
3000
2000
0
1
2
3
4
1e6

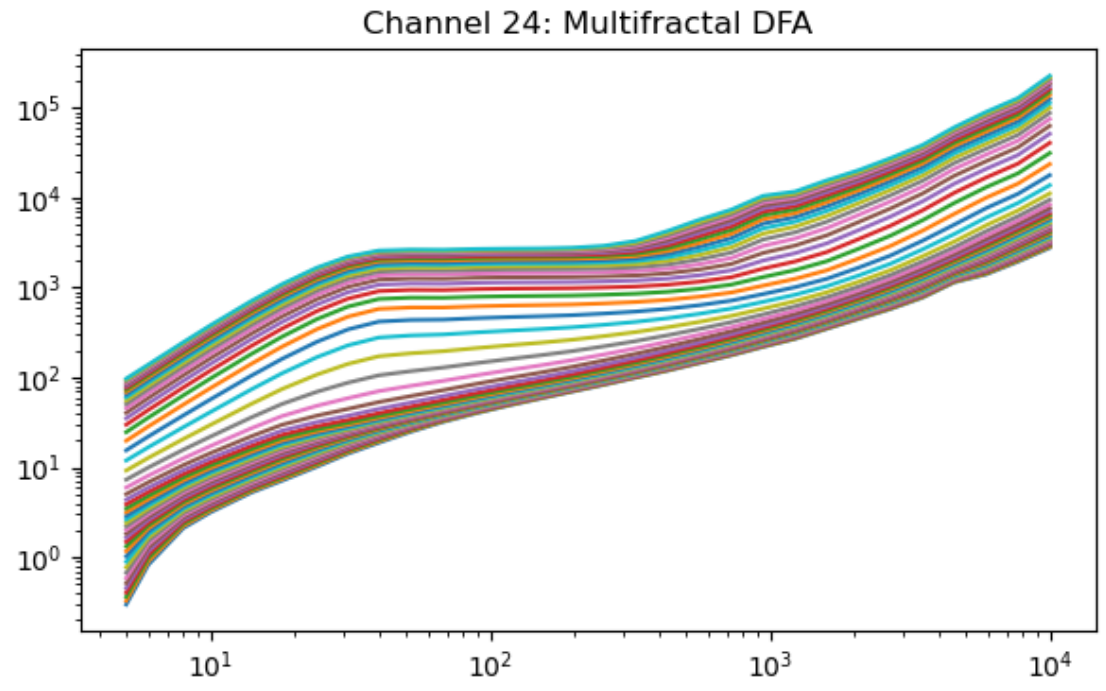
Channel 24: Multifractal DFA
$10^5$
$10^4$
$10^3$
$10^2$
$10^1$
$10^0$
$10^1$
$10^2$
$10^3$
$10^4$

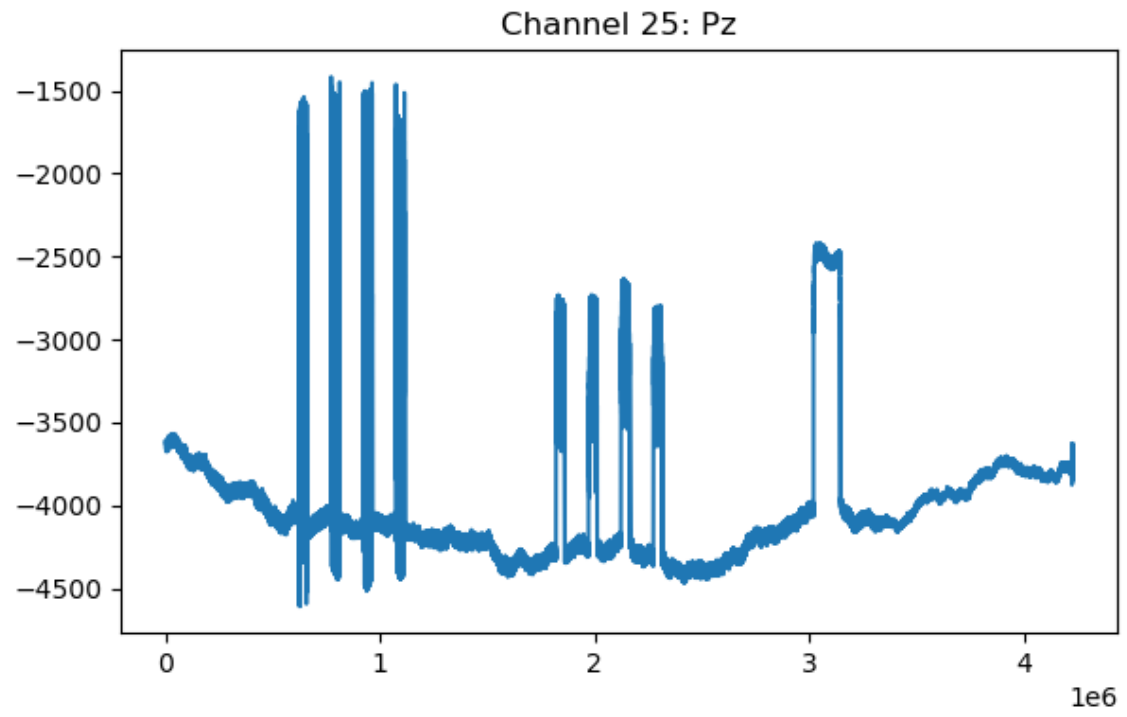
Channel 25: Pz
−1500
−2000
−2500
−3000
−3500
−4000
−4500
0
1
2
3
4
1e6

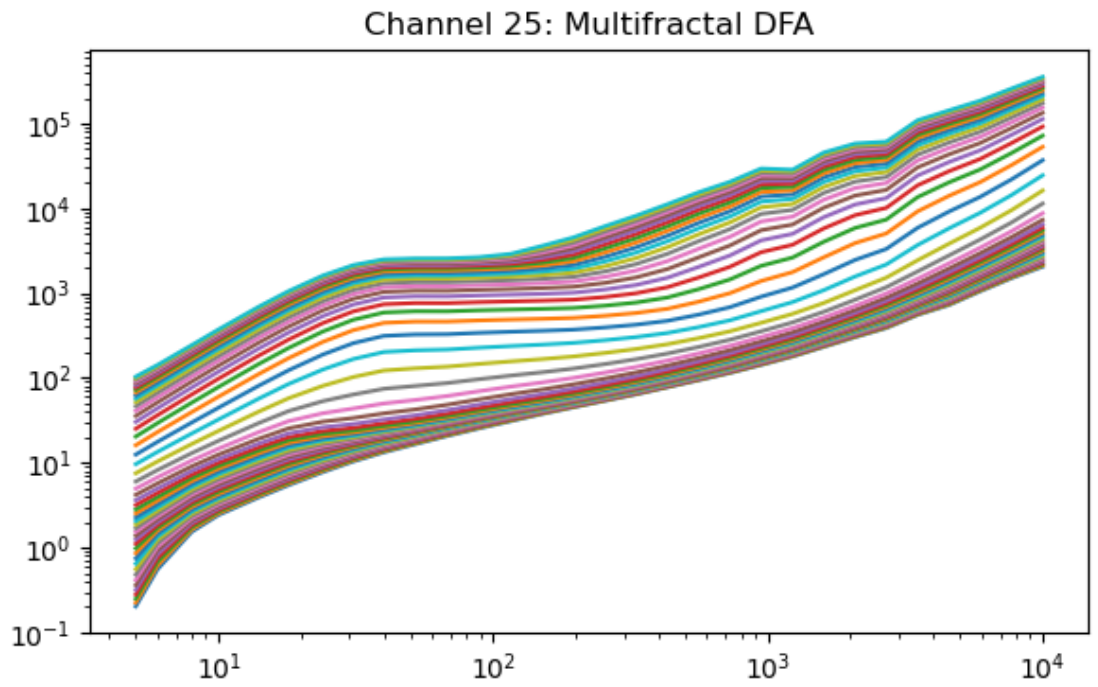
Channel 25: Multifractal DFA
$10^5$
$10^4$
$10^3$
$10^2$
$10^1$
$10^0$
$10^{-1}$
$10^1$
$10^2$
$10^3$
$10^4$

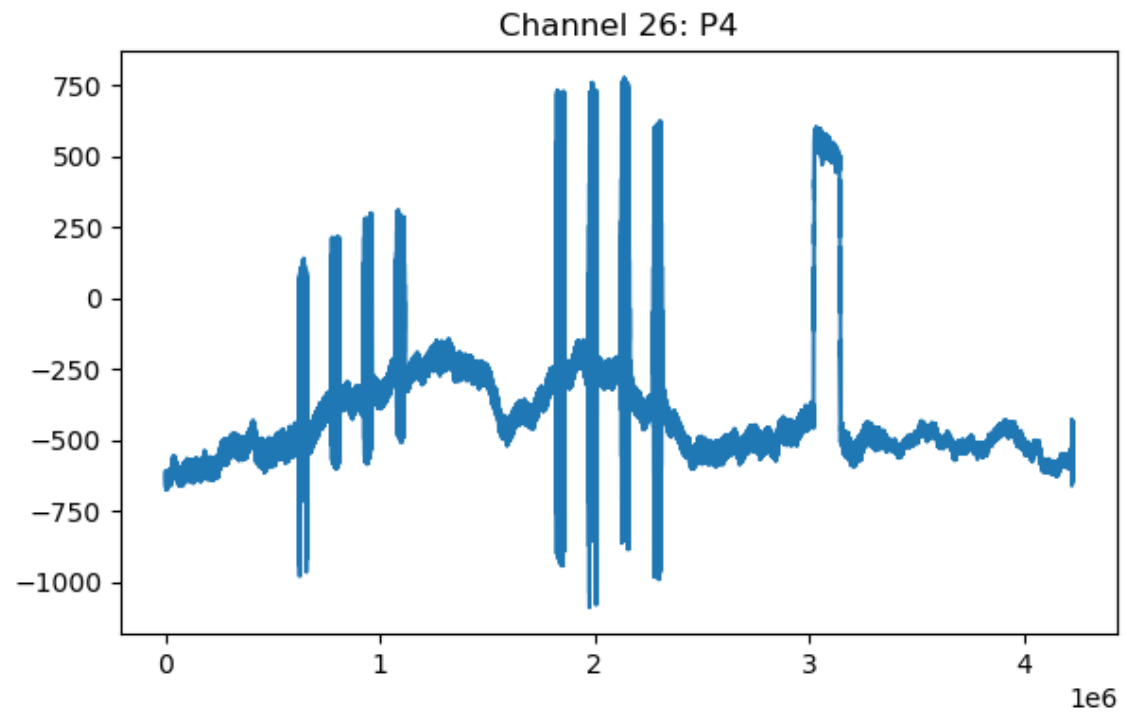
Channel 26: P4
750
500
250
0
−250
−500
−750
−1000
0
1
2
3
4
1e6

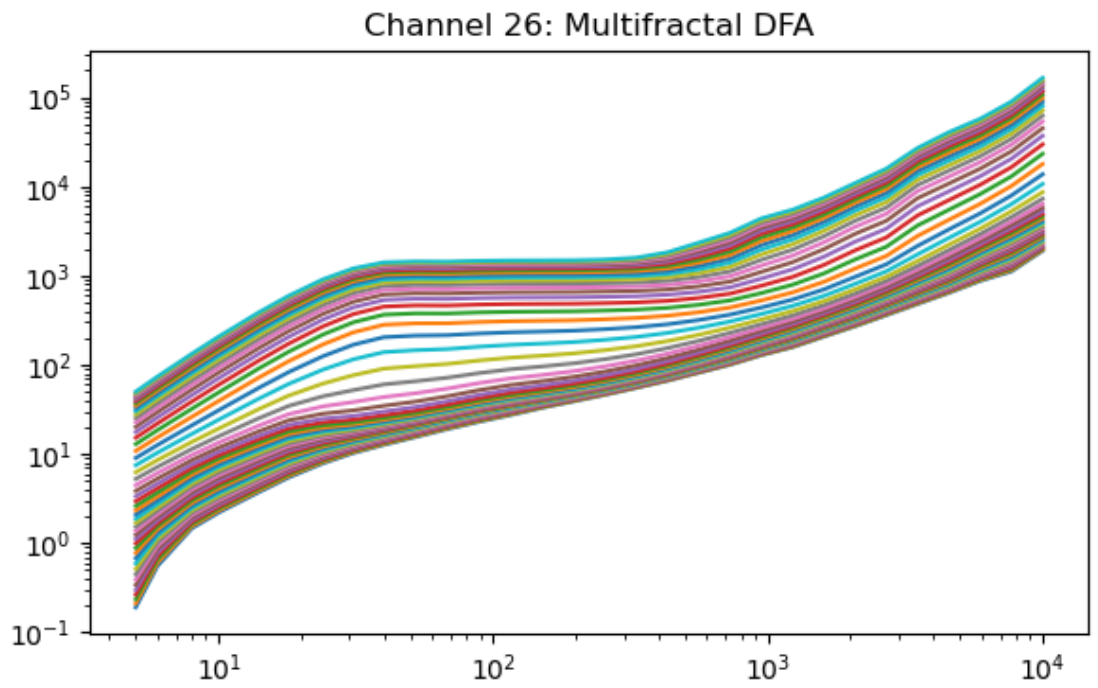
Channel 26: Multifractal DFA
$10^5$
$10^4$
$10^3$
$10^2$
$10^1$
$10^0$
$10^{-1}$
$10^1$
$10^2$
$10^3$
$10^4$

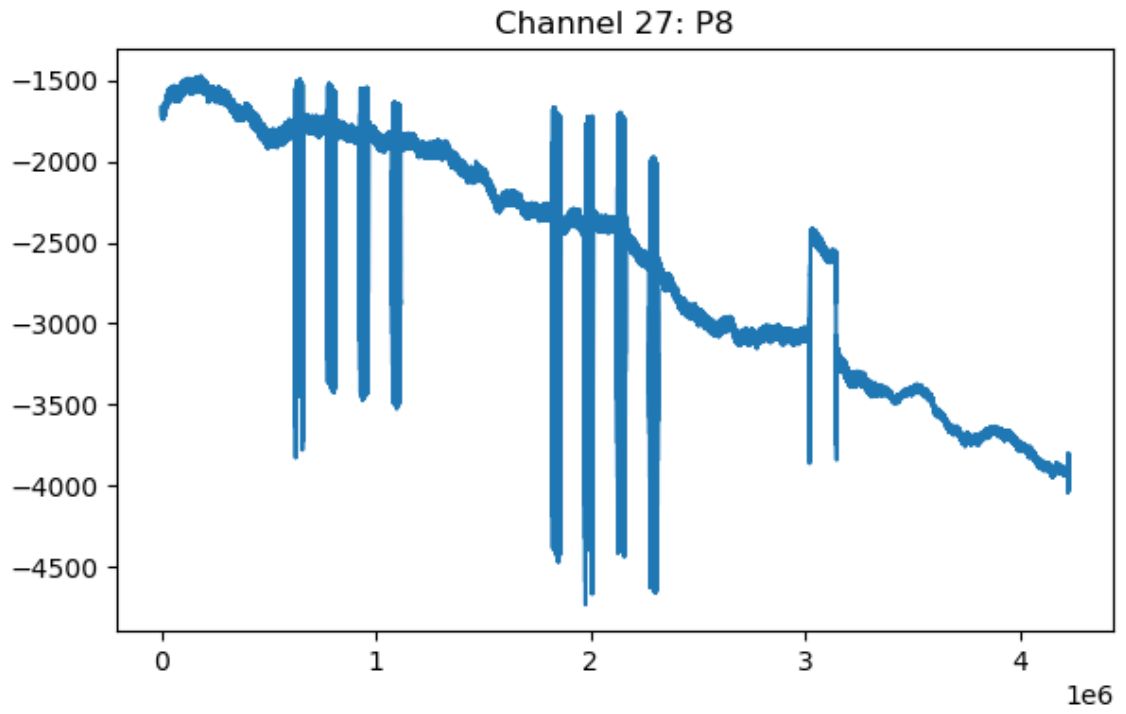
Channel 27: P8
−1500
−2000
−2500
−3000
−3500
−4000
−4500
0
1
2
3
4
1e6

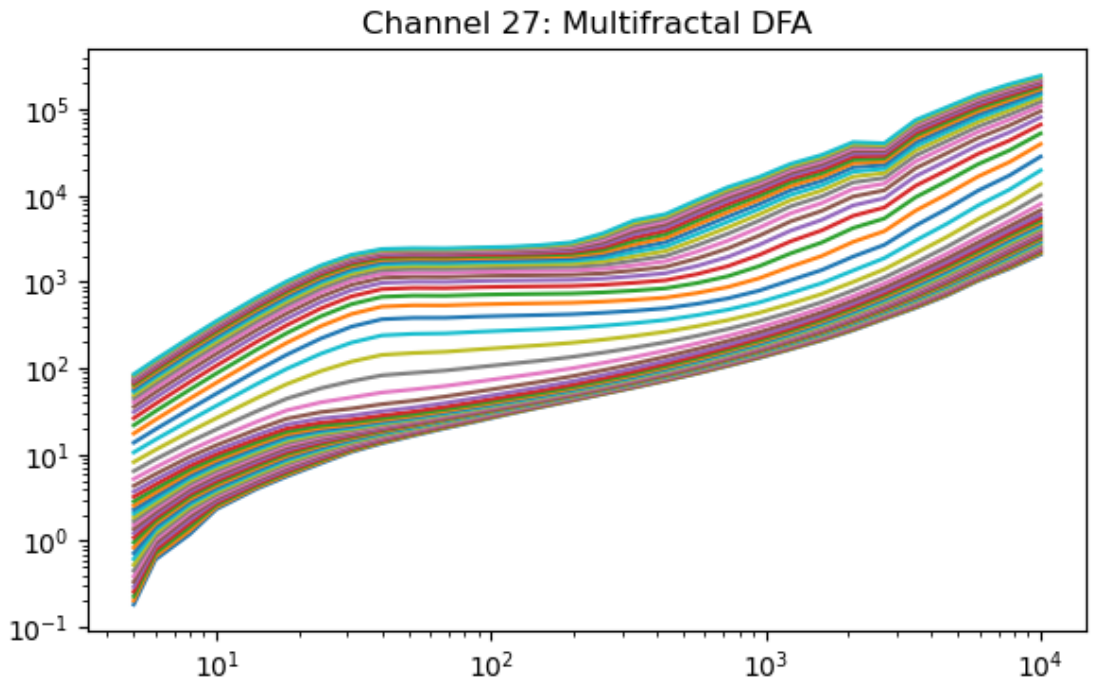
Channel 27: Multifractal DFA

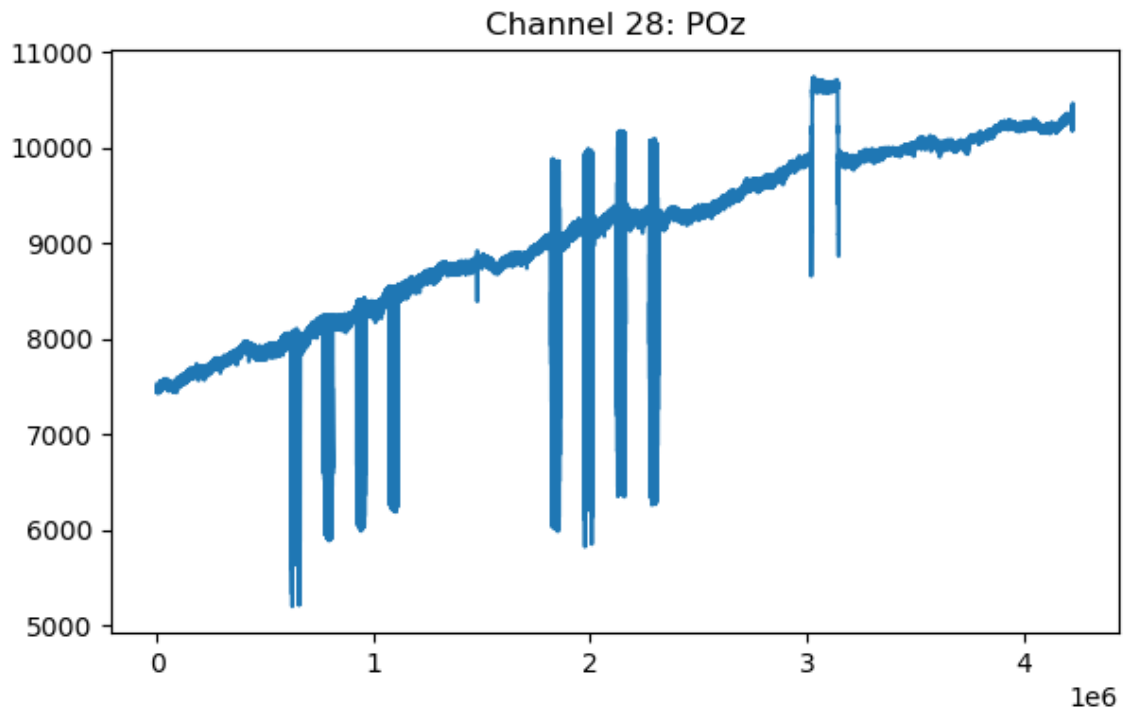
Channel 28: POz
11000
10000
9000
8000
7000
6000
5000
0
1
2
3
4
1e6

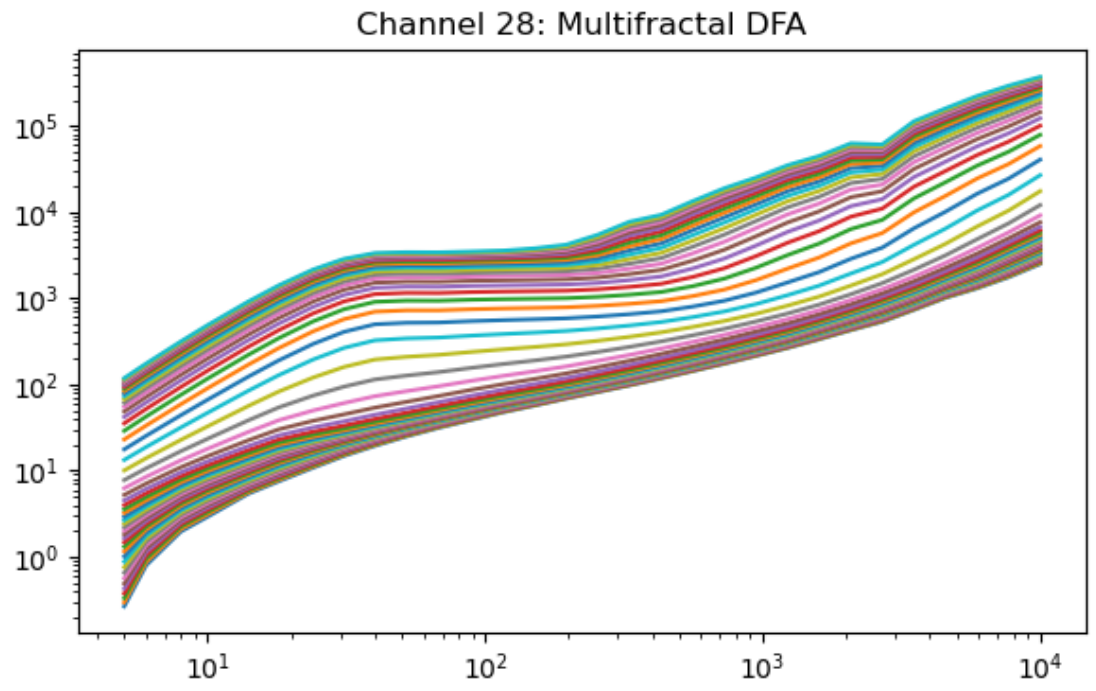
Channel 28: Multifractal DFA

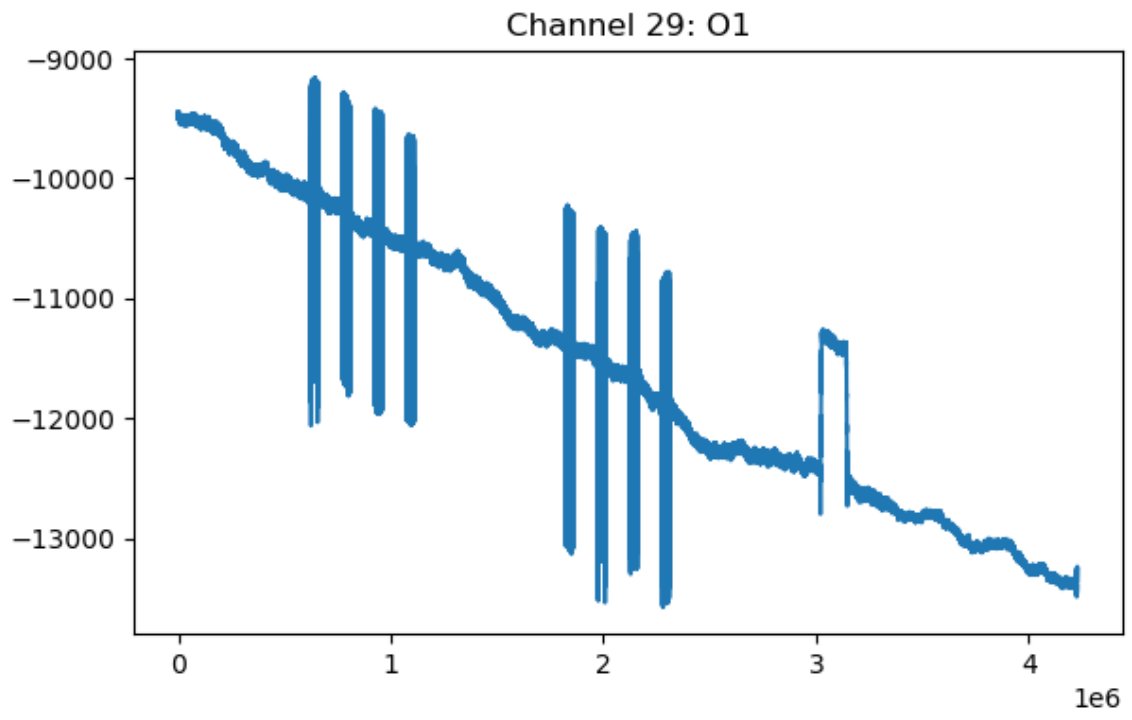
Channel 29: O1
−9000
−10000
−11000
−12000
−13000
0
1
2
3
4
1e6

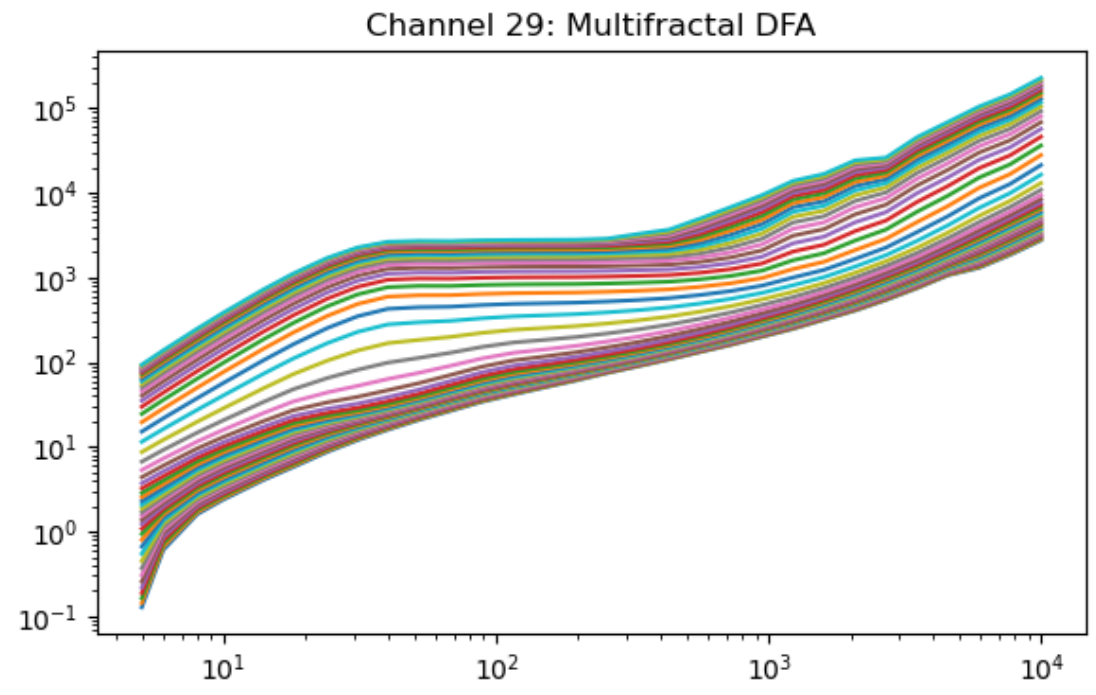
Channel 29: Multifractal DFA

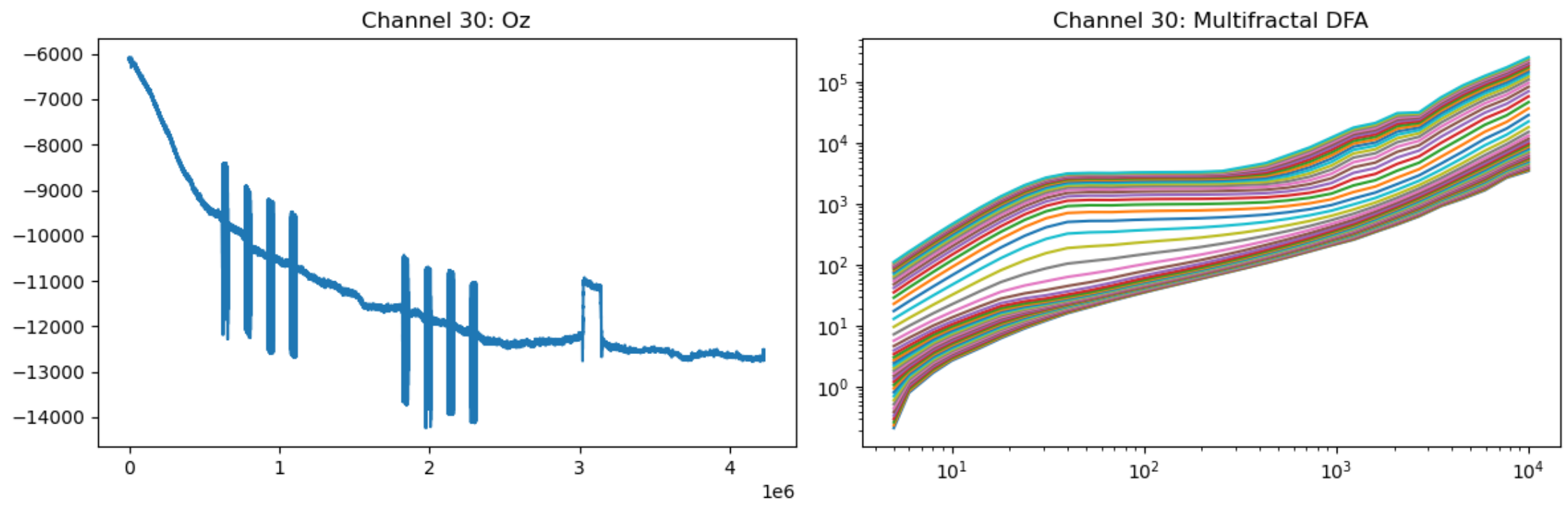

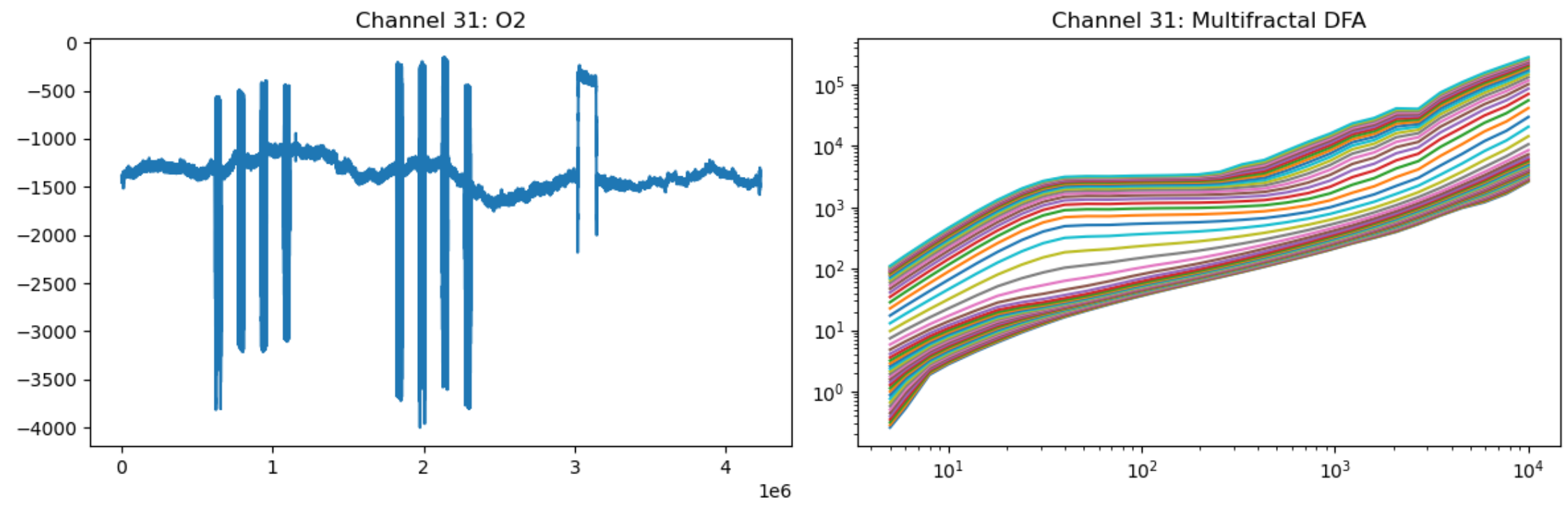

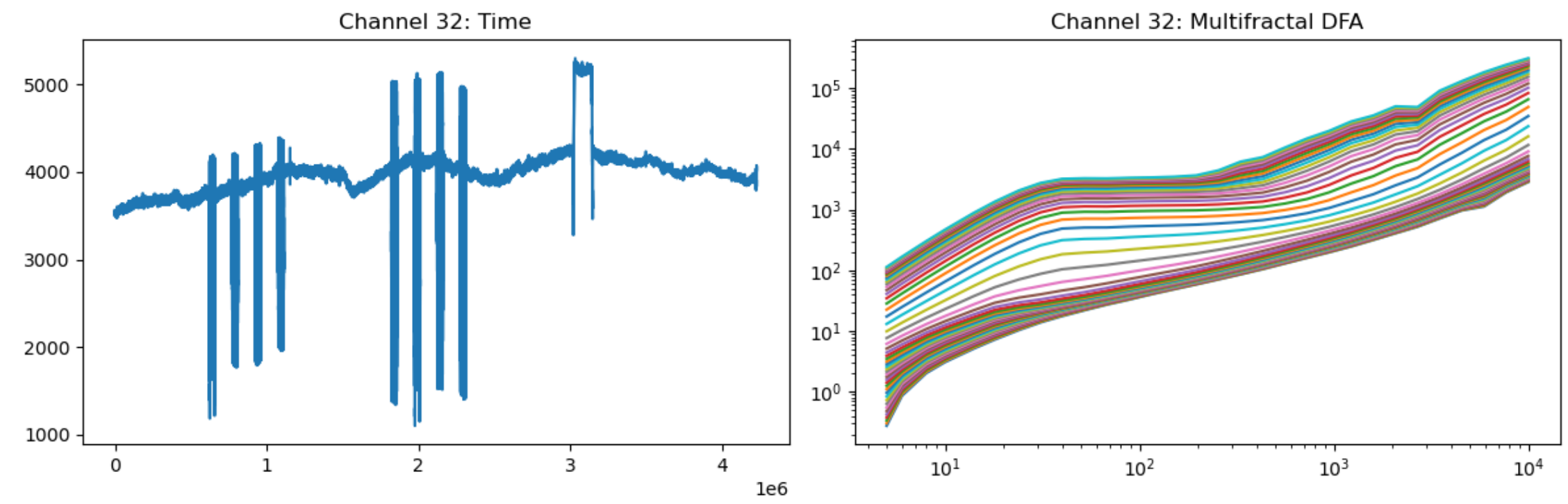

[5]:

```python
# Save mfdfa_results to a file
mfdfa_results_dict = {f'channel_{i}': result for i, result in enumerate(mfdfa_results)}

# Save the results to a file
save_dir = '/home/vincent/AAA_projects/MVCS/Neuroscience/MFDFA/'
np.save(save_dir + 'MFDFA_results.npy', mfdfa_results_dict)
```

## 3 Extract and save features

[11]:
```
import numpy as np
import os

# 1. Load MFDFA results
mfdfa_results_path = '/home/vincent/AAA_projects/MVCS/Neuroscience/Analysis/
↪MFDFA/MFDFA_results.npy'
mfdfa_results_dict = np.load(mfdfa_results_path, allow_pickle=True).item()

# Convert the results into an array for ease of processing
mfdfa_data = [mfdfa_results_dict[key][1] for key in sorted(mfdfa_results_dict.
↪keys())]  # extracting the fluctuation part
mfdfa_data = np.array(mfdfa_data)  # shape (num_channels, q_size, lag_size)

# 2. Extract CNN and RNN features

# CNN Features
window_size = 10
step_size = window_size // 2
n_windows = (mfdfa_data.shape[2] - window_size) // step_size + 1
cnn_features = np.zeros((n_windows, mfdfa_data.shape[0], window_size))

for i in range(n_windows):
    start = i * step_size
    end = start + window_size
    cnn_features[i] = mfdfa_data[:, :, start:end].mean(axis=1)  # Taking mean␣
↪over q to reduce dimension

# RNN Features
rnn_features = mfdfa_data.mean(axis=1).reshape(mfdfa_data.shape[0], mfdfa_data.
↪shape[2], 1)

# 3. Save the features and print their heads and shapes

save_path = "/home/vincent/AAA_projects/MVCS/Neuroscience/Analysis/MFDFA"
np.save(os.path.join(save_path, 'cnn_mfdfa.npy'), cnn_features)
np.save(os.path.join(save_path, 'rnn_mfdfa.npy'), rnn_features)

print(f"CNN Features Shape: {cnn_features.shape}")
print(f"RNN Features Shape: {rnn_features.shape}")

print("CNN Feature Head:")
print(cnn_features[0, :, :5])

print("RNN Feature Head:")
print(rnn_features[:, :5, :])
```

```
CNN Features Shape: (9, 32, 10)
RNN Features Shape: (32, 50, 1)
CNN Feature Head:
[[808.78941721 831.00442732 854.7410567  880.12387655 907.28505507]
 [702.95878259 720.23441807 738.61648119 758.18501829 779.01974983]
 [340.18842364 347.04416282 354.25879743 361.85748201 369.8677704 ]
 [347.94740111 355.05768471 362.535901   370.40556346 378.69189862]
 [464.54112682 474.24992283 484.48328065 495.27454899 506.65882269]
 [425.92907677 435.25576806 445.16549063 455.70478323 466.92387353]
 [295.19338221 302.11861158 309.50598242 317.39657168 325.83563606]
 [230.52373467 236.32210387 242.48060533 249.02668904 255.98996496]
 [249.51939669 256.38016509 263.67960321 271.44548593 279.70636752]
 [345.53220487 353.13214059 361.19101425 369.74267459 378.8229611 ]
 [491.41899525 501.49106744 512.10885031 523.3087109  535.12952201]
 [292.33557652 298.98846364 306.01958873 313.45573303 321.32578637]
 [747.1362778  767.99189053 790.12370047 813.60498171 838.51150602]
 [205.70463205 211.25815164 217.19656933 223.55441627 230.36951101]
 [197.19228356 202.70113502 208.5923901  214.89692728 221.64764109]
 [281.1320621  287.84611077 294.94984705 302.47229619 310.44507526]
 [362.7548123  370.44662105 378.58658508 387.20904262 396.35112986]
 [258.74566443 264.82415982 271.31486184 278.25396736 285.68030413]
 [223.00574234 228.35527715 234.05646592 240.13855249 246.63316765]
 [249.84875076 256.27334974 263.07662292 270.27954655 277.90431733]
 [334.96964821 342.33936032 350.0923249  358.25427266 366.8531522 ]
 [342.49171032 350.63392889 359.27998569 368.4669541  378.23438147]
 [431.36730363 442.54896252 454.43352574 467.07296938 480.52357021]
 [486.36037792 496.773076   507.72036538 519.23666113 531.35930219]
 [325.5337415  333.63843199 342.16008617 351.11335335 360.51340386]
 [338.06945632 345.10713958 352.53489632 360.38134016 368.67760805]
 [414.07400572 422.69823946 431.79658924 441.40679474 451.57107341]
 [460.21909111 469.99734993 480.29037808 491.1290055  502.54639224]
 [405.32873856 413.87515683 422.88104757 432.37318659 442.37942467]
 [437.48631526 446.52941929 456.04418083 466.06320766 476.62247274]
 [353.81174556 361.67465794 369.9755212  378.74510873 388.01715142]
 [319.75306088 326.72902532 334.08520288 341.84816312 350.04688261]]
RNN Feature Head:
[[[808.78941721]
  [831.00442732]
  [854.7410567 ]
  [880.12387655]
  [907.28505507]]

 [[702.95878259]
  [720.23441807]
  [738.61648119]
  [758.18501829]
  [779.01974983]]
```

```
[[340.18842364]
 [347.04416282]
 [354.25879743]
 [361.85748201]
 [369.8677704 ]]

[[347.94740111]
 [355.05768471]
 [362.535901  ]
 [370.40556346]
 [378.69189862]]

[[464.54112682]
 [474.24992283]
 [484.48328065]
 [495.27454899]
 [506.65882269]]

[[425.92907677]
 [435.25576806]
 [445.16549063]
 [455.70478323]
 [466.92387353]]

[[295.19338221]
 [302.11861158]
 [309.50598242]
 [317.39657168]
 [325.83563606]]

[[230.52373467]
 [236.32210387]
 [242.48060533]
 [249.02668904]
 [255.98996496]]

[[249.51939669]
 [256.38016509]
 [263.67960321]
 [271.44548593]
 [279.70636752]]

[[345.53220487]
 [353.13214059]
 [361.19101425]
 [369.74267459]
 [378.8229611 ]]
```

```
[[491.41899525]
 [501.49106744]
 [512.10885031]
 [523.3087109 ]
 [535.12952201]]

[[292.33557652]
 [298.98846364]
 [306.01958873]
 [313.45573303]
 [321.32578637]]

[[747.1362778 ]
 [767.99189053]
 [790.12370047]
 [813.60498171]
 [838.51150602]]

[[205.70463205]
 [211.25815164]
 [217.19656933]
 [223.55441627]
 [230.36951101]]

[[197.19228356]
 [202.70113502]
 [208.5923901 ]
 [214.89692728]
 [221.64764109]]

[[281.1320621 ]
 [287.84611077]
 [294.94984705]
 [302.47229619]
 [310.44507526]]

[[362.7548123 ]
 [370.44662105]
 [378.58658508]
 [387.20904262]
 [396.35112986]]

[[258.74566443]
 [264.82415982]
 [271.31486184]
 [278.25396736]
 [285.68030413]]
```

```
[[223.00574234]
 [228.35527715]
 [234.05646592]
 [240.13855249]
 [246.63316765]]

[[249.84875076]
 [256.27334974]
 [263.07662292]
 [270.27954655]
 [277.90431733]]

[[334.96964821]
 [342.33936032]
 [350.0923249 ]
 [358.25427266]
 [366.8531522 ]]

[[342.49171032]
 [350.63392889]
 [359.27998569]
 [368.4669541 ]
 [378.23438147]]

[[431.36730363]
 [442.54896252]
 [454.43352574]
 [467.07296938]
 [480.52357021]]

[[486.36037792]
 [496.773076  ]
 [507.72036538]
 [519.23666113]
 [531.35930219]]

[[325.5337415 ]
 [333.63843199]
 [342.16008617]
 [351.11335335]
 [360.51340386]]

[[338.06945632]
 [345.10713958]
 [352.53489632]
 [360.38134016]
 [368.67760805]]
```

```
[[414.07400572]
 [422.69823946]
 [431.79658924]
 [441.40679474]
 [451.57107341]]

[[460.21909111]
 [469.99734993]
 [480.29037808]
 [491.1290055 ]
 [502.54639224]]

[[405.32873856]
 [413.87515683]
 [422.88104757]
 [432.37318659]
 [442.37942467]]

[[437.48631526]
 [446.52941929]
 [456.04418083]
 [466.06320766]
 [476.62247274]]

[[353.81174556]
 [361.67465794]
 [369.9755212 ]
 [378.74510873]
 [388.01715142]]

[[319.75306088]
 [326.72902532]
 [334.08520288]
 [341.84816312]
 [350.04688261]]]
```

## 4 Hurst Exponents

[115]:
```
# Initialize an array to store the Hurst exponent for each channel
num_channels = eeg_df_values.shape[1]
hurst_exponents = np.zeros(num_channels)

# Check if mfdfa_results has data for each channel
if len(mfdfa_results) != num_channels:
    print("Error: MFDFA results do not have data for each channel.")
else:
    # Loop over each channel
```

```
    for ch in range(num_channels):
        # Extract the scale and fluctuation parameters for the current channel
        scale, fluct = mfdfa_results[ch]

        # Calculate the mean fluctuation over all time scales
        mean_fluct = np.mean(fluct, axis=0)

        # Print the head (first 10 elements) of scale and mean fluctuation␣
↪values for the current channel
        print(f"Channel {ch+1} - Scale:", scale[:10])
        print(f"Channel {ch+1} - Mean Fluctuation:", mean_fluct[:10])

        # Fit a line in log-log space to get the Hurst exponent
        H_hat = np.polyfit(np.log(scale)[4:20], np.log(mean_fluct)[4:20], 1)[0]

        # Store the Hurst exponent for the current channel
        hurst_exponents[ch] = H_hat.item()  # Extract the scalar value

    # Specify the file path for saving the Hurst exponents
    file_path = '/home/vincent/AAA_projects/MVCS/Neuroscience/HurstExponents/
↪hurst_exponents.npy'

    # Save the Hurst exponents to the file
    np.save(file_path, hurst_exponents)

    # Print the head (first 10 elements) of the Hurst exponents array
    print("Hurst exponents:", hurst_exponents)

    print("Hurst exponents saved to:", file_path)
```

```
Channel 1 - Scale: [ 5  6  8 10 14 18 24 31 40 52]
Channel 1 - Mean Fluctuation: [ 808.78941721  831.00442732  854.7410567
880.12387655  907.28505507
  936.36302415  967.50039245 1000.84106592 1036.52671978 1074.6930805 ]
Channel 2 - Scale: [ 5  6  8 10 14 18 24 31 40 52]
Channel 2 - Mean Fluctuation: [702.95878259 720.23441807 738.61648119
758.18501829 779.01974983
 801.19698379 824.78610711 849.84637455 876.42514008 904.55888951]
Channel 3 - Scale: [ 5  6  8 10 14 18 24 31 40 52]
Channel 3 - Mean Fluctuation: [747.1362778  767.99189053 790.12370047
813.60498171 838.51150602
 864.9218368  892.91779162 922.58511433 954.01441952 987.30251307]
Channel 4 - Scale: [ 5  6  8 10 14 18 24 31 40 52]
Channel 4 - Mean Fluctuation: [486.36037792 496.773076   507.72036538
519.23666113 531.35930219
 544.12894623 557.59006258 571.79156228 586.78762672 602.63883684]
Channel 5 - Scale: [ 5  6  8 10 14 18 24 31 40 52]
```

```
Channel 5 - Mean Fluctuation: [414.07400572 422.69823946 431.79658924
441.40679474 451.57107341
 462.33686534 473.75774219 485.89456497 498.81704696 512.60592194]
Channel 6 - Scale: [ 5  6  8 10 14 18 24 31 40 52]
Channel 6 - Mean Fluctuation: [460.21909111 469.99734993 480.29037808
491.1290055  502.54639224
 514.57840159 527.26406349 540.64616475 554.77203566 569.69462631]
Channel 7 - Scale: [ 5  6  8 10 14 18 24 31 40 52]
Channel 7 - Mean Fluctuation: [405.32873856 413.87515683 422.88104757
432.37318659 442.37942467
 452.92871757 464.05120508 475.77836517 488.14327653 501.18103558]
Channel 8 - Scale: [ 5  6  8 10 14 18 24 31 40 52]
Channel 8 - Mean Fluctuation: [437.48631526 446.52941929 456.04418083
466.06320766 476.62247274
 487.76172183 499.52490294 511.96062724 525.12268042 539.07061749]
Channel 9 - Scale: [ 5  6  8 10 14 18 24 31 40 52]
Channel 9 - Mean Fluctuation: [353.81174556 361.67465794 369.9755212
378.74510873 388.01715142
 397.82877423 408.22103452 419.23960662 430.93567971 443.36717441]
Channel 10 - Scale: [ 5  6  8 10 14 18 24 31 40 52]
Channel 10 - Mean Fluctuation: [319.75306088 326.72902532 334.08520288
341.84816312 350.04688261
 358.71304226 367.88137871 377.59010898 387.88145871 398.80234301]
Channel 11 - Scale: [ 5  6  8 10 14 18 24 31 40 52]
Channel 11 - Mean Fluctuation: [340.18842364 347.04416282 354.25879743
361.85748201 369.8677704
 378.31988452 387.24700468 396.68558385 406.67569295 417.26141563]
Channel 12 - Scale: [ 5  6  8 10 14 18 24 31 40 52]
Channel 12 - Mean Fluctuation: [347.94740111 355.05768471 362.535901
370.40556346 378.69189862
 387.42198907 396.62493684 406.33205656 416.57711299 427.39662958]
Channel 13 - Scale: [ 5  6  8 10 14 18 24 31 40 52]
Channel 13 - Mean Fluctuation: [464.54112682 474.24992283 484.48328065
495.27454899 506.65882269
 518.6729496  531.35554827 544.74705364 558.8898249  573.82837718]
Channel 14 - Scale: [ 5  6  8 10 14 18 24 31 40 52]
Channel 14 - Mean Fluctuation: [425.92907677 435.25576806 445.16549063
455.70478323 466.92387353
 478.87682375 491.62165783 505.22048413 519.73964868 535.2499907 ]
Channel 15 - Scale: [ 5  6  8 10 14 18 24 31 40 52]
Channel 15 - Mean Fluctuation: [295.19338221 302.11861158 309.50598242
317.39657168 325.83563606
 334.8730631  344.56387889 354.96882774 366.15504876 378.19689543]
Channel 16 - Scale: [ 5  6  8 10 14 18 24 31 40 52]
Channel 16 - Mean Fluctuation: [230.52373467 236.32210387 242.48060533
249.02668904 255.98996496
 263.40233187 271.298107   279.71416033 288.69006208 298.26826166]
Channel 17 - Scale: [ 5  6  8 10 14 18 24 31 40 52]
```

```
Channel 17 - Mean Fluctuation: [249.51939669 256.38016509 263.67960321
271.44548593 279.70636752
 288.49180896 297.83280065 307.76238154 318.31641617 329.53445484]
Channel 18 - Scale: [ 5  6  8 10 14 18 24 31 40 52]
Channel 18 - Mean Fluctuation: [345.53220487 353.13214059 361.19101425
369.74267459 378.8229611
 388.46967529 398.72256154 409.62332528 421.2157284  433.54581716]
Channel 19 - Scale: [ 5  6  8 10 14 18 24 31 40 52]
Channel 19 - Mean Fluctuation: [491.41899525 501.49106744 512.10885031
523.3087109  535.12952201
 547.61288373 560.80341176 574.74912107 589.50194478 605.11844824]
Channel 20 - Scale: [ 5  6  8 10 14 18 24 31 40 52]
Channel 20 - Mean Fluctuation: [292.33557652 298.98846364 306.01958873
313.45573303 321.32578637
 329.66091781 338.49476299 347.86363749 357.80679573 368.36677563]
Channel 21 - Scale: [ 5  6  8 10 14 18 24 31 40 52]
Channel 21 - Mean Fluctuation: [205.70463205 211.25815164 217.19656933
223.55441627 230.36951101
 237.68320103 245.54060558 253.99086097 263.08737367 272.88809617]
Channel 22 - Scale: [ 5  6  8 10 14 18 24 31 40 52]
Channel 22 - Mean Fluctuation: [197.19228356 202.70113502 208.5923901
214.89692728 221.64764109
 228.87939722 236.62895233 244.93484487 253.83727335 263.37799464]
Channel 23 - Scale: [ 5  6  8 10 14 18 24 31 40 52]
Channel 23 - Mean Fluctuation: [281.1320621  287.84611077 294.94984705
302.47229619 310.44507526
 318.90261242 327.88237562 337.42510568 347.57504601 358.38016663]
Channel 24 - Scale: [ 5  6  8 10 14 18 24 31 40 52]
Channel 24 - Mean Fluctuation: [362.7548123  370.44662105 378.58658508
387.20904262 396.35112986
 406.05289903 416.35740402 427.3107469  438.96208312 451.36359659]
Channel 25 - Scale: [ 5  6  8 10 14 18 24 31 40 52]
Channel 25 - Mean Fluctuation: [258.74566443 264.82415982 271.31486184
278.25396736 285.68030413
 293.63519448 302.16222571 311.30692955 321.11639116 331.63883874]
Channel 26 - Scale: [ 5  6  8 10 14 18 24 31 40 52]
Channel 26 - Mean Fluctuation: [223.00574234 228.35527715 234.05646592
240.13855249 246.63316765
 253.57443084 260.99902958 268.9462676  277.45807321 286.57896366]
Channel 27 - Scale: [ 5  6  8 10 14 18 24 31 40 52]
Channel 27 - Mean Fluctuation: [249.84875076 256.27334974 263.07662292
270.27954655 277.90431733
 285.97493034 294.51790902 303.56313367 313.14470093 323.30176143]
Channel 28 - Scale: [ 5  6  8 10 14 18 24 31 40 52]
Channel 28 - Mean Fluctuation: [334.96964821 342.33936032 350.0923249
358.25427266 366.8531522
 375.91934423 385.48589646 395.58878765 406.26723666 417.56408839]
Channel 29 - Scale: [ 5  6  8 10 14 18 24 31 40 52]
```

```
Channel 29 - Mean Fluctuation: [342.49171032 350.63392889 359.27998569
368.4669541  378.23438147
 388.6244378  399.68210322 411.45540169 423.99568433 437.35796508]
Channel 30 - Scale: [ 5  6  8 10 14 18 24 31 40 52]
Channel 30 - Mean Fluctuation: [431.36730363 442.54896252 454.43352574
467.07296938 480.52357021
 494.84650029 510.10853654 526.38288187 543.75008657 562.29905725]
Channel 31 - Scale: [ 5  6  8 10 14 18 24 31 40 52]
Channel 31 - Mean Fluctuation: [325.5337415  333.63843199 342.16008617
351.11335335 360.51340386
 370.37664356 380.72157466 391.56975619 402.94682099 414.88352232]
Channel 32 - Scale: [ 5  6  8 10 14 18 24 31 40 52]
Channel 32 - Mean Fluctuation: [338.06945632 345.10713958 352.53489632
360.38134016 368.67760805
 377.45761355 386.75833248 396.6201316  407.08715926 418.20783212]
Hurst exponents: [0.14766394 0.12747417 0.13936538 0.11284491 0.12113714
0.11161206
 0.10870372 0.11140832 0.12256221 0.11815558 0.10900003 0.10714402
 0.10927929 0.12524573 0.14115001 0.13666074 0.14234257 0.12094061
 0.11011225 0.12241291 0.15583857 0.15274446 0.1286373  0.11745562
 0.13577366 0.13653462 0.13278782 0.11488157 0.12993843 0.14124633
 0.11902633 0.11474097]
Hurst exponents saved to:
/home/vincent/AAA_projects/MVCS/Neuroscience/HurstExponents/hurst_exponents.npy
```

# 5 Montage Setup for EEG Channels

[104]:

```
scalp_positions_1010 = {
    'Fp1': (-0.9511, -0.3090),
    'Fpz': (-1.0000, 0.0000),
    'Fp2': (-0.9511, 0.3090),
    'F7': (-0.5878, -0.8090),
    'F3': (-0.8090, -0.5878),
    'Fz': (-0.8090, 0.0000),
    'F4': (-0.8090, 0.5878),
    'F8': (-0.5878, 0.8090),
    'FC5': (-0.4045, -0.8090),
    'FC1': (-0.6545, -0.3090),
    'FC2': (-0.6545, 0.3090),
    'FC6': (-0.4045, 0.8090),
    'M1': (0.1545, -1.0000),
    'T7': (-0.0000, -1.0000),
    'C3': (-0.3090, -0.9511),
    'Cz': (0.0000, -0.8090),
    'C4': (-0.3090, -0.9511),
    'T8': (0.0000, 1.0000),
    'M2': (0.1545, 1.0000),
```

```python
    'CP5': (0.4045, -0.8090),
    'CP1': (0.6545, -0.3090),
    'CP2': (0.6545, 0.3090),
    'CP6': (0.4045, 0.8090),
    'P7': (0.5878, -0.8090),
    'P3': (0.8090, -0.5878),
    'Pz': (0.8090, 0.0000),
    'P4': (0.8090, 0.5878),
    'P8': (0.5878, 0.8090),
    'POz': (0.9511, 0.0000),
    'O1': (1.0000, -0.3090),
    'Oz': (1.0000, 0.0000),
    'O2': (1.0000, 0.3090)
}

# Create an MNE info object with channel names and sampling rate
info = mne.create_info(ch_names=list(channel_names), sfreq=sampling_rate, ch_types='eeg')

# Create the montage
montage = mne.channels.make_dig_montage(ch_pos=dict(zip(channel_names, scalp_positions)), coord_frame='head')

# Set the montage to the Info object
info.set_montage(montage)
```

[104]:
```
<Info | 8 non-empty values
 bads: []
 ch_names: Fp1, Fpz, Fp2, F7, F3, Fz, F4, F8, FC5, FC1, FC2, FC6, M1, T7, …
 chs: 32 EEG
 custom_ref_applied: False
 dig: 35 items (3 Cardinal, 32 EEG)
 highpass: 0.0 Hz
 lowpass: 500.0 Hz
 meas_date: unspecified
 nchan: 32
 projs: []
 sfreq: 1000.0 Hz
>
```

## 6 Visualizing Mean Hurst Exponent Topographical Map

[129]:
```python
import numpy as np
import mne

# Replace the following variables with your actual data
```

```
num_channels = 32  # Number of EEG channels
mean_hurst_exponent = hurst_exponents  # Array with mean Hurst exponent value␣
↪for each channel
grid_size = 10

# Generate random scalp positions with a small random offset to avoid␣
↪overlapping
offset = 0.1
scalp_positions = np.random.uniform(0, grid_size - offset, size=(num_channels,␣
↪2))
scalp_positions += np.random.uniform(0, offset, size=(num_channels, 2))
scalp_positions_3d = [(*pos, 0) for pos in scalp_positions]

# Step 1: Organize mean Hurst exponent values into a 2D grid corresponding to␣
↪scalp positions
# Initialize the mean_hurst_exponent_map with NaN values to avoid index out of␣
↪bounds error
mean_hurst_exponent_map = np.full((grid_size, grid_size), np.nan)

for i in range(num_channels):
    x, y = scalp_positions[i]
    mean_hurst_exponent_map[int(x), int(y)] = mean_hurst_exponent[i]

# Step 2: Plot the topographical map using MNE-Python
# Create an MNE info object with channel names and montage information
channel_names = [f'EEG {i+1}' for i in range(num_channels)]
sampling_rate = 1000  # Replace with your actual sampling rate
info = mne.create_info(ch_names=channel_names, sfreq=sampling_rate,␣
↪ch_types='eeg')

# Create a digitization object and add electrode locations to it
dig = mne.channels.make_dig_montage(ch_pos=dict(zip(channel_names,␣
↪scalp_positions_3d)), coord_frame='head')

# Add the digitization to the info object
info.set_montage(dig)

# Create the EvokedArray with your mean Hurst exponent data
data = mean_hurst_exponent[:, np.newaxis]
evoked = mne.EvokedArray(data, info)

# Plot the topographical map of mean Hurst exponent values
evoked.plot_topomap(times=0, scalings=1.0, cmap='viridis')

# Consider performing statistical tests (e.g., ANOVA, t-test) to assess group␣
↪differences or condition-related variations in the mean Hurst exponent values
```

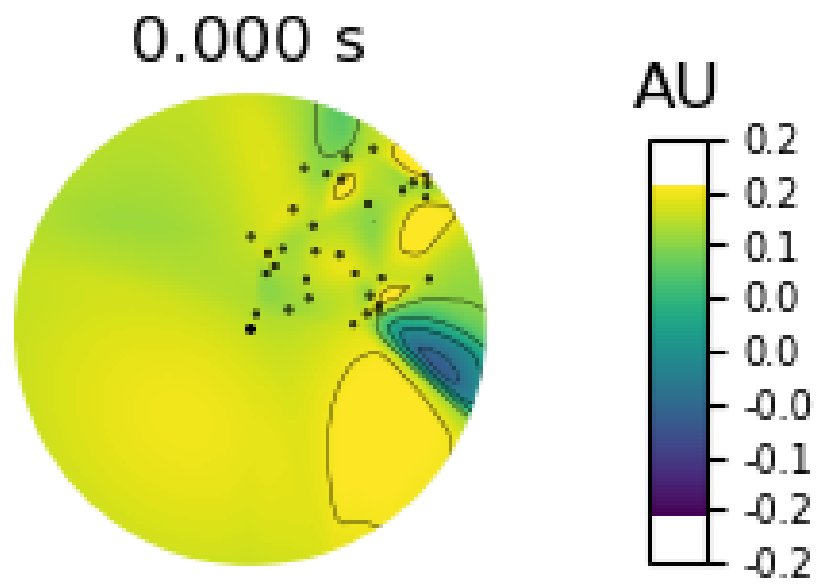

[129]:

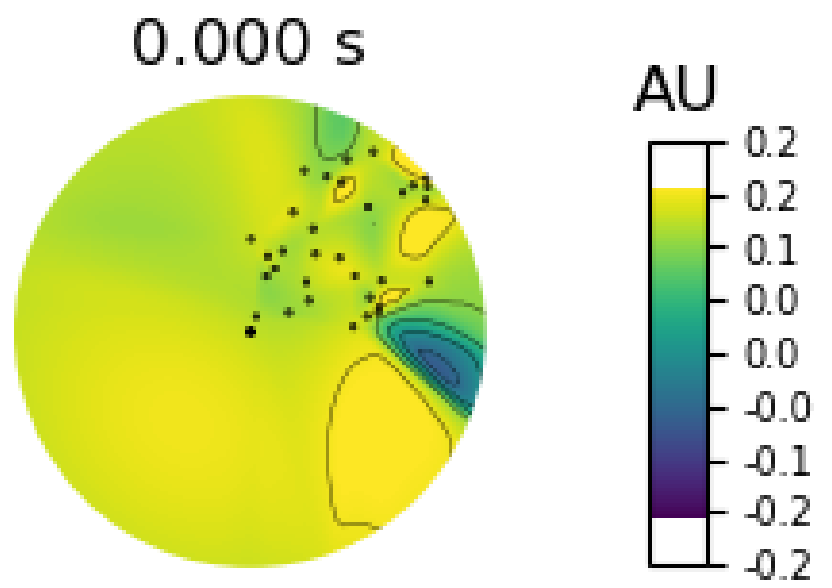

# 7 Load saved MFDFA results and plot them

[16]:

```
# Load the MFDFA results from the file
save_dir = '/home/vincent/AAA_projects/MVCS/Neuroscience/Analysis/MFDFA/'  #␣
↪the directory path where you saved the file
mfdfa_results = np.load(save_dir + 'MFDFA_results.npy', allow_pickle=True)
```

[17]:

```
# Check if mfdfa_results has data for each channel
num_channels = mfdfa_results.shape[0]
if num_channels == 0:
    print("Error: MFDFA results do not have data for any channel.")
else:
    # Initialize an array to store the Hurst exponent for each channel
    hurst_exponents = np.zeros(num_channels)

    # Loop over each channel
    for ch in range(num_channels):
        # Extract the scale and fluctuation parameters for the current channel
        scale, fluct, _, _ = saved_mfdfa_results[ch]

        # Fit a line in log-log space to get the Hurst exponent
        H_hat = np.polyfit(np.log(scale)[4:20], np.log(fluct)[4:20], 1)[0]
```

```
        # Store the Hurst exponent for the current channel
        hurst_exponents[ch] = H_hat

    # Visualize the results
    plt.figure(figsize=(10, 6))
    plt.bar(range(1, num_channels + 1), hurst_exponents)
    plt.xlabel('Channel')
    plt.ylabel('Hurst Exponent')
    plt.title('Hurst Exponent for Each EEG Channel')
    plt.show()
```

```
---------------------------------------------------------------------------
IndexError                                Traceback (most recent call last)
Cell In[17], line 2
      1 # Check if mfdfa_results has data for each channel
----> 2 num_channels = mfdfa_results.shape[0]
      3 if num_channels == 0:
      4     print("Error: MFDFA results do not have data for any channel.")

IndexError: tuple index out of range
```

# 8 MFDFA Feature Extraction and Data Formatting

[19]:
```
import numpy as np
import os

# Load the MFDFA results from the file
save_dir = '/home/vincent/AAA_projects/MVCS/Neuroscience/Analysis/MFDFA/'
mfdfa_results = np.load(save_dir + 'MFDFA_results.npy', allow_pickle=True)


# Define the number of timesteps in each input sequence
timesteps = 100

# Create continuous data by concatenating all scale and fluctuation arrays
continuous_scale = np.concatenate([scale for scale, _ in mfdfa_results])
continuous_fluct = np.concatenate([fluct for _, fluct in mfdfa_results])

# Initialize X data
cnn_X_data = []
rnn_X_data = []

# Create overlapping sequences
for i in range(len(continuous_scale) - timesteps):
```

```
    scale_seq = continuous_scale[i : i + timesteps]
    fluct_seq = continuous_fluct[i : i + timesteps]
    seq = np.column_stack([scale_seq, fluct_seq])

    cnn_X_data.append(seq.flatten())  # Flatten sequence for CNN
    rnn_X_data.append(seq)  # Keep sequence shape for RNN

# Convert lists to numpy arrays
cnn_X_data = np.array(cnn_X_data)
rnn_X_data = np.array(rnn_X_data)

print("Shape of cnn_X_data:", cnn_X_data.shape)
print("Shape of rnn_X_data:", rnn_X_data.shape)

save_path = "/home/vincent/AAA_projects/MVCS/Neuroscience/Analysis/MFDFA"
np.save(os.path.join(save_path, 'cnn_mfdfa_concatenated.npy'), cnn_X_data)
np.save(os.path.join(save_path, 'rnn_mfdfa_concatenated.npy'), rnn_X_data)
```

```
---------------------------------------------------------------------------
TypeError                                 Traceback (most recent call last)
Cell In[19], line 13
     10 timesteps = 100
     12 # Create continuous data by concatenating all scale and fluctuation␣
 ↪arrays
---> 13 continuous_scale = np.concatenate([scale for scale, _ in mfdfa_results]
     14 continuous_fluct = np.concatenate([fluct for _, fluct in mfdfa_results]
     16 # Initialize X data

TypeError: iteration over a 0-d array
```

[24]:

```
print(mfdfa_results)
```

```
{'channel_0': (array([    5,     6,     8,    10,    14,    18,    24,    31,
40,
          52,    68,    89,   116,   151,   196,   255,   331,   431,
         560,   727,   945,  1229,  1597,  2076,  2697,  3506,  4556,
        5921,  7694, 10000]), array([[3.17812754e-01, 3.54225702e-01,
3.97791564e-01, …,
        1.55199608e+02, 1.67307625e+02, 1.79692663e+02],
       [1.06771347e+00, 1.15658587e+00, 1.25650014e+00, …,
        1.88711853e+02, 2.02926240e+02, 2.17466373e+02],
       [2.91624019e+00, 3.06278577e+00, 3.21751270e+00, …,
        2.61666838e+02, 2.81624459e+02, 3.02396051e+02],
       …,
       [2.76022917e+03, 2.87026143e+03, 2.99107366e+03, …,
        5.74617372e+05, 6.00112829e+05, 6.24710145e+05],
       [5.00978504e+03, 5.16203350e+03, 5.32575195e+03, …,
```

```
        7.54650370e+05, 7.85482412e+05, 8.15147759e+05],
       [8.75979602e+03, 8.94817355e+03, 9.14550353e+03, …,
        9.71676625e+05, 1.00678134e+06, 1.04030819e+06]])), 'channel_1':
(array([    5,     6,     8,    10,    14,    18,    24,    31,    40,
          52,    68,    89,   116,   151,   196,   255,   331,   431,
         560,   727,   945,  1229,  1597,  2076,  2697,  3506,  4556,
        5921,  7694, 10000]), array([[2.81270381e-01, 3.10134330e-01,
3.44182802e-01, …,
        1.26342561e+02, 1.32079735e+02, 1.37875462e+02],
       [6.96043943e-01, 7.66790377e-01, 8.47495814e-01, …,
        1.79325945e+02, 1.86538122e+02, 1.93597071e+02],
       [2.10381441e+00, 2.19954124e+00, 2.30175144e+00, …,
        3.15971774e+02, 3.28109883e+02, 3.39819755e+02],
       …,
       [2.15318244e+03, 2.25751020e+03, 2.37261446e+03, …,
        3.93015538e+05, 4.10217026e+05, 4.26794170e+05],
       [4.71312119e+03, 4.81992136e+03, 4.93232219e+03, …,
        5.35807663e+05, 5.58049166e+05, 5.79474914e+05],
       [7.07570350e+03, 7.19509597e+03, 7.32006231e+03, …,
        7.25535109e+05, 7.53610770e+05, 7.80724247e+05]])), 'channel_2':
(array([    5,     6,     8,    10,    14,    18,    24,    31,    40,
          52,    68,    89,   116,   151,   196,   255,   331,   431,
         560,   727,   945,  1229,  1597,  2076,  2697,  3506,  4556,
        5921,  7694, 10000]), array([[2.54252149e-01, 2.82512926e-01,
3.16297456e-01, …,
        7.63228862e+01, 8.22335307e+01, 8.84349232e+01],
       [9.53329883e-01, 1.01997687e+00, 1.09324267e+00, …,
        9.68483814e+01, 1.03473301e+02, 1.10428071e+02],
       [2.24894751e+00, 2.35925640e+00, 2.47688837e+00, …,
        1.47537411e+02, 1.56043177e+02, 1.65047571e+02],
       …,
       [3.17059259e+03, 3.25761431e+03, 3.35108801e+03, …,
        3.21919199e+05, 3.36205124e+05, 3.49992672e+05],
       [4.55748381e+03, 4.68983513e+03, 4.83162763e+03, …,
        4.25024760e+05, 4.42434199e+05, 4.59186583e+05],
       [7.43498795e+03, 7.65310000e+03, 7.88100885e+03, …,
        5.43994232e+05, 5.63481430e+05, 5.82094947e+05]])), 'channel_3':
(array([    5,     6,     8,    10,    14,    18,    24,    31,    40,
          52,    68,    89,   116,   151,   196,   255,   331,   431,
         560,   727,   945,  1229,  1597,  2076,  2697,  3506,  4556,
        5921,  7694, 10000]), array([[4.93478758e-01, 5.44183062e-01,
6.04032230e-01, …,
        1.74136353e+03, 1.80760181e+03, 1.87096524e+03],
       [1.45956720e+00, 1.57712303e+00, 1.70691312e+00, …,
        2.57855882e+03, 2.67661479e+03, 2.77041476e+03],
       [3.44619200e+00, 3.60797689e+00, 3.77631475e+00, …,
        4.65806429e+03, 4.83500104e+03, 5.00426047e+03],
       …,
```

```
        [1.92408249e+03, 1.96699115e+03, 2.01210914e+03, …,
         1.88967436e+06, 1.99888665e+06, 2.10593814e+06],
        [2.74460444e+03, 2.80279747e+03, 2.86416353e+03, …,
         2.75559494e+06, 2.91170015e+06, 3.06428909e+06],
        [3.53613208e+03, 3.63309428e+03, 3.73584665e+03, …,
         4.51984320e+06, 4.74832148e+06, 4.96960409e+06]])), 'channel_4':
(array([    5,     6,     8,    10,    14,    18,    24,    31,    40,
           52,    68,    89,   116,   151,   196,   255,   331,   431,
          560,   727,   945,  1229,  1597,  2076,  2697,  3506,  4556,
         5921,  7694, 10000]), array([[4.54945936e-01, 4.99850018e-01,
5.52411839e-01, …,
         5.34521902e+02, 5.53675608e+02, 5.72071683e+02],
        [7.15138720e-01, 8.02329729e-01, 9.07202361e-01, …,
         7.88285338e+02, 8.16470819e+02, 8.43517460e+02],
        [2.99846181e+00, 3.12453249e+00, 3.25732794e+00, …,
         1.42085180e+03, 1.47173635e+03, 1.52056331e+03],
        …,
        [1.75603708e+03, 1.79076573e+03, 1.82750673e+03, …,
         8.91385043e+05, 9.29627946e+05, 9.66433675e+05],
        [2.38961902e+03, 2.43957619e+03, 2.49239817e+03, …,
         1.20425333e+06, 1.25255502e+06, 1.29890340e+06],
        [3.25314067e+03, 3.32397805e+03, 3.39922193e+03, …,
         1.62671780e+06, 1.68379287e+06, 1.73824289e+06]])), 'channel_5':
(array([    5,     6,     8,    10,    14,    18,    24,    31,    40,
           52,    68,    89,   116,   151,   196,   255,   331,   431,
          560,   727,   945,  1229,  1597,  2076,  2697,  3506,  4556,
         5921,  7694, 10000]), array([[2.36645013e-01, 2.62856884e-01,
2.93751156e-01, …,
         3.31976805e+02, 3.44613913e+02, 3.56707521e+02],
        [3.99284317e-01, 4.47963408e-01, 5.06555268e-01, …,
         4.91427641e+02, 5.10138096e+02, 5.28042705e+02],
        [1.68300302e+00, 1.75956355e+00, 1.84156393e+00, …,
         8.87513626e+02, 9.21269988e+02, 9.53570152e+02],
        …,
        [1.97752958e+03, 2.02037378e+03, 2.06538709e+03, …,
         4.20339706e+05, 4.40838279e+05, 4.60768809e+05],
        [2.80967039e+03, 2.86312522e+03, 2.91951803e+03, …,
         6.55528935e+05, 6.87956238e+05, 7.19428945e+05],
        [3.69323932e+03, 3.77948773e+03, 3.87046656e+03, …,
         1.17710345e+06, 1.23878207e+06, 1.29870265e+06]])), 'channel_6':
(array([    5,     6,     8,    10,    14,    18,    24,    31,    40,
           52,    68,    89,   116,   151,   196,   255,   331,   431,
          560,   727,   945,  1229,  1597,  2076,  2697,  3506,  4556,
         5921,  7694, 10000]), array([[2.47746155e-01, 2.73611214e-01,
3.04239746e-01, …,
         5.86189030e+01, 6.33723387e+01, 6.83582030e+01],
        [6.58173825e-01, 7.23727315e-01, 7.98309785e-01, …,
         7.36873355e+01, 7.90941128e+01, 8.47809410e+01],
```

[1.96805222e+00, 2.06861692e+00, 2.17642111e+00, …,
 1.09608732e+02, 1.16598639e+02, 1.24091864e+02],
…,
[1.73698806e+03, 1.77105703e+03, 1.80693889e+03, …,
 2.52749649e+05, 2.63884318e+05, 2.74619120e+05],
[2.33583302e+03, 2.38586135e+03, 2.43877252e+03, …,
 3.36738943e+05, 3.50507706e+05, 3.63748151e+05],
[3.26064393e+03, 3.33452925e+03, 3.41278852e+03, …,
 4.31687974e+05, 4.47100794e+05, 4.61817375e+05]])), 'channel_7':
(array([    5,     6,     8,    10,    14,    18,    24,    31,    40,
    52,    68,    89,   116,   151,   196,   255,   331,   431,
   560,   727,   945,  1229,  1597,  2076,  2697,  3506,  4556,
  5921,  7694, 10000]), array([[1.61238990e-01, 1.80779759e-01,
2.04626577e-01, …,
 9.10139188e+01, 9.43929775e+01, 9.77360130e+01],
[9.07124866e-01, 9.81491317e-01, 1.06388612e+00, …,
 1.32748204e+02, 1.37367146e+02, 1.41838310e+02],
[2.10147785e+00, 2.24843405e+00, 2.40634759e+00, …,
 2.37505684e+02, 2.45572860e+02, 2.53322408e+02],
…,
[1.88590006e+03, 1.91839521e+03, 1.95281838e+03, …,
 2.59347251e+05, 2.70973887e+05, 2.82214520e+05],
[2.42918083e+03, 2.48228665e+03, 2.53850844e+03, …,
 3.41680501e+05, 3.55980999e+05, 3.69784857e+05],
[3.26930382e+03, 3.34721478e+03, 3.42986974e+03, …,
 4.38880158e+05, 4.54788828e+05, 4.70008082e+05]])), 'channel_8':
(array([    5,     6,     8,    10,    14,    18,    24,    31,    40,
    52,    68,    89,   116,   151,   196,   255,   331,   431,
   560,   727,   945,  1229,  1597,  2076,  2697,  3506,  4556,
  5921,  7694, 10000]), array([[5.18070125e-01, 5.64200639e-01,
6.17443654e-01, …,
 1.26062014e+03, 1.30049543e+03, 1.33847439e+03],
[1.12426114e+00, 1.22239328e+00, 1.33201633e+00, …,
 1.86641849e+03, 1.92543165e+03, 1.98163908e+03],
[2.73385248e+00, 2.88291365e+00, 3.03926472e+00, …,
 3.37116980e+03, 3.47762981e+03, 3.57903344e+03],
…,
[1.35178762e+03, 1.38749860e+03, 1.42526339e+03, …,
 1.16844180e+06, 1.23287091e+06, 1.29605319e+06],
[1.91884415e+03, 1.97106483e+03, 2.02598954e+03, …,
 1.69957704e+06, 1.79256568e+06, 1.88357162e+06],
[2.69013711e+03, 2.75563989e+03, 2.82561913e+03, …,
 2.72413625e+06, 2.85886845e+06, 2.98955940e+06]])), 'channel_9':
(array([    5,     6,     8,    10,    14,    18,    24,    31,    40,
    52,    68,    89,   116,   151,   196,   255,   331,   431,
   560,   727,   945,  1229,  1597,  2076,  2697,  3506,  4556,
  5921,  7694, 10000]), array([[1.51377489e-01, 1.68710553e-01,
1.89656736e-01, …,

```
         3.75733505e+02, 3.88830401e+02, 4.01390513e+02],
        [6.21089785e-01, 6.74994988e-01, 7.35209393e-01, …,
         5.56303701e+02, 5.75689734e+02, 5.94281595e+02],
        [1.43273652e+00, 1.52582835e+00, 1.62504687e+00, …,
         1.00479921e+03, 1.03977290e+03, 1.07331499e+03],
        …,
        [1.30892803e+03, 1.33454790e+03, 1.36173963e+03, …,
         4.12253437e+05, 4.35574970e+05, 4.58342685e+05],
        [1.78647504e+03, 1.82811295e+03, 1.87225604e+03, …,
         6.44545385e+05, 6.78606476e+05, 7.11576188e+05],
        [2.61051809e+03, 2.67628695e+03, 2.74609197e+03, …,
         1.19526232e+06, 1.25856390e+06, 1.31985246e+06]])), 'channel_10':
(array([    5,     6,     8,    10,    14,    18,    24,    31,    40,
           52,    68,    89,   116,   151,   196,   255,   331,   431,
          560,   727,   945,  1229,  1597,  2076,  2697,  3506,  4556,
         5921,  7694, 10000]), array([[2.09663791e-01, 2.31473156e-01,
2.57197123e-01, …,
         6.46384466e+01, 6.75768090e+01, 7.05462479e+01],
        [5.90453699e-01, 6.40030349e-01, 6.95367409e-01, …,
         9.20942730e+01, 9.58240303e+01, 9.94740275e+01],
        [1.43629652e+00, 1.51734700e+00, 1.60463631e+00, …,
         1.62675263e+02, 1.68961870e+02, 1.75018385e+02],
        …,
        [1.47148038e+03, 1.49728011e+03, 1.52452191e+03, …,
         2.06120297e+05, 2.15009910e+05, 2.23564555e+05],
        [1.92939664e+03, 1.96582368e+03, 2.00456650e+03, …,
         2.85339456e+05, 2.97122268e+05, 3.08471451e+05],
        [2.88178832e+03, 2.94869280e+03, 3.01915555e+03, …,
         3.93210604e+05, 4.08566766e+05, 4.23413877e+05]])), 'channel_11':
(array([    5,     6,     8,    10,    14,    18,    24,    31,    40,
           52,    68,    89,   116,   151,   196,   255,   331,   431,
          560,   727,   945,  1229,  1597,  2076,  2697,  3506,  4556,
         5921,  7694, 10000]), array([[2.08499594e-01, 2.30290003e-01,
2.56278783e-01, …,
         7.86812361e+01, 8.15885778e+01, 8.44758046e+01],
        [7.03756759e-01, 7.63779584e-01, 8.31287258e-01, …,
         1.14568113e+02, 1.18496341e+02, 1.22301863e+02],
        [1.84969017e+00, 1.95868743e+00, 2.07510875e+00, …,
         2.04731610e+02, 2.11548126e+02, 2.18091214e+02],
        …,
        [1.38776874e+03, 1.41430723e+03, 1.44234506e+03, …,
         2.28161980e+05, 2.38400003e+05, 2.48297997e+05],
        [1.91617624e+03, 1.95986500e+03, 2.00592566e+03, …,
         3.00323784e+05, 3.12910599e+05, 3.25061014e+05],
        [2.84995495e+03, 2.92118284e+03, 2.99652326e+03, …,
         3.85253073e+05, 3.99257641e+05, 4.12656267e+05]])), 'channel_12':
(array([    5,     6,     8,    10,    14,    18,    24,    31,    40,
           52,    68,    89,   116,   151,   196,   255,   331,   431,
```

```
          560,   727,   945,  1229,  1597,  2076,  2697,  3506,  4556,
         5921,  7694, 10000]), array([[2.65263841e-01, 2.95621367e-01,
3.31742163e-01, …,
         1.22714312e+02, 1.27340707e+02, 1.31781097e+02],
        [8.32974948e-01, 9.01913918e-01, 9.78984883e-01, …,
         1.81450653e+02, 1.88263388e+02, 1.94795775e+02],
        [1.80135735e+00, 1.94706582e+00, 2.10458097e+00, …,
         3.27462931e+02, 3.39728102e+02, 3.51485477e+02],
        …,
        [1.88940116e+03, 1.92306579e+03, 1.95879492e+03, …,
         2.40289080e+05, 2.51103721e+05, 2.61570274e+05],
        [2.47900537e+03, 2.53951242e+03, 2.60368841e+03, …,
         3.20044312e+05, 3.33632821e+05, 3.46762183e+05],
        [3.66228807e+03, 3.75419825e+03, 3.85159484e+03, …,
         4.08578282e+05, 4.23062020e+05, 4.36922545e+05]])), 'channel_13':
(array([    5,     6,     8,    10,    14,    18,    24,    31,    40,
           52,    68,    89,   116,   151,   196,   255,   331,   431,
          560,   727,   945,  1229,  1597,  2076,  2697,  3506,  4556,
         5921,  7694, 10000]), array([[3.09638891e-01, 3.42703812e-01,
3.81913234e-01, …,
         7.54983414e+02, 7.83791309e+02, 8.11357253e+02],
        [1.03462499e+00, 1.11203796e+00, 1.19783438e+00, …,
         1.11790192e+03, 1.16052925e+03, 1.20131833e+03],
        [2.42383292e+00, 2.56005987e+00, 2.70557838e+00, …,
         2.01904051e+03, 2.09589647e+03, 2.16943606e+03],
        …,
        [1.66551292e+03, 1.70491569e+03, 1.74686129e+03, …,
         1.77586605e+05, 1.85603849e+05, 1.93358657e+05],
        [2.40018181e+03, 2.45943898e+03, 2.52265141e+03, …,
         2.55314987e+05, 2.66698216e+05, 2.77699746e+05],
        [3.54388517e+03, 3.63102818e+03, 3.72413614e+03, …,
         3.95045794e+05, 4.14101699e+05, 4.32695569e+05]])), 'channel_14':
(array([    5,     6,     8,    10,    14,    18,    24,    31,    40,
           52,    68,    89,   116,   151,   196,   255,   331,   431,
          560,   727,   945,  1229,  1597,  2076,  2697,  3506,  4556,
         5921,  7694, 10000]), array([[1.48497813e-01, 1.66525787e-01,
1.88360952e-01, …,
         6.21539784e+02, 6.44535918e+02, 6.66603624e+02],
        [7.04025267e-01, 7.52853941e-01, 8.06794463e-01, …,
         9.20191468e+02, 9.54207681e+02, 9.86849502e+02],
        [1.58323964e+00, 1.65978457e+00, 1.74263618e+00, …,
         1.66159866e+03, 1.72288625e+03, 1.78169347e+03],
        …,
        [1.12771336e+03, 1.15553492e+03, 1.18537063e+03, …,
         3.06214715e+05, 3.24971361e+05, 3.43267191e+05],
        [1.64727038e+03, 1.69532468e+03, 1.74690227e+03, …,
         4.55869478e+05, 4.81704448e+05, 5.06796468e+05],
        [2.78260237e+03, 2.84693358e+03, 2.91597706e+03, …,
```

```
        8.28274930e+05, 8.71684841e+05, 9.13623941e+05]])), 'channel_15':
(array([    5,     6,     8,    10,    14,    18,    24,    31,    40,
          52,    68,    89,   116,   151,   196,   255,   331,   431,
         560,   727,   945,  1229,  1597,  2076,  2697,  3506,  4556,
        5921,  7694, 10000]), array([[2.03070855e-01, 2.24430359e-01,
2.49558385e-01, …,
        8.29205949e+01, 8.60614658e+01, 8.90701250e+01],
       [6.12446169e-01, 6.58820225e-01, 7.10266667e-01, …,
        1.22665418e+02, 1.27312263e+02, 1.31762745e+02],
       [1.46935768e+00, 1.54313278e+00, 1.62306265e+00, …,
        2.21424427e+02, 2.29802807e+02, 2.37825523e+02],
       …,
       [9.15818572e+02, 9.37716814e+02, 9.61142369e+02, …,
        9.64001133e+04, 1.01866903e+05, 1.07202778e+05],
       [1.33089189e+03, 1.37057242e+03, 1.41274599e+03, …,
        1.51052872e+05, 1.59049257e+05, 1.66788114e+05],
       [2.00909698e+03, 2.06476788e+03, 2.12447445e+03, …,
        2.80270073e+05, 2.95112070e+05, 3.09484638e+05]])), 'channel_16':
(array([    5,     6,     8,    10,    14,    18,    24,    31,    40,
          52,    68,    89,   116,   151,   196,   255,   331,   431,
         560,   727,   945,  1229,  1597,  2076,  2697,  3506,  4556,
        5921,  7694, 10000]), array([[2.15400635e-01, 2.38201914e-01,
2.64987102e-01, …,
        5.73289175e+01, 5.95119208e+01, 6.17147042e+01],
       [5.79988779e-01, 6.28695310e-01, 6.83566838e-01, …,
        8.28308371e+01, 8.56499744e+01, 8.83979961e+01],
       [1.60412969e+00, 1.68058470e+00, 1.76332921e+00, …,
        1.47187866e+02, 1.51947316e+02, 1.56514810e+02],
       …,
       [1.03135446e+03, 1.05665321e+03, 1.08355385e+03, …,
        1.76543988e+05, 1.84482583e+05, 1.92158184e+05],
       [1.47181921e+03, 1.51477349e+03, 1.56060548e+03, …,
        2.31834656e+05, 2.41542883e+05, 2.50912517e+05],
       [1.98462620e+03, 2.06452300e+03, 2.15032478e+03, …,
        2.97136086e+05, 3.07960038e+05, 3.18314034e+05]])), 'channel_17':
(array([    5,     6,     8,    10,    14,    18,    24,    31,    40,
          52,    68,    89,   116,   151,   196,   255,   331,   431,
         560,   727,   945,  1229,  1597,  2076,  2697,  3506,  4556,
        5921,  7694, 10000]), array([[3.44022133e-01, 3.75063624e-01,
4.11005622e-01, …,
        1.07699015e+02, 1.11527956e+02, 1.15218847e+02],
       [8.12645414e-01, 8.76622335e-01, 9.48008581e-01, …,
        1.58844092e+02, 1.64432487e+02, 1.69800459e+02],
       [1.99679749e+00, 2.11380157e+00, 2.23850807e+00, …,
        2.86199105e+02, 2.96238466e+02, 3.05873670e+02],
       …,
       [1.34881971e+03, 1.37606371e+03, 1.40517084e+03, …,
        2.32736626e+05, 2.43127942e+05, 2.53176691e+05],
```

```
       [1.89623480e+03, 1.94231631e+03, 1.99139155e+03, …,
        3.07957038e+05, 3.20818384e+05, 3.33231855e+05],
       [2.54340003e+03, 2.62163076e+03, 2.70535551e+03, …,
        3.94764742e+05, 4.08791884e+05, 4.22206772e+05]])), 'channel_18':
(array([    5,     6,     8,    10,    14,    18,    24,    31,    40,
          52,    68,    89,   116,   151,   196,   255,   331,   431,
         560,   727,   945,  1229,  1597,  2076,  2697,  3506,  4556,
        5921,  7694, 10000]), array([[2.48575847e-01, 2.76153709e-01,
3.09279495e-01, …,
        1.33341376e+02, 1.38203958e+02, 1.42878113e+02],
       [9.60914554e-01, 1.03609996e+00, 1.11946007e+00, …,
        1.97043043e+02, 2.04199171e+02, 2.11069182e+02],
       [2.29711256e+00, 2.42633664e+00, 2.56435931e+00, …,
        3.55484023e+02, 3.68376332e+02, 3.80749589e+02],
       …,
       [1.97447963e+03, 2.01349250e+03, 2.05486475e+03, …,
        2.54281401e+05, 2.65596749e+05, 2.76535201e+05],
       [2.98119741e+03, 3.03750485e+03, 3.09688964e+03, …,
        3.38382479e+05, 3.52417201e+05, 3.65945845e+05],
       [3.82238030e+03, 3.91686780e+03, 4.01717968e+03, …,
        4.37125477e+05, 4.52459502e+05, 4.67103273e+05]])), 'channel_19':
(array([    5,     6,     8,    10,    14,    18,    24,    31,    40,
          52,    68,    89,   116,   151,   196,   255,   331,   431,
         560,   727,   945,  1229,  1597,  2076,  2697,  3506,  4556,
        5921,  7694, 10000]), array([[2.68749309e-01, 2.97058275e-01,
3.30269259e-01, …,
        1.15491376e+03, 1.19901325e+03, 1.24120996e+03],
       [7.03984393e-01, 7.65114734e-01, 8.33746451e-01, …,
        1.71001432e+03, 1.77526448e+03, 1.83769810e+03],
       [1.77608106e+00, 1.88085444e+00, 1.99167621e+00, …,
        3.08829055e+03, 3.20591784e+03, 3.31846273e+03],
       …,
       [1.08381744e+03, 1.11566231e+03, 1.14927129e+03, …,
        9.66233479e+04, 1.01077962e+05, 1.05414554e+05],
       [1.68581164e+03, 1.73020092e+03, 1.77698749e+03, …,
        1.29841760e+05, 1.35210912e+05, 1.40390472e+05],
       [2.45192931e+03, 2.51214076e+03, 2.57648545e+03, …,
        1.81706572e+05, 1.88417588e+05, 1.94856145e+05]])), 'channel_20':
(array([    5,     6,     8,    10,    14,    18,    24,    31,    40,
          52,    68,    89,   116,   151,   196,   255,   331,   431,
         560,   727,   945,  1229,  1597,  2076,  2697,  3506,  4556,
        5921,  7694, 10000]), array([[1.73895183e-01, 1.94266441e-01,
2.18640975e-01, …,
        9.58312902e+01, 9.89841730e+01, 1.02011542e+02],
       [6.33286920e-01, 6.80742857e-01, 7.33254649e-01, …,
        1.40258795e+02, 1.44760557e+02, 1.49059825e+02],
       [1.46656746e+00, 1.54691846e+00, 1.63357740e+00, …,
        2.51458620e+02, 2.59477138e+02, 2.67126804e+02],
```

```
       …,
       [7.87152395e+02, 8.09402036e+02, 8.33339088e+02, …,
        1.67261049e+05, 1.74663909e+05, 1.81812601e+05],
       [1.23524011e+03, 1.27259579e+03, 1.31266005e+03, …,
        2.19566494e+05, 2.28578160e+05, 2.37259507e+05],
       [1.94228576e+03, 2.00066421e+03, 2.06345158e+03, …,
        2.85338142e+05, 2.95458396e+05, 3.05125156e+05]])), 'channel_21':
(array([    5,     6,     8,    10,    14,    18,    24,    31,    40,
          52,    68,    89,   116,   151,   196,   255,   331,   431,
         560,   727,   945,  1229,  1597,  2076,  2697,  3506,  4556,
        5921,  7694, 10000]), array([[1.68712519e-01, 1.87741279e-01,
2.10428858e-01, …,
        3.84305859e+01, 3.97323635e+01, 4.09940328e+01],
       [5.78025062e-01, 6.23494008e-01, 6.74089239e-01, …,
        5.62085546e+01, 5.80525401e+01, 5.98213495e+01],
       [1.46035257e+00, 1.53520155e+00, 1.61638902e+00, …,
        1.00617562e+02, 1.03879598e+02, 1.06998818e+02],
       …,
       [8.15606941e+02, 8.39498143e+02, 8.65026937e+02, …,
        9.28593059e+04, 9.71055116e+04, 1.01218407e+05],
       [1.23049647e+03, 1.26886951e+03, 1.30997583e+03, …,
        1.21427030e+05, 1.26595998e+05, 1.31593298e+05],
       [1.88491909e+03, 1.94348045e+03, 2.00631643e+03, …,
        1.52294434e+05, 1.57807092e+05, 1.63086365e+05]])), 'channel_22':
(array([    5,     6,     8,    10,    14,    18,    24,    31,    40,
          52,    68,    89,   116,   151,   196,   255,   331,   431,
         560,   727,   945,  1229,  1597,  2076,  2697,  3506,  4556,
        5921,  7694, 10000]), array([[2.04597047e-01, 2.28431152e-01,
2.56922184e-01, …,
        9.17374631e+01, 9.49200265e+01, 9.79829186e+01],
       [6.69612558e-01, 7.28555452e-01, 7.94892237e-01, …,
        1.35336692e+02, 1.39989341e+02, 1.44454012e+02],
       [1.78091651e+00, 1.87946302e+00, 1.98539624e+00, …,
        2.43876043e+02, 2.52237652e+02, 2.60256047e+02],
       …,
       [1.12593186e+03, 1.15099664e+03, 1.17776423e+03, …,
        1.90975608e+05, 1.99591148e+05, 2.07930392e+05],
       [1.38921839e+03, 1.43719127e+03, 1.48849674e+03, …,
        2.51556815e+05, 2.62150167e+05, 2.72382526e+05],
       [2.50343493e+03, 2.56714556e+03, 2.63488176e+03, …,
        3.21129501e+05, 3.32552300e+05, 3.43481880e+05]])), 'channel_23':
(array([    5,     6,     8,    10,    14,    18,    24,    31,    40,
          52,    68,    89,   116,   151,   196,   255,   331,   431,
         560,   727,   945,  1229,  1597,  2076,  2697,  3506,  4556,
        5921,  7694, 10000]), array([[2.98734966e-01, 3.28880695e-01,
3.64441182e-01, …,
        9.10523417e+01, 9.42593499e+01, 9.73590307e+01],
       [8.28866769e-01, 9.01113864e-01, 9.82137371e-01, …,
```

```
         1.31010588e+02, 1.35362425e+02, 1.39533223e+02],
        [2.11709690e+00, 2.23361122e+00, 2.35743976e+00, …,
         2.32306901e+02, 2.39856421e+02, 2.47074666e+02],
        …,
        [1.40688187e+03, 1.43741385e+03, 1.46979087e+03, …,
         8.17343708e+04, 8.60047783e+04, 9.01853978e+04],
        [1.94606373e+03, 1.99559679e+03, 2.04838486e+03, …,
         1.16557083e+05, 1.22527261e+05, 1.28350442e+05],
        [2.77796883e+03, 2.85499303e+03, 2.93698988e+03, …,
         2.07370923e+05, 2.17796737e+05, 2.27904353e+05]])), 'channel_24':
(array([    5,     6,     8,    10,    14,    18,    24,    31,    40,
           52,    68,    89,   116,   151,   196,   255,   331,   431,
          560,   727,   945,  1229,  1597,  2076,  2697,  3506,  4556,
         5921,  7694, 10000]), array([[2.02968780e-01, 2.24581422e-01,
2.50248138e-01, …,
         9.52111665e+01, 9.95363862e+01, 1.03791582e+02],
        [5.54786642e-01, 6.12397867e-01, 6.78068069e-01, …,
         1.31800053e+02, 1.37120854e+02, 1.42267864e+02],
        [1.53948359e+00, 1.63462794e+00, 1.73620106e+00, …,
         2.27172637e+02, 2.35822843e+02, 2.44118272e+02],
        …,
        [1.05318210e+03, 1.07721946e+03, 1.10285243e+03, …,
         1.71086794e+05, 1.78519725e+05, 1.85674905e+05],
        [1.50140874e+03, 1.53761911e+03, 1.57660100e+03, …,
         2.39117055e+05, 2.48722240e+05, 2.57937576e+05],
        [2.05493308e+03, 2.12126508e+03, 2.19286323e+03, …,
         3.31543687e+05, 3.43678385e+05, 3.55285205e+05]])), 'channel_25':
(array([    5,     6,     8,    10,    14,    18,    24,    31,    40,
           52,    68,    89,   116,   151,   196,   255,   331,   431,
          560,   727,   945,  1229,  1597,  2076,  2697,  3506,  4556,
         5921,  7694, 10000]), array([[1.91002991e-01, 2.11871454e-01,
2.36492171e-01, …,
         4.66719885e+01, 4.83848676e+01, 5.00300065e+01],
        [5.56308532e-01, 6.04072661e-01, 6.57713935e-01, …,
         6.89303814e+01, 7.14577932e+01, 7.38843081e+01],
        [1.46908924e+00, 1.54841047e+00, 1.63431001e+00, …,
         1.24208419e+02, 1.28752886e+02, 1.33114486e+02],
        …,
        [8.73076420e+02, 8.95217014e+02, 9.18954569e+02, …,
         5.19212983e+04, 5.46641786e+04, 5.73422726e+04],
        [1.11513430e+03, 1.15313519e+03, 1.19391407e+03, …,
         8.12230813e+04, 8.53977433e+04, 8.94453353e+04],
        [1.90425513e+03, 1.95230145e+03, 2.00404619e+03, …,
         1.49794200e+05, 1.57618976e+05, 1.65212829e+05]])), 'channel_26':
(array([    5,     6,     8,    10,    14,    18,    24,    31,    40,
           52,    68,    89,   116,   151,   196,   255,   331,   431,
          560,   727,   945,  1229,  1597,  2076,  2697,  3506,  4556,
         5921,  7694, 10000]), array([[1.82792841e-01, 2.03487803e-01,
```

2.28148789e-01, …,
        8.00566638e+01, 8.27785331e+01, 8.53890431e+01],
       [6.08073140e-01, 6.61013152e-01, 7.20426189e-01, …,
        1.18288587e+02, 1.22297720e+02, 1.26139757e+02],
       [1.19443218e+00, 1.31565625e+00, 1.45104699e+00, …,
        2.13346306e+02, 2.20565351e+02, 2.27482246e+02],
       …,
       [1.03435551e+03, 1.05672958e+03, 1.08064712e+03, …,
        1.37736093e+05, 1.44035135e+05, 1.50136376e+05],
       [1.42939440e+03, 1.47072302e+03, 1.51496538e+03, …,
        1.81186198e+05, 1.88962821e+05, 1.96479208e+05],
       [2.08774044e+03, 2.16088620e+03, 2.23825232e+03, …,
        2.30231930e+05, 2.38459868e+05, 2.46341563e+05]])), 'channel_27':
(array([    5,     6,     8,    10,    14,    18,    24,    31,    40,
          52,    68,    89,   116,   151,   196,   255,   331,   431,
         560,   727,   945,  1229,  1597,  2076,  2697,  3506,  4556,
        5921,  7694, 10000]), array([[2.65721870e-01, 2.96143156e-01,
3.32338608e-01, …,
        1.11067882e+02, 1.14978426e+02, 1.18735724e+02],
       [7.96864171e-01, 8.68915068e-01, 9.49684561e-01, …,
        1.64129239e+02, 1.69885736e+02, 1.75410415e+02],
       [1.96400203e+00, 2.09285919e+00, 2.23029622e+00, …,
        2.96089163e+02, 3.06457862e+02, 3.16406563e+02],
       …,
       [1.30763497e+03, 1.33499503e+03, 1.36378862e+03, …,
        2.05690657e+05, 2.14928373e+05, 2.23865775e+05],
       [1.76004151e+03, 1.80762462e+03, 1.85776348e+03, …,
        2.72421818e+05, 2.83869943e+05, 2.94918933e+05],
       [2.50694651e+03, 2.57826438e+03, 2.65396423e+03, …,
        3.49243055e+05, 3.61554731e+05, 3.73322809e+05]])), 'channel_28':
(array([    5,     6,     8,    10,    14,    18,    24,    31,    40,
          52,    68,    89,   116,   151,   196,   255,   331,   431,
         560,   727,   945,  1229,  1597,  2076,  2697,  3506,  4556,
        5921,  7694, 10000]), array([[1.27204578e-01, 1.43316208e-01,
1.63039499e-01, …,
        8.79506461e+01, 9.07588433e+01, 9.34364889e+01],
       [6.07167057e-01, 6.61465826e-01, 7.22484905e-01, …,
        1.30105411e+02, 1.34253012e+02, 1.38207176e+02],
       [1.60445440e+00, 1.69287598e+00, 1.78775763e+00, …,
        2.34824948e+02, 2.42291789e+02, 2.49409495e+02],
       …,
       [1.30153888e+03, 1.33524636e+03, 1.37117281e+03, …,
        9.52156704e+04, 9.99581543e+04, 1.04564511e+05],
       [1.84813247e+03, 1.90101098e+03, 1.95758776e+03, …,
        1.32682626e+05, 1.39410901e+05, 1.45955215e+05],
       [2.74890895e+03, 2.83064742e+03, 2.91809703e+03, …,
        2.08467308e+05, 2.18605533e+05, 2.28434104e+05]])), 'channel_29':
(array([    5,     6,     8,    10,    14,    18,    24,    31,    40,

```
          52,     68,     89,    116,    151,    196,    255,    331,    431,
         560,    727,    945,   1229,   1597,   2076,   2697,   3506,   4556,
        5921,   7694, 10000]), array([[2.14486687e-01, 2.38639527e-01,
2.67435494e-01, …,
        1.06862547e+02, 1.10243455e+02, 1.13464167e+02],
       [8.16185351e-01, 8.70552348e-01, 9.30456546e-01, …,
        1.58091714e+02, 1.63085856e+02, 1.67842624e+02],
       [1.72655595e+00, 1.83545365e+00, 1.95143334e+00, …,
        2.85340345e+02, 2.94331062e+02, 3.02893232e+02],
       …,
       [1.70436749e+03, 1.75169548e+03, 1.80224325e+03, …,
        1.17136877e+05, 1.22732824e+05, 1.28156387e+05],
       [2.74214074e+03, 2.81024111e+03, 2.88297532e+03, …,
        1.61624569e+05, 1.69422000e+05, 1.76988226e+05],
       [3.53168232e+03, 3.65107173e+03, 3.77845786e+03, …,
        2.37376642e+05, 2.48298079e+05, 2.58902064e+05]])), 'channel_30':
(array([    5,      6,      8,     10,     14,     18,     24,     31,     40,
          52,     68,     89,    116,    151,    196,    255,    331,    431,
         560,    727,    945,   1229,   1597,   2076,   2697,   3506,   4556,
        5921,   7694, 10000]), array([[2.58563870e-01, 2.84549974e-01,
3.15120991e-01, …,
        1.04157350e+02, 1.07558161e+02, 1.10807813e+02],
       [5.23003476e-01, 5.81676348e-01, 6.50574904e-01, …,
        1.54097072e+02, 1.59120843e+02, 1.63920516e+02],
       [1.87155199e+00, 1.96919888e+00, 2.07416299e+00, …,
        2.78172402e+02, 2.87222237e+02, 2.95867715e+02],
       …,
       [1.20950190e+03, 1.24929607e+03, 1.29093945e+03, …,
        1.42680026e+05, 1.49268180e+05, 1.55642362e+05],
       [1.66355524e+03, 1.72249297e+03, 1.78464795e+03, …,
        1.91874483e+05, 2.00483929e+05, 2.08811001e+05],
       [2.62252953e+03, 2.68775240e+03, 2.75724266e+03, …,
        2.58662051e+05, 2.69088296e+05, 2.79173504e+05]])), 'channel_31':
(array([    5,      6,      8,     10,     14,     18,     24,     31,     40,
          52,     68,     89,    116,    151,    196,    255,    331,    431,
         560,    727,    945,   1229,   1597,   2076,   2697,   3506,   4556,
        5921,   7694, 10000]), array([[2.74939784e-01, 3.03374050e-01,
3.36975450e-01, …,
        1.06561927e+02, 1.10207821e+02, 1.13703693e+02],
       [8.58400427e-01, 9.20205480e-01, 9.88991849e-01, …,
        1.57600998e+02, 1.62983153e+02, 1.68142035e+02],
       [2.02721178e+00, 2.13763096e+00, 2.25681506e+00, …,
        2.84443969e+02, 2.94141590e+02, 3.03436084e+02],
       …,
       [1.12757918e+03, 1.16050570e+03, 1.19554544e+03, …,
        1.69291141e+05, 1.76975221e+05, 1.84410714e+05],
       [1.93089516e+03, 1.97414854e+03, 2.01990029e+03, …,
        2.25288963e+05, 2.34978986e+05, 2.44337345e+05],
```

```
       [2.85795921e+03, 2.91646458e+03, 2.97860453e+03, …,
        2.92578002e+05, 3.03281848e+05, 3.13555297e+05]]))}
```

[31]:
```python
# Print shape and first elements of CNN data
print("Shape of cnn_X_data:", cnn_X_data.shape)
print("First element of cnn_X_data:\n", cnn_X_data[0])

# Print shape and first elements of RNN data
print("Shape of rnn_X_data:", rnn_X_data.shape)
print("First element of rnn_X_data:\n", rnn_X_data[0])
```

```
Shape of cnn_X_data: (860, 5100)
First element of cnn_X_data:
 [0.00000000e+00 0.00000000e+00 3.50020883e-08 … 1.02874780e-02
 1.06212100e-02 1.09375542e-02]
Shape of rnn_X_data: (860, 100, 51)
First element of rnn_X_data:
 [[0.00000000e+00 0.00000000e+00 3.50020883e-08 … 1.48880730e-04
  1.60519608e-04 1.72424775e-04]
 [1.00050025e-04 7.20844984e-07 8.06273925e-07 … 1.81094506e-04
  1.94758141e-04 2.08734901e-04]
 [3.00150075e-04 2.49774851e-06 2.63861602e-06 … 2.51222770e-04
  2.70407111e-04 2.90373884e-04]
 …
 [2.60130065e-03 3.42822519e-06 3.53222373e-06 … 8.57237580e-03
  8.87304395e-03 9.16003688e-03]
 [3.50175088e-03 4.44041983e-06 4.55638833e-06 … 1.00475252e-02
  1.03789221e-02 1.06936306e-02]
 [4.70235118e-03 5.62150246e-06 5.74783045e-06 … 1.02874780e-02
  1.06212100e-02 1.09375542e-02]]
```

[10]:
```python
 print("Shape of cnn_X_data:", cnn_X_data.shape)
print(cnn_X_data.head())
```

```
Shape of cnn_X_data: (860, 5100)
        0             1             2             3             4  \
0  0.0000  0.000000e+00  3.500209e-08  7.687994e-08  1.273251e-07
1  0.0001  7.208450e-07  8.062739e-07  9.023169e-07  1.009805e-06
2  0.0003  2.497749e-06  2.638616e-06  2.787348e-06  2.943969e-06
3  0.0005  4.023184e-06  4.207777e-06  4.401253e-06  4.603944e-06
4  0.0009  6.903312e-06  7.153461e-06  7.418168e-06  7.697934e-06

              5             6             7             8             9  …  \
0  1.884323e-07  2.627076e-07  3.529606e-07  4.619666e-07  5.917862e-07  …
1  1.129261e-06  1.260744e-06  1.403757e-06  1.557257e-06  1.719825e-06  …
2  3.108491e-06  3.280934e-06  3.461341e-06  3.649793e-06  3.846420e-06  …
3  4.816177e-06  5.038258e-06  5.270445e-06  5.512933e-06  5.765839e-06  …
4  7.993139e-06  8.304015e-06  8.630621e-06  8.972824e-06  9.330300e-06  …
```

```
       5090      5091      5092      5093      5094      5095      5096  \
0  0.007344  0.007841  0.008309  0.008751  0.009168  0.009563  0.009935
1  0.007306  0.007800  0.008267  0.008707  0.009123  0.009515  0.009887
2  0.007470  0.007975  0.008451  0.008900  0.009324  0.009725  0.010103
3  0.007542  0.008050  0.008530  0.008983  0.009411  0.009814  0.010196
4  0.007562  0.008072  0.008553  0.009006  0.009434  0.009838  0.010220

       5097      5098      5099
0  0.010287  0.010621  0.010938
1  0.010238  0.010571  0.010886
2  0.010462  0.010801  0.011122
3  0.010557  0.010898  0.011222
4  0.010581  0.010923  0.011248

[5 rows x 5100 columns]
```

## 9 Save MFDFA X data for CNN and RNN as numpy

[3]:
```
import pandas as pd
import numpy as np

# Path to the CSV file
csv_file_path = '/home/vincent/AAA_projects/MVCS/Neuroscience/CNN_data/cnn_X_df.
 ↪csv'

# Load the CSV file into a DataFrame
cnn_X_data = pd.read_csv(csv_file_path)

cnn_save_dir = '/home/vincent/AAA_projects/MVCS/Neuroscience/CNN_data/'
# Save CNN data
np.save(cnn_save_dir + 'cnn_mfdfa.npy', cnn_X_data)
```

[16]:
```
# Define the directory paths where you want to save the data
cnn_save_dir = '/home/vincent/AAA_projects/MVCS/Neuroscience/CNN_data/'
rnn_save_dir = '/home/vincent/AAA_projects/MVCS/Neuroscience/RNN_data/'

# Save CNN data
np.save(cnn_save_dir + 'cnn_mfdfa_X_df.npy', cnn_X_data)

# Save RNN data
np.save(rnn_save_dir + 'rnn_mfdfa_X_df.npy', rnn_X_data)
```

[ ]:
```
# Define the directory paths where you want to save the data
cnn_save_dir = '/home/vincent/AAA_projects/MVCS/Neuroscience/CNN_data/'
```

```
# Save CNN data
np.save(cnn_save_dir + 'cnn_mfdfa_X_df.npy', cnn_X_data
```

## 10 Load Hurst exponents as numpy

[18]:
```
# Load the saved Hurst exponents from the file
hurst_exponents_npy = np.load('/home/vincent/AAA_projects/MVCS/Neuroscience/
↪HurstExponents/hurst_exponents.npy')
```

## 11 Load MFDFA RNN X data as numpy

[22]:
```
# Load data
rnn_mfdfa_X_df_path = '/home/vincent/AAA_projects/MVCS/Neuroscience/RNN_data/'

# Load RNN data
rnn_mfdfa_X_df_npy = np.load(rnn_mfdfa_X_df_path + 'rnn_mfdfa_X_df.npy')
```

## 12 Load MFDFA CNN X data as numpy

[ ]:
```
# Load data
cnn_save_dir = '/home/vincent/AAA_projects/MVCS/Neuroscience/CNN_data/'

# Load CNN data
rnn_mfdfa_X_df_npy = np.load(rnn_mfdfa_X_df_path + 'cnn_mfdfa_X_df.npy')
```

## 13 Convert RNN, CNN X data, and Hurst exp data from numpy to df

[38]:
```
# Display the shapes
print("Shape of cnn_mfdfa_X_df:", cnn_mfdfa_X_df.npy.shape)
print("Shape of rnn_mfdfa_X_df.npy:", rnn_mfdfa_X_df.npy.shape)

# Convert the 3D CNN data to a DataFrame
cnn_X_df = pd.DataFrame(cnn_mfdfa_X_df)

# Convert the 3D RNN data to a list of DataFrames with generic column labels
rnn_X_df = [pd.DataFrame(time_series) for time_series in rnn_mfdfa_X_df.npy]

# Display the head of the first DataFrame in the list
print("cnn_X_df:")
print(cnn_X_df.head())
print()
```

```python
# Display the head of the first 5 DataFrames in the list
print("rnn_X_df:")
for i, df in enumerate(rnn_X_df[:5]):
    print(f"DataFrame {i+1}:")
    print(df.head())
    print()

# Assuming hurst_exponents_npy has shape (32,)
# Convert Numpy array to DataFrame
hurst_exponents_df = pd.DataFrame(hurst_exponents_npy)

print("hurst_exponents:")
print(hurst_exponents_df.head())
```

```
Shape of cnn_X_data: (860, 5100)
Shape of rnn_X_data: (860, 100, 51)
cnn_X_df:
        0             1             2             3             4     \
0  0.0000  0.000000e+00  3.500209e-08  7.687994e-08  1.273251e-07
1  0.0001  7.208450e-07  8.062739e-07  9.023169e-07  1.009805e-06
2  0.0003  2.497749e-06  2.638616e-06  2.787348e-06  2.943969e-06
3  0.0005  4.023184e-06  4.207777e-06  4.401253e-06  4.603944e-06
4  0.0009  6.903312e-06  7.153461e-06  7.418168e-06  7.697934e-06

           5             6             7             8             9     …   \
0  1.884323e-07  2.627076e-07  3.529606e-07  4.619666e-07  5.917862e-07  …
1  1.129261e-06  1.260744e-06  1.403757e-06  1.557257e-06  1.719825e-06  …
2  3.108491e-06  3.280934e-06  3.461341e-06  3.649793e-06  3.846420e-06  …
3  4.816177e-06  5.038258e-06  5.270445e-06  5.512933e-06  5.765839e-06  …
4  7.993139e-06  8.304015e-06  8.630621e-06  8.972824e-06  9.330300e-06  …

       5090      5091      5092      5093      5094      5095      5096  \
0  0.007344  0.007841  0.008309  0.008751  0.009168  0.009563  0.009935
1  0.007306  0.007800  0.008267  0.008707  0.009123  0.009515  0.009887
2  0.007470  0.007975  0.008451  0.008900  0.009324  0.009725  0.010103
3  0.007542  0.008050  0.008530  0.008983  0.009411  0.009814  0.010196
4  0.007562  0.008072  0.008553  0.009006  0.009434  0.009838  0.010220

       5097      5098      5099
0  0.010287  0.010621  0.010938
1  0.010238  0.010571  0.010886
2  0.010462  0.010801  0.011122
3  0.010557  0.010898  0.011222
4  0.010581  0.010923  0.011248

[5 rows x 5100 columns]
```

```
rnn_X_df:
DataFrame 1:
        0             1             2             3             4   \
0  0.0000  0.000000e+00  3.500209e-08  7.687994e-08  1.273251e-07
1  0.0001  7.208450e-07  8.062739e-07  9.023169e-07  1.009805e-06
2  0.0003  2.497749e-06  2.638616e-06  2.787348e-06  2.943969e-06
3  0.0005  4.023184e-06  4.207777e-06  4.401253e-06  4.603944e-06
4  0.0009  6.903312e-06  7.153461e-06  7.418168e-06  7.697934e-06

             5             6             7             8             9   …  \
0  1.884323e-07  2.627076e-07  3.529606e-07  4.619666e-07  5.917862e-07  …
1  1.129261e-06  1.260744e-06  1.403757e-06  1.557257e-06  1.719825e-06  …
2  3.108491e-06  3.280934e-06  3.461341e-06  3.649793e-06  3.846420e-06  …
3  4.816177e-06  5.038258e-06  5.270445e-06  5.512933e-06  5.765839e-06  …
4  7.993139e-06  8.304015e-06  8.630621e-06  8.972824e-06  9.330300e-06  …

         41        42        43        44        45        46        47  \
0  0.000075  0.000085  0.000095  0.000105  0.000116  0.000126  0.000138
1  0.000095  0.000106  0.000118  0.000130  0.000142  0.000155  0.000168
2  0.000138  0.000152  0.000167  0.000182  0.000198  0.000215  0.000233
3  0.000188  0.000205  0.000222  0.000240  0.000259  0.000278  0.000298
4  0.000315  0.000340  0.000365  0.000391  0.000418  0.000445  0.000475

         48        49        50
0  0.000149  0.000161  0.000172
1  0.000181  0.000195  0.000209
2  0.000251  0.000270  0.000290
3  0.000319  0.000340  0.000362
4  0.000505  0.000538  0.000572

[5 rows x 51 columns]

DataFrame 2:
          0             1             2             3         4         5   \
0  0.000100  7.208450e-07  8.062739e-07  9.023169e-07  0.000001  0.000001
1  0.000300  2.497749e-06  2.638616e-06  2.787348e-06  0.000003  0.000003
2  0.000500  4.023184e-06  4.207777e-06  4.401253e-06  0.000005  0.000005
3  0.000900  6.903312e-06  7.153461e-06  7.418168e-06  0.000008  0.000008
4  0.001301  9.928155e-06  1.028365e-05  1.065725e-05  0.000011  0.000011

          6         7         8         9  …        41        42        43  \
0  0.000001  0.000001  0.000002  0.000002  …  0.000095  0.000106  0.000118
1  0.000003  0.000003  0.000004  0.000004  …  0.000138  0.000152  0.000167
2  0.000005  0.000005  0.000006  0.000006  …  0.000188  0.000205  0.000222
3  0.000008  0.000009  0.000009  0.000009  …  0.000315  0.000340  0.000365
4  0.000012  0.000012  0.000013  0.000013  …  0.000464  0.000498  0.000533

         44        45        46        47        48        49        50
```

```
0  0.000130  0.000142  0.000155  0.000168  0.000181  0.000195  0.000209
1  0.000182  0.000198  0.000215  0.000233  0.000251  0.000270  0.000290
2  0.000240  0.000259  0.000278  0.000298  0.000319  0.000340  0.000362
3  0.000391  0.000418  0.000445  0.000475  0.000505  0.000538  0.000572
4  0.000567  0.000601  0.000637  0.000673  0.000710  0.000748  0.000789

[5 rows x 51 columns]

DataFrame 3:
         0         1         2         3         4         5         6   \
0  0.000300  0.000002  0.000003  0.000003  0.000003  0.000003  0.000003
1  0.000500  0.000004  0.000004  0.000004  0.000005  0.000005  0.000005
2  0.000900  0.000007  0.000007  0.000007  0.000008  0.000008  0.000008
3  0.001301  0.000010  0.000010  0.000011  0.000011  0.000011  0.000012
4  0.001901  0.000015  0.000015  0.000015  0.000016  0.000016  0.000017

         7         8         9   …        41        42        43        44  \
0  0.000003  0.000004  0.000004  …  0.000138  0.000152  0.000167  0.000182
1  0.000005  0.000006  0.000006  …  0.000188  0.000205  0.000222  0.000240
2  0.000009  0.000009  0.000009  …  0.000315  0.000340  0.000365  0.000391
3  0.000012  0.000013  0.000013  …  0.000464  0.000498  0.000533  0.000567
4  0.000017  0.000018  0.000018  …  0.000698  0.000748  0.000797  0.000846

         45        46        47        48        49        50
0  0.000198  0.000215  0.000233  0.000251  0.000270  0.000290
1  0.000259  0.000278  0.000298  0.000319  0.000340  0.000362
2  0.000418  0.000445  0.000475  0.000505  0.000538  0.000572
3  0.000601  0.000637  0.000673  0.000710  0.000748  0.000789
4  0.000895  0.000944  0.000994  0.001045  0.001097  0.001152

[5 rows x 51 columns]

DataFrame 4:
         0         1         2         3         4         5         6   \
0  0.000500  0.000004  0.000004  0.000004  0.000005  0.000005  0.000005
1  0.000900  0.000007  0.000007  0.000007  0.000008  0.000008  0.000008
2  0.001301  0.000010  0.000010  0.000011  0.000011  0.000011  0.000012
3  0.001901  0.000015  0.000015  0.000015  0.000016  0.000016  0.000017
4  0.002601  0.000019  0.000020  0.000020  0.000021  0.000021  0.000022

         7         8         9   …        41        42        43        44  \
0  0.000005  0.000006  0.000006  …  0.000188  0.000205  0.000222  0.000240
1  0.000009  0.000009  0.000009  …  0.000315  0.000340  0.000365  0.000391
2  0.000012  0.000013  0.000013  …  0.000464  0.000498  0.000533  0.000567
3  0.000017  0.000018  0.000018  …  0.000698  0.000748  0.000797  0.000846
4  0.000022  0.000023  0.000023  …  0.000938  0.001003  0.001067  0.001130

         45        46        47        48        49        50
```

```
0  0.000259  0.000278  0.000298  0.000319  0.000340  0.000362
1  0.000418  0.000445  0.000475  0.000505  0.000538  0.000572
2  0.000601  0.000637  0.000673  0.000710  0.000748  0.000789
3  0.000895  0.000944  0.000994  0.001045  0.001097  0.001152
4  0.001193  0.001255  0.001318  0.001382  0.001448  0.001516

[5 rows x 51 columns]

DataFrame 5:
         0         1         2         3         4         5         6   \
0  0.000900  0.000007  0.000007  0.000007  0.000008  0.000008  0.000008
1  0.001301  0.000010  0.000010  0.000011  0.000011  0.000011  0.000012
2  0.001901  0.000015  0.000015  0.000015  0.000016  0.000016  0.000017
3  0.002601  0.000019  0.000020  0.000020  0.000021  0.000021  0.000022
4  0.003502  0.000025  0.000026  0.000026  0.000026  0.000027  0.000027

         7         8         9   …         41        42        43        44  \
0  0.000009  0.000009  0.000009  …  0.000315  0.000340  0.000365  0.000391
1  0.000012  0.000013  0.000013  …  0.000464  0.000498  0.000533  0.000567
2  0.000017  0.000018  0.000018  …  0.000698  0.000748  0.000797  0.000846
3  0.000022  0.000023  0.000023  …  0.000938  0.001003  0.001067  0.001130
4  0.000028  0.000028  0.000029  …  0.001134  0.001215  0.001295  0.001377

        45        46        47        48        49        50
0  0.000418  0.000445  0.000475  0.000505  0.000538  0.000572
1  0.000601  0.000637  0.000673  0.000710  0.000748  0.000789
2  0.000895  0.000944  0.000994  0.001045  0.001097  0.001152
3  0.001193  0.001255  0.001318  0.001382  0.001448  0.001516
4  0.001459  0.001545  0.001633  0.001725  0.001822  0.001923

[5 rows x 51 columns]

hurst_exponents:
          0
0  0.147664
1  0.127474
2  0.139365
3  0.112845
4  0.121137
```

## 14 Load MFDFA CNN X data as numpy

[28]:
```python
# Define file paths
cnn_save_dir = '/home/vincent/AAA_projects/MVCS/Neuroscience/CNN_data/'
cnn_mfdfa_X_data_path = cnn_save_dir + 'cnn_mfdfa_X_data.npy'

# Load the data as numpy arrays
```

```python
cnn_mfdfa_X_data = np.load(cnn_mfdfa_X_data_path)
cnn_mfdfa_Y_data = np.load(cnn_mfdfa_Y_data_path)

# Now you have the data as numpy arrays
print("cnn_mfdfa_X_data:")
print(cnn_mfdfa_X_data)

print("cnn_mfdfa_Y_data:")
print(cnn_mfdfa_Y_data)
```

```
cnn_mfdfa_X_data:
[[[0.00000000e+00 0.00000000e+00 3.50020883e-08 … 1.48880730e-04
   1.60519608e-04 1.72424775e-04]
  [1.00050025e-04 7.20844984e-07 8.06273925e-07 … 1.81094506e-04
   1.94758141e-04 2.08734901e-04]
  [3.00150075e-04 2.49774851e-06 2.63861602e-06 … 2.51222770e-04
   2.70407111e-04 2.90373884e-04]
  …
  [5.91895948e-01 2.65297556e-03 2.75874450e-03 … 5.52352885e-01
   5.76860492e-01 6.00504758e-01]
  [7.69284642e-01 4.81536991e-03 4.96171933e-03 … 7.25410306e-01
   7.55047727e-01 7.83563658e-01]
  [1.00000000e+00 8.42008260e-03 8.60116123e-03 … 9.34027641e-01
   9.67772183e-01 1.00000000e+00]]

 [[0.00000000e+00 0.00000000e+00 3.69707482e-08 … 1.61467172e-04
   1.68815702e-04 1.76239232e-04]
  [1.00050025e-04 5.31267875e-07 6.21884324e-07 … 2.29331598e-04
   2.38569405e-04 2.47610948e-04]
  [3.00150075e-04 2.33442819e-06 2.45704109e-06 … 4.04356107e-04
   4.19903355e-04 4.34902090e-04]
  …
  [5.91895948e-01 2.75757024e-03 2.89119975e-03 … 5.03398479e-01
   5.25431218e-01 5.46664259e-01]
  [7.69284642e-01 6.03649961e-03 6.17329594e-03 … 6.86295547e-01
   7.14783853e-01 7.42227289e-01]
  [1.00000000e+00 9.06264255e-03 9.21556787e-03 … 9.29310306e-01
   9.65271366e-01 1.00000000e+00]]

 [[0.00000000e+00 0.00000000e+00 4.85501352e-08 … 1.30680858e-04
   1.40834953e-04 1.51488533e-04]
  [1.00050025e-04 1.20096909e-06 1.31546419e-06 … 1.65942295e-04
   1.77323467e-04 1.89271299e-04]
  [3.00150075e-04 3.42675408e-06 3.61625741e-06 … 2.53022679e-04
   2.67635021e-04 2.83103971e-04]
  …
  [5.91895948e-01 5.44643058e-03 5.59592812e-03 … 5.53035353e-01
```

```
  5.77577624e-01 6.01263715e-01]
 [7.69284642e-01 7.82901754e-03 8.05638831e-03 … 7.30163856e-01
  7.60072115e-01 7.88851598e-01]
 [1.00000000e+00 1.27723784e-02 1.31470804e-02 … 9.34545504e-01
  9.68023215e-01 1.00000000e+00]]

…

[[0.00000000e+00 0.00000000e+00 9.32895630e-08 … 4.11924674e-04
  4.24983324e-04 4.37423218e-04]
 [1.00050025e-04 2.32404158e-06 2.53403235e-06 … 6.09795670e-04
  6.29085384e-04 6.47458247e-04]
 [3.00150075e-04 5.84031849e-06 6.26093234e-06 … 1.10128938e-03
  1.13601574e-03 1.16908684e-03]
 …
 [5.91895948e-01 6.58223571e-03 6.76503854e-03 … 4.52436559e-01
  4.74050726e-01 4.94999059e-01]
 [7.69284642e-01 1.05906013e-02 1.08536367e-02 … 6.24268830e-01
  6.54386154e-01 6.83610455e-01]
 [1.00000000e+00 1.36401800e-02 1.41013177e-02 … 9.16858757e-01
  9.59042452e-01 1.00000000e+00]]

[[0.00000000e+00 0.00000000e+00 9.30823589e-08 … 3.72165987e-04
  3.84347708e-04 3.95987978e-04]
 [1.00050025e-04 9.47224025e-07 1.15739056e-06 … 5.51050325e-04
  5.69045500e-04 5.86237952e-04]
 [3.00150075e-04 5.77773174e-06 6.12750338e-06 … 9.95488794e-04
  1.02790535e-03 1.05887350e-03]
 …
 [5.91895948e-01 4.33151584e-03 4.47405875e-03 … 5.11079660e-01
  5.34678460e-01 5.57510812e-01]
 [7.69284642e-01 5.95793723e-03 6.16905248e-03 … 6.87294459e-01
  7.18133536e-01 7.47961151e-01]
 [1.00000000e+00 9.39298807e-03 9.62661672e-03 … 9.26527868e-01
  9.63874732e-01 1.00000000e+00]]

[[0.00000000e+00 0.00000000e+00 9.06834985e-08 … 3.38973958e-04
  3.50601564e-04 3.61750714e-04]
 [1.00050025e-04 1.86079189e-06 2.05790260e-06 … 5.01749445e-04
  5.18914392e-04 5.35367267e-04]
 [3.00150075e-04 5.58840355e-06 5.94055602e-06 … 9.06281224e-04
  9.37209196e-04 9.66851503e-04]
 …
 [5.91895948e-01 3.59523580e-03 3.70024615e-03 … 5.39908003e-01
  5.64414325e-01 5.88127844e-01]
 [7.69284642e-01 6.15719756e-03 6.29514266e-03 … 7.18498101e-01
  7.49401841e-01 7.79247828e-01]
 [1.00000000e+00 9.11382077e-03 9.30040802e-03 … 9.33098521e-01
```

```
   9.67235578e-01 1.00000000e+00]]]
cnn_mfdfa_Y_data:
[[[0.00000000e+00 0.00000000e+00]
  [0.00000000e+00 3.69707483e-08]
  [0.00000000e+00 8.05821572e-08]
  …
  [1.00000000e+00 9.29310306e-01]
  [1.00000000e+00 9.65271366e-01]
  [1.00000000e+00 1.00000000e+00]]

 [[0.00000000e+00 0.00000000e+00]
  [0.00000000e+00 4.85501327e-08]
  [0.00000000e+00 1.06589707e-07]
  …
  [1.00000000e+00 9.34545504e-01]
  [1.00000000e+00 9.68023215e-01]
  [1.00000000e+00 1.00000000e+00]]

 [[0.00000000e+00 0.00000000e+00]
  [0.00000000e+00 1.02028870e-08]
  [0.00000000e+00 2.22459336e-08]
  …
  [1.00000000e+00 9.09497632e-01]
  [1.00000000e+00 9.55472783e-01]
  [1.00000000e+00 1.00000000e+00]]

 …

 [[0.00000000e+00 0.00000000e+00]
  [0.00000000e+00 9.30823591e-08]
  [0.00000000e+00 2.02587898e-07]
  …
  [1.00000000e+00 9.26527868e-01]
  [1.00000000e+00 9.63874732e-01]
  [1.00000000e+00 1.00000000e+00]]

 [[0.00000000e+00 0.00000000e+00]
  [0.00000000e+00 9.06834983e-08]
  [0.00000000e+00 1.97846188e-07]
  …
  [1.00000000e+00 9.33098521e-01]
  [1.00000000e+00 9.67235578e-01]
  [1.00000000e+00 1.00000000e+00]]

 [[0.00000000e+00 0.00000000e+00]
  [0.00000000e+00 3.41868871e-17]
  [0.00000000e+00 6.83737741e-17]
  …
```

```
[1.00000000e+00 1.00000000e+00]
[1.00000000e+00 1.00000000e+00]
[1.00000000e+00 1.00000000e+00]]]
```

## 15 Save MFDFA CNN X data as df

[40]:
```
# Define the save directory
cnn_save_path = '/home/vincent/AAA_projects/MVCS/Neuroscience/CNN_data/'

# Save the CNN X data
cnn_X_df.to_csv(cnn_save_path + 'cnn_X_df.csv', index=False)
```

## 16 Save MFDFA RNN X data and Hurst exp as DF

[47]:
```
# Save rnn_mfdfa_X_df
rnn_data_path = '/home/vincent/AAA_projects/MVCS/Neuroscience/RNN_data/'
for i, df in enumerate(rnn_X_df):
    df.to_csv(f"{rnn_data_path}rnn_X_df_{i}.csv", index=False)

# Save hurst_exponents_df
hurst_exponents_path = '/home/vincent/AAA_projects/MVCS/Neuroscience/
↪HurstExponents/'
hurst_exponents_df.to_csv(f"{hurst_exponents_path}hurst_exponents_df.csv",␣
↪index=False)
```

## 17 Make the MFDFA RNN X DF 1 file instead of 860

[48]:
```
# Combine the list of DataFrames into a single 3D DataFrame
rnn_X_df_combined = np.stack(rnn_X_data)

# The shape of rnn_X_df_combined should be (860, 100, 51)
print("Shape of rnn_X_df_combined:", rnn_X_df_combined.shape)

# Save the combined 3D DataFrame to a CSV file
rnn_data_path = '/home/vincent/AAA_projects/MVCS/Neuroscience/RNN_data/'
np.save(f"{rnn_data_path}rnn_X_data_combined.npy", rnn_X_df_combined)
```

```
Shape of rnn_X_df_combined: (860, 100, 51)
```

# Phase Space

September 8, 2023

## 1 Phase Space Plot

## 2 False Nearest Neighbors

[4]:
```
import numpy as np
import matplotlib.pyplot as plt
from sklearn.neighbors import NearestNeighbors

# Load EEG data
EEG_data = np.load('/home/vincent/AAA_projects/MVCS/Neuroscience/
↪eeg_data_with_channels.npy', allow_pickle=True)

# Extract EEG channel names
eeg_channel_names = ['Fp1', 'Fpz', 'Fp2', 'F7', 'F3', 'Fz', 'F4', 'F8', 'FC5',␣
↪'FC1', 'FC2', 'FC6',
                     'M1', 'T7', 'C3', 'Cz', 'C4', 'T8', 'M2', 'CP5', 'CP1',␣
↪'CP2', 'CP6',
                     'P7', 'P3', 'Pz', 'P4', 'P8', 'POz', 'O1', 'Oz', 'O2']

max_dim = 20  # Maximum embedding dimension to consider

def delay_embedding(data, emb_dim, delay):
    N = len(data)
    return np.array([data[i:i+emb_dim*delay:delay].flatten() for i in range(N -␣
↪emb_dim * delay + 1)])

def false_nearest_neighbors(data, emb_dim, delay, R=10):
    N = len(data)
    false_neighbors = np.zeros(emb_dim)

    for d in range(1, emb_dim + 1):
        emb_data = delay_embedding(data, d, delay)
        nbrs = NearestNeighbors(n_neighbors=2).fit(emb_data[:-delay])
        distances, indices = nbrs.kneighbors(emb_data[:-delay])
        neighbor_index = indices[:, 1]
        neighbor_distance = np.abs(data[neighbor_index + delay] - data[np.
↪arange(N - d * delay) + delay])
```

```python
        false_neighbors[d - 1] = np.mean((neighbor_distance / distances[:, 1])␣
↪> R)

    return false_neighbors

# Calculate FNN for different embedding dimensions for each channel
fnn_data = {}
for channel_idx, channel_name in enumerate(eeg_channel_names):
    channel_data = EEG_data[channel_idx]
    channel_data_flat = channel_data.flatten()
    fnn = false_nearest_neighbors(channel_data_flat, emb_dim=max_dim, delay=1)
    fnn_data[channel_name] = fnn
    print(f'Channel: {channel_name}')
    for dim, fnn_value in enumerate(fnn, start=1):
        print(f'Embedding Dimension {dim}: Fraction of FNN = {fnn_value:.4f}')

# Plot the FNN as a function of embedding dimension for each channel
plt.figure()
for channel_name in eeg_channel_names:
    fnn = fnn_data[channel_name]
    plt.plot(np.arange(1, max_dim+1), fnn, label=channel_name)
plt.xlabel('Embedding Dimension')
plt.ylabel('Fraction of False Nearest Neighbors')
plt.title('Estimation of Embedding Dimension using FNN Method')
plt.legend()
plt.show()

# Save the FNN data to a file
output_filename = '/home/vincent/AAA_projects/MVCS/Neuroscience/
↪false_nearest_neighbors.npy'
np.save(output_filename, fnn_data)
print(f'FNN data saved to {output_filename}')
```

```
Channel: Fp1
Embedding Dimension 1: Fraction of FNN = 0.5161
Embedding Dimension 2: Fraction of FNN = 0.0000
Embedding Dimension 3: Fraction of FNN = 0.0000
Embedding Dimension 4: Fraction of FNN = 0.0000
Embedding Dimension 5: Fraction of FNN = 0.0000
Embedding Dimension 6: Fraction of FNN = 0.0000
Embedding Dimension 7: Fraction of FNN = 0.0000
Embedding Dimension 8: Fraction of FNN = 0.0000
Embedding Dimension 9: Fraction of FNN = 0.0000
Embedding Dimension 10: Fraction of FNN = 0.0000
Embedding Dimension 11: Fraction of FNN = 0.0000
Embedding Dimension 12: Fraction of FNN = 0.0000
Embedding Dimension 13: Fraction of FNN = 0.0000
```

Embedding Dimension 14: Fraction of FNN = 0.0000
Embedding Dimension 15: Fraction of FNN = 0.0000
Embedding Dimension 16: Fraction of FNN = 0.0000
Embedding Dimension 17: Fraction of FNN = 0.0000
Embedding Dimension 18: Fraction of FNN = 0.0000
Embedding Dimension 19: Fraction of FNN = 0.0000
Embedding Dimension 20: Fraction of FNN = 0.0000
Channel: Fpz
Embedding Dimension 1: Fraction of FNN = 0.5161
Embedding Dimension 2: Fraction of FNN = 0.0000
Embedding Dimension 3: Fraction of FNN = 0.0000
Embedding Dimension 4: Fraction of FNN = 0.0000
Embedding Dimension 5: Fraction of FNN = 0.0000
Embedding Dimension 6: Fraction of FNN = 0.0000
Embedding Dimension 7: Fraction of FNN = 0.0000
Embedding Dimension 8: Fraction of FNN = 0.0000
Embedding Dimension 9: Fraction of FNN = 0.0000
Embedding Dimension 10: Fraction of FNN = 0.0000
Embedding Dimension 11: Fraction of FNN = 0.0000
Embedding Dimension 12: Fraction of FNN = 0.0000
Embedding Dimension 13: Fraction of FNN = 0.0000
Embedding Dimension 14: Fraction of FNN = 0.0000
Embedding Dimension 15: Fraction of FNN = 0.0000
Embedding Dimension 16: Fraction of FNN = 0.0000
Embedding Dimension 17: Fraction of FNN = 0.0000
Embedding Dimension 18: Fraction of FNN = 0.0000
Embedding Dimension 19: Fraction of FNN = 0.0000
Embedding Dimension 20: Fraction of FNN = 0.0000
Channel: Fp2
Embedding Dimension 1: Fraction of FNN = 0.5161
Embedding Dimension 2: Fraction of FNN = 0.0000
Embedding Dimension 3: Fraction of FNN = 0.0000
Embedding Dimension 4: Fraction of FNN = 0.0000
Embedding Dimension 5: Fraction of FNN = 0.0000
Embedding Dimension 6: Fraction of FNN = 0.0000
Embedding Dimension 7: Fraction of FNN = 0.0000
Embedding Dimension 8: Fraction of FNN = 0.0000
Embedding Dimension 9: Fraction of FNN = 0.0000
Embedding Dimension 10: Fraction of FNN = 0.0000
Embedding Dimension 11: Fraction of FNN = 0.0000
Embedding Dimension 12: Fraction of FNN = 0.0000
Embedding Dimension 13: Fraction of FNN = 0.0000
Embedding Dimension 14: Fraction of FNN = 0.0000
Embedding Dimension 15: Fraction of FNN = 0.0000
Embedding Dimension 16: Fraction of FNN = 0.0000
Embedding Dimension 17: Fraction of FNN = 0.0000
Embedding Dimension 18: Fraction of FNN = 0.0000
Embedding Dimension 19: Fraction of FNN = 0.0000

```
Embedding Dimension 20: Fraction of FNN = 0.0000
Channel: F7
Embedding Dimension 1: Fraction of FNN = 0.5161
Embedding Dimension 2: Fraction of FNN = 0.0000
Embedding Dimension 3: Fraction of FNN = 0.0000
Embedding Dimension 4: Fraction of FNN = 0.0000
Embedding Dimension 5: Fraction of FNN = 0.0000
Embedding Dimension 6: Fraction of FNN = 0.0000
Embedding Dimension 7: Fraction of FNN = 0.0000
Embedding Dimension 8: Fraction of FNN = 0.0000
Embedding Dimension 9: Fraction of FNN = 0.0000
Embedding Dimension 10: Fraction of FNN = 0.0000
Embedding Dimension 11: Fraction of FNN = 0.0000
Embedding Dimension 12: Fraction of FNN = 0.0000
Embedding Dimension 13: Fraction of FNN = 0.0000
Embedding Dimension 14: Fraction of FNN = 0.0000
Embedding Dimension 15: Fraction of FNN = 0.0000
Embedding Dimension 16: Fraction of FNN = 0.0000
Embedding Dimension 17: Fraction of FNN = 0.0000
Embedding Dimension 18: Fraction of FNN = 0.0000
Embedding Dimension 19: Fraction of FNN = 0.0000
Embedding Dimension 20: Fraction of FNN = 0.0000
Channel: F3
Embedding Dimension 1: Fraction of FNN = 0.5161
Embedding Dimension 2: Fraction of FNN = 0.0000
Embedding Dimension 3: Fraction of FNN = 0.0000
Embedding Dimension 4: Fraction of FNN = 0.0000
Embedding Dimension 5: Fraction of FNN = 0.0000
Embedding Dimension 6: Fraction of FNN = 0.0000
Embedding Dimension 7: Fraction of FNN = 0.0000
Embedding Dimension 8: Fraction of FNN = 0.0000
Embedding Dimension 9: Fraction of FNN = 0.0000
Embedding Dimension 10: Fraction of FNN = 0.0000
Embedding Dimension 11: Fraction of FNN = 0.0000
Embedding Dimension 12: Fraction of FNN = 0.0000
Embedding Dimension 13: Fraction of FNN = 0.0000
Embedding Dimension 14: Fraction of FNN = 0.0000
Embedding Dimension 15: Fraction of FNN = 0.0000
Embedding Dimension 16: Fraction of FNN = 0.0000
Embedding Dimension 17: Fraction of FNN = 0.0000
Embedding Dimension 18: Fraction of FNN = 0.0000
Embedding Dimension 19: Fraction of FNN = 0.0000
Embedding Dimension 20: Fraction of FNN = 0.0000
Channel: Fz
Embedding Dimension 1: Fraction of FNN = 0.5161
Embedding Dimension 2: Fraction of FNN = 0.0000
Embedding Dimension 3: Fraction of FNN = 0.0000
Embedding Dimension 4: Fraction of FNN = 0.0000
```

```
Embedding Dimension 5: Fraction of FNN = 0.0000
Embedding Dimension 6: Fraction of FNN = 0.0000
Embedding Dimension 7: Fraction of FNN = 0.0000
Embedding Dimension 8: Fraction of FNN = 0.0000
Embedding Dimension 9: Fraction of FNN = 0.0000
Embedding Dimension 10: Fraction of FNN = 0.0000
Embedding Dimension 11: Fraction of FNN = 0.0000
Embedding Dimension 12: Fraction of FNN = 0.0000
Embedding Dimension 13: Fraction of FNN = 0.0000
Embedding Dimension 14: Fraction of FNN = 0.0000
Embedding Dimension 15: Fraction of FNN = 0.0000
Embedding Dimension 16: Fraction of FNN = 0.0000
Embedding Dimension 17: Fraction of FNN = 0.0000
Embedding Dimension 18: Fraction of FNN = 0.0000
Embedding Dimension 19: Fraction of FNN = 0.0000
Embedding Dimension 20: Fraction of FNN = 0.0000
Channel: F4
Embedding Dimension 1: Fraction of FNN = 0.5161
Embedding Dimension 2: Fraction of FNN = 0.0000
Embedding Dimension 3: Fraction of FNN = 0.0000
Embedding Dimension 4: Fraction of FNN = 0.0000
Embedding Dimension 5: Fraction of FNN = 0.0000
Embedding Dimension 6: Fraction of FNN = 0.0000
Embedding Dimension 7: Fraction of FNN = 0.0000
Embedding Dimension 8: Fraction of FNN = 0.0000
Embedding Dimension 9: Fraction of FNN = 0.0000
Embedding Dimension 10: Fraction of FNN = 0.0000
Embedding Dimension 11: Fraction of FNN = 0.0000
Embedding Dimension 12: Fraction of FNN = 0.0000
Embedding Dimension 13: Fraction of FNN = 0.0000
Embedding Dimension 14: Fraction of FNN = 0.0000
Embedding Dimension 15: Fraction of FNN = 0.0000
Embedding Dimension 16: Fraction of FNN = 0.0000
Embedding Dimension 17: Fraction of FNN = 0.0000
Embedding Dimension 18: Fraction of FNN = 0.0000
Embedding Dimension 19: Fraction of FNN = 0.0000
Embedding Dimension 20: Fraction of FNN = 0.0000
Channel: F8
Embedding Dimension 1: Fraction of FNN = 0.5161
Embedding Dimension 2: Fraction of FNN = 0.0000
Embedding Dimension 3: Fraction of FNN = 0.0000
Embedding Dimension 4: Fraction of FNN = 0.0000
Embedding Dimension 5: Fraction of FNN = 0.0000
Embedding Dimension 6: Fraction of FNN = 0.0000
Embedding Dimension 7: Fraction of FNN = 0.0000
Embedding Dimension 8: Fraction of FNN = 0.0000
Embedding Dimension 9: Fraction of FNN = 0.0000
Embedding Dimension 10: Fraction of FNN = 0.0000
```

```
Embedding Dimension 11: Fraction of FNN = 0.0000
Embedding Dimension 12: Fraction of FNN = 0.0000
Embedding Dimension 13: Fraction of FNN = 0.0000
Embedding Dimension 14: Fraction of FNN = 0.0000
Embedding Dimension 15: Fraction of FNN = 0.0000
Embedding Dimension 16: Fraction of FNN = 0.0000
Embedding Dimension 17: Fraction of FNN = 0.0000
Embedding Dimension 18: Fraction of FNN = 0.0000
Embedding Dimension 19: Fraction of FNN = 0.0000
Embedding Dimension 20: Fraction of FNN = 0.0000
Channel: FC5
Embedding Dimension 1: Fraction of FNN = 0.5161
Embedding Dimension 2: Fraction of FNN = 0.0000
Embedding Dimension 3: Fraction of FNN = 0.0000
Embedding Dimension 4: Fraction of FNN = 0.0000
Embedding Dimension 5: Fraction of FNN = 0.0000
Embedding Dimension 6: Fraction of FNN = 0.0000
Embedding Dimension 7: Fraction of FNN = 0.0000
Embedding Dimension 8: Fraction of FNN = 0.0000
Embedding Dimension 9: Fraction of FNN = 0.0000
Embedding Dimension 10: Fraction of FNN = 0.0000
Embedding Dimension 11: Fraction of FNN = 0.0000
Embedding Dimension 12: Fraction of FNN = 0.0000
Embedding Dimension 13: Fraction of FNN = 0.0000
Embedding Dimension 14: Fraction of FNN = 0.0000
Embedding Dimension 15: Fraction of FNN = 0.0000
Embedding Dimension 16: Fraction of FNN = 0.0000
Embedding Dimension 17: Fraction of FNN = 0.0000
Embedding Dimension 18: Fraction of FNN = 0.0000
Embedding Dimension 19: Fraction of FNN = 0.0000
Embedding Dimension 20: Fraction of FNN = 0.0000
Channel: FC1
Embedding Dimension 1: Fraction of FNN = 0.5161
Embedding Dimension 2: Fraction of FNN = 0.0000
Embedding Dimension 3: Fraction of FNN = 0.0000
Embedding Dimension 4: Fraction of FNN = 0.0000
Embedding Dimension 5: Fraction of FNN = 0.0000
Embedding Dimension 6: Fraction of FNN = 0.0000
Embedding Dimension 7: Fraction of FNN = 0.0000
Embedding Dimension 8: Fraction of FNN = 0.0000
Embedding Dimension 9: Fraction of FNN = 0.0000
Embedding Dimension 10: Fraction of FNN = 0.0000
Embedding Dimension 11: Fraction of FNN = 0.0000
Embedding Dimension 12: Fraction of FNN = 0.0000
Embedding Dimension 13: Fraction of FNN = 0.0000
Embedding Dimension 14: Fraction of FNN = 0.0000
Embedding Dimension 15: Fraction of FNN = 0.0000
Embedding Dimension 16: Fraction of FNN = 0.0000
```

```
Embedding Dimension 17: Fraction of FNN = 0.0000
Embedding Dimension 18: Fraction of FNN = 0.0000
Embedding Dimension 19: Fraction of FNN = 0.0000
Embedding Dimension 20: Fraction of FNN = 0.0000
Channel: FC2
Embedding Dimension 1: Fraction of FNN = 0.5161
Embedding Dimension 2: Fraction of FNN = 0.0000
Embedding Dimension 3: Fraction of FNN = 0.0000
Embedding Dimension 4: Fraction of FNN = 0.0000
Embedding Dimension 5: Fraction of FNN = 0.0000
Embedding Dimension 6: Fraction of FNN = 0.0000
Embedding Dimension 7: Fraction of FNN = 0.0000
Embedding Dimension 8: Fraction of FNN = 0.0000
Embedding Dimension 9: Fraction of FNN = 0.0000
Embedding Dimension 10: Fraction of FNN = 0.0000
Embedding Dimension 11: Fraction of FNN = 0.0000
Embedding Dimension 12: Fraction of FNN = 0.0000
Embedding Dimension 13: Fraction of FNN = 0.0000
Embedding Dimension 14: Fraction of FNN = 0.0000
Embedding Dimension 15: Fraction of FNN = 0.0000
Embedding Dimension 16: Fraction of FNN = 0.0000
Embedding Dimension 17: Fraction of FNN = 0.0000
Embedding Dimension 18: Fraction of FNN = 0.0000
Embedding Dimension 19: Fraction of FNN = 0.0000
Embedding Dimension 20: Fraction of FNN = 0.0000
Channel: FC6
Embedding Dimension 1: Fraction of FNN = 0.5161
Embedding Dimension 2: Fraction of FNN = 0.0000
Embedding Dimension 3: Fraction of FNN = 0.0000
Embedding Dimension 4: Fraction of FNN = 0.0000
Embedding Dimension 5: Fraction of FNN = 0.0000
Embedding Dimension 6: Fraction of FNN = 0.0000
Embedding Dimension 7: Fraction of FNN = 0.0000
Embedding Dimension 8: Fraction of FNN = 0.0000
Embedding Dimension 9: Fraction of FNN = 0.0000
Embedding Dimension 10: Fraction of FNN = 0.0000
Embedding Dimension 11: Fraction of FNN = 0.0000
Embedding Dimension 12: Fraction of FNN = 0.0000
Embedding Dimension 13: Fraction of FNN = 0.0000
Embedding Dimension 14: Fraction of FNN = 0.0000
Embedding Dimension 15: Fraction of FNN = 0.0000
Embedding Dimension 16: Fraction of FNN = 0.0000
Embedding Dimension 17: Fraction of FNN = 0.0000
Embedding Dimension 18: Fraction of FNN = 0.0000
Embedding Dimension 19: Fraction of FNN = 0.0000
Embedding Dimension 20: Fraction of FNN = 0.0000
Channel: M1
Embedding Dimension 1: Fraction of FNN = 0.5161
```

```
Embedding Dimension 2: Fraction of FNN = 0.0000
Embedding Dimension 3: Fraction of FNN = 0.0000
Embedding Dimension 4: Fraction of FNN = 0.0000
Embedding Dimension 5: Fraction of FNN = 0.0000
Embedding Dimension 6: Fraction of FNN = 0.0000
Embedding Dimension 7: Fraction of FNN = 0.0000
Embedding Dimension 8: Fraction of FNN = 0.0000
Embedding Dimension 9: Fraction of FNN = 0.0000
Embedding Dimension 10: Fraction of FNN = 0.0000
Embedding Dimension 11: Fraction of FNN = 0.0000
Embedding Dimension 12: Fraction of FNN = 0.0000
Embedding Dimension 13: Fraction of FNN = 0.0000
Embedding Dimension 14: Fraction of FNN = 0.0000
Embedding Dimension 15: Fraction of FNN = 0.0000
Embedding Dimension 16: Fraction of FNN = 0.0000
Embedding Dimension 17: Fraction of FNN = 0.0000
Embedding Dimension 18: Fraction of FNN = 0.0000
Embedding Dimension 19: Fraction of FNN = 0.0000
Embedding Dimension 20: Fraction of FNN = 0.0000
Channel: T7
Embedding Dimension 1: Fraction of FNN = 0.5161
Embedding Dimension 2: Fraction of FNN = 0.0000
Embedding Dimension 3: Fraction of FNN = 0.0000
Embedding Dimension 4: Fraction of FNN = 0.0000
Embedding Dimension 5: Fraction of FNN = 0.0000
Embedding Dimension 6: Fraction of FNN = 0.0000
Embedding Dimension 7: Fraction of FNN = 0.0000
Embedding Dimension 8: Fraction of FNN = 0.0000
Embedding Dimension 9: Fraction of FNN = 0.0000
Embedding Dimension 10: Fraction of FNN = 0.0000
Embedding Dimension 11: Fraction of FNN = 0.0000
Embedding Dimension 12: Fraction of FNN = 0.0000
Embedding Dimension 13: Fraction of FNN = 0.0000
Embedding Dimension 14: Fraction of FNN = 0.0000
Embedding Dimension 15: Fraction of FNN = 0.0000
Embedding Dimension 16: Fraction of FNN = 0.0000
Embedding Dimension 17: Fraction of FNN = 0.0000
Embedding Dimension 18: Fraction of FNN = 0.0000
Embedding Dimension 19: Fraction of FNN = 0.0000
Embedding Dimension 20: Fraction of FNN = 0.0000
Channel: C3
Embedding Dimension 1: Fraction of FNN = 0.5161
Embedding Dimension 2: Fraction of FNN = 0.0000
Embedding Dimension 3: Fraction of FNN = 0.0000
Embedding Dimension 4: Fraction of FNN = 0.0000
Embedding Dimension 5: Fraction of FNN = 0.0000
Embedding Dimension 6: Fraction of FNN = 0.0000
Embedding Dimension 7: Fraction of FNN = 0.0000
```

```
Embedding Dimension 8: Fraction of FNN = 0.0000
Embedding Dimension 9: Fraction of FNN = 0.0000
Embedding Dimension 10: Fraction of FNN = 0.0000
Embedding Dimension 11: Fraction of FNN = 0.0000
Embedding Dimension 12: Fraction of FNN = 0.0000
Embedding Dimension 13: Fraction of FNN = 0.0000
Embedding Dimension 14: Fraction of FNN = 0.0000
Embedding Dimension 15: Fraction of FNN = 0.0000
Embedding Dimension 16: Fraction of FNN = 0.0000
Embedding Dimension 17: Fraction of FNN = 0.0000
Embedding Dimension 18: Fraction of FNN = 0.0000
Embedding Dimension 19: Fraction of FNN = 0.0000
Embedding Dimension 20: Fraction of FNN = 0.0000
Channel: Cz
Embedding Dimension 1: Fraction of FNN = 0.5161
Embedding Dimension 2: Fraction of FNN = 0.0000
Embedding Dimension 3: Fraction of FNN = 0.0000
Embedding Dimension 4: Fraction of FNN = 0.0000
Embedding Dimension 5: Fraction of FNN = 0.0000
Embedding Dimension 6: Fraction of FNN = 0.0000
Embedding Dimension 7: Fraction of FNN = 0.0000
Embedding Dimension 8: Fraction of FNN = 0.0000
Embedding Dimension 9: Fraction of FNN = 0.0000
Embedding Dimension 10: Fraction of FNN = 0.0000
Embedding Dimension 11: Fraction of FNN = 0.0000
Embedding Dimension 12: Fraction of FNN = 0.0000
Embedding Dimension 13: Fraction of FNN = 0.0000
Embedding Dimension 14: Fraction of FNN = 0.0000
Embedding Dimension 15: Fraction of FNN = 0.0000
Embedding Dimension 16: Fraction of FNN = 0.0000
Embedding Dimension 17: Fraction of FNN = 0.0000
Embedding Dimension 18: Fraction of FNN = 0.0000
Embedding Dimension 19: Fraction of FNN = 0.0000
Embedding Dimension 20: Fraction of FNN = 0.0000
Channel: C4
Embedding Dimension 1: Fraction of FNN = 0.5161
Embedding Dimension 2: Fraction of FNN = 0.0000
Embedding Dimension 3: Fraction of FNN = 0.0000
Embedding Dimension 4: Fraction of FNN = 0.0000
Embedding Dimension 5: Fraction of FNN = 0.0000
Embedding Dimension 6: Fraction of FNN = 0.0000
Embedding Dimension 7: Fraction of FNN = 0.0000
Embedding Dimension 8: Fraction of FNN = 0.0000
Embedding Dimension 9: Fraction of FNN = 0.0000
Embedding Dimension 10: Fraction of FNN = 0.0000
Embedding Dimension 11: Fraction of FNN = 0.0000
Embedding Dimension 12: Fraction of FNN = 0.0000
Embedding Dimension 13: Fraction of FNN = 0.0000
```

```
Embedding Dimension 14: Fraction of FNN = 0.0000
Embedding Dimension 15: Fraction of FNN = 0.0000
Embedding Dimension 16: Fraction of FNN = 0.0000
Embedding Dimension 17: Fraction of FNN = 0.0000
Embedding Dimension 18: Fraction of FNN = 0.0000
Embedding Dimension 19: Fraction of FNN = 0.0000
Embedding Dimension 20: Fraction of FNN = 0.0000
Channel: T8
Embedding Dimension 1: Fraction of FNN = 0.4839
Embedding Dimension 2: Fraction of FNN = 0.0000
Embedding Dimension 3: Fraction of FNN = 0.0000
Embedding Dimension 4: Fraction of FNN = 0.0000
Embedding Dimension 5: Fraction of FNN = 0.0000
Embedding Dimension 6: Fraction of FNN = 0.0000
Embedding Dimension 7: Fraction of FNN = 0.0000
Embedding Dimension 8: Fraction of FNN = 0.0000
Embedding Dimension 9: Fraction of FNN = 0.0000
Embedding Dimension 10: Fraction of FNN = 0.0000
Embedding Dimension 11: Fraction of FNN = 0.0000
Embedding Dimension 12: Fraction of FNN = 0.0000
Embedding Dimension 13: Fraction of FNN = 0.0000
Embedding Dimension 14: Fraction of FNN = 0.0000
Embedding Dimension 15: Fraction of FNN = 0.0000
Embedding Dimension 16: Fraction of FNN = 0.0000
Embedding Dimension 17: Fraction of FNN = 0.0000
Embedding Dimension 18: Fraction of FNN = 0.0000
Embedding Dimension 19: Fraction of FNN = 0.0000
Embedding Dimension 20: Fraction of FNN = 0.0000
Channel: M2
Embedding Dimension 1: Fraction of FNN = 0.4839
Embedding Dimension 2: Fraction of FNN = 0.0000
Embedding Dimension 3: Fraction of FNN = 0.0000
Embedding Dimension 4: Fraction of FNN = 0.0000
Embedding Dimension 5: Fraction of FNN = 0.0000
Embedding Dimension 6: Fraction of FNN = 0.0000
Embedding Dimension 7: Fraction of FNN = 0.0000
Embedding Dimension 8: Fraction of FNN = 0.0000
Embedding Dimension 9: Fraction of FNN = 0.0000
Embedding Dimension 10: Fraction of FNN = 0.0000
Embedding Dimension 11: Fraction of FNN = 0.0000
Embedding Dimension 12: Fraction of FNN = 0.0000
Embedding Dimension 13: Fraction of FNN = 0.0000
Embedding Dimension 14: Fraction of FNN = 0.0000
Embedding Dimension 15: Fraction of FNN = 0.0000
Embedding Dimension 16: Fraction of FNN = 0.0000
Embedding Dimension 17: Fraction of FNN = 0.0000
Embedding Dimension 18: Fraction of FNN = 0.0000
Embedding Dimension 19: Fraction of FNN = 0.0000
```

```
Embedding Dimension 20: Fraction of FNN = 0.0000
Channel: CP5
Embedding Dimension 1: Fraction of FNN = 0.4839
Embedding Dimension 2: Fraction of FNN = 0.0000
Embedding Dimension 3: Fraction of FNN = 0.0000
Embedding Dimension 4: Fraction of FNN = 0.0000
Embedding Dimension 5: Fraction of FNN = 0.0000
Embedding Dimension 6: Fraction of FNN = 0.0000
Embedding Dimension 7: Fraction of FNN = 0.0000
Embedding Dimension 8: Fraction of FNN = 0.0000
Embedding Dimension 9: Fraction of FNN = 0.0000
Embedding Dimension 10: Fraction of FNN = 0.0000
Embedding Dimension 11: Fraction of FNN = 0.0000
Embedding Dimension 12: Fraction of FNN = 0.0000
Embedding Dimension 13: Fraction of FNN = 0.0000
Embedding Dimension 14: Fraction of FNN = 0.0000
Embedding Dimension 15: Fraction of FNN = 0.0000
Embedding Dimension 16: Fraction of FNN = 0.0000
Embedding Dimension 17: Fraction of FNN = 0.0000
Embedding Dimension 18: Fraction of FNN = 0.0000
Embedding Dimension 19: Fraction of FNN = 0.0000
Embedding Dimension 20: Fraction of FNN = 0.0000
Channel: CP1
Embedding Dimension 1: Fraction of FNN = 0.4839
Embedding Dimension 2: Fraction of FNN = 0.0000
Embedding Dimension 3: Fraction of FNN = 0.0000
Embedding Dimension 4: Fraction of FNN = 0.0000
Embedding Dimension 5: Fraction of FNN = 0.0000
Embedding Dimension 6: Fraction of FNN = 0.0000
Embedding Dimension 7: Fraction of FNN = 0.0000
Embedding Dimension 8: Fraction of FNN = 0.0000
Embedding Dimension 9: Fraction of FNN = 0.0000
Embedding Dimension 10: Fraction of FNN = 0.0000
Embedding Dimension 11: Fraction of FNN = 0.0000
Embedding Dimension 12: Fraction of FNN = 0.0000
Embedding Dimension 13: Fraction of FNN = 0.0000
Embedding Dimension 14: Fraction of FNN = 0.0000
Embedding Dimension 15: Fraction of FNN = 0.0000
Embedding Dimension 16: Fraction of FNN = 0.0000
Embedding Dimension 17: Fraction of FNN = 0.0000
Embedding Dimension 18: Fraction of FNN = 0.0000
Embedding Dimension 19: Fraction of FNN = 0.0000
Embedding Dimension 20: Fraction of FNN = 0.0000
Channel: CP2
Embedding Dimension 1: Fraction of FNN = 0.5161
Embedding Dimension 2: Fraction of FNN = 0.0000
Embedding Dimension 3: Fraction of FNN = 0.0000
Embedding Dimension 4: Fraction of FNN = 0.0000
```

```
Embedding Dimension 5: Fraction of FNN = 0.0000
Embedding Dimension 6: Fraction of FNN = 0.0000
Embedding Dimension 7: Fraction of FNN = 0.0000
Embedding Dimension 8: Fraction of FNN = 0.0000
Embedding Dimension 9: Fraction of FNN = 0.0000
Embedding Dimension 10: Fraction of FNN = 0.0000
Embedding Dimension 11: Fraction of FNN = 0.0000
Embedding Dimension 12: Fraction of FNN = 0.0000
Embedding Dimension 13: Fraction of FNN = 0.0000
Embedding Dimension 14: Fraction of FNN = 0.0000
Embedding Dimension 15: Fraction of FNN = 0.0000
Embedding Dimension 16: Fraction of FNN = 0.0000
Embedding Dimension 17: Fraction of FNN = 0.0000
Embedding Dimension 18: Fraction of FNN = 0.0000
Embedding Dimension 19: Fraction of FNN = 0.0000
Embedding Dimension 20: Fraction of FNN = 0.0000
Channel: CP6
Embedding Dimension 1: Fraction of FNN = 0.5161
Embedding Dimension 2: Fraction of FNN = 0.0000
Embedding Dimension 3: Fraction of FNN = 0.0000
Embedding Dimension 4: Fraction of FNN = 0.0000
Embedding Dimension 5: Fraction of FNN = 0.0000
Embedding Dimension 6: Fraction of FNN = 0.0000
Embedding Dimension 7: Fraction of FNN = 0.0000
Embedding Dimension 8: Fraction of FNN = 0.0000
Embedding Dimension 9: Fraction of FNN = 0.0000
Embedding Dimension 10: Fraction of FNN = 0.0000
Embedding Dimension 11: Fraction of FNN = 0.0000
Embedding Dimension 12: Fraction of FNN = 0.0000
Embedding Dimension 13: Fraction of FNN = 0.0000
Embedding Dimension 14: Fraction of FNN = 0.0000
Embedding Dimension 15: Fraction of FNN = 0.0000
Embedding Dimension 16: Fraction of FNN = 0.0000
Embedding Dimension 17: Fraction of FNN = 0.0000
Embedding Dimension 18: Fraction of FNN = 0.0000
Embedding Dimension 19: Fraction of FNN = 0.0000
Embedding Dimension 20: Fraction of FNN = 0.0000
Channel: P7
Embedding Dimension 1: Fraction of FNN = 0.5161
Embedding Dimension 2: Fraction of FNN = 0.0000
Embedding Dimension 3: Fraction of FNN = 0.0000
Embedding Dimension 4: Fraction of FNN = 0.0000
Embedding Dimension 5: Fraction of FNN = 0.0000
Embedding Dimension 6: Fraction of FNN = 0.0000
Embedding Dimension 7: Fraction of FNN = 0.0000
Embedding Dimension 8: Fraction of FNN = 0.0000
Embedding Dimension 9: Fraction of FNN = 0.0000
Embedding Dimension 10: Fraction of FNN = 0.0000
```

```
Embedding Dimension 11: Fraction of FNN = 0.0000
Embedding Dimension 12: Fraction of FNN = 0.0000
Embedding Dimension 13: Fraction of FNN = 0.0000
Embedding Dimension 14: Fraction of FNN = 0.0000
Embedding Dimension 15: Fraction of FNN = 0.0000
Embedding Dimension 16: Fraction of FNN = 0.0000
Embedding Dimension 17: Fraction of FNN = 0.0000
Embedding Dimension 18: Fraction of FNN = 0.0000
Embedding Dimension 19: Fraction of FNN = 0.0000
Embedding Dimension 20: Fraction of FNN = 0.0000
Channel: P3
Embedding Dimension 1: Fraction of FNN = 0.4839
Embedding Dimension 2: Fraction of FNN = 0.0000
Embedding Dimension 3: Fraction of FNN = 0.0000
Embedding Dimension 4: Fraction of FNN = 0.0000
Embedding Dimension 5: Fraction of FNN = 0.0000
Embedding Dimension 6: Fraction of FNN = 0.0000
Embedding Dimension 7: Fraction of FNN = 0.0000
Embedding Dimension 8: Fraction of FNN = 0.0000
Embedding Dimension 9: Fraction of FNN = 0.0000
Embedding Dimension 10: Fraction of FNN = 0.0000
Embedding Dimension 11: Fraction of FNN = 0.0000
Embedding Dimension 12: Fraction of FNN = 0.0000
Embedding Dimension 13: Fraction of FNN = 0.0000
Embedding Dimension 14: Fraction of FNN = 0.0000
Embedding Dimension 15: Fraction of FNN = 0.0000
Embedding Dimension 16: Fraction of FNN = 0.0000
Embedding Dimension 17: Fraction of FNN = 0.0000
Embedding Dimension 18: Fraction of FNN = 0.0000
Embedding Dimension 19: Fraction of FNN = 0.0000
Embedding Dimension 20: Fraction of FNN = 0.0000
Channel: Pz
Embedding Dimension 1: Fraction of FNN = 0.4839
Embedding Dimension 2: Fraction of FNN = 0.0000
Embedding Dimension 3: Fraction of FNN = 0.0000
Embedding Dimension 4: Fraction of FNN = 0.0000
Embedding Dimension 5: Fraction of FNN = 0.0000
Embedding Dimension 6: Fraction of FNN = 0.0000
Embedding Dimension 7: Fraction of FNN = 0.0000
Embedding Dimension 8: Fraction of FNN = 0.0000
Embedding Dimension 9: Fraction of FNN = 0.0000
Embedding Dimension 10: Fraction of FNN = 0.0000
Embedding Dimension 11: Fraction of FNN = 0.0000
Embedding Dimension 12: Fraction of FNN = 0.0000
Embedding Dimension 13: Fraction of FNN = 0.0000
Embedding Dimension 14: Fraction of FNN = 0.0000
Embedding Dimension 15: Fraction of FNN = 0.0000
Embedding Dimension 16: Fraction of FNN = 0.0000
```

```
Embedding Dimension 17: Fraction of FNN = 0.0000
Embedding Dimension 18: Fraction of FNN = 0.0000
Embedding Dimension 19: Fraction of FNN = 0.0000
Embedding Dimension 20: Fraction of FNN = 0.0000
Channel: P4
Embedding Dimension 1: Fraction of FNN = 0.5161
Embedding Dimension 2: Fraction of FNN = 0.0000
Embedding Dimension 3: Fraction of FNN = 0.0000
Embedding Dimension 4: Fraction of FNN = 0.0000
Embedding Dimension 5: Fraction of FNN = 0.0000
Embedding Dimension 6: Fraction of FNN = 0.0000
Embedding Dimension 7: Fraction of FNN = 0.0000
Embedding Dimension 8: Fraction of FNN = 0.0000
Embedding Dimension 9: Fraction of FNN = 0.0000
Embedding Dimension 10: Fraction of FNN = 0.0000
Embedding Dimension 11: Fraction of FNN = 0.0000
Embedding Dimension 12: Fraction of FNN = 0.0000
Embedding Dimension 13: Fraction of FNN = 0.0000
Embedding Dimension 14: Fraction of FNN = 0.0000
Embedding Dimension 15: Fraction of FNN = 0.0000
Embedding Dimension 16: Fraction of FNN = 0.0000
Embedding Dimension 17: Fraction of FNN = 0.0000
Embedding Dimension 18: Fraction of FNN = 0.0000
Embedding Dimension 19: Fraction of FNN = 0.0000
Embedding Dimension 20: Fraction of FNN = 0.0000
Channel: P8
Embedding Dimension 1: Fraction of FNN = 0.4839
Embedding Dimension 2: Fraction of FNN = 0.0000
Embedding Dimension 3: Fraction of FNN = 0.0000
Embedding Dimension 4: Fraction of FNN = 0.0000
Embedding Dimension 5: Fraction of FNN = 0.0000
Embedding Dimension 6: Fraction of FNN = 0.0000
Embedding Dimension 7: Fraction of FNN = 0.0000
Embedding Dimension 8: Fraction of FNN = 0.0000
Embedding Dimension 9: Fraction of FNN = 0.0000
Embedding Dimension 10: Fraction of FNN = 0.0000
Embedding Dimension 11: Fraction of FNN = 0.0000
Embedding Dimension 12: Fraction of FNN = 0.0000
Embedding Dimension 13: Fraction of FNN = 0.0000
Embedding Dimension 14: Fraction of FNN = 0.0000
Embedding Dimension 15: Fraction of FNN = 0.0000
Embedding Dimension 16: Fraction of FNN = 0.0000
Embedding Dimension 17: Fraction of FNN = 0.0000
Embedding Dimension 18: Fraction of FNN = 0.0000
Embedding Dimension 19: Fraction of FNN = 0.0000
Embedding Dimension 20: Fraction of FNN = 0.0000
Channel: POz
Embedding Dimension 1: Fraction of FNN = 0.5161
```

```
Embedding Dimension 2: Fraction of FNN = 0.0000
Embedding Dimension 3: Fraction of FNN = 0.0000
Embedding Dimension 4: Fraction of FNN = 0.0000
Embedding Dimension 5: Fraction of FNN = 0.0000
Embedding Dimension 6: Fraction of FNN = 0.0000
Embedding Dimension 7: Fraction of FNN = 0.0000
Embedding Dimension 8: Fraction of FNN = 0.0000
Embedding Dimension 9: Fraction of FNN = 0.0000
Embedding Dimension 10: Fraction of FNN = 0.0000
Embedding Dimension 11: Fraction of FNN = 0.0000
Embedding Dimension 12: Fraction of FNN = 0.0000
Embedding Dimension 13: Fraction of FNN = 0.0000
Embedding Dimension 14: Fraction of FNN = 0.0000
Embedding Dimension 15: Fraction of FNN = 0.0000
Embedding Dimension 16: Fraction of FNN = 0.0000
Embedding Dimension 17: Fraction of FNN = 0.0000
Embedding Dimension 18: Fraction of FNN = 0.0000
Embedding Dimension 19: Fraction of FNN = 0.0000
Embedding Dimension 20: Fraction of FNN = 0.0000
Channel: O1
Embedding Dimension 1: Fraction of FNN = 0.4839
Embedding Dimension 2: Fraction of FNN = 0.0000
Embedding Dimension 3: Fraction of FNN = 0.0000
Embedding Dimension 4: Fraction of FNN = 0.0000
Embedding Dimension 5: Fraction of FNN = 0.0000
Embedding Dimension 6: Fraction of FNN = 0.0000
Embedding Dimension 7: Fraction of FNN = 0.0000
Embedding Dimension 8: Fraction of FNN = 0.0000
Embedding Dimension 9: Fraction of FNN = 0.0000
Embedding Dimension 10: Fraction of FNN = 0.0000
Embedding Dimension 11: Fraction of FNN = 0.0000
Embedding Dimension 12: Fraction of FNN = 0.0000
Embedding Dimension 13: Fraction of FNN = 0.0000
Embedding Dimension 14: Fraction of FNN = 0.0000
Embedding Dimension 15: Fraction of FNN = 0.0000
Embedding Dimension 16: Fraction of FNN = 0.0000
Embedding Dimension 17: Fraction of FNN = 0.0000
Embedding Dimension 18: Fraction of FNN = 0.0000
Embedding Dimension 19: Fraction of FNN = 0.0000
Embedding Dimension 20: Fraction of FNN = 0.0000
Channel: Oz
Embedding Dimension 1: Fraction of FNN = 0.4839
Embedding Dimension 2: Fraction of FNN = 0.0000
Embedding Dimension 3: Fraction of FNN = 0.0000
Embedding Dimension 4: Fraction of FNN = 0.0000
Embedding Dimension 5: Fraction of FNN = 0.0000
Embedding Dimension 6: Fraction of FNN = 0.0000
Embedding Dimension 7: Fraction of FNN = 0.0000
```

```
Embedding Dimension 8: Fraction of FNN = 0.0000
Embedding Dimension 9: Fraction of FNN = 0.0000
Embedding Dimension 10: Fraction of FNN = 0.0000
Embedding Dimension 11: Fraction of FNN = 0.0000
Embedding Dimension 12: Fraction of FNN = 0.0000
Embedding Dimension 13: Fraction of FNN = 0.0000
Embedding Dimension 14: Fraction of FNN = 0.0000
Embedding Dimension 15: Fraction of FNN = 0.0000
Embedding Dimension 16: Fraction of FNN = 0.0000
Embedding Dimension 17: Fraction of FNN = 0.0000
Embedding Dimension 18: Fraction of FNN = 0.0000
Embedding Dimension 19: Fraction of FNN = 0.0000
Embedding Dimension 20: Fraction of FNN = 0.0000
Channel: O2
Embedding Dimension 1: Fraction of FNN = 0.4839
Embedding Dimension 2: Fraction of FNN = 0.0000
Embedding Dimension 3: Fraction of FNN = 0.0000
Embedding Dimension 4: Fraction of FNN = 0.0000
Embedding Dimension 5: Fraction of FNN = 0.0000
Embedding Dimension 6: Fraction of FNN = 0.0000
Embedding Dimension 7: Fraction of FNN = 0.0000
Embedding Dimension 8: Fraction of FNN = 0.0000
Embedding Dimension 9: Fraction of FNN = 0.0000
Embedding Dimension 10: Fraction of FNN = 0.0000
Embedding Dimension 11: Fraction of FNN = 0.0000
Embedding Dimension 12: Fraction of FNN = 0.0000
Embedding Dimension 13: Fraction of FNN = 0.0000
Embedding Dimension 14: Fraction of FNN = 0.0000
Embedding Dimension 15: Fraction of FNN = 0.0000
Embedding Dimension 16: Fraction of FNN = 0.0000
Embedding Dimension 17: Fraction of FNN = 0.0000
Embedding Dimension 18: Fraction of FNN = 0.0000
Embedding Dimension 19: Fraction of FNN = 0.0000
Embedding Dimension 20: Fraction of FNN = 0.0000
```

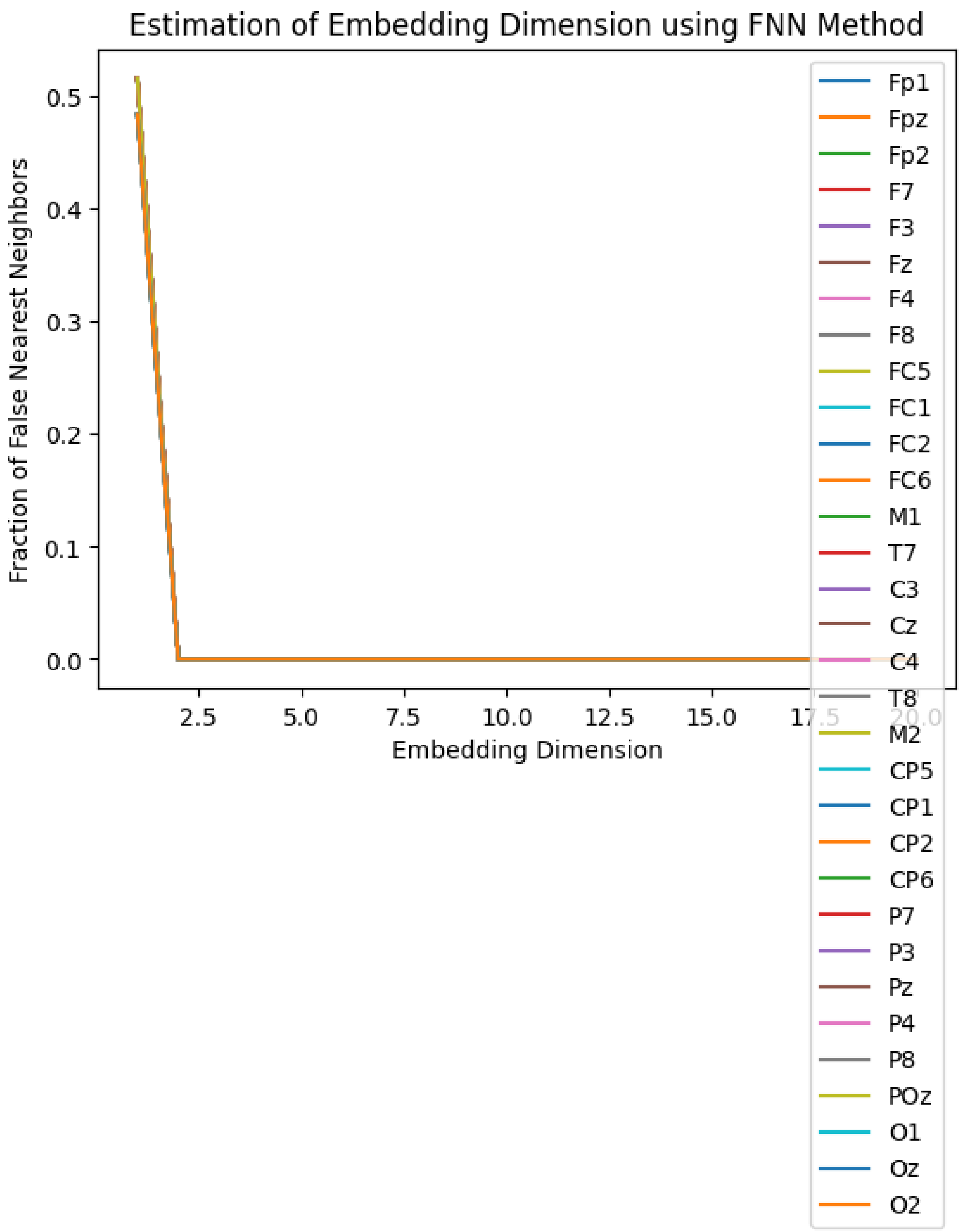

FNN data saved to
/home/vincent/AAA_projects/MVCS/Neuroscience/false_nearest_neighbors.npy

# 3 2D Phase Space Plot

[1]:
```
import numpy as np
import matplotlib.pyplot as plt
from minepy import MINE
import multiprocessing

def mutual_info_worker(args):
    data1, data2 = args
    mine = MINE()
    mine.compute_score(data1, data2)
    return mine.mic()

def determine_delay(data, max_delay=100, subsample_factor=10):
    subsampled_data = data[::subsample_factor]
    with multiprocessing.Pool() as pool:
        args_list = [(subsampled_data[:-i], subsampled_data[i:]) for i in␣
 ↪range(1, max_delay+1)]
        mi_values = pool.map(mutual_info_worker, args_list)
    min_index = np.argmin(mi_values)
    return min_index + 1

def delay_embedding(data, emb_dim, delay):
    N = len(data)
    embedded_data = np.zeros((N - (emb_dim - 1) * delay, emb_dim))
    for i in range(N - (emb_dim - 1) * delay):
        embedded_data[i] = [data[i + j * delay] for j in range(emb_dim)]
    return embedded_data

# Load EEG data
EEG_data = np.load('/home/vincent/AAA_projects/MVCS/Neuroscience/
 ↪eeg_data_with_channels.npy', allow_pickle=True)

# Adjust the list to match your data's channels
eeg_channels = ['Fp1', 'Fpz', 'Fp2', 'F7', 'F3', 'Fz', 'F4', 'F8', 'FC5',␣
 ↪'FC1', 'FC2', 'FC6',
                'M1', 'T7', 'C3', 'Cz', 'C4', 'T8', 'M2', 'CP5', 'CP1', 'CP2',␣
 ↪'CP6',
                'P7', 'P3', 'Pz', 'P4', 'P8', 'POz', 'O1', 'Oz', 'O2']


# Create directory for saving the plots if it doesn't exist
plots_directory = "/home/vincent/AAA_projects/MVCS/Neuroscience/Analysis/Phase␣
 ↪Space/2dembedding_data/Plots"
if not os.path.exists(plots_directory):
    os.makedirs(plots_directory)
```

```python
# Loop through each channel
for selected_channel in eeg_channels:
    channel_index = eeg_channels.index(selected_channel)
    channel_data = EEG_data[:, channel_index]

    # Determine optimal delay using mutual information with subsampling
    optimal_delay = determine_delay(channel_data, subsample_factor=50)

    # Embedding dimension
    emb_dim = 2

    # Perform delay embedding
    embedded_channel_data = delay_embedding(channel_data, emb_dim=emb_dim,␣
 ↪delay=optimal_delay)

    # Save the embedded data
    np.save(f'/home/vincent/AAA_projects/MVCS/Neuroscience/Analysis/Phase Space/
 ↪2dembedding_data/2dembedded_{selected_channel}_data.npy',␣
 ↪embedded_channel_data)

    # Create 2D scatter plot with black background
    plt.figure(figsize=(8, 6), facecolor='black')
    plt.scatter(embedded_channel_data[:, 0], embedded_channel_data[:, 1],␣
 ↪color='red', s=0.5)
    plt.title(f'Phase Space Plot for Channel {selected_channel}', color='white')
    plt.xlabel('Embedding Dimension 1', color='grey')
    plt.ylabel('Embedding Dimension 2', color='grey')
    plt.xticks(color='grey')
    plt.yticks(color='grey')
    plt.gca().spines['left'].set_color('grey')
    plt.gca().spines['right'].set_color('grey')
    plt.gca().spines['bottom'].set_color('grey')
    plt.gca().spines['top'].set_color('grey')

    # Save the figure to the specified directory
    plt.savefig(os.path.join(plots_directory, f'PhaseSpace_{selected_channel}.
 ↪png'), facecolor='black', dpi=300)
    plt.close()
```

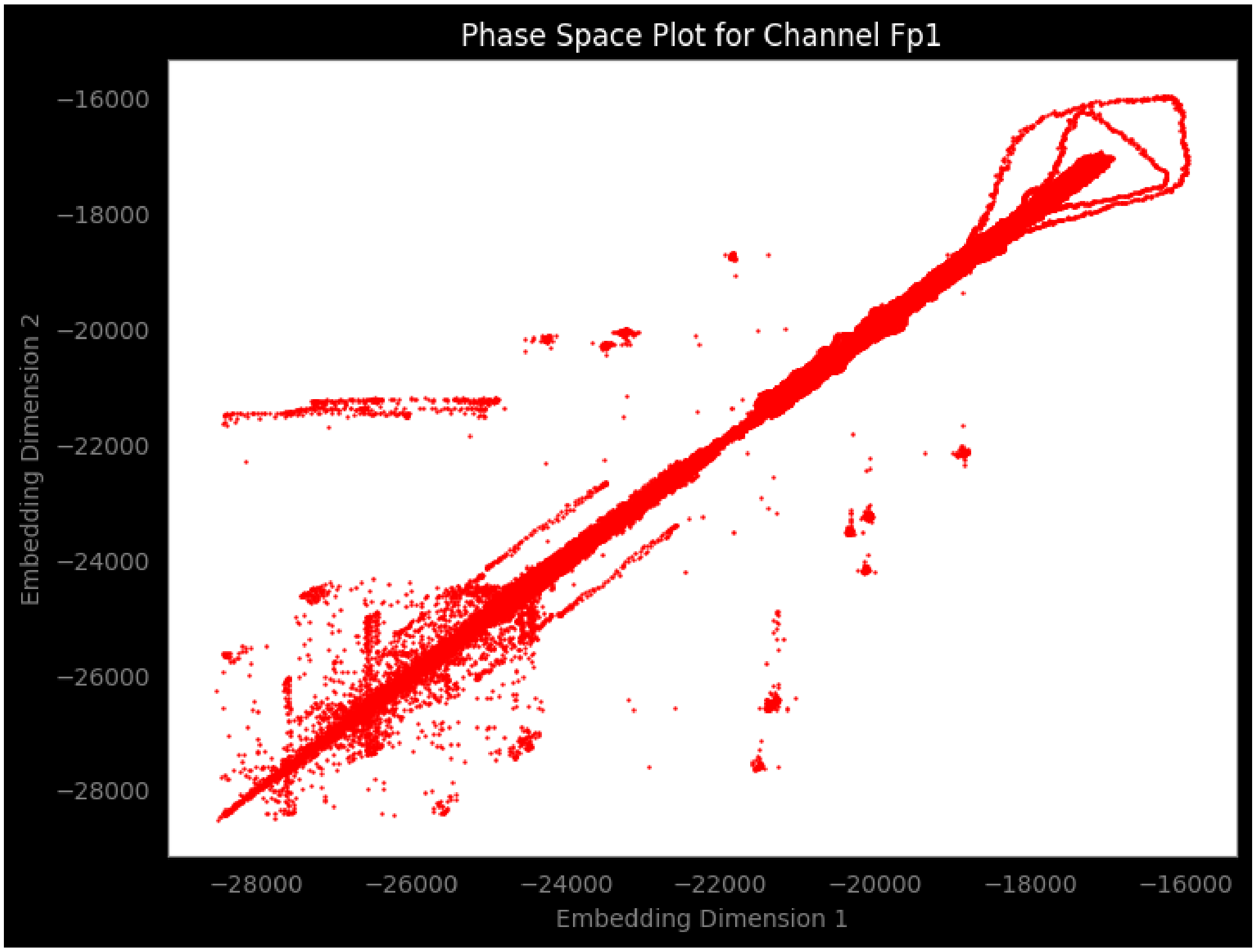
Phase Space Plot for Channel Fp1
Embedding Dimension 2
−16000
−18000
−20000
−22000
−24000
−26000
−28000
−28000
−26000
−24000
−22000
−20000
−18000
−16000
Embedding Dimension 1

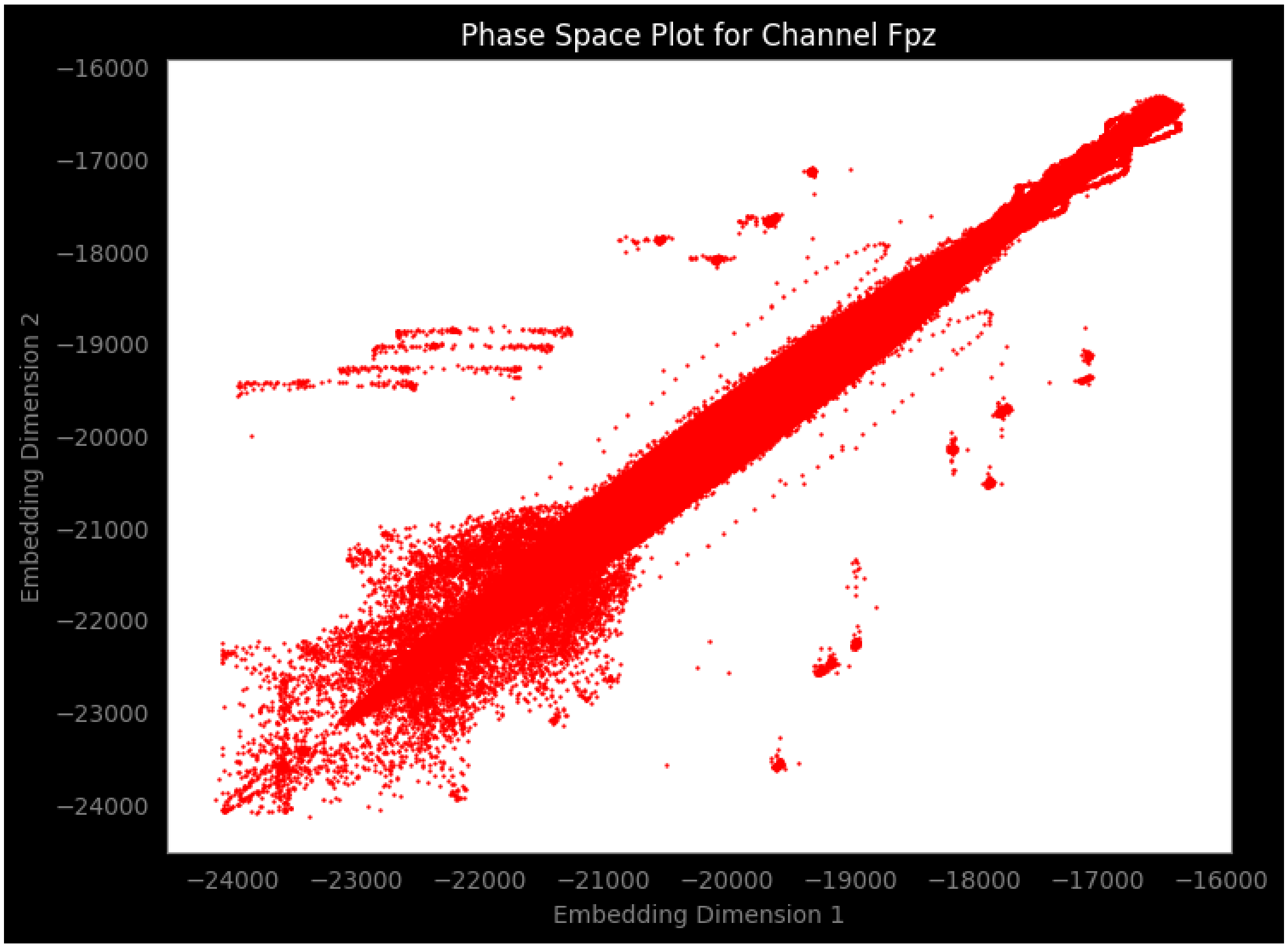

Phase Space Plot for Channel Fpz
−16000
−17000
−18000
−19000
−20000
−21000
−22000
−23000
−24000
Embedding Dimension 2
−24000
−23000
−22000
−21000
−20000
−19000
−18000
−17000
−16000
Embedding Dimension 1

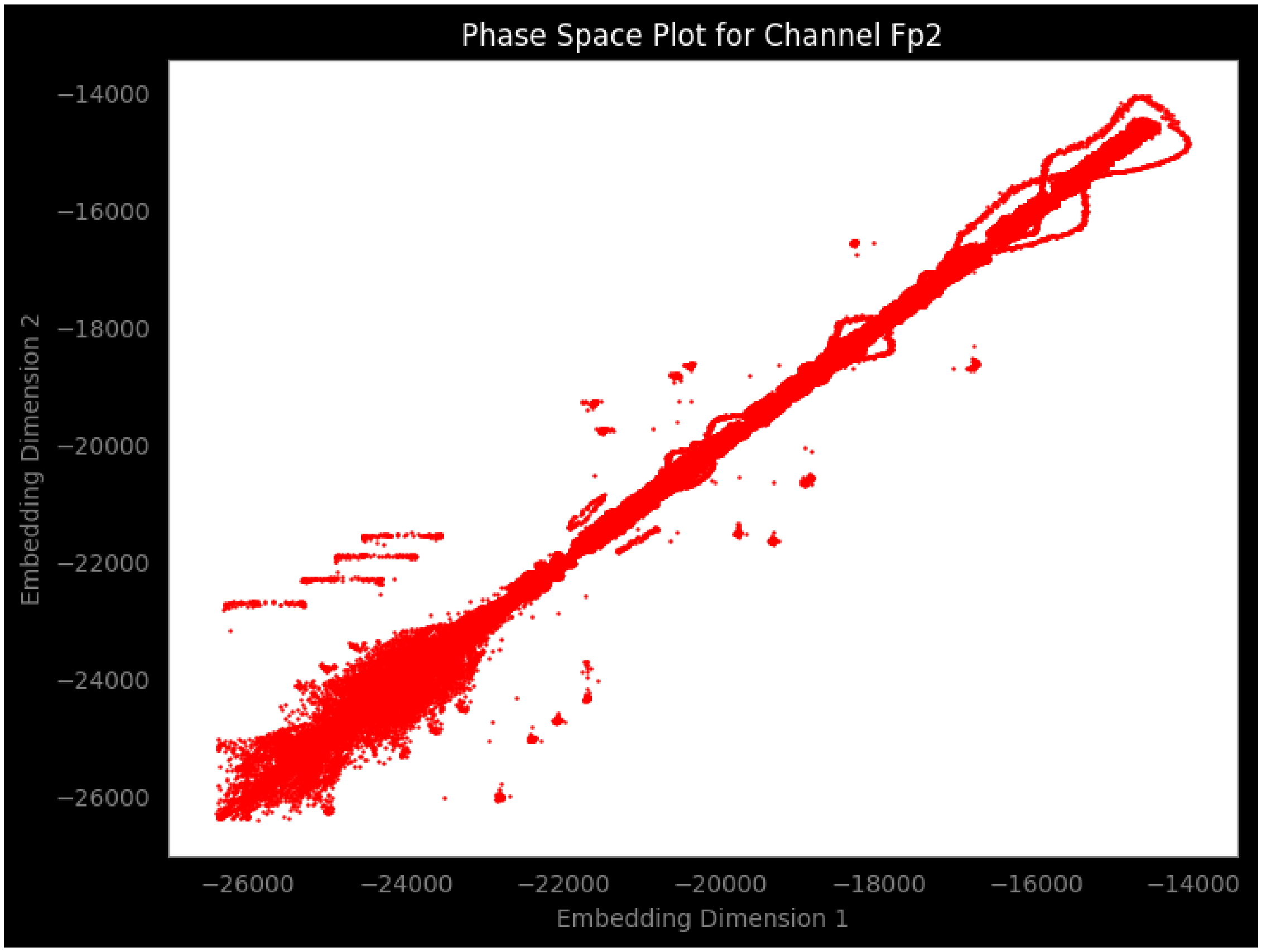
Phase Space Plot for Channel Fp2
−14000
−16000
−18000
−20000
−22000
−24000
−26000
Embedding Dimension 2
−26000
−24000
−22000
−20000
−18000
−16000
−14000
Embedding Dimension 1

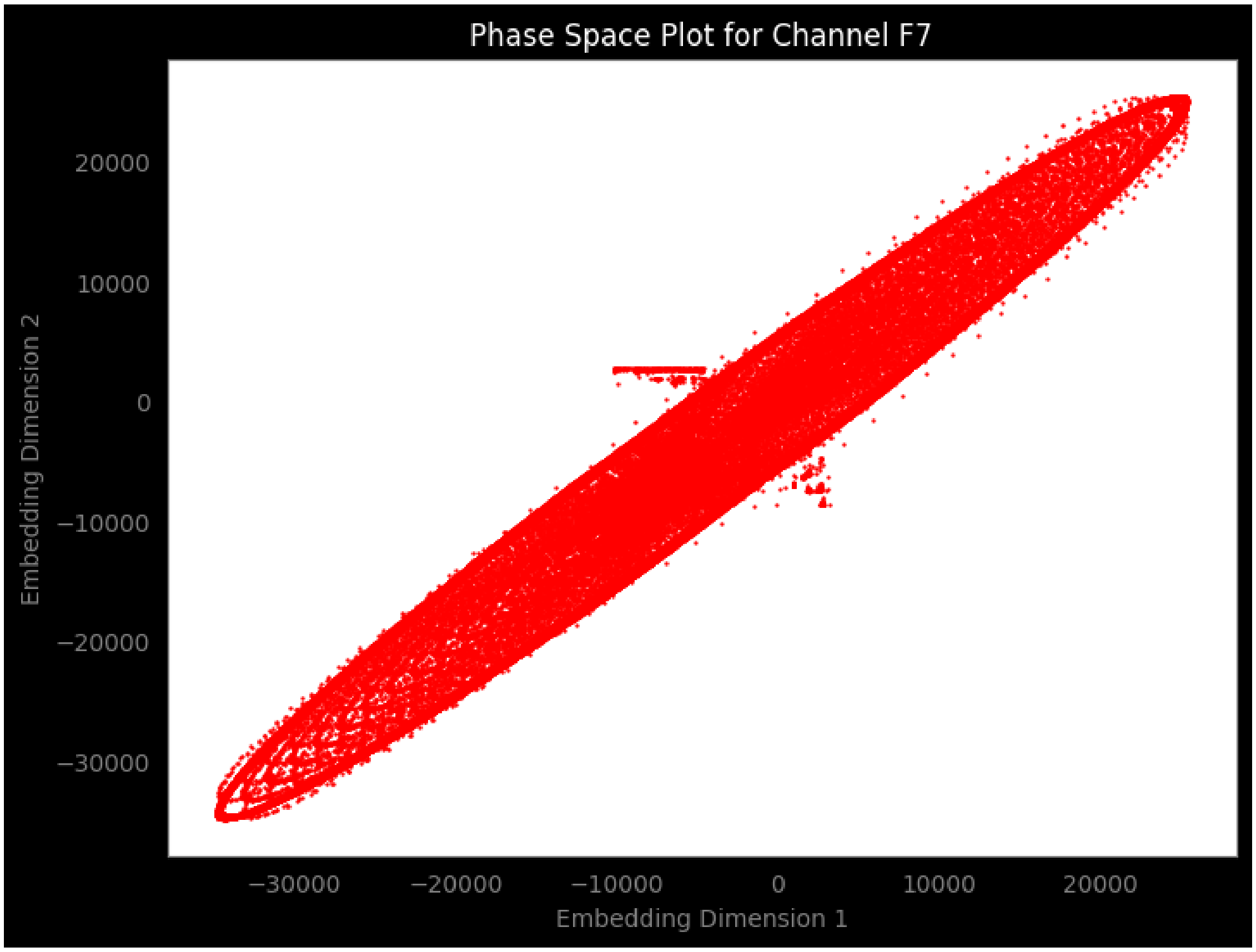
Phase Space Plot for Channel F7
20000
10000
0
−10000
−20000
−30000
Embedding Dimension 2
−30000
−20000
−10000
0
10000
20000
Embedding Dimension 1

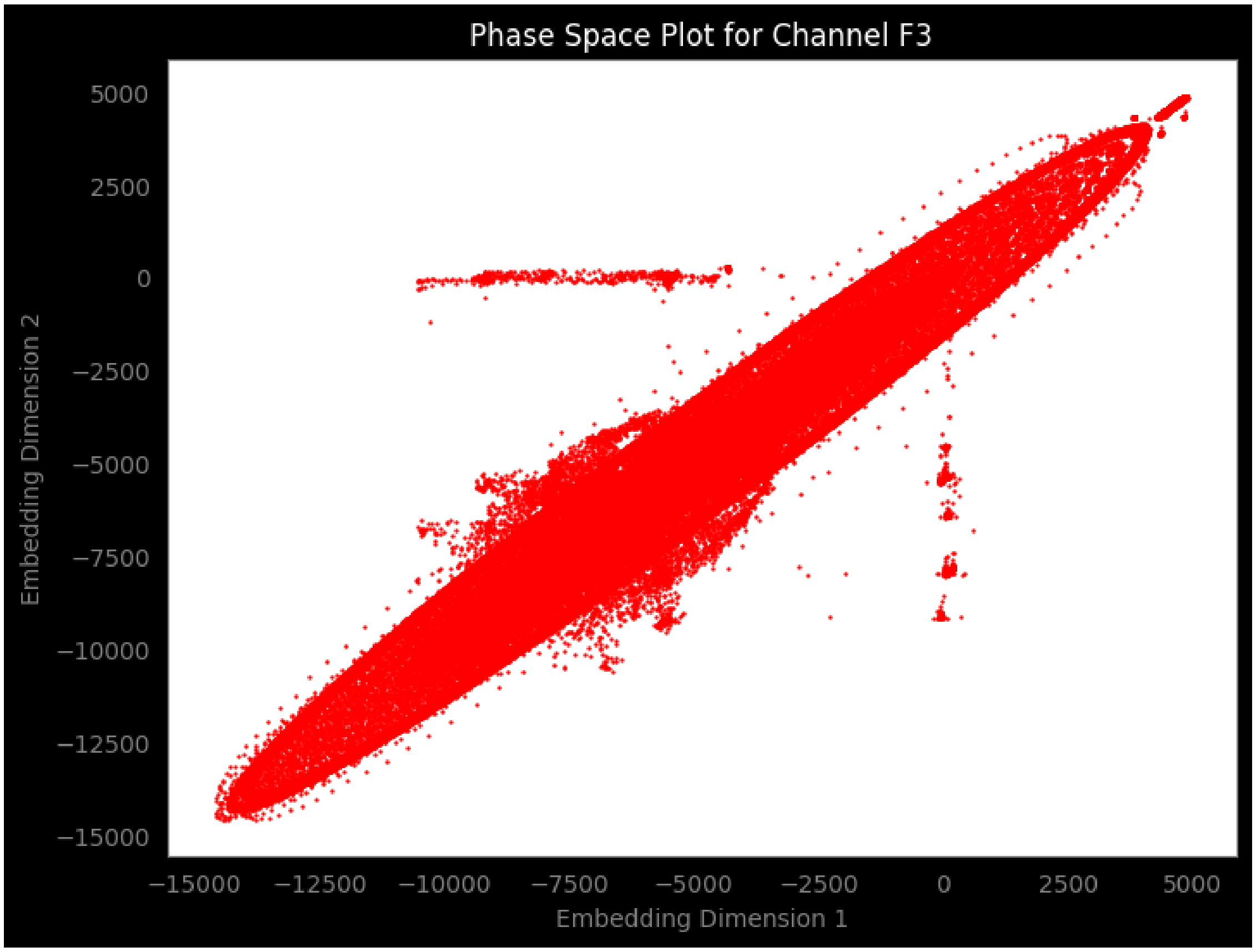
Phase Space Plot for Channel F3
Embedding Dimension 2
Embedding Dimension 1
5000
2500
0
−2500
−5000
−7500
−10000
−12500
−15000
−15000
−12500
−10000
−7500
−5000
−2500
0
2500
5000

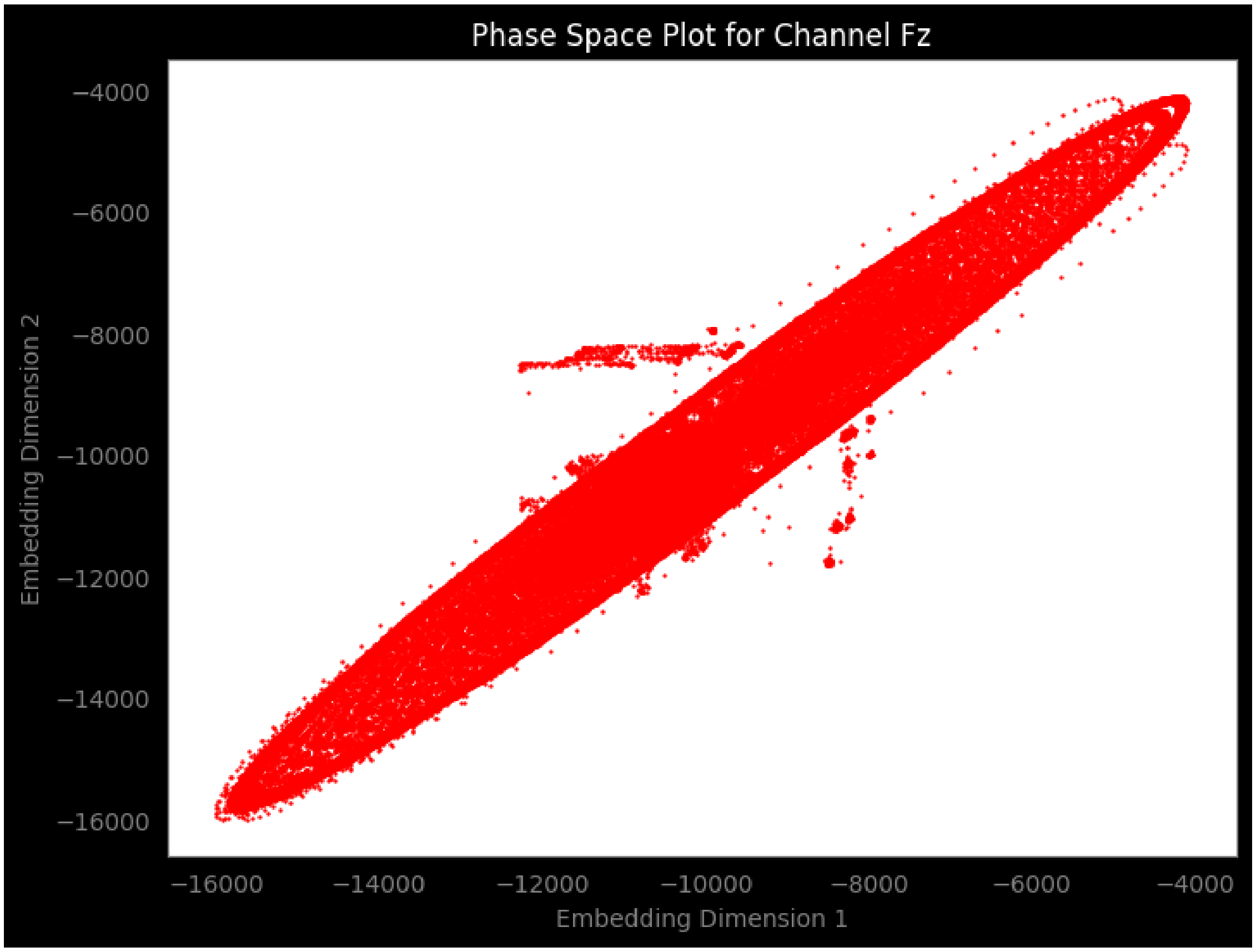
Phase Space Plot for Channel Fz
Embedding Dimension 2
−4000
−6000
−8000
−10000
−12000
−14000
−16000
−16000
−14000
−12000
−10000
−8000
−6000
−4000
Embedding Dimension 1

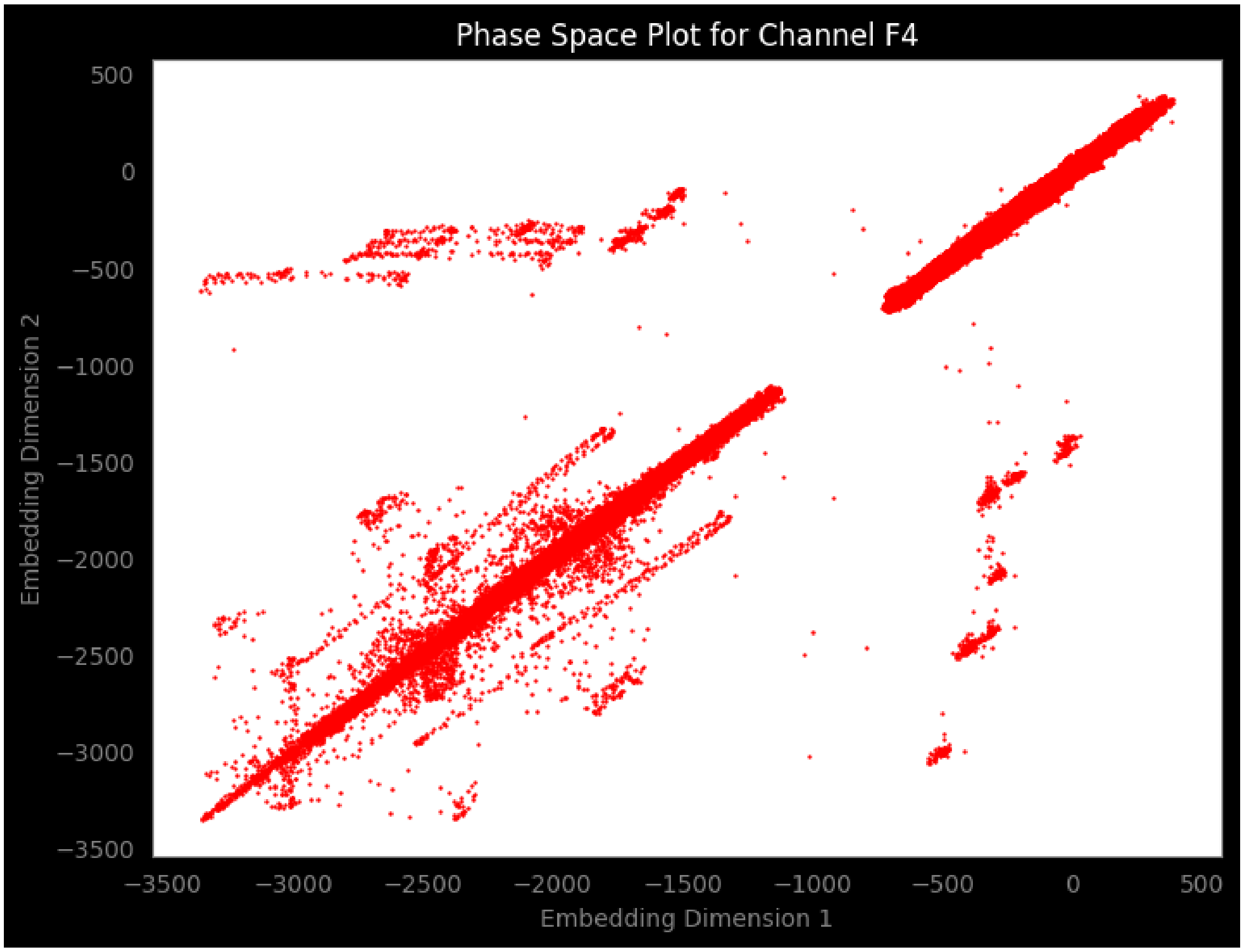

Phase Space Plot for Channel F4
Embedding Dimension 2
500
0
−500
−1000
−1500
−2000
−2500
−3000
−3500
−3500
−3000
−2500
−2000
−1500
−1000
−500
0
500
Embedding Dimension 1

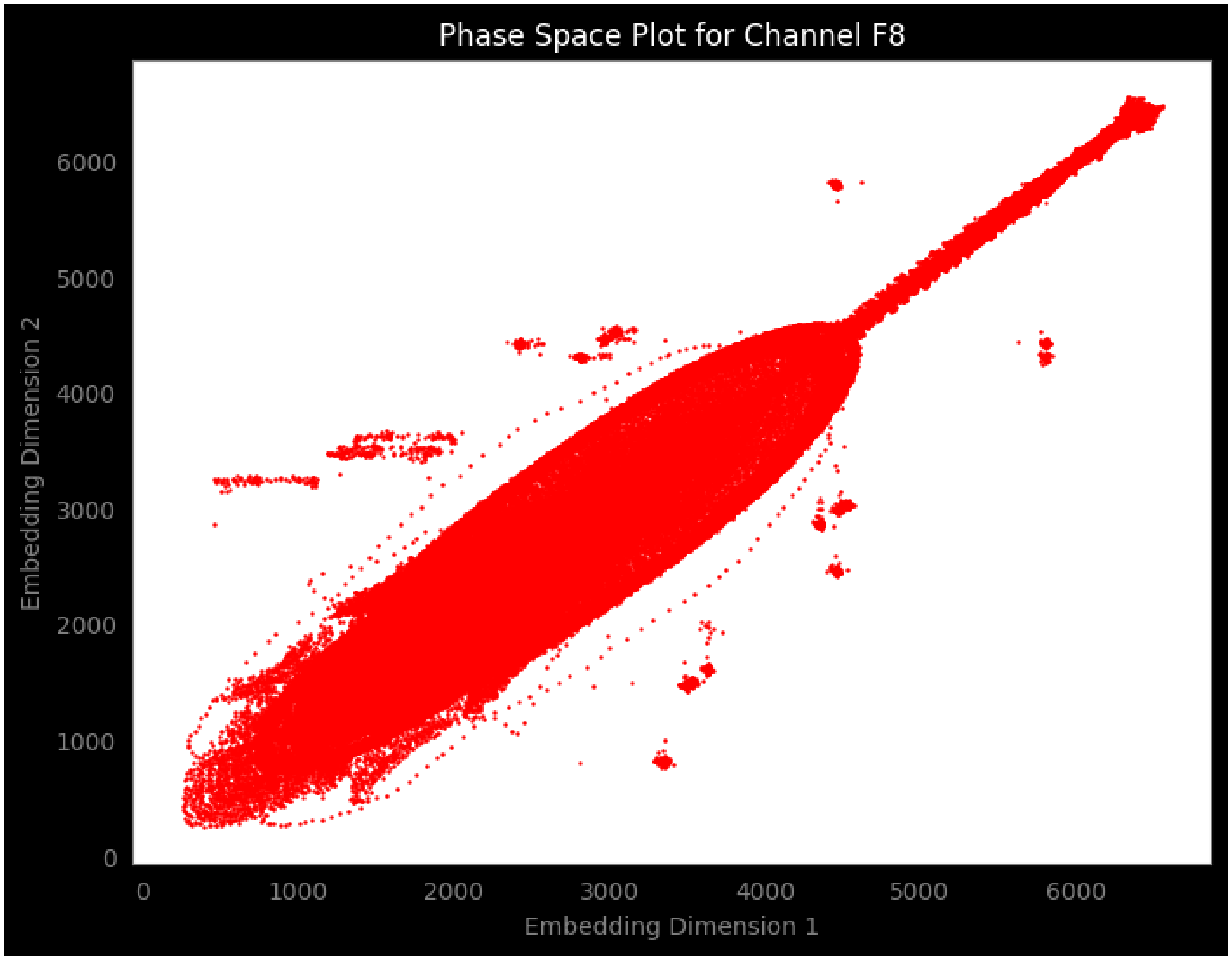
Phase Space Plot for Channel F8
6000
5000
4000
3000
2000
1000
0
Embedding Dimension 2
0
1000
2000
3000
4000
5000
6000
Embedding Dimension 1

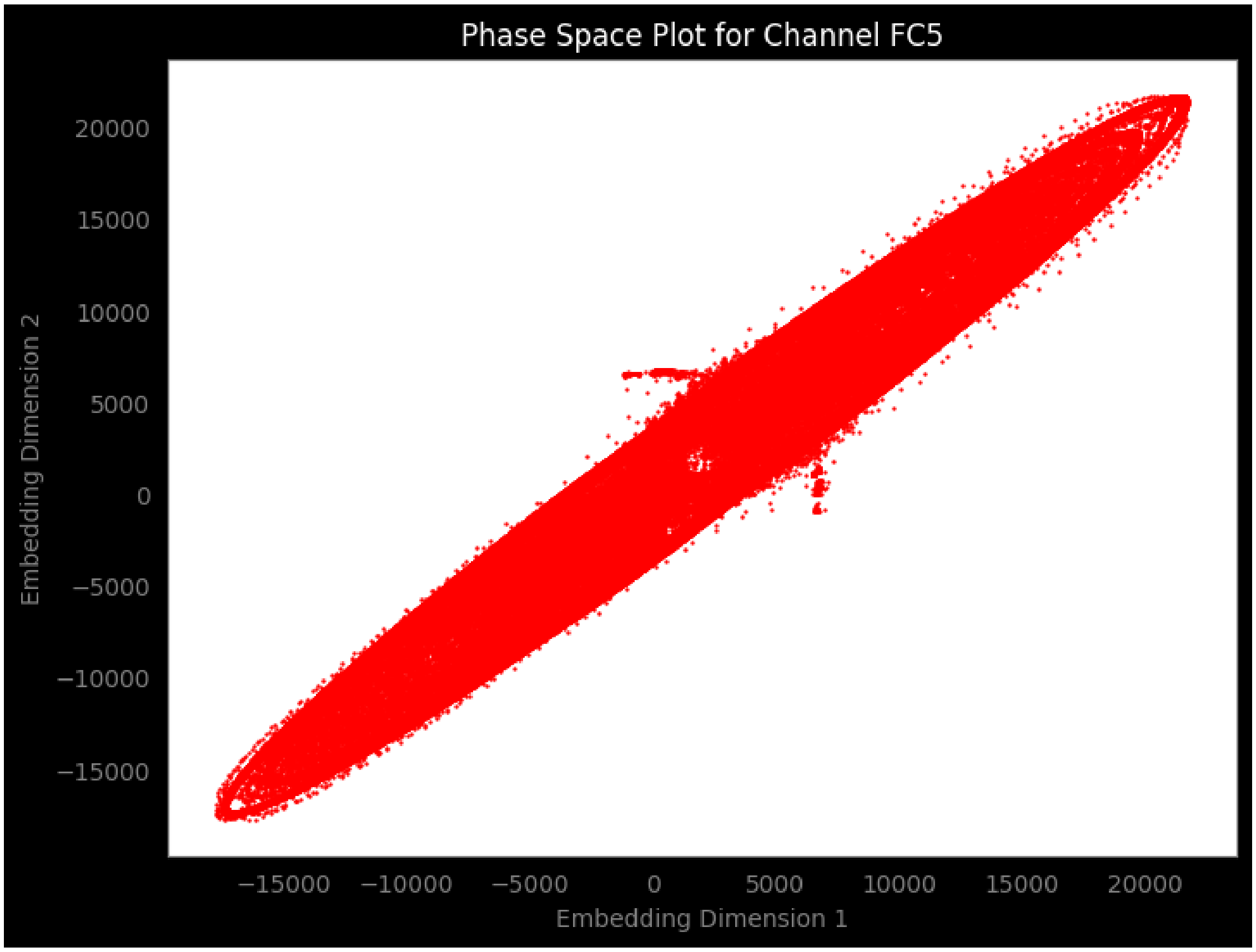
Phase Space Plot for Channel FC5
Embedding Dimension 2
20000
15000
10000
5000
0
−5000
−10000
−15000
−15000
−10000
−5000
0
5000
10000
15000
20000
Embedding Dimension 1

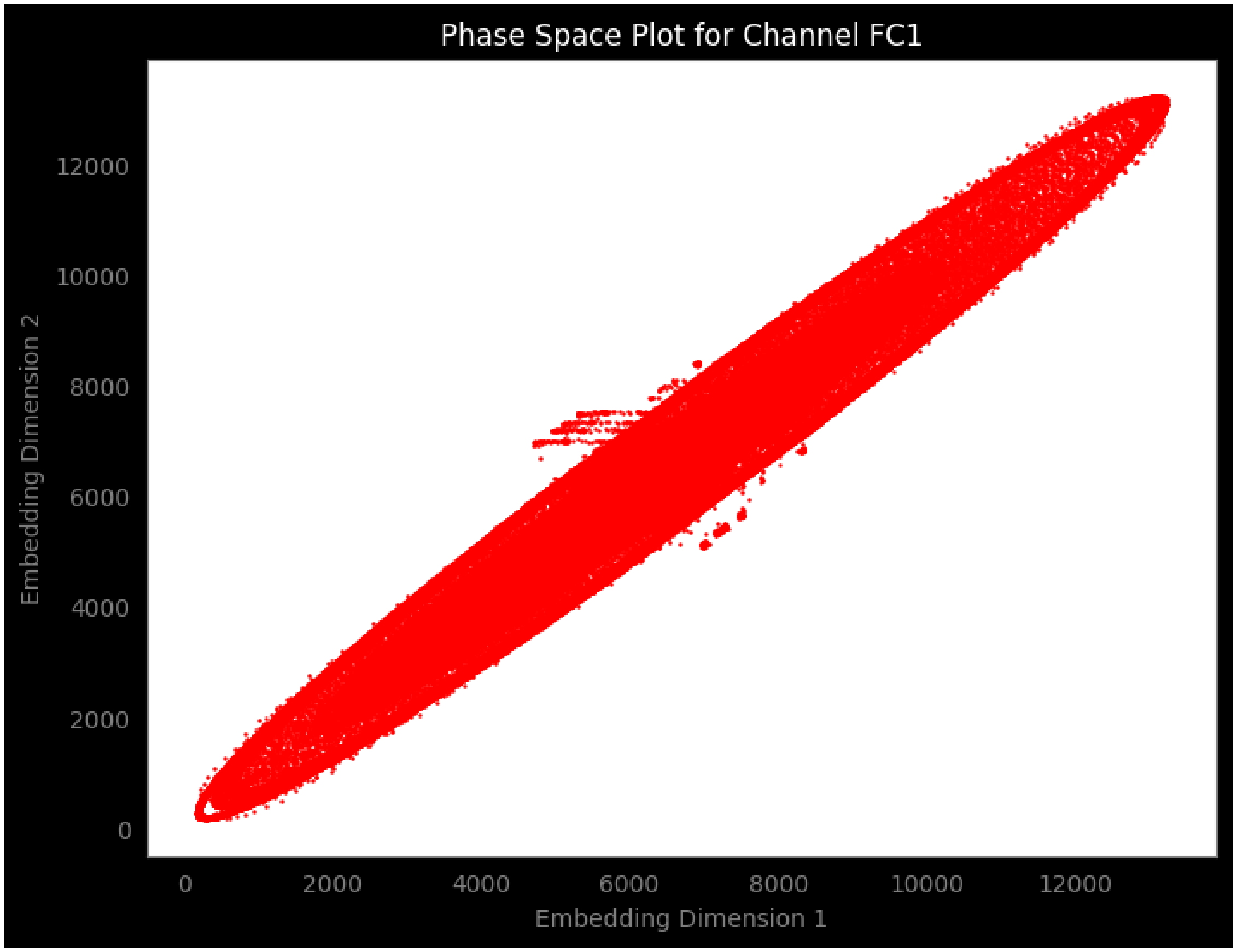
Phase Space Plot for Channel FC1
Embedding Dimension 2
Embedding Dimension 1
0
2000
4000
6000
8000
10000
12000
0
2000
4000
6000
8000
10000
12000

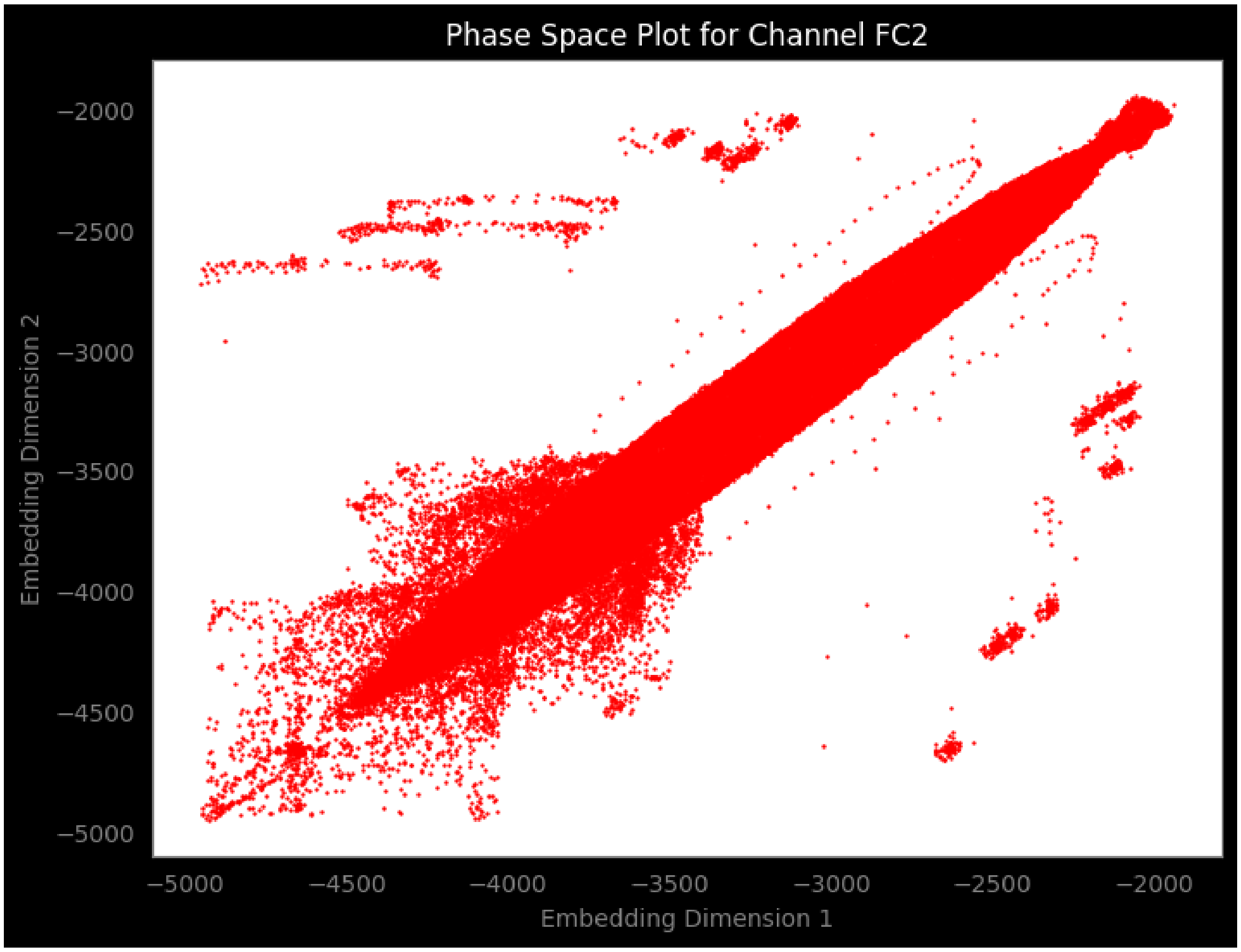
Phase Space Plot for Channel FC2
Embedding Dimension 2
−2000
−2500
−3000
−3500
−4000
−4500
−5000
−5000
−4500
−4000
−3500
−3000
−2500
−2000
Embedding Dimension 1

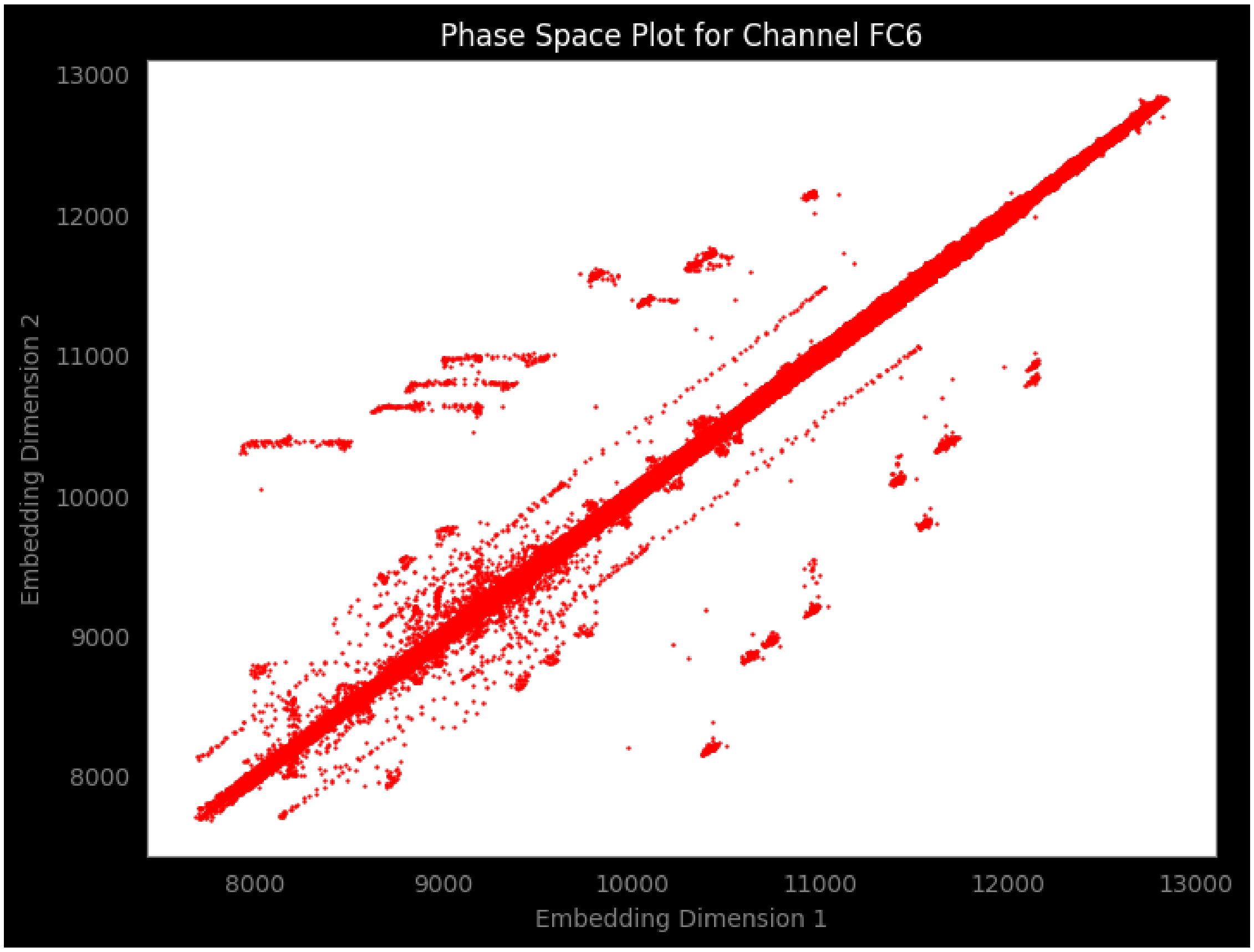

Phase Space Plot for Channel FC6
Embedding Dimension 2
Embedding Dimension 1
13000
12000
11000
10000
9000
8000
8000
9000
10000
11000
12000
13000

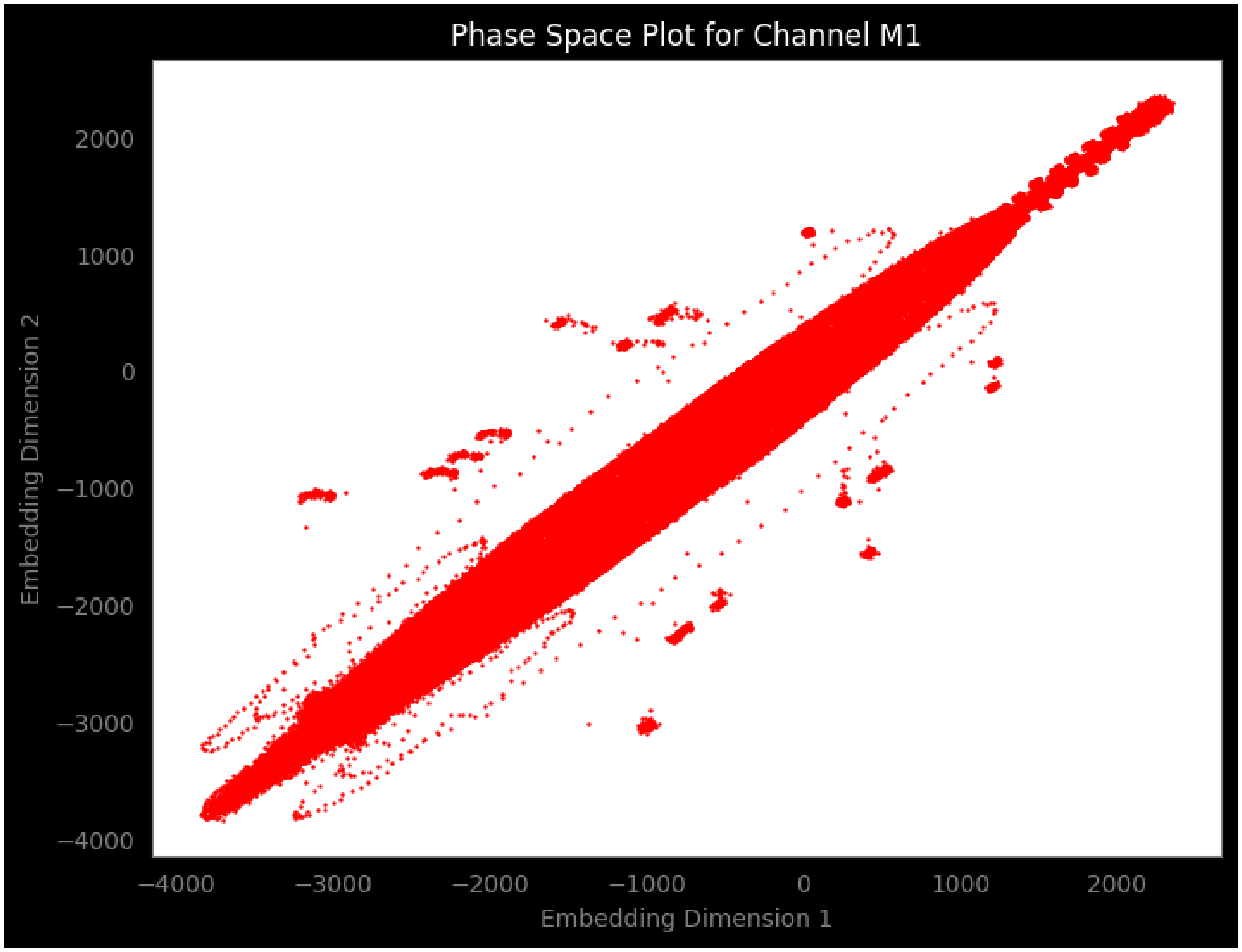

Phase Space Plot for Channel M1
Embedding Dimension 2
Embedding Dimension 1
2000
1000
0
−1000
−2000
−3000
−4000
−4000
−3000
−2000
−1000
0
1000
2000

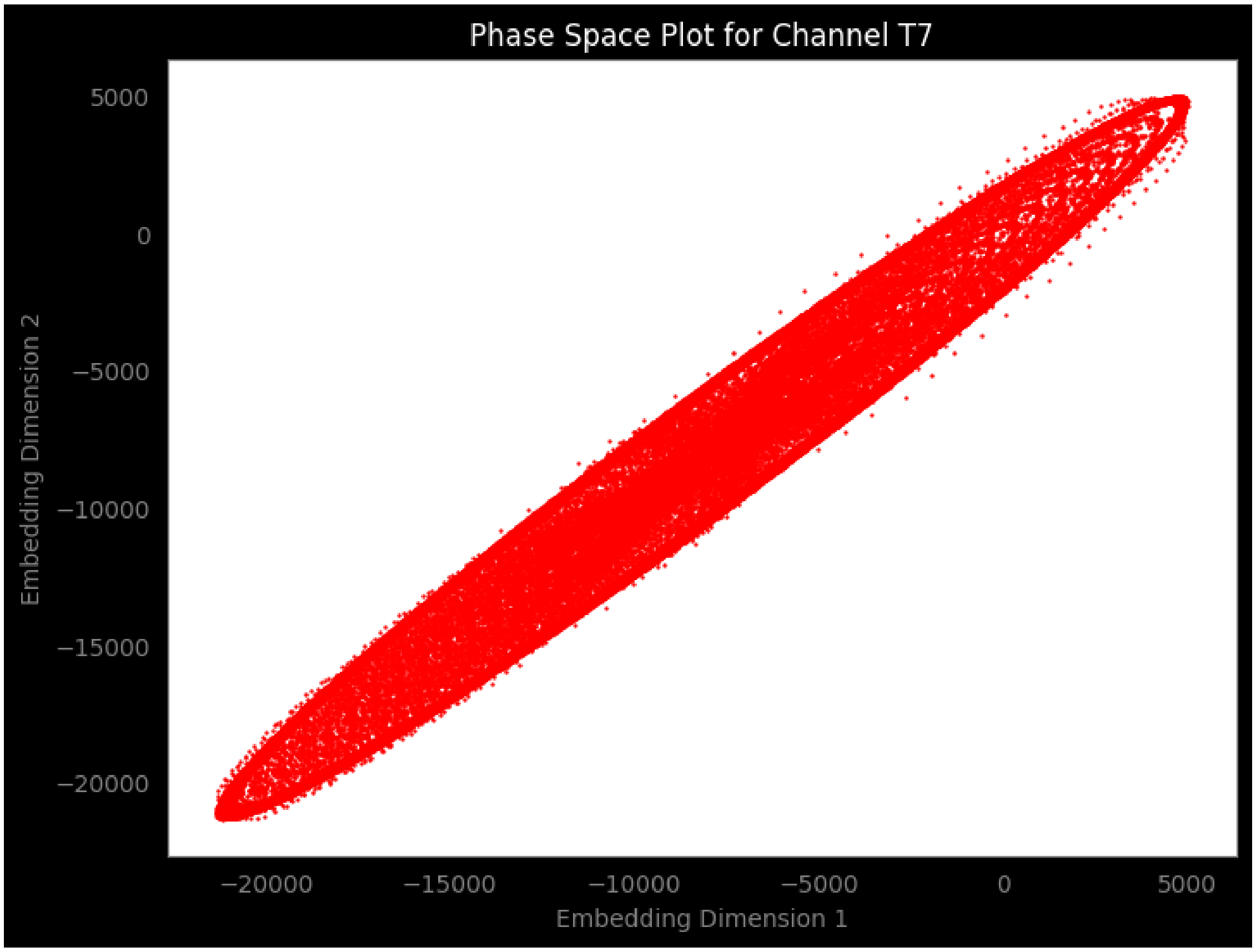
Phase Space Plot for Channel T7
5000
0
−5000
−10000
−15000
−20000
Embedding Dimension 2
−20000
−15000
−10000
−5000
0
5000
Embedding Dimension 1

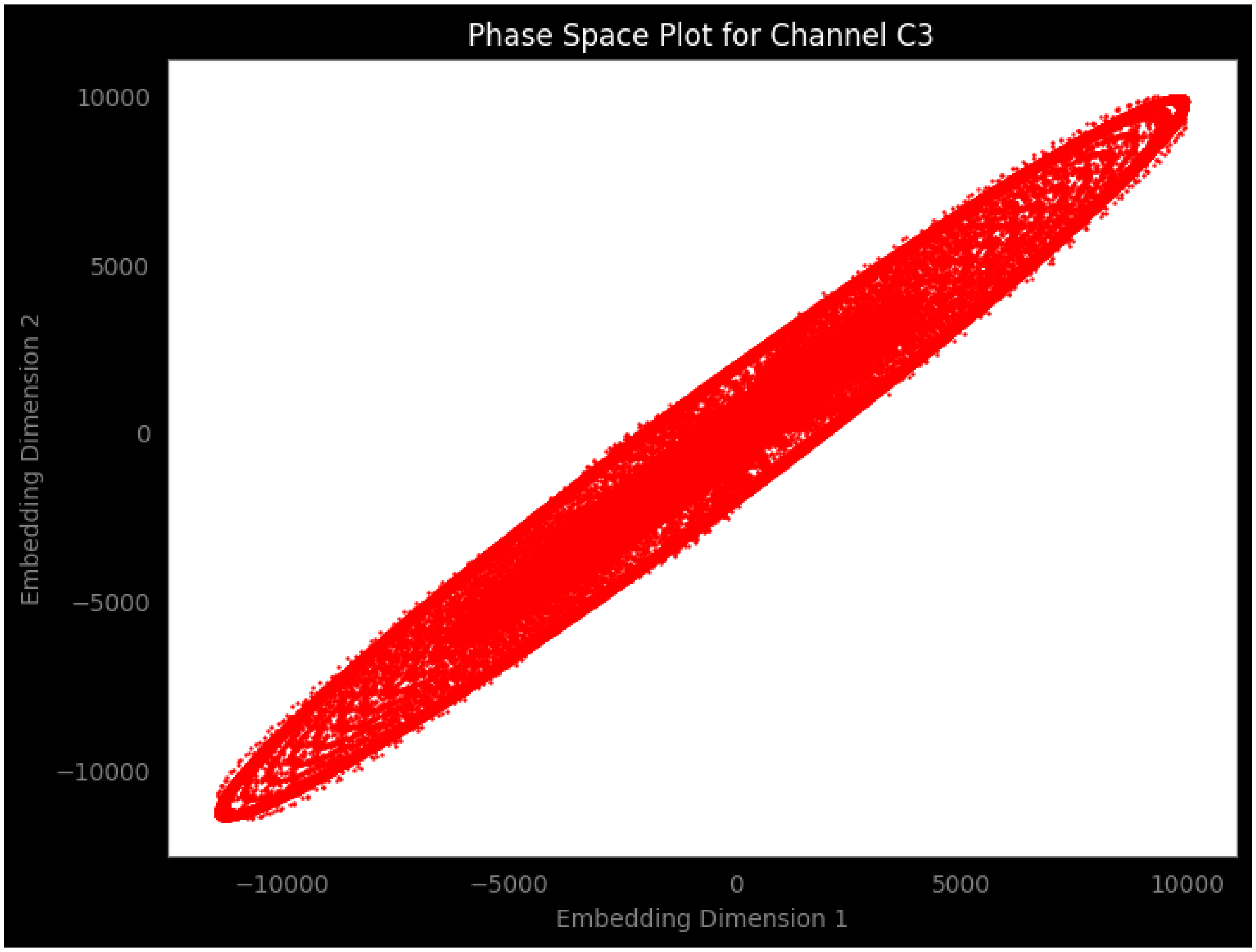
Phase Space Plot for Channel C3
10000
5000
0
−5000
−10000
Embedding Dimension 2
−10000
−5000
0
5000
10000
Embedding Dimension 1

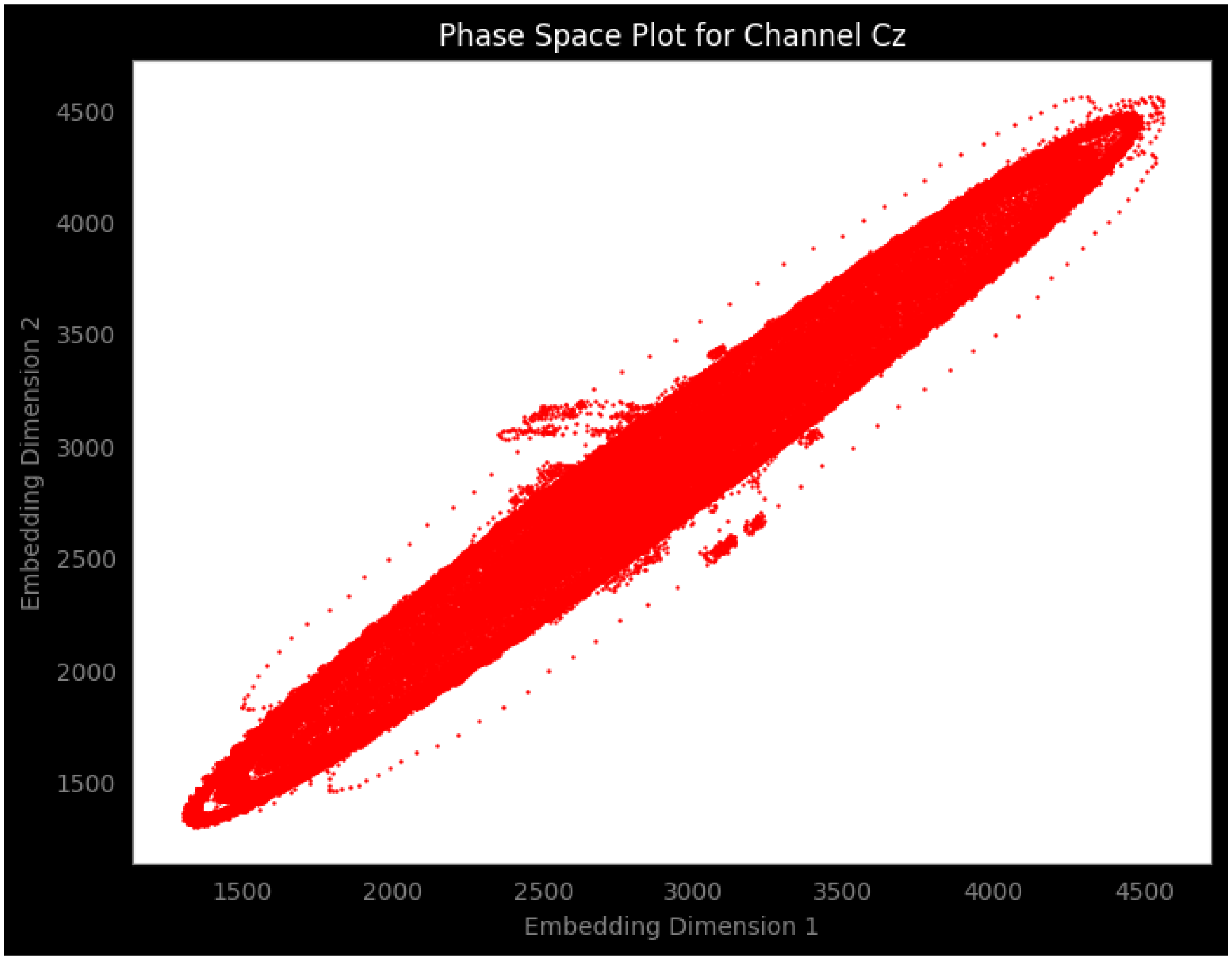
Phase Space Plot for Channel Cz
Embedding Dimension 2
Embedding Dimension 1
4500
4000
3500
3000
2500
2000
1500
1500
2000
2500
3000
3500
4000
4500

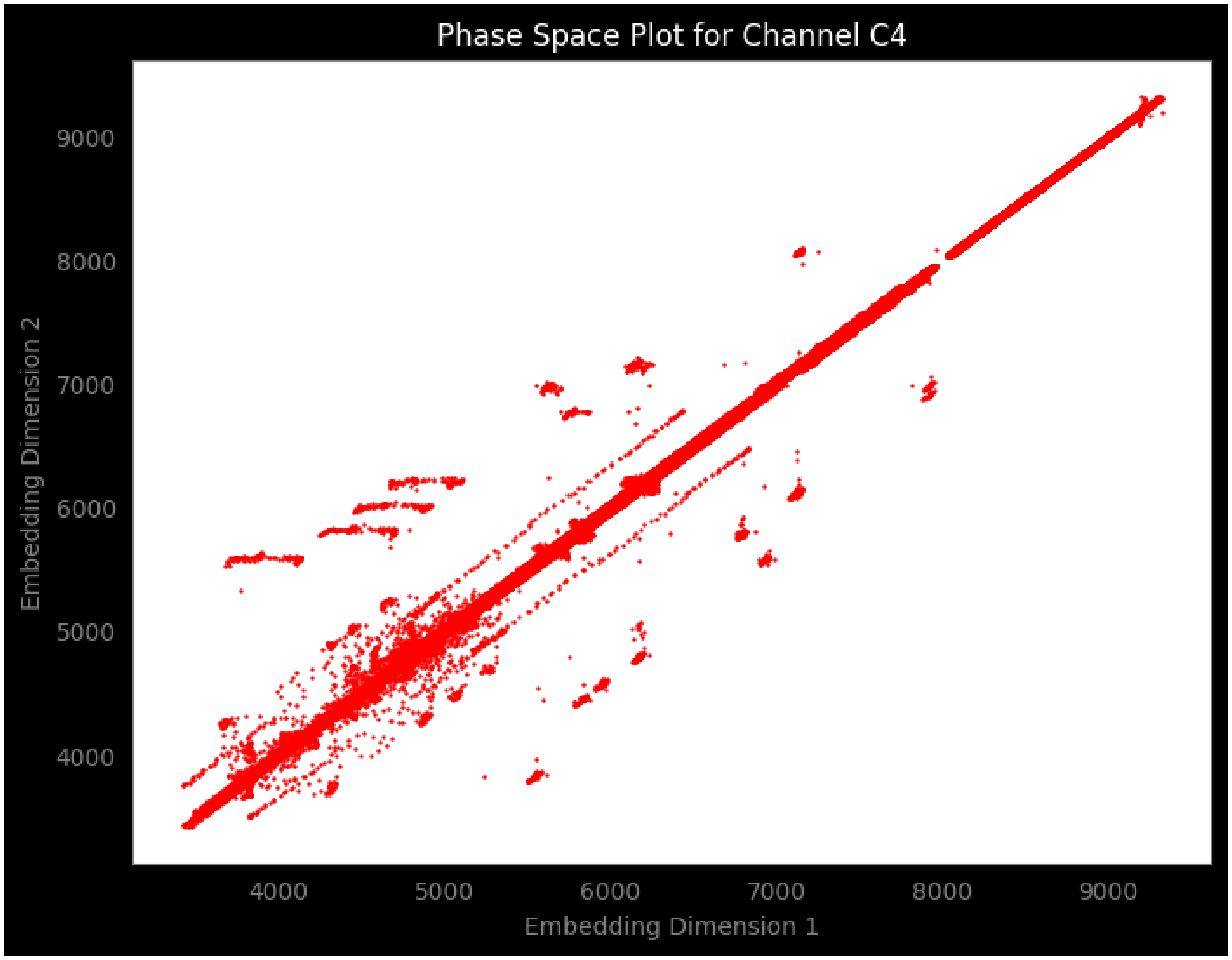
Phase Space Plot for Channel C4
Embedding Dimension 2
Embedding Dimension 1
9000
8000
7000
6000
5000
4000
4000
5000
6000
7000
8000
9000

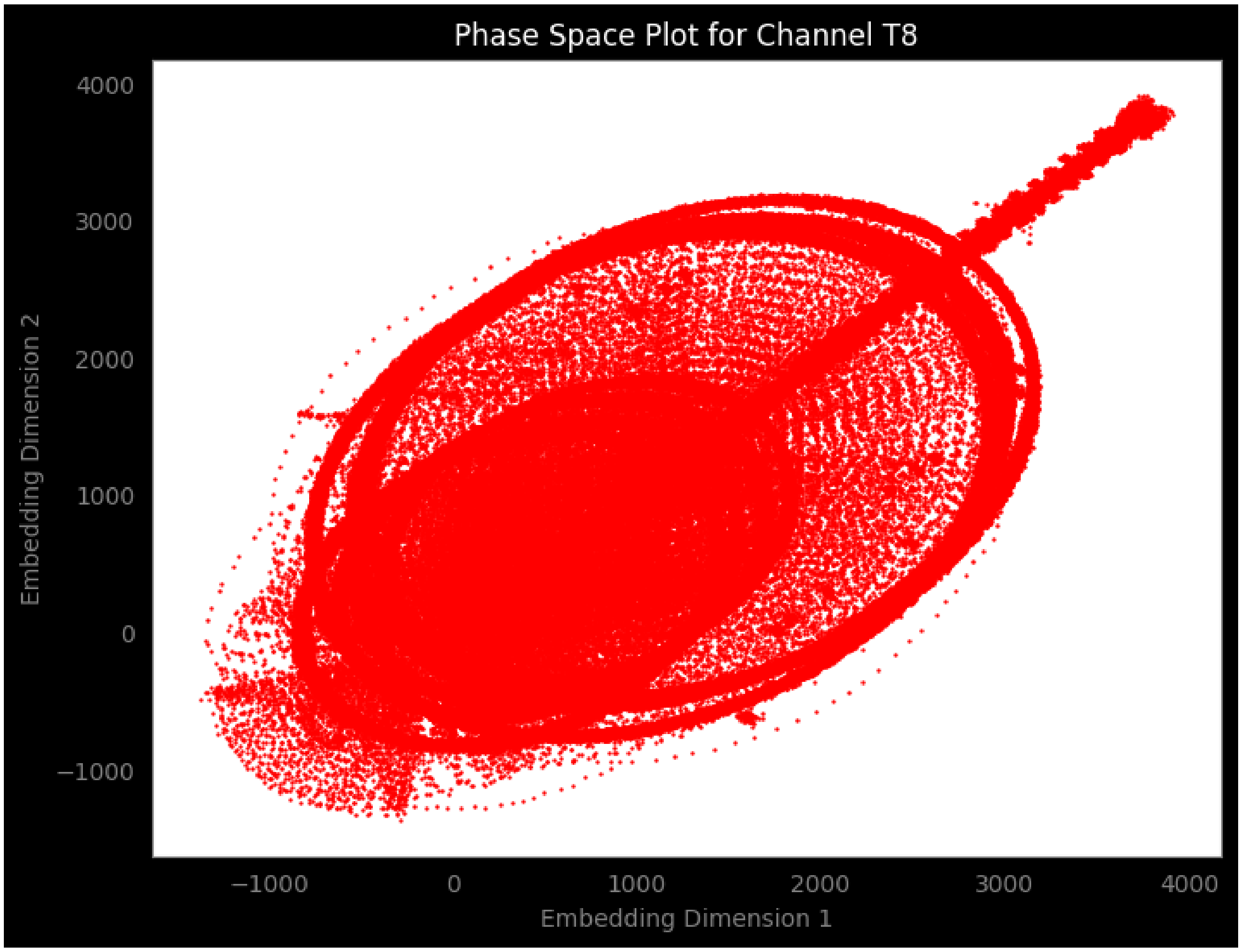
Phase Space Plot for Channel T8
4000
3000
2000
1000
0
−1000
−1000
0
1000
2000
3000
4000
Embedding Dimension 1
Embedding Dimension 2

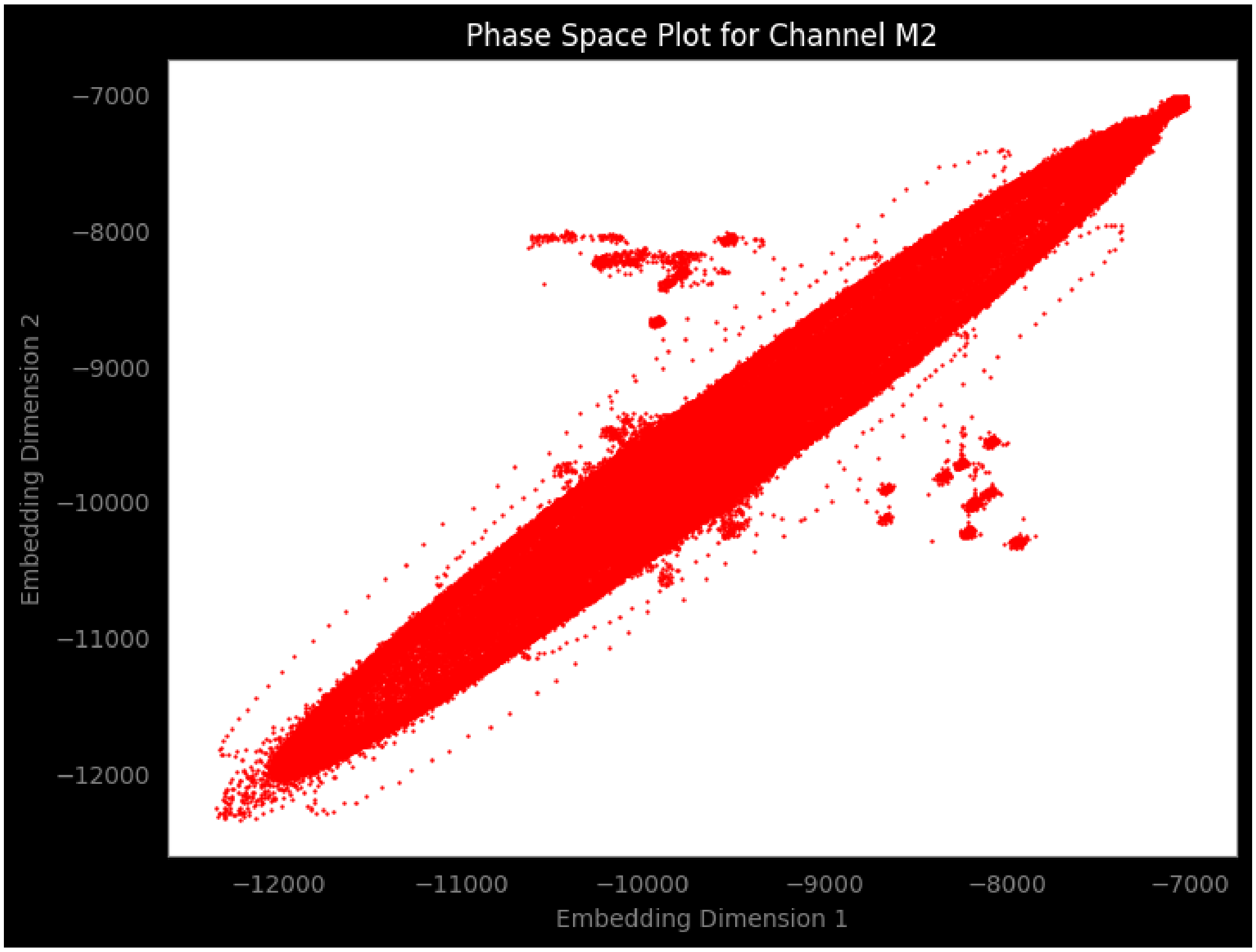
Phase Space Plot for Channel M2
−7000
−8000
−9000
−10000
−11000
−12000
Embedding Dimension 2
−12000
−11000
−10000
−9000
−8000
−7000
Embedding Dimension 1

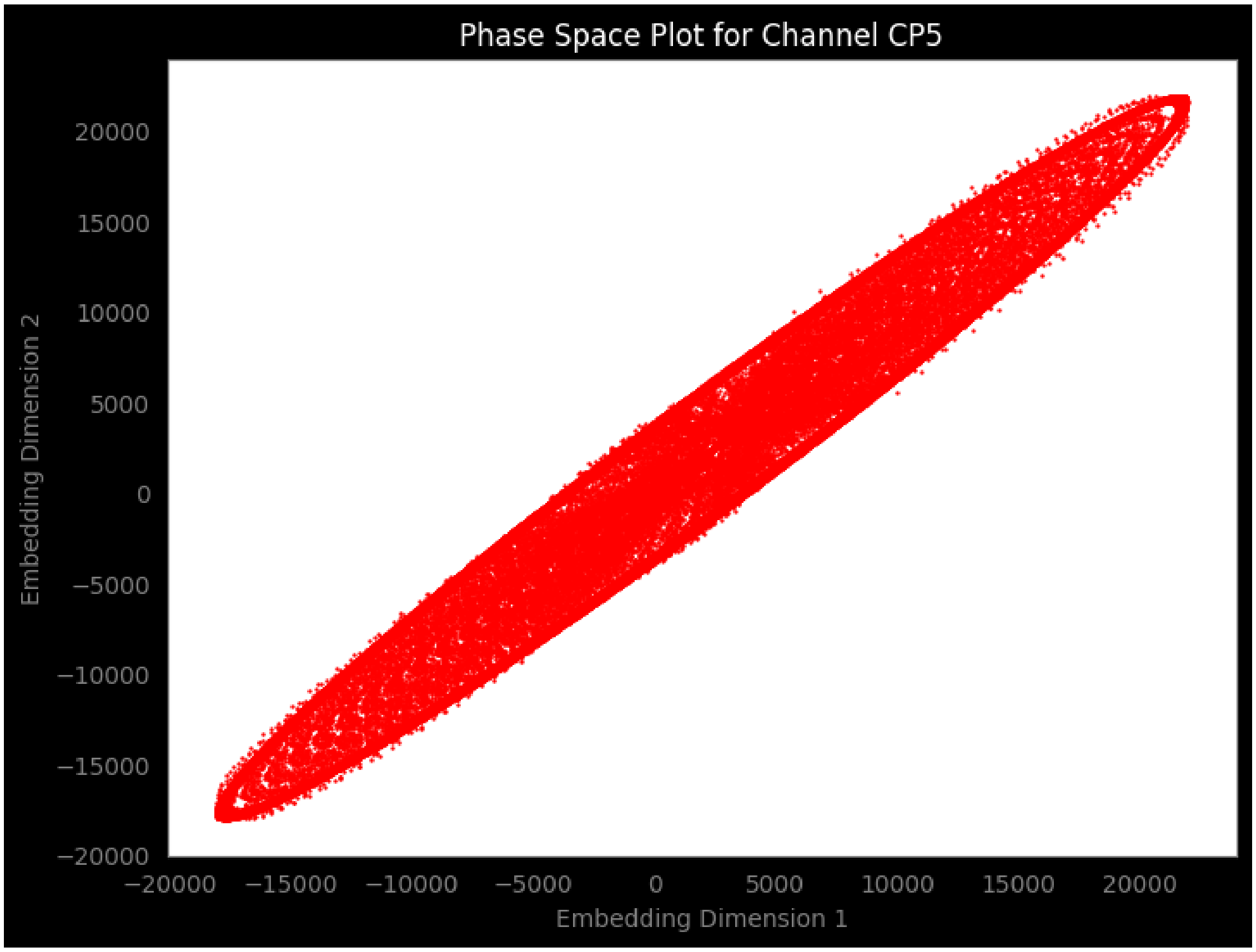

Phase Space Plot for Channel CP5
Embedding Dimension 2
20000
15000
10000
5000
0
−5000
−10000
−15000
−20000
−20000
−15000
−10000
−5000
0
5000
10000
15000
20000
Embedding Dimension 1

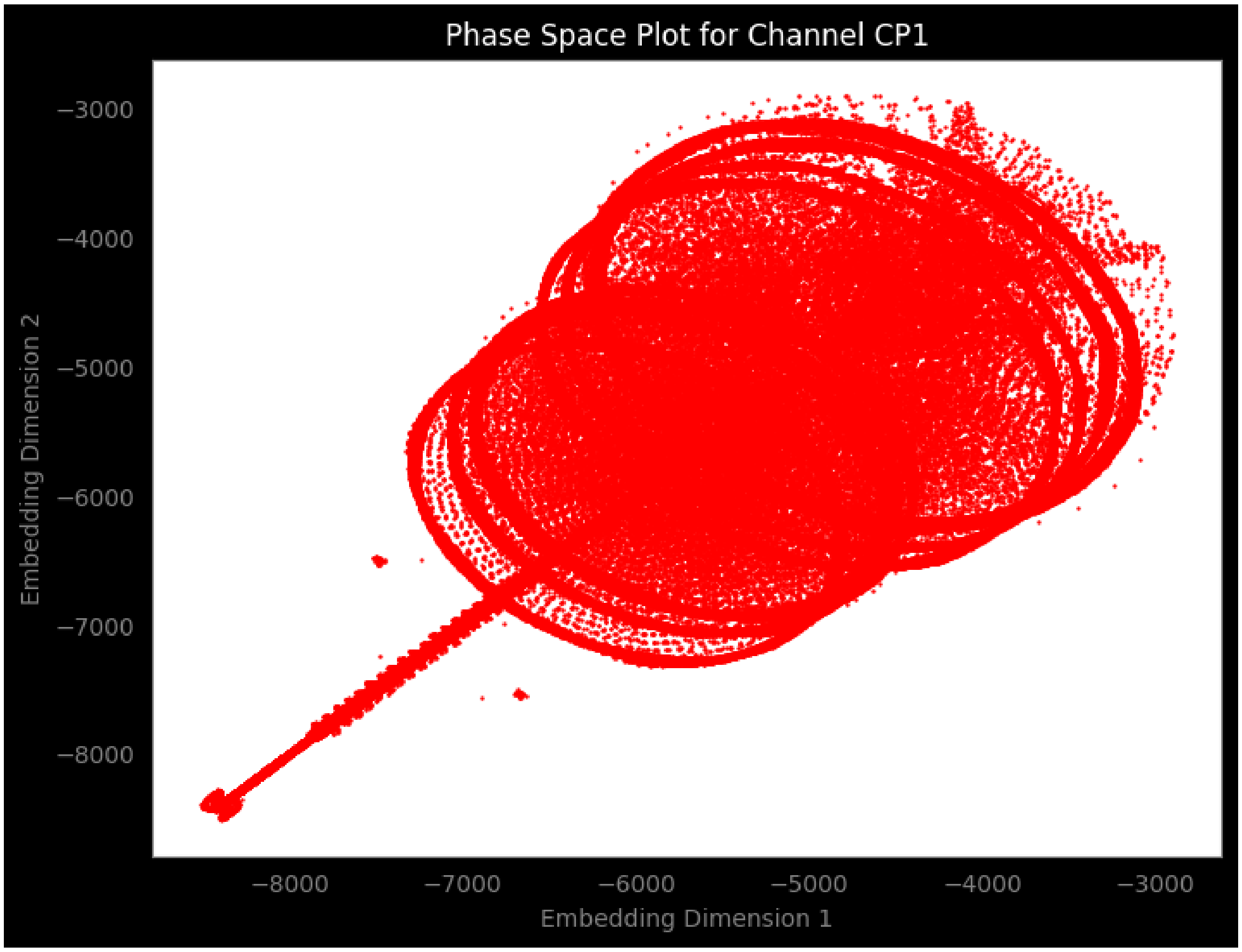
Phase Space Plot for Channel CP1
−3000
−4000
−5000
−6000
−7000
−8000
Embedding Dimension 2
−8000
−7000
−6000
−5000
−4000
−3000
Embedding Dimension 1

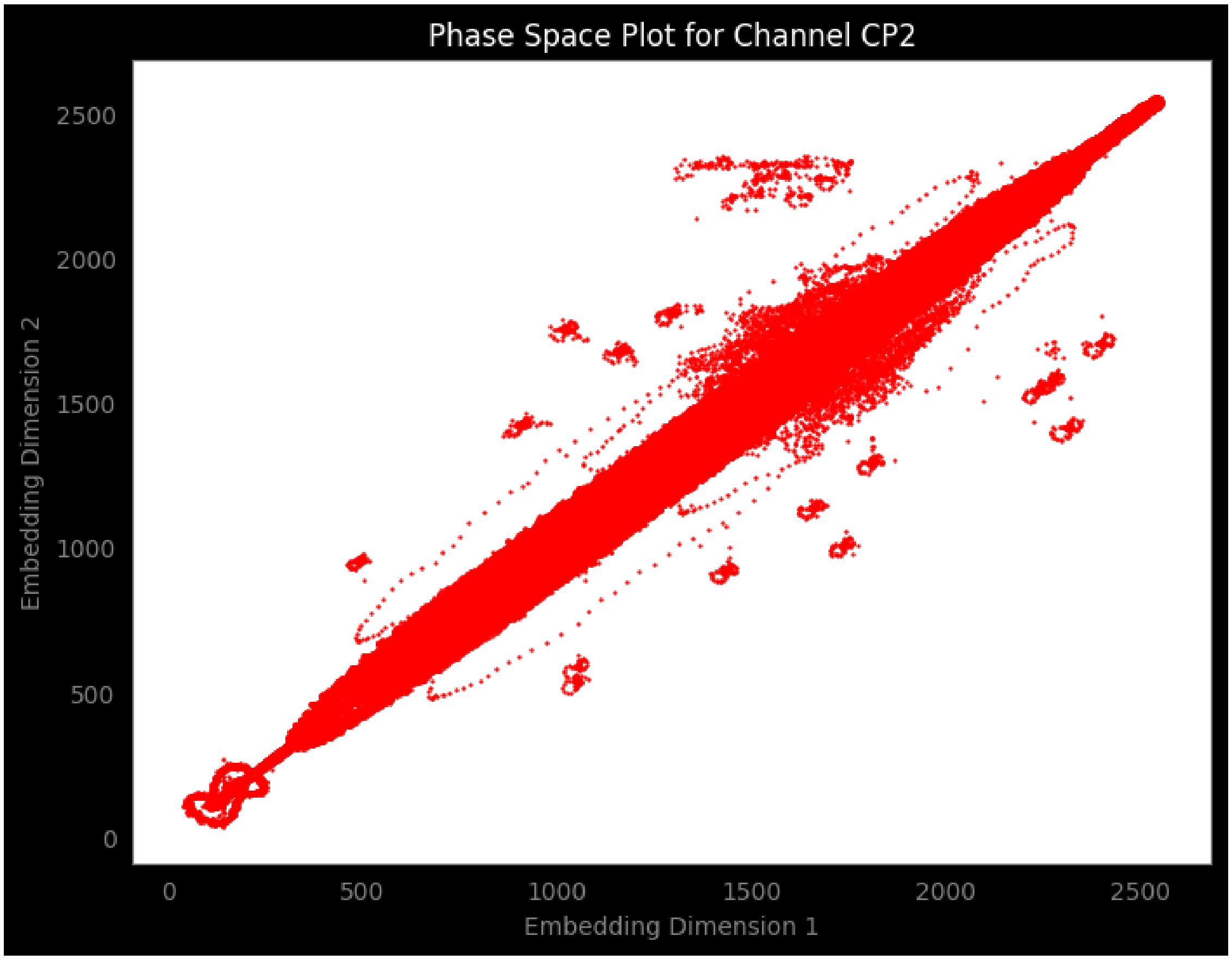
Phase Space Plot for Channel CP2
Embedding Dimension 2
Embedding Dimension 1
0
500
1000
1500
2000
2500
0
500
1000
1500
2000
2500

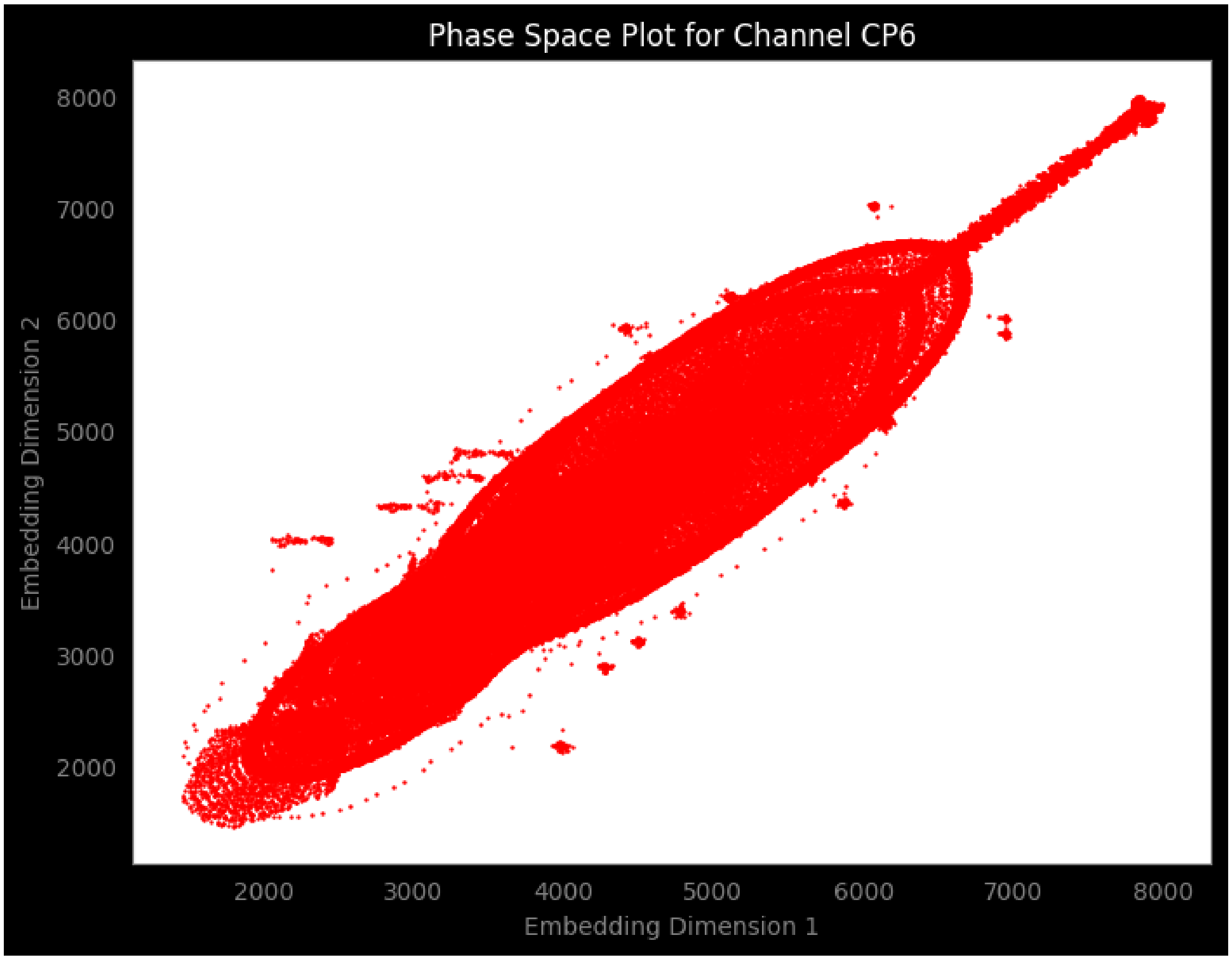

Phase Space Plot for Channel CP6
Embedding Dimension 2
8000
7000
6000
5000
4000
3000
2000
2000
3000
4000
5000
6000
7000
8000
Embedding Dimension 1

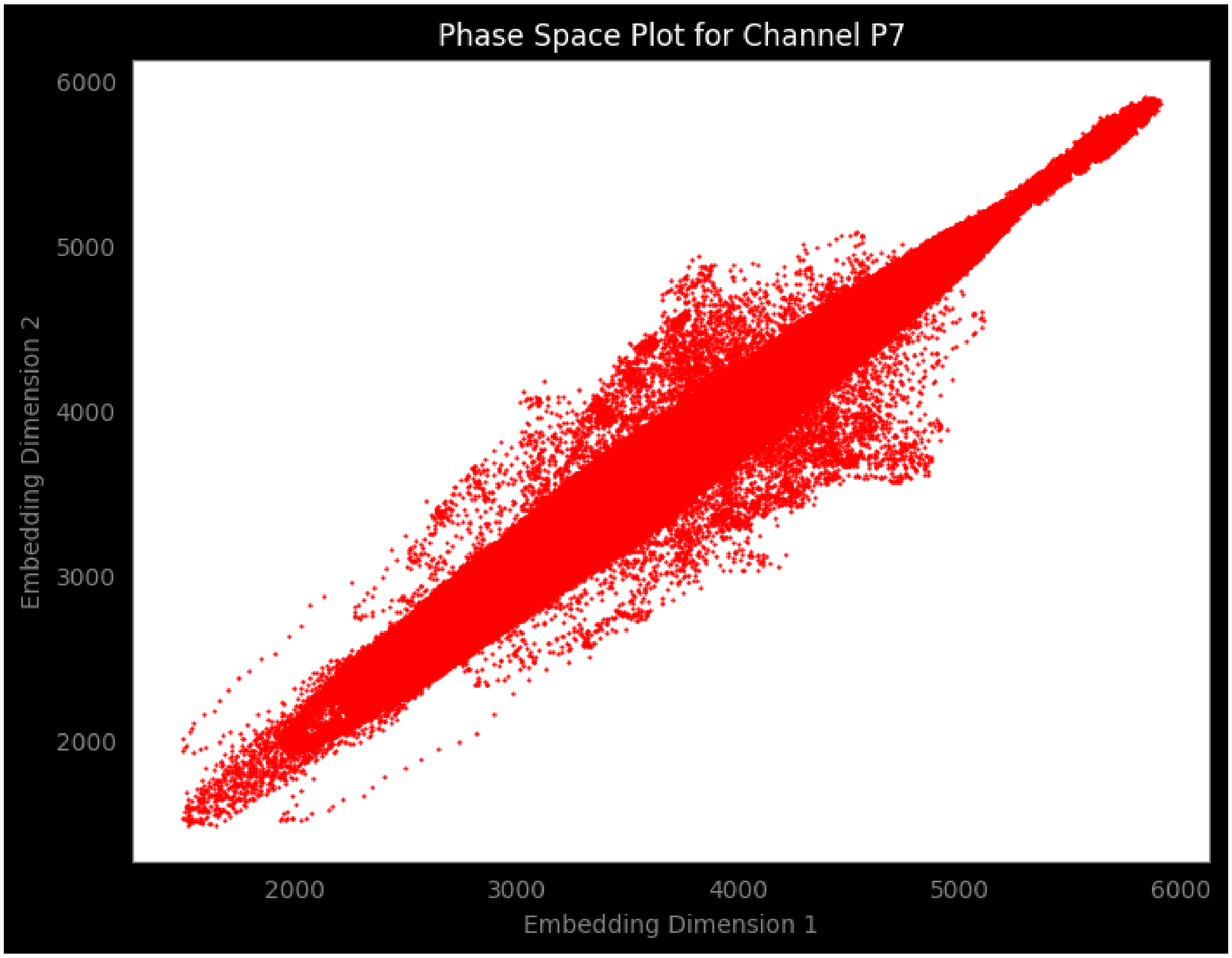
Phase Space Plot for Channel P7
6000
5000
4000
3000
2000
Embedding Dimension 2
2000
3000
4000
5000
6000
Embedding Dimension 1

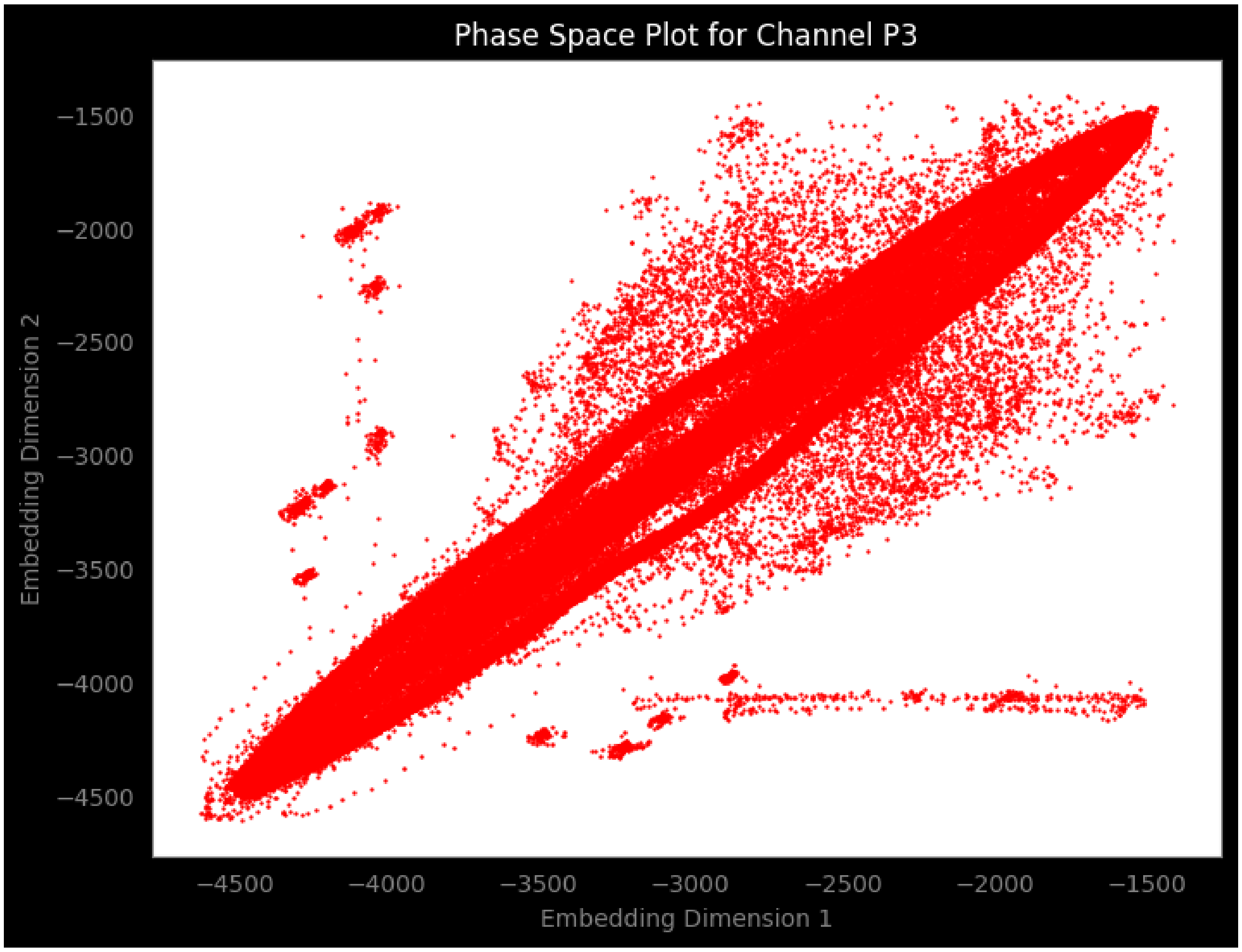
Phase Space Plot for Channel P3
Embedding Dimension 2
Embedding Dimension 1
−1500
−2000
−2500
−3000
−3500
−4000
−4500
−4500
−4000
−3500
−3000
−2500
−2000
−1500

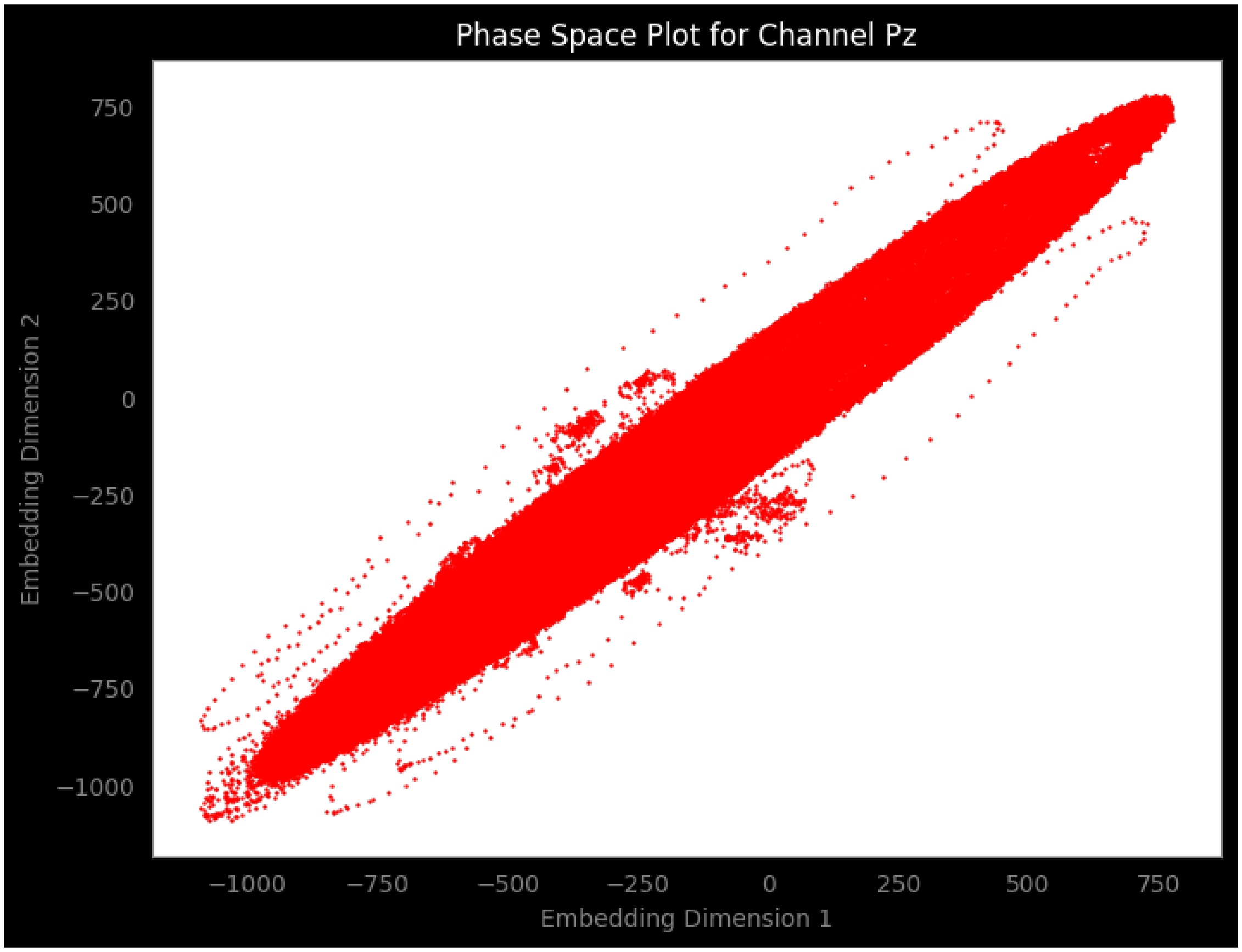
Phase Space Plot for Channel Pz
Embedding Dimension 2
Embedding Dimension 1
750
500
250
0
−250
−500
−750
−1000
−1000
−750
−500
−250
0
250
500
750

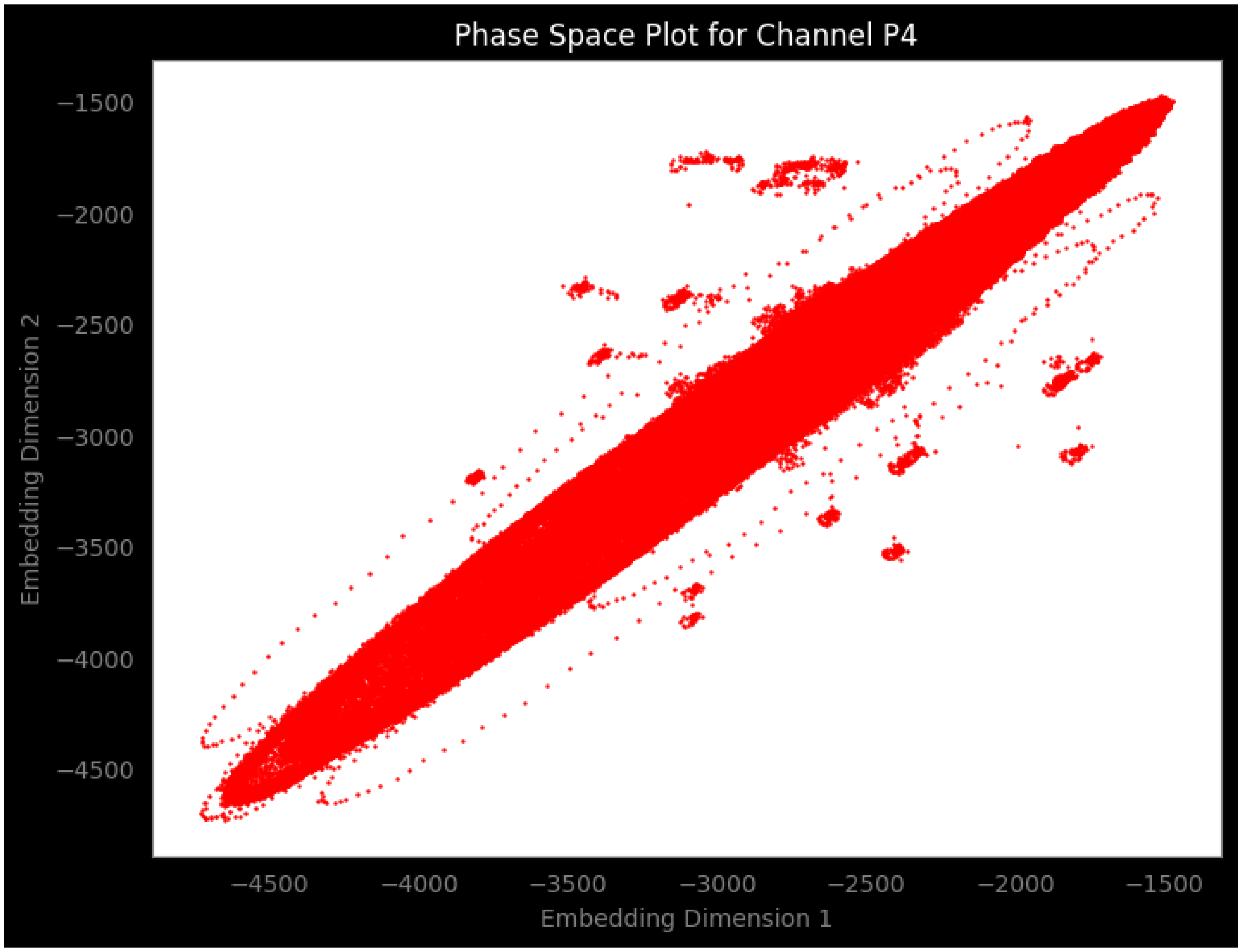
Phase Space Plot for Channel P4
−1500
−2000
−2500
−3000
−3500
−4000
−4500
Embedding Dimension 2
−4500
−4000
−3500
−3000
−2500
−2000
−1500
Embedding Dimension 1

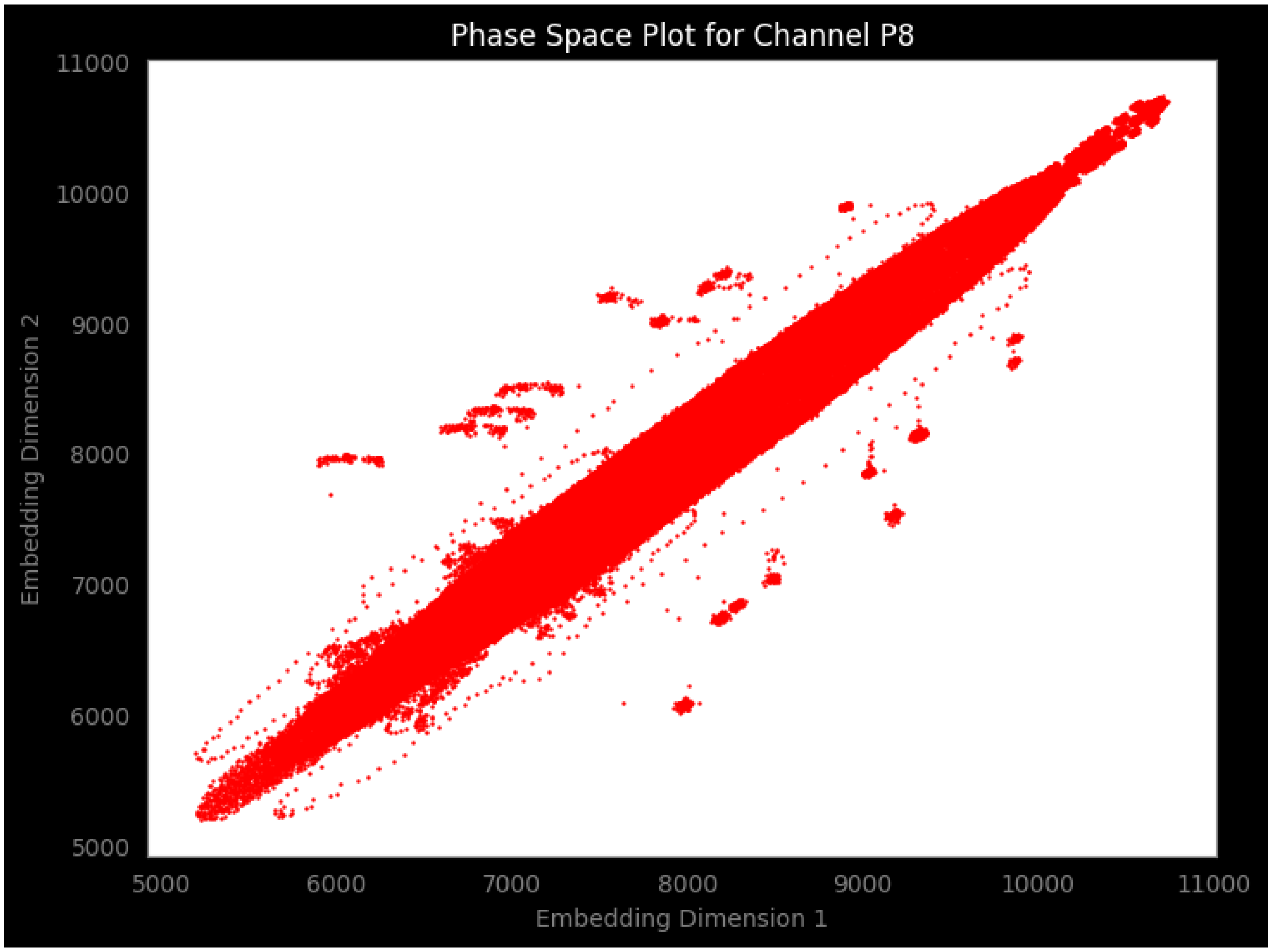
Phase Space Plot for Channel P8
Embedding Dimension 2
11000
10000
9000
8000
7000
6000
5000
5000
6000
7000
8000
9000
10000
11000
Embedding Dimension 1

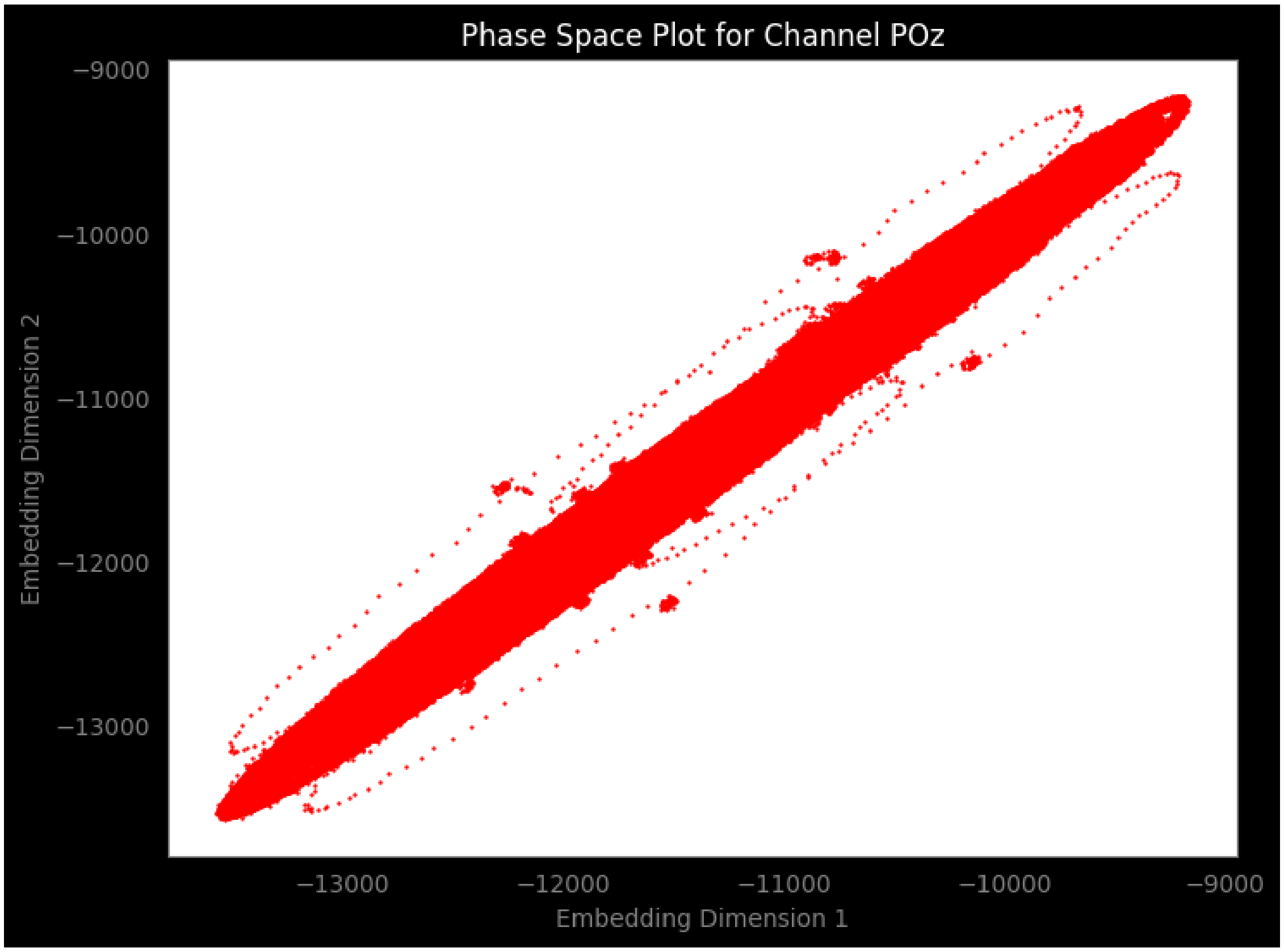
Phase Space Plot for Channel POz
−9000
−10000
−11000
−12000
−13000
Embedding Dimension 2
−13000
−12000
−11000
−10000
−9000
Embedding Dimension 1

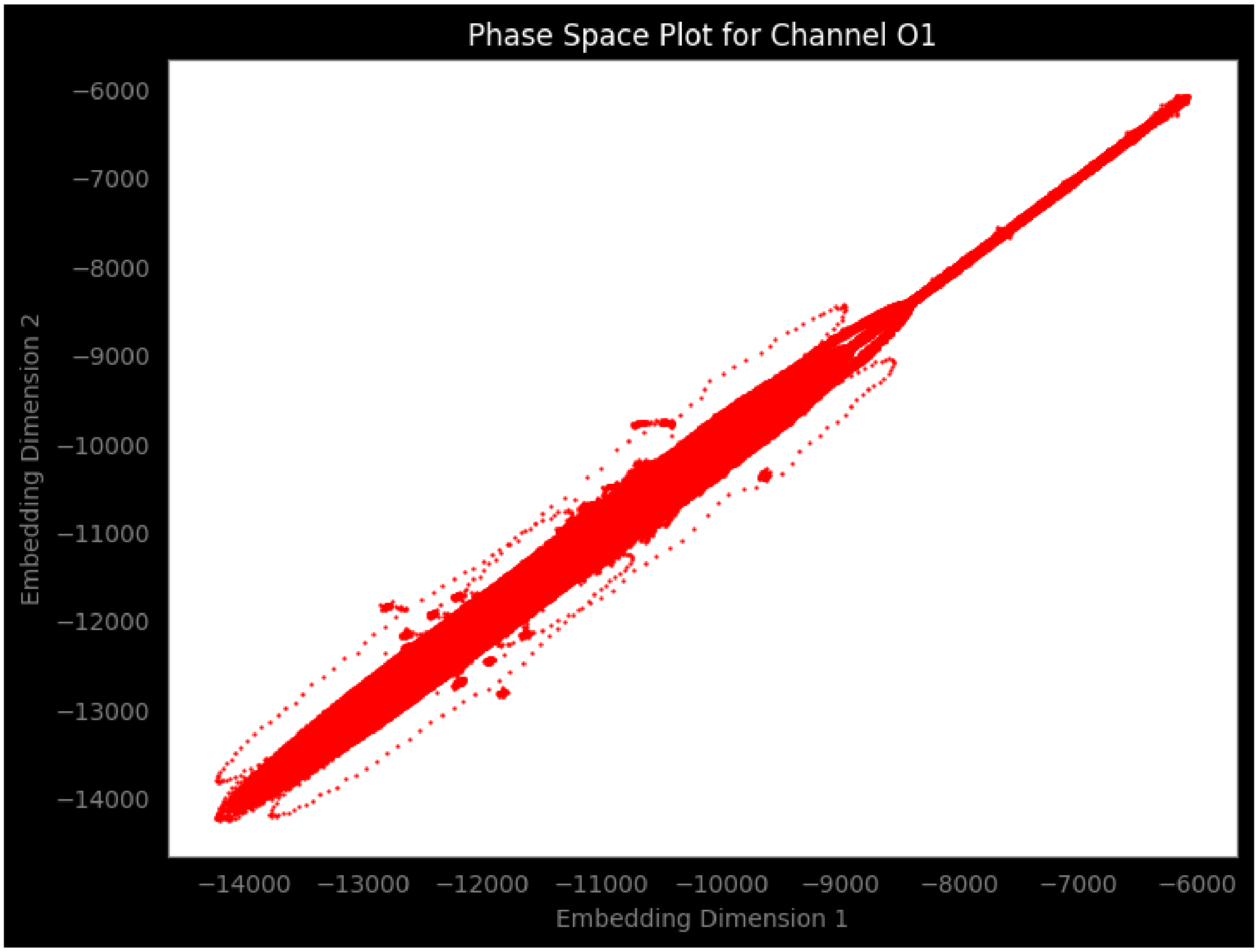
Phase Space Plot for Channel O1
Embedding Dimension 2
Embedding Dimension 1
−6000
−7000
−8000
−9000
−10000
−11000
−12000
−13000
−14000
−14000
−13000
−12000
−11000
−10000
−9000
−8000
−7000
−6000

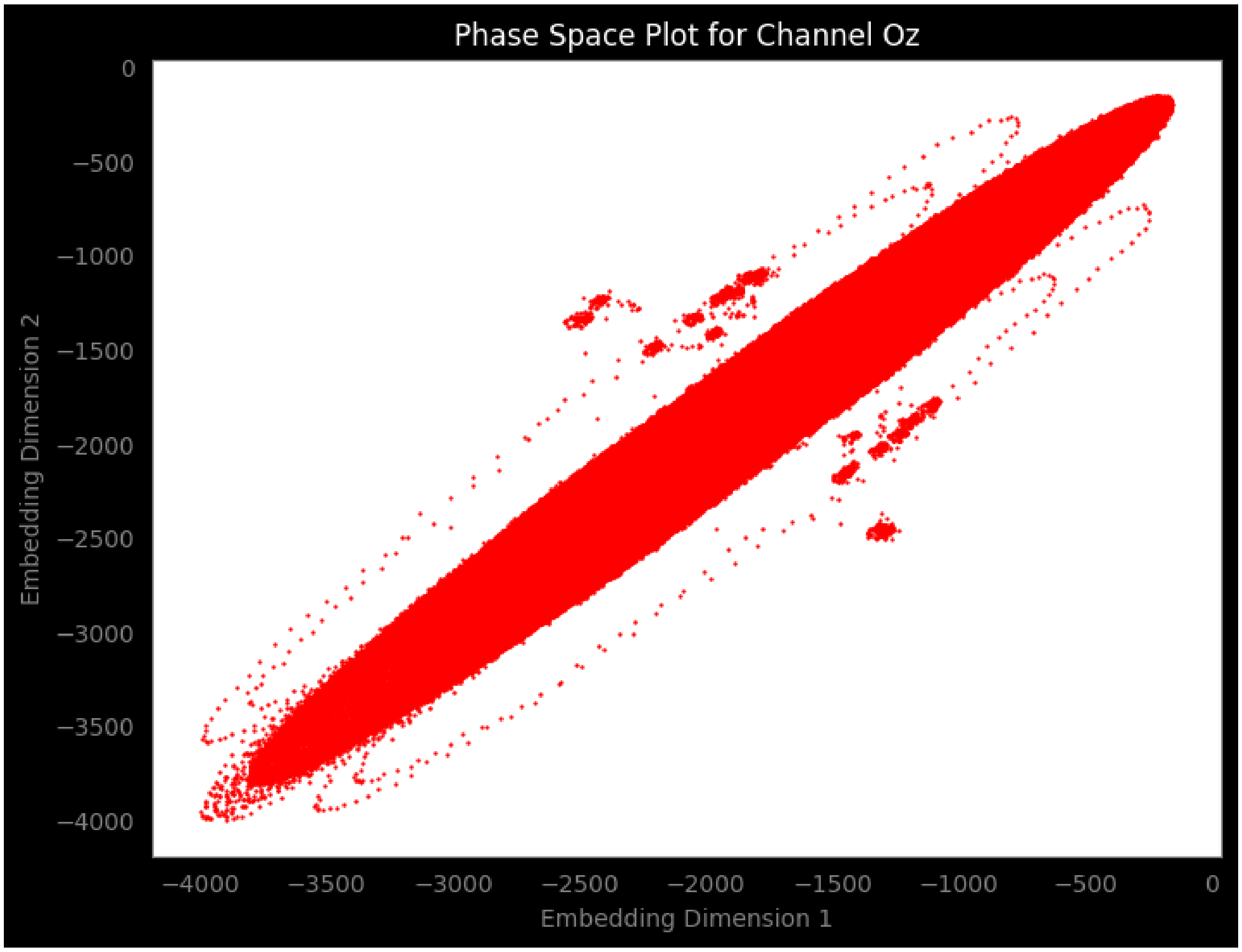
Phase Space Plot for Channel Oz
Embedding Dimension 2
Embedding Dimension 1
0
−500
−1000
−1500
−2000
−2500
−3000
−3500
−4000
−4000
−3500
−3000
−2500
−2000
−1500
−1000
−500
0

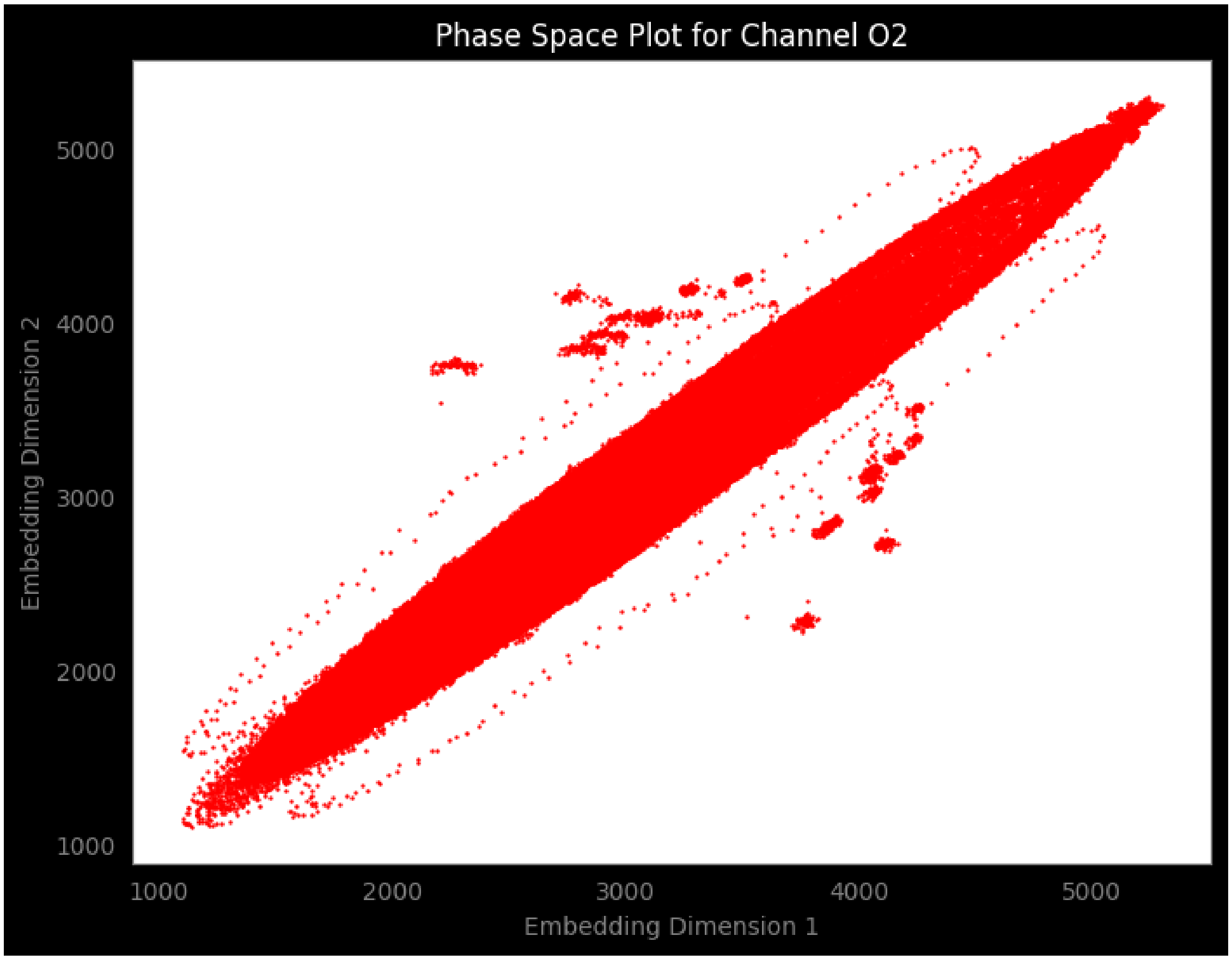

# 4 Create a zipfile for the results

[5]:
```
import zipfile
import os

# Directory containing the embedded data files
embedded_data_dir = '/home/vincent/AAA_projects/MVCS/Neuroscience/Analysis/
↪Phase Space/2dembedding_data'

# List of file paths to the embedded data files
embedded_data_files = [os.path.join(embedded_data_dir,␣
↪f'2dembedded_{channel}_data.npy') for channel in eeg_channels]

zip_file_path = '/home/vincent/AAA_projects/MVCS/Neuroscience/Analysis/Phase␣
↪Space/2dembedded_data.zip'

# Create a zipped file containing all embedded data files
with zipfile.ZipFile(zip_file_path, 'w') as zipf:
    for data_file in embedded_data_files:
```

```
        zipf.write(data_file, os.path.basename(data_file))  # Add the file to␣
↪the zip with its original name
```

## 5 3D Phase Space Plot

[6]:
```
import numpy as np
import matplotlib.pyplot as plt
from minepy import MINE
import multiprocessing
import zipfile
import os

# Load EEG data
EEG_data = np.load('/home/vincent/AAA_projects/MVCS/Neuroscience/
↪eeg_data_with_channels.npy', allow_pickle=True)

# Extract EEG channel names
eeg_channels = ['Fp1', 'Fpz', 'Fp2', 'F7', 'F3', 'Fz', 'F4', 'F8', 'FC5',␣
↪'FC1', 'FC2', 'FC6',
                'M1', 'T7', 'C3', 'Cz', 'C4', 'T8', 'M2', 'CP5', 'CP1',␣
↪'CP2', 'CP6',
                'P7', 'P3', 'Pz', 'P4', 'P8', 'POz', 'O1', 'Oz', 'O2']

def mutual_info_worker(args):
    data1, data2 = args
    mine = MINE()
    mine.compute_score(data1, data2)
    return mine.mic()

def determine_delay(data, max_delay=100, subsample_factor=10):
    subsampled_data = data[::subsample_factor]
    with multiprocessing.Pool() as pool:
        args_list = [(subsampled_data[:-i], subsampled_data[i:]) for i in␣
↪range(1, max_delay+1)]
        mi_values = pool.map(mutual_info_worker, args_list)
    min_index = np.argmin(mi_values)
    return min_index + 1

def delay_embedding(data, emb_dim, delay):
    N = len(data)
    embedded_data = np.zeros((N - (emb_dim - 1) * delay, emb_dim))
    for i in range(N - (emb_dim - 1) * delay):
        embedded_data[i] = [data[i + j * delay] for j in range(emb_dim)]
    return embedded_data

embedded_data_dict = {}
```

```
# Load EEG data
EEG_data = np.load('/home/vincent/AAA_projects/MVCS/Neuroscience/
↪eeg_data_with_channels.npy', allow_pickle=True)

# Assuming EEG_data is of shape (num_samples, num_channels)
for channel_index, channel_name in enumerate(eeg_channels):
    channel_data = EEG_data[:, channel_index]

    # Normalize the data
    channel_data = (channel_data - np.mean(channel_data)) / np.std(channel_data)

    # Determine optimal delay using mutual information with subsampling
    optimal_delay = determine_delay(channel_data, subsample_factor=50)  #␣
↪Adjust subsample_factor as needed

    # Embedding dimension
    emb_dim = 3

    # Perform delay embedding
    embedded_channel_data = delay_embedding(channel_data, emb_dim=emb_dim,␣
↪delay=optimal_delay)
    embedded_data_dict[channel_name] = embedded_channel_data  # store the␣
↪embedded data in the dictionary

    # Create 3D scatter plot with black background
    fig = plt.figure(figsize=(10,8), facecolor='black')
    ax = fig.add_subplot(111, projection='3d', frame_on=False)
    ax.scatter(embedded_channel_data[:, 0], embedded_channel_data[:, 1],␣
↪embedded_channel_data[:, 2], color='red', s=0.2)
    ax.set_facecolor('black')
    ax.set_title(f'Phase Space Plot for Channel {channel_name}', color='white')
    ax.set_xlabel('Embedding Dimension 1', color='grey')
    ax.set_ylabel('Embedding Dimension 2', color='grey')
    ax.set_zlabel('Embedding Dimension 3', color='grey')
    ax.spines['left'].set_color('grey')
    ax.spines['right'].set_color('grey')
    ax.spines['bottom'].set_color('grey')
    ax.spines['top'].set_color('grey')
    ax.xaxis.label.set_color('grey')
    ax.yaxis.label.set_color('grey')
    ax.zaxis.label.set_color('grey')
    ax.tick_params(axis='both', colors='grey')
    plt.show()

# Path to save the numpy files before zipping
```

```python
temp_save_path = '/home/vincent/AAA_projects/MVCS/Neuroscience/Analysis/Phase␣
↪Space/3dembedding_data/temp'

# Check if the temp directory exists. If not, create it
if not os.path.exists(temp_save_path):
    os.makedirs(temp_save_path)

# Save the embedded data for each channel as separate numpy files
for channel_name, data in embedded_data_dict.items():
    file_path = os.path.join(temp_save_path, f'3dembedded_{channel_name}.npy')
    np.save(file_path, data)

# Create a zipped file containing all embedded data files
with zipfile.ZipFile('/home/vincent/AAA_projects/MVCS/Neuroscience/Analysis/
↪Phase Space/3dembedded_data.zip', 'w') as zipf:
    for channel_name in eeg_channels:
        data_file_name = f'3dembedded_{channel_name}.npy'
        file_path = os.path.join(temp_save_path, data_file_name)
        zipf.write(file_path, arcname=data_file_name)
```

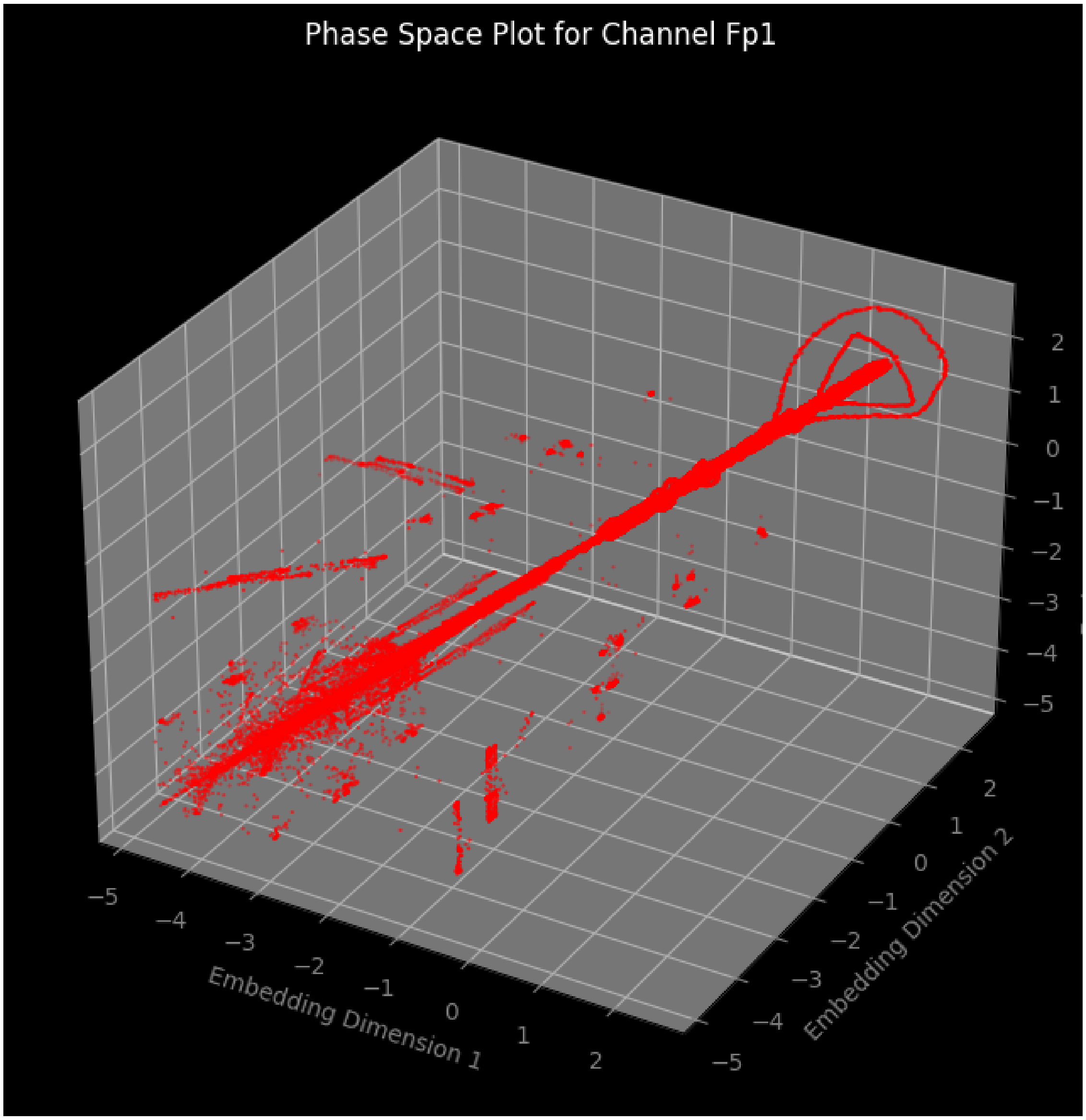

Phase Space Plot for Channel Fp1
Embedding Dimension 1
Embedding Dimension 2
−5
−4
−3
−2
−1
0
1
2

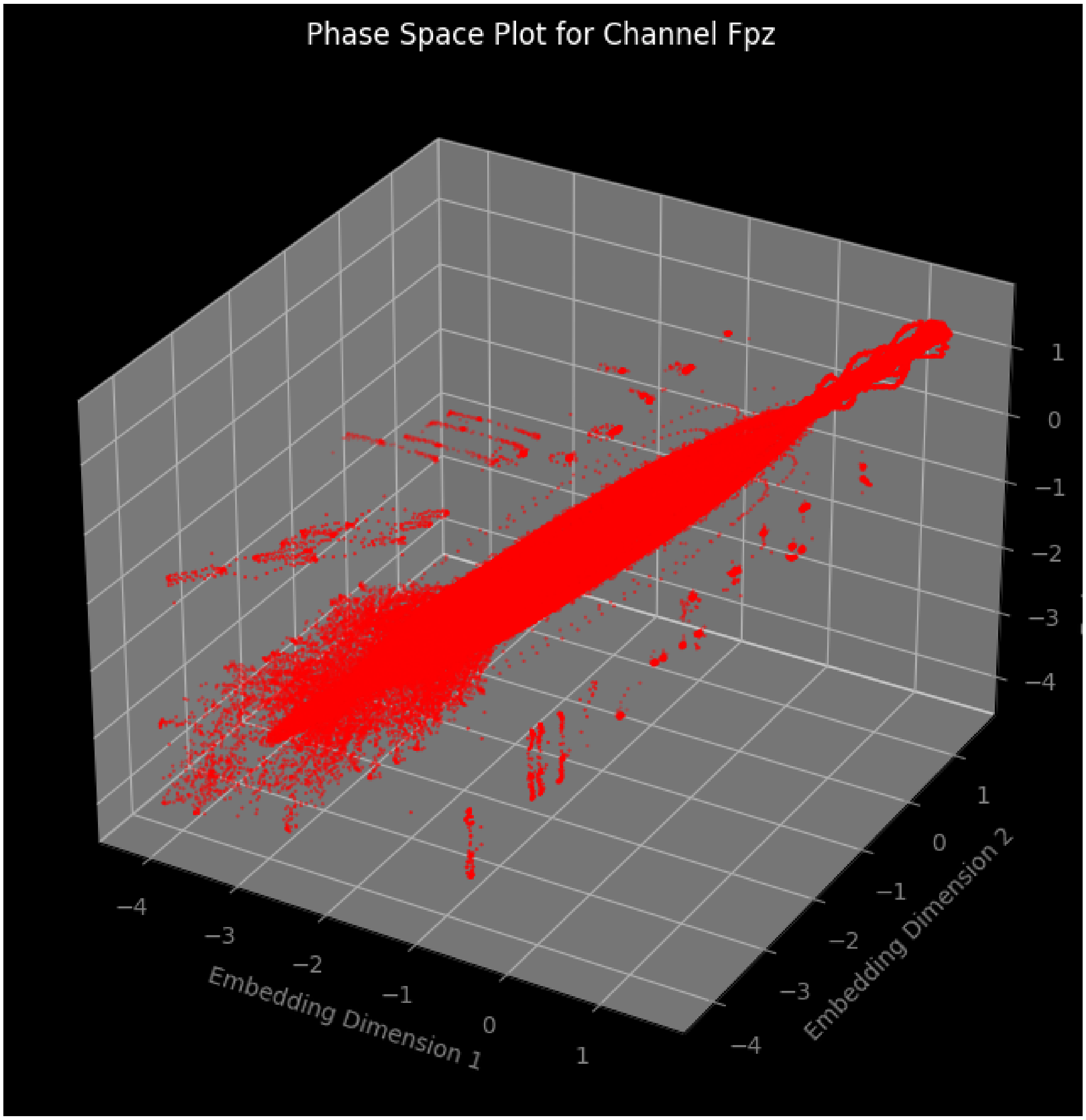
Phase Space Plot for Channel Fpz
Embedding Dimension 1
Embedding Dimension 2
−4
−3
−2
−1
0
1

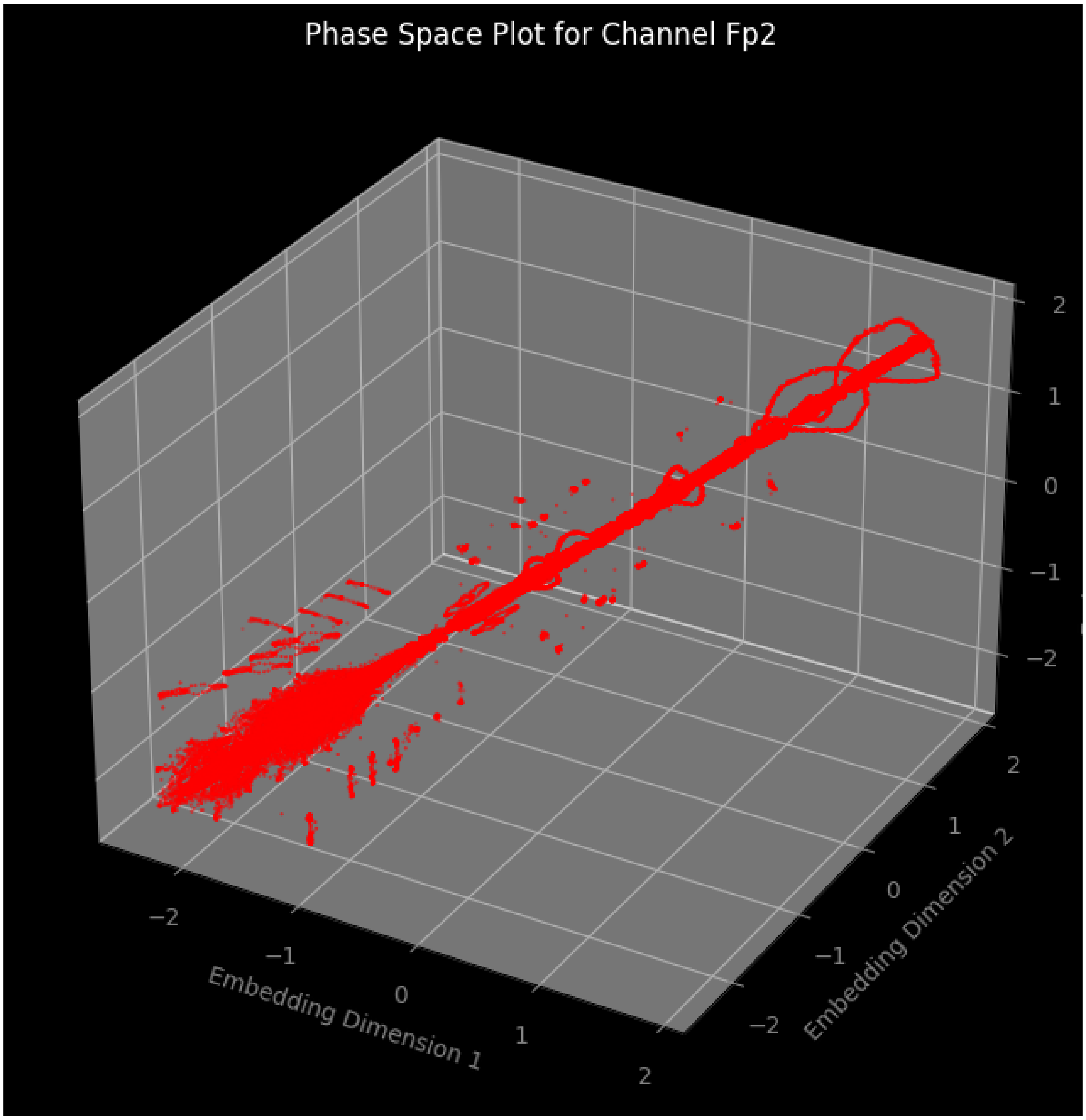
Phase Space Plot for Channel Fp2
Embedding Dimension 1
Embedding Dimension 2
−2
−1
0
1
2
−2
−1
0
1
2
−2
−1
0
1
2

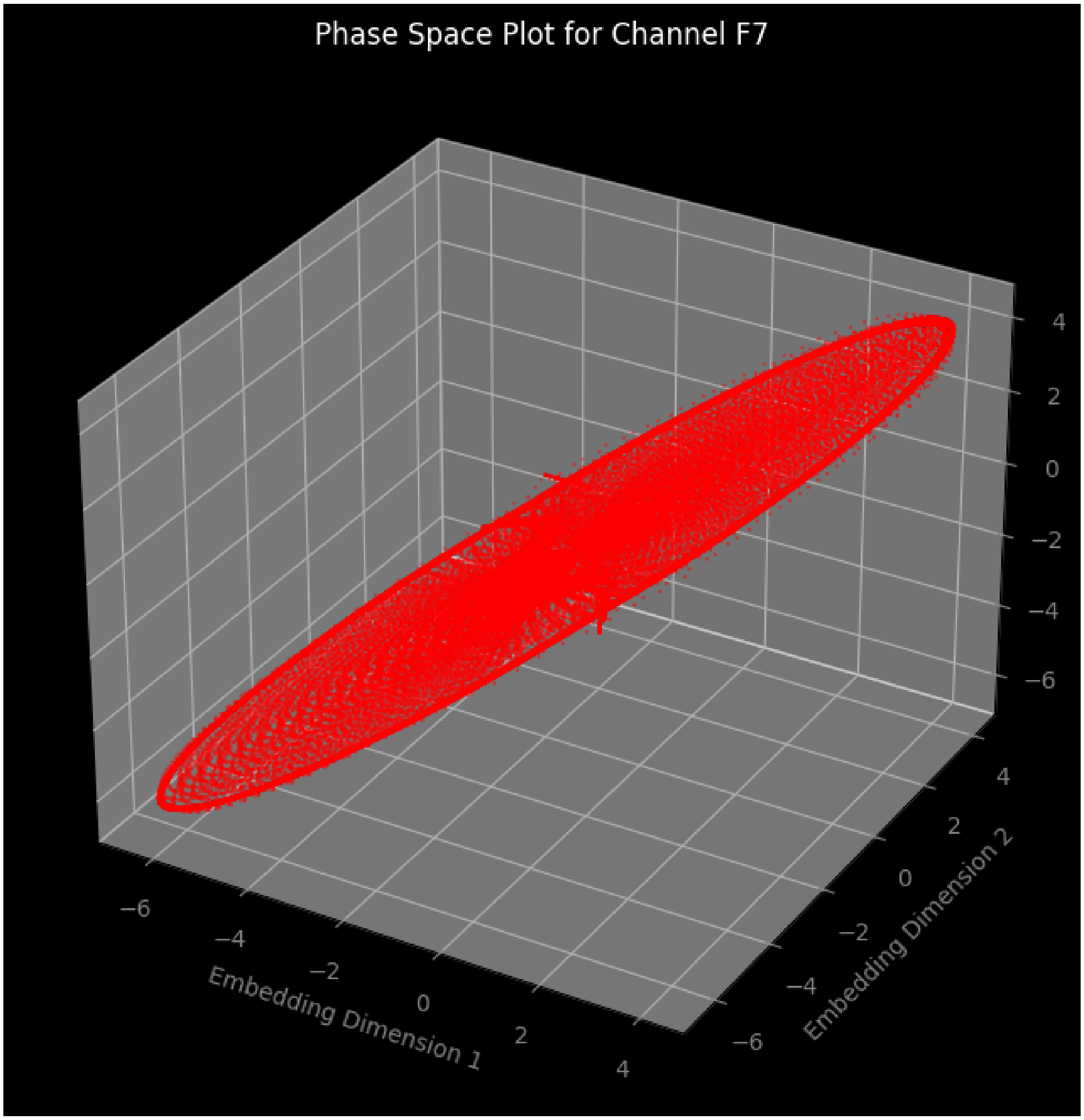
Phase Space Plot for Channel F7
Embedding Dimension 1
Embedding Dimension 2
−6
−4
−2
0
2
4

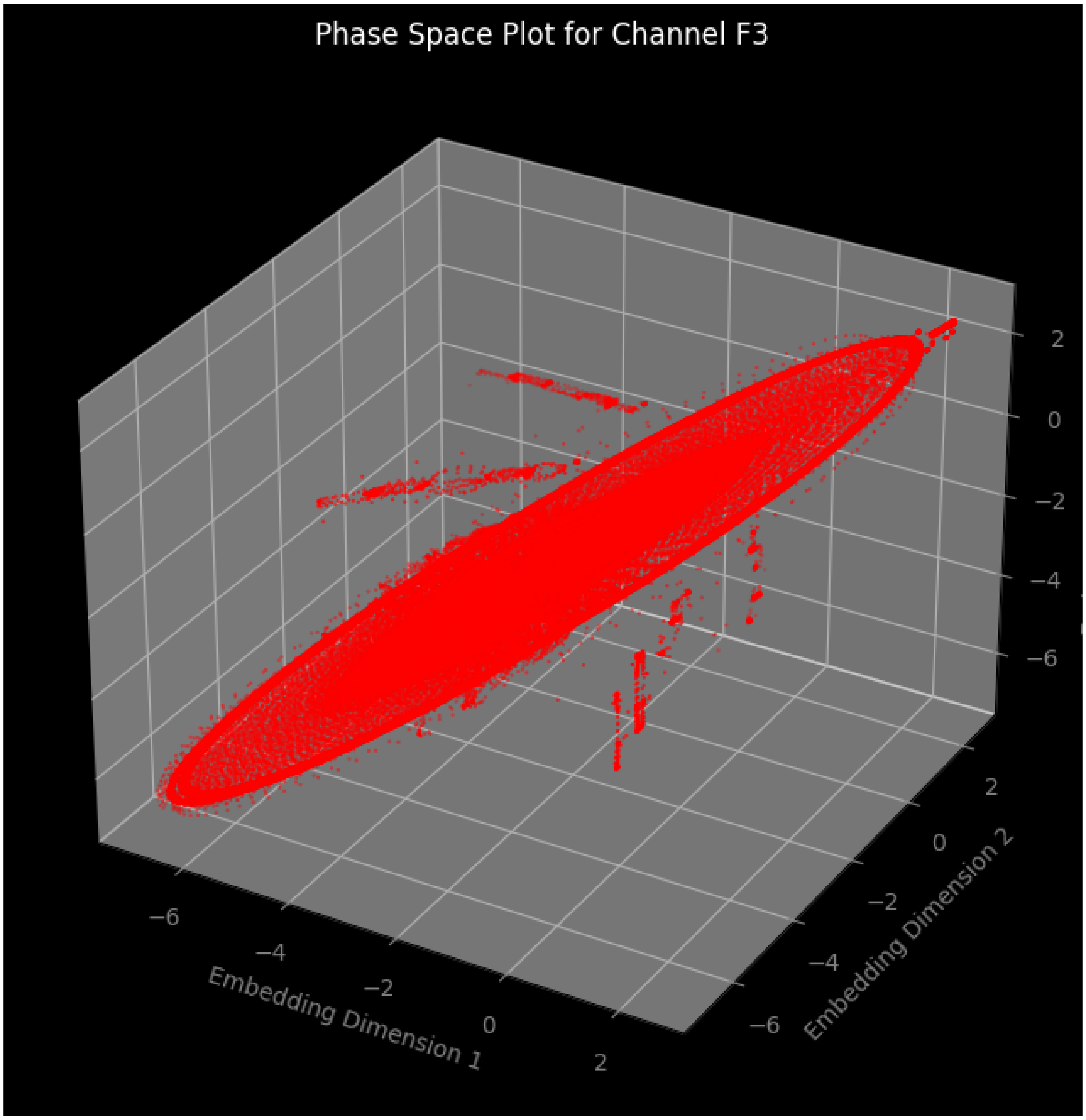
Phase Space Plot for Channel F3
Embedding Dimension 1
Embedding Dimension 2
−6
−4
−2
0
2

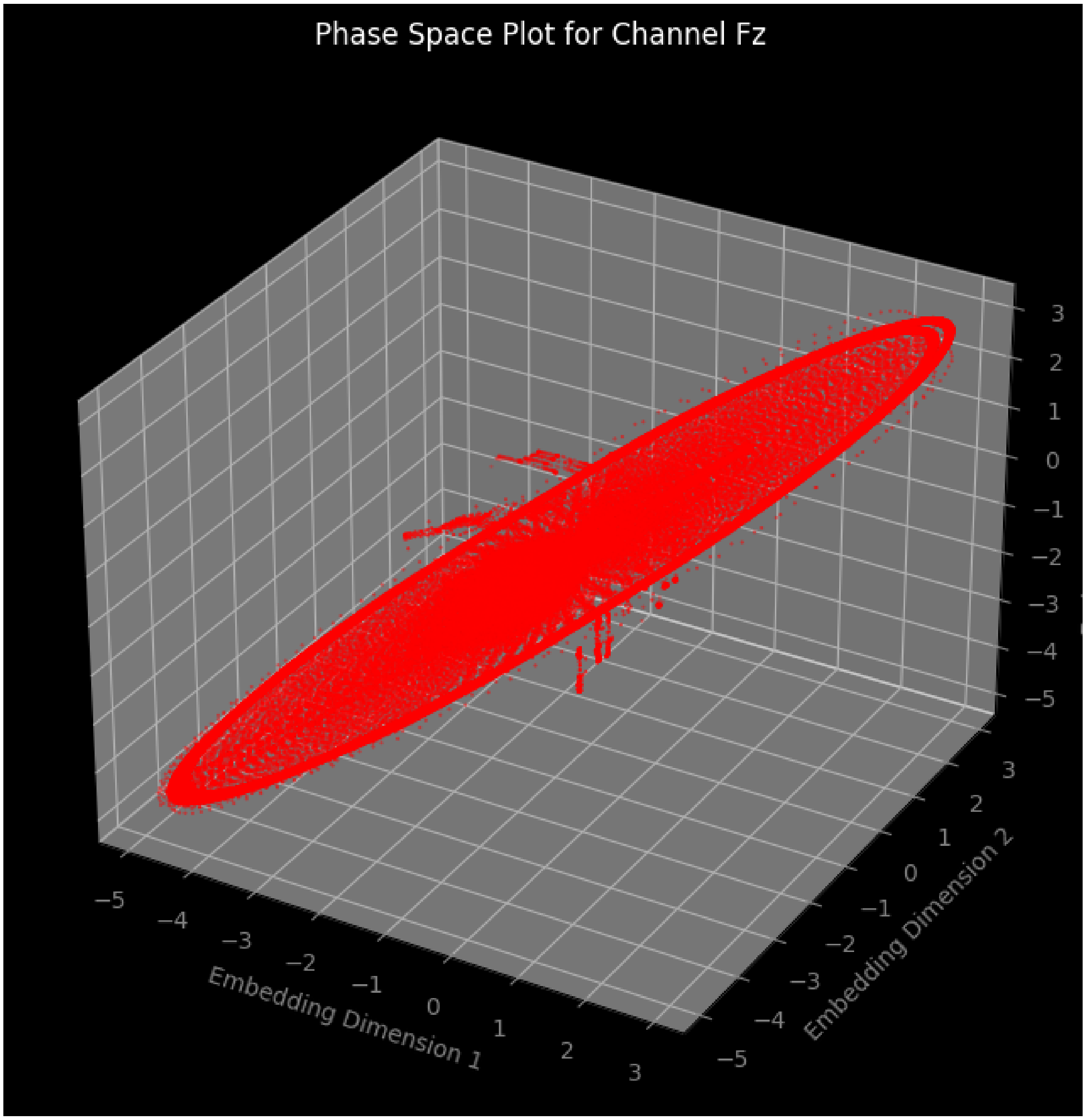
Phase Space Plot for Channel Fz
Embedding Dimension 1
Embedding Dimension 2
−5
−4
−3
−2
−1
0
1
2
3

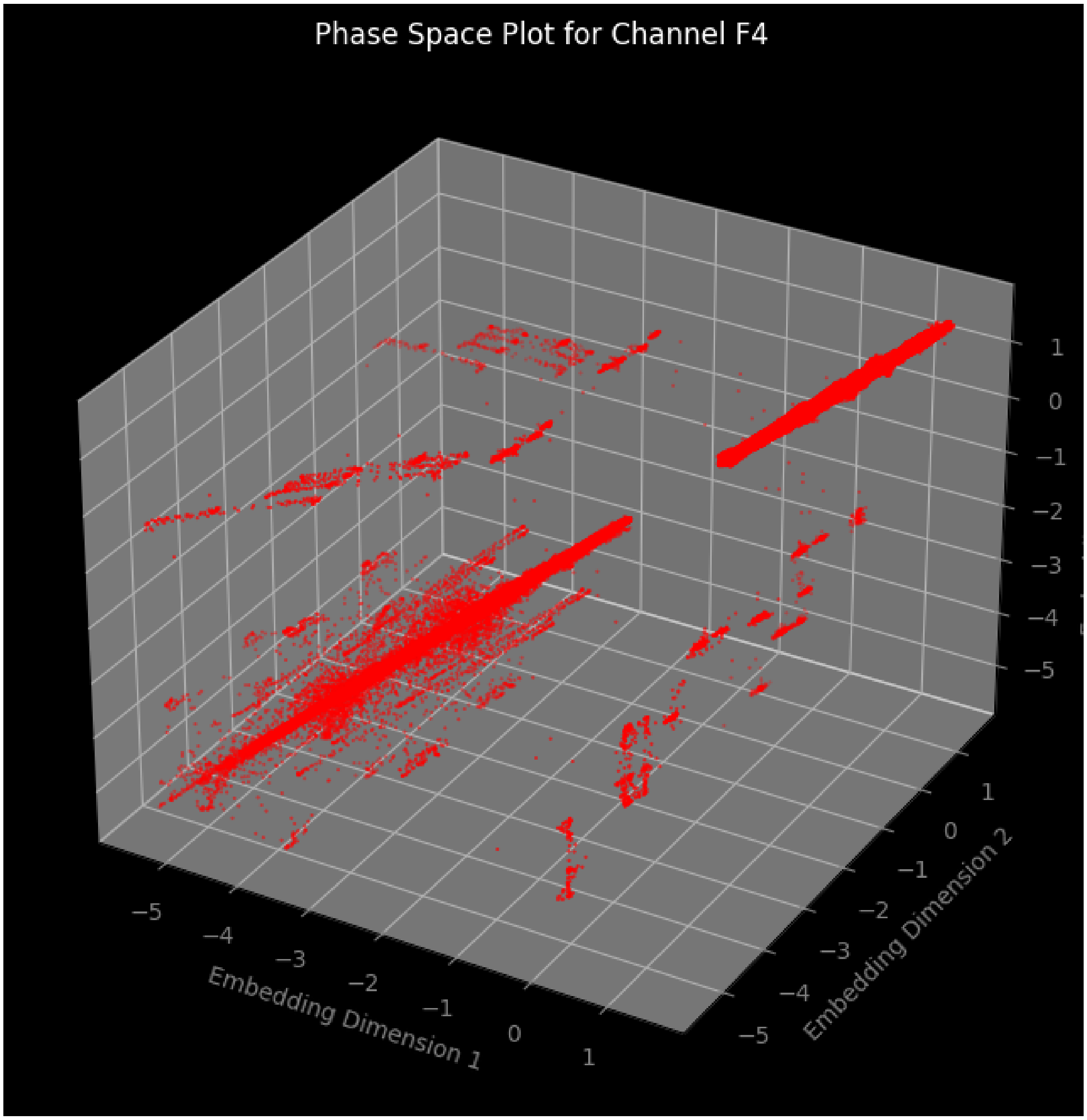

Phase Space Plot for Channel F4
Embedding Dimension 1
Embedding Dimension 2
−5
−4
−3
−2
−1
0
1

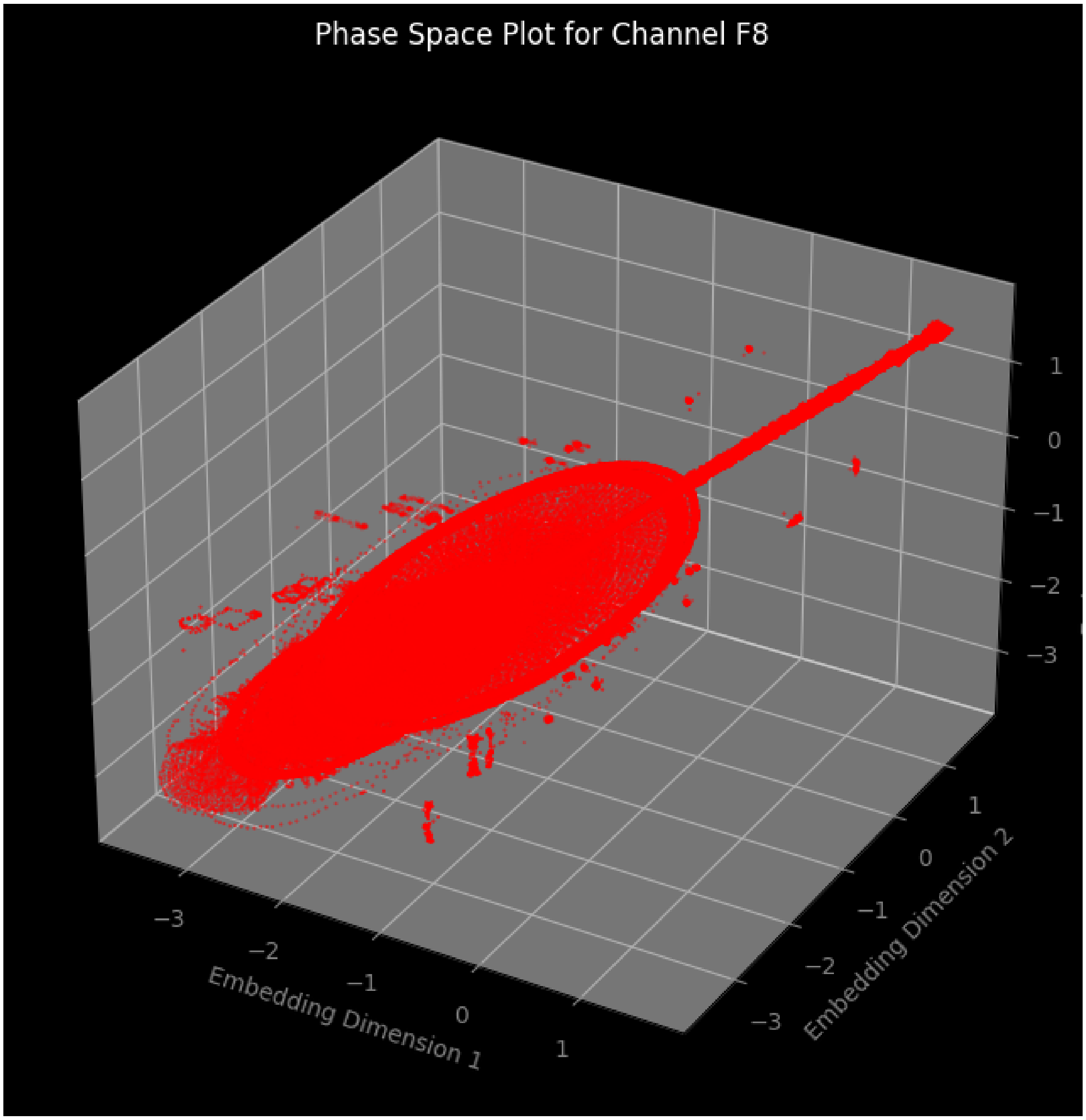

Phase Space Plot for Channel F8
Embedding Dimension 1
Embedding Dimension 2
−3
−2
−1
0
1

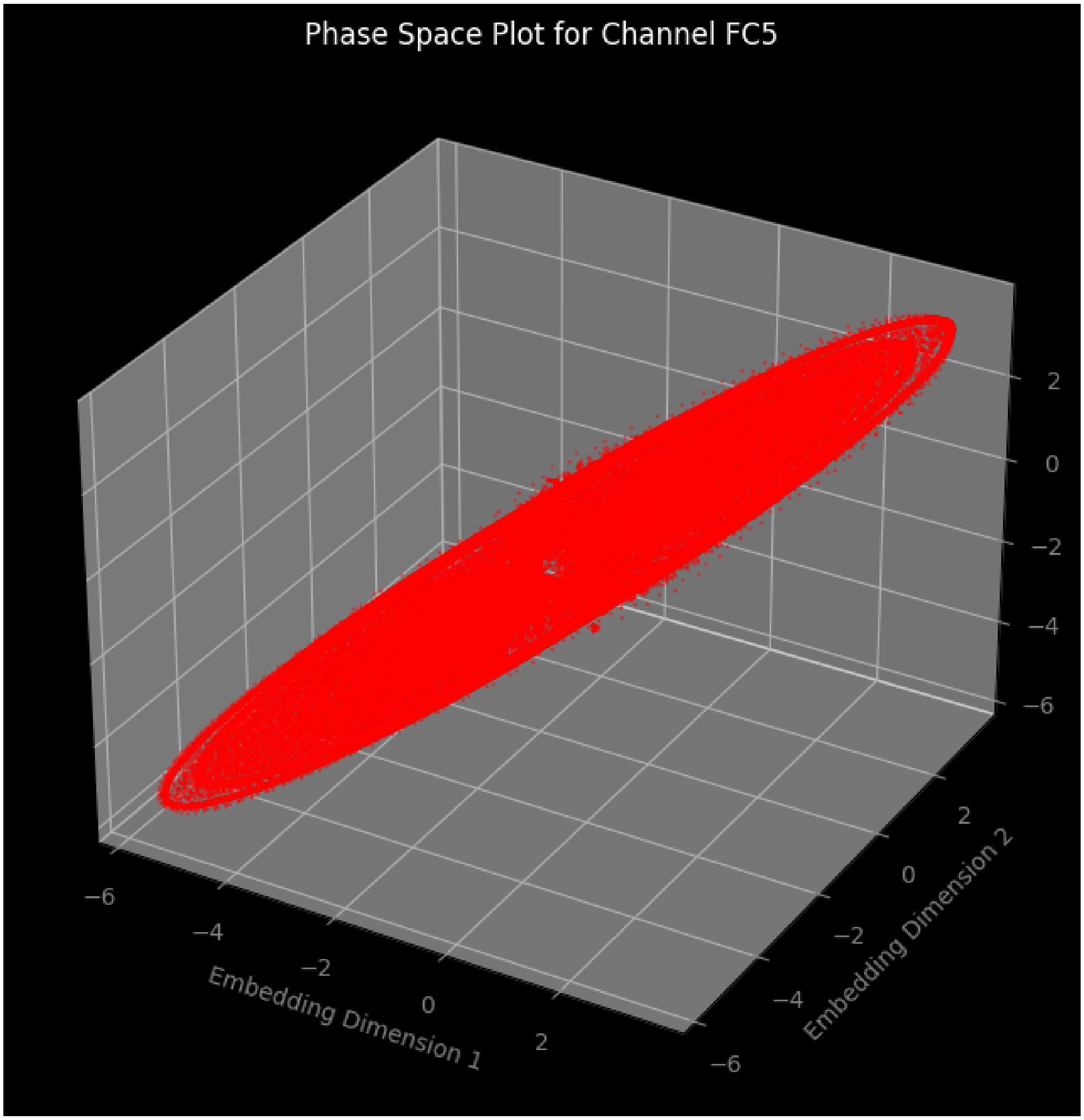
Phase Space Plot for Channel FC5
Embedding Dimension 1
Embedding Dimension 2
−6
−4
−2
0
2

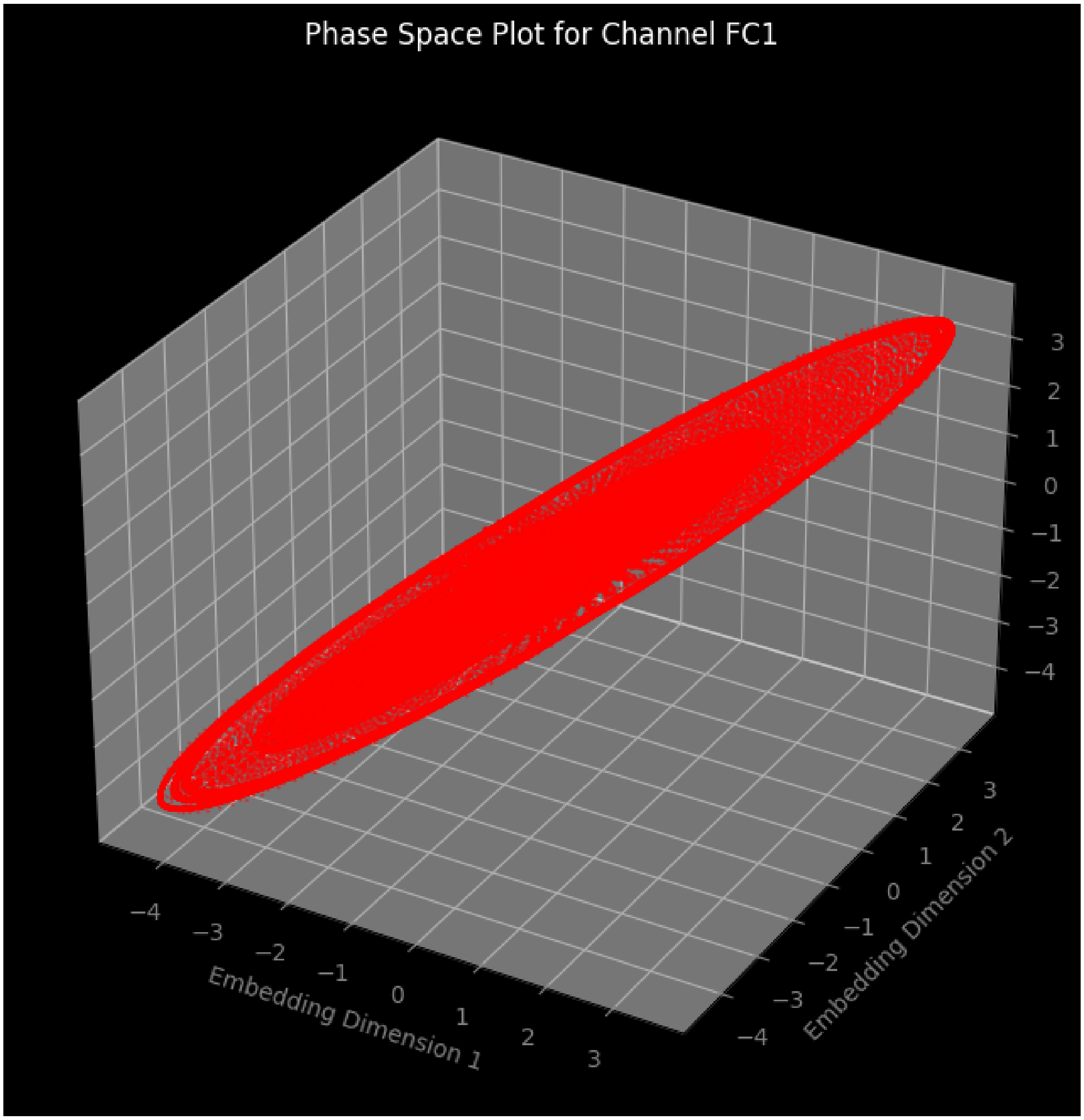

Phase Space Plot for Channel FC1
Embedding Dimension 1
Embedding Dimension 2
−4
−3
−2
−1
0
1
2
3

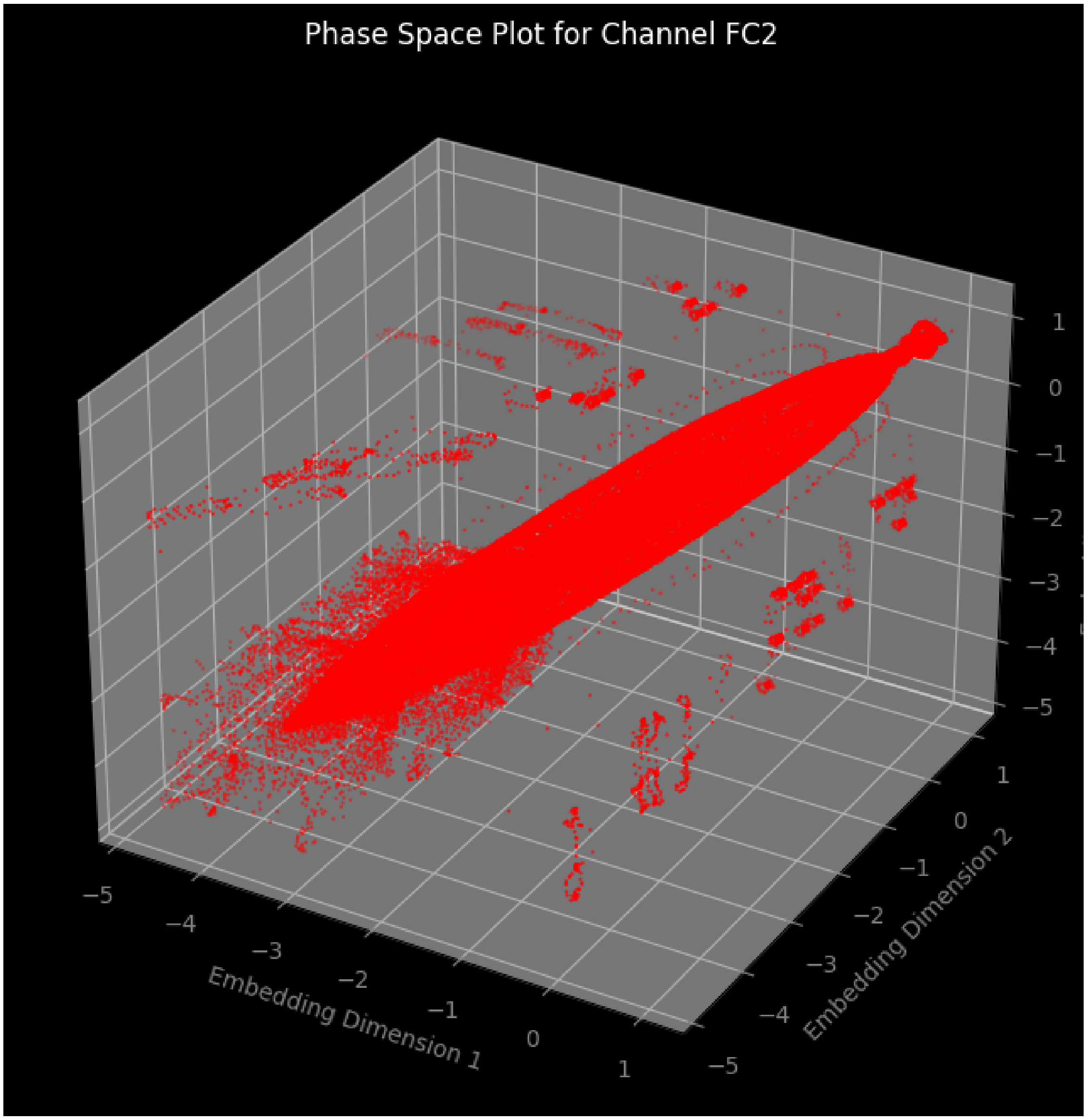
Phase Space Plot for Channel FC2
Embedding Dimension 1
Embedding Dimension 2
−5
−4
−3
−2
−1
0
1

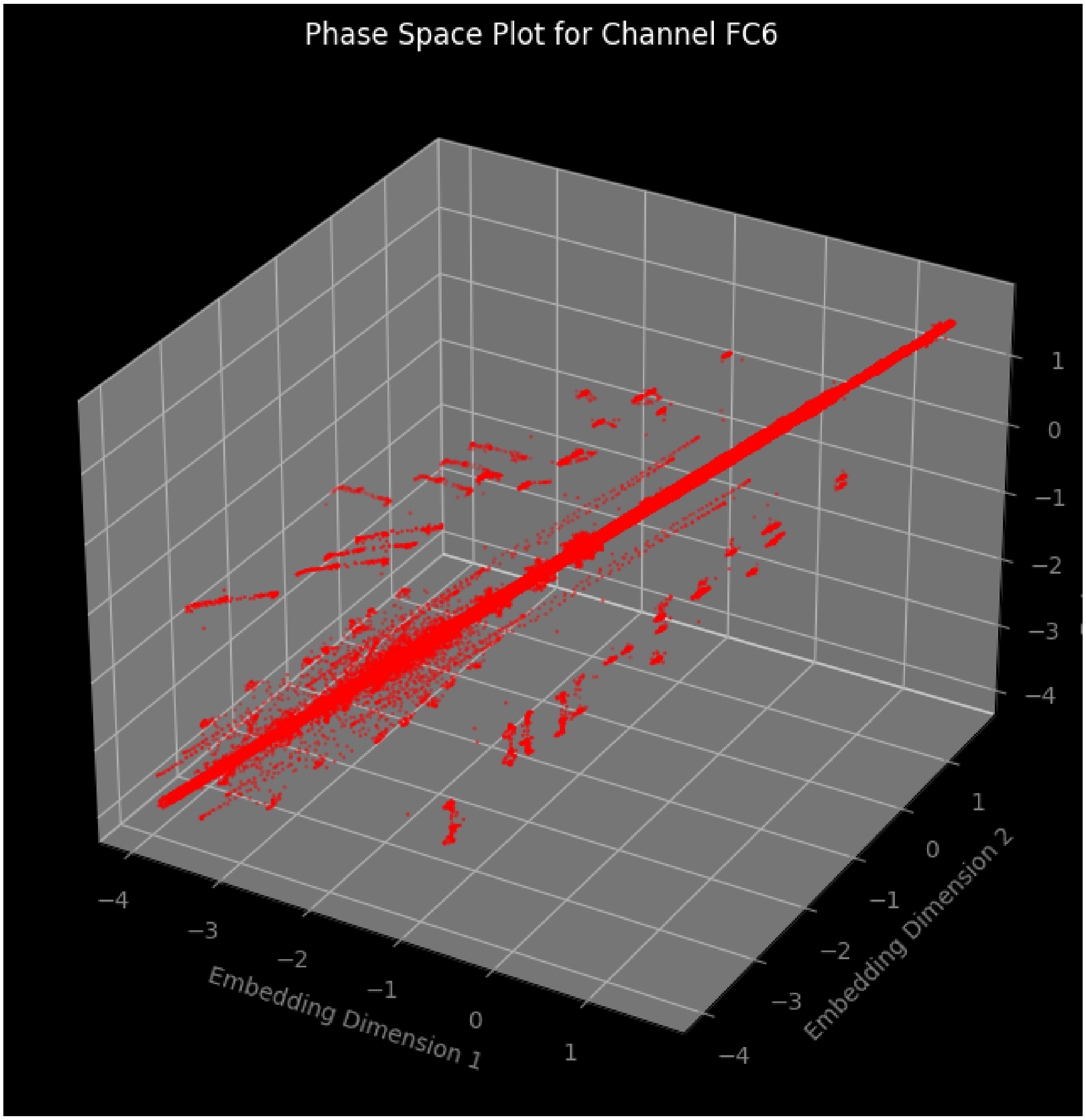

Phase Space Plot for Channel FC6
Embedding Dimension 1
Embedding Dimension 2
−4
−3
−2
−1
0
1

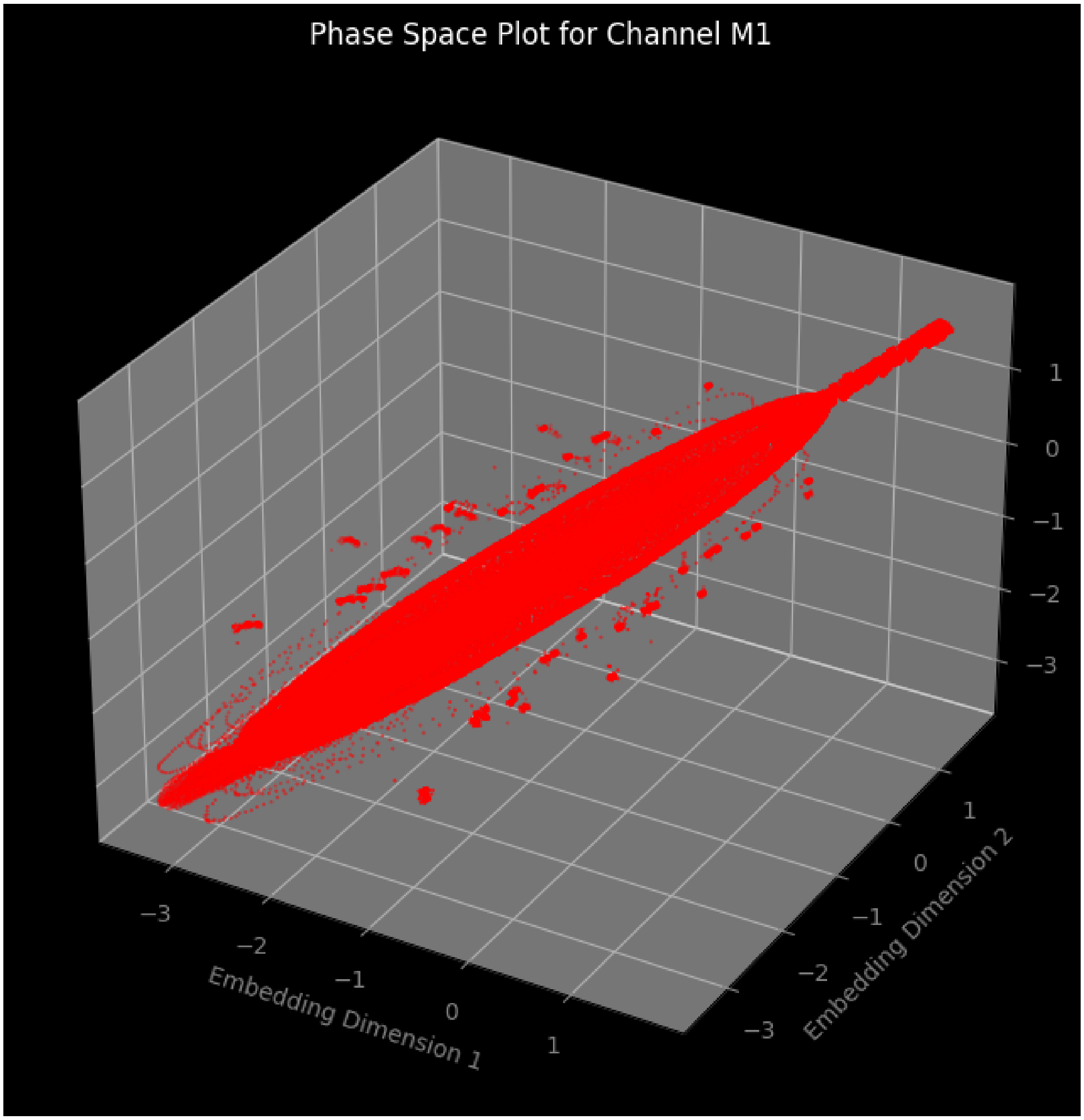

Phase Space Plot for Channel M1
Embedding Dimension 1
Embedding Dimension 2
−3
−2
−1
0
1

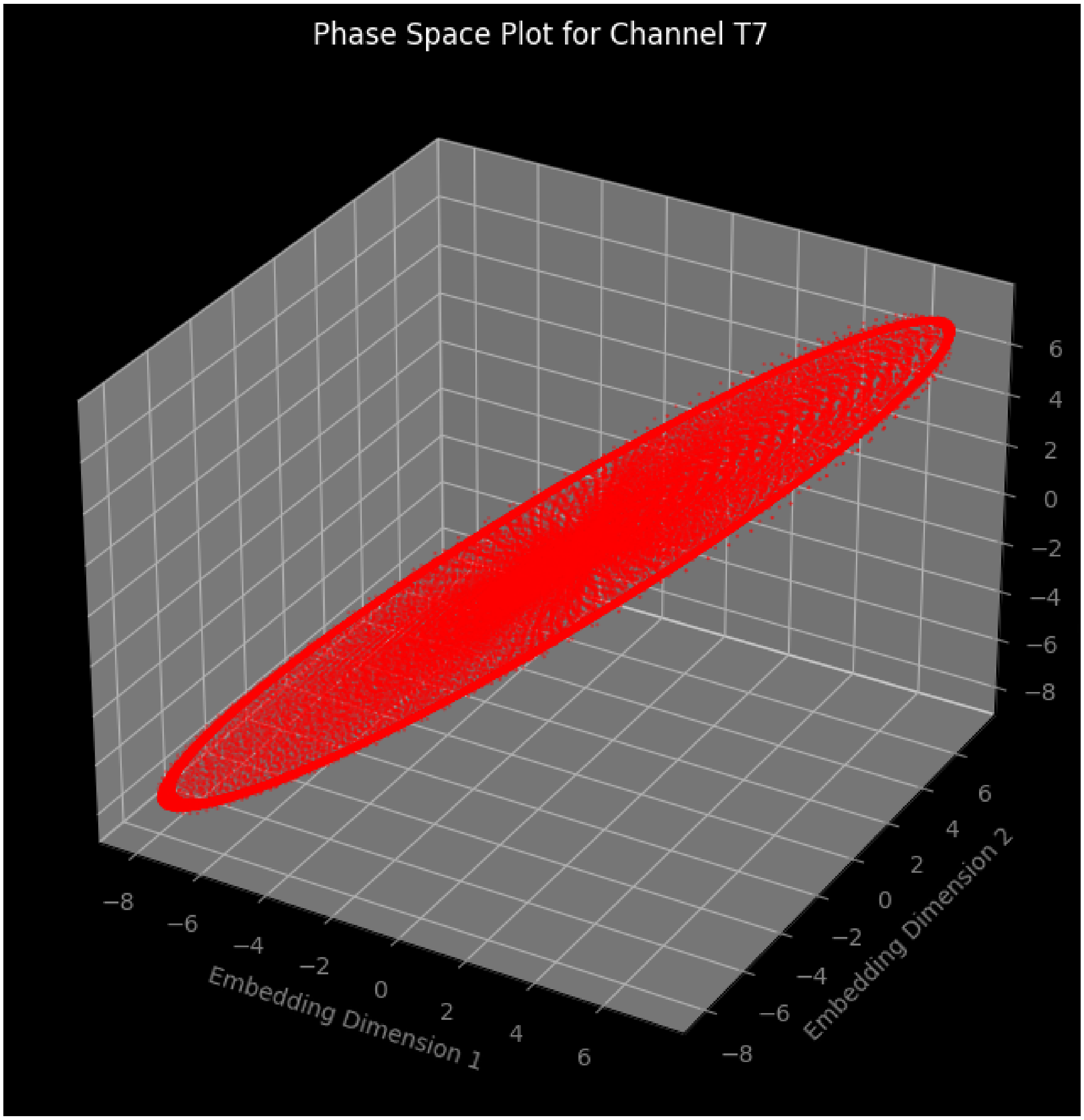

Phase Space Plot for Channel T7
−8
−6
−4
−2
0
2
4
6
Embedding Dimension 1
−8
−6
−4
−2
0
2
4
6
Embedding Dimension 2
6
4
2
0
−2
−4
−6
−8

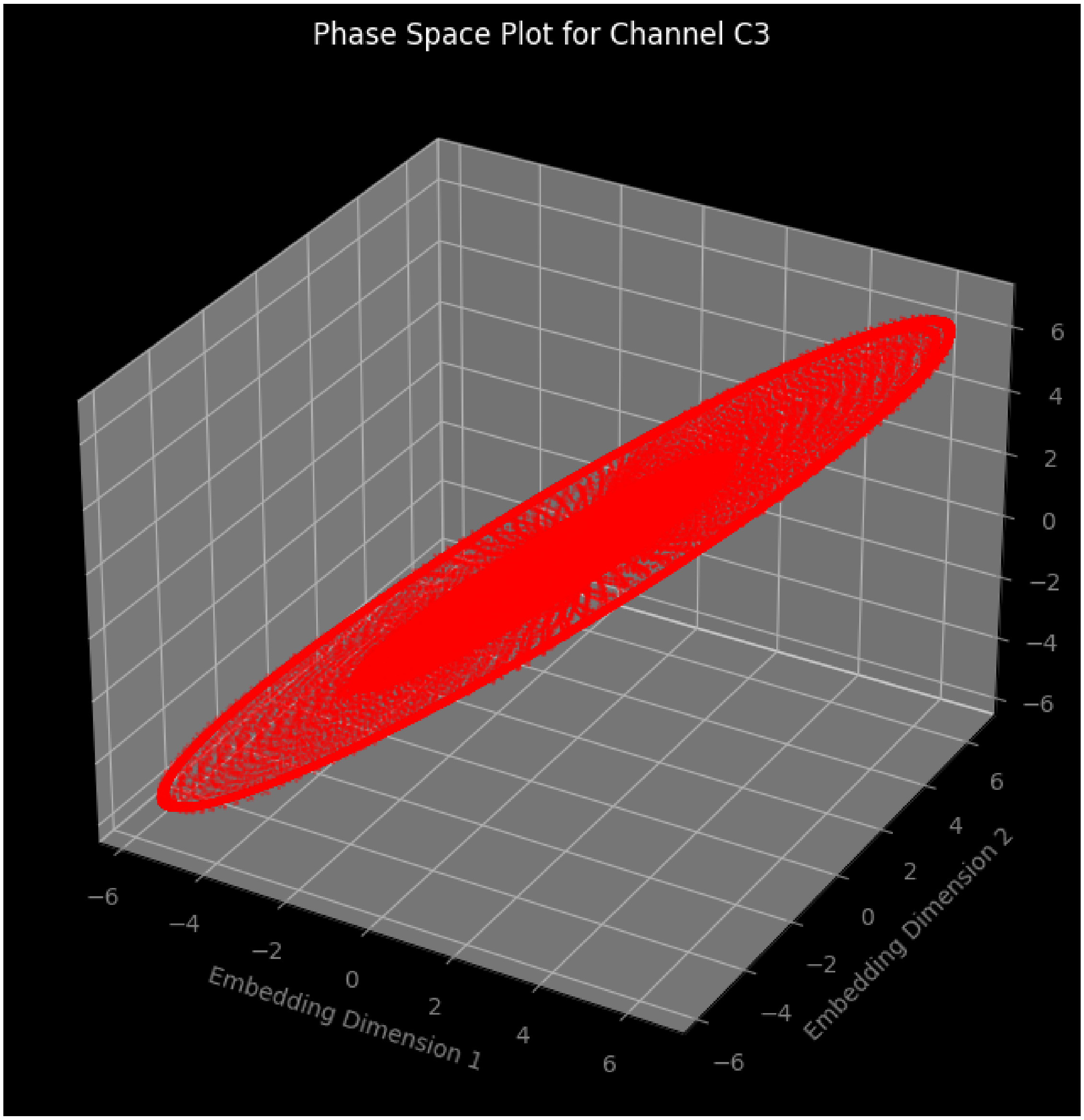
Phase Space Plot for Channel C3
Embedding Dimension 1
Embedding Dimension 2
−6
−4
−2
0
2
4
6

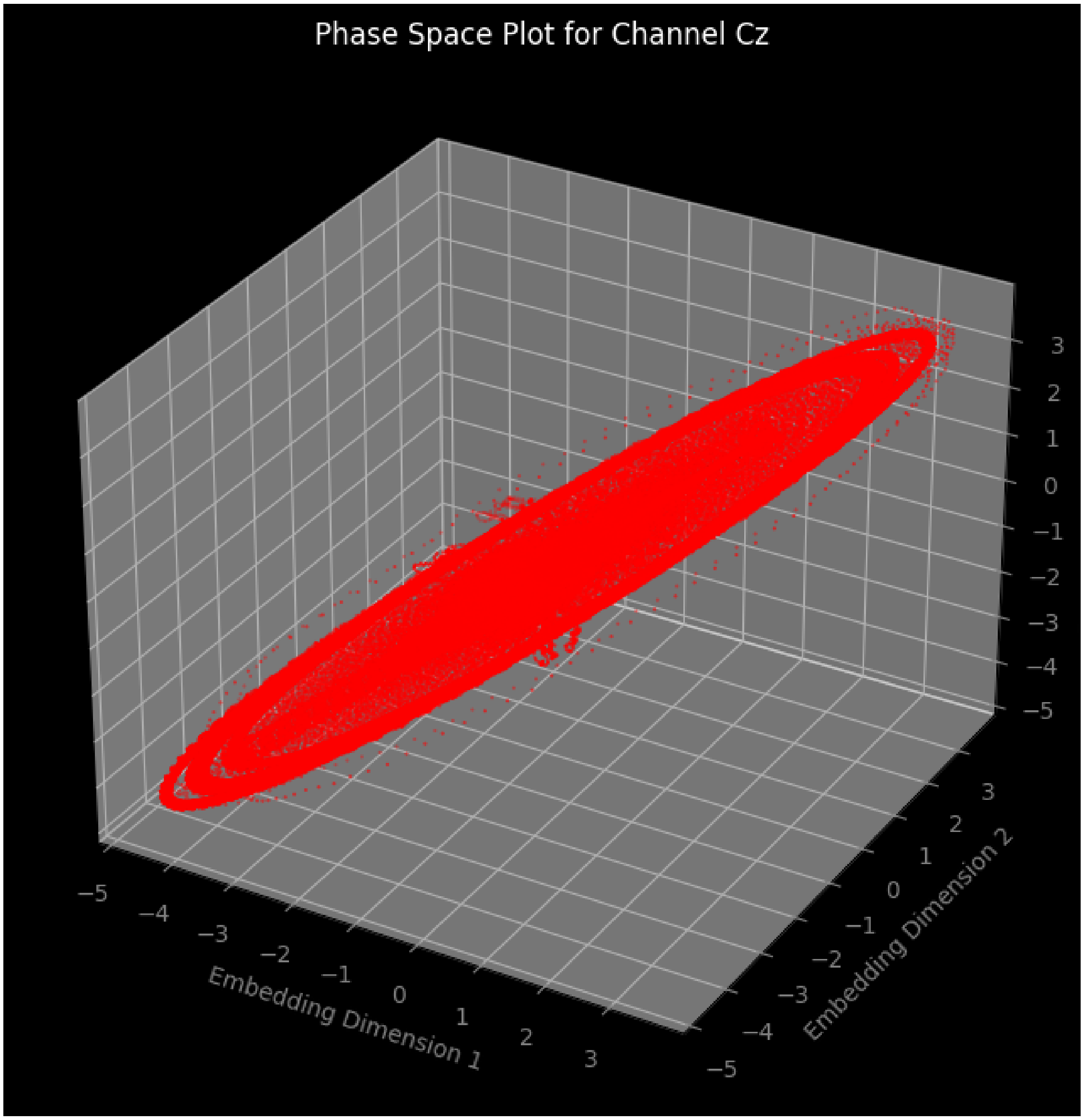

Phase Space Plot for Channel Cz
−5
−4
−3
−2
−1
0
1
2
3
Embedding Dimension 1
Embedding Dimension 2

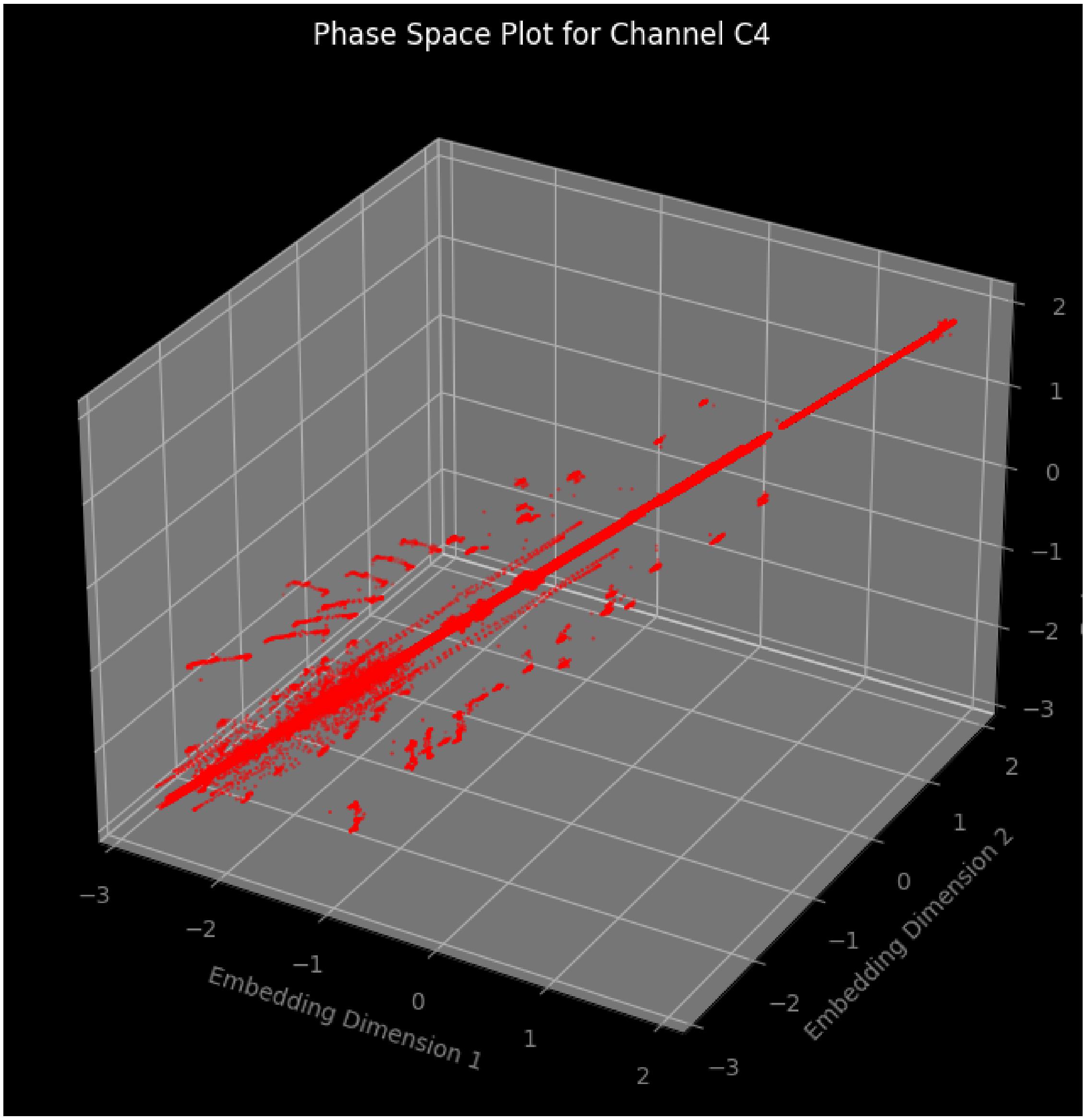

Phase Space Plot for Channel C4
Embedding Dimension 1
Embedding Dimension 2
−3
−2
−1
0
1
2

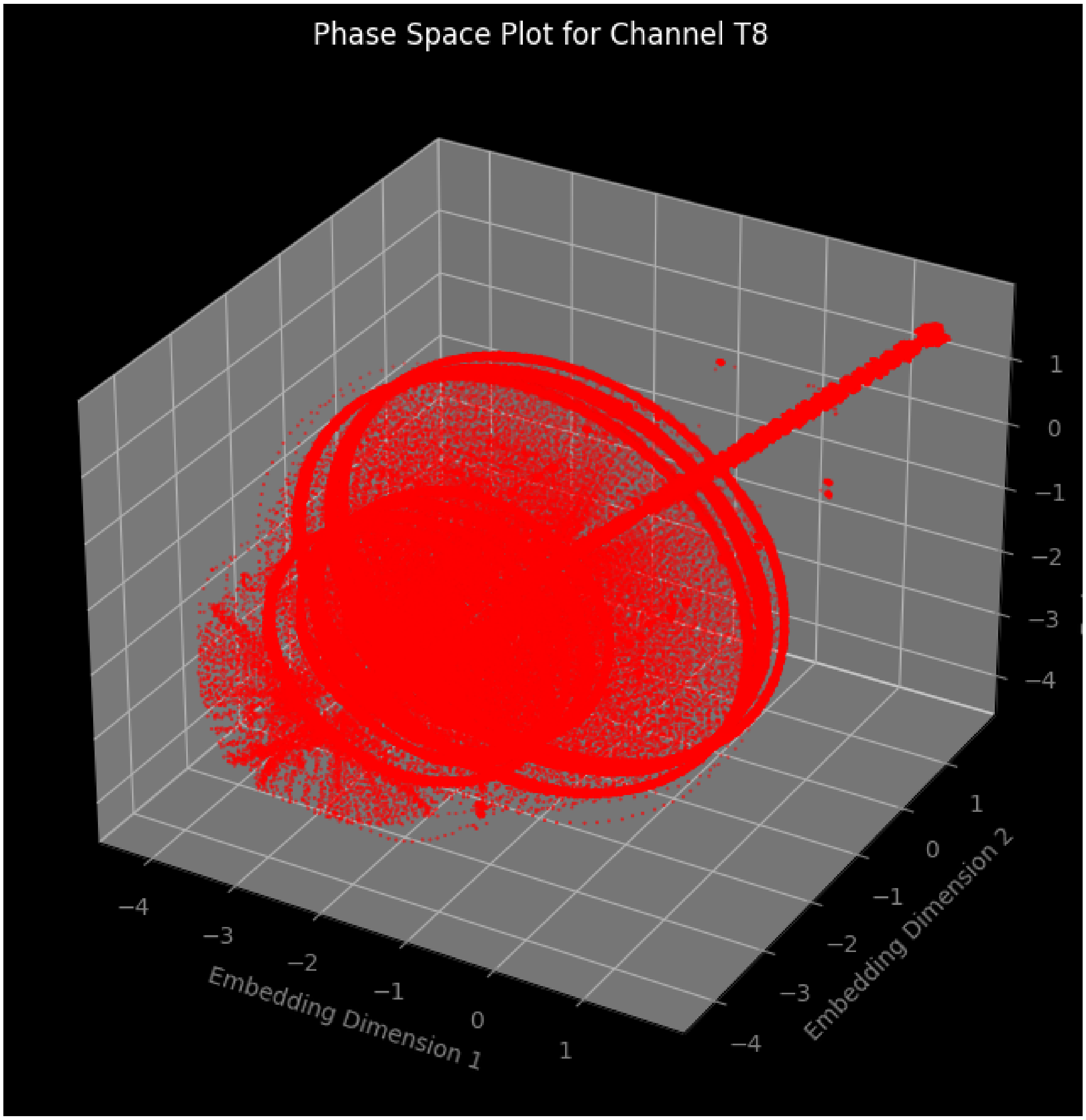

Phase Space Plot for Channel T8
Embedding Dimension 1
Embedding Dimension 2
−4
−3
−2
−1
0
1

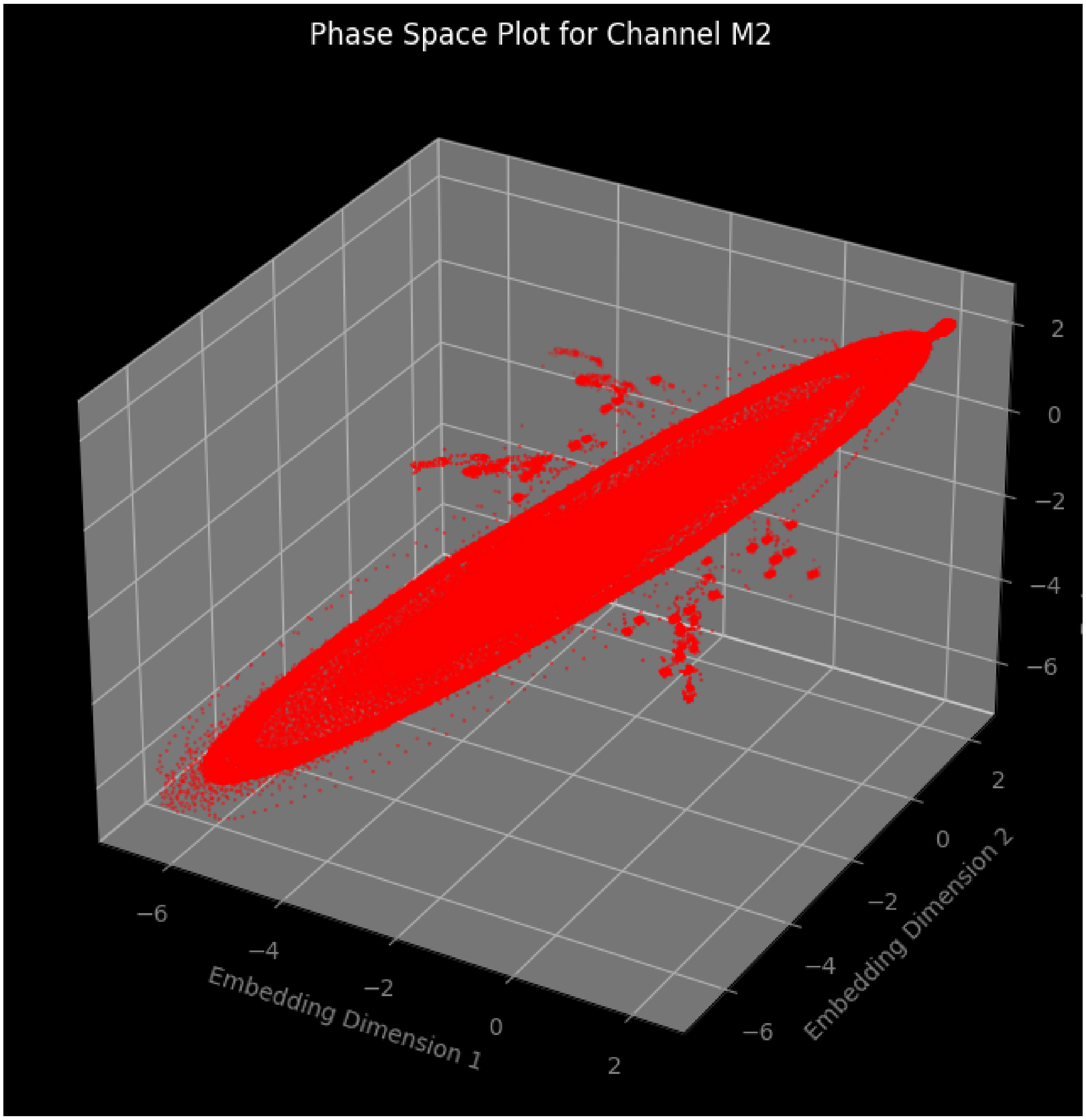

Phase Space Plot for Channel M2
Embedding Dimension 1
Embedding Dimension 2
−6
−4
−2
0
2

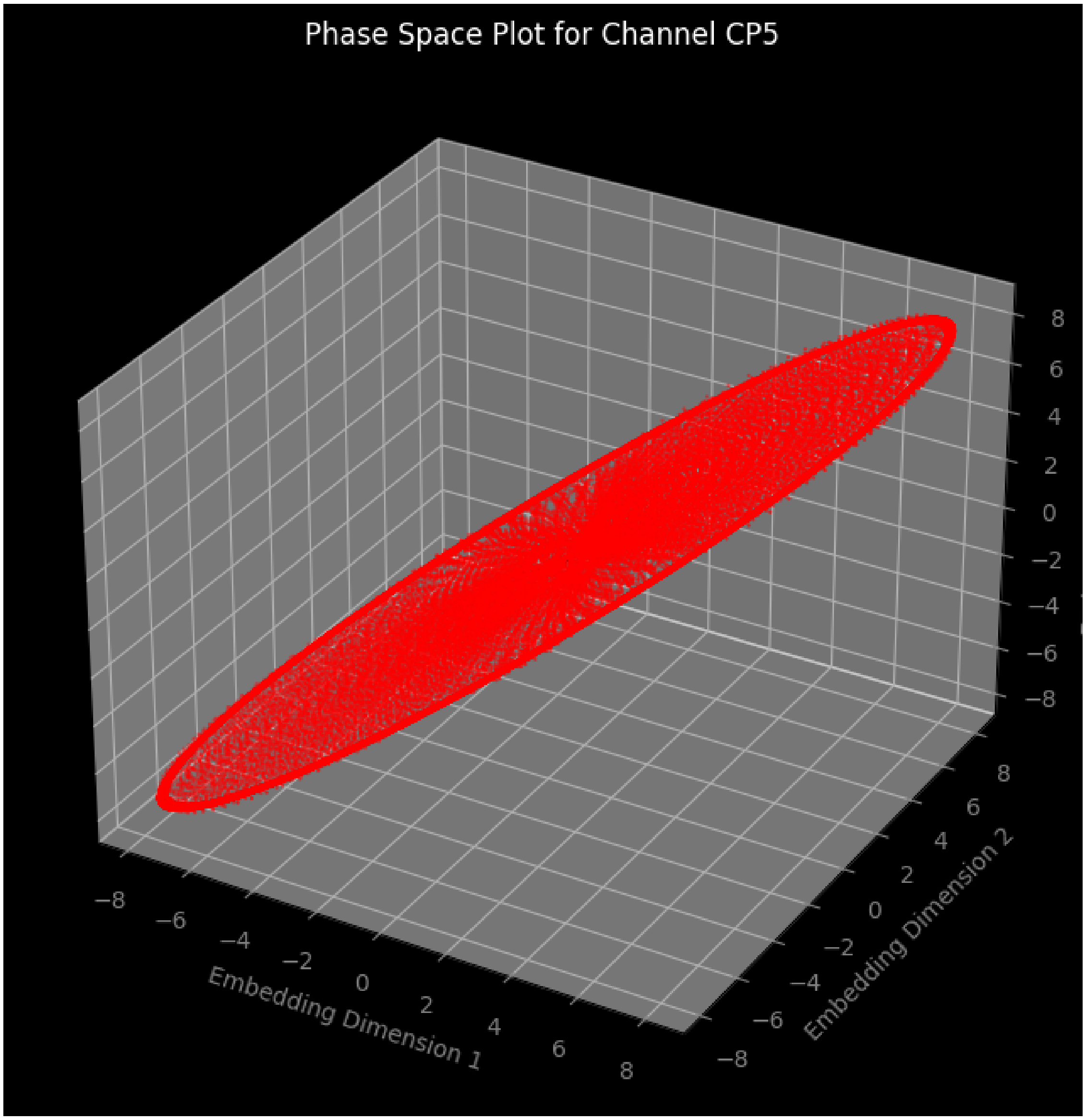
Phase Space Plot for Channel CP5
Embedding Dimension 1
Embedding Dimension 2
−8
−6
−4
−2
0
2
4
6
8

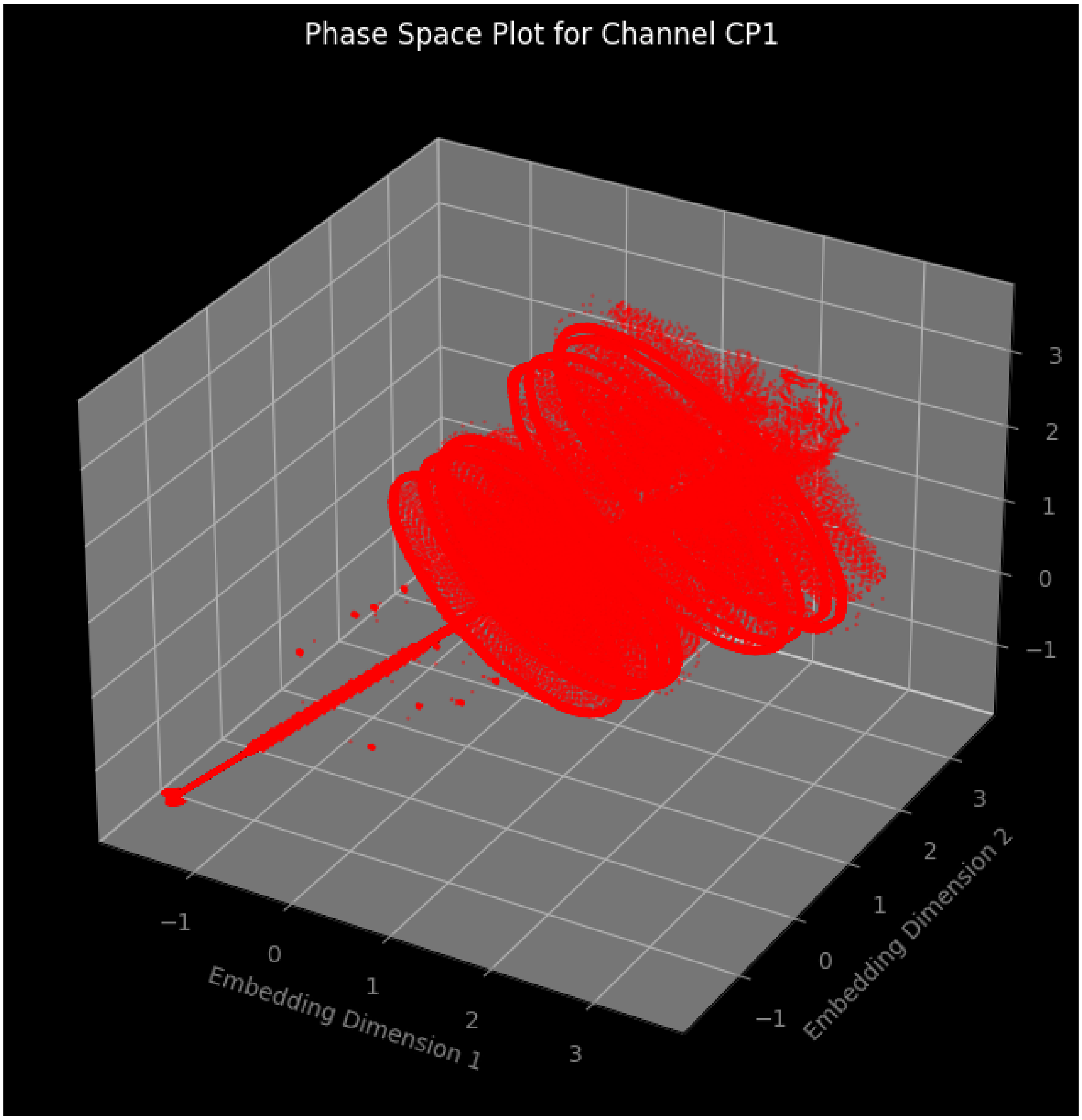
Phase Space Plot for Channel CP1
Embedding Dimension 1
Embedding Dimension 2
−1
0
1
2
3

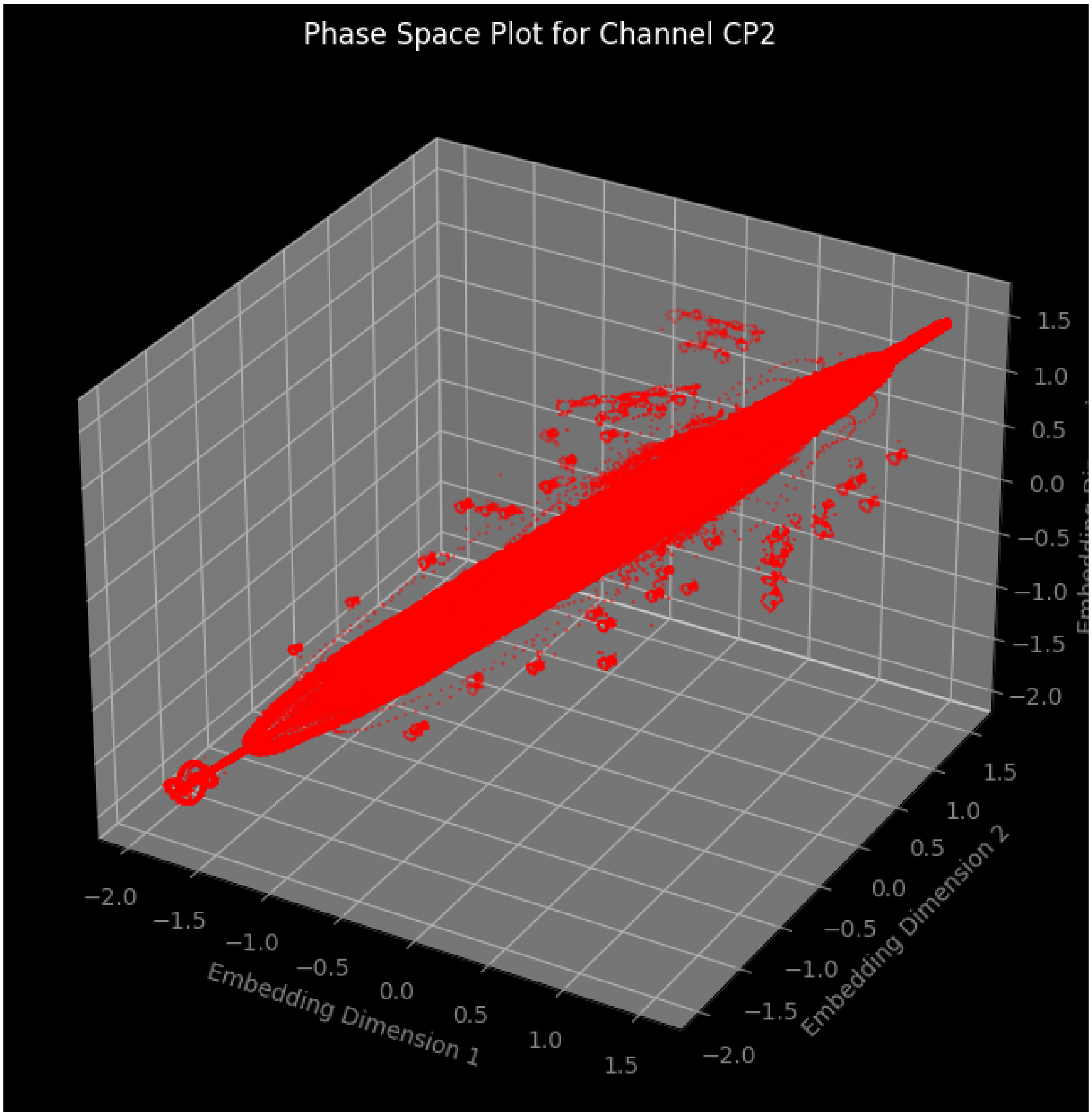

Phase Space Plot for Channel CP2
Embedding Dimension 1
Embedding Dimension 2
−2.0
−1.5
−1.0
−0.5
0.0
0.5
1.0
1.5

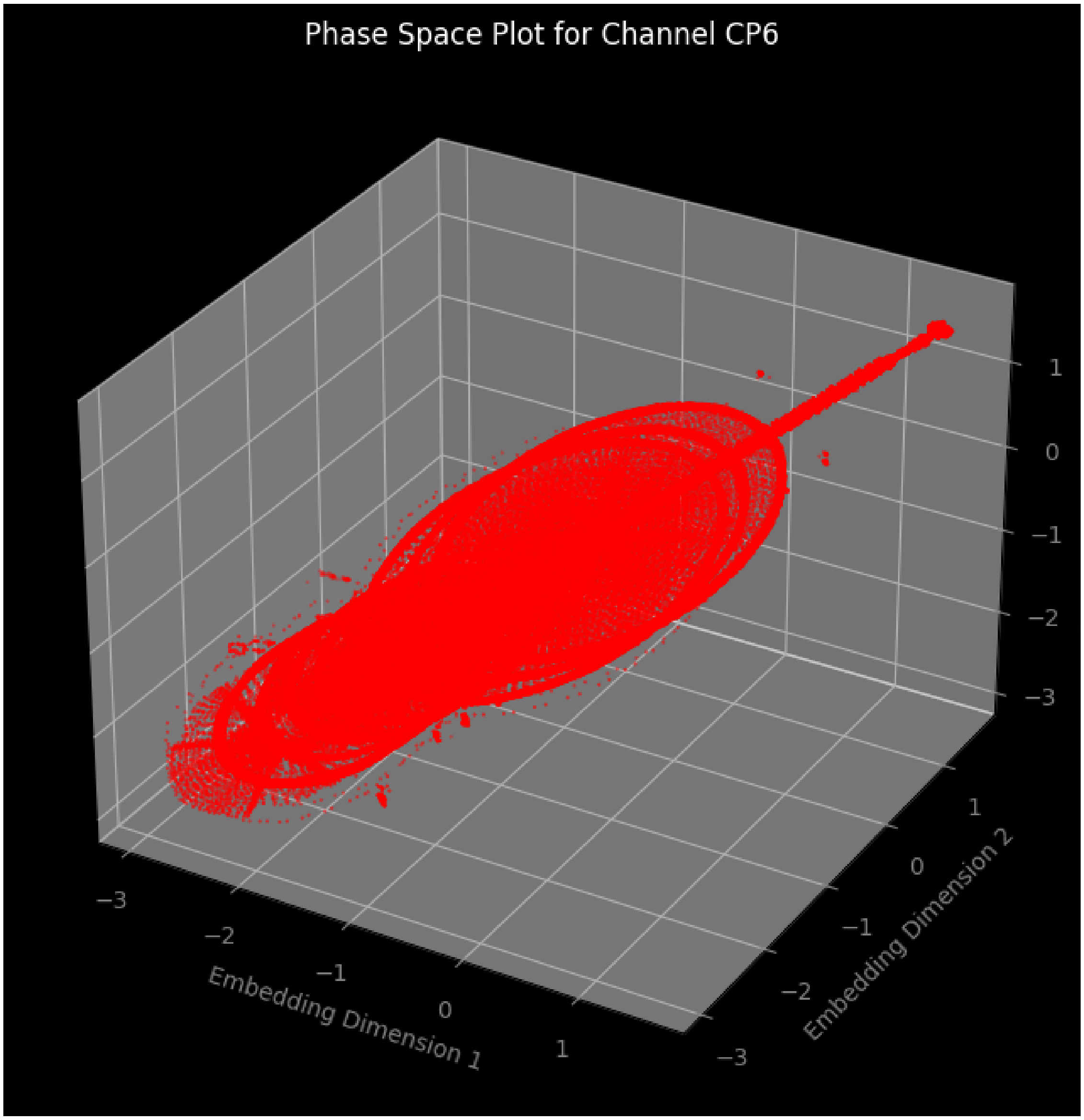
Phase Space Plot for Channel CP6
Embedding Dimension 1
Embedding Dimension 2
−3
−2
−1
0
1

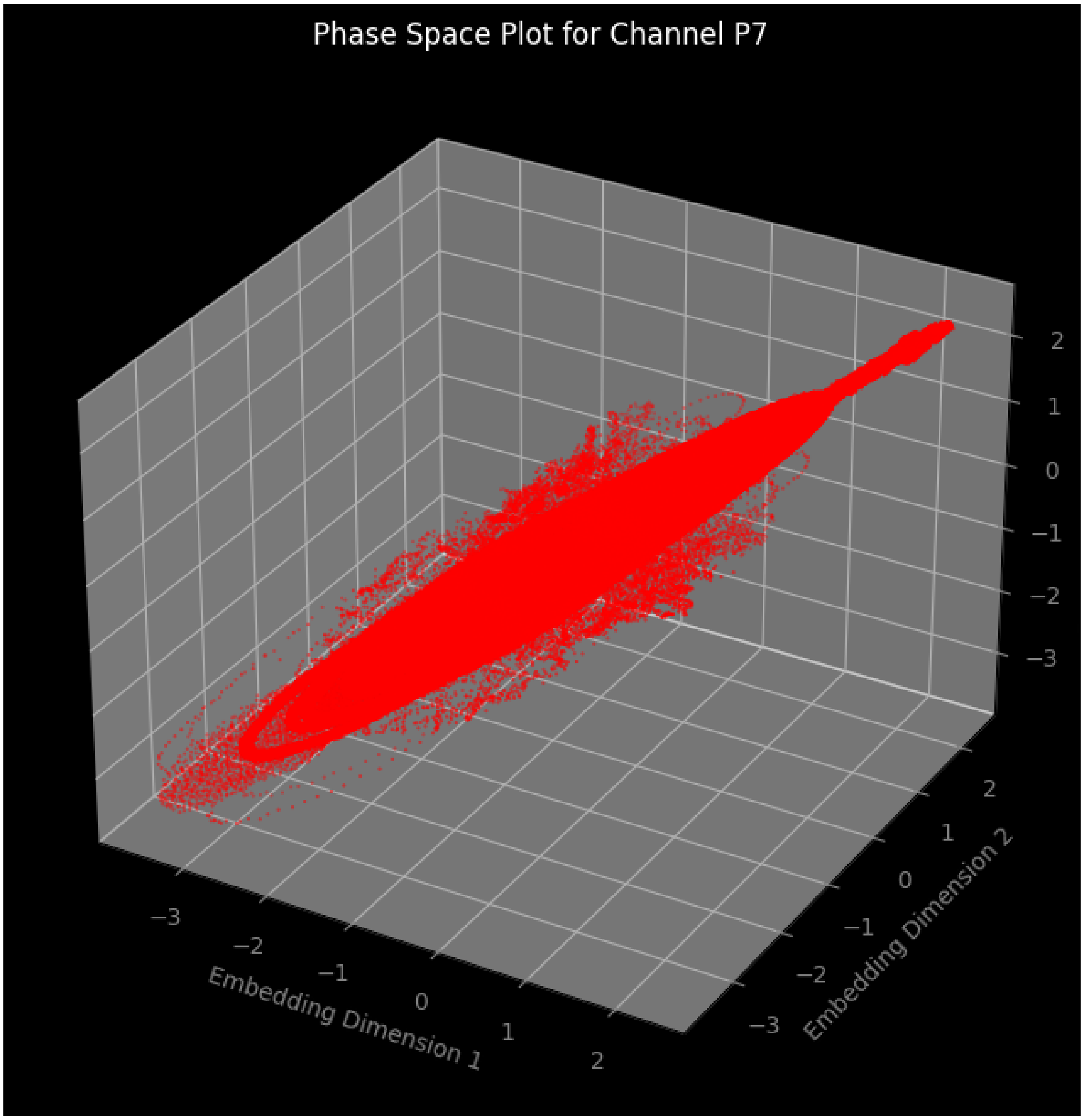
Phase Space Plot for Channel P7
Embedding Dimension 1
Embedding Dimension 2
−3
−2
−1
0
1
2

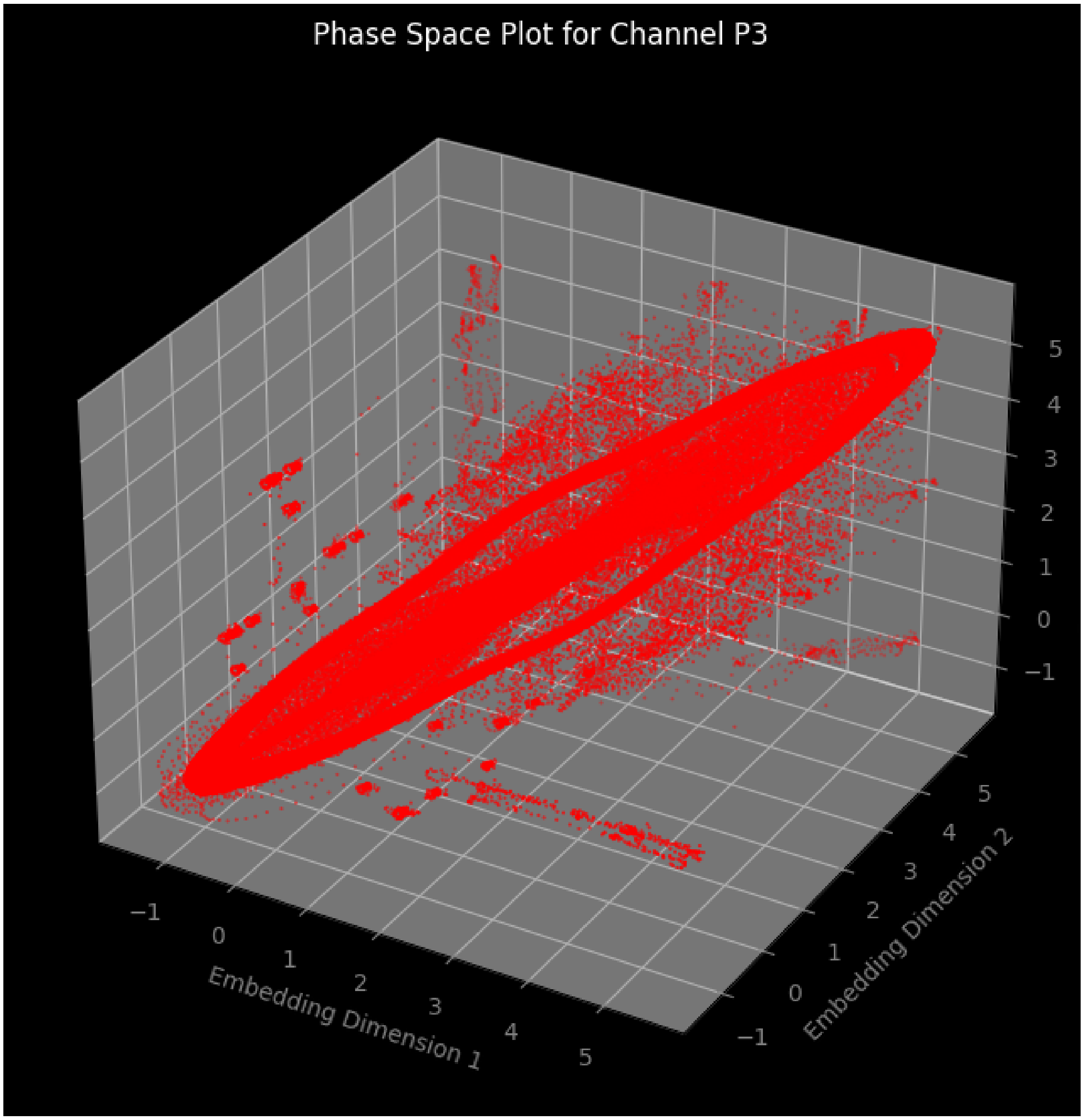
Phase Space Plot for Channel P3
Embedding Dimension 1
Embedding Dimension 2
−1
0
1
2
3
4
5

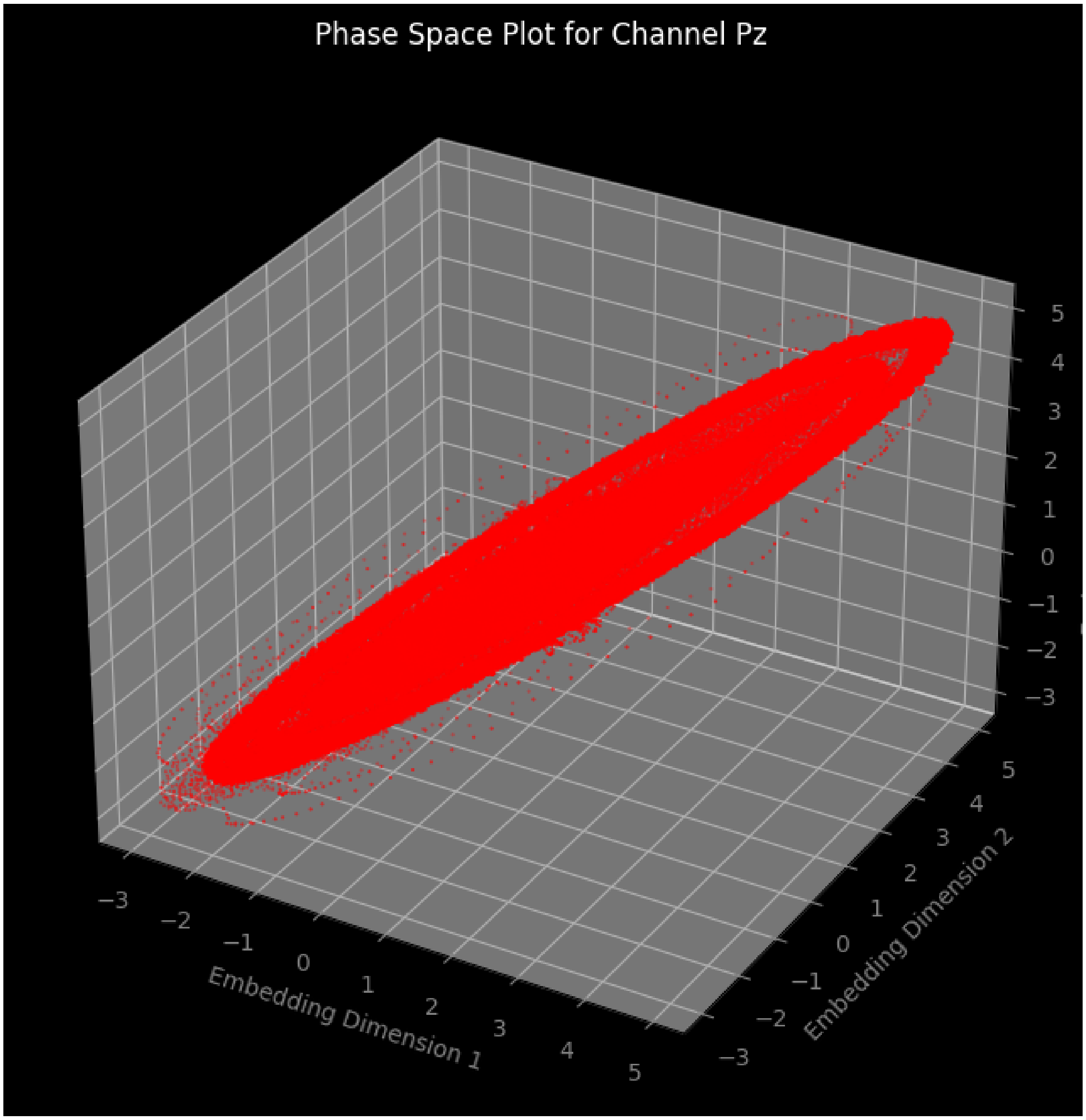
Phase Space Plot for Channel Pz
Embedding Dimension 1
Embedding Dimension 2
−3
−2
−1
0
1
2
3
4
5

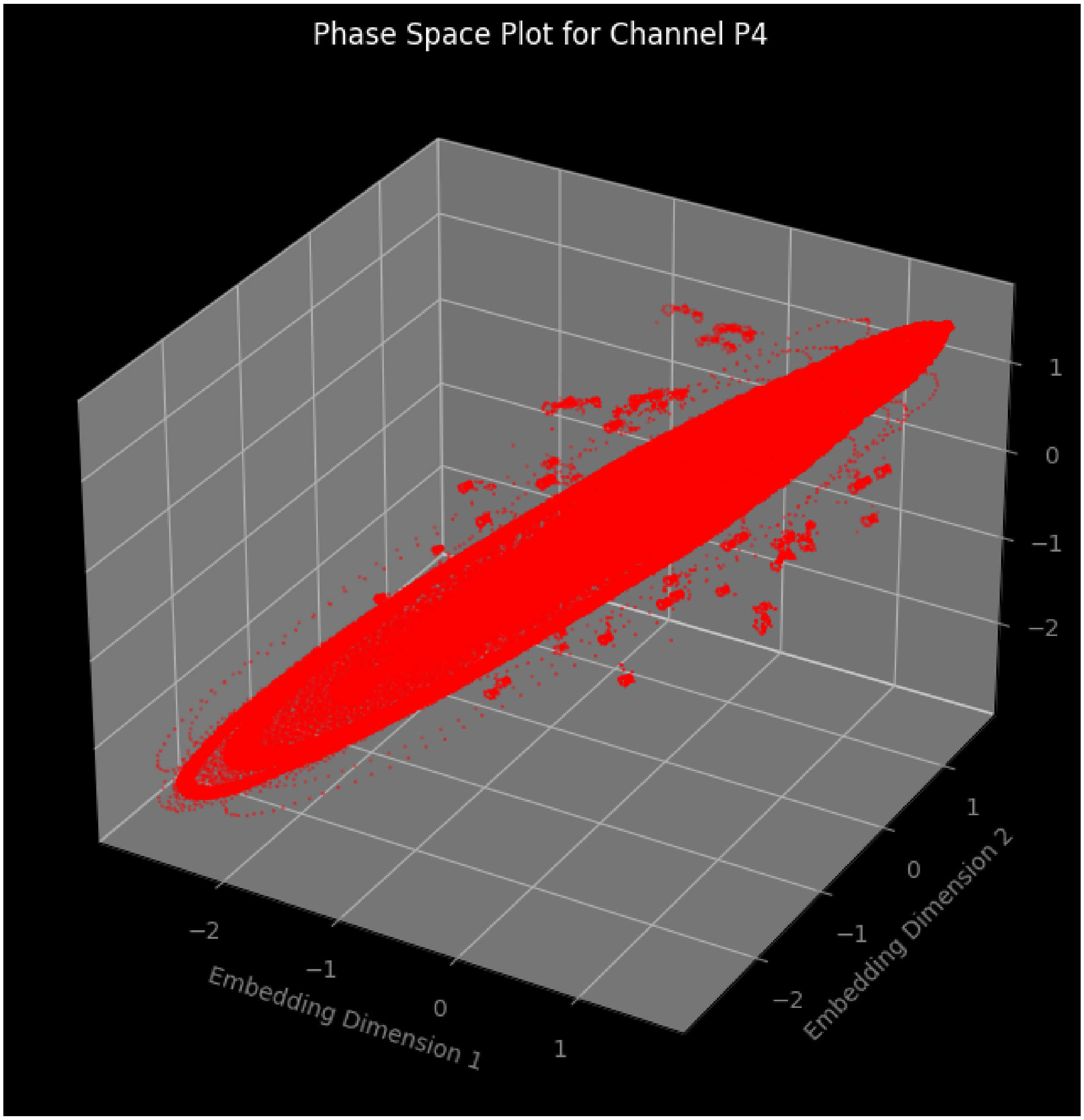

Phase Space Plot for Channel P4
Embedding Dimension 1
Embedding Dimension 2
−2
−1
0
1

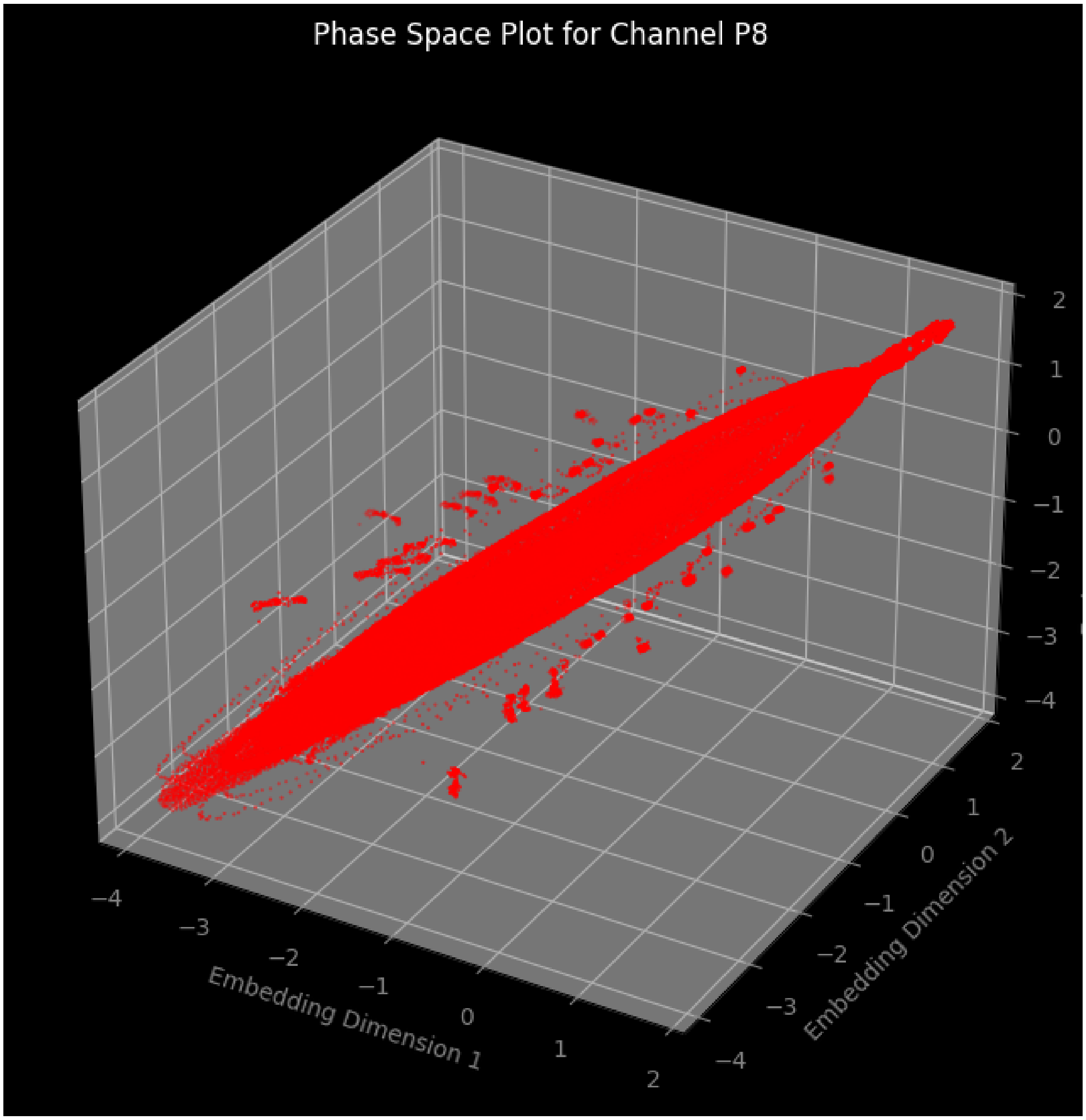
Phase Space Plot for Channel P8
Embedding Dimension 1
Embedding Dimension 2
−4
−3
−2
−1
0
1
2

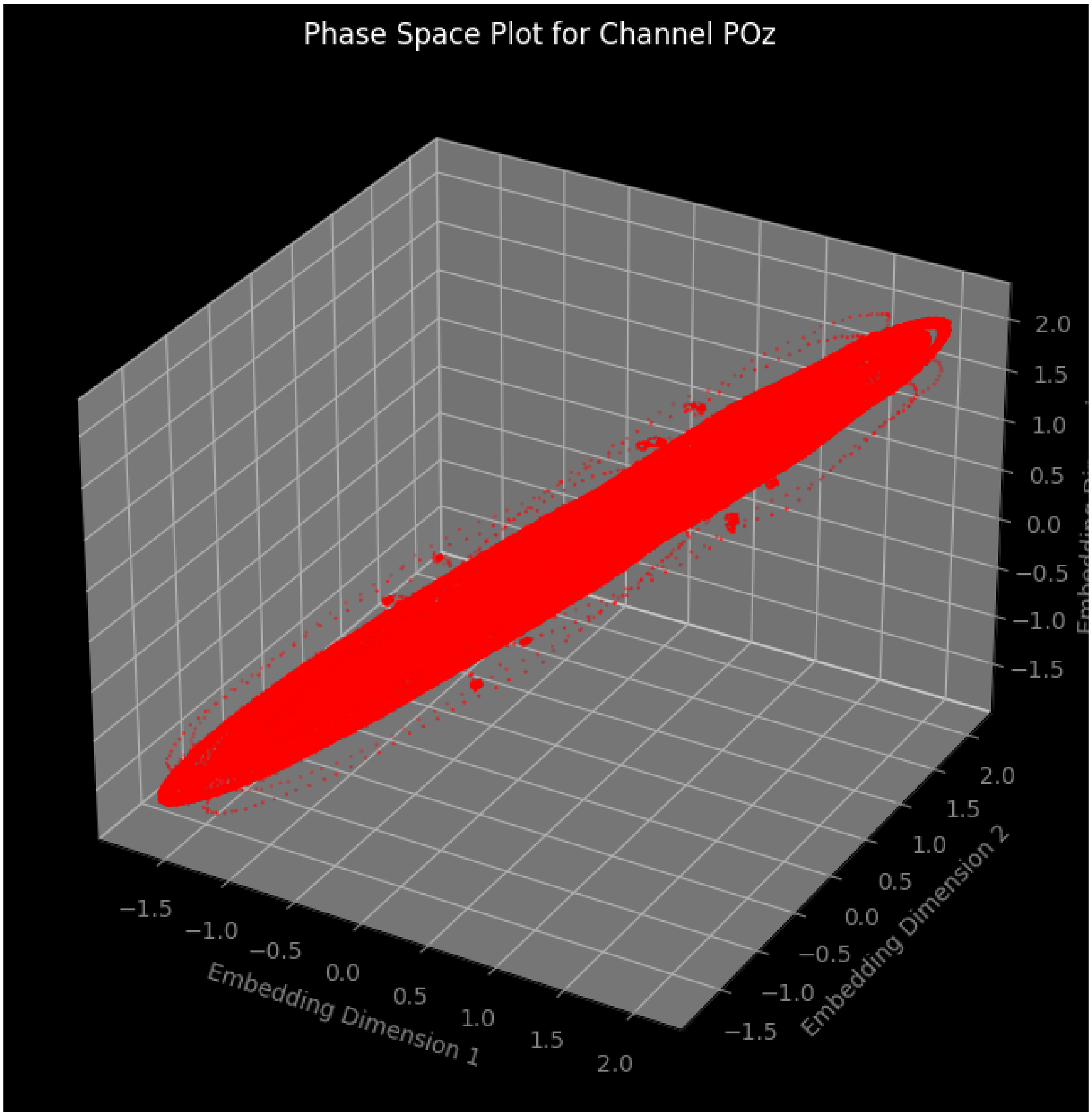

Phase Space Plot for Channel POz
Embedding Dimension 1
Embedding Dimension 2
−1.5
−1.0
−0.5
0.0
0.5
1.0
1.5
2.0

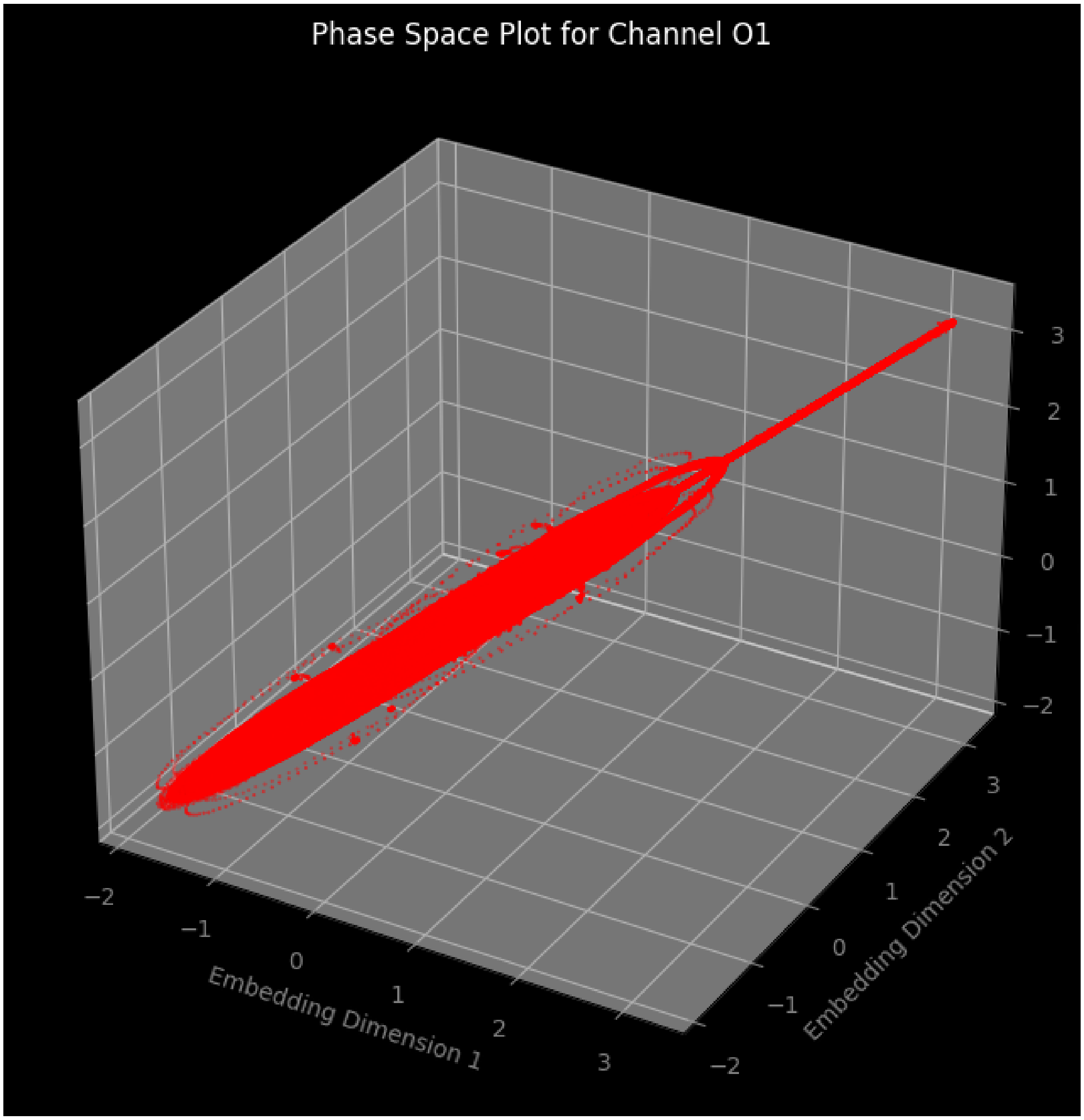
Phase Space Plot for Channel O1
Embedding Dimension 1
Embedding Dimension 2
−2
−1
0
1
2
3

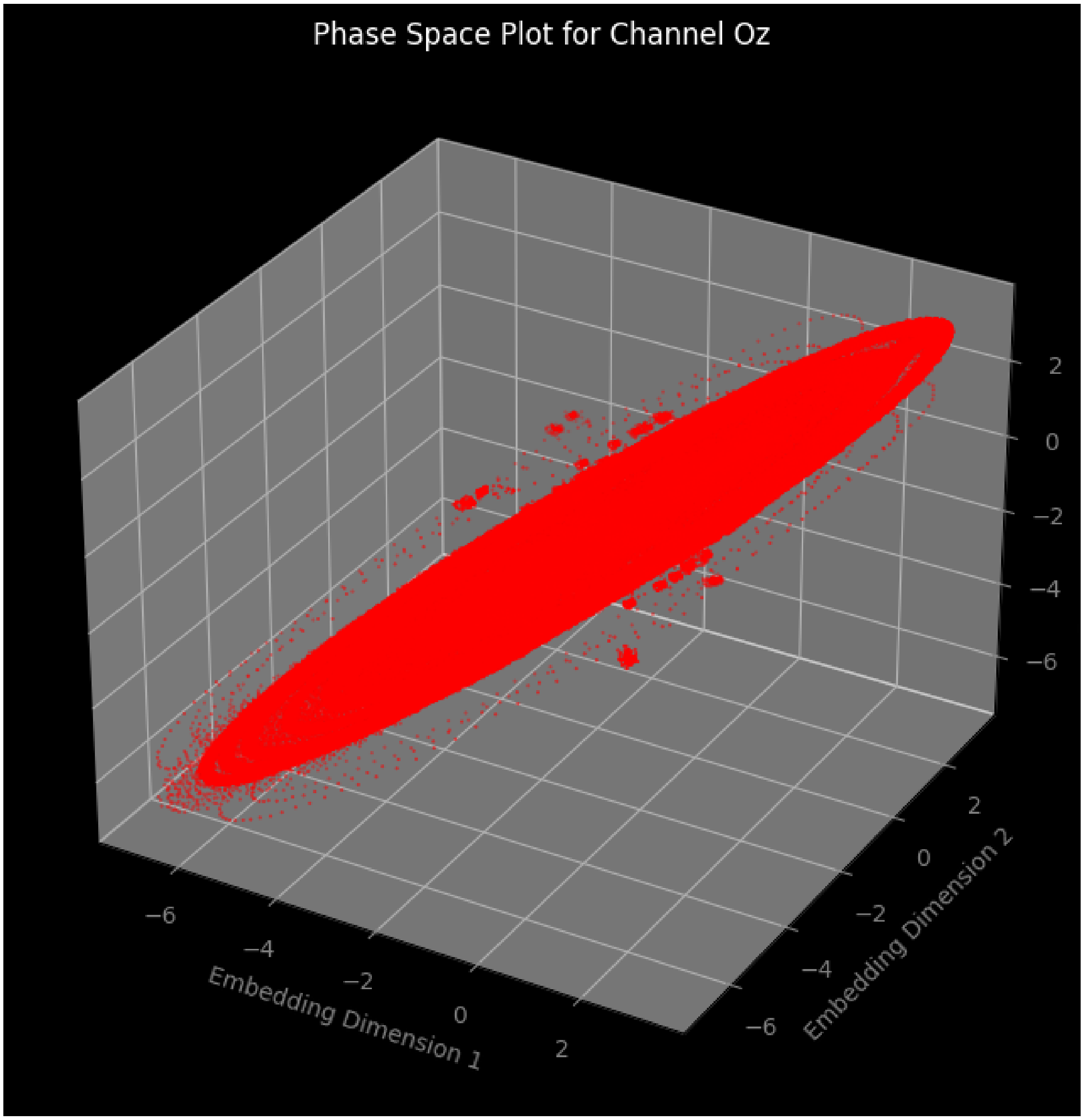
Phase Space Plot for Channel Oz
Embedding Dimension 1
Embedding Dimension 2
−6
−4
−2
0
2

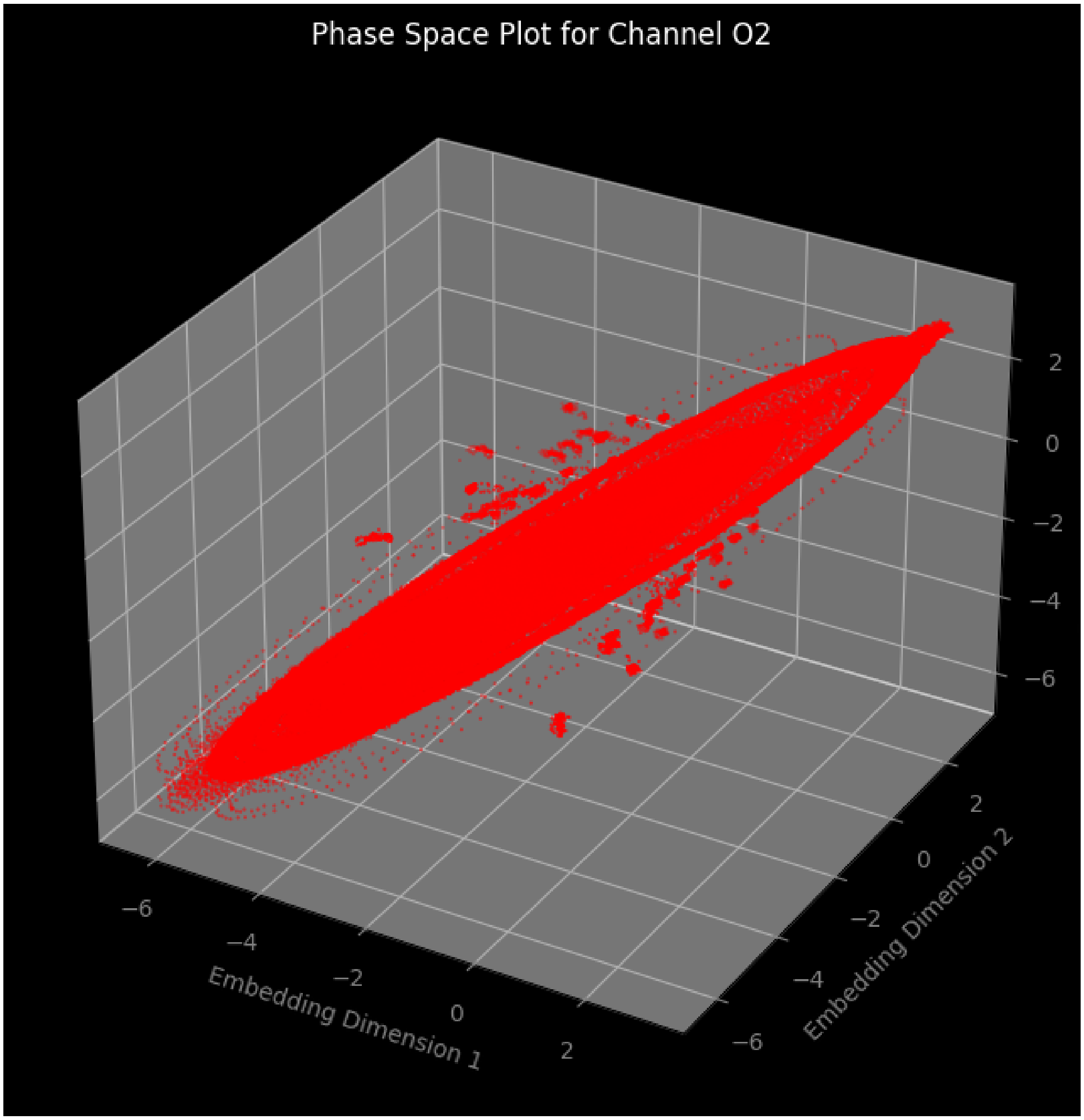
Phase Space Plot for Channel O2
Embedding Dimension 1
Embedding Dimension 2
−6
−4
−2
0
2

Phase Syncronization Analysis

September 8, 2023

# 1 Phase Synchonization Analysis

# 2 Phase Locking Values

[3]:
```
import numpy as np
import itertools
import matplotlib.pyplot as plt
from scipy.signal import hilbert

def compute_phase_locking_value(signal1, signal2):
    phase1 = np.angle(hilbert(signal1))
    phase2 = np.angle(hilbert(signal2))
    phase_diff = phase1 - phase2
    PLV = abs(np.sum(np.exp(1j * phase_diff))) / len(phase_diff)
    return PLV

# Load EEG data
EEG_data = np.load('/home/vincent/AAA_projects/MVCS/Neuroscience/
 ↪eeg_data_with_channels.npy', allow_pickle=True)

eeg_channel_names = ['Fp1', 'Fpz', 'Fp2', 'F7', 'F3', 'Fz', 'F4', 'F8', 'FC5',␣
 ↪'FC1', 'FC2', 'FC6',
                     'M1', 'T7', 'C3', 'Cz', 'C4', 'T8', 'M2', 'CP5', 'CP1',␣
 ↪'CP2', 'CP6',
                     'P7', 'P3', 'Pz', 'P4', 'P8', 'POz', 'O1', 'Oz', 'O2']

num_channels = len(eeg_channel_names)
plv_matrix = np.zeros((num_channels, num_channels))

for i, j in itertools.combinations(range(num_channels), 2):
    plv_matrix[i, j] = compute_phase_locking_value(EEG_data[:, i], EEG_data[:,␣
 ↪j])
    plv_matrix[j, i] = plv_matrix[i, j]

# Save the PLV matrix
save_path = "/home/vincent/AAA_projects/MVCS/Neuroscience/Analysis/Phase␣
 ↪Syncronization/plv_matrix.npy"
```

```python
np.save(save_path, plv_matrix)

# Plot the PLV matrix
plt.figure(figsize=(10, 10))
plt.imshow(plv_matrix, cmap="viridis", interpolation="none")
plt.colorbar(label="PLV")
plt.title("Phase Locking Value (PLV) between EEG channels")
plt.xticks(np.arange(num_channels), eeg_channel_names, rotation=90)
plt.yticks(np.arange(num_channels), eeg_channel_names)

# Save the plot
plot_save_path = "/home/vincent/AAA_projects/MVCS/Neuroscience/Analysis/Phase␣
↪Syncronization/plots/plv_plot.png"
plt.tight_layout()
plt.savefig(plot_save_path, dpi=300)
plt.show()
```

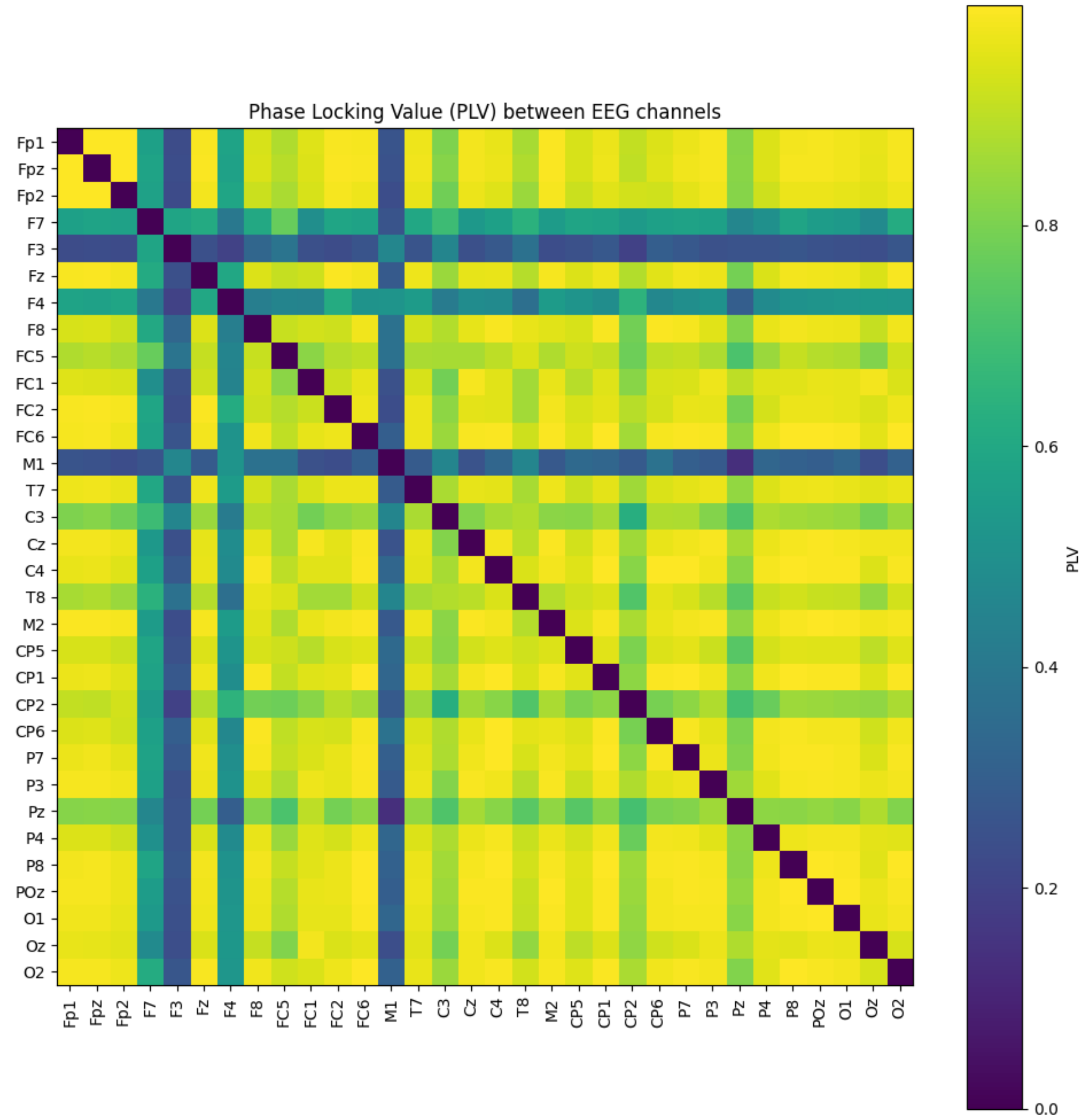
Phase Locking Value (PLV) between EEG channels
Fp1
Fpz
Fp2
F7
F3
Fz
F4
F8
FC5
FC1
FC2
FC6
M1
T7
C3
Cz
C4
T8
M2
CP5
CP1
CP2
CP6
P7
P3
Pz
P4
P8
POz
O1
Oz
O2
PLV
0.0
0.2
0.4
0.6
0.8

```
[ ]:
```

```
[ ]:
```

# Recurrence Quantification Analysis

September 8, 2023

## 1 Recurrence Quantification Analysis

#### 1.0.1 2D Recurrence Quantification

[ ]:
```
import numpy as np
import os
import pandas as pd
from pyrqa.settings import Settings
from pyrqa.neighbourhood import FixedRadius
from pyrqa.computation import RQAComputation
from pyrqa.time_series import TimeSeries
from pyrqa.metric import EuclideanMetric
import zipfile


eeg_channels = ['Fp1', 'Fpz', 'Fp2', 'F7', 'F3', 'Fz', 'F4', 'F8', 'FC5',␣
 ↪'FC1', 'FC2', 'FC6',
                'M1', 'T7', 'C3', 'Cz', 'C4', 'T8', 'M2', 'CP5', 'CP1', 'CP2',␣
 ↪'CP6',
                'P7', 'P3', 'Pz', 'P4', 'P8', 'POz', 'O1', 'Oz', 'O2']

zipped_file_path = '/home/vincent/AAA_projects/MVCS/Neuroscience/Analysis/Phase␣
 ↪Space/2dembedded_data.zip'
output_directory = '/home/vincent/AAA_projects/MVCS/Neuroscience/Analysis/
 ↪Recurrence Quantification'

def apply_channel_computation(args):
    return compute_rqa_for_channel(*args)

def compute_rqa_for_channel(channel_data, channel_name):
    time_series = TimeSeries(channel_data, embedding_dimension=2, time_delay=1)
    settings = Settings(time_series, neighbourhood=FixedRadius(0.1),␣
 ↪similarity_measure=EuclideanMetric(), theiler_corrector=1)
    computation = RQAComputation.create(settings)
    rqa_result = computation.run()

    metrics = {
```

```
        "RR": rqa_result.recurrence_rate,
        "DET": rqa_result.determinism,
        "LAM": rqa_result.laminarity,
        "L": rqa_result.average_diagonal_line,
        "Lmax": rqa_result.longest_diagonal_line,
        "L_entr": rqa_result.entropy_diagonal_lines,
        "TT": rqa_result.trapping_time,
        "RT": rqa_result.recurrence_time,
        "TREND": rqa_result.trend,
        "DET/RR": rqa_result.determinism / rqa_result.recurrence_rate if␣
↪rqa_result.recurrence_rate != 0 else 0
    }

    return channel_name, metrics

embedded_data = {}
with zipfile.ZipFile(zipped_file_path, 'r') as zipf:
    for channel in eeg_channels:
        data_file_name = f'2dembedded_{channel}_data.npy'
        with zipf.open(data_file_name) as data_file:
            embedded_data[channel] = np.load(data_file)

channels_with_data = [(embedded_data[channel], channel) for channel in␣
↪eeg_channels]

results = []
for channel_data, channel_name in [(embedded_data[channel], channel) for␣
↪channel in eeg_channels]:
    try:
        results.append(compute_rqa_for_channel(channel_data, channel_name))
    except Exception as e:
        print(f"An error occurred with channel {channel_name}: {e}")

if results:
    flattened_results = [(channel, *metrics.values()) for channel, metrics in␣
↪results]
    df_columns = ['Channel', 'RR', 'DET', 'LAM', 'L', 'Lmax', 'L_entr', 'TT',␣
↪'RT', 'TREND', 'DET/RR']
    df = pd.DataFrame(flattened_results, columns=df_columns)
    csv_path = os.path.join(output_directory, "rqa_results.csv")
    df.to_csv(csv_path, index=False)
    print("RQA analysis results saved in", csv_path)
else:
    print("No results to save.")
```

```
An error occurred with channel Fp1: module 'numpy' has no attribute 'float'.
`np.float` was a deprecated alias for the builtin `float`. To avoid this error
```

in existing code, use `float` by itself. Doing this will not modify any behavior and is safe. If you specifically wanted the numpy scalar type, use `np.float64` here.
The aliases was originally deprecated in NumPy 1.20; for more details and guidance see the original release note at:
    https://numpy.org/devdocs/release/1.20.0-notes.html#deprecations

### 1.0.2 Load 3D embedded from zip

[ ]:
```
import zipfile
import numpy as np
import os

# Define the path to the zipped file
zip_file_path = '/home/vincent/AAA_projects/MVCS/Neuroscience/Analysis/Phase␣
 ↪Space/3dembedded_data.zip'

# Directory to temporarily extract the numpy files
temp_extract_dir = '/home/vincent/temp_embedded_data/'

# Check if the temp directory exists, if not create it
if not os.path.exists(temp_extract_dir):
    os.mkdir(temp_extract_dir)

# Extract the contents of the zip file
with zipfile.ZipFile(zip_file_path, 'r') as zip_ref:
    zip_ref.extractall(temp_extract_dir)

# Load each numpy file and store them in a dictionary for easy access
embedded_data_dict = {}

for channel in eeg_channels:
    data_file_path = os.path.join(temp_extract_dir, f'3dembedded_{channel}.npy')
    embedded_data_dict[channel] = np.load(data_file_path)

# Optionally, remove the extracted .npy files to clean up after loading
for channel_file in os.listdir(temp_extract_dir):
    os.remove(os.path.join(temp_extract_dir, channel_file))

# Remove the temporary directory
os.rmdir(temp_extract_dir)

# Now, embedded_data_dict contains the loaded numpy arrays for each channel
```

### 1.0.3 3D Recurrence Quantification

[ ]:
```python
import numpy as np
from pyrqa.settings import Settings
from pyrqa.neighbourhood import FixedRadius
from pyrqa.computation import RQAComputation
from pyrqa.image_generator import ImageGenerator
from pyrqa.result import RQAResult
from pyrqa.time_series import TimeSeries
from pyrqa.metric import EuclideanMetric
from concurrent.futures import ThreadPoolExecutor

# Load the 3D embedding data for all EEG channels
eeg_channels = ['Fp1', 'Fpz', 'Fp2', 'F7', 'F3', 'Fz', 'F4', 'F8', 'FC5',␣
↪'FC1', 'FC2', 'FC6',
                'M1', 'T7', 'C3', 'Cz', 'C4', 'T8', 'M2', 'CP5', 'CP1', 'CP2',␣
↪'CP6',
                'P7', 'P3', 'Pz', 'P4', 'P8', 'POz', 'O1', 'Oz', 'O2']

embedding_data_directory = '/home/vincent/AAA_projects/MVCS/Neuroscience/
↪Analysis/Phase Space/3dembedding_data'
embedding_data_list = []

for channel in eeg_channels:
    embedded_channel_data = np.load(f'{embedding_data_directory}/
↪3dembedded_{channel}_data.npy')
    embedding_data_list.append(embedded_channel_data)

# Convert the embedded data into TimeSeries type for each channel
time_series_list = []

for channel_data in embedding_data_list:
    time_series_x = TimeSeries(channel_data[:, 0], embedding_dimension=1)
    time_series_y = TimeSeries(channel_data[:, 1], embedding_dimension=1)
    time_series_z = TimeSeries(channel_data[:, 2], embedding_dimension=1)
    time_series_list.append((time_series_x, time_series_y, time_series_z))

# RQA settings
radius = 0.5  # Adjust this based on your data's characteristics

def compute_rqa_for_channel(channel_time_series_data):
    channel, time_series_tuple = channel_time_series_data

    settings = Settings(time_series_tuple,
                        neighbourhood=FixedRadius(radius),
                        similarity_measure=EuclideanMetric(),
                        theiler_corrector=1)
```

```
    # Compute RQA
    computation = RQAComputation.create(settings)
    result = computation.run()

    # Print RQA results
    print(f"RQA results for {channel}:\n{result}")

    # Generate a recurrence plot
    ImageGenerator.save_recurrence_plot(result.recurrence_matrix_reverse,
                                        f'{embedding_data_directory}/plots/
↪recurrence_plot_{channel}.png')
    return result

# Using thread-based parallel processing to compute RQA for each channel
with ThreadPoolExecutor() as executor:
    rqa_results = list(executor.map(compute_rqa_for_channel, zip(eeg_channels,␣
↪time_series_list)))

print("RQA analysis and recurrence plots saved for all EEG channels.")
```

[ ]:

[ ]:

[6]:
```
import mne
import numpy as np

# Load EEG data
EEGdata = np.load('/home/vincent/AAA_projects/MVCS/Neuroscience/
↪eeg_data_with_channels.npy', allow_pickle=True)

# EEG channel names (hardcoded from previous code snippet)
eeg_channel_names = ['Fp1', 'Fpz', 'Fp2', 'F7', 'F3', 'Fz', 'F4', 'F8', 'FC5',␣
↪'FC1', 'FC2', 'FC6',
                     'M1', 'T7', 'C3', 'Cz', 'C4', 'T8', 'M2', 'CP5', 'CP1',␣
↪'CP2', 'CP6',
                     'P7', 'P3', 'Pz', 'P4', 'P8', 'POz', 'O1', 'Oz', 'O2']

# Transpose the EEG data if necessary
if EEGdata.shape[0] != len(eeg_channel_names):
    EEGdata = EEGdata.T

sfreq = 1000  # Your sampling frequency

ch_types = ['eeg'] * len(eeg_channel_names)
```

```python
info = mne.create_info(ch_names=eeg_channel_names, sfreq=sfreq,␣
 ↪ch_types=ch_types)

# Create Raw object
raw = mne.io.RawArray(EEGdata, info)

# Apply band-pass filter
raw.filter(l_freq=1., h_freq=50.)

# Standardize data (mean 0, variance 1)
raw_standardized = (raw.get_data() - np.mean(raw.get_data())) / np.std(raw.
 ↪get_data())

# If your data have multiple channels and you want to use RQA on each channel␣
 ↪separately,
# you should split your data here
eeg_data_split = np.split(raw_standardized, len(eeg_channel_names), axis=0)

# Now 'eeg_data_split' is a list of 1D numpy arrays, each one representing a␣
 ↪channel
# You can now use these arrays as inputs for your RQA computations
```

```
Creating RawArray with float64 data, n_channels=32, n_times=4227788
    Range : 0 … 4227787 =      0.000 …  4227.787 secs
Ready.
Filtering raw data in 1 contiguous segment
Setting up band-pass filter from 1 - 50 Hz

FIR filter parameters
---------------------
Designing a one-pass, zero-phase, non-causal bandpass filter:
- Windowed time-domain design (firwin) method
- Hamming window with 0.0194 passband ripple and 53 dB stopband attenuation
- Lower passband edge: 1.00
- Lower transition bandwidth: 1.00 Hz (-6 dB cutoff frequency: 0.50 Hz)
- Upper passband edge: 50.00 Hz
- Upper transition bandwidth: 12.50 Hz (-6 dB cutoff frequency: 56.25 Hz)
- Filter length: 3301 samples (3.301 s)


[Parallel(n_jobs=1)]: Using backend SequentialBackend with 1 concurrent workers.
[Parallel(n_jobs=1)]: Done   1 out of   1 | elapsed:    0.1s remaining:    0.0s
[Parallel(n_jobs=1)]: Done   2 out of   2 | elapsed:    0.1s remaining:    0.0s
[Parallel(n_jobs=1)]: Done   3 out of   3 | elapsed:    0.2s remaining:    0.0s
[Parallel(n_jobs=1)]: Done   4 out of   4 | elapsed:    0.2s remaining:    0.0s
[Parallel(n_jobs=1)]: Done  32 out of  32 | elapsed:    1.9s finished
```

[25]:
```
import matplotlib.pyplot as plt
from pyts.image import RecurrencePlot

# Downsample factor
downsample_factor = 10  # Further downsample the data

# Select a segment of the data
start_time = 1  # Adjust these to your needs
end_time = 10000

# Downsample the data and select a segment
downsampled_channel_data = EEGdata[0, start_time:end_time:downsample_factor]

# Create a RecurrencePlot object
rp = RecurrencePlot(dimension=embedding_dimension, time_delay=time_delay,␣
↪threshold='distance')

# Transform the time series into a Recurrence Plot
X_rp = rp.fit_transform(downsampled_channel_data.reshape(1, -1))

# Plot the result
fig, ax = plt.subplots(figsize=(8, 8))

# Set black background
fig.patch.set_facecolor('black')
ax.set_facecolor('black')

# Draw the recurrence plot
im = ax.imshow(X_rp[0], cmap='binary', origin='lower')

# Set title and labels with red color
ax.set_title('Recurrence Plot', fontsize=14, color='red')
ax.set_xlabel('Time Steps', color='red')
ax.set_ylabel('Time Steps', color='red')

# Change ticks color to dark grey
ax.tick_params(colors='darkgrey')

# Show the plot
plt.show()
```

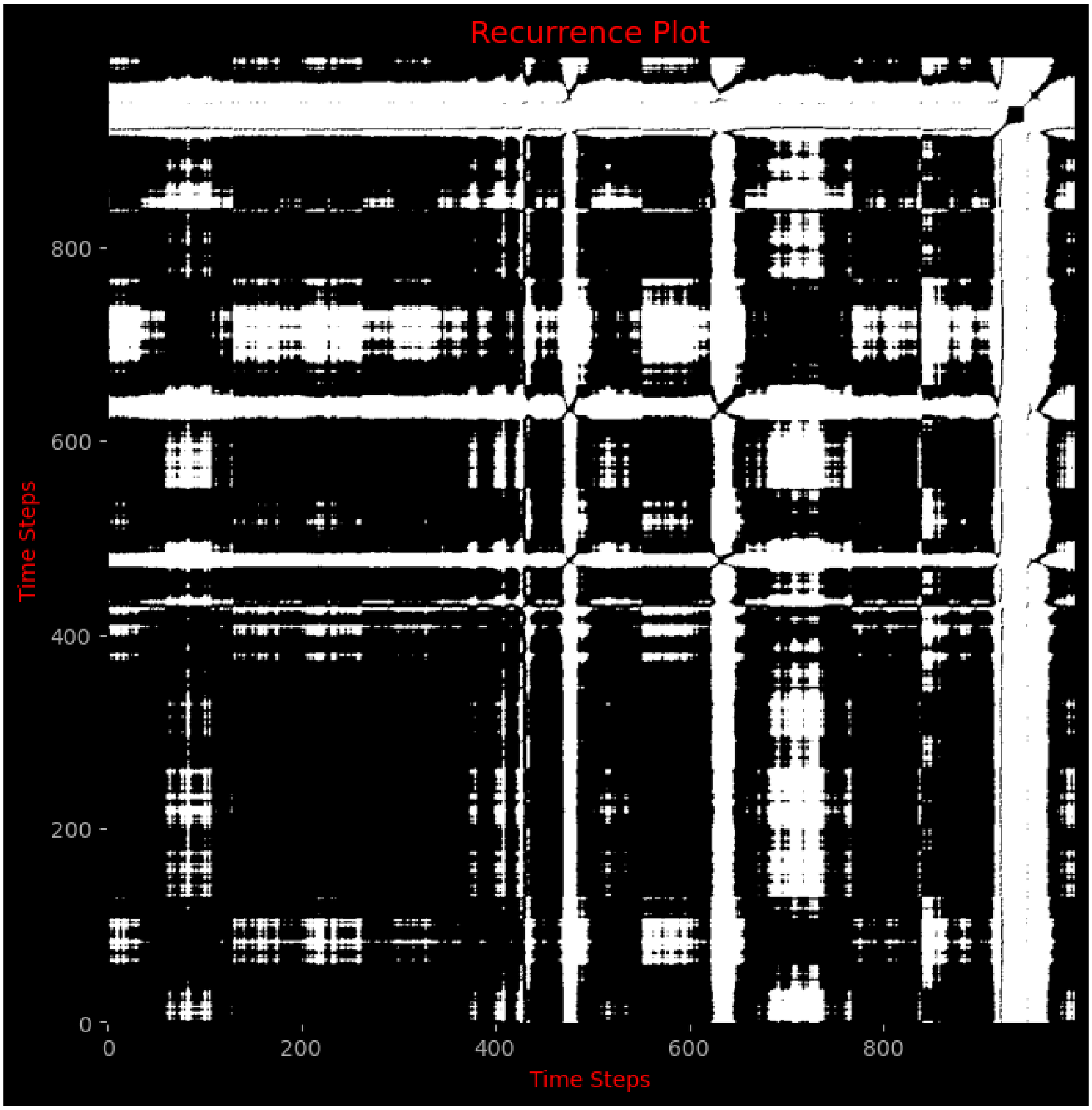
Recurrence Plot
Time Steps
Time Steps
0
200
400
600
800
0
200
400
600
800

# Arnold Tongue

September 8, 2023

## 1 Arnold Tongue

[3]:
```
import zipfile
import numpy as np
import os
# Path to the zipped 2D embedded data
zip_file_path = '/home/vincent/AAA_projects/MVCS/Neuroscience/Analysis/Phase␣
↪Space/2dembedded_data.zip'
extraction_dir = '/home/vincent/AAA_projects/MVCS/Neuroscience/Analysis/Phase␣
↪Space/temp_extraction'

# Extract zipped data to a temporary directory
with zipfile.ZipFile(zip_file_path, 'r') as zipf:
    zipf.extractall(extraction_dir)

eeg_channels = ['Fp1', 'Fpz', 'Fp2', 'F7', 'F3', 'Fz', 'F4', 'F8', 'FC5',␣
↪'FC1', 'FC2', 'FC6',
                'M1', 'T7', 'C3', 'Cz', 'C4', 'T8', 'M2', 'CP5', 'CP1', 'CP2',␣
↪'CP6',
                'P7', 'P3', 'Pz', 'P4', 'P8', 'POz', 'O1', 'Oz', 'O2']
```

## 2 ODEINT

## 3 BDF

[ ]:
```
import numpy as np
import matplotlib.pyplot as plt
from scipy.integrate import solve_ivp

def kuramoto_model(t, theta, omega, K, N, y):
    """ Kuramoto model to describe synchronization phenomena.
    y is the driving force with amplitude a and frequency b.
    """
    dydt = omega + (K/N) * np.sum(np.sin(theta - theta[:, np.newaxis]), axis=1)␣
↪+ y(t)
    return dydt
```

```python
def driving_force(t, a, b):
    """ Periodic driving force with amplitude 'a' and frequency 'b'. """
    return a * np.cos(2 * np.pi * b * t)

# Parameters
N = len(eeg_channels)  # Number of oscillators (one for each EEG channel)
K = 1.0  # Coupling strength
omega = np.random.uniform(-1, 1, N)  # Natural frequencies

# Ranges for amplitude 'a' and frequency 'b' of the driving force
a_values = np.linspace(0.1, 2.0, 50)
b_values = np.linspace(0.1, 2.0, 50)

# Store results
synchronization_array = np.zeros((len(a_values), len(b_values)))

# Iterate over values of a and b
for i, a in enumerate(a_values):
    for j, b in enumerate(b_values):
        y = lambda t: driving_force(t, a, b)

        # Integrate the Kuramoto model with the given driving force using BDF␣
↪method
        theta_0 = np.random.uniform(0, 2 * np.pi, N)
        T = [0, 100]
        solution = solve_ivp(kuramoto_model, T, theta_0, args=(omega, K, N, y),␣
↪method='BDF', t_eval=np.linspace(0, 100, 1000))

        # Measure synchronization
        r = np.abs(np.sum(np.exp(1j * solution.y), axis=0)) / N
        synchronization_array[i, j] = np.mean(r)

# Plot
plt.imshow(synchronization_array, extent=[a_values.min(), a_values.max(),␣
↪b_values.min(), b_values.max()], origin='lower', aspect='auto')
plt.colorbar(label="Synchronization")
plt.xlabel('Amplitude a')
plt.ylabel('Frequency b')
plt.title('Arnold Tongues')
plt.show()
```

## 4 Extract and save features

[ ]:
```
import numpy as np
import pandas as pd

# CNN Feature
cnn_feature = (synchronization_array - synchronization_array.min()) /␣
↪(synchronization_array.max() - synchronization_array.min())

# RNN Feature
rnn_feature = cnn_feature  # We can directly use the same normalized data;␣
↪during training, we would treat each row/column as a sequence.

# Save the features
save_path = '/home/vincent/AAA_projects/MVCS/Neuroscience/Features'
np.save(f"{save_path}/cnn_synch_array.npy", cnn_feature)
np.save(f"{save_path}/rnn_synch_array.npy", rnn_feature)

# Print shape and head
# For CNN feature
cnn_df = pd.DataFrame(cnn_feature)
print("\nCNN Feature Shape:")
print(cnn_df.shape)
print("\nCNN Feature Head:")
print(cnn_df.head())

# For RNN feature
rnn_df = pd.DataFrame(rnn_feature)
print("\nRNN Feature Shape:")
print(rnn_df.shape)
print("\nRNN Feature Head:")
print(rnn_df.head())
```

## 5 Mode Locked

[ ]:
```
import numpy as np
import matplotlib.pyplot as plt
from scipy.signal import hilbert
from matplotlib.colors import LinearSegmentedColormap
from multiprocessing import Pool

# Load EEG data
EEG_data = np.load('/home/vincent/AAA_projects/MVCS/Neuroscience/
↪eeg_data_with_channels.npy', allow_pickle=True)

# Channel labels
```

```
eeg_channels = ['Fp1', 'Fpz', 'Fp2', 'F7', 'F3', 'Fz', 'F4', 'F8', 'FC5',␣
↪'FC1', 'FC2', 'FC6',
                'M1', 'T7', 'C3', 'Cz', 'C4', 'T8', 'M2', 'CP5', 'CP1', 'CP2',␣
↪'CP6',
                'P7', 'P3', 'Pz', 'P4', 'P8', 'POz', 'O1', 'Oz', 'O2']

# Convert the EEG data to a dictionary
EEG_data_dict = {ch: EEG_data[:, i] for i, ch in enumerate(eeg_channels)}

# Extract the instantaneous phase from EEG data for each channel
instantaneous_phase_dict = {ch: np.angle(hilbert(data)) for ch, data in␣
↪EEG_data_dict.items()}

# Compute average phase across channels for each time point
avg_phase = np.mean(list(instantaneous_phase_dict.values()), axis=0)

# Define the circle map
def circle_map(theta, Omega, K):
    return theta + Omega - K / (2 * np.pi) * np.sin(2 * np.pi * theta)


# Reduce the number of iterations for mode-locking checking for speed-up (can␣
↪be adjusted)
ITERATIONS = 100

# Function to check mode-locking (revised to vectorize with numpy)
def is_mode_locked(theta, Omega, K, iterations=ITERATIONS, tol=1e-6):
    for _ in range(iterations):
        theta_next = circle_map(theta, Omega, K)
        if np.all(np.abs(theta_next - theta) < tol):
            return True
        theta = theta_next
    return False

# Function for mode-locking for a pair of Omega and K values
def mode_locking_for_pair(params):
    omega, K, phases = params
    locked_sum = 0
    for phase in phases:
        if is_mode_locked(phase, omega, K):
            locked_sum += 1
    return locked_sum / len(phases)

# Parameters
omegas = np.linspace(0, 1, 300)  # Omega values
K_values = np.linspace(0, 4 * np.pi, 300)  # K values
```

```
# Flatten the omega and K grid for parallel processing
omegas_grid, K_grid = np.meshgrid(omegas, K_values)
omegas_flat = omegas_grid.flatten()
K_flat = K_grid.flatten()

# Replicate the average phase for each pair of omega and K
phases_list = [avg_phase] * len(omegas_flat)

# Use multiprocessing pool to parallelize the calculation
with Pool() as pool:
    locked_flat = pool.map(mode_locking_for_pair, zip(omegas_flat, K_flat,␣
↪phases_list))

# Using all available CPU cores for multiprocessing
num_cpus = cpu_count()

# Use multiprocessing pool to parallelize the calculation
with Pool(num_cpus) as pool:
    locked_flat = pool.map(mode_locking_for_pair, zip(omegas_flat, K_flat,␣
↪phases_list))

# Reshape the locked_flat back to the grid shape
locked = np.array(locked_flat).reshape(omegas_grid.shape)

# Custom colormap
colors = [(1, 0, 1), (0, 0, 1)]  # R -> G -> B; Purple to Blue
cm = LinearSegmentedColormap.from_list('custom_cmap', colors, N=256)

# Plotting
plt.figure(figsize=(10, 7))
plt.imshow(locked, extent=(omegas.min(), omegas.max(), K_values.min(), K_values.
↪max()), aspect='auto', origin='lower', cmap=cm)
plt.colorbar(label='Proportion Mode-Locked')
plt.title('Arnold Tongues with EEG data')
plt.xlabel('Omega (Ω)')
plt.ylabel('K')

# Save the plot
plot_save_path = "/home/vincent/AAA_projects/MVCS/Neuroscience/Analysis/Arnold␣
↪Tongues/arnold_tongues_plot.png"
plt.savefig(plot_save_path)
print(f"Plot saved to {plot_save_path}")

# Show the plot
plt.show()

# Save the results to a .npy file
```

```
results_save_path = "/home/vincent/AAA_projects/MVCS/Neuroscience/Analysis/
 ↪Arnold Tongues/arnold_tongues_results.npy"
np.save(results_save_path, locked)
print(f"Results saved to {results_save_path}")
```

## 6 Extract and save features

[ ]:
```
import numpy as np
import pandas as pd

# Load the saved locked results
locked = np.load("/home/vincent/AAA_projects/MVCS/Neuroscience/Analysis/Arnold␣
 ↪Tongues/arnold_tongues_results.npy")

# Create normalized features from the locked results
# Normalize the locked data between 0 and 1 for neural network compatibility
normalized_locked = (locked - locked.min()) / (locked.max() - locked.min())

# CNN Feature
cnn_feature = normalized_locked

# RNN Feature
# Here we'll consider each row as a time series.
# So the RNN feature would be a collection of time series.
rnn_feature = normalized_locked

# Save the features
save_path = '/home/vincent/AAA_projects/MVCS/Neuroscience/Features'
np.save(f"{save_path}/cnn_modelocked.npy", cnn_feature)
np.save(f"{save_path}/rnn_modelocked.npy", rnn_feature)

# Print shape and head
# For CNN feature
cnn_df = pd.DataFrame(cnn_feature)
print("\nCNN Feature Shape:")
print(cnn_df.shape)
print("\nCNN Feature Head:")
print(cnn_df.head())

# For RNN feature
rnn_df = pd.DataFrame(rnn_feature)
print("\nRNN Feature Shape:")
print(rnn_df.shape)
print("\nRNN Feature Head:")
print(rnn_df.head())
```

## 7 Arnold tongues rotations

[20]:
```
import numpy as np
import matplotlib.pyplot as plt
import matplotlib.colors as mcolors

# Load EEG data
EEG_data = np.load('/home/vincent/AAA_projects/MVCS/Neuroscience/
↪eeg_data_with_channels.npy', allow_pickle=True)

# Channel labels
eeg_channels = ['Fp1', 'Fpz', 'Fp2', 'F7', 'F3', 'Fz', 'F4', 'F8', 'FC5',␣
↪'FC1', 'FC2', 'FC6',
                'M1', 'T7', 'C3', 'Cz', 'C4', 'T8', 'M2', 'CP5', 'CP1', 'CP2',␣
↪'CP6',
                'P7', 'P3', 'Pz', 'P4', 'P8', 'POz', 'O1', 'Oz', 'O2']

# Convert the EEG data to a dictionary
EEG_data_dict = {ch: EEG_data[:, i] for i, ch in enumerate(eeg_channels)}

# Define the circle map
def circle_map(theta, Omega, K):
    return theta + Omega - K / (2 * np.pi) * np.sin(2 * np.pi * theta)

# Compute average rotation number across all EEG channels
def average_rotation_number(omega, K, ch_data, iterations=1000, transient=100):
    theta = np.mean(ch_data)  # Use mean EEG value for initial condition
    for _ in range(transient):  # discard transient iterations
        theta = circle_map(theta, omega, K)

    diffs = []
    for _ in range(iterations):
        theta_next = circle_map(theta, omega, K)
        diffs.append(theta_next - theta)
        theta = theta_next
    return np.mean(diffs) % 1

# Parameters
omegas = np.linspace(0, 1, 300)
K_values = np.linspace(0, 2 * np.pi, 300)
rotation_numbers_dict = {}

# Compute rotation numbers for each channel
for ch, data in EEG_data_dict.items():
    rotation_numbers = np.zeros((len(K_values), len(omegas)))
    for i, K in enumerate(K_values):
        for j, omega in enumerate(omegas):
```

```
            rotation_numbers[i, j] = average_rotation_number(omega, K, data)
    rotation_numbers_dict[ch] = rotation_numbers

# Save rotation numbers
save_path = '/home/vincent/AAA_projects/MVCS/Neuroscience/Analysis/Arnold␣
 ↪Tongues/'
np.save(save_path + 'rotation_numbers.npy', rotation_numbers_dict)

# Define custom colormap
colors = [(0, 0, 0), (0, 1, 0), (1, 0, 0)]
cmap = mcolors.LinearSegmentedColormap.from_list('custom', colors, N=256)

# Plotting
rows = 4
cols = 8
fig, axs = plt.subplots(rows, cols, figsize=(20, 15), constrained_layout=True)

for ax, (ch, rotation_numbers) in zip(axs.flat, rotation_numbers_dict.items()):
    ax.imshow(rotation_numbers, extent=(omegas.min(), omegas.max(), K_values.
 ↪min(), K_values.max()),
              aspect='auto', origin='lower', cmap=cmap, clim=(0, 1))
    ax.set_title(ch)

# Remove unused subplots (if any)
for idx in range(len(eeg_channels), rows * cols):
    axs.flat[idx].axis('off')

# Save the plot
plt.savefig(save_path + 'arnold_tongues_plot.png', dpi=300)
plt.show()
```

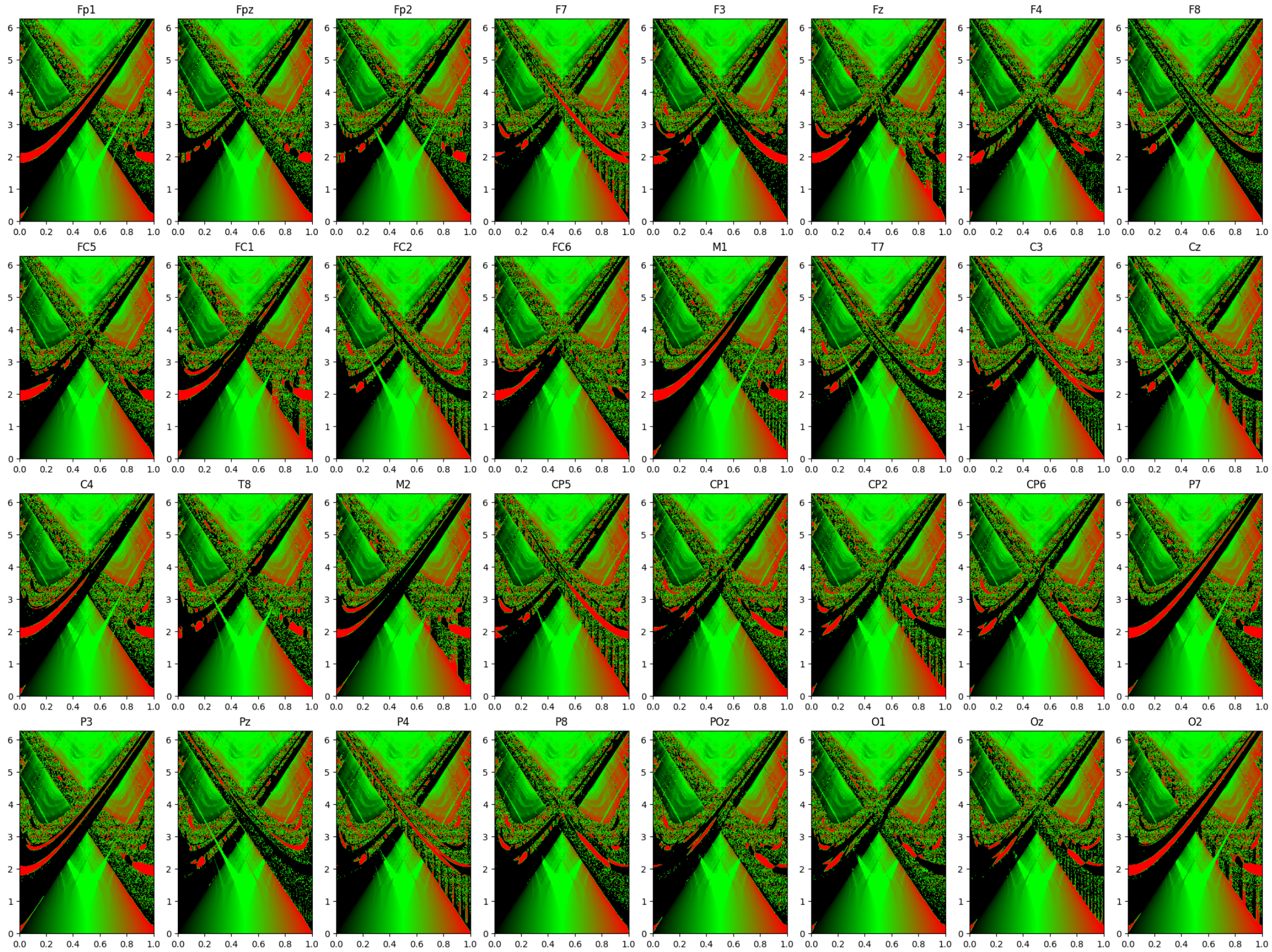

# 8 Extract and save features

[8]:
```python
import numpy as np
import pandas as pd

# Load rotation numbers
rotation_numbers_dict = np.load('/home/vincent/AAA_projects/MVCS/Neuroscience/
↪Analysis/Arnold Tongues/rotation_numbers.npy', allow_pickle=True).item()

# CNN Feature
# Convert the dictionary to a 3D array (channels x K_values x omegas)
cnn_features = np.array([rotation_numbers_dict[ch] for ch in eeg_channels])

# RNN Feature
# We consider each column in the rotation number for a channel as a time␣
↪series, and we transpose for RNN input.
rnn_features = np.transpose(cnn_features, (0, 2, 1))

# Save the features
save_path = '/home/vincent/AAA_projects/MVCS/Neuroscience/Features'
```

```
np.save(f"{save_path}/cnn_arnold_tongues_rotations.npy", cnn_features)
np.save(f"{save_path}/rnn_arnold_tongues_rotations.npy", rnn_features)

# Print shape and head using pandas
# For CNN feature
cnn_df = pd.DataFrame(cnn_features.reshape(cnn_features.shape[0], -1))
print("\nCNN Feature Shape:")
print(cnn_df.shape)
print("\nCNN Feature Head:")
print(cnn_df.head())

# For RNN feature
rnn_df = pd.DataFrame(rnn_features.reshape(rnn_features.shape[0], -1))
print("\nRNN Feature Shape:")
print(rnn_df.shape)
print("\nRNN Feature Head:")
print(rnn_df.head())
```

```
CNN Feature Shape:
(32, 90000)

CNN Feature Head:
   0         1         2         3         4         5         6      \
0    0.0  0.003344  0.006689  0.010033  0.013378  0.016722  0.020067
1    0.0  0.003344  0.006689  0.010033  0.013378  0.016722  0.020067
2    0.0  0.003344  0.006689  0.010033  0.013378  0.016722  0.020067
3    0.0  0.003344  0.006689  0.010033  0.013378  0.016722  0.020067
4    0.0  0.003344  0.006689  0.010033  0.013378  0.016722  0.020067

      7         8        9      …  89990  89991  89992  89993  89994  89995  \
0  0.023411  0.026756  0.0301  …    0.0    0.0    0.0    0.0    0.0    0.0
1  0.023411  0.026756  0.0301  …    0.0    0.0    0.0    0.0    0.0    0.0
2  0.023411  0.026756  0.0301  …    0.0    0.0    0.0    0.0    0.0    0.0
3  0.023411  0.026756  0.0301  …    0.0    0.0    0.0    0.0    0.0    0.0
4  0.023411  0.026756  0.0301  …    0.0    0.0    0.0    0.0    0.0    0.0

   89996  89997  89998         89999
0    0.0    0.0    0.0  6.915963e-07
1    0.0    0.0    0.0  9.999994e-01
2    0.0    0.0    0.0  4.546754e-07
3    0.0    0.0    0.0  5.371592e-07
4    0.0    0.0    0.0  4.784017e-07

[5 rows x 90000 columns]

RNN Feature Shape:
```

```
(32, 90000)

RNN Feature Head:
   0         1         2             3             4             5      \
0    0.0  0.999975  0.999997  9.999997e-01  1.000000e+00  1.000000e+00
1    0.0  0.999659  0.999930  9.999917e-01  9.999991e-01  9.999999e-01
2    0.0  0.999555  0.999800  9.999724e-01  9.999969e-01  9.999997e-01
3    0.0  0.000049  0.000006  6.322868e-07  6.621423e-08  6.581701e-09
4    0.0  0.000061  0.000007  7.965284e-07  8.370595e-08  8.350443e-09

          6             7             8             9      …     89990  \
0  1.000000e+00  1.000000e+00  1.000000e+00  1.000000e+00  …  0.997571
1  1.000000e+00  1.000000e+00  1.000000e+00  1.000000e+00  …  0.966209
2  1.000000e+00  1.000000e+00  1.000000e+00  1.000000e+00  …  0.971405
3  6.194421e-10  5.505308e-11  4.607045e-12  3.612968e-13  …  0.005906
4  7.888418e-10  7.037966e-11  5.913449e-12  4.664571e-13  …  0.961757

      89991     89992     89993     89994     89995     89996     89997  \
0  0.019064  0.011107  0.016453  0.036061  0.016125  0.012074  0.066612
1  0.968837  0.945844  0.979108  0.949785  0.976870  0.012531  0.034475
2  0.956600  0.992731  0.990966  0.000794  0.004616  0.013527  0.969350
3  0.974935  0.997710  0.995251  0.962983  0.965301  0.007396  0.926366
4  0.992619  0.953266  0.979591  0.022200  0.092059  0.013020  0.023107

      89998         89999
0  0.973163  6.915963e-07
1  0.930762  9.999994e-01
2  0.019783  4.546754e-07
3  0.025825  5.371592e-07
4  0.972908  4.784017e-07

[5 rows x 90000 columns]
```

[ ]:

[9]:
```python
import numpy as np
import matplotlib.pyplot as plt

# Circle map function
def circle_map(theta, Omega, K):
    return theta + Omega - K / (2 * np.pi) * np.sin(2 * np.pi * theta)

# Parameters
Omega = 1/3
K_values = np.linspace(0, 4 * np.pi, 300)
iterations = 50
```

```
# For each channel, apply the circle map and store the results
all_hist_data = []

for channel in range(EEG_data.shape[0]):
    theta_values_init = EEG_data[channel]
    theta_values_init = (theta_values_init - np.min(theta_values_init)) / (np.max(theta_values_init) - np.min(theta_values_init))

    hist_data = []

    for K in K_values:
        theta_values = theta_values_init.copy()
        for _ in range(iterations):
            theta_values = circle_map(theta_values, Omega, K)
        hist_data.append(theta_values)

    # Convert to numpy array and adjust for desired plotting range
    hist_data = np.array(hist_data)
    hist_data = np.mod(hist_data, 1)
    all_hist_data.append(hist_data)

# Now, you can plot for each channel or derive some aggregate metric across all channels.
# As a simplistic approach, we'll just average across all channels for plotting.
avg_hist_data = np.mean(all_hist_data, axis=0)

# Directory to save the results and plots
save_directory = "/home/vincent/AAA_projects/MVCS/Neuroscience/Analysis/Arnold Tongues"

# Save the averaged histogram data
np.save(os.path.join(save_directory, "avg_hist_data.npy"), avg_hist_data)

# Plotting
plt.figure(figsize=(10, 7))

# We can use the histogram approach to better visualize mode-locking behavior for each K value
for idx, K in enumerate(K_values):
    plt.hist(avg_hist_data[idx], bins=np.linspace(0, 1, 51), range=(0, 1), color='gray', alpha=0.3)

plt.title('Density of Converged Values for EEG data with Ω = 1/3')
plt.xlabel('Converged Values')
plt.ylabel('K')
# Save the plot
plt.savefig(os.path.join(save_directory, "Arnold_Tongues_Plot.png"))
```

```
plt.show()
```

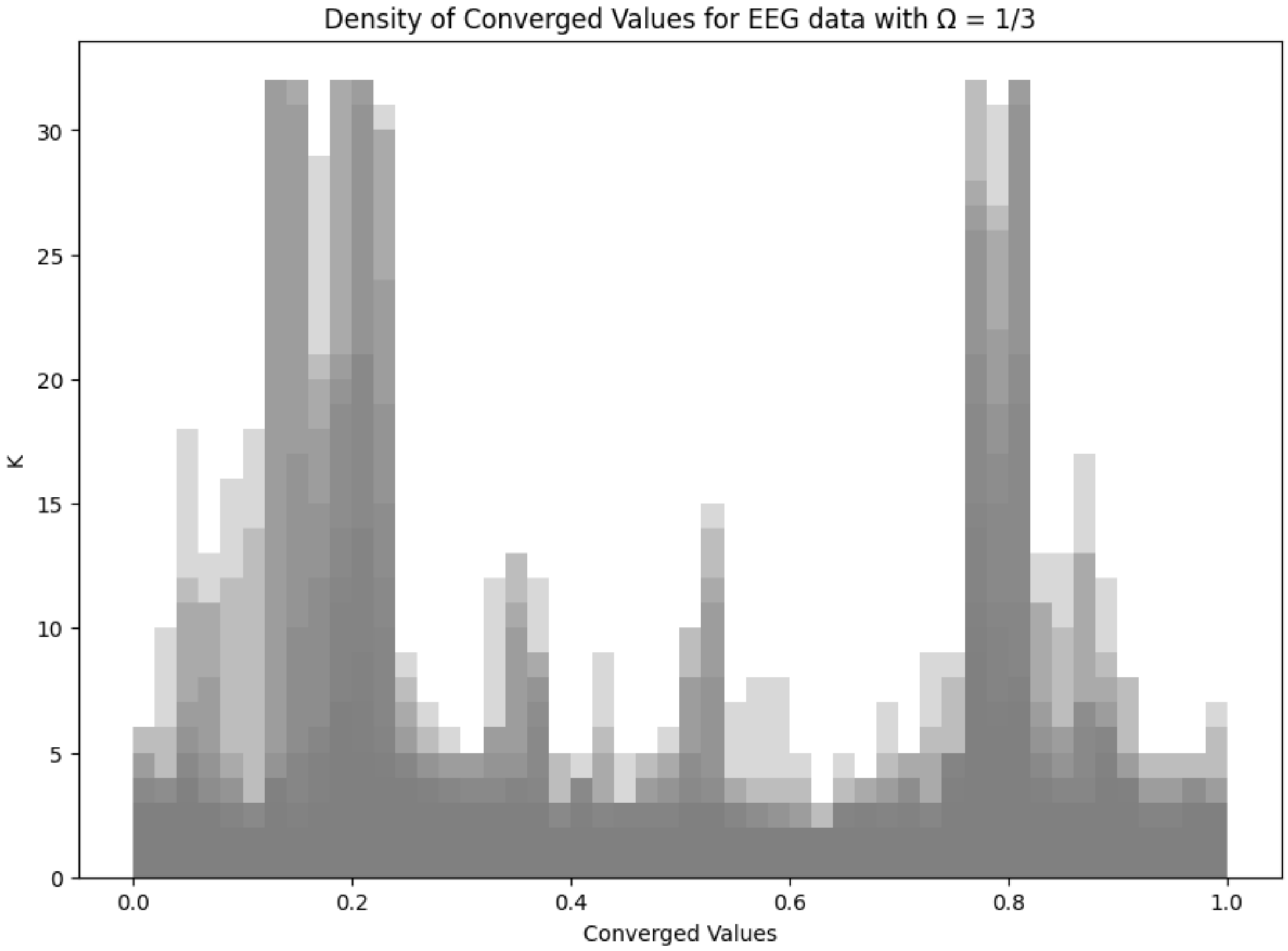

Density of Converged Values for EEG data with Ω = 1/3
K
Converged Values
0
5
10
15
20
25
30
0.0
0.2
0.4
0.6
0.8
1.0

[ ]:

```

```

Transfer Entropy

September 8, 2023

# 1 Transfer Entropy

# 2 Hemispheric

[17]:
```
import zipfile
import numpy as np
import os
import shutil
from pyinform import transfer_entropy

# Path to the zipped 2D embedded data
zip_file_path = '/home/vincent/AAA_projects/MVCS/Neuroscience/Analysis/Phase␣
↪Space/2dembedded_data.zip'
extraction_dir = '/home/vincent/AAA_projects/MVCS/Neuroscience/Analysis/Phase␣
↪Space/temp_extraction'

embedding_data_list_2D = []

# Extract zipped data to a temporary directory
with zipfile.ZipFile(zip_file_path, 'r') as zipf:
    zipf.extractall(extraction_dir)

eeg_channels = ['Fp1', 'Fpz', 'Fp2', 'F7', 'F3', 'Fz', 'F4', 'F8', 'FC5',␣
↪'FC1', 'FC2', 'FC6',
                'M1', 'T7', 'C3', 'Cz', 'C4', 'T8', 'M2', 'CP5', 'CP1', 'CP2',␣
↪'CP6',
                'P7', 'P3', 'Pz', 'P4', 'P8', 'POz', 'O1', 'Oz', 'O2']

for channel in eeg_channels:
    file_path = os.path.join(extraction_dir, f'2dembedded_{channel}_data.npy')
    embedding_data = np.load(file_path)
    embedding_data_list_2D.append(embedding_data)


# Set the desired length
desired_length = 4227688
```

```python
# Trim each array in embedding_data_list_2D to desired_length
embedding_data_list_2D_trimmed = [data[:desired_length] for data in␣
 ↪embedding_data_list_2D]

left_channels = ['Fp1', 'F7', 'F3', 'FC5', 'M1', 'T7', 'C3', 'CP5', 'P7', 'P3',␣
 ↪'O1']
right_channels = ['Fp2', 'F8', 'F4', 'FC6', 'M2', 'T8', 'C4', 'CP6', 'P8',␣
 ↪'P4', 'O2']
central_channels = ['Fpz', 'Fz', 'FC1', 'FC2', 'Cz', 'CP1', 'CP2', 'Pz', 'POz',␣
 ↪'Oz']

# Cleanup: Optionally remove the extraction directory after loading the data
# shutil.rmtree(extraction_dir)

# Group channels by hemisphere
left_indices = [eeg_channels.index(ch) for ch in left_channels]
right_indices = [eeg_channels.index(ch) for ch in right_channels]

# Compute average embeddings for left and right channels
left_embedding_avg = np.mean(np.array([embedding_data_list_2D_trimmed[i] for i␣
 ↪in left_indices]), axis=0)
right_embedding_avg = np.mean(np.array([embedding_data_list_2D_trimmed[i] for i␣
 ↪in right_indices]), axis=0)

# Use one of the dimensions (e.g., x-coordinate) for Transfer Entropy␣
 ↪calculation
left_data = left_embedding_avg[:, 0]
right_data = right_embedding_avg[:, 0]

def bin_data(data, num_bins):
    hist, bins = np.histogram(data, bins=num_bins)
    binned_data = np.digitize(data, bins[:-1]) - 1  # subtract 1 to start␣
 ↪binning from 0
    return binned_data

left_data_binned = bin_data(left_data, 1000)
right_data_binned = bin_data(right_data, 1000)

# Compute Transfer Entropy
k, l = 1, 1
try:
    TE_left_to_right = transfer_entropy(left_data_binned, right_data_binned, k)
    TE_right_to_left = transfer_entropy(right_data_binned, left_data_binned, k)

    print(f"Transfer Entropy from Left to Right (2D embeddings):␣
 ↪{TE_left_to_right}")
```

```
    print(f"Transfer Entropy from Right to Left (2D embeddings):␣
 ↪{TE_right_to_left}")
except Exception as e:
    print(f"Error computing Transfer Entropy: {e}")
```

```
Transfer Entropy from Left to Right (2D embeddings): 0.5814364131726154
Transfer Entropy from Right to Left (2D embeddings): 0.5016624825390452
```

## 3 Feature extraction

[27]:

```
WINDOW_SIZE = 1000  # Size of each window
STEP_SIZE = 500     # Size of each step between windows

def compute_transfer_entropy_sequence(left_data, right_data,␣
 ↪window_size=WINDOW_SIZE, step_size=STEP_SIZE):
    """
    Compute Transfer Entropy for sliding windows of the data.

    :param left_data: Left hemisphere data
    :param right_data: Right hemisphere data
    :param window_size: Size of each window
    :param step_size: Size of each step between windows
    :return: Sequence of Transfer Entropy values
    """
    te_sequence = []

    for start in range(0, len(left_data) - window_size + 1, step_size):
        window_left_data = left_data[start:start+window_size]
        window_right_data = right_data[start:start+window_size]

        window_left_binned = bin_data(window_left_data, 1000)
        window_right_binned = bin_data(window_right_data, 1000)

        te_left_to_right = transfer_entropy(window_left_binned,␣
 ↪window_right_binned, 1)
        te_sequence.append(te_left_to_right)

    return np.array(te_sequence)

# Compute the TE sequence for the averaged hemispheres
te_sequence = compute_transfer_entropy_sequence(left_data_binned,␣
 ↪right_data_binned)

# For CNNs, you may reshape this sequence to suit the model's input shape.
# Get the nearest square greater than or equal to len(te_sequence)
next_square = int(np.ceil(np.sqrt(len(te_sequence)))) ** 2
```

```
# Create a zero-filled square matrix of shape (sqrt(next_square),␣
↪sqrt(next_square))
cnn_input = np.zeros((int(np.sqrt(next_square)), int(np.sqrt(next_square))))

# Fill in the matrix with the values from te_sequence
cnn_input.flat[:len(te_sequence)] = te_sequence

# For RNNs, you can directly use te_sequence as a feature.
rnn_input = te_sequence

print(f"Shape for CNN input: {cnn_input.shape}")
print(f"Shape for RNN input: {rnn_input.shape}")
```

```
Shape for CNN input: (92, 92)
Shape for RNN input: (8454,)
```

## 4 Save features

[29]:
```
import os
import numpy as np

# Define the directory where you want to save the results
output_dir = "/home/vincent/AAA_projects/MVCS/Neuroscience/Analysis/Transfer␣
↪Entropy"

# Ensure the directory exists, if not, create it
if not os.path.exists(output_dir):
    os.makedirs(output_dir)

# Save the computed TE sequence
np.save(os.path.join(output_dir, 'transfer_entropy_hemispheric_avg.npy'),␣
↪te_sequence)

# Save the CNN input
np.save(os.path.join(output_dir, 'cnn_transfer_entropy_hemispheric_avg_input.
↪npy'), cnn_input)

# Save the RNN input (which is the same as the TE sequence in this case)
np.save(os.path.join(output_dir, 'rnn_transfer_entropy_hemispheric_avg_input.
↪npy'), rnn_input)

print(f"Data saved in {output_dir}")
```

```
Data saved in /home/vincent/AAA_projects/MVCS/Neuroscience/Analysis/Transfer
Entropy
```

## 5 Hemispherical Pairs

[24]:
```python
def compute_transfer_entropy_for_hemispheres(data, left_channels,␣
↪right_channels, window_size=1000, step_size=500):
    """
    Compute Transfer Entropy for pairs of channels between left and right␣
↪hemispheres over sliding windows of the data.

    :param data: EEG data
    :param left_channels: List of channels from the left hemisphere
    :param right_channels: List of channels from the right hemisphere
    :param window_size: Size of each window
    :param step_size: Size of each step between windows
    :return: Matrix of Transfer Entropy values; each row is a time window and␣
↪each column a channel pair
    """
    te_matrix = []

    for ch1 in left_channels:
        for ch2 in right_channels:
            te_values = []
            for start in range(0, len(data[0]) - window_size + 1, step_size):
                ch1_data = data[eeg_channels.index(ch1)][start:
↪start+window_size]
                ch2_data = data[eeg_channels.index(ch2)][start:
↪start+window_size]

                ch1_data_binned = bin_data(ch1_data[:, 0], 1000)
                ch2_data_binned = bin_data(ch2_data[:, 0], 1000)

                te_ch1_to_ch2 = transfer_entropy(ch1_data_binned,␣
↪ch2_data_binned, 1)
                te_values.append(te_ch1_to_ch2)

            te_matrix.append(te_values)

    return np.array(te_matrix)

# Compute transfer entropy for each channel pair between left and right␣
↪hemispheres
te_matrix =␣
↪compute_transfer_entropy_for_hemispheres(embedding_data_list_2D_trimmed,␣
↪left_channels, right_channels)

# This te_matrix is suitable for CNN as it's 2D.
```

```
# For RNN, you can consider each row of the te_matrix as a sequence of TE␣
↪values for a particular channel pairing over time.
```

```
---------------------------------------------------------------------------
KeyboardInterrupt                         Traceback (most recent call last)
Cell In[24], line 32
     29     return np.array(te_matrix)
     31 # Compute transfer entropy for each channel pair between left and right␣
↪hemispheres
---> 32 te_matrix =␣
↪compute_transfer_entropy_for_hemispheres(embedding_data_list_2D_trimmed, left_channels, rig
     34 # This te_matrix is suitable for CNN as it's 2D.
     35 # For RNN, you can consider each row of the te_matrix as a sequence of␣
↪TE values for a particular channel pairing over time.

Cell In[24], line 24, in compute_transfer_entropy_for_hemispheres(data,␣
↪left_channels, right_channels, window_size, step_size)
     21     ch1_data_binned = bin_data(ch1_data[:, 0], 1000)
     22     ch2_data_binned = bin_data(ch2_data[:, 0], 1000)
---> 24     te_ch1_to_ch2 = transfer_entropy(ch1_data_binned, ch2_data_binned,  )
     25     te_values.append(te_ch1_to_ch2)
     27 te_matrix.append(te_values)

File ~/miniconda3/lib/python3.10/site-packages/pyinform/transferentropy.py:220,␣
↪in transfer_entropy(source, target, k, condition, local)
    218     _local_transfer_entropy(ydata, xdata, cdata, c_ulong(z), c_ulong(n) ␣
↪c_ulong(m), c_int(b), c_ulong(k), out, byref(e))
    219 else:
--> 220     te =␣
↪_transfer_entropy(ydata, xdata, cdata, c_ulong(z), c_ulong(n), c_ulong(m), c_int(b), c_ulor
    222 error_guard(e)
    224 return te

KeyboardInterrupt:
```

[ ]:

## 6 Regional

[3]:

```
import zipfile
import numpy as np
import os
import shutil
from pyinform import transfer_entropy
```

```python
# Path to the zipped 2D embedded data
zip_file_path = '/home/vincent/AAA_projects/MVCS/Neuroscience/Analysis/Phase␣
↪Space/2dembedded_data.zip'
extraction_dir = '/home/vincent/AAA_projects/MVCS/Neuroscience/Analysis/Phase␣
↪Space/temp_extraction'

embedding_data_list_2D = []

# Extract zipped data to a temporary directory
with zipfile.ZipFile(zip_file_path, 'r') as zipf:
    zipf.extractall(extraction_dir)

eeg_channels = ['Fp1', 'Fpz', 'Fp2', 'F7', 'F3', 'Fz', 'F4', 'F8', 'FC5',␣
↪'FC1', 'FC2', 'FC6',
                'M1', 'T7', 'C3', 'Cz', 'C4', 'T8', 'M2', 'CP5', 'CP1', 'CP2',␣
↪'CP6',
                'P7', 'P3', 'Pz', 'P4', 'P8', 'POz', 'O1', 'Oz', 'O2']

for channel in eeg_channels:
    file_path = os.path.join(extraction_dir, f'2dembedded_{channel}_data.npy')
    embedding_data = np.load(file_path)
    embedding_data_list_2D.append(embedding_data)

desired_length = 4227688
embedding_data_list_2D_trimmed = [data[:desired_length] for data in␣
↪embedding_data_list_2D]

# Brain regions
frontal_channels = ['Fp1', 'Fpz', 'Fp2', 'F7', 'F3', 'Fz', 'F4', 'F8']
temporal_channels = ['T7', 'T8']
parietal_channels = ['CP5', 'CP1', 'CP2', 'CP6', 'P7', 'P3', 'Pz', 'P4', 'P8']
occipital_channels = ['O1', 'Oz', 'O2']

frontal_indices = [eeg_channels.index(ch) for ch in frontal_channels]
temporal_indices = [eeg_channels.index(ch) for ch in temporal_channels]
parietal_indices = [eeg_channels.index(ch) for ch in parietal_channels]
occipital_indices = [eeg_channels.index(ch) for ch in occipital_channels]

# Getting average for each region
frontal_data_avg = np.mean([embedding_data_list_2D_trimmed[i] for i in␣
↪frontal_indices], axis=0)
temporal_data_avg = np.mean([embedding_data_list_2D_trimmed[i] for i in␣
↪temporal_indices], axis=0)
parietal_data_avg = np.mean([embedding_data_list_2D_trimmed[i] for i in␣
↪parietal_indices], axis=0)
```

```python
occipital_data_avg = np.mean([embedding_data_list_2D_trimmed[i] for i in␣
↪occipital_indices], axis=0)

# Use one of the dimensions for Transfer Entropy calculation
frontal_data = frontal_data_avg[:, 0]
temporal_data = temporal_data_avg[:, 0]
parietal_data = parietal_data_avg[:, 0]
occipital_data = occipital_data_avg[:, 0]

def bin_data(data, num_bins):
    hist, bins = np.histogram(data, bins=num_bins)
    binned_data = np.digitize(data, bins[:-1]) - 1
    return binned_data

frontal_data_binned = bin_data(frontal_data, 1000)
temporal_data_binned = bin_data(temporal_data, 1000)
parietal_data_binned = bin_data(parietal_data, 1000)
occipital_data_binned = bin_data(occipital_data, 1000)

# Store the regional data in a dictionary for easier processing
region_data_binned = {
    "Frontal": frontal_data_binned,
    "Temporal": temporal_data_binned,
    "Parietal": parietal_data_binned,
    "Occipital": occipital_data_binned
}

k = 1


# Directory to save the transfer entropy results
results_dir = '/home/vincent/AAA_projects/MVCS/Neuroscience/Analysis/Transfer␣
↪Entropy'
if not os.path.exists(results_dir):
    os.makedirs(results_dir)  # Create the directory if it doesn't exist

# Dictionary to store results
TE_results = {}

# Compute and print the transfer entropy from each region to every other region
for source_region, source_data in region_data_binned.items():
    for target_region, target_data in region_data_binned.items():
        if source_region != target_region:
            try:
                TE_value = transfer_entropy(source_data, target_data, k)
                key = f"{source_region}_to_{target_region}"
                TE_results[key] = TE_value
```

```
                print(f"Transfer Entropy from {source_region} to␣
 ↪{target_region} (2D embeddings): {TE_value}")
            except Exception as e:
                print(f"Error computing Transfer Entropy from {source_region}␣
 ↪to {target_region}: {e}")

# Save the computed Transfer Entropy results
results_file_path = os.path.join(results_dir,␣
 ↪'regional_transfer_entropy_results.npy')
np.save(results_file_path, TE_results)

print(f"Transfer entropy results saved to: {results_file_path}")
```

```
Transfer Entropy from Frontal to Temporal (2D embeddings): 0.25110661733698914
Transfer Entropy from Frontal to Parietal (2D embeddings): 0.5773050773854287
Transfer Entropy from Frontal to Occipital (2D embeddings): 0.6741032253719645
Transfer Entropy from Temporal to Frontal (2D embeddings): 0.33752969048400794
Transfer Entropy from Temporal to Parietal (2D embeddings): 0.4711875861147353
Transfer Entropy from Temporal to Occipital (2D embeddings): 0.6126095442734832
Transfer Entropy from Parietal to Frontal (2D embeddings): 0.38813154044068165
Transfer Entropy from Parietal to Temporal (2D embeddings): 0.22015981791417624
Transfer Entropy from Parietal to Occipital (2D embeddings): 0.6353775260357438
Transfer Entropy from Occipital to Frontal (2D embeddings): 0.42216948662291265
Transfer Entropy from Occipital to Temporal (2D embeddings): 0.2811912780352561
Transfer Entropy from Occipital to Parietal (2D embeddings): 0.5849692342942411
Transfer entropy results saved to:
/home/vincent/AAA_projects/MVCS/Neuroscience/Analysis/Transfer
Entropy/regional_transfer_entropy_results.npy
```

## 7 Full Granularity

[4]:

```
import zipfile
import numpy as np
import os
from pyinform import transfer_entropy
import shutil

# Paths and constants
zip_file_path = '/home/vincent/AAA_projects/MVCS/Neuroscience/Analysis/Phase␣
 ↪Space/2dembedded_data.zip'
extraction_dir = '/home/vincent/AAA_projects/MVCS/Neuroscience/Analysis/Phase␣
 ↪Space/temp_extraction'
results_dir = '/home/vincent/AAA_projects/MVCS/Neuroscience/Analysis/Transfer␣
 ↪Entropy'
results_file_name = 'full_granularity_transfer_entropy_results.npy'
```

```python
eeg_channels = ['Fp1', 'Fpz', 'Fp2', 'F7', 'F3', 'Fz', 'F4', 'F8', 'FC5',␣
↪'FC1', 'FC2', 'FC6',
                'M1', 'T7', 'C3', 'Cz', 'C4', 'T8', 'M2', 'CP5', 'CP1', 'CP2',␣
↪'CP6',
                'P7', 'P3', 'Pz', 'P4', 'P8', 'POz', 'O1', 'Oz', 'O2']

desired_length = 4227688
num_bins = 1000
k = 1

# Helper functions
def extract_data(zip_path, extraction_path):
    with zipfile.ZipFile(zip_path, 'r') as zipf:
        zipf.extractall(extraction_path)
    return extraction_path

def load_embedded_data(channels, extraction_path):
    data = []
    for channel in channels:
        file_path = os.path.join(extraction_path, f'2dembedded_{channel}_data.
↪npy')
        embedding_data = np.load(file_path)
        data.append(embedding_data)
    return data

def bin_data(data, bins):
    hist, bin_edges = np.histogram(data, bins=bins)
    binned_data = np.digitize(data, bin_edges[:-1]) - 1  # subtract 1 to start␣
↪binning from 0
    return binned_data

# Create results directory if it doesn't exist
if not os.path.exists(results_dir):
    os.makedirs(results_dir)

# Main execution
extract_data(zip_file_path, extraction_dir)
embedding_data_list_2D = load_embedded_data(eeg_channels, extraction_dir)

# Trimming and binning the data
embedding_data_list_2D_trimmed = [data[:desired_length, 0] for data in␣
↪embedding_data_list_2D]
embedding_data_list_2D_binned = [bin_data(data, num_bins) for data in␣
↪embedding_data_list_2D_trimmed]

# Dictionary to store transfer entropy results
```

```python
TE_results = {}

# Compute Transfer Entropy for all pairs
for i, source_channel in enumerate(eeg_channels):
    for j, target_channel in enumerate(eeg_channels):
        if i != j:  # To avoid computing Transfer Entropy for the same channel
            try:
                TE = transfer_entropy(embedding_data_list_2D_binned[i],␣
 ↪embedding_data_list_2D_binned[j], k)
                key = f"{source_channel}_to_{target_channel}"
                TE_results[key] = TE
                print(f"Transfer Entropy from {source_channel} to␣
 ↪{target_channel}: {TE}")
            except Exception as e:
                print(f"Error computing Transfer Entropy from {source_channel}␣
 ↪to {target_channel}: {e}")

# Save the computed Transfer Entropy results to the specified directory
results_file_path = os.path.join(results_dir, results_file_name)
np.save(results_file_path, TE_results)

print(f"Transfer entropy results saved to: {results_file_path}")

# Optionally remove the extraction directory after loading the data
shutil.rmtree(extraction_dir)
```

```
Transfer Entropy from Fp1 to Fpz: 0.40323706303475343
Transfer Entropy from Fp1 to Fp2: 0.2818704697781004
Transfer Entropy from Fp1 to F7: 0.25450640673035146
Transfer Entropy from Fp1 to F3: 0.3920366504025133
Transfer Entropy from Fp1 to Fz: 0.340570439690493
Transfer Entropy from Fp1 to F4: 0.36684320006698923
Transfer Entropy from Fp1 to F8: 0.3814174869152535
Transfer Entropy from Fp1 to FC5: 0.4008950684194118
Transfer Entropy from Fp1 to FC1: 0.4868985959943641
Transfer Entropy from Fp1 to FC2: 0.5637154724066433
Transfer Entropy from Fp1 to FC6: 0.49389085353899803
Transfer Entropy from Fp1 to M1: 0.47221756489424793
Transfer Entropy from Fp1 to T7: 0.32894960730094375
Transfer Entropy from Fp1 to C3: 0.38167164803444587
Transfer Entropy from Fp1 to Cz: 0.6011588901131365
Transfer Entropy from Fp1 to C4: 0.4697287943283873
Transfer Entropy from Fp1 to T8: 0.5130112097503553
Transfer Entropy from Fp1 to M2: 0.5672168204033003
Transfer Entropy from Fp1 to CP5: 0.23824754062478157
Transfer Entropy from Fp1 to CP1: 0.5927575243094223
Transfer Entropy from Fp1 to CP2: 0.7304516712278878
```

Transfer Entropy from Fp1 to CP6: 0.5333527182744573
Transfer Entropy from Fp1 to P7: 0.5848131842038433
Transfer Entropy from Fp1 to P3: 0.6223911000280086
Transfer Entropy from Fp1 to Pz: 0.8419868187140546
Transfer Entropy from Fp1 to P4: 0.741978782470686
Transfer Entropy from Fp1 to P8: 0.5263928153052267
Transfer Entropy from Fp1 to POz: 0.7199603925232134
Transfer Entropy from Fp1 to O1: 0.6084761791049695
Transfer Entropy from Fp1 to Oz: 0.6292160960469199
Transfer Entropy from Fp1 to O2: 0.6162789021465184
Transfer Entropy from Fpz to Fp1: 0.22548421233814356
Transfer Entropy from Fpz to Fp2: 0.29535056198551324
Transfer Entropy from Fpz to F7: 0.2728162785141356
Transfer Entropy from Fpz to F3: 0.45271738327135497
Transfer Entropy from Fpz to Fz: 0.3705970672961796
Transfer Entropy from Fpz to F4: 0.39432191077563183
Transfer Entropy from Fpz to F8: 0.4504715834492651
Transfer Entropy from Fpz to FC5: 0.44204977680335245
Transfer Entropy from Fpz to FC1: 0.4992650572589718
Transfer Entropy from Fpz to FC2: 0.5845975781662247
Transfer Entropy from Fpz to FC6: 0.5319088200449946
Transfer Entropy from Fpz to M1: 0.5159233442783275
Transfer Entropy from Fpz to T7: 0.34237604979236935
Transfer Entropy from Fpz to C3: 0.425376402632431
Transfer Entropy from Fpz to Cz: 0.6280055415339607
Transfer Entropy from Fpz to C4: 0.4962428042428292
Transfer Entropy from Fpz to T8: 0.5601906232950797
Transfer Entropy from Fpz to M2: 0.619271715497946
Transfer Entropy from Fpz to CP5: 0.25440289170605873
Transfer Entropy from Fpz to CP1: 0.6628801471616039
Transfer Entropy from Fpz to CP2: 0.8008384697272103
Transfer Entropy from Fpz to CP6: 0.5650311929970476
Transfer Entropy from Fpz to P7: 0.6536962363897626
Transfer Entropy from Fpz to P3: 0.6977601985654825
Transfer Entropy from Fpz to Pz: 0.889685555984976
Transfer Entropy from Fpz to P4: 0.7971261643508482
Transfer Entropy from Fpz to P8: 0.5727492123229471
Transfer Entropy from Fpz to POz: 0.7609620316667555
Transfer Entropy from Fpz to O1: 0.6206092451968968
Transfer Entropy from Fpz to Oz: 0.6929789947962747
Transfer Entropy from Fpz to O2: 0.6671739801660761
Transfer Entropy from Fp2 to Fp1: 0.22253144619849255
Transfer Entropy from Fp2 to Fpz: 0.4151031562687186
Transfer Entropy from Fp2 to F7: 0.2959589968258271
Transfer Entropy from Fp2 to F3: 0.4708117467388301
Transfer Entropy from Fp2 to Fz: 0.3988820131932256
Transfer Entropy from Fp2 to F4: 0.44835964986756977
Transfer Entropy from Fp2 to F8: 0.4599383774702956

Transfer Entropy from Fp2 to FC5: 0.4721761008201325
Transfer Entropy from Fp2 to FC1: 0.549185894698739
Transfer Entropy from Fp2 to FC2: 0.6349349586694903
Transfer Entropy from Fp2 to FC6: 0.5384525072092683
Transfer Entropy from Fp2 to M1: 0.5263852605331477
Transfer Entropy from Fp2 to T7: 0.3703422497183083
Transfer Entropy from Fp2 to C3: 0.4440234007863593
Transfer Entropy from Fp2 to Cz: 0.6793640737691301
Transfer Entropy from Fp2 to C4: 0.494082674954434
Transfer Entropy from Fp2 to T8: 0.5652986856228639
Transfer Entropy from Fp2 to M2: 0.623616768434703
Transfer Entropy from Fp2 to CP5: 0.2972795221100576
Transfer Entropy from Fp2 to CP1: 0.6766718698004434
Transfer Entropy from Fp2 to CP2: 0.8371871287177642
Transfer Entropy from Fp2 to CP6: 0.562561096852585
Transfer Entropy from Fp2 to P7: 0.6631875126940144
Transfer Entropy from Fp2 to P3: 0.7026918646703283
Transfer Entropy from Fp2 to Pz: 0.9608386698956521
Transfer Entropy from Fp2 to P4: 0.8017619570836667
Transfer Entropy from Fp2 to P8: 0.5746023873839303
Transfer Entropy from Fp2 to POz: 0.752846871809526
Transfer Entropy from Fp2 to O1: 0.6268262361141601
Transfer Entropy from Fp2 to Oz: 0.7097136754151835
Transfer Entropy from Fp2 to O2: 0.6845591879800277
Transfer Entropy from F7 to Fp1: 0.209224736645039
Transfer Entropy from F7 to Fpz: 0.427079720994545
Transfer Entropy from F7 to Fp2: 0.3118092032414955
Transfer Entropy from F7 to F3: 0.33288594898698215
Transfer Entropy from F7 to Fz: 0.3000542930570356
Transfer Entropy from F7 to F4: 0.34980948427680597
Transfer Entropy from F7 to F8: 0.35892417857510517
Transfer Entropy from F7 to FC5: 0.2899615436500479
Transfer Entropy from F7 to FC1: 0.4174725539041163
Transfer Entropy from F7 to FC2: 0.5107229461509603
Transfer Entropy from F7 to FC6: 0.4911189664837978
Transfer Entropy from F7 to M1: 0.4673645712077154
Transfer Entropy from F7 to T7: 0.27471370091822567
Transfer Entropy from F7 to C3: 0.29553752616052786
Transfer Entropy from F7 to Cz: 0.5057262221139318
Transfer Entropy from F7 to C4: 0.4899128913209771
Transfer Entropy from F7 to T8: 0.48247863762864207
Transfer Entropy from F7 to M2: 0.46220881046253354
Transfer Entropy from F7 to CP5: 0.17243649607056385
Transfer Entropy from F7 to CP1: 0.49161749487330897
Transfer Entropy from F7 to CP2: 0.5457029981636722
Transfer Entropy from F7 to CP6: 0.5310947742973297
Transfer Entropy from F7 to P7: 0.5367681550726636
Transfer Entropy from F7 to P3: 0.5201898680210068

Transfer Entropy from F7 to Pz: 0.7026634818914608
Transfer Entropy from F7 to P4: 0.5676616328450815
Transfer Entropy from F7 to P8: 0.5021918223810214
Transfer Entropy from F7 to POz: 0.5865800437226907
Transfer Entropy from F7 to O1: 0.5083477071393259
Transfer Entropy from F7 to Oz: 0.5238700528348819
Transfer Entropy from F7 to O2: 0.5156330051788834
Transfer Entropy from F3 to Fp1: 0.2651447541394184
Transfer Entropy from F3 to Fpz: 0.5259369841598357
Transfer Entropy from F3 to Fp2: 0.39939512893641066
Transfer Entropy from F3 to F7: 0.23154387513362276
Transfer Entropy from F3 to Fz: 0.3441846498135728
Transfer Entropy from F3 to F4: 0.4141160298990385
Transfer Entropy from F3 to F8: 0.4171993976843795
Transfer Entropy from F3 to FC5: 0.3218769419099474
Transfer Entropy from F3 to FC1: 0.4712629740370795
Transfer Entropy from F3 to FC2: 0.5399011048476986
Transfer Entropy from F3 to FC6: 0.5711178492240735
Transfer Entropy from F3 to M1: 0.545520262200558
Transfer Entropy from F3 to T7: 0.28283591535388636
Transfer Entropy from F3 to C3: 0.3522619247972086
Transfer Entropy from F3 to Cz: 0.5620882102184691
Transfer Entropy from F3 to C4: 0.5755898623520428
Transfer Entropy from F3 to T8: 0.5633592210358906
Transfer Entropy from F3 to M2: 0.5405512905318697
Transfer Entropy from F3 to CP5: 0.1876820610957885
Transfer Entropy from F3 to CP1: 0.619760519107217
Transfer Entropy from F3 to CP2: 0.7202108894519004
Transfer Entropy from F3 to CP6: 0.6295818062477966
Transfer Entropy from F3 to P7: 0.6091414061649228
Transfer Entropy from F3 to P3: 0.554641906902627
Transfer Entropy from F3 to Pz: 0.726373859113473
Transfer Entropy from F3 to P4: 0.7049984174036981
Transfer Entropy from F3 to P8: 0.5874718064291378
Transfer Entropy from F3 to POz: 0.7445814986336887
Transfer Entropy from F3 to O1: 0.6234449285247656
Transfer Entropy from F3 to Oz: 0.5695312892033086
Transfer Entropy from F3 to O2: 0.545099083772171
Transfer Entropy from Fz to Fp1: 0.24221807468398338
Transfer Entropy from Fz to Fpz: 0.45796095840297535
Transfer Entropy from Fz to Fp2: 0.3491674607300556
Transfer Entropy from Fz to F7: 0.24413404282676826
Transfer Entropy from Fz to F3: 0.38998596556393417
Transfer Entropy from Fz to F4: 0.36822219654546046
Transfer Entropy from Fz to F8: 0.4181566754429746
Transfer Entropy from Fz to FC5: 0.3703491567190001
Transfer Entropy from Fz to FC1: 0.42485324784106754
Transfer Entropy from Fz to FC2: 0.5350318296342667

Transfer Entropy from Fz to FC6: 0.5443053672251921
Transfer Entropy from Fz to M1: 0.5160736364728813
Transfer Entropy from Fz to T7: 0.2833177855486371
Transfer Entropy from Fz to C3: 0.35944006231316306
Transfer Entropy from Fz to Cz: 0.4692669166953484
Transfer Entropy from Fz to C4: 0.5460395670150509
Transfer Entropy from Fz to T8: 0.5396732732487385
Transfer Entropy from Fz to M2: 0.5360380747508289
Transfer Entropy from Fz to CP5: 0.20463758349428982
Transfer Entropy from Fz to CP1: 0.6233771166028202
Transfer Entropy from Fz to CP2: 0.6843582561045939
Transfer Entropy from Fz to CP6: 0.6001780862726105
Transfer Entropy from Fz to P7: 0.5750892965159602
Transfer Entropy from Fz to P3: 0.5938263122327581
Transfer Entropy from Fz to Pz: 0.7586462746385891
Transfer Entropy from Fz to P4: 0.7008879119460192
Transfer Entropy from Fz to P8: 0.5632205190987256
Transfer Entropy from Fz to POz: 0.6784001143749826
Transfer Entropy from Fz to O1: 0.5967238422253116
Transfer Entropy from Fz to Oz: 0.5954800395156108
Transfer Entropy from Fz to O2: 0.5793375696001052
Transfer Entropy from F4 to Fp1: 0.2192756988143739
Transfer Entropy from F4 to Fpz: 0.4209312090152685
Transfer Entropy from F4 to Fp2: 0.34292640494827303
Transfer Entropy from F4 to F7: 0.2620720374093887
Transfer Entropy from F4 to F3: 0.4099273245907356
Transfer Entropy from F4 to Fz: 0.3268706759704917
Transfer Entropy from F4 to F8: 0.3974203835713979
Transfer Entropy from F4 to FC5: 0.4178700169859713
Transfer Entropy from F4 to FC1: 0.49261977837303084
Transfer Entropy from F4 to FC2: 0.5574090256285634
Transfer Entropy from F4 to FC6: 0.5464305450548195
Transfer Entropy from F4 to M1: 0.5119359419839761
Transfer Entropy from F4 to T7: 0.304108794748545
Transfer Entropy from F4 to C3: 0.39117567192768576
Transfer Entropy from F4 to Cz: 0.5470985817570448
Transfer Entropy from F4 to C4: 0.5415503069297042
Transfer Entropy from F4 to T8: 0.5333846750024308
Transfer Entropy from F4 to M2: 0.5830639341582593
Transfer Entropy from F4 to CP5: 0.2133684270845234
Transfer Entropy from F4 to CP1: 0.6431847991622369
Transfer Entropy from F4 to CP2: 0.7704642801713107
Transfer Entropy from F4 to CP6: 0.6012495368219137
Transfer Entropy from F4 to P7: 0.6093803436505312
Transfer Entropy from F4 to P3: 0.5963866941007402
Transfer Entropy from F4 to Pz: 0.789305865221836
Transfer Entropy from F4 to P4: 0.772906097520076
Transfer Entropy from F4 to P8: 0.5559533564921287

Transfer Entropy from F4 to POz: 0.768061333099793
Transfer Entropy from F4 to O1: 0.6302728069486067
Transfer Entropy from F4 to Oz: 0.629919967112253
Transfer Entropy from F4 to O2: 0.5874266530822311
Transfer Entropy from F8 to Fp1: 0.22607537767458188
Transfer Entropy from F8 to Fpz: 0.4724746462576873
Transfer Entropy from F8 to Fp2: 0.34817209626635187
Transfer Entropy from F8 to F7: 0.2506342934769467
Transfer Entropy from F8 to F3: 0.3941614672378394
Transfer Entropy from F8 to Fz: 0.355483108910025
Transfer Entropy from F8 to F4: 0.38873398698112566
Transfer Entropy from F8 to FC5: 0.411445005410226
Transfer Entropy from F8 to FC1: 0.49762255381462794
Transfer Entropy from F8 to FC2: 0.6067761080867932
Transfer Entropy from F8 to FC6: 0.43028824032416946
Transfer Entropy from F8 to M1: 0.449084666553153
Transfer Entropy from F8 to T7: 0.330972143584159
Transfer Entropy from F8 to C3: 0.4008503861471955
Transfer Entropy from F8 to Cz: 0.5964801538506499
Transfer Entropy from F8 to C4: 0.49260626591346146
Transfer Entropy from F8 to T8: 0.4400142448346192
Transfer Entropy from F8 to M2: 0.5441394925933131
Transfer Entropy from F8 to CP5: 0.2437786861228755
Transfer Entropy from F8 to CP1: 0.60088116800224
Transfer Entropy from F8 to CP2: 0.763606907917181
Transfer Entropy from F8 to CP6: 0.5698072304881495
Transfer Entropy from F8 to P7: 0.5728444933785435
Transfer Entropy from F8 to P3: 0.6603182351271305
Transfer Entropy from F8 to Pz: 0.8707034282684714
Transfer Entropy from F8 to P4: 0.7527762501637311
Transfer Entropy from F8 to P8: 0.49280671228224576
Transfer Entropy from F8 to POz: 0.7562746722759742
Transfer Entropy from F8 to O1: 0.6242284253413856
Transfer Entropy from F8 to Oz: 0.6173839736505841
Transfer Entropy from F8 to O2: 0.5751129716709564
Transfer Entropy from FC5 to Fp1: 0.2744206689739513
Transfer Entropy from FC5 to Fpz: 0.507377247240566
Transfer Entropy from FC5 to Fp2: 0.3891431508384015
Transfer Entropy from FC5 to F7: 0.20309841850488092
Transfer Entropy from FC5 to F3: 0.3235433671666623
Transfer Entropy from FC5 to Fz: 0.3354941809952175
Transfer Entropy from FC5 to F4: 0.40721054447816757
Transfer Entropy from FC5 to F8: 0.43516365610998575
Transfer Entropy from FC5 to FC1: 0.4509404614195442
Transfer Entropy from FC5 to FC2: 0.5417885848010565
Transfer Entropy from FC5 to FC6: 0.5664585940534665
Transfer Entropy from FC5 to M1: 0.5328288880358402
Transfer Entropy from FC5 to T7: 0.2824350489402219

Transfer Entropy from FC5 to C3: 0.3139761237203531
Transfer Entropy from FC5 to Cz: 0.5505130485605755
Transfer Entropy from FC5 to C4: 0.558403361979837
Transfer Entropy from FC5 to T8: 0.5479714050108613
Transfer Entropy from FC5 to M2: 0.5519878041012428
Transfer Entropy from FC5 to CP5: 0.18411069289979728
Transfer Entropy from FC5 to CP1: 0.6409745166981834
Transfer Entropy from FC5 to CP2: 0.7222924979017571
Transfer Entropy from FC5 to CP6: 0.6049106124336661
Transfer Entropy from FC5 to P7: 0.6053678419999592
Transfer Entropy from FC5 to P3: 0.5584981208094395
Transfer Entropy from FC5 to Pz: 0.7194732906750475
Transfer Entropy from FC5 to P4: 0.700297947104999
Transfer Entropy from FC5 to P8: 0.5715212927638625
Transfer Entropy from FC5 to POz: 0.7467573852213278
Transfer Entropy from FC5 to O1: 0.6265307573415986
Transfer Entropy from FC5 to Oz: 0.5431474149512052
Transfer Entropy from FC5 to O2: 0.5482844878207151
Transfer Entropy from FC1 to Fp1: 0.2808880493602157
Transfer Entropy from FC1 to Fpz: 0.4858918042913545
Transfer Entropy from FC1 to Fp2: 0.38369708964922816
Transfer Entropy from FC1 to F7: 0.25986257554970843
Transfer Entropy from FC1 to F3: 0.42525898411532703
Transfer Entropy from FC1 to Fz: 0.31400960864444977
Transfer Entropy from FC1 to F4: 0.43870320509696437
Transfer Entropy from FC1 to F8: 0.4607621704658753
Transfer Entropy from FC1 to FC5: 0.4242960078989762
Transfer Entropy from FC1 to FC2: 0.5837206621022315
Transfer Entropy from FC1 to FC6: 0.5488606438948963
Transfer Entropy from FC1 to M1: 0.5437530094706433
Transfer Entropy from FC1 to T7: 0.33163429107339604
Transfer Entropy from FC1 to C3: 0.381342498278394
Transfer Entropy from FC1 to Cz: 0.5778924105476672
Transfer Entropy from FC1 to C4: 0.5505372796617737
Transfer Entropy from FC1 to T8: 0.5566677146184026
Transfer Entropy from FC1 to M2: 0.5807701125320506
Transfer Entropy from FC1 to CP5: 0.23541848104565302
Transfer Entropy from FC1 to CP1: 0.7087686882160183
Transfer Entropy from FC1 to CP2: 0.7935257515650826
Transfer Entropy from FC1 to CP6: 0.6065033484155081
Transfer Entropy from FC1 to P7: 0.6101733579249651
Transfer Entropy from FC1 to P3: 0.647368987552461
Transfer Entropy from FC1 to Pz: 0.8329630289734958
Transfer Entropy from FC1 to P4: 0.7653903840605324
Transfer Entropy from FC1 to P8: 0.5794370610120535
Transfer Entropy from FC1 to POz: 0.7718258087860932
Transfer Entropy from FC1 to O1: 0.646324634845309
Transfer Entropy from FC1 to Oz: 0.649722768281883

Transfer Entropy from FC1 to O2: 0.6436136793265926
Transfer Entropy from FC2 to Fp1: 0.2581282749611645
Transfer Entropy from FC2 to Fpz: 0.4673143714611598
Transfer Entropy from FC2 to Fp2: 0.3625051592895219
Transfer Entropy from FC2 to F7: 0.2655752058203292
Transfer Entropy from FC2 to F3: 0.4230332505453134
Transfer Entropy from FC2 to Fz: 0.3526079092233726
Transfer Entropy from FC2 to F4: 0.42442705275287973
Transfer Entropy from FC2 to F8: 0.47524539833901486
Transfer Entropy from FC2 to FC5: 0.4204362547625147
Transfer Entropy from FC2 to FC1: 0.5108896027505884
Transfer Entropy from FC2 to FC6: 0.5983969054688704
Transfer Entropy from FC2 to M1: 0.5658290875490554
Transfer Entropy from FC2 to T7: 0.3041310860045622
Transfer Entropy from FC2 to C3: 0.4153756404837343
Transfer Entropy from FC2 to Cz: 0.6048296472894954
Transfer Entropy from FC2 to C4: 0.594206606352386
Transfer Entropy from FC2 to T8: 0.6067456442666673
Transfer Entropy from FC2 to M2: 0.6158604586160105
Transfer Entropy from FC2 to CP5: 0.23235873355815445
Transfer Entropy from FC2 to CP1: 0.6953476482405847
Transfer Entropy from FC2 to CP2: 0.8187040476825275
Transfer Entropy from FC2 to CP6: 0.650305089710449
Transfer Entropy from FC2 to P7: 0.6645718106272651
Transfer Entropy from FC2 to P3: 0.7039992526205677
Transfer Entropy from FC2 to Pz: 0.8289935987858127
Transfer Entropy from FC2 to P4: 0.8236403212256016
Transfer Entropy from FC2 to P8: 0.6229087667747264
Transfer Entropy from FC2 to POz: 0.8089796669984013
Transfer Entropy from FC2 to O1: 0.6612533531438027
Transfer Entropy from FC2 to Oz: 0.6372086425936588
Transfer Entropy from FC2 to O2: 0.6147472573983395
Transfer Entropy from FC6 to Fp1: 0.24001588645879307
Transfer Entropy from FC6 to Fpz: 0.46586700310665996
Transfer Entropy from FC6 to Fp2: 0.3321860185328585
Transfer Entropy from FC6 to F7: 0.28252231313360143
Transfer Entropy from FC6 to F3: 0.44764502373053294
Transfer Entropy from FC6 to Fz: 0.3893231218508609
Transfer Entropy from FC6 to F4: 0.4499423535729028
Transfer Entropy from FC6 to F8: 0.3420531757869358
Transfer Entropy from FC6 to FC5: 0.47152023756452505
Transfer Entropy from FC6 to FC1: 0.5167628259182291
Transfer Entropy from FC6 to FC2: 0.6444400780203419
Transfer Entropy from FC6 to M1: 0.4285857089447552
Transfer Entropy from FC6 to T7: 0.36026119848140564
Transfer Entropy from FC6 to C3: 0.4402211299962277
Transfer Entropy from FC6 to Cz: 0.654068412676228
Transfer Entropy from FC6 to C4: 0.4462629134411483

```
Transfer Entropy from FC6 to T8: 0.4442412420759182
Transfer Entropy from FC6 to M2: 0.5868951280792212
Transfer Entropy from FC6 to CP5: 0.2866576443176058
Transfer Entropy from FC6 to CP1: 0.662040601301077
Transfer Entropy from FC6 to CP2: 0.8392118334850389
Transfer Entropy from FC6 to CP6: 0.5114632503423725
Transfer Entropy from FC6 to P7: 0.5938451883044574
Transfer Entropy from FC6 to P3: 0.7105037956598554
Transfer Entropy from FC6 to Pz: 0.9151925560938446
Transfer Entropy from FC6 to P4: 0.8000260441670335
Transfer Entropy from FC6 to P8: 0.46372480884918943
Transfer Entropy from FC6 to POz: 0.7999453687613175
Transfer Entropy from FC6 to O1: 0.6555571387673637
Transfer Entropy from FC6 to Oz: 0.6769463605789564
Transfer Entropy from FC6 to O2: 0.6226202797377951
Transfer Entropy from M1 to Fp1: 0.24224604348833373
Transfer Entropy from M1 to Fpz: 0.4667125319018863
Transfer Entropy from M1 to Fp2: 0.3403045599619948
Transfer Entropy from M1 to F7: 0.286165304805969
Transfer Entropy from M1 to F3: 0.43989471566785937
Transfer Entropy from M1 to Fz: 0.38046254006051783
Transfer Entropy from M1 to F4: 0.438379869746892
Transfer Entropy from M1 to F8: 0.3840361509393835
Transfer Entropy from M1 to FC5: 0.4376068372189374
Transfer Entropy from M1 to FC1: 0.5151564922734129
Transfer Entropy from M1 to FC2: 0.6339474793478506
Transfer Entropy from M1 to FC6: 0.4446750821342816
Transfer Entropy from M1 to T7: 0.3714765275064403
Transfer Entropy from M1 to C3: 0.4260250946035193
Transfer Entropy from M1 to Cz: 0.6519110103559751
Transfer Entropy from M1 to C4: 0.45179213863778594
Transfer Entropy from M1 to T8: 0.43323330276118277
Transfer Entropy from M1 to M2: 0.5783357064805341
Transfer Entropy from M1 to CP5: 0.27880461935593864
Transfer Entropy from M1 to CP1: 0.6394202324423525
Transfer Entropy from M1 to CP2: 0.8301337714823536
Transfer Entropy from M1 to CP6: 0.5225083680229148
Transfer Entropy from M1 to P7: 0.6180791486891902
Transfer Entropy from M1 to P3: 0.7141066530638719
Transfer Entropy from M1 to Pz: 0.909821512406858
Transfer Entropy from M1 to P4: 0.8021909788631344
Transfer Entropy from M1 to P8: 0.4252346696563039
Transfer Entropy from M1 to POz: 0.7890174990898806
Transfer Entropy from M1 to O1: 0.6604098464460043
Transfer Entropy from M1 to Oz: 0.6822375190117876
Transfer Entropy from M1 to O2: 0.6391347400673207
Transfer Entropy from T7 to Fp1: 0.22761504089207293
Transfer Entropy from T7 to Fpz: 0.4239460049755076
```

Transfer Entropy from T7 to Fp2: 0.33814743630504857
Transfer Entropy from T7 to F7: 0.1944597504923592
Transfer Entropy from T7 to F3: 0.3029863691161947
Transfer Entropy from T7 to Fz: 0.2525203355964807
Transfer Entropy from T7 to F4: 0.332396543779861
Transfer Entropy from T7 to F8: 0.376867563582715
Transfer Entropy from T7 to FC5: 0.2933285144784581
Transfer Entropy from T7 to FC1: 0.4092611806935111
Transfer Entropy from T7 to FC2: 0.4249328244537932
Transfer Entropy from T7 to FC6: 0.4969823896291912
Transfer Entropy from T7 to M1: 0.49071973547447745
Transfer Entropy from T7 to C3: 0.3143682061613503
Transfer Entropy from T7 to Cz: 0.47534586101104254
Transfer Entropy from T7 to C4: 0.5013964653762745
Transfer Entropy from T7 to T8: 0.5038427748820057
Transfer Entropy from T7 to M2: 0.4989677214183821
Transfer Entropy from T7 to CP5: 0.16183906581020063
Transfer Entropy from T7 to CP1: 0.55509837768748
Transfer Entropy from T7 to CP2: 0.6180650371935951
Transfer Entropy from T7 to CP6: 0.5532742739744586
Transfer Entropy from T7 to P7: 0.524902325888267
Transfer Entropy from T7 to P3: 0.488016750183168
Transfer Entropy from T7 to Pz: 0.589082732651706
Transfer Entropy from T7 to P4: 0.6357775037585331
Transfer Entropy from T7 to P8: 0.521378108959349
Transfer Entropy from T7 to POz: 0.6470938548125686
Transfer Entropy from T7 to O1: 0.5654459309283346
Transfer Entropy from T7 to Oz: 0.4816908775485304
Transfer Entropy from T7 to O2: 0.4962223139530741
Transfer Entropy from C3 to Fp1: 0.26599422030334374
Transfer Entropy from C3 to Fpz: 0.5075861043761491
Transfer Entropy from C3 to Fp2: 0.39021067272664656
Transfer Entropy from C3 to F7: 0.21736618208897066
Transfer Entropy from C3 to F3: 0.37269497146156705
Transfer Entropy from C3 to Fz: 0.3293137441282195
Transfer Entropy from C3 to F4: 0.4233258532616056
Transfer Entropy from C3 to F8: 0.4448526163233863
Transfer Entropy from C3 to FC5: 0.3275768729100352
Transfer Entropy from C3 to FC1: 0.45662390264475117
Transfer Entropy from C3 to FC2: 0.5853391658383275
Transfer Entropy from C3 to FC6: 0.5826314024478404
Transfer Entropy from C3 to M1: 0.5460034116483873
Transfer Entropy from C3 to T7: 0.30525794505866244
Transfer Entropy from C3 to Cz: 0.5719632729081359
Transfer Entropy from C3 to C4: 0.5884527266755075
Transfer Entropy from C3 to T8: 0.574715755189436
Transfer Entropy from C3 to M2: 0.5706733166095102
Transfer Entropy from C3 to CP5: 0.2063820812401189

Transfer Entropy from C3 to CP1: 0.6174597784107029
Transfer Entropy from C3 to CP2: 0.6730648642222391
Transfer Entropy from C3 to CP6: 0.6286615396568008
Transfer Entropy from C3 to P7: 0.6031867586756696
Transfer Entropy from C3 to P3: 0.6222820805900595
Transfer Entropy from C3 to Pz: 0.8029063632371876
Transfer Entropy from C3 to P4: 0.6773142640121943
Transfer Entropy from C3 to P8: 0.5960733699661078
Transfer Entropy from C3 to POz: 0.7063683729220696
Transfer Entropy from C3 to O1: 0.6109749947659162
Transfer Entropy from C3 to Oz: 0.5900569885917403
Transfer Entropy from C3 to O2: 0.6092954200234817
Transfer Entropy from Cz to Fp1: 0.28480675910703523
Transfer Entropy from Cz to Fpz: 0.5071831978079726
Transfer Entropy from Cz to Fp2: 0.38647272497545315
Transfer Entropy from Cz to F7: 0.2500469951653287
Transfer Entropy from Cz to F3: 0.41482459542364397
Transfer Entropy from Cz to Fz: 0.30539702255159423
Transfer Entropy from Cz to F4: 0.42264600503894373
Transfer Entropy from Cz to F8: 0.46955272876065457
Transfer Entropy from Cz to FC5: 0.4124846419838271
Transfer Entropy from Cz to FC1: 0.48128874888437345
Transfer Entropy from Cz to FC2: 0.6044827204247961
Transfer Entropy from Cz to FC6: 0.6018637538042874
Transfer Entropy from Cz to M1: 0.5772639752124228
Transfer Entropy from Cz to T7: 0.31885978893068434
Transfer Entropy from Cz to C3: 0.3983733094020566
Transfer Entropy from Cz to C4: 0.606091637796723
Transfer Entropy from Cz to T8: 0.5989982568302028
Transfer Entropy from Cz to M2: 0.6020879266861191
Transfer Entropy from Cz to CP5: 0.22715511753514855
Transfer Entropy from Cz to CP1: 0.7121875114604178
Transfer Entropy from Cz to CP2: 0.8050172284026301
Transfer Entropy from Cz to CP6: 0.6515702820310094
Transfer Entropy from Cz to P7: 0.6315034163606156
Transfer Entropy from Cz to P3: 0.6713260541069318
Transfer Entropy from Cz to Pz: 0.8305610713509728
Transfer Entropy from Cz to P4: 0.8031952400601028
Transfer Entropy from Cz to P8: 0.6165286498197367
Transfer Entropy from Cz to POz: 0.7780858889704905
Transfer Entropy from Cz to O1: 0.627452817472712
Transfer Entropy from Cz to Oz: 0.6598440347199691
Transfer Entropy from Cz to O2: 0.6440152411518153
Transfer Entropy from C4 to Fp1: 0.2275191774530697
Transfer Entropy from C4 to Fpz: 0.4431219343979519
Transfer Entropy from C4 to Fp2: 0.3037208314855347
Transfer Entropy from C4 to F7: 0.2820416525303275
Transfer Entropy from C4 to F3: 0.4532265199022479

Transfer Entropy from C4 to Fz: 0.39723446494561176
Transfer Entropy from C4 to F4: 0.45957377547439576
Transfer Entropy from C4 to F8: 0.4103631097284736
Transfer Entropy from C4 to FC5: 0.4672289780711388
Transfer Entropy from C4 to FC1: 0.5275102141848488
Transfer Entropy from C4 to FC2: 0.6561561573247182
Transfer Entropy from C4 to FC6: 0.4581155555191046
Transfer Entropy from C4 to M1: 0.44791833470872894
Transfer Entropy from C4 to T7: 0.36449265438272377
Transfer Entropy from C4 to C3: 0.44443186092731285
Transfer Entropy from C4 to Cz: 0.6779679052624765
Transfer Entropy from C4 to T8: 0.4959258602576995
Transfer Entropy from C4 to M2: 0.6016009362276266
Transfer Entropy from C4 to CP5: 0.29928942395079783
Transfer Entropy from C4 to CP1: 0.688447336698057
Transfer Entropy from C4 to CP2: 0.8462868393474471
Transfer Entropy from C4 to CP6: 0.46341425984702805
Transfer Entropy from C4 to P7: 0.6137324712653139
Transfer Entropy from C4 to P3: 0.7249921612162684
Transfer Entropy from C4 to Pz: 0.9399977256806578
Transfer Entropy from C4 to P4: 0.8138322710724541
Transfer Entropy from C4 to P8: 0.5003839377526326
Transfer Entropy from C4 to POz: 0.8040994363792496
Transfer Entropy from C4 to O1: 0.6679406462325422
Transfer Entropy from C4 to Oz: 0.6875079624632604
Transfer Entropy from C4 to O2: 0.6412873600007891
Transfer Entropy from T8 to Fp1: 0.24964982460632099
Transfer Entropy from T8 to Fpz: 0.4797653226930973
Transfer Entropy from T8 to Fp2: 0.3395873911293969
Transfer Entropy from T8 to F7: 0.2646837701840812
Transfer Entropy from T8 to F3: 0.4333858901330164
Transfer Entropy from T8 to Fz: 0.37874566418292643
Transfer Entropy from T8 to F4: 0.42830455306832677
Transfer Entropy from T8 to F8: 0.34845808086859764
Transfer Entropy from T8 to FC5: 0.4387354914158649
Transfer Entropy from T8 to FC1: 0.501159264435174
Transfer Entropy from T8 to FC2: 0.6352359029593639
Transfer Entropy from T8 to FC6: 0.42391791931223965
Transfer Entropy from T8 to M1: 0.398864907633702
Transfer Entropy from T8 to T7: 0.3583212314863955
Transfer Entropy from T8 to C3: 0.4265676378963282
Transfer Entropy from T8 to Cz: 0.6326928576466867
Transfer Entropy from T8 to C4: 0.4643395519887053
Transfer Entropy from T8 to M2: 0.5721742189669775
Transfer Entropy from T8 to CP5: 0.27289662497741407
Transfer Entropy from T8 to CP1: 0.6429466750665006
Transfer Entropy from T8 to CP2: 0.8192865734814008
Transfer Entropy from T8 to CP6: 0.5403926456217284

Transfer Entropy from T8 to P7: 0.5886061053303752
Transfer Entropy from T8 to P3: 0.6991534839179641
Transfer Entropy from T8 to Pz: 0.8994952265377982
Transfer Entropy from T8 to P4: 0.7819066174712515
Transfer Entropy from T8 to P8: 0.45416272287548126
Transfer Entropy from T8 to POz: 0.7861547877249202
Transfer Entropy from T8 to O1: 0.6467492868565516
Transfer Entropy from T8 to Oz: 0.6590203798493345
Transfer Entropy from T8 to O2: 0.5964088154544178
Transfer Entropy from M2 to Fp1: 0.2729832346653959
Transfer Entropy from M2 to Fpz: 0.5094704346604599
Transfer Entropy from M2 to Fp2: 0.37520690003344453
Transfer Entropy from M2 to F7: 0.2555019245965073
Transfer Entropy from M2 to F3: 0.38492955702625553
Transfer Entropy from M2 to Fz: 0.3538606256699939
Transfer Entropy from M2 to F4: 0.4540385109661914
Transfer Entropy from M2 to F8: 0.42449246572650545
Transfer Entropy from M2 to FC5: 0.38256722097920126
Transfer Entropy from M2 to FC1: 0.48762928873033723
Transfer Entropy from M2 to FC2: 0.6229090171362787
Transfer Entropy from M2 to FC6: 0.5524168297441244
Transfer Entropy from M2 to M1: 0.5198949360159302
Transfer Entropy from M2 to T7: 0.329828916619646
Transfer Entropy from M2 to C3: 0.4016949675770862
Transfer Entropy from M2 to Cz: 0.6155832434842541
Transfer Entropy from M2 to C4: 0.5590706314417696
Transfer Entropy from M2 to T8: 0.5522477580895141
Transfer Entropy from M2 to CP5: 0.23464230512466072
Transfer Entropy from M2 to CP1: 0.6189841060124309
Transfer Entropy from M2 to CP2: 0.7479228310415068
Transfer Entropy from M2 to CP6: 0.619640816584136
Transfer Entropy from M2 to P7: 0.6330427457217864
Transfer Entropy from M2 to P3: 0.6506398339563948
Transfer Entropy from M2 to Pz: 0.8150277015405445
Transfer Entropy from M2 to P4: 0.7432592726656267
Transfer Entropy from M2 to P8: 0.5912004387734271
Transfer Entropy from M2 to POz: 0.7364247205878687
Transfer Entropy from M2 to O1: 0.5990944908408338
Transfer Entropy from M2 to Oz: 0.6129759834623539
Transfer Entropy from M2 to O2: 0.5695889107487881
Transfer Entropy from CP5 to Fp1: 0.20276766588279074
Transfer Entropy from CP5 to Fpz: 0.3950494742723641
Transfer Entropy from CP5 to Fp2: 0.31675680383413013
Transfer Entropy from CP5 to F7: 0.18125419196131184
Transfer Entropy from CP5 to F3: 0.30033438325435846
Transfer Entropy from CP5 to Fz: 0.2595997420269019
Transfer Entropy from CP5 to F4: 0.288273190389089
Transfer Entropy from CP5 to F8: 0.3467223722672132

Transfer Entropy from CP5 to FC5: 0.30435119789689213
Transfer Entropy from CP5 to FC1: 0.374315445816313
Transfer Entropy from CP5 to FC2: 0.4357363305945459
Transfer Entropy from CP5 to FC6: 0.44153014082768777
Transfer Entropy from CP5 to M1: 0.4432265679399791
Transfer Entropy from CP5 to T7: 0.238148429520541
Transfer Entropy from CP5 to C3: 0.2928430158398981
Transfer Entropy from CP5 to Cz: 0.44856004239163566
Transfer Entropy from CP5 to C4: 0.4543071879274041
Transfer Entropy from CP5 to T8: 0.4516290057986561
Transfer Entropy from CP5 to M2: 0.471179603354463
Transfer Entropy from CP5 to CP1: 0.5499837565918735
Transfer Entropy from CP5 to CP2: 0.5987868153301504
Transfer Entropy from CP5 to CP6: 0.501745868302056
Transfer Entropy from CP5 to P7: 0.5041637668744706
Transfer Entropy from CP5 to P3: 0.453689229483312
Transfer Entropy from CP5 to Pz: 0.6069875823953674
Transfer Entropy from CP5 to P4: 0.60859394921269
Transfer Entropy from CP5 to P8: 0.4737066160339837
Transfer Entropy from CP5 to POz: 0.616158248955714
Transfer Entropy from CP5 to O1: 0.5347324545223567
Transfer Entropy from CP5 to Oz: 0.4989838148113529
Transfer Entropy from CP5 to O2: 0.5007355176586465
Transfer Entropy from CP1 to Fp1: 0.28085265404163934
Transfer Entropy from CP1 to Fpz: 0.5513275944351442
Transfer Entropy from CP1 to Fp2: 0.4123930425903522
Transfer Entropy from CP1 to F7: 0.3120042086253545
Transfer Entropy from CP1 to F3: 0.45123019953972016
Transfer Entropy from CP1 to Fz: 0.4211768211961787
Transfer Entropy from CP1 to F4: 0.5238125327518869
Transfer Entropy from CP1 to F8: 0.4666262030194362
Transfer Entropy from CP1 to FC5: 0.466388403689516
Transfer Entropy from CP1 to FC1: 0.5844991287950965
Transfer Entropy from CP1 to FC2: 0.7241289172717942
Transfer Entropy from CP1 to FC6: 0.6218656436586346
Transfer Entropy from CP1 to M1: 0.570665767090662
Transfer Entropy from CP1 to T7: 0.3639420147576637
Transfer Entropy from CP1 to C3: 0.4452804042215395
Transfer Entropy from CP1 to Cz: 0.7415446385160949
Transfer Entropy from CP1 to C4: 0.6406168219160892
Transfer Entropy from CP1 to T8: 0.6113897442736007
Transfer Entropy from CP1 to M2: 0.6162727077013641
Transfer Entropy from CP1 to CP5: 0.27994393144847773
Transfer Entropy from CP1 to CP2: 0.7681985098593306
Transfer Entropy from CP1 to CP6: 0.6758437872381273
Transfer Entropy from CP1 to P7: 0.697370575484575
Transfer Entropy from CP1 to P3: 0.73420658784119
Transfer Entropy from CP1 to Pz: 0.9919956638205839

Transfer Entropy from CP1 to P4: 0.7735496026495331
Transfer Entropy from CP1 to P8: 0.6138433575184544
Transfer Entropy from CP1 to POz: 0.7394211992317483
Transfer Entropy from CP1 to O1: 0.6560182533045447
Transfer Entropy from CP1 to Oz: 0.7115763909329005
Transfer Entropy from CP1 to O2: 0.6809660858900355
Transfer Entropy from CP2 to Fp1: 0.30421781335650794
Transfer Entropy from CP2 to Fpz: 0.5646288759772858
Transfer Entropy from CP2 to Fp2: 0.4291230429293488
Transfer Entropy from CP2 to F7: 0.28842373566430163
Transfer Entropy from CP2 to F3: 0.46298405185763025
Transfer Entropy from CP2 to Fz: 0.39380031283329625
Transfer Entropy from CP2 to F4: 0.5396443739655108
Transfer Entropy from CP2 to F8: 0.519020172997948
Transfer Entropy from CP2 to FC5: 0.45873626265639583
Transfer Entropy from CP2 to FC1: 0.5662926163521093
Transfer Entropy from CP2 to FC2: 0.7319134378071182
Transfer Entropy from CP2 to FC6: 0.6735540127553125
Transfer Entropy from CP2 to M1: 0.6425461387006969
Transfer Entropy from CP2 to T7: 0.34973186853930105
Transfer Entropy from CP2 to C3: 0.4272977356808945
Transfer Entropy from CP2 to Cz: 0.7166377731465874
Transfer Entropy from CP2 to C4: 0.6700458419963367
Transfer Entropy from CP2 to T8: 0.67028654163871
Transfer Entropy from CP2 to M2: 0.6375280369295905
Transfer Entropy from CP2 to CP5: 0.26974241780418323
Transfer Entropy from CP2 to CP1: 0.6644306610759935
Transfer Entropy from CP2 to CP6: 0.7236152380464264
Transfer Entropy from CP2 to P7: 0.705102261658833
Transfer Entropy from CP2 to P3: 0.7338189565156642
Transfer Entropy from CP2 to Pz: 1.009245230553781
Transfer Entropy from CP2 to P4: 0.7385680535688937
Transfer Entropy from CP2 to P8: 0.6851745937716947
Transfer Entropy from CP2 to POz: 0.7167524266958696
Transfer Entropy from CP2 to O1: 0.6161716761825685
Transfer Entropy from CP2 to Oz: 0.7198698617889328
Transfer Entropy from CP2 to O2: 0.7308687917076059
Transfer Entropy from CP6 to Fp1: 0.23712503359213838
Transfer Entropy from CP6 to Fpz: 0.45560025014485045
Transfer Entropy from CP6 to Fp2: 0.31190751480795037
Transfer Entropy from CP6 to F7: 0.2752122988952096
Transfer Entropy from CP6 to F3: 0.45701127072376785
Transfer Entropy from CP6 to Fz: 0.40007241304909325
Transfer Entropy from CP6 to F4: 0.47087543330620324
Transfer Entropy from CP6 to F8: 0.4393704419564172
Transfer Entropy from CP6 to FC5: 0.4563405619653145
Transfer Entropy from CP6 to FC1: 0.5217609529753696
Transfer Entropy from CP6 to FC2: 0.6614396073811792

Transfer Entropy from CP6 to FC6: 0.4675166577801748
Transfer Entropy from CP6 to M1: 0.46124207713040677
Transfer Entropy from CP6 to T7: 0.36914362442736065
Transfer Entropy from CP6 to C3: 0.4364730751454995
Transfer Entropy from CP6 to Cz: 0.6740313565292575
Transfer Entropy from CP6 to C4: 0.4079575264607737
Transfer Entropy from CP6 to T8: 0.5169216441989501
Transfer Entropy from CP6 to M2: 0.6117063910724522
Transfer Entropy from CP6 to CP5: 0.3004210782934484
Transfer Entropy from CP6 to CP1: 0.6856097618073231
Transfer Entropy from CP6 to CP2: 0.8669077150109646
Transfer Entropy from CP6 to P7: 0.6153711492488189
Transfer Entropy from CP6 to P3: 0.7344500950373174
Transfer Entropy from CP6 to Pz: 0.9450782689708108
Transfer Entropy from CP6 to P4: 0.8153349840700354
Transfer Entropy from CP6 to P8: 0.4929453408429554
Transfer Entropy from CP6 to POz: 0.7958692753280208
Transfer Entropy from CP6 to O1: 0.6575521358705486
Transfer Entropy from CP6 to Oz: 0.6888025376108107
Transfer Entropy from CP6 to O2: 0.6404597743222343
Transfer Entropy from P7 to Fp1: 0.25119918165370947
Transfer Entropy from P7 to Fpz: 0.49637217298958447
Transfer Entropy from P7 to Fp2: 0.368857763155908
Transfer Entropy from P7 to F7: 0.259174338237262
Transfer Entropy from P7 to F3: 0.42015797018986556
Transfer Entropy from P7 to Fz: 0.34259467995486975
Transfer Entropy from P7 to F4: 0.42916081642256265
Transfer Entropy from P7 to F8: 0.40550129806769014
Transfer Entropy from P7 to FC5: 0.41377534441489117
Transfer Entropy from P7 to FC1: 0.47506229940989103
Transfer Entropy from P7 to FC2: 0.6212094437343963
Transfer Entropy from P7 to FC6: 0.5097007478952168
Transfer Entropy from P7 to M1: 0.516060964414289
Transfer Entropy from P7 to T7: 0.31634397652769036
Transfer Entropy from P7 to C3: 0.3877339723979816
Transfer Entropy from P7 to Cz: 0.5848997256436373
Transfer Entropy from P7 to C4: 0.5169786786107995
Transfer Entropy from P7 to T8: 0.5209513395138157
Transfer Entropy from P7 to M2: 0.5865202941979875
Transfer Entropy from P7 to CP5: 0.2473366970773057
Transfer Entropy from P7 to CP1: 0.6555025823465721
Transfer Entropy from P7 to CP2: 0.7741693405813315
Transfer Entropy from P7 to CP6: 0.5743459486735876
Transfer Entropy from P7 to P3: 0.6618620342612336
Transfer Entropy from P7 to Pz: 0.8580709032558461
Transfer Entropy from P7 to P4: 0.768404418320693
Transfer Entropy from P7 to P8: 0.5295737324510357
Transfer Entropy from P7 to POz: 0.7599441026378368

Transfer Entropy from P7 to O1: 0.6344779857514274
Transfer Entropy from P7 to Oz: 0.6069663370639066
Transfer Entropy from P7 to O2: 0.5812729857541896
Transfer Entropy from P3 to Fp1: 0.28277505134572395
Transfer Entropy from P3 to Fpz: 0.5468056980820039
Transfer Entropy from P3 to Fp2: 0.40481585648822593
Transfer Entropy from P3 to F7: 0.27275276262810133
Transfer Entropy from P3 to F3: 0.39360880341621357
Transfer Entropy from P3 to Fz: 0.3797184546363368
Transfer Entropy from P3 to F4: 0.44324428605691263
Transfer Entropy from P3 to F8: 0.4958044442460554
Transfer Entropy from P3 to FC5: 0.39928962330612344
Transfer Entropy from P3 to FC1: 0.5186131255242833
Transfer Entropy from P3 to FC2: 0.6788775775045697
Transfer Entropy from P3 to FC6: 0.632285503275481
Transfer Entropy from P3 to M1: 0.6120931599791238
Transfer Entropy from P3 to T7: 0.32878705199843633
Transfer Entropy from P3 to C3: 0.42214121217628126
Transfer Entropy from P3 to Cz: 0.6510652227808342
Transfer Entropy from P3 to C4: 0.6336933596905726
Transfer Entropy from P3 to T8: 0.6366797941091709
Transfer Entropy from P3 to M2: 0.6182120038156498
Transfer Entropy from P3 to CP5: 0.250619686117817
Transfer Entropy from P3 to CP1: 0.6909293884206953
Transfer Entropy from P3 to CP2: 0.7970164614954279
Transfer Entropy from P3 to CP6: 0.6900603001271559
Transfer Entropy from P3 to P7: 0.6729583642534109
Transfer Entropy from P3 to Pz: 0.8573290952156406
Transfer Entropy from P3 to P4: 0.8003444092054056
Transfer Entropy from P3 to P8: 0.6562966457988416
Transfer Entropy from P3 to POz: 0.7913375828186647
Transfer Entropy from P3 to O1: 0.6729057694689432
Transfer Entropy from P3 to Oz: 0.6714088271134114
Transfer Entropy from P3 to O2: 0.6859573135512539
Transfer Entropy from Pz to Fp1: 0.28611226103862686
Transfer Entropy from Pz to Fpz: 0.5197412275855209
Transfer Entropy from Pz to Fp2: 0.409060351153753
Transfer Entropy from Pz to F7: 0.2639465087968371
Transfer Entropy from Pz to F3: 0.4056213572197862
Transfer Entropy from Pz to Fz: 0.3695983243439643
Transfer Entropy from Pz to F4: 0.4438572612792324
Transfer Entropy from Pz to F8: 0.5026368844368214
Transfer Entropy from Pz to FC5: 0.4013818709186775
Transfer Entropy from Pz to FC1: 0.5133665993942269
Transfer Entropy from Pz to FC2: 0.6085302533436973
Transfer Entropy from Pz to FC6: 0.6198221509954254
Transfer Entropy from Pz to M1: 0.5929499161768697
Transfer Entropy from Pz to T7: 0.3018269518329596

Transfer Entropy from Pz to C3: 0.4283781835243477
Transfer Entropy from Pz to Cz: 0.6070932522211424
Transfer Entropy from Pz to C4: 0.6234678265284159
Transfer Entropy from Pz to T8: 0.6276026058632869
Transfer Entropy from Pz to M2: 0.6008866153268025
Transfer Entropy from Pz to CP5: 0.22875700687687325
Transfer Entropy from Pz to CP1: 0.7189017051277292
Transfer Entropy from Pz to CP2: 0.8508620913248183
Transfer Entropy from Pz to CP6: 0.6762782792070746
Transfer Entropy from Pz to P7: 0.6648833073233149
Transfer Entropy from Pz to P3: 0.6599618828306203
Transfer Entropy from Pz to P4: 0.8355532916484941
Transfer Entropy from Pz to P8: 0.6436348879513077
Transfer Entropy from Pz to POz: 0.8346202051154836
Transfer Entropy from Pz to O1: 0.6635358250622138
Transfer Entropy from Pz to Oz: 0.6312024138992479
Transfer Entropy from Pz to O2: 0.6205787123304894
Transfer Entropy from P4 to Fp1: 0.3089480021931509
Transfer Entropy from P4 to Fpz: 0.5592582419993397
Transfer Entropy from P4 to Fp2: 0.41097035257941833
Transfer Entropy from P4 to F7: 0.27244901412699835
Transfer Entropy from P4 to F3: 0.43316991485828543
Transfer Entropy from P4 to Fz: 0.39658439170282744
Transfer Entropy from P4 to F4: 0.5292397852573394
Transfer Entropy from P4 to F8: 0.499327177435332
Transfer Entropy from P4 to FC5: 0.42067429148983576
Transfer Entropy from P4 to FC1: 0.5342203466656243
Transfer Entropy from P4 to FC2: 0.7217821126333386
Transfer Entropy from P4 to FC6: 0.6359979014064775
Transfer Entropy from P4 to M1: 0.6138674625879365
Transfer Entropy from P4 to T7: 0.3423126676404655
Transfer Entropy from P4 to C3: 0.39867722622352686
Transfer Entropy from P4 to Cz: 0.7036578040056441
Transfer Entropy from P4 to C4: 0.6417440480400238
Transfer Entropy from P4 to T8: 0.633372759018917
Transfer Entropy from P4 to M2: 0.6193631521241213
Transfer Entropy from P4 to CP5: 0.26020121072894253
Transfer Entropy from P4 to CP1: 0.6509452060772606
Transfer Entropy from P4 to CP2: 0.7211052575621919
Transfer Entropy from P4 to CP6: 0.6880168050335022
Transfer Entropy from P4 to P7: 0.690167851933255
Transfer Entropy from P4 to P3: 0.7203292134508612
Transfer Entropy from P4 to Pz: 0.9758305381264579
Transfer Entropy from P4 to P8: 0.6468967789760403
Transfer Entropy from P4 to POz: 0.7225874460437725
Transfer Entropy from P4 to O1: 0.5953697436524638
Transfer Entropy from P4 to Oz: 0.6614312125686215
Transfer Entropy from P4 to O2: 0.6753239086754647

Transfer Entropy from P8 to Fp1: 0.24392259407346656
Transfer Entropy from P8 to Fpz: 0.4731676160157628
Transfer Entropy from P8 to Fp2: 0.3345029484405739
Transfer Entropy from P8 to F7: 0.2618976474825331
Transfer Entropy from P8 to F3: 0.43198080290834356
Transfer Entropy from P8 to Fz: 0.37908782915650147
Transfer Entropy from P8 to F4: 0.4342890385820715
Transfer Entropy from P8 to F8: 0.37730105494787547
Transfer Entropy from P8 to FC5: 0.4317311901181148
Transfer Entropy from P8 to FC1: 0.5011716902398708
Transfer Entropy from P8 to FC2: 0.6384685140523935
Transfer Entropy from P8 to FC6: 0.42657333467501796
Transfer Entropy from P8 to M1: 0.3730038077135971
Transfer Entropy from P8 to T7: 0.354700397684711
Transfer Entropy from P8 to C3: 0.42371387525390997
Transfer Entropy from P8 to Cz: 0.633499996611392
Transfer Entropy from P8 to C4: 0.45143627143975223
Transfer Entropy from P8 to T8: 0.43964209881932526
Transfer Entropy from P8 to M2: 0.596212379553361
Transfer Entropy from P8 to CP5: 0.2708579860269619
Transfer Entropy from P8 to CP1: 0.6282375645095174
Transfer Entropy from P8 to CP2: 0.8227482704275855
Transfer Entropy from P8 to CP6: 0.4999007868742137
Transfer Entropy from P8 to P7: 0.5796121090329287
Transfer Entropy from P8 to P3: 0.7052043815467225
Transfer Entropy from P8 to Pz: 0.9018078768683981
Transfer Entropy from P8 to P4: 0.7843588835460187
Transfer Entropy from P8 to POz: 0.7799399217482365
Transfer Entropy from P8 to O1: 0.6446870259333928
Transfer Entropy from P8 to Oz: 0.6602865129903335
Transfer Entropy from P8 to O2: 0.5901483811461566
Transfer Entropy from POz to Fp1: 0.3058488325837628
Transfer Entropy from POz to Fpz: 0.5438917286473721
Transfer Entropy from POz to Fp2: 0.3974323709017269
Transfer Entropy from POz to F7: 0.29261419265048794
Transfer Entropy from POz to F3: 0.46006444790186235
Transfer Entropy from POz to Fz: 0.38118453143887054
Transfer Entropy from POz to F4: 0.5392110846544154
Transfer Entropy from POz to F8: 0.517246041727831
Transfer Entropy from POz to FC5: 0.45140823779034733
Transfer Entropy from POz to FC1: 0.5452654216586684
Transfer Entropy from POz to FC2: 0.7233900643450631
Transfer Entropy from POz to FC6: 0.6504106888222768
Transfer Entropy from POz to M1: 0.6188130446030056
Transfer Entropy from POz to T7: 0.3463809327790957
Transfer Entropy from POz to C3: 0.421442434052833
Transfer Entropy from POz to Cz: 0.695085368850095
Transfer Entropy from POz to C4: 0.6437068168960972

Transfer Entropy from POz to T8: 0.6524473948146415
Transfer Entropy from POz to M2: 0.6343606692360518
Transfer Entropy from POz to CP5: 0.26792201204451355
Transfer Entropy from POz to CP1: 0.6258598498988869
Transfer Entropy from POz to CP2: 0.7079979881496614
Transfer Entropy from POz to CP6: 0.6842725039586182
Transfer Entropy from POz to P7: 0.6988512841269896
Transfer Entropy from POz to P3: 0.7235996628810841
Transfer Entropy from POz to Pz: 1.0033988374843419
Transfer Entropy from POz to P4: 0.7315188567910448
Transfer Entropy from POz to P8: 0.6621032447982088
Transfer Entropy from POz to O1: 0.5040522550485206
Transfer Entropy from POz to Oz: 0.7050951202710743
Transfer Entropy from POz to O2: 0.7164956509334204
Transfer Entropy from O1 to Fp1: 0.2879157257824763
Transfer Entropy from O1 to Fpz: 0.5000748663326027
Transfer Entropy from O1 to Fp2: 0.3780831236749345
Transfer Entropy from O1 to F7: 0.2813700837033936
Transfer Entropy from O1 to F3: 0.4327106868384678
Transfer Entropy from O1 to Fz: 0.3816143580749602
Transfer Entropy from O1 to F4: 0.497881365433139
Transfer Entropy from O1 to F8: 0.47828007722228916
Transfer Entropy from O1 to FC5: 0.4283644234505236
Transfer Entropy from O1 to FC1: 0.5211604477493811
Transfer Entropy from O1 to FC2: 0.6767893874608424
Transfer Entropy from O1 to FC6: 0.607622135647792
Transfer Entropy from O1 to M1: 0.583110582310506
Transfer Entropy from O1 to T7: 0.3533451294030663
Transfer Entropy from O1 to C3: 0.41189099924103
Transfer Entropy from O1 to Cz: 0.6585049028156075
Transfer Entropy from O1 to C4: 0.6175644498271423
Transfer Entropy from O1 to T8: 0.6133454756795951
Transfer Entropy from O1 to M2: 0.5823056835436806
Transfer Entropy from O1 to CP5: 0.2700300914126598
Transfer Entropy from O1 to CP1: 0.6512620700340936
Transfer Entropy from O1 to CP2: 0.7312113750695021
Transfer Entropy from O1 to CP6: 0.6498583240445425
Transfer Entropy from O1 to P7: 0.6633442503574516
Transfer Entropy from O1 to P3: 0.7051634770828585
Transfer Entropy from O1 to Pz: 0.9278486806470021
Transfer Entropy from O1 to P4: 0.7048952120425446
Transfer Entropy from O1 to P8: 0.6226052335977479
Transfer Entropy from O1 to POz: 0.5889830924424602
Transfer Entropy from O1 to Oz: 0.6508313953265039
Transfer Entropy from O1 to O2: 0.6706090000917205
Transfer Entropy from Oz to Fp1: 0.27895411465641
Transfer Entropy from Oz to Fpz: 0.5284562406015565
Transfer Entropy from Oz to Fp2: 0.3942450961859365

```
Transfer Entropy from Oz to F7: 0.2463407700134964
Transfer Entropy from Oz to F3: 0.3929228169603419
Transfer Entropy from Oz to Fz: 0.3612123355721033
Transfer Entropy from Oz to F4: 0.44625056886118014
Transfer Entropy from Oz to F8: 0.439282774939853
Transfer Entropy from Oz to FC5: 0.3578672074697291
Transfer Entropy from Oz to FC1: 0.4983413538762678
Transfer Entropy from Oz to FC2: 0.5758763327463812
Transfer Entropy from Oz to FC6: 0.5820479867705555
Transfer Entropy from Oz to M1: 0.5671435707731142
Transfer Entropy from Oz to T7: 0.30359123572497193
Transfer Entropy from Oz to C3: 0.3760892011127713
Transfer Entropy from Oz to Cz: 0.6024649545973955
Transfer Entropy from Oz to C4: 0.5812707964908721
Transfer Entropy from Oz to T8: 0.5796809363948214
Transfer Entropy from Oz to M2: 0.5605332259628545
Transfer Entropy from Oz to CP5: 0.2415387737510631
Transfer Entropy from Oz to CP1: 0.6491334300118347
Transfer Entropy from Oz to CP2: 0.7571980012909698
Transfer Entropy from Oz to CP6: 0.6342426495795793
Transfer Entropy from Oz to P7: 0.6001838356093723
Transfer Entropy from Oz to P3: 0.6444603825421643
Transfer Entropy from Oz to Pz: 0.7835963027587441
Transfer Entropy from Oz to P4: 0.7205077166560746
Transfer Entropy from Oz to P8: 0.5973476297530016
Transfer Entropy from Oz to POz: 0.7555103256179122
Transfer Entropy from Oz to O1: 0.618781422982721
Transfer Entropy from Oz to O2: 0.5491675015094409
Transfer Entropy from O2 to Fp1: 0.2680962760415231
Transfer Entropy from O2 to Fpz: 0.5032716643608136
Transfer Entropy from O2 to Fp2: 0.37258731619218394
Transfer Entropy from O2 to F7: 0.2402288791445957
Transfer Entropy from O2 to F3: 0.38220653960286394
Transfer Entropy from O2 to Fz: 0.34699060053854974
Transfer Entropy from O2 to F4: 0.39920834827980745
Transfer Entropy from O2 to F8: 0.3984273366845345
Transfer Entropy from O2 to FC5: 0.37571885790177195
Transfer Entropy from O2 to FC1: 0.49889110740516707
Transfer Entropy from O2 to FC2: 0.5424604971940453
Transfer Entropy from O2 to FC6: 0.5282512540448491
Transfer Entropy from O2 to M1: 0.5238390919159943
Transfer Entropy from O2 to T7: 0.31403881978301934
Transfer Entropy from O2 to C3: 0.39140910309991866
Transfer Entropy from O2 to Cz: 0.5877067028384575
Transfer Entropy from O2 to C4: 0.5352647446042443
Transfer Entropy from O2 to T8: 0.5177069517265268
Transfer Entropy from O2 to M2: 0.5120614298723932
Transfer Entropy from O2 to CP5: 0.2362900554214557
```

```
Transfer Entropy from O2 to CP1: 0.6251003317790271
Transfer Entropy from O2 to CP2: 0.7816889949624142
Transfer Entropy from O2 to CP6: 0.5882846398934183
Transfer Entropy from O2 to P7: 0.570606202572885
Transfer Entropy from O2 to P3: 0.6614497142756367
Transfer Entropy from O2 to Pz: 0.7636914586781992
Transfer Entropy from O2 to P4: 0.7378666245416413
Transfer Entropy from O2 to P8: 0.5321677892014791
Transfer Entropy from O2 to POz: 0.7681231418244422
Transfer Entropy from O2 to O1: 0.6342821531304308
Transfer Entropy from O2 to Oz: 0.5444472315129361
Transfer entropy results saved to:
/home/vincent/AAA_projects/MVCS/Neuroscience/Analysis/Transfer
Entropy/full_granularity_transfer_entropy_results.npy
```

[ ]:

Kuramoto Model

September 8, 2023

# 1 Kuramoto Model

# 2 Regional Avg

[ ]:
```
import numpy as np
from scipy.integrate import solve_ivp
from scipy.signal import hilbert
import matplotlib.pyplot as plt
import os

# Set random seed for consistent results
np.random.seed(42)

# Load EEG data
EEG_data = np.load('/home/vincent/AAA_projects/MVCS/Neuroscience/
↪eeg_data_with_channels.npy', allow_pickle=True)

# EEG channel names
eeg_channel_names = ['Fp1', 'Fpz', 'Fp2', 'F7', 'F3', 'Fz', 'F4', 'F8', 'FC5',␣
↪'FC1', 'FC2', 'FC6',
                     'M1', 'T7', 'C3', 'Cz', 'C4', 'T8', 'M2', 'CP5', 'CP1',␣
↪'CP2', 'CP6',
                     'P7', 'P3', 'Pz', 'P4', 'P8', 'POz', 'O1', 'Oz', 'O2']

# Broad regions and corresponding channels
regions = {
    "frontal": ['Fp1', 'Fpz', 'Fp2', 'F7', 'F3', 'Fz', 'F4', 'F8'],
    "temporal": ['T7', 'T8'],
    "parietal": ['CP5', 'CP1', 'CP2', 'CP6', 'P7', 'P3', 'Pz', 'P4', 'P8'],
    "occipital": ['O1', 'Oz', 'O2']
}

# Load PLV matrix
plv_matrix = np.load("/home/vincent/AAA_projects/MVCS/Neuroscience/Analysis/
↪Phase Syncronization/plv_matrix.npy")

N = len(regions)  # Number of regions
```

```
# Calculate average PLV per region
plv_regions = np.zeros((N, N))
for i, region1 in enumerate(regions.keys()):
    indices1 = [eeg_channel_names.index(ch) for ch in regions[region1]]
    for j, region2 in enumerate(regions.keys()):
        indices2 = [eeg_channel_names.index(ch) for ch in regions[region2]]
        plv_regions[i, j] = np.mean(plv_matrix[np.ix_(indices1, indices2)])

# Set natural frequencies proportional to mean PLV of each region
omega = np.mean(plv_regions, axis=1)

# Phase bias based on mean phase difference between regions
phase_diff_matrix = np.zeros((N, N))
for i, region1 in enumerate(regions.keys()):
    indices1 = [eeg_channel_names.index(ch) for ch in regions[region1]]
    for j, region2 in enumerate(regions.keys()):
        if i != j:
            indices2 = [eeg_channel_names.index(ch) for ch in regions[region2]]

            # Compute pairwise phase differences for each combination of␣
↪channels across the two regions
            pairwise_diffs = []
            for idx1 in indices1:
                for idx2 in indices2:
                    phase_diff = np.angle(hilbert(EEG_data[:, idx1])) - np.
↪angle(hilbert(EEG_data[:, idx2]))
                    pairwise_diffs.append(phase_diff)

            # Average over all pairwise differences
            phase_diff_matrix[i, j] = np.mean(pairwise_diffs)

# Modify the Kuramoto function to include weighted coupling and phase bias
def kuramoto_weighted_bias(t, y, omega, K):
    weighted_sin = plv_regions * np.sin(y - y[:, np.newaxis] -␣
↪phase_diff_matrix)
    dydt = omega + K/N * np.sum(weighted_sin, axis=1)
    return dydt

# Hilbert transform to get analytical signal
analytic_signal = hilbert(EEG_data)
phases = np.angle(analytic_signal)

# Average phase for the initial time point for each region
avg_phases = {}
for region, channels in regions.items():
    indices = [eeg_channel_names.index(ch) for ch in channels]
```

```python
    avg_phases[region] = np.mean(phases[0, indices])

initial_phases = np.array(list(avg_phases.values()))

K = 5.0  # Coupling strength

# Time span for the simulation
t_span = (0, 300)
t_eval = np.linspace(t_span[0], t_span[1], 300)

# Solve differential equations using the modified model
solution = solve_ivp(kuramoto_weighted_bias, t_span, initial_phases,␣
 ↪t_eval=t_eval, args=(omega, K))

# The simulated Kuramoto phases
kuramoto_phases = solution.y

# Create a time axis for plotting
time_axis = np.linspace(t_span[0], t_span[1], 2 * len(EEG_data))

# Interpolating the Kuramoto phases to match the EEG data length
from scipy.interpolate import interp1d
interp_funcs = [interp1d(t_eval, kuramoto_phases[i, :]) for i in range(N)]
kuramoto_phases_interp = np.array([f(time_axis) for f in interp_funcs])

# Plot
plt.figure(figsize=(14, 7))
for i, region in enumerate(regions.keys()):
    plt.plot(time_axis, kuramoto_phases_interp[i, :], label=region, linewidth=1.
 ↪5)

plt.legend()  # Place legend call here to ensure one entry per region
plt.xlabel('Time')
plt.ylabel('Phase')
plt.title('Simulated Kuramoto Phases for Brain Regions')

save_directory = "/home/vincent/AAA_projects/MVCS/Neuroscience/Modelling/
 ↪Kuramoto"

# Check if the directory exists and create it if not
if not os.path.exists(save_directory):
    os.makedirs(save_directory)

plot_path = os.path.join(save_directory, "Kuramoto_Brain_Regions_Plot_Extended.
 ↪png")
plt.savefig(plot_path, dpi=300)  # Increased dpi for better resolution
```

```
# Save the simulated Kuramoto phases
kuramoto_data_path = os.path.join(save_directory, "kuramoto_phases_extended.
↪npy")
np.save(kuramoto_data_path, kuramoto_phases)

print(f"Plot saved at {plot_path}")
print(f"Kuramoto phases saved at {kuramoto_data_path}")

plt.show()  # Show the plot at the end
```

## 3 Granular

[22]:
```
import numpy as np
from scipy.integrate import solve_ivp
from scipy.signal import hilbert
import matplotlib.pyplot as plt
import os

# Set random seed for consistent results
np.random.seed(42)

# Load EEG data
EEG_data = np.load('/home/vincent/AAA_projects/MVCS/Neuroscience/
↪eeg_data_with_channels.npy', allow_pickle=True)

# EEG channel names
eeg_channel_names = ['Fp1', 'Fpz', 'Fp2', 'F7', 'F3', 'Fz', 'F4', 'F8', 'FC5',␣
↪'FC1', 'FC2', 'FC6',
                     'M1', 'T7', 'C3', 'Cz', 'C4', 'T8', 'M2', 'CP5', 'CP1',␣
↪'CP2', 'CP6',
                     'P7', 'P3', 'Pz', 'P4', 'P8', 'POz', 'O1', 'Oz', 'O2']

# Broad regions and corresponding channels
regions = {
    "frontal": ['Fp1', 'Fpz', 'Fp2', 'F7', 'F3', 'Fz', 'F4', 'F8'],
    "temporal": ['T7', 'T8'],
    "parietal": ['CP5', 'CP1', 'CP2', 'CP6', 'P7', 'P3', 'Pz', 'P4', 'P8'],
    "occipital": ['O1', 'Oz', 'O2']
}

# Load PLV matrix
plv_matrix = np.load("/home/vincent/AAA_projects/MVCS/Neuroscience/Analysis/
↪Phase Syncronization/plv_matrix.npy")

# Number of channels
N = len(eeg_channel_names)
```

```python
# Set natural frequencies proportional to mean PLV of each channel
omega = np.mean(plv_matrix, axis=1)

# Phase bias based on mean phase difference between channels
phase_diff_matrix = np.zeros((N, N))
for i in range(N):
    for j in range(N):
        if i != j:
            phase_diff = np.mean(np.angle(hilbert(EEG_data[:, i])) - np.
 ↪angle(hilbert(EEG_data[:, j])))
            phase_diff_matrix[i, j] = phase_diff

# Use initial phases from the EEG data directly, not regions
initial_phases = np.angle(hilbert(EEG_data))[0, :]

# Modify the Kuramoto function to include weighted coupling and phase bias
def kuramoto_weighted_bias(t, y, omega, K):
    weighted_sin = plv_matrix * np.sin(y - y[:, np.newaxis] - phase_diff_matrix)
    dydt = omega + K/N * np.sum(weighted_sin, axis=1)
    return dydt

# Hilbert transform to get analytical signal
analytic_signal = hilbert(EEG_data)
phases = np.angle(analytic_signal)

# Average phase for the initial time point for each region
avg_phases = {}
for region, channels in regions.items():
    indices = [eeg_channel_names.index(ch) for ch in channels]
    avg_phases[region] = np.mean(phases[0, indices])

K = 5.0  # Coupling strength

# Time span for the simulation
t_span = (0, 300)
t_eval = np.linspace(t_span[0], t_span[1], 300)

# Solve differential equations using the modified model
solution = solve_ivp(kuramoto_weighted_bias, t_span, initial_phases,␣
 ↪t_eval=t_eval, args=(omega, K))

# The simulated Kuramoto phases
kuramoto_phases = solution.y

# Create a time axis for plotting
time_axis = np.linspace(t_span[0], t_span[1], 2 * len(EEG_data))
```

```python
# Interpolating the Kuramoto phases to match the EEG data length
from scipy.interpolate import interp1d
interp_funcs = [interp1d(t_eval, kuramoto_phases[i, :]) for i in range(N)]
kuramoto_phases_interp = np.array([f(time_axis) for f in interp_funcs])

# Plot
plt.figure(figsize=(14, 7))
for i, region in enumerate(regions.keys()):
    plt.plot(time_axis, kuramoto_phases_interp[i, :], label=region, linewidth=1.
 ↪5)

plt.legend()  # Place legend call here to ensure one entry per region
plt.xlabel('Time')
plt.ylabel('Phase')
plt.title('Simulated Kuramoto Phases for Brain Regions')

save_directory = "/home/vincent/AAA_projects/MVCS/Neuroscience/Modelling/
 ↪Kuramoto"

# Check if the directory exists and create it if not
if not os.path.exists(save_directory):
    os.makedirs(save_directory)

plot_path = os.path.join(save_directory, "Kuramoto_EEG_Channels_Plot.png")
plt.savefig(plot_path, dpi=300)  # Increased dpi for better resolution

# Save the simulated Kuramoto phases
kuramoto_data_path = os.path.join(save_directory, "kuramoto_phases_channels.
 ↪npy")
np.save(kuramoto_data_path, kuramoto_phases)

print(f"Plot saved at {plot_path}")
print(f"Kuramoto phases saved at {kuramoto_data_path}")

plt.show()  # Show the plot at the end
```

```
Plot saved at /home/vincent/AAA_projects/MVCS/Neuroscience/Modelling/Kuramoto/Ku
ramoto_EEG_Channels_Plot.png
Kuramoto phases saved at /home/vincent/AAA_projects/MVCS/Neuroscience/Modelling/
Kuramoto/kuramoto_phases_channels.npy
```

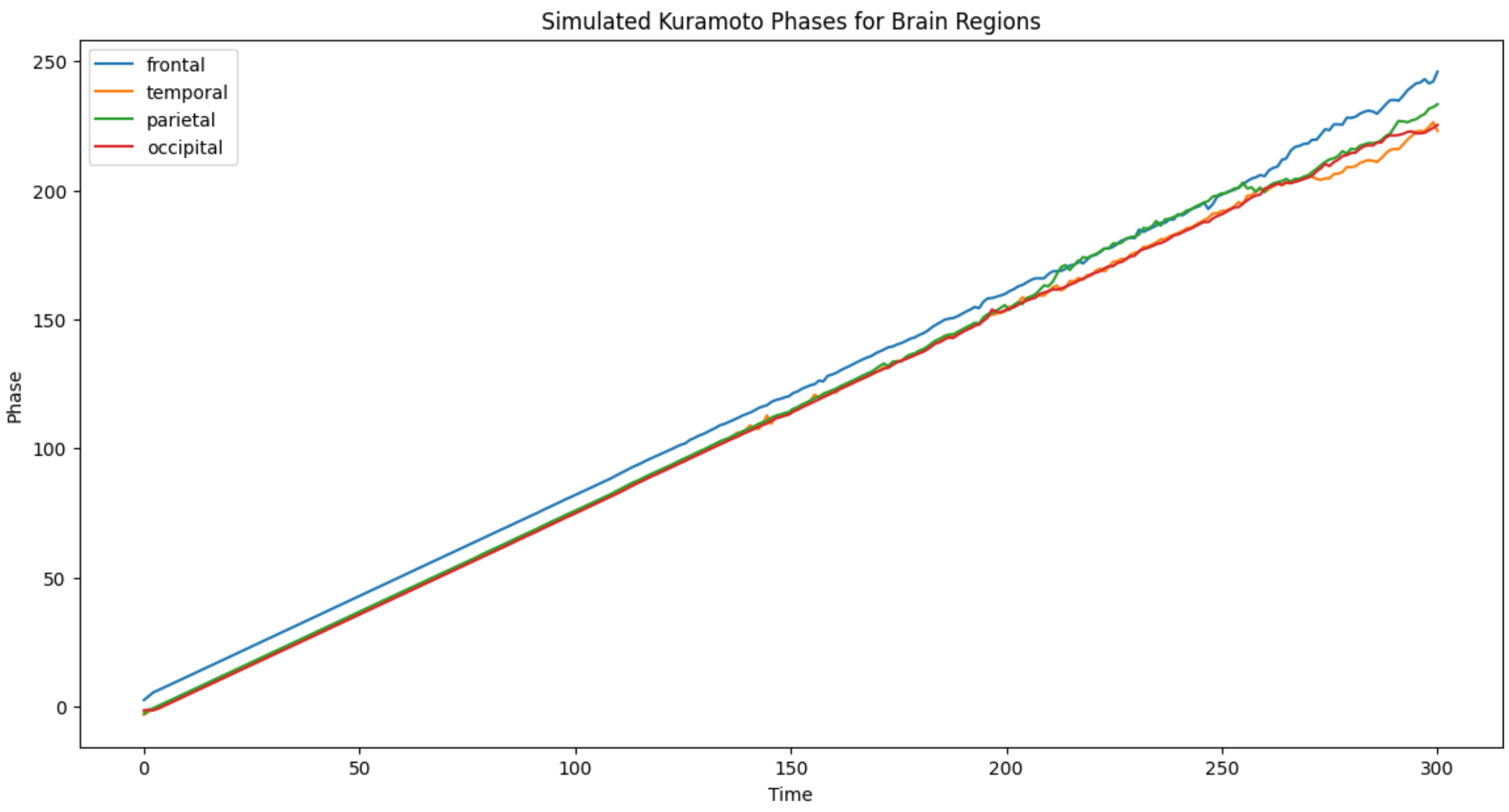

[23]:
```
plt.hist(kuramoto_phases.ravel(), bins=100, density=True)
plt.xlabel('Phase')
plt.ylabel('Frequency')
plt.title('Distribution of Kuramoto Phases')
plt.show()
```

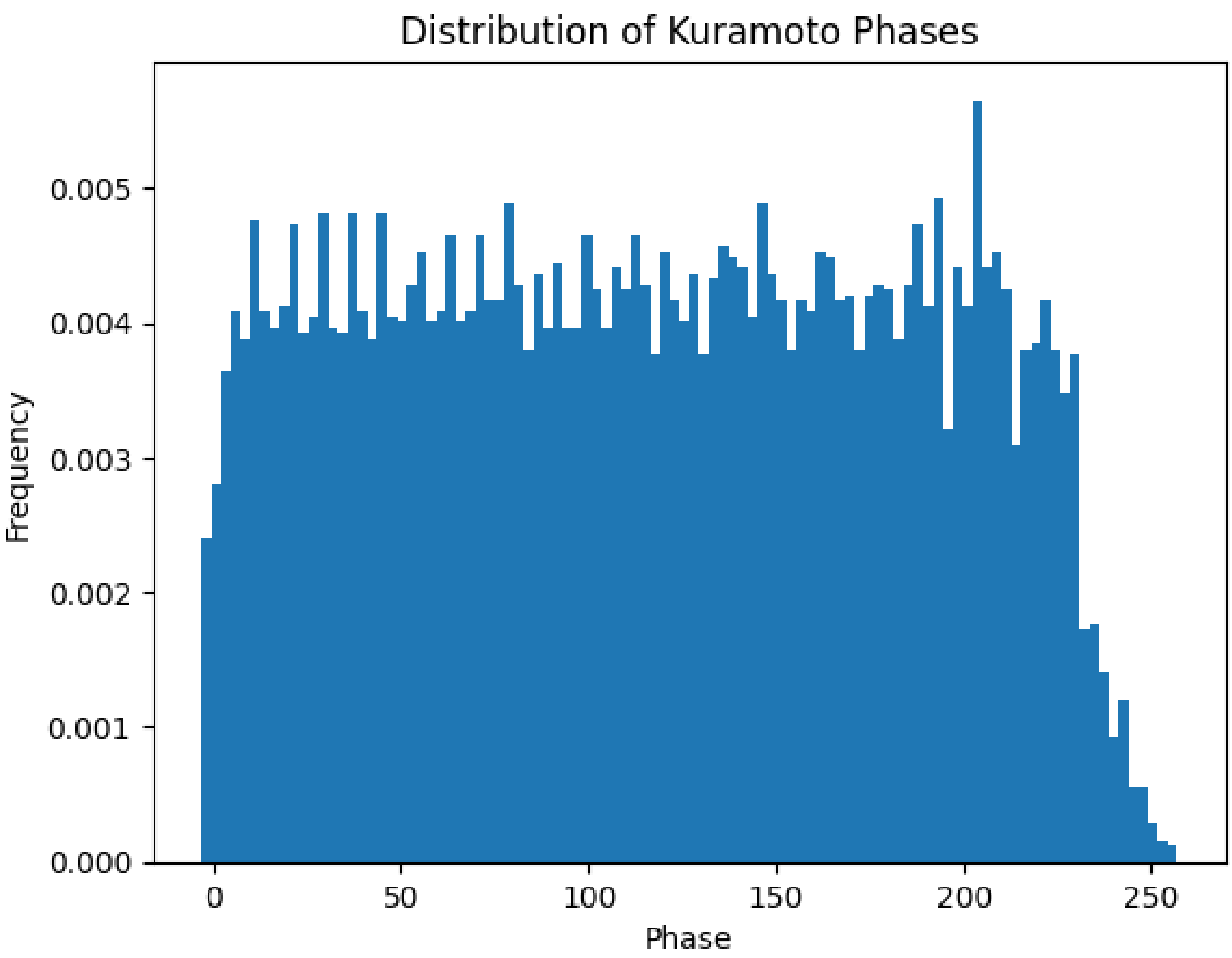

## 4 Compare with EEG data

[24]:
```
import numpy as np
import matplotlib.pyplot as plt
from scipy.signal import hilbert

def compute_phase_locking_value(signal1, signal2):
    phase1 = np.angle(hilbert(signal1))
    phase2 = np.angle(hilbert(signal2))
    phase_diff = phase1 - phase2
    PLV = abs(np.sum(np.exp(1j * phase_diff))) / len(phase_diff)
    return PLV


# List of EEG channels
eeg_channel_names = ['Fp1', 'Fpz', 'Fp2', 'F7', 'F3', 'Fz', 'F4', 'F8', 'FC5',␣
 ↪'FC1', 'FC2', 'FC6',
                    'M1', 'T7', 'C3', 'Cz', 'C4', 'T8', 'M2', 'CP5', 'CP1',␣
 ↪'CP2', 'CP6',
```

```
                    'P7', 'P3', 'Pz', 'P4', 'P8', 'POz', 'O1', 'Oz', 'O2']

num_channels = len(eeg_channel_names)

# Load the precomputed EEG PLVs
eeg_plv_path = "/home/vincent/AAA_projects/MVCS/Neuroscience/Analysis/Phase␣
↪Syncronization/plv_matrix.npy"
plv_matrix = np.load(eeg_plv_path)

# Compute Kuramoto model channel PLVs
kuramoto_channel_phases = ...  # Assuming you have Kuramoto phase data for each␣
↪channel
kuramoto_channel_plvs = np.zeros((num_channels, num_channels))
for i in range(num_channels):
    for j in range(num_channels):
        if i != j:
            kuramoto_channel_plvs[i, j] =␣
↪compute_phase_locking_value(kuramoto_channel_phases[:, i],␣
↪kuramoto_channel_phases[:, j])

# Visualize channel-wise EEG PLVs and channel-wise Kuramoto PLVs
plt.figure(figsize=(20, 10))

# Plot for real EEG data
plt.subplot(2, 2, 1)
plt.imshow(plv_matrix, cmap="viridis", interpolation="none")
plt.colorbar(label="PLV")
plt.title("EEG Channel PLV")
plt.xticks(np.arange(num_channels), eeg_channel_names, rotation=90)
plt.yticks(np.arange(num_channels), eeg_channel_names)

# Plot for Kuramoto model channel-wise
plt.subplot(2, 2, 2)
plt.imshow(kuramoto_channel_plvs, cmap="viridis", interpolation="none")
plt.colorbar(label="PLV")
plt.title("Kuramoto Channel PLV")
plt.xticks(np.arange(num_channels), eeg_channel_names, rotation=90)
plt.yticks(np.arange(num_channels), eeg_channel_names)

# Assuming you've computed regional Kuramoto PLVs as kuramoto_plv_matrix
plt.subplot(2, 2, 3)
plt.imshow(kuramoto_plv_matrix, cmap="viridis", interpolation="none")
plt.colorbar(label="PLV")
plt.title("Kuramoto Regional PLV")
plt.xticks(np.arange(N), regions.keys(), rotation=90)
plt.yticks(np.arange(N), regions.keys())
```

```
plt.tight_layout()
plt.show()
```

```
---------------------------------------------------------------------------
TypeError                                 Traceback (most recent call last)
Cell In[24], line 30
     28     for j in range(num_channels):
     29         if i != j:
---> 30             kuramoto_channel_plvs[i, j] =␣
 ↪compute_phase_locking_value(kuramoto_channel_phases[:, i],␣
 ↪kuramoto_channel_phases[:, j])
     32 # Visualize channel-wise EEG PLVs and channel-wise Kuramoto PLVs
     33 plt.figure(figsize=(20, 10))

TypeError: 'ellipsis' object is not subscriptable
```

## 5 Feature Extraction

```
[ ]: # Normalization, since phases from the Kuramoto model will be between 0 and␣
      ↪2*pi,
     normalized_phases = kuramoto_phases / (2 * np.pi)

     # Feature Creation
     instantaneous_frequency = np.vstack([np.zeros(normalized_phases.shape[0]),
                                          np.diff(normalized_phases, axis=1)])

     # Combining both the normalized phases and instantaneous frequency for a richer␣
      ↪feature set
     features = np.vstack([normalized_phases, instantaneous_frequency]).T

     # Printing the head (first 5 rows)
     print(features[:5])

     # Save the features as a .npy file for future use
     np.save('/home/vincent/AAA_projects/MVCS/Neuroscience/Modelling/Kuramoto/
      ↪kuramoto_features.npy', features)
```

## 6 Bifurcation

```
[ ]: def compute_order_parameter(phases):
         """Compute the order parameter r from the phases."""
         return np.abs(np.mean(np.exp(1j * phases), axis=0))

     # Bifurcation analysis
```

```
K_values = np.linspace(0, 5, 50)  # Change the range and number of points as␣
↪needed
r_values = []

for K in K_values:
    solution = solve_ivp(kuramoto, t_span, initial_phases, t_eval=t_eval,␣
↪args=(omega, K))
    r = compute_order_parameter(solution.y)
    r_values.append(np.mean(r[-int(desired_length * 0.1):]))  # average of the␣
↪last 10% to ensure steady-state

# Plot bifurcation diagram
plt.figure(figsize=(12, 6))
plt.plot(K_values, r_values, '-o')
plt.xlabel('Coupling Strength K')
plt.ylabel('Order Parameter r')
plt.title('Bifurcation Diagram of Kuramoto Model')
plt.tight_layout()

bifurcation_plot_path = os.path.join(save_directory, "Kuramoto_Bifurcation_Plot.
↪png")
plt.savefig(bifurcation_plot_path)

print(f"Bifurcation plot saved at {bifurcation_plot_path}")
```

# CNN Tensor Creation

September 8, 2023

# 1 CNN Tensor Creation

# 2 EEG

[3]:
```
import torch
from sklearn.preprocessing import StandardScaler
import numpy as np

# Load EEG data
EEG_data = np.load('/home/vincent/AAA_projects/MVCS/Neuroscience/
 ↪eeg_data_with_channels.npy', allow_pickle=True)

# Standardize the EEG data
scaler = StandardScaler()
EEG_data_standardized = scaler.fit_transform(EEG_data)

# Convert the data to a PyTorch tensor
eeg_tensor = torch.tensor(EEG_data_standardized, dtype=torch.float32)

eeg_tensor = eeg_tensor.T  # Transpose to make channels the second dimension
eeg_tensor = eeg_tensor.unsqueeze(0).unsqueeze(2)  # Add batch and height␣
 ↪dimensions
# Now eeg_tensor should have shape [1, num_channels, 1, num_timepoints]

print(f"The reshaped tensor has shape: {eeg_tensor.shape}")

# Step 4: Save the tensor
save_path = "/home/vincent/AAA_projects/MVCS/Neuroscience/Models/CNN/EEG_tensor.
 ↪pth"
torch.save(eeg_tensor, save_path)
```

```
The reshaped tensor has shape: torch.Size([1, 32, 1, 4227788])
```

## 3 Band Powers

[9]:
```
import numpy as np
import torch

# Load the band powers data from the .npy file
band_powers_data = np.load('/home/vincent/AAA_projects/MVCS/Neuroscience/
↪Analysis/Spectral Analysis/BandPowers_x.npy', allow_pickle=True).item()

# Print the populated keys in band_powers_data (channels)
print("Populated keys in band_powers_data:", list(band_powers_data.keys()))

# Extract eeg_channels from the data (assuming they're the top-level keys)
eeg_channels = list(band_powers_data.keys())

# Extract frequency_bands from the data (assuming they're the second-level keys␣
↪for any channel)
frequency_bands = list(next(iter(band_powers_data.values())).keys())

# Assuming band_powers_data is in the format {channel: {band: values}}
# Convert the band powers dictionary to a tensor
num_time_steps = len(next(iter(next(iter(band_powers_data.values())).values())))
tensor_shape = (num_time_steps, len(eeg_channels), len(frequency_bands))
band_power_tensor = torch.empty(tensor_shape)

for i, channel in enumerate(eeg_channels):
    for j, band in enumerate(frequency_bands):
        band_power_tensor[:, i, j] = torch.
↪tensor(band_powers_data[channel][band])

# Saving path for the band power tensor
save_path = "/home/vincent/AAA_projects/MVCS/Neuroscience/Models/CNN/
↪band_power_tensor.pth"

# Save the tensor
torch.save(band_power_tensor, save_path)
```

```
Populated keys in band_powers_data: ['Fp1', 'Fpz', 'Fp2', 'F7', 'F3', 'Fz',
'F4', 'F8', 'FC5', 'FC1', 'FC2', 'FC6', 'M1', 'T7', 'C3', 'Cz', 'C4', 'T8',
'M2', 'CP5', 'CP1', 'CP2', 'CP6', 'P7', 'P3', 'Pz', 'P4', 'P8', 'POz', 'O1',
'Oz', 'O2']
```

## 4 Fast Fourier Transform

[14]:
```
import numpy as np
import scipy.signal
import torch

# Load the FFT PSD data
load_path = '/home/vincent/AAA_projects/MVCS/Neuroscience/Analysis/Spectral␣
↪Analysis/combined_fft_psd_x.npy'
combined_fft_psd_data = np.load(load_path, allow_pickle=True).item()

eeg_channels = ['Fp1', 'Fpz', 'Fp2', 'F7', 'F3', 'Fz', 'F4', 'F8', 'FC5',␣
↪'FC1', 'FC2', 'FC6',
                'M1', 'T7', 'C3', 'Cz', 'C4', 'T8', 'M2', 'CP5', 'CP1', 'CP2',␣
↪'CP6',
                'P7', 'P3', 'Pz', 'P4', 'P8', 'POz', 'O1', 'Oz', 'O2']

num_channels = len(combined_fft_psd_data.keys())
# Assuming all PSD sequences have the same length
num_frequencies = len(next(iter(combined_fft_psd_data.values())))
# Reshape the data
fft_psd_tensor = torch.empty((num_channels, num_frequencies))

for i, channel in enumerate(eeg_channels):
    fft_psd_tensor[i] = torch.tensor(combined_fft_psd_data[channel])

# Saving path for the CNN tensor
save_path = "/home/vincent/AAA_projects/MVCS/Neuroscience/Models/CNN/
↪fast_fourier_transform_psd_tensor.pth"

# Save the tensor
torch.save(fft_psd_tensor, save_path)
```

## 5 Short Time Fourier Transform

[17]:
```
import numpy as np
import torch

# Load the STFT data
stft_data_dict = np.load('/home/vincent/AAA_projects/MVCS/Neuroscience/Analysis/
↪Spectral Analysis/STFT_x.npy', allow_pickle=True).item()

# Get number of channels, frequencies, and time intervals from the data
num_channels = len(stft_data_dict)
num_frequencies, num_time_intervals = stft_data_dict[next(iter(stft_data_dict.
↪keys()))].shape
```

```
# Initialize a tensor to store the STFT data
stft_tensor = torch.empty((num_channels, num_frequencies, num_time_intervals))

# Populate the tensor
for i, channel_data in enumerate(stft_data_dict.values()):
    stft_tensor[i] = torch.tensor(channel_data)


# Saving path for the CNN tensor
save_path = "/home/vincent/AAA_projects/MVCS/Neuroscience/Models/CNN/
 ↪short_time_fourier_transform_tensor.pth"

# Save the tensor
torch.save(stft_tensor, save_path)
```

## 6 3D Embedding data EEG

[26]:
```
import numpy as np
import zipfile
import os

eeg_channels = ['Fp1', 'Fpz', 'Fp2', 'F7', 'F3', 'Fz', 'F4', 'F8', 'FC5',␣
 ↪'FC1', 'FC2', 'FC6',
                'M1', 'T7', 'C3', 'Cz', 'C4', 'T8', 'M2', 'CP5', 'CP1', 'CP2',␣
 ↪'CP6',
                'P7', 'P3', 'Pz', 'P4', 'P8', 'POz', 'O1', 'Oz', 'O2']

# Extract zip file
zip_path = '/home/vincent/AAA_projects/MVCS/Neuroscience/Analysis/Phase Space/
 ↪3dembedded_data.zip'
extract_path = '/home/vincent/AAA_projects/MVCS/Neuroscience/Analysis/Phase␣
 ↪Space/3dembedding_data/temp'
with zipfile.ZipFile(zip_path, 'r') as zip_ref:
    zip_ref.extractall(extract_path)

# Load the first channel to determine the number of data points
first_channel_data = np.load(os.path.join(extract_path,␣
 ↪f'3dembedded_{eeg_channels[0]}.npy'))
num_data_points, emb_dim = first_channel_data.shape
num_channels = len(eeg_channels)

# Initialize a 4D tensor to store embeddings
eeg_tensor = np.zeros((num_data_points, emb_dim, num_channels, 1))
```

```
# Populate the tensor
for idx, channel in enumerate(eeg_channels):
    eeg_tensor[:, :, idx, 0] = np.load(os.path.join(extract_path,␣
 ↪f'3dembedded_{channel}.npy'))

# Saving path for the CNN tensor
save_path = "/home/vincent/AAA_projects/MVCS/Neuroscience/Models/CNN/
 ↪3D_embedding_EEG_tensor.pth"

# Save the tensor
torch.save(eeg_tensor, save_path)
```

```
---------------------------------------------------------------------------
ValueError                                Traceback (most recent call last)
Cell In[26], line 25
     23 # Populate the tensor
     24 for idx, channel in enumerate(eeg_channels):
---> 25     eeg_tensor[:, :, idx, 0] = np.load(os.path.join(extract_path,␣
 ↪f'3dembedded_{channel}.npy'))
     27 # Saving path for the CNN tensor
     28 save_path = "/home/vincent/AAA_projects/MVCS/Neuroscience/Models/CNN/
 ↪3D_embedding_EEG_tensor.pth"

ValueError: could not broadcast input array from shape (4227590,3) into shape␣
 ↪(4227588,3)
```

# 7 Hurst Exponents

[69]:
```
import numpy as np
import torch

# Load Data
hurst_exponents = np.load('/home/vincent/AAA_projects/MVCS/Neuroscience/
 ↪HurstExponents/hurst_exponents.npy')

# Reshape and normalize data
hurst_exponents = hurst_exponents.reshape(-1, 1)  # Reshape to (num_channels, 1)
normalized_data = (hurst_exponents - np.mean(hurst_exponents)) / np.
 ↪std(hurst_exponents)

# Convert to PyTorch tensor and reshape for CNN
hurst_tensor = torch.tensor(normalized_data).float().unsqueeze(0).unsqueeze(1) ␣
 ↪# Shape: (1, 1, num_channels, 1)

print(hurst_tensor.shape)  # should print torch.Size([1, 1, 32, 1])
```

```
# Saving path for the CNN tensor
save_path = "/home/vincent/AAA_projects/MVCS/Neuroscience/Models/CNN/
↪Hurst_tensor.pth"

# Save the tensor
torch.save(hurst_tensor, save_path)
```

```
torch.Size([1, 1, 32, 1])
```

# 8 MFDFA

[68]:
```
import numpy as np
import torch

# Load the saved numpy features
cnn_features = np.load('/home/vincent/AAA_projects/MVCS/Neuroscience/Analysis/
↪MFDFA/cnn_mfdfa.npy')

# Reshape the Data to include a channel dimension
cnn_features_tensor = cnn_features[..., np.newaxis]  # Adding a channel␣
↪dimension

# Convert to PyTorch tensor
cnn_features_tensor_torch = torch.tensor(cnn_features_tensor, dtype=torch.
↪float32)

print(cnn_features_tensor_torch.shape)

# Save the tensor using torch.save
save_path = "/home/vincent/AAA_projects/MVCS/Neuroscience/Models/CNN/
↪mfdfa_tensor.pth"
torch.save(cnn_features_tensor_torch, save_path)
```

```
torch.Size([9, 32, 10, 1])
```

# 9 MFDFA Concatenated

[67]:
```
import torch
import numpy as np

# Load the MFDFA results from the file
save_dir = '/home/vincent/AAA_projects/MVCS/Neuroscience/Analysis/MFDFA/'
mfdfa_results_np = np.load(save_dir + 'MFDFA_results.npy', allow_pickle=True)

# Extract the dictionary from the numpy scalar
```

```
mfdfa_results = mfdfa_results_np.item()

# Extract the channels data from your dictionary
channels_data = [mfdfa_results[key] for key in sorted(mfdfa_results.keys())]

# List to store the data for each channel
channel_images = []

for channel_data in channels_data:
    scales, fluctuations = channel_data
    # Reshape the scales and fluctuations into column vectors
    scales = scales.reshape(-1, 1)
    fluctuations = fluctuations.mean(axis=1).reshape(-1, 1)  # Take the mean of␣
 ↪fluctuations across its scales for simplicity

    # Concatenate the scales and fluctuations horizontally to form an "image"
    channel_image = np.hstack([scales, fluctuations])
    channel_images.append(channel_image)

# Convert the list of channel images to a tensor
tensor = torch.from_numpy(np.array(channel_images)).float()

# Reshape the tensor to [batch_size, channels, height, width]
mfdfa_concatd_tensor = tensor.unsqueeze(1)  # Adding a channel dimension

print(mfdfa_concatd_tensor.shape)

# Save the tensor using torch.save
save_path = "/home/vincent/AAA_projects/MVCS/Neuroscience/Models/CNN/
 ↪mfdfa_concatd_tensor.pth"
torch.save(mfdfa_concatd_tensor, save_path)
```

```
torch.Size([32, 1, 30, 2])
```

## 10 Arnold Tongues Rotation Numbers

[66]:
```
import numpy as np
import torch

# Load the saved rotation numbers
rotation_numbers_path = '/home/vincent/AAA_projects/MVCS/Neuroscience/Analysis/
 ↪Arnold Tongues/rotation_numbers.npy'
rotation_numbers_dict = np.load(rotation_numbers_path, allow_pickle=True).item()

# EEG channels definition (as per your previous code)
```

```
eeg_channels = ['Fp1', 'Fpz', 'Fp2', 'F7', 'F3', 'Fz', 'F4', 'F8', 'FC5',␣
↪'FC1', 'FC2', 'FC6',
                'M1', 'T7', 'C3', 'Cz', 'C4', 'T8', 'M2', 'CP5', 'CP1', 'CP2',␣
↪'CP6',
                'P7', 'P3', 'Pz', 'P4', 'P8', 'POz', 'O1', 'Oz', 'O2']

# Omegas and K_values are derived from your earlier code
omegas = np.linspace(0, 1, 300)
K_values = np.linspace(0, 2 * np.pi, 300)

tensor_shape = (len(eeg_channels), len(K_values), len(omegas))
rotation_numbers_tensor = np.zeros(tensor_shape)

for idx, ch in enumerate(eeg_channels):
    rotation_numbers_tensor[idx] = rotation_numbers_dict[ch]

# Convert the numpy tensor to a PyTorch tensor
rotation_numbers_torch_tensor = torch.tensor(rotation_numbers_tensor)

print(rotation_numbers_torch_tensor.shape)

# Save the PyTorch tensor to disk
torch_save_path = '/home/vincent/AAA_projects/MVCS/Neuroscience/Analysis/Arnold␣
↪Tongues/rotation_numbers_tensor.pt'
torch.save(rotation_numbers_torch_tensor, torch_save_path)
```

```
torch.Size([32, 300, 300])
```

## 11 Transfer Entropy Hemispheric

[65]:
```
import torch
import numpy as np

# Load the saved CNN input data
cnn_input_path = '/home/vincent/AAA_projects/MVCS/Neuroscience/Features/CNN/
↪cnn_transfer_entropy_hemispheric_avg_input.npy'
cnn_input_data = np.load(cnn_input_path)

# Convert the numpy array to a PyTorch tensor
cnn_input_tensor = torch.tensor(cnn_input_data)

print(cnn_input_tensor.shape)

# Save the PyTorch tensor to disk
torch_save_path = '/home/vincent/AAA_projects/MVCS/Neuroscience/Models/CNN/
↪transfer_entropy_hemispheric_avg_input_tensor.pt'
```

```
torch.save(cnn_input_tensor, torch_save_path)
```

```
torch.Size([92, 92])
```

## 12 Transfer Entropy Regional

[64]:
```
import numpy as np

# Load the saved CNN input data
cnn_input_path = '/home/vincent/AAA_projects/MVCS/Neuroscience/Analysis/
↪Transfer Entropy/regional_transfer_entropy_results.npy'
cnn_input_data = np.load(cnn_input_path, allow_pickle=True).item()

regions = ["Frontal", "Temporal", "Parietal", "Occipital"]
TE_matrix = np.zeros((4, 4))

for i, source in enumerate(regions):
    for j, target in enumerate(regions):
        if i != j:
            key = f"{source}_to_{target}"
            TE_matrix[i, j] = cnn_input_data[key]

print(TE_matrix)

# Convert the numpy array to a PyTorch tensor
cnn_input_tensor = torch.tensor(TE_matrix)

print(cnn_input_tensor.shape)

# Save the PyTorch tensor to disk
torch_save_path = '/home/vincent/AAA_projects/MVCS/Neuroscience/Models/CNN/
↪transfer_entropy_regional_tensor.pt'
torch.save(cnn_input_tensor, torch_save_path)
```

```
[[0.         0.25110662 0.57730508 0.67410323]
 [0.33752969 0.         0.47118759 0.61260954]
 [0.38813154 0.22015982 0.         0.63537753]
 [0.42216949 0.28119128 0.58496923 0.        ]]
torch.Size([4, 4])
```

## 13 Transfer Entropy Granular

[ ]:
```
import numpy as np
import torch

# Load the numpy results file
```

```
results_file_path = '/home/vincent/AAA_projects/MVCS/Neuroscience/Analysis/
↪Transfer Entropy/full_granularity_transfer_entropy_results.npy'
TE_results = np.load(results_file_path, allow_pickle=True).item()

# Create a 4x4 matrix with zeros
TE_matrix = np.zeros((32, 32))

# Fill the matrix using the provided results
for i, source_region in enumerate(regions):
    for j, target_region in enumerate(regions):
        if i != j:
            key = f"{source_region}_to_{target_region}"
            TE_matrix[i, j] = TE_results[key]

# Convert the numpy matrix to PyTorch tensor
TE_tensor = torch.tensor(TE_matrix)

print(TE_tensor.shape)

# Save the PyTorch tensor to disk
torch_save_path = '/home/vincent/AAA_projects/MVCS/Neuroscience/Models/CNN/
↪transfer_entropy_granular_tensor.pt'
torch.save(TE_tensor, torch_save_path)
```

# 14 DSPM

[77]:
```
import numpy as np
import torch

# Load the numpy data
cnn_data_np = np.load('/home/vincent/AAA_projects/MVCS/Neuroscience/Features/
↪CNN/cnn_dspm.npy')

# Convert the numpy array to a PyTorch tensor
cnn_data_tensor = torch.tensor(cnn_data_np)

print(cnn_data_tensor.shape)

# Save the tensor using PyTorch's save function
torch.save(cnn_data_tensor, '/home/vincent/AAA_projects/MVCS/Neuroscience/
↪Models/CNN/dspm_tensor.pt')
```

```
torch.Size([19, 18840, 10])
```

## 15 Higuchi Fractal Dimension

[60]:
```
import torch
import numpy as np

# Load the CNN feature
cnn_features_path = '/home/vincent/AAA_projects/MVCS/Neuroscience/Features/CNN/
↪cnn_fractal.npy'
cnn_features = np.load(cnn_features_path)

# Reshape for CNN (batch_size, channels, height, width)
cnn_tensor = torch.tensor(cnn_features).unsqueeze(0).unsqueeze(0)

print(cnn_tensor.shape)

# Save the PyTorch tensor to disk
torch_save_path = '/home/vincent/AAA_projects/MVCS/Neuroscience/Models/CNN/
↪higuchi_fractal_dimensions_tensor.pt'
torch.save(cnn_tensor, torch_save_path)
```

```
torch.Size([1, 1, 4, 8])
```

## 16 Spectral Entropy

[59]:
```
import numpy as np
import torch

# 1. Load the spectral entropy values
results_folder_path = '/home/vincent/AAA_projects/MVCS/Neuroscience/Analysis/
↪Spectral Analysis/'
results_file = "SpectralEntropy_x.npy"
spectral_entropy_dict = np.load(results_folder_path + results_file,␣
↪allow_pickle=True).item()

# Extract only the values from the dictionary
entropy_values = list(spectral_entropy_dict.values())

# 2. Convert the spectral entropy values into a tensor
entropy_tensor = torch.Tensor(entropy_values)

# 3. Reshape the tensor for CNN (1 sample, 1 channel, 32 height (EEG channels),␣
↪1 width (spectral entropy))
cnn_tensor = entropy_tensor.view(1, 1, 32, 1)

print(cnn_tensor.shape)  # should print torch.Size([1, 1, 32, 1])
```

```
# Save the PyTorch tensor to disk
torch_save_path = '/home/vincent/AAA_projects/MVCS/Neuroscience/Models/CNN/
↪spectral_entropy_tensor.pt'
torch.save(cnn_tensor, torch_save_path)
```

```
torch.Size([1, 1, 32, 1])
```

# 17 Spectral Centroids

[58]:
```
import numpy as np
import torch

# 1. Load the spectral centroid values
results_path = '/home/vincent/AAA_projects/MVCS/Neuroscience/Analysis/Spectral␣
↪Analysis/'
results_file = "SpectralCentroids_x.npy"
spectral_centroids_dict = np.load(results_path + results_file,␣
↪allow_pickle=True).item()

# Extract only the values from the dictionary
centroid_values = list(spectral_centroids_dict.values())

# 2. Convert the spectral centroid values into a tensor
centroid_tensor = torch.Tensor(centroid_values)

# 3. Reshape the tensor for CNN (1 sample, 1 channel, 32 height (EEG channels),␣
↪1 width (spectral centroid))
cnn_tensor = centroid_tensor.view(1, 1, 32, 1)

print(cnn_tensor.shape)  # should print torch.Size([1, 1, 32, 1])

# Save the PyTorch tensor to disk
torch_save_path = '/home/vincent/AAA_projects/MVCS/Neuroscience/Models/CNN/
↪spectral_centroids_tensor.pt'
torch.save(cnn_tensor, torch_save_path)
```

```
torch.Size([1, 1, 32, 1])
```

# 18 Frequency of Max Power

[71]:
```
import numpy as np
import torch

# Load the peak frequency data
```

```
results_path = '/home/vincent/AAA_projects/MVCS/Neuroscience/Analysis/Spectral␣
 ↪Analysis/'
results_file = "PeakFrequencies_x.npy"
peak_frequencies_dict = np.load(results_path + results_file, allow_pickle=True).
 ↪item()

# Extract only the values from the dictionary
peak_frequency_values = list(peak_frequencies_dict.values())

# Convert the peak frequency values into a tensor
peak_frequency_tensor = torch.Tensor(peak_frequency_values)

# Reshape the tensor for CNN (1 sample, 1 channel, 32 height (EEG channels), 1␣
 ↪width (peak frequency))
cnn_tensor = peak_frequency_tensor.view(1, 1, 32, 1)

print(cnn_tensor.shape)  # should print torch.Size([1, 1, 32, 1])

# Save the PyTorch tensor to disk
torch_save_path = '/home/vincent/AAA_projects/MVCS/Neuroscience/Models/CNN/
 ↪freq_max_power_tensor.pt'
torch.save(cnn_tensor, torch_save_path)
```

```
torch.Size([1, 1, 32, 1])
```

## 19 Spectral Edge Frequencies

[72]:

```
import numpy as np
import torch

# Load the spectral edge density data
results_path = '/home/vincent/AAA_projects/MVCS/Neuroscience/Analysis/Spectral␣
 ↪Analysis/'
results_file = "SpectralEdgeDensities_x.npy"
spectral_edge_densities_dict = np.load(results_path + results_file,␣
 ↪allow_pickle=True).item()

# Extract only the values from the dictionary
spectral_edge_density_values = list(spectral_edge_densities_dict.values())

# Convert the spectral edge density values into a tensor
spectral_edge_density_tensor = torch.Tensor(spectral_edge_density_values)

# Reshape the tensor for CNN (1 sample, 1 channel, 32 height (EEG channels), 1␣
 ↪width (spectral edge density))
cnn_tensor = spectral_edge_density_tensor.view(1, 1, 32, 1)
```

```python
print(cnn_tensor.shape)  # should print torch.Size([1, 1, 32, 1])

# Save the PyTorch tensor to disk
torch_save_path = '/home/vincent/AAA_projects/MVCS/Neuroscience/Models/CNN/
 ↪spectral_edge_freqs_tensor.pt'
torch.save(cnn_tensor, torch_save_path)
```

```
torch.Size([1, 1, 32, 1])
```

[ ]:

```python

```

# Neural Net

September 21, 2023

## 1 Load Tensors

[66]:
```
import torch

# List of tensor paths
tensor_paths = [
    "/home/vincent/AAA_projects/MVCS/Neuroscience/Models/CNN/
 ↪arnold_tongues_rotation_numbers_tensor.pt",
    "/home/vincent/AAA_projects/MVCS/Neuroscience/Models/CNN/dspm_tensor.pt",
    "/home/vincent/AAA_projects/MVCS/Neuroscience/Models/CNN/
 ↪higuchi_fractal_dimensions_tensor.pt",
    "/home/vincent/AAA_projects/MVCS/Neuroscience/Models/CNN/Hurst_tensor.pth",
    "/home/vincent/AAA_projects/MVCS/Neuroscience/Models/CNN/
 ↪mfdfa_concatd_tensor.pth",
    "/home/vincent/AAA_projects/MVCS/Neuroscience/Models/CNN/mfdfa_tensor.pth",
    "/home/vincent/AAA_projects/MVCS/Neuroscience/Models/CNN/
 ↪short_time_fourier_transform_tensor.pth",
    "/home/vincent/AAA_projects/MVCS/Neuroscience/Models/CNN/
 ↪transfer_entropy_granular_tensor.pt",
    "/home/vincent/AAA_projects/MVCS/Neuroscience/Models/CNN/
 ↪transfer_entropy_hemispheric_avg_input_tensor.pt",
    "/home/vincent/AAA_projects/MVCS/Neuroscience/Models/CNN/
 ↪transfer_entropy_regional_tensor.pt",
    "/home/vincent/AAA_projects/MVCS/Neuroscience/Models/CNN/
 ↪spectral_entropy_tensor.pt",
    "/home/vincent/AAA_projects/MVCS/Neuroscience/Models/CNN/
 ↪spectral_centroids_tensor.pt",
    "/home/vincent/AAA_projects/MVCS/Neuroscience/Models/CNN/
 ↪freq_max_power_tensor.pt",
    "/home/vincent/AAA_projects/MVCS/Neuroscience/Models/CNN/
 ↪spectral_edge_freqs_tensor.pt",
]

# Initialize an empty dictionary to store the tensors and another for their␣
 ↪shapes
tensors = {}
tensor_shapes = {}
```

```
# Load the tensors into a dictionary and collect their shapes
for path in tensor_paths:
    tensor_name = path.split('/')[-1].replace('.pt', '').replace('.pth', '')

    # Remove the 'h' from the end, if it exists
    if tensor_name.endswith("h"):
        tensor_name = tensor_name[:-1]

    # Load the tensor
    data = torch.load(path)
    tensors[tensor_name] = data

    # Check the type of the loaded data
    if isinstance(data, torch.Tensor):
        tensor_shapes[tensor_name] = data.shape
    elif isinstance(data, dict):  # Likely a state_dict
        tensor_shapes[tensor_name] = "state_dict (model parameters)"
    else:
        tensor_shapes[tensor_name] = "unknown type"

# Print the shapes of all loaded tensors
for name, shape in tensor_shapes.items():
    print(f"{name}: {shape}")
```

```
arnold_tongues_rotation_numbers_tensor: torch.Size([32, 300, 300])
dspm_tensor: torch.Size([19, 18840, 10])
higuchi_fractal_dimensions_tensor: torch.Size([1, 1, 4, 8])
Hurst_tensor: torch.Size([1, 1, 32, 1])
mfdfa_concatd_tensor: torch.Size([32, 1, 30, 2])
mfdfa_tensor: torch.Size([9, 32, 10, 1])
short_time_fourier_transform_tensor: torch.Size([32, 1001, 4229])
transfer_entropy_granular_tensor: torch.Size([4, 4])
transfer_entropy_hemispheric_avg_input_tensor: torch.Size([92, 92])
transfer_entropy_regional_tensor: torch.Size([4, 4])
spectral_entropy_tensor: torch.Size([1, 1, 32, 1])
spectral_centroids_tensor: torch.Size([1, 1, 32, 1])
freq_max_power_tensor: torch.Size([1, 1, 32, 1])
spectral_edge_freqs_tensor: torch.Size([1, 1, 32, 1])
```

[67]:
```
# Loop through the loaded tensors to check for NaNs and Infs
for tensor_name, tensor_data in tensors.items():
    # Only perform the checks if the data is a tensor
    if isinstance(tensor_data, torch.Tensor):
        total_elements = torch.numel(tensor_data)

        nans_count = torch.sum(torch.isnan(tensor_data)).item()
```

```
        infs_count = torch.sum(torch.isinf(tensor_data)).item()

        nans_percentage = (nans_count / total_elements) * 100
        infs_percentage = (infs_count / total_elements) * 100

        print(f"Percentage of NaNs in {tensor_name}: {nans_percentage}%")
        print(f"Percentage of Infs in {tensor_name}: {infs_percentage}%")
    else:
        print(f"{tensor_name} is not a tensor, skipping...")
```

```
Percentage of NaNs in arnold_tongues_rotation_numbers_tensor: 0.0%
Percentage of Infs in arnold_tongues_rotation_numbers_tensor: 0.0%
Percentage of NaNs in dspm_tensor: 0.0%
Percentage of Infs in dspm_tensor: 0.0%
Percentage of NaNs in higuchi_fractal_dimensions_tensor: 0.0%
Percentage of Infs in higuchi_fractal_dimensions_tensor: 0.0%
Percentage of NaNs in Hurst_tensor: 0.0%
Percentage of Infs in Hurst_tensor: 0.0%
Percentage of NaNs in mfdfa_concatd_tensor: 0.0%
Percentage of Infs in mfdfa_concatd_tensor: 0.0%
Percentage of NaNs in mfdfa_tensor: 0.0%
Percentage of Infs in mfdfa_tensor: 0.0%
Percentage of NaNs in short_time_fourier_transform_tensor: 0.0%
Percentage of Infs in short_time_fourier_transform_tensor: 0.0%
Percentage of NaNs in transfer_entropy_granular_tensor: 0.0%
Percentage of Infs in transfer_entropy_granular_tensor: 0.0%
Percentage of NaNs in transfer_entropy_hemispheric_avg_input_tensor: 0.0%
Percentage of Infs in transfer_entropy_hemispheric_avg_input_tensor: 0.0%
Percentage of NaNs in transfer_entropy_regional_tensor: 0.0%
Percentage of Infs in transfer_entropy_regional_tensor: 0.0%
Percentage of NaNs in spectral_entropy_tensor: 0.0%
Percentage of Infs in spectral_entropy_tensor: 0.0%
Percentage of NaNs in spectral_centroids_tensor: 0.0%
Percentage of Infs in spectral_centroids_tensor: 0.0%
Percentage of NaNs in freq_max_power_tensor: 0.0%
Percentage of Infs in freq_max_power_tensor: 0.0%
Percentage of NaNs in spectral_edge_freqs_tensor: 0.0%
Percentage of Infs in spectral_edge_freqs_tensor: 0.0%
```

## 2 Match dimensions, reshape, and normalize

[2]:
```
import torch.nn.functional as F

def preprocess_and_resize_tensor(tensor, target_shape):
    # Add missing batch and channel dimensions
    while len(tensor.shape) < 4:
```

```python
        tensor = tensor.unsqueeze(0)

    # Reduce the channel dimension to 1 by taking the mean along that axis
    tensor = torch.mean(tensor, dim=1, keepdim=True)

    # Normalize
    mean = tensor.mean()
    std = tensor.std()
    if std != 0:
        tensor = (tensor - mean) / std

    # Reshape/resize to target_shape
    tensor = F.interpolate(tensor, size=target_shape[2:], mode='bilinear',␣
 ↪align_corners=True)

    return tensor

target_shape = [1, 1, 32, 32]

# List of all your tensors
all_tensors = [tensors[key] for key in tensors]

# Preprocess all tensors
processed_tensors = [preprocess_and_resize_tensor(tensor, target_shape) for␣
 ↪tensor in all_tensors]

# Store preprocessed tensors back into the `tensors` dictionary
for key, tensor in zip(tensors.keys(), processed_tensors):
    tensors[key] = tensor

# Print out the new shapes
for key in tensors.keys():
    print(f"Processed tensor {key} shape: {tensors[key].shape}")
```

```
Processed tensor arnold_tongues_rotation_numbers_tensor shape: torch.Size([1, 1,
32, 32])
Processed tensor dspm_tensor shape: torch.Size([1, 1, 32, 32])
Processed tensor higuchi_fractal_dimensions_tensor shape: torch.Size([1, 1, 32,
32])
Processed tensor Hurst_tensor shape: torch.Size([1, 1, 32, 32])
Processed tensor mfdfa_concatd_tensor shape: torch.Size([32, 1, 32, 32])
Processed tensor mfdfa_tensor shape: torch.Size([9, 1, 32, 32])
Processed tensor short_time_fourier_transform_tensor shape: torch.Size([1, 1,
32, 32])
Processed tensor transfer_entropy_granular_tensor shape: torch.Size([1, 1, 32,
32])
Processed tensor transfer_entropy_hemispheric_avg_input_tensor shape:
```

```
torch.Size([1, 1, 32, 32])
Processed tensor transfer_entropy_regional_tensor shape: torch.Size([1, 1, 32,
32])
Processed tensor spectral_entropy_tensor shape: torch.Size([1, 1, 32, 32])
Processed tensor spectral_centroids_tensor shape: torch.Size([1, 1, 32, 32])
Processed tensor freq_max_power_tensor shape: torch.Size([1, 1, 32, 32])
Processed tensor spectral_edge_freqs_tensor shape: torch.Size([1, 1, 32, 32])
```

# 3 CNN

[3]:
```
import torch
import torch.nn as nn
from torch.utils.data import Dataset, DataLoader

tensor_names = [
    'arnold_tongues_rotation_numbers_tensor',
    'dspm_tensor',
    'higuchi_fractal_dimensions_tensor',
    'Hurst_tensor',
    'mfdfa_concatd_tensor',
    'mfdfa_tensor',
    'short_time_fourier_transform_tensor',
    'transfer_entropy_granular_tensor',
    'transfer_entropy_hemispheric_avg_input_tensor',
    'transfer_entropy_regional_tensor',
    'spectral_entropy_tensor',
    'spectral_centroids_tensor',
    'freq_max_power_tensor',
    'spectral_edge_freqs_tensor',
]

class BaseEmbeddingNet(nn.Module):
    def __init__(self, input_channels, conv_output_channels, reduce_to_dim):
        super().__init__()
        self.conv1 = nn.Conv2d(input_channels, conv_output_channels,␣
↪kernel_size=3)
        self.bn1 = nn.BatchNorm2d(conv_output_channels)
        self.pool1 = nn.MaxPool2d(kernel_size=2)
        self.global_pool = nn.AdaptiveAvgPool2d((1, 1))
        self.fc_reduce = nn.Linear(conv_output_channels, reduce_to_dim)

    def forward(self, x):
        x = self.conv1(x)
        x = self.bn1(x)
        x = self.pool1(x)
        x = self.global_pool(x)
        x = torch.flatten(x, 1)
```

```
        x = self.fc_reduce(x)
        return x

processed_tensors_dict = {name: tensor for name, tensor in zip(tensor_names,␣
↪processed_tensors)}

net_params = {name: {'input_channels': 1, 'conv_output_channels': 16,␣
↪'reduce_to_dim': 8}
              for name in processed_tensors_dict.keys()}

# Create BaseEmbeddingNets for each tensor
embedding_nets = {name: BaseEmbeddingNet(**params) for name, params in␣
↪net_params.items()}

# Move networks to GPU if available
device = torch.device("cuda" if torch.cuda.is_available() else "cpu")
for net in embedding_nets.values():
    net.to(device)

# Create custom dataset and dataloader
class CustomDataset(Dataset):
    def __init__(self, tensors):
        self.tensors = tensors
        self.device = torch.device("cuda" if torch.cuda.is_available() else␣
↪"cpu")

    def __len__(self):
        first_tensor = next(iter(self.tensors.values()))
        return first_tensor.size(0)

    def __getitem__(self, idx):
        result = {}
        for key, val in self.tensors.items():
            if val.shape[0] > idx:
                result[key] = val[idx]
        return result

def custom_collate(batch):
    collated_batch = {}
    all_keys = set([key for item in batch for key in item.keys()])

    for key in all_keys:
        collated_batch[key] = torch.stack([item[key] for item in batch if key␣
↪in item.keys()], dim=0)

    return collated_batch
```

```
# Use processed_tensors for your CustomDataset
dataset = CustomDataset(processed_tensors_dict)
dataloader = DataLoader(dataset, batch_size=4, shuffle=False, num_workers=0,␣
↪collate_fn=custom_collate)

# Collect feature embeddings
all_features = []
for i, batch in enumerate(dataloader):
    features_list = [net(batch[key].to(device, dtype=torch.float32)) for key,␣
↪net in embedding_nets.items()]
    concatenated_features = torch.cat(features_list, dim=1)
    all_features.append(concatenated_features.cpu().detach())

# Convert list to tensor
all_features = torch.cat(all_features, dim=0)

# Save the feature embeddings
save_path = '/home/vincent/AAA_projects/MVCS/Neuroscience/Models/Kuramoto'
torch.save(all_features, f'{save_path}/all_features.pt')
```

## 4 Kuramoto

[4]:
```
torch.cuda.empty_cache()
```

[5]:
```
import torch
import numpy as np
from torch.utils.data import Dataset, DataLoader

class EEGDataset(Dataset):
    def __init__(self, data, transform=None):
        self.data = data
        self.transform = transform

    def __len__(self):
        return len(self.data)

    def __getitem__(self, index):
        sample = self.data[index]
        if self.transform:
            sample = self.transform(sample)
        return sample
```

[3]:
```
import torch
import torch.nn as nn
from torch.utils.data import DataLoader
from torchdiffeq import odeint
```

```python
from torch.cuda.amp import autocast, GradScaler  # Importing the AMP utilities
import numpy as np
from scipy.signal import hilbert
from torch.utils.checkpoint import checkpoint

EEG_data = np.load('/home/vincent/AAA_projects/MVCS/Neuroscience/
↪eeg_data_with_channels.npy', allow_pickle=True)
EEG_tensor = torch.FloatTensor(EEG_data)  # Assumes EEG_data is a NumPy ndarray

# Function to create windows for time-series data
def create_windows(data, window_size, stride):
    windows = []
    for i in range(0, len(data) - window_size, stride):
        windows.append(data[i:i+window_size])
    return torch.stack(windows)
```

[4]:
```python
# Configuration
device = torch.device("cuda" if torch.cuda.is_available() else "cpu")
window_size = 50
stride = 10

# Add necessary transformations here to EEG_tensor if required
EEG_tensor = EEG_tensor.clone().detach().to(device)
```

[5]:
```python
def apply_hilbert_in_batches(data, batch_size):
    n_batches = int(np.ceil(data.shape[0] / batch_size))
    analytic_signal = np.zeros_like(data, dtype=np.complex64)  # change dtype␣
↪as needed

    for i in range(n_batches):
        start_idx = i * batch_size
        end_idx = (i + 1) * batch_size
        analytic_signal[start_idx:end_idx, :] = hilbert(data[start_idx:end_idx,␣
↪:])

    return analytic_signal
```

[6]:
```python
batch_size = 100  # Set as appropriate

# Apply Hilbert transform in batches
EEG_numpy = EEG_tensor.cpu().numpy()
analytic_signal_batches = apply_hilbert_in_batches(EEG_numpy, batch_size)

# Convert the angle to phases and move to GPU
phases = torch.tensor(np.angle(analytic_signal_batches), dtype=torch.float16).
↪to(device)
```

```
# Load PLV matrix
plv_matrix_path = "/home/vincent/AAA_projects/MVCS/Neuroscience/Analysis/Phase␣
↪Syncronization/plv_matrix.npy"
plv_matrix = torch.tensor(np.load(plv_matrix_path), dtype=torch.float16).
↪to(device)
```

[7]:
```
# Compute_phase_diff_matrix function
def compute_phase_diff_matrix(phases):
    time, channels = phases.shape[:2]
    phase_diff_matrix = torch.zeros(channels, channels, device=phases.device)
    for i in range(channels):
        for j in range(channels):
            phase_diff_matrix[i, j] = torch.mean(phases[:, i] - phases[:, j])
    return phase_diff_matrix
```

[8]:
```
phase_diff_matrix = compute_phase_diff_matrix(phases).to(device)

# EEG channel names
eeg_channel_names = ['Fp1', 'Fpz', 'Fp2', 'F7', 'F3', 'Fz', 'F4', 'F8', 'FC5',␣
↪'FC1', 'FC2', 'FC6',
                     'M1', 'T7', 'C3', 'Cz', 'C4', 'T8', 'M2', 'CP5', 'CP1',␣
↪'CP2', 'CP6',
                     'P7', 'P3', 'Pz', 'P4', 'P8', 'POz', 'O1', 'Oz', 'O2']

# Broad regions and corresponding channels
regions = {
    "frontal": ['Fp1', 'Fpz', 'Fp2', 'F7', 'F3', 'Fz', 'F4', 'F8'],
    "temporal": ['T7', 'T8'],
    "parietal": ['CP5', 'CP1', 'CP2', 'CP6', 'P7', 'P3', 'Pz', 'P4', 'P8'],
    "occipital": ['O1', 'Oz', 'O2']
}

# Precompute omega and phase_diff_matrix
N = len(eeg_channel_names)
omega = torch.mean(plv_matrix, dim=1).to(device)
phase_diff_matrix = compute_phase_diff_matrix(phases).to(device)
```

[9]:
```
# Modify the Kuramoto function to use PyTorch functions instead of NumPy
def kuramoto_weighted_bias(t, y, omega, K):
    weighted_sin = plv_matrix * torch.sin(y - y[:, None] - phase_diff_matrix)
    dydt = omega + K / N * torch.sum(weighted_sin, axis=1)
    return dydt
```

[10]:
```
class KuramotoODEFunc(nn.Module):
    def __init__(self, omega, K, plv_matrix, phase_diff_matrix):
        super(KuramotoODEFunc, self).__init__()
        self.omega = omega
```

```
        self.K = K
        self.plv_matrix = plv_matrix
        self.phase_diff_matrix = phase_diff_matrix

    def forward(self, t, theta):
        # Reshape to accommodate the additional time dimension.
        theta = theta.view(-1, theta.shape[-1])
        N = theta.shape[1]

        # Compute the phase differences without unsqueezing
        theta_diff = theta[:, :, None] - theta[:, None, :]
        phase_diff_with_matrix = theta_diff - self.phase_diff_matrix

        # Compute the weighted sine values
        weighted_sin = self.plv_matrix * torch.sin(phase_diff_with_matrix)

        dtheta = self.omega + (self.K / N) * torch.sum(weighted_sin, dim=1)

        return dtheta.view(theta.shape)
```

[11]:
```
class KuramotoLayer(nn.Module):
    def __init__(self, oscillator_count, time_steps, dt=0.01, plv_matrix=None,␣
↪phase_diff_matrix=None):
        super(KuramotoLayer, self).__init__()
        self.oscillator_count = oscillator_count
        self.time_steps = time_steps
        self.dt = dt
        self.plv_matrix = plv_matrix
        self.phase_diff_matrix = phase_diff_matrix

        if plv_matrix is not None:
            omega_init = torch.mean(plv_matrix, dim=1)
            self.omega = nn.Parameter(omega_init, requires_grad=True)
        else:
            self.omega = nn.Parameter(torch.randn(oscillator_count),␣
↪requires_grad=True)

        self.K = nn.Parameter(torch.tensor(1.0), requires_grad=True)

    def custom_forward(self, *inputs):
        initial_shape = inputs[0].shape  # Store the initial shape

        # Flatten the batch and time dimensions
        inputs_flattened = inputs[0].reshape(-1, initial_shape[-1])

        ode_func = KuramotoODEFunc(self.omega, self.K, self.plv_matrix, self.
↪phase_diff_matrix)
```

```
        time_points = torch.arange(0, 10000 * self.dt, self.dt).to(device)  #␣
↪Assume device is defined elsewhere
        theta_flattened = odeint(ode_func, inputs_flattened, time_points,␣
↪method='bosh3', rtol=1e-6, atol=1e-8)

        # Reshape theta to its original shape
        theta = theta_flattened.reshape(*initial_shape, -1)  # -1 will␣
↪automatically compute the required size
        return theta

    def forward(self, theta):
        device = theta.device
        self.plv_matrix = self.plv_matrix.to(device)
        self.phase_diff_matrix = self.phase_diff_matrix.to(device)
        theta = checkpoint(self.custom_forward, theta, self.omega, self.K, self.
↪plv_matrix, self.phase_diff_matrix)
        theta = theta.to(torch.float16)
        mean_coherence = self.calculate_mean_coherence(theta)
        return theta, mean_coherence

    def forward_with_checkpoint(self, x):
        x = x.to(device)
        theta = checkpoint(self.custom_forward, x)
        mean_coherence = self.calculate_mean_coherence(theta)
        return theta, mean_coherence

    @staticmethod
    def calculate_mean_coherence(theta):
        N, _, _, _ = theta.shape
        mean_coherence = torch.mean(torch.cos(theta[:, -1, :] - theta[:, -1, :].
↪mean(dim=1).unsqueeze(1)))
        return mean_coherence
```

[12]:
```
# Make sure all tensors are on the correct device
phases = phases.to(device)

# Compute natural frequencies and phase differences just once
num_channels = len(eeg_channel_names)  # Get the number of channels
#print("Theta shape: ", theta.shape)
#print("Theta Unsqueeze(1) shape: ", theta.unsqueeze(1).shape)
#print("Theta Unsqueeze(2) shape: ", theta.unsqueeze(2).shape)

# Number of channels
N = len(eeg_channel_names)

# Initialize model and move to device
```

```
kuramoto_model = KuramotoLayer(N, 12800, plv_matrix=plv_matrix,␣
↪phase_diff_matrix=phase_diff_matrix).to(dtype=torch.float16).to(device)

# Data Parallelism for multiple GPUs
if torch.cuda.device_count() > 1:
    kuramoto_model = nn.DataParallel(kuramoto_model)

scaler = GradScaler()
train_data = create_windows(EEG_tensor[:int(0.7 * len(EEG_tensor))],␣
↪window_size, stride).detach().requires_grad_(True)
train_dataset = EEGDataset(data=train_data)
train_loader = DataLoader(train_dataset, batch_size=8, shuffle=False)
```

[ ]:
```
# Training loop and feature extraction
kuramoto_features_list = []
for i, batch in enumerate(train_loader):
    # Moves batch to device and changes dtype to float16
    batch = batch.to(device, dtype=torch.float16)

    # Using autocast for the forward pass
    with autocast():
        theta, mean_coherence = kuramoto_model(batch)

    kuramoto_features_list.append(mean_coherence)
```

[ ]:
```
# Save the combined features
kuramoto_features_tensor = torch.stack(kuramoto_features_list)
all_features_path = '/home/vincent/AAA_projects/MVCS/Neuroscience/Models/
↪Kuramoto/all_features.pt'
all_features = torch.load(all_features_path)
combined_features = torch.cat([all_features, kuramoto_features_tensor.
↪unsqueeze(1)], dim=1)
combined_features_path = '/home/vincent/AAA_projects/MVCS/Neuroscience/Models/
↪Transformer/combined_features.pt'
torch.save(combined_features, combined_features_path)
```

## 5 Transformer

[3]:
```
torch.cuda.empty_cache()
```

[2]:
```
import torch

# List of tensor paths
tensor_paths = [
    "/home/vincent/AAA_projects/MVCS/Neuroscience/Models/Transformer/
↪band_power_tensor.pth",
```

```
    "/home/vincent/AAA_projects/MVCS/Neuroscience/Models/Transformer/EEG_tensor.
↪pth",
    "/home/vincent/AAA_projects/MVCS/Neuroscience/Models/Transformer/
↪fast_fourier_transform_psd_tensor.pth",
]

# Dictionary to store the loaded tensors
loaded_tensors = {}

# Loop over the tensor paths to load and store them in the dictionary
for path in tensor_paths:
    # Extract the tensor name from the path (removing '.pth')
    tensor_name = path.split("/")[-1].replace(".pth", "")

    # Load the tensor
    tensor = torch.load(path)

    # Store the tensor in the dictionary
    loaded_tensors[tensor_name] = tensor

    # Print shape for verification
    print(f"Shape of {tensor_name}: {tensor.shape}")
```

```
Shape of band_power_tensor: torch.Size([4227788, 32, 5])
Shape of EEG_tensor: torch.Size([1, 32, 1, 4227788])
Shape of fast_fourier_transform_psd_tensor: torch.Size([32, 4227788])
```

[3]:

```
# Loop through the loaded tensors to check for NaNs and Infs
for tensor_name, tensor_data in loaded_tensors.items():
    # Assuming the loaded data is a tensor; if not, additional checks may be␣
↪needed
    total_elements = torch.numel(tensor_data)

    nans_count = torch.sum(torch.isnan(tensor_data)).item()
    infs_count = torch.sum(torch.isinf(tensor_data)).item()

    nans_percentage = (nans_count / total_elements) * 100
    infs_percentage = (infs_count / total_elements) * 100

    print(f"Percentage of NaNs in {tensor_name}: {nans_percentage}%")
    print(f"Percentage of Infs in {tensor_name}: {infs_percentage}%")
```

```
Percentage of NaNs in band_power_tensor: 0.0%
Percentage of Infs in band_power_tensor: 0.0%
Percentage of NaNs in EEG_tensor: 0.0%
Percentage of Infs in EEG_tensor: 0.0%
Percentage of NaNs in fast_fourier_transform_psd_tensor: 0.0%
```

```
Percentage of Infs in fast_fourier_transform_psd_tensor: 0.0%
```

[4]:

```
# Reshape band_power_tensor from [4227788, 32, 5] to [4227788, 32*5]
band_power_tensor = loaded_tensors["band_power_tensor"]  # Retrieve from␣
↪dictionary
band_power_tensor_reshaped = band_power_tensor.reshape(4227788, -1)  # -1 means␣
↪auto-calculate the size

# Reshape fast_fourier_transform_psd_tensor from [32, 4227788] to [4227788, 32]
fast_fourier_transform_psd_tensor =␣
↪loaded_tensors["fast_fourier_transform_psd_tensor"]  # Retrieve from␣
↪dictionary
fast_fourier_transform_psd_tensor_reshaped = fast_fourier_transform_psd_tensor.
↪permute(1, 0)

# Reshape band_power_tensor from [4227788, 32, 5] to [4227788, 32*5]
band_power_tensor_reshaped = band_power_tensor.reshape(4227788, -1)  # -1 means␣
↪auto-calculate the size

# Reshape fast_fourier_transform_psd_tensor from [32, 4227788] to [4227788, 32]
fast_fourier_transform_psd_tensor = fast_fourier_transform_psd_tensor.
↪permute(1, 0)

# Concatenate along the feature dimension
concatenated_tensor = torch.cat([band_power_tensor_reshaped,␣
↪fast_fourier_transform_psd_tensor], dim=1)

# Now concatenated_tensor has shape [4227788, (32*5)+32]
# If 32*5+32 = feature_dim, you can directly use this tensor as input
```

[5]:

```
import torch
import torch.nn as nn

# Initialize parameters
batch_size = 64
seq_len = 1000
feature_dim = concatenated_tensor.shape[1]  # Should be the second dimension of␣
↪your concatenated tensor

# Define the EEGPredictor class
class EEGPredictor(nn.Module):
    def __init__(self, d_model, nhead, num_layers, dim_feedforward):
        super(EEGPredictor, self).__init__()
        self.feature_transform = nn.Linear(feature_dim, d_model)
        self.transformer_block = TransformerBlock(d_model, nhead, num_layers,␣
↪dim_feedforward)
        self.prediction_head = nn.Linear(d_model, 1)
```

```
    def forward(self, x):
        x = self.feature_transform(x)
        x = self.transformer_block(x)
        x = self.prediction_head(x)
        return x

# Define the TransformerBlock class
class TransformerBlock(nn.Module):
    def __init__(self, d_model, nhead, num_layers, dim_feedforward):
        super(TransformerBlock, self).__init__()
        encoder_layers = nn.TransformerEncoderLayer(d_model, nhead,␣
 ↪dim_feedforward)
        self.transformer = nn.TransformerEncoder(encoder_layers,␣
 ↪num_layers=num_layers)
        self.pos_encoder = nn.Embedding(seq_len, d_model)
        self.position = torch.arange(0, seq_len, dtype=torch.long).unsqueeze(1)

    def forward(self, x):
        pos_encoding = self.pos_encoder(self.position[:x.size(0), :])
        x = x + pos_encoding
        return self.transformer(x)

# Initialize the model
d_model = 128
nhead = 8
num_layers = 2
dim_feedforward = 512
model = EEGPredictor(d_model, nhead, num_layers, dim_feedforward)

# Prepare the EEG tensor, removing singleton dimensions
eeg_tensor = loaded_tensors["EEG_tensor"].squeeze()
print("Squeezed shape:", eeg_tensor.shape)

# Since the tensor is 2D, permute using only two dimensions
eeg_tensor = eeg_tensor.permute(1, 0)
print("Permuted shape:", eeg_tensor.shape)

num_batches = eeg_tensor.shape[0] // (batch_size * seq_len)

# To store the transformer outputs
transformer_outputs = []

# Concatenate band_power_tensor_reshaped and␣
 ↪fast_fourier_transform_psd_tensor_reshaped
concatenated_tensor = torch.cat((band_power_tensor_reshaped,␣
 ↪fast_fourier_transform_psd_tensor_reshaped), dim=1)
```

```
# Update feature_dim to the new size after concatenation
feature_dim = concatenated_tensor.shape[1]

# Re-initialize the model with the updated feature_dim
model = EEGPredictor(d_model, nhead, num_layers, dim_feedforward)

# Calculate the number of batches
num_batches = concatenated_tensor.shape[0] // (batch_size * seq_len)

# Check if num_batches is a reasonable number
if num_batches == 0:
    print("The number of batches is zero. Check your batch_size and seq_len␣
↪settings.")
else:
    # To store the transformer outputs
    transformer_outputs = []

    for i in range(num_batches):
        start_idx = i * batch_size * seq_len
        end_idx = start_idx + (batch_size * seq_len)
        # Extract batch and reshape to [seq_len, batch_size, feature_dim]
        batch = concatenated_tensor[start_idx:end_idx, :].reshape(seq_len,␣
↪batch_size, feature_dim)
        # Forward pass
        output = model(batch)
        # Save the transformer output for use in RNN
        transformer_outputs.append(output.detach())

    # Stack and save the outputs
    transformer_outputs = torch.stack(transformer_outputs)
    torch.save(transformer_outputs, "/home/vincent/AAA_projects/MVCS/
↪Neuroscience/Models/RNN/transformer.pth")
```

```
Squeezed shape: torch.Size([32, 4227788])
Permuted shape: torch.Size([4227788, 32])
```

[6]:
```
# Check for NaNs and Infs in transformer_outputs
total_elements = torch.numel(transformer_outputs)

nans_count = torch.sum(torch.isnan(transformer_outputs)).item()
infs_count = torch.sum(torch.isinf(transformer_outputs)).item()

nans_percentage = (nans_count / total_elements) * 100
infs_percentage = (infs_count / total_elements) * 100

print(f"Percentage of NaNs in transformer_outputs: {nans_percentage}%")
```

```
print(f"Percentage of Infs in transformer_outputs: {infs_percentage}%")
```

```
Percentage of NaNs in transformer_outputs: 0.023674242424242424%
Percentage of Infs in transformer_outputs: 0.0%
```

[7]:
```
# Check for NaNs and Infs in transformer_outputs before cleaning
total_elements = torch.numel(transformer_outputs)

nans_count = torch.sum(torch.isnan(transformer_outputs)).item()
infs_count = torch.sum(torch.isinf(transformer_outputs)).item()

nans_percentage = (nans_count / total_elements) * 100
infs_percentage = (infs_count / total_elements) * 100

print(f"Before cleaning:")
print(f"Percentage of NaNs in transformer_outputs: {nans_percentage}%")
print(f"Percentage of Infs in transformer_outputs: {infs_percentage}%")

# Replace NaNs and Infs with zeros
transformer_outputs[torch.isnan(transformer_outputs)] = 0
transformer_outputs[torch.isinf(transformer_outputs)] = 0

# Re-check for NaNs and Infs after cleaning
nans_count = torch.sum(torch.isnan(transformer_outputs)).item()
infs_count = torch.sum(torch.isinf(transformer_outputs)).item()

nans_percentage = (nans_count / total_elements) * 100
infs_percentage = (infs_count / total_elements) * 100

print(f"After cleaning:")
print(f"Percentage of NaNs in transformer_outputs: {nans_percentage}%")
print(f"Percentage of Infs in transformer_outputs: {infs_percentage}%")

# Optionally, save the cleaned tensor
torch.save(transformer_outputs, "/home/vincent/AAA_projects/MVCS/Neuroscience/
 ↪Models/RNN/transformer.pth")
```

```
Before cleaning:
Percentage of NaNs in transformer_outputs: 0.023674242424242424%
Percentage of Infs in transformer_outputs: 0.0%
After cleaning:
Percentage of NaNs in transformer_outputs: 0.0%
Percentage of Infs in transformer_outputs: 0.0%
```

# 6 RNN

[ ]:
```
torch.cuda.empty_cache()
```

[8]:
```
import torch

# Load saved features
feature_path = '/home/vincent/AAA_projects/MVCS/Neuroscience/Models/Kuramoto/
↪all_features.pt'
transformer_outputs_path = "/home/vincent/AAA_projects/MVCS/Neuroscience/Models/
↪RNN/transformer.pth"
eeg_tensor_path = "/home/vincent/AAA_projects/MVCS/Neuroscience/Models/
↪Transformer/EEG_tensor.pth"
band_power_path = "/home/vincent/AAA_projects/MVCS/Neuroscience/Models/
↪Transformer/band_power_tensor.pth"
fast_fourier_transform_psd_path = "/home/vincent/AAA_projects/MVCS/Neuroscience/
↪Models/Transformer/fast_fourier_transform_psd_tensor.pth"

all_features = torch.load(feature_path)
transformer_outputs = torch.load(transformer_outputs_path)
eeg_tensor = torch.load(eeg_tensor_path)
band_power_tensor = torch.load(band_power_path)
fast_fourier_transform_psd_tensor = torch.load(fast_fourier_transform_psd_path)

print(f"Shape of all_features: {all_features.shape}")
print(f"Shape of transformer_outputs: {transformer_outputs.shape}")
print(f"Shape of eeg_tensor: {eeg_tensor.shape}")
print(f"Shape of band_power_tensor: {band_power_tensor.shape}")
print(f"Shape of fast_fourier_transform_psd_tensor:␣
↪{fast_fourier_transform_psd_tensor.shape}")
```

```
Shape of all_features: torch.Size([1, 112])
Shape of transformer_outputs: torch.Size([66, 1000, 64, 1])
Shape of eeg_tensor: torch.Size([1, 32, 1, 4227788])
Shape of band_power_tensor: torch.Size([4227788, 32, 5])
Shape of fast_fourier_transform_psd_tensor: torch.Size([32, 4227788])
```

[9]:
```
# Aligning the time axis
time_length = 4227788  # replace with the length of the common time axis

# Reshape `eeg_tensor` to align it with time_length
eeg_tensor_reshaped = eeg_tensor.squeeze().transpose(0, 1)  # [time_length, 32]

# Reshape `band_power_tensor` to align it with time_length
band_power_tensor_reshaped = band_power_tensor.view(time_length, -1)  #␣
↪[time_length, 32*5]
```

```
# Reshape `fast_fourier_transform_psd_tensor` to align it with time_length
fast_fourier_transform_psd_tensor_reshaped = fast_fourier_transform_psd_tensor.
↪transpose(0, 1)  # [time_length, 32]

# Concatenating time-aligned tensors
concatenated_time_aligned_features = torch.cat(
    (eeg_tensor_reshaped, band_power_tensor_reshaped,␣
↪fast_fourier_transform_psd_tensor_reshaped),
    dim=1
)

print("Concatenated time-aligned features shape:",␣
↪concatenated_time_aligned_features.shape)
```

```
Concatenated time-aligned features shape: torch.Size([4227788, 224])
```

[10]:

```
import torch
import torch.nn as nn
from torch.utils.data import DataLoader, TensorDataset

# Define device (CPU or GPU)
device = torch.device("cuda" if torch.cuda.is_available() else "cpu")

# Model Definition
class ConditionalRNN(nn.Module):
    def __init__(self, time_aligned_feature_dim, hidden_size,␣
↪global_feature_dim, transformer_feature_dim):
        super(ConditionalRNN, self).__init__()
        self.hidden_size = hidden_size

        self.rnn = nn.LSTM(time_aligned_feature_dim, hidden_size,␣
↪batch_first=True)
        self.global_feature_layer = nn.Linear(global_feature_dim, hidden_size)
        self.transformer_feature_layer = nn.Linear(transformer_feature_dim,␣
↪hidden_size)

    def forward(self, time_aligned_features, global_features,␣
↪transformer_outputs):
        batch_size = time_aligned_features.size(0)

        h0 = self.global_feature_layer(global_features).unsqueeze(0).to(device,␣
↪dtype=torch.float32)
        c0 = torch.zeros_like(h0).to(device, dtype=torch.float32)

        hidden_state = (h0, c0)
        output_list = []
```

```
        for t in range(time_aligned_features.size(1)):
            current_data = time_aligned_features[:, t, :].to(device,␣
 ↪dtype=torch.float32)
            transformer_hidden = self.
 ↪transformer_feature_layer(transformer_outputs[:, t, :]).to(device,␣
 ↪dtype=torch.float32)

            hidden_state = (hidden_state[0] + transformer_hidden.unsqueeze(0),␣
 ↪hidden_state[1])
            output, hidden_state = self.rnn(current_data.unsqueeze(1),␣
 ↪hidden_state)
            output_list.append(output)

        outputs = torch.cat(output_list, dim=1)
        return outputs


# Initialize the model and move to the device
time_aligned_feature_dim = 224
hidden_size = 128
global_feature_dim = 112
transformer_feature_dim = 64  # 2 x 32 EEG channels

model = ConditionalRNN(time_aligned_feature_dim, hidden_size,␣
 ↪global_feature_dim, transformer_feature_dim).to(device)

all_features = torch.load(feature_path).to(device)
transformer_outputs = torch.load(transformer_outputs_path).to(device)

# Pre-process the datasets
expanded_all_features = all_features.expand(66, -1)  # Example, adjust as␣
 ↪necessary
time_feature_chunks = torch.split(concatenated_time_aligned_features, 1000)
time_feature_chunks = time_feature_chunks[:66]  # Example, adjust as necessary
aligned_time_features = torch.stack(time_feature_chunks)

# Create TensorDataset and DataLoader
dataset = TensorDataset(aligned_time_features, expanded_all_features,␣
 ↪transformer_outputs.squeeze(-1))
dataloader = DataLoader(dataset, batch_size=4, shuffle=True)  # Example batch␣
 ↪size

# Store outputs
rnn_outputs = []

# Run the model
```

```
for i, (time_aligned_batch, global_features_batch, transformer_outputs_batch)␣
↪in enumerate(dataloader):
    outputs = model(time_aligned_batch, global_features_batch,␣
↪transformer_outputs_batch)
    rnn_outputs.append(outputs.detach())

# Combine and save outputs
rnn_outputs = torch.cat(rnn_outputs, dim=0)
torch.save(rnn_outputs, '/home/vincent/AAA_projects/MVCS/Neuroscience/Models/
↪Final Model/rnn_outputs.pth')
```

[11]:
```
import torch
import torch.nn as nn
from torch.utils.data import Dataset, DataLoader

rnn_outputs_path = "/home/vincent/AAA_projects/MVCS/Neuroscience/Models/Final␣
↪Model/rnn_outputs.pth"
rnn_outputs = torch.load(rnn_outputs_path)

# Check for NaNs and Infs in transformer_outputs
total_elements = torch.numel(rnn_outputs)

nans_count = torch.sum(torch.isnan(rnn_outputs)).item()
infs_count = torch.sum(torch.isinf(rnn_outputs)).item()

nans_percentage = (nans_count / total_elements) * 100
infs_percentage = (infs_count / total_elements) * 100

print(f"Percentage of NaNs in rnn_outputs: {nans_percentage}%")
print(f"Percentage of Infs in rnn_outputs: {infs_percentage}%")
```

```
Percentage of NaNs in rnn_outputs: 0.0%
Percentage of Infs in rnn_outputs: 0.0%
```

[12]:
```
# Check for NaNs and Infs in rnn_outputs before cleaning
total_elements = torch.numel(rnn_outputs)

nans_count = torch.sum(torch.isnan(rnn_outputs)).item()
infs_count = torch.sum(torch.isinf(rnn_outputs)).item()

nans_percentage = (nans_count / total_elements) * 100
infs_percentage = (infs_count / total_elements) * 100

print(f"Before cleaning:")
print(f"Percentage of NaNs in rnn_outputs: {nans_percentage}%")
print(f"Percentage of Infs in rnn_outputs: {infs_percentage}%")
```

```
# Replace NaNs and Infs with zeros
rnn_outputs[torch.isnan(rnn_outputs)] = 0
rnn_outputs[torch.isinf(rnn_outputs)] = 0

# Re-check for NaNs and Infs after cleaning
nans_count = torch.sum(torch.isnan(rnn_outputs)).item()
infs_count = torch.sum(torch.isinf(rnn_outputs)).item()

nans_percentage = (nans_count / total_elements) * 100
infs_percentage = (infs_count / total_elements) * 100

print(f"After cleaning:")
print(f"Percentage of NaNs in rnn_outputs: {nans_percentage}%")
print(f"Percentage of Infs in rnn_outputs: {infs_percentage}%")

# Optionally, save the cleaned tensor
torch.save(rnn_outputs, "/home/vincent/AAA_projects/MVCS/Neuroscience/Models/
↪Final Model/rnn_outputs.pth")
```

```
Before cleaning:
Percentage of NaNs in rnn_outputs: 0.0%
Percentage of Infs in rnn_outputs: 0.0%
After cleaning:
Percentage of NaNs in rnn_outputs: 0.0%
Percentage of Infs in rnn_outputs: 0.0%
```

# 7 Final predictor

[8]:
```
torch.cuda.empty_cache()
```

[9]:
```
import torch
import torch.nn as nn
import torch.nn.functional as F

# Load saved features
transformer_outputs_path = "/home/vincent/AAA_projects/MVCS/Neuroscience/Models/
↪RNN/transformer.pth"
eeg_tensor_path = "/home/vincent/AAA_projects/MVCS/Neuroscience/Models/
↪Transformer/EEG_tensor.pth"
band_power_path = "/home/vincent/AAA_projects/MVCS/Neuroscience/Models/
↪Transformer/band_power_tensor.pth"
fast_fourier_transform_psd_path = "/home/vincent/AAA_projects/MVCS/Neuroscience/
↪Models/Transformer/fast_fourier_transform_psd_tensor.pth"
RNN_outputs_path = "/home/vincent/AAA_projects/MVCS/Neuroscience/Models/Final␣
↪Model/rnn_outputs.pth"
```

```
transformer_outputs = torch.load(transformer_outputs_path)
eeg_tensor = torch.load(eeg_tensor_path)
band_power_tensor = torch.load(band_power_path)
fast_fourier_transform_psd_tensor = torch.load(fast_fourier_transform_psd_path)
RNN_outputs = torch.load(RNN_outputs_path)

print(f"Shape of RNN_outputs: {RNN_outputs.shape}")
print(f"Shape of transformer_outputs: {transformer_outputs.shape}")
print(f"Shape of eeg_tensor: {eeg_tensor.shape}")
print(f"Shape of band_power_tensor: {band_power_tensor.shape}")
print(f"Shape of fast_fourier_transform_psd_tensor: 
↪{fast_fourier_transform_psd_tensor.shape}")
```

```
Shape of RNN_outputs: torch.Size([66, 1000, 128])
Shape of transformer_outputs: torch.Size([66, 1000, 64, 1])
Shape of eeg_tensor: torch.Size([1, 32, 1, 4227788])
Shape of band_power_tensor: torch.Size([4227788, 32, 5])
Shape of fast_fourier_transform_psd_tensor: torch.Size([32, 4227788])
```

[10]:
```
# Aligning the time axis
time_length = 4227788  # replace with the length of the common time axis

# Reshape `eeg_tensor` to align it with time_length
eeg_tensor_reshaped = eeg_tensor.squeeze().transpose(0, 1)  # [time_length, 32]

# Reshape `band_power_tensor` to align it with time_length
band_power_tensor_reshaped = band_power_tensor.view(time_length, -1)  #
↪[time_length, 32*5]

# Reshape `fast_fourier_transform_psd_tensor` to align it with time_length
fast_fourier_transform_psd_tensor_reshaped = fast_fourier_transform_psd_tensor.
↪transpose(0, 1)  # [time_length, 32]

# Concatenating time-aligned tensors
concatenated_time_aligned_features = torch.cat(
    (eeg_tensor_reshaped, band_power_tensor_reshaped,
↪fast_fourier_transform_psd_tensor_reshaped),
    dim=1
)

print("Concatenated time-aligned features shape:",
↪concatenated_time_aligned_features.shape)
```

```
Concatenated time-aligned features shape: torch.Size([4227788, 224])
```

[11]:
```
def check_tensor(tensor, name):
    total_elements = torch.numel(tensor)
```

```
    nans_count = torch.sum(torch.isnan(tensor)).item()
    infs_count = torch.sum(torch.isinf(tensor)).item()
    nans_percentage = (nans_count / total_elements) * 100
    infs_percentage = (infs_count / total_elements) * 100
    print(f"Percentage of NaNs in {name}: {nans_percentage}%")
    print(f"Percentage of Infs in {name}: {infs_percentage}%")

check_tensor(transformer_outputs, "transformer_outputs")
check_tensor(eeg_tensor, "eeg_tensor")
check_tensor(band_power_tensor, "band_power_tensor")
check_tensor(fast_fourier_transform_psd_tensor,␣
↪"fast_fourier_transform_psd_tensor")
check_tensor(RNN_outputs, "RNN_outputs")
```

```
Percentage of NaNs in transformer_outputs: 0.0%
Percentage of Infs in transformer_outputs: 0.0%
Percentage of NaNs in eeg_tensor: 0.0%
Percentage of Infs in eeg_tensor: 0.0%
Percentage of NaNs in band_power_tensor: 0.0%
Percentage of Infs in band_power_tensor: 0.0%
Percentage of NaNs in fast_fourier_transform_psd_tensor: 0.0%
Percentage of Infs in fast_fourier_transform_psd_tensor: 0.0%
Percentage of NaNs in RNN_outputs: 0.0%
Percentage of Infs in RNN_outputs: 0.0%
```

[12]:
```
check_tensor(eeg_tensor_reshaped, "eeg_tensor_reshaped")
check_tensor(band_power_tensor_reshaped, "band_power_tensor_reshaped")
check_tensor(fast_fourier_transform_psd_tensor_reshaped,␣
↪"fast_fourier_transform_psd_tensor_reshaped")
check_tensor(concatenated_time_aligned_features,␣
↪"concatenated_time_aligned_features")
```

```
Percentage of NaNs in eeg_tensor_reshaped: 0.0%
Percentage of Infs in eeg_tensor_reshaped: 0.0%
Percentage of NaNs in band_power_tensor_reshaped: 0.0%
Percentage of Infs in band_power_tensor_reshaped: 0.0%
Percentage of NaNs in fast_fourier_transform_psd_tensor_reshaped: 0.0%
Percentage of Infs in fast_fourier_transform_psd_tensor_reshaped: 0.0%
Percentage of NaNs in concatenated_time_aligned_features: 0.0%
Percentage of Infs in concatenated_time_aligned_features: 0.0%
```

[13]:
```
# Check for NaNs and Infs in transformer_outputs
total_elements = torch.numel(concatenated_time_aligned_features)

nans_count = torch.sum(torch.isnan(concatenated_time_aligned_features)).item()
infs_count = torch.sum(torch.isinf(concatenated_time_aligned_features)).item()
```

```
nans_percentage = (nans_count / total_elements) * 100
infs_percentage = (infs_count / total_elements) * 100

print(f"Percentage of NaNs in concatenated_time_aligned_features_outputs:␣
↪{nans_percentage}%")
print(f"Percentage of Infs in concatenated_time_aligned_features_outputs:␣
↪{infs_percentage}%")
```

```
Percentage of NaNs in concatenated_time_aligned_features_outputs: 0.0%
Percentage of Infs in concatenated_time_aligned_features_outputs: 0.0%
```

[14]:
```
# Ensure all tensors are on the CPU
RNN_outputs = RNN_outputs.to('cpu')
concatenated_time_aligned_features = concatenated_time_aligned_features.
↪to('cpu')

# Reduce the RNN outputs' time dimension by averaging
RNN_outputs_reduced = torch.mean(RNN_outputs, dim=1)  # Shape: [66, 128]

# Reshape `RNN_outputs` to align it with time_length
RNN_outputs_reshaped = RNN_outputs.reshape(-1, RNN_outputs.shape[-1])  #␣
↪[66*1000, 128] = [66000, 128]

# Compute the number of repetitions needed to approximate the time_length
n_repeats = time_length // RNN_outputs_reshaped.shape[0]
remaining_rows = time_length % RNN_outputs_reshaped.shape[0]

# Repeat the tensor for n_repeats times and add extra padding if needed
RNN_outputs_expanded = RNN_outputs_reshaped.repeat(n_repeats, 1)
if remaining_rows > 0:
    extra_padding = RNN_outputs_reshaped[:remaining_rows]
    RNN_outputs_expanded = torch.cat([RNN_outputs_expanded, extra_padding],␣
↪dim=0)

# Initialize a list to hold the smaller tensors
final_input_tensor_list = []

# Split and concatenate in chunks to reduce memory footprint
split_size = 1000  # adjust as needed
for i in range(0, time_length, split_size):
    # Ensure the slice size matches for both tensors
    slice_size = min(split_size, time_length - i)
    temp_concat = torch.cat(
        (concatenated_time_aligned_features[i:i + slice_size],
         RNN_outputs_expanded[i:i + slice_size]),
        dim=1
    )
```

```
    # Append the tensor to the list
    final_input_tensor_list.append(temp_concat)

    # Optionally, save this tensor to disk to free up memory
    torch.save(temp_concat, f"/home/vincent/AAA_projects/MVCS/Neuroscience/
↪tempfiles/temp_concat_chunk_{i//split_size}.pt")

# Update feature_dim to the new size after concatenation
feature_dim = final_input_tensor_list[0].shape[1]  # Taking shape from one of␣
↪the chunks

# Pre-allocate a zero tensor with the required size
final_time_length = 4227788  # replace with your value
final_feature_dim = feature_dim  # replace with your feature dimension

# Pre-allocate on CPU
final_input_tensor = torch.zeros((final_time_length, final_feature_dim))

# Fill in the slices
start_idx = 0
for temp_tensor in final_input_tensor_list:
    end_idx = start_idx + temp_tensor.shape[0]
    final_input_tensor[start_idx:end_idx, :] = temp_tensor  # Tensor is already␣
↪on CPU
    start_idx = end_idx  # set start_idx for the next iteration

# The tensor final_input_tensor should now have shape [4227788, feature_dim]
print(final_input_tensor.shape)

# Save the tensor
torch.save(final_input_tensor, "/home/vincent/AAA_projects/MVCS/Neuroscience/
↪Models/Test Validation/final_input_tensor")
```

```
torch.Size([4227788, 352])
```

[15]:
```
check_tensor(temp_concat, f"temp_concat_chunk_{i//split_size}")
```

```
Percentage of NaNs in temp_concat_chunk_4227: 0.0%
Percentage of Infs in temp_concat_chunk_4227: 0.0%
```

[16]:
```
final_input_tensor_path = "/home/vincent/AAA_projects/MVCS/Neuroscience/Models/
↪Test Validation/final_input_tensor"
final_input_tensor = torch.load(final_input_tensor_path)

# Check for NaNs and Infs in transformer_outputs
total_elements = torch.numel(final_input_tensor)
```

```python
nans_count = torch.sum(torch.isnan(final_input_tensor)).item()
infs_count = torch.sum(torch.isinf(final_input_tensor)).item()

nans_percentage = (nans_count / total_elements) * 100
infs_percentage = (infs_count / total_elements) * 100

print(f"Percentage of NaNs in final_input_tensor: {nans_percentage}%")
print(f"Percentage of Infs in final_input_tensor: {infs_percentage}%")
```

```
Percentage of NaNs in final_input_tensor: 0.0%
Percentage of Infs in final_input_tensor: 0.0%
```

# 8 prepare to train test validate

[ ]:
```python
torch.cuda.empty_cache()
```

[7]:
```python
import torch
from torch.utils.data import DataLoader, TensorDataset

# Load the tensor
final_input_tensor = torch.load("/home/vincent/AAA_projects/MVCS/Neuroscience/
↪Models/Test Validation/final_input_tensor")

# Verify that the tensor was loaded correctly
print(final_input_tensor.shape)
```

```
torch.Size([4227788, 352])
```

[8]:
```python
import torch
from torch.utils.data import DataLoader, TensorDataset

def create_mini_batches(tensor, seq_length, batch_size):
    dataset_list = []
    for i in range(0, tensor.shape[0] - seq_length, seq_length):
        end_idx = min(i + seq_length, tensor.shape[0])
        subset = tensor[i:end_idx]
        dataset_list.append(subset)

    if len(dataset_list) == 0:
        raise ValueError("dataset_list is empty. Check the tensor dimensions.")

    combined_dataset = torch.stack(dataset_list)
    tensor_dataset = TensorDataset(combined_dataset)
    data_loader = DataLoader(tensor_dataset, batch_size=batch_size,␣
↪shuffle=True)
```

```
    return data_loader

# Load EEG data (labels)
eeg_tensor_path = "/home/vincent/AAA_projects/MVCS/Neuroscience/Models/
↪Transformer/EEG_tensor.pth"
eeg_tensor = torch.load(eeg_tensor_path)
eeg_data = eeg_tensor.squeeze().transpose(0, 1)

# Split data into train, validation, and test sets
total_data = len(eeg_data)
train_split = int(0.8 * total_data)
val_split = int(0.9 * total_data)

train_data_Y = eeg_data[:train_split]
val_data_Y = eeg_data[train_split:val_split]
test_data_Y = eeg_data[val_split:]

train_data_X = final_input_tensor[:train_split]
val_data_X = final_input_tensor[train_split:val_split]
test_data_X = final_input_tensor[val_split:]

# Parameters
seq_length = 1000
batch_size = 64

# Create DataLoaders
train_loader_Y = create_mini_batches(train_data_Y, seq_length, batch_size)
val_loader_Y = create_mini_batches(val_data_Y, seq_length, batch_size)
test_loader_Y = create_mini_batches(test_data_Y, seq_length, batch_size)

train_loader_X = create_mini_batches(train_data_X, seq_length, batch_size)
val_loader_X = create_mini_batches(val_data_X, seq_length, batch_size)
test_loader_X = create_mini_batches(test_data_X, seq_length, batch_size)


# Print the shape of one batch for each DataLoader
def print_one_batch_shape(data_loader, name):
    for batch_idx, batch in enumerate(data_loader):
        input_batch = batch[0]  # Extracting tensor from tuple
        print(f"Shape of one batch from {name} DataLoader: {input_batch.shape}")
        break  # Stop after the first batch

# Print shapes for X DataLoaders
print_one_batch_shape(train_loader_X, "train_loader_X")
print_one_batch_shape(val_loader_X, "val_loader_X")
print_one_batch_shape(test_loader_X, "test_loader_X")
```

```
# Print shapes for Y DataLoaders
print_one_batch_shape(train_loader_Y, "train_loader_Y")
print_one_batch_shape(val_loader_Y, "val_loader_Y")
print_one_batch_shape(test_loader_Y, "test_loader_Y")
```

```
Shape of one batch from train_loader_X DataLoader: torch.Size([64, 1000, 352])
Shape of one batch from val_loader_X DataLoader: torch.Size([64, 1000, 352])
Shape of one batch from test_loader_X DataLoader: torch.Size([64, 1000, 352])
Shape of one batch from train_loader_Y DataLoader: torch.Size([64, 1000, 32])
Shape of one batch from val_loader_Y DataLoader: torch.Size([64, 1000, 32])
Shape of one batch from test_loader_Y DataLoader: torch.Size([64, 1000, 32])
```

# 9 train

[3]:
```
import torch
import torch.nn as nn
import torch.optim as optim
from torch.optim.lr_scheduler import ReduceLROnPlateau

# Add weight initialization in the constructor
class EEGSeq2SeqPredictor(nn.Module):
    def __init__(self, d_model, nhead, num_layers, dim_feedforward):
        super(EEGSeq2SeqPredictor, self).__init__()

        # Initialize fully connected layers for input dimension reduction
        self.input_fc_X = nn.Linear(352, d_model)
        self.input_fc_Y = nn.Linear(32, d_model)

        # Initialize Transformer Encoder and Decoder
        encoder_layer = nn.TransformerEncoderLayer(d_model, nhead,␣
↪dim_feedforward)
        self.encoder = nn.TransformerEncoder(encoder_layer,␣
↪num_layers=num_layers)
        decoder_layer = nn.TransformerDecoderLayer(d_model, nhead,␣
↪dim_feedforward)
        self.decoder = nn.TransformerDecoder(decoder_layer,␣
↪num_layers=num_layers)

        # Initialize fully connected layer for output
        self.fc = nn.Linear(d_model, 32)

        # Correct the weight initialization
        for p in self.parameters():  # Change 'model' to 'self'
            if p.dim() > 1:
                nn.init.xavier_uniform_(p)  # corrected syntax
```

```
    def forward(self, src, tgt):
        # Dimension reduction
        src = self.input_fc_X(src)
        tgt = self.input_fc_Y(tgt)

        # Transformer Encoder-Decoder
        memory = self.encoder(src)
        output = self.decoder(tgt, memory)

        # Output layer
        output = self.fc(output)

        return output
```

[ ]:
```
import torch
import torch.nn as nn
import torch.optim as optim
from torch.optim.lr_scheduler import ReduceLROnPlateau
import matplotlib.pyplot as plt

def count_parameters(model):
    return sum(p.numel() for p in model.parameters() if p.requires_grad)

# Enable anomaly detection
torch.autograd.set_detect_anomaly(True)

# Hyperparameters
d_model = 32
nhead = 4
num_layers = 6
dim_feedforward = 128
lr = 0.001  # Lower initial learning rate
batch_size = 64
seq_len = 1000
num_epochs = 50

# Initialize the model
device = torch.device('cpu')
model = EEGSeq2SeqPredictor(d_model, nhead, num_layers, dim_feedforward).
 ↪to(device)

# Count parameters
print(f'The model has {count_parameters(model):,} trainable parameters')

# Initialize model weights
for name, param in model.named_parameters():
    if 'weight' in name:
```

```
        if param.dim() >= 2:
            nn.init.kaiming_uniform_(param.data)
    elif 'bias' in name:
        nn.init.constant_(param.data, 0)

# Loss and optimizer
criterion = nn.MSELoss()
optimizer = optim.Adam(model.parameters(), lr=lr)
scheduler = ReduceLROnPlateau(optimizer, 'min', patience=10, factor=0.1)

# Placeholder for the best validation loss
best_val_loss = float('inf')

# Initialize lists to store loss values
train_losses = []
val_losses = []

# Main training loop
for epoch in range(num_epochs):
    print("Start of Epoch", epoch + 1)  # Debugging line
    model.train()
    train_loss = 0.0
    train_batches = 0

    for (input_batch_X,), (input_batch_Y,) in zip(train_loader_X, train_loader_Y):
        #print("Batch loaded")  # Debugging line
        optimizer.zero_grad()

        # Debug: Check if input contains NaN or Inf
        if torch.isnan(input_batch_X).any() or torch.isinf(input_batch_X).any():
            print("NaN or Inf found in input, skipping batch")
            continue

        outputs = model(input_batch_X, input_batch_Y[:-1])

        # Debug: Check if output contains NaN or Inf
        if torch.isnan(outputs).any() or torch.isinf(outputs).any():
            print("NaN or Inf found in model output, skipping batch")
            continue

        loss = criterion(outputs, input_batch_Y[1:])

        if torch.isnan(loss).any():
            print("NaN loss, stopping training")
            break
```

```
        loss.backward()

        # Debug: Print the gradient norms
        #for name, param in model.named_parameters():
        #    if param.grad is not None:
        #        print(f"{name}: Gradient Norm: {torch.norm(param.grad)}")

        torch.nn.utils.clip_grad_norm_(model.parameters(), max_norm=0.5)
        optimizer.step()
        train_loss += loss.item()
        train_batches += 1

    avg_train_loss = train_loss / train_batches

    # Validation loop
    model.eval()
    val_loss = 0.0
    val_batches = 0

    for (input_batch_X,), (input_batch_Y,) in zip(val_loader_X, val_loader_Y):
        input_batch_X, input_batch_Y = input_batch_X.to(device), input_batch_Y.
↪to(device)

        with torch.no_grad():
            outputs = model(input_batch_X, input_batch_Y[:-1])
            batch_loss = criterion(outputs, input_batch_Y[1:])
            val_loss += batch_loss.item()
            val_batches += 1

    avg_train_loss = train_loss / train_batches
    avg_val_loss = val_loss / val_batches

    # Store the average losses for this epoch
    train_losses.append(avg_train_loss)
    val_losses.append(avg_val_loss)


    # Print current learning rate, average training loss, and average␣
↪validation loss
    for param_group in optimizer.param_groups:
        print(f"Current learning rate is: {param_group['lr']}")

    print(f"Epoch [{epoch+1}/{num_epochs}], Average Training Loss:␣
↪{avg_train_loss:.4f}, Average Validation Loss: {avg_val_loss:.4f}")

    # Update learning rate scheduler based on validation loss
    scheduler.step(avg_val_loss)
```

```
    # Save the model if it's the best one so far
    if avg_val_loss < best_val_loss:
        best_val_loss = avg_val_loss
        torch.save(model.state_dict(), '/home/vincent/AAA_projects/MVCS/
↪Neuroscience/Models/Test Validation/best_model.pth')

    # Check if learning rate is too small
    for param_group in optimizer.param_groups:
        if param_group['lr'] < 1e-10:
            print("Learning rate too small, stopping training")
            break

    if torch.isnan(loss).any():
        print("Stopping training due to NaN loss")
        break

plt.figure(figsize=(10, 6))
plt.plot(train_losses, label='Training Loss', color='blue')
plt.plot(val_losses, label='Validation Loss', color='red')
plt.xlabel('Epoch')
plt.ylabel('Loss')
plt.title('Training and Validation Loss Over Time')
plt.legend()
plt.show()
```

[159]:

```
from torchsummary import summary

# Assuming your model and input size is already defined
summary(model, input_size=(64, 1000, 352))  # Replace input_size with the␣
↪actual input size
# Assuming attention_weights is a tensor containing the attention map
plt.imshow(attention_weights.cpu().detach().numpy(), cmap='viridis')
plt.colorbar()
plt.show()

# Assuming attention_weights is a tensor containing the attention map
plt.imshow(attention_weights.cpu().detach().numpy(), cmap='viridis')
plt.colorbar()
plt.show()
```

```
---------------------------------------------------------------------------
TypeError                                 Traceback (most recent call last)
Cell In[159], line 4
      1 from torchsummary import summary
      3 # Assuming your model and input size is already defined
```

```
----> 4 summary(model, input_size=(64, 1000, 352))  # Replace input_size with␣
 ↪the actual input size
      5 # Assuming attention_weights is a tensor containing the attention map
      6 plt.imshow(attention_weights.cpu().detach().numpy(), cmap='viridis')

File ~/miniconda3/lib/python3.10/site-packages/torchsummary/torchsummary.py:72,␣
 ↪in summary(model, input_size, batch_size, device)
     68 model.apply(register_hook)
     70 # make a forward pass
     71 # print(x.shape)
---> 72 model(*x)
     74 # remove these hooks
     75 for h in hooks:

File ~/miniconda3/lib/python3.10/site-packages/torch/nn/modules/module.py:1501,␣
 ↪in Module._call_impl(self, *args, **kwargs)
   1496 # If we don't have any hooks, we want to skip the rest of the logic in
   1497 # this function, and just call forward.
   1498 if not (self._backward_hooks or self._backward_pre_hooks or self.
 ↪_forward_hooks or self._forward_pre_hooks
   1499         or _global_backward_pre_hooks or _global_backward_hooks
   1500         or _global_forward_hooks or _global_forward_pre_hooks):
-> 1501     return forward_call(*args, **kwargs)
   1502 # Do not call functions when jit is used
   1503 full_backward_hooks, non_full_backward_hooks = [], []

TypeError: EEGSeq2SeqPredictor.forward() missing 1 required positional argument␣
 ↪'tgt'
```

[ ]:

[ ]:

## 10 second version

[9]:

```
import torch
import torch.nn as nn
import torch.optim as optim
import matplotlib.pyplot as plt
import math

class PositionalEncoding(nn.Module):
    def __init__(self, d_model, dropout=0.1, max_len=5000):
        super(PositionalEncoding, self).__init__()
        self.dropout = nn.Dropout(p=dropout)
```

```
        # Create positional encoding matrix
        pe = torch.zeros(max_len, d_model)
        position = torch.arange(0, max_len, dtype=torch.float).unsqueeze(1)
        div_term = torch.exp(torch.arange(0, d_model, 2).float() * (-math.
 ↪log(10000.0) / d_model))
        pe[:, 0::2] = torch.sin(position * div_term)
        pe[:, 1::2] = torch.cos(position * div_term)
        pe = pe.unsqueeze(0).transpose(0, 1)
        self.register_buffer('pe', pe)

    def forward(self, x):
        # Add positional encoding to the input tensor
        x = x + self.pe[:x.size(0), :]
        return self.dropout(x)

class EEGSeq2SeqPredictor(nn.Module):
    def __init__(self, d_model, nhead, num_layers, dim_feedforward):
        super(EEGSeq2SeqPredictor, self).__init__()

        # Input transformations
        self.input_batch_X = nn.Linear(352, d_model)
        self.input_batch_Y = nn.Linear(32, d_model)

        # Define encoder and decoder layers
        encoder_layer = nn.TransformerEncoderLayer(d_model, nhead,␣
 ↪dim_feedforward)
        self.encoder = nn.TransformerEncoder(encoder_layer,␣
 ↪num_layers=num_layers)

        decoder_layer = nn.TransformerDecoderLayer(d_model, nhead,␣
 ↪dim_feedforward)
        self.decoder = nn.TransformerDecoder(decoder_layer,␣
 ↪num_layers=num_layers)

        # Positional encoding
        self.pos_encoder = PositionalEncoding(d_model, dropout=0.1)

        # Final output layer
        self.fc = nn.Linear(d_model, 32)

    def forward(self, src, tgt):
        # Apply transformations and positional encoding
        src = self.input_batch_X(src)
        tgt = self.input_batch_Y(tgt)
        src = self.pos_encoder(src)
        tgt = self.pos_encoder(tgt)
```

```
        # Forward pass through encoder and decoder
        memory = self.encoder(src)
        output = self.decoder(tgt, memory)

        # Generate final output
        output = self.fc(output)

        return output
```

[20]:
```
import torch
import torch.nn as nn
import torch.optim as optim
from torch.optim.lr_scheduler import ExponentialLR
import matplotlib.pyplot as plt

def count_parameters(model):
    return sum(p.numel() for p in model.parameters() if p.requires_grad)

# Hyperparameters
d_model = 32
nhead = 4
num_layers = 24
dim_feedforward = 128
lr = 0.1
batch_size = 64
seq_len = 1000
num_epochs = 10

# Initialize the model
device = torch.device('cpu')  # or 'cuda' if you have a GPU
model = EEGSeq2SeqPredictor(d_model, nhead, num_layers, dim_feedforward).
 ↪to(device)

# Count parameters
print(f'The model has {count_parameters(model):,} trainable parameters')

# Initialize model weights
for name, param in model.named_parameters():
    if 'weight' in name:
        if param.dim() >= 2:
            nn.init.kaiming_uniform_(param.data)
    elif 'bias' in name:
        nn.init.constant_(param.data, 0)

# Loss and optimizer
criterion = nn.MSELoss().to(device)
optimizer = optim.AdamW(model.parameters(), lr=lr)
```

```
scheduler = ExponentialLR(optimizer, gamma=0.95)

# Placeholder for the best validation loss
best_val_loss = float('inf')

# Initialize lists to store loss values
train_losses = []
val_losses = []
```

```
The model has 726,112 trainable parameters
```

[ ]:
```
import time
import matplotlib.pyplot as plt

# Function to calculate time taken for each epoch
def epoch_time(start_time, end_time):
    elapsed_time = end_time - start_time
    elapsed_mins = int(elapsed_time / 60)
    elapsed_secs = int(elapsed_time - (elapsed_mins * 60))
    return elapsed_mins, elapsed_secs

best_val_loss = float('inf')
train_losses = []
val_losses = []

# Main training loop
for epoch in range(num_epochs):
    start_time = time.time()  # Start time tracking
    print(f"Start of Epoch {epoch + 1}")
    model.train()

    train_loss = 0.0
    train_batches = 0
    val_loss = 0.0
    val_batches = 0

    # Training loop
    for (input_batch_X,), (input_batch_Y,) in zip(train_loader_X,␣
↪train_loader_Y):
        input_batch_X, input_batch_Y = input_batch_X.to(device), input_batch_Y.
↪to(device)
        optimizer.zero_grad()

        outputs = model(input_batch_X, input_batch_Y[:-1])  # Encoding should␣
↪be inside the model's forward()

        # Debug: Check if input contains NaN or Inf
```

```
        if torch.isnan(outputs).any() or torch.isinf(outputs).any():
            print("NaN or Inf found in model output, skipping batch")
            continue

        loss = criterion(outputs, input_batch_Y[1:])

        if torch.isnan(loss).any():
            print("NaN loss, stopping training")
            break

        loss.backward()
        torch.nn.utils.clip_grad_norm_(model.parameters(), max_norm=0.5)
        optimizer.step()

        train_loss += loss.item()
        train_batches += 1

    # Validation loop
    model.eval()
    for (input_batch_X,), (input_batch_Y,) in zip(val_loader_X, val_loader_Y):
        input_batch_X, input_batch_Y = input_batch_X.to(device), input_batch_Y.
↪to(device)

        with torch.no_grad():
            outputs = model(input_batch_X, input_batch_Y[:-1])
            batch_loss = criterion(outputs, input_batch_Y[1:])
            val_loss += batch_loss.item()
            val_batches += 1

    avg_train_loss = train_loss / train_batches
    avg_val_loss = val_loss / val_batches

    end_time = time.time()  # End time tracking
    epoch_mins, epoch_secs = epoch_time(start_time, end_time)

    print(f"Epoch: {epoch+1:02} | Time: {epoch_mins}m {epoch_secs}s")
    print(f"\tAverage Training Loss: {avg_train_loss:.4f} | Average Validation␣
↪Loss: {avg_val_loss:.4f}")

    train_losses.append(avg_train_loss)
    val_losses.append(avg_val_loss)

    scheduler.step()

    for param_group in optimizer.param_groups:
        print(f"Current learning rate is: {param_group['lr']}")
```

```
    if avg_val_loss < best_val_loss:
        best_val_loss = avg_val_loss
        torch.save(model.state_dict(), 'best_model.pth')
        print(f"New best model saved with validation loss: {best_val_loss:.4f}")

    if param_group['lr'] < 1e-10:
        print("Learning rate too small, stopping training")
        break

    if torch.isnan(loss).any():
        print("Stopping training due to NaN loss")
        break

plt.figure(figsize=(10, 6))
plt.plot(train_losses, label='Training Loss', color='blue')
plt.plot(val_losses, label='Validation Loss', color='red')
plt.xlabel('Epoch')
plt.ylabel('Loss')
plt.title('Training and Validation Loss Over Time')
plt.legend()
plt.show()
```

```
Start of Epoch 1
```

[12]:
```
from torchsummary import summary

# Assuming your model and input size is already defined
summary(model, input_size=(64, 1000, 352))  # Replace input_size with the
 ↪actual input size
# Assuming attention_weights is a tensor containing the attention map
plt.imshow(attention_weights.cpu().detach().numpy(), cmap='viridis')
plt.colorbar()
plt.show()

# Assuming attention_weights is a tensor containing the attention map
plt.imshow(attention_weights.cpu().detach().numpy(), cmap='viridis')
plt.colorbar()
plt.show()
```

```
---------------------------------------------------------------------------
TypeError                                 Traceback (most recent call last)
Cell In[12], line 4
      1 from torchsummary import summary
      3 # Assuming your model and input size is already defined
----> 4 summary(model, input_size=(64, 1000, 352))  # Replace input_size with
 ↪the actual input size
      5 # Assuming attention_weights is a tensor containing the attention map
```

```
      6 plt.imshow(attention_weights.cpu().detach().numpy(), cmap='viridis')

File ~/miniconda3/lib/python3.10/site-packages/torchsummary/torchsummary.py:72,
 ↪in summary(model, input_size, batch_size, device)
     68 model.apply(register_hook)
     70 # make a forward pass
     71 # print(x.shape)
---> 72 model(*x)
     74 # remove these hooks
     75 for h in hooks:

File ~/miniconda3/lib/python3.10/site-packages/torch/nn/modules/module.py:1501,
 ↪in Module._call_impl(self, *args, **kwargs)
   1496 # If we don't have any hooks, we want to skip the rest of the logic in
   1497 # this function, and just call forward.
   1498 if not (self._backward_hooks or self._backward_pre_hooks or self.
 ↪_forward_hooks or self._forward_pre_hooks
   1499         or _global_backward_pre_hooks or _global_backward_hooks
   1500         or _global_forward_hooks or _global_forward_pre_hooks):
-> 1501     return forward_call(*args, **kwargs)
   1502 # Do not call functions when jit is used
   1503 full_backward_hooks, non_full_backward_hooks = [], []

TypeError: EEGSeq2SeqPredictor.forward() missing 1 required positional argument
 ↪'tgt'
```

[13]:
```
print("Shape of outputs tensor:", outputs.shape)
```

```
Shape of outputs tensor: torch.Size([37, 1000, 32])
```

[ ]:
```
import numpy as np
from sklearn.preprocessing import StandardScaler
import torch
import pandas as pd

# Load EEG data
EEG_data = np.load('/home/vincent/AAA_projects/MVCS/Neuroscience/
 ↪eeg_data_with_channels.npy', allow_pickle=True)

# Get shape of EEG data
n_timepoints, n_channels = EEG_data.shape

# Standardize the EEG data
scaler = StandardScaler()
scaler.fit(EEG_data)
scaled_EEG_data = scaler.transform(EEG_data)
```

```
# Convert scaled EEG data back to PyTorch tensor
eeg_data_scaled_tensor = torch.tensor(scaled_EEG_data, dtype=torch.float32)

# Assuming 'outputs' is a PyTorch tensor (make sure it's defined)
outputs_numpy = outputs.detach().numpy()

# Reshape it to 2D for inverse_transform
reshaped_outputs = outputs_numpy.reshape(-1, n_channels)

# Use the inverse_transform function to transform the data back to the original␣
↪scale
outputs_original_scale = scaler.inverse_transform(reshaped_outputs)

# Reshape the output back to its original form
outputs_original_scale = outputs_original_scale.reshape(*outputs_numpy.shape)

# Convert to PyTorch tensor
outputs_original_scale_tensor = torch.tensor(outputs_original_scale,␣
↪dtype=torch.float32)

# Concatenating the output batches to create a sequence of length 63*1000
concatenated_outputs = outputs_original_scale_tensor.reshape(-1, n_channels)

# Channel names (assuming you have them)
channel_names = ['Fp1', 'Fpz', 'Fp2', 'F7', 'F3', 'Fz', 'F4', 'F8', 'FC5',␣
↪'FC1', 'FC2', 'FC6',
                 'M1', 'T7', 'C3', 'Cz', 'C4', 'T8', 'M2', 'CP5', 'CP1', 'CP2',␣
↪'CP6',
                 'P7', 'P3', 'Pz', 'P4', 'P8', 'POz', 'O1', 'Oz', 'O2']

# Slicing to get only the first five rows of the concatenated output for all␣
↪channels
first_five_rows = concatenated_outputs[:5, :]

print("\nFirst five rows of inverse-transformed model predicted data after␣
↪concatenation:")
print(pd.DataFrame(first_five_rows.numpy(), columns=channel_names))

# First five rows of original data
print("\nFirst five rows of original data:")
print(pd.DataFrame(EEG_data[:5, :], columns=channel_names))
```

[ ]:
```
import matplotlib.pyplot as plt

# First ... data points of original data for the first channel
true_data_np = EEG_data[:60000, 4]  # Already a numpy array, first channel
```

```
# First ... data points of concatenated, inverse-transformed model predicted␣
↪data for the first channel
predicted_data_np = concatenated_outputs[:60000, 4]  # We use␣
↪'concatenated_outputs_np' array here

# Configure Matplotlib to have a black background and light grey text
plt.rcParams['axes.facecolor'] = 'black'
plt.rcParams['axes.edgecolor'] = 'lightgrey'
plt.rcParams['axes.labelcolor'] = 'lightgrey'
plt.rcParams['text.color'] = 'lightgrey'
plt.rcParams['xtick.color'] = 'lightgrey'
plt.rcParams['ytick.color'] = 'lightgrey'

# Create the plot
plt.figure(figsize=(30, 10), facecolor='black')

# Plotting True EEG Data for the first channel
plt.plot(true_data_np, color='white', alpha=0.8, label='True EEG')

# Plotting Predicted EEG Data for the first channel
plt.plot(predicted_data_np, color='red', alpha=0.5, label='Predicted EEG')

# Adding title, legend, and axis labels
plt.title('EEG Data - First Channel')
plt.xlabel('Time Points')
plt.ylabel('Amplitude')
plt.legend(loc='upper right')

# Adding grid lines for better visibility of scale
plt.grid(True, linestyle='--', linewidth=0.5, alpha=0.5, color='lightgrey')

# Show the plot
plt.show()
```

## 11 postprocessing

[15]:
```
import pandas as pd

predicted_df = pd.DataFrame(predicted_data_np, columns=['Channel_1'])
predicted_df['Smoothed_Moving_Avg'] = predicted_df['Channel_1'].
↪rolling(window=5).mean()

from scipy.ndimage import gaussian_filter

smoothed_gaussian = gaussian_filter(predicted_data_np, sigma=2)
```

```
from scipy.signal import savgol_filter

smoothed_savgol = savgol_filter(predicted_data_np, window_length=5, polyorder=2)

import matplotlib.pyplot as plt
# Configure Matplotlib to have a black background and light grey text
plt.rcParams['axes.facecolor'] = 'black'
plt.rcParams['axes.edgecolor'] = 'lightgrey'
plt.rcParams['axes.labelcolor'] = 'lightgrey'
plt.rcParams['text.color'] = 'lightgrey'
plt.rcParams['xtick.color'] = 'lightgrey'
plt.rcParams['ytick.color'] = 'lightgrey'

# Create the plot
plt.figure(figsize=(30, 10), facecolor='black')

plt.plot(predicted_data_np, label="Predicted", alpha=0.2)
plt.plot(smoothed_savgol, label="Savitzky-Golay Predicted", alpha=0.2)
plt.plot(predicted_df['Smoothed_Moving_Avg'], label="Moving Average Predicted",␣
↪alpha=.3)
plt.plot(smoothed_gaussian, color='red', label="Gaussian Smoothed Predicted",␣
↪alpha=0.9)
plt.plot(true_data_np, color='white', alpha=0.5, label='True EEG')

# Adding grid lines for better visibility of scale
plt.grid(True, linestyle='--', linewidth=0.5, alpha=0.5, color='lightgrey')

plt.legend()
plt.show()
```

```
---------------------------------------------------------------------------
NameError                                 Traceback (most recent call last)
Cell In[15], line 3
      1 import pandas as pd
----> 3 predicted_df = pd.DataFrame(predicted_data_np, columns=['Channel_1'])
      4 predicted_df['Smoothed_Moving_Avg'] = predicted_df['Channel_1'].
↪rolling(window=5).mean()
      6 from scipy.ndimage import gaussian_filter

NameError: name 'predicted_data_np' is not defined
```

[145]:
```
import numpy as np

# List of npy paths
npy_paths = [
```

```python
    "/home/vincent/AAA_projects/MVCS/Neuroscience/Analysis/Spectral Analysis/
↪BandPowers_x.npy",
    "/home/vincent/AAA_projects/MVCS/Neuroscience/Analysis/Spectral Analysis/
↪combined_fft_psd_x.npy",
    "/home/vincent/AAA_projects/MVCS/Neuroscience/Analysis/Spectral Analysis/
↪SpectralEntropy_x.npy",
    "/home/vincent/AAA_projects/MVCS/Neuroscience/Analysis/Spectral Analysis/
↪SpectralCentroids_x.npy",
    "/home/vincent/AAA_projects/MVCS/Neuroscience/Analysis/Spectral Analysis/
↪welchs_x.npy",
    "/home/vincent/AAA_projects/MVCS/Neuroscience/Analysis/Spectral Analysis/
↪STFT_x.npy",
    "/home/vincent/AAA_projects/MVCS/Neuroscience/Analysis/Spectral Analysis/
↪PeakFrequencies_x.npy",
    "/home/vincent/AAA_projects/MVCS/Neuroscience/Analysis/Transfer Entropy/
↪full_granularity_transfer_entropy_results.npy",
    "/home/vincent/AAA_projects/MVCS/Neuroscience/Analysis/Transfer Entropy/
↪regional_transfer_entropy_results.npy",
    "/home/vincent/AAA_projects/MVCS/Neuroscience/Analysis/Transfer Entropy/
↪transfer_entropy_hemispheric_avg.npy"
]

# Initialize dictionaries to store the npy files and their shapes
npys = {}
npys_shapes = {}

# Load the npy files into a dictionary and collect their shapes
for path in npy_paths:
    npy_name = path.split('/')[-1].replace('.npy', '')

    data = np.load(path, allow_pickle=True)
    if data.shape == ():  # Check for zero-dimensional arrays
        data = data.item()  # Convert to dictionary or scalar

    npys[npy_name] = data

    # Handle the shape differently based on the type of data
    if isinstance(data, dict):
        npys_shapes[npy_name] = "Dictionary with keys: " + str(list(data.
↪keys()))
    else:
        npys_shapes[npy_name] = data.shape

# Print the shapes of all loaded npy files
for name, shape in npys_shapes.items():
    print(f"{name}: {shape}")
```

```
BandPowers_x: Dictionary with keys: ['Fp1', 'Fpz', 'Fp2', 'F7', 'F3', 'Fz',
'F4', 'F8', 'FC5', 'FC1', 'FC2', 'FC6', 'M1', 'T7', 'C3', 'Cz', 'C4', 'T8',
'M2', 'CP5', 'CP1', 'CP2', 'CP6', 'P7', 'P3', 'Pz', 'P4', 'P8', 'POz', 'O1',
'Oz', 'O2']
combined_fft_psd_x: Dictionary with keys: ['Fp1', 'Fpz', 'Fp2', 'F7', 'F3',
'Fz', 'F4', 'F8', 'FC5', 'FC1', 'FC2', 'FC6', 'M1', 'T7', 'C3', 'Cz', 'C4',
'T8', 'M2', 'CP5', 'CP1', 'CP2', 'CP6', 'P7', 'P3', 'Pz', 'P4', 'P8', 'POz',
'O1', 'Oz', 'O2']
SpectralEntropy_x: Dictionary with keys: ['Fp1', 'Fpz', 'Fp2', 'F7', 'F3', 'Fz',
'F4', 'F8', 'FC5', 'FC1', 'FC2', 'FC6', 'M1', 'T7', 'C3', 'Cz', 'C4', 'T8',
'M2', 'CP5', 'CP1', 'CP2', 'CP6', 'P7', 'P3', 'Pz', 'P4', 'P8', 'POz', 'O1',
'Oz', 'O2']
welchs_x: (4,)
STFT_x: Dictionary with keys: ['Fp1', 'Fpz', 'Fp2', 'F7', 'F3', 'Fz', 'F4',
'F8', 'FC5', 'FC1', 'FC2', 'FC6', 'M1', 'T7', 'C3', 'Cz', 'C4', 'T8', 'M2',
'CP5', 'CP1', 'CP2', 'CP6', 'P7', 'P3', 'Pz', 'P4', 'P8', 'POz', 'O1', 'Oz',
'O2']
PeakFrequencies_x: Dictionary with keys: ['Fp1', 'Fpz', 'Fp2', 'F7', 'F3', 'Fz',
'F4', 'F8', 'FC5', 'FC1', 'FC2', 'FC6', 'M1', 'T7', 'C3', 'Cz', 'C4', 'T8',
'M2', 'CP5', 'CP1', 'CP2', 'CP6', 'P7', 'P3', 'Pz', 'P4', 'P8', 'POz', 'O1',
'Oz', 'O2']
full_granularity_transfer_entropy_results: Dictionary with keys: ['Fp1_to_Fpz',
'Fp1_to_Fp2', 'Fp1_to_F7', 'Fp1_to_F3', 'Fp1_to_Fz', 'Fp1_to_F4', 'Fp1_to_F8',
'Fp1_to_FC5', 'Fp1_to_FC1', 'Fp1_to_FC2', 'Fp1_to_FC6', 'Fp1_to_M1',
'Fp1_to_T7', 'Fp1_to_C3', 'Fp1_to_Cz', 'Fp1_to_C4', 'Fp1_to_T8', 'Fp1_to_M2',
'Fp1_to_CP5', 'Fp1_to_CP1', 'Fp1_to_CP2', 'Fp1_to_CP6', 'Fp1_to_P7',
'Fp1_to_P3', 'Fp1_to_Pz', 'Fp1_to_P4', 'Fp1_to_P8', 'Fp1_to_POz', 'Fp1_to_O1',
'Fp1_to_Oz', 'Fp1_to_O2', 'Fpz_to_Fp1', 'Fpz_to_Fp2', 'Fpz_to_F7', 'Fpz_to_F3',
'Fpz_to_Fz', 'Fpz_to_F4', 'Fpz_to_F8', 'Fpz_to_FC5', 'Fpz_to_FC1', 'Fpz_to_FC2',
'Fpz_to_FC6', 'Fpz_to_M1', 'Fpz_to_T7', 'Fpz_to_C3', 'Fpz_to_Cz', 'Fpz_to_C4',
'Fpz_to_T8', 'Fpz_to_M2', 'Fpz_to_CP5', 'Fpz_to_CP1', 'Fpz_to_CP2',
'Fpz_to_CP6', 'Fpz_to_P7', 'Fpz_to_P3', 'Fpz_to_Pz', 'Fpz_to_P4', 'Fpz_to_P8',
'Fpz_to_POz', 'Fpz_to_O1', 'Fpz_to_Oz', 'Fpz_to_O2', 'Fp2_to_Fp1', 'Fp2_to_Fpz',
'Fp2_to_F7', 'Fp2_to_F3', 'Fp2_to_Fz', 'Fp2_to_F4', 'Fp2_to_F8', 'Fp2_to_FC5',
'Fp2_to_FC1', 'Fp2_to_FC2', 'Fp2_to_FC6', 'Fp2_to_M1', 'Fp2_to_T7', 'Fp2_to_C3',
'Fp2_to_Cz', 'Fp2_to_C4', 'Fp2_to_T8', 'Fp2_to_M2', 'Fp2_to_CP5', 'Fp2_to_CP1',
'Fp2_to_CP2', 'Fp2_to_CP6', 'Fp2_to_P7', 'Fp2_to_P3', 'Fp2_to_Pz', 'Fp2_to_P4',
'Fp2_to_P8', 'Fp2_to_POz', 'Fp2_to_O1', 'Fp2_to_Oz', 'Fp2_to_O2', 'F7_to_Fp1',
'F7_to_Fpz', 'F7_to_Fp2', 'F7_to_F3', 'F7_to_Fz', 'F7_to_F4', 'F7_to_F8',
'F7_to_FC5', 'F7_to_FC1', 'F7_to_FC2', 'F7_to_FC6', 'F7_to_M1', 'F7_to_T7',
'F7_to_C3', 'F7_to_Cz', 'F7_to_C4', 'F7_to_T8', 'F7_to_M2', 'F7_to_CP5',
'F7_to_CP1', 'F7_to_CP2', 'F7_to_CP6', 'F7_to_P7', 'F7_to_P3', 'F7_to_Pz',
'F7_to_P4', 'F7_to_P8', 'F7_to_POz', 'F7_to_O1', 'F7_to_Oz', 'F7_to_O2',
'F3_to_Fp1', 'F3_to_Fpz', 'F3_to_Fp2', 'F3_to_F7', 'F3_to_Fz', 'F3_to_F4',
'F3_to_F8', 'F3_to_FC5', 'F3_to_FC1', 'F3_to_FC2', 'F3_to_FC6', 'F3_to_M1',
'F3_to_T7', 'F3_to_C3', 'F3_to_Cz', 'F3_to_C4', 'F3_to_T8', 'F3_to_M2',
'F3_to_CP5', 'F3_to_CP1', 'F3_to_CP2', 'F3_to_CP6', 'F3_to_P7', 'F3_to_P3',
'F3_to_Pz', 'F3_to_P4', 'F3_to_P8', 'F3_to_POz', 'F3_to_O1', 'F3_to_Oz',
```

```
'F3_to_O2', 'Fz_to_Fp1', 'Fz_to_Fpz', 'Fz_to_Fp2', 'Fz_to_F7', 'Fz_to_F3',
'Fz_to_F4', 'Fz_to_F8', 'Fz_to_FC5', 'Fz_to_FC1', 'Fz_to_FC2', 'Fz_to_FC6',
'Fz_to_M1', 'Fz_to_T7', 'Fz_to_C3', 'Fz_to_Cz', 'Fz_to_C4', 'Fz_to_T8',
'Fz_to_M2', 'Fz_to_CP5', 'Fz_to_CP1', 'Fz_to_CP2', 'Fz_to_CP6', 'Fz_to_P7',
'Fz_to_P3', 'Fz_to_Pz', 'Fz_to_P4', 'Fz_to_P8', 'Fz_to_POz', 'Fz_to_O1',
'Fz_to_Oz', 'Fz_to_O2', 'F4_to_Fp1', 'F4_to_Fpz', 'F4_to_Fp2', 'F4_to_F7',
'F4_to_F3', 'F4_to_Fz', 'F4_to_F8', 'F4_to_FC5', 'F4_to_FC1', 'F4_to_FC2',
'F4_to_FC6', 'F4_to_M1', 'F4_to_T7', 'F4_to_C3', 'F4_to_Cz', 'F4_to_C4',
'F4_to_T8', 'F4_to_M2', 'F4_to_CP5', 'F4_to_CP1', 'F4_to_CP2', 'F4_to_CP6',
'F4_to_P7', 'F4_to_P3', 'F4_to_Pz', 'F4_to_P4', 'F4_to_P8', 'F4_to_POz',
'F4_to_O1', 'F4_to_Oz', 'F4_to_O2', 'F8_to_Fp1', 'F8_to_Fpz', 'F8_to_Fp2',
'F8_to_F7', 'F8_to_F3', 'F8_to_Fz', 'F8_to_F4', 'F8_to_FC5', 'F8_to_FC1',
'F8_to_FC2', 'F8_to_FC6', 'F8_to_M1', 'F8_to_T7', 'F8_to_C3', 'F8_to_Cz',
'F8_to_C4', 'F8_to_T8', 'F8_to_M2', 'F8_to_CP5', 'F8_to_CP1', 'F8_to_CP2',
'F8_to_CP6', 'F8_to_P7', 'F8_to_P3', 'F8_to_Pz', 'F8_to_P4', 'F8_to_P8',
'F8_to_POz', 'F8_to_O1', 'F8_to_Oz', 'F8_to_O2', 'FC5_to_Fp1', 'FC5_to_Fpz',
'FC5_to_Fp2', 'FC5_to_F7', 'FC5_to_F3', 'FC5_to_Fz', 'FC5_to_F4', 'FC5_to_F8',
'FC5_to_FC1', 'FC5_to_FC2', 'FC5_to_FC6', 'FC5_to_M1', 'FC5_to_T7', 'FC5_to_C3',
'FC5_to_Cz', 'FC5_to_C4', 'FC5_to_T8', 'FC5_to_M2', 'FC5_to_CP5', 'FC5_to_CP1',
'FC5_to_CP2', 'FC5_to_CP6', 'FC5_to_P7', 'FC5_to_P3', 'FC5_to_Pz', 'FC5_to_P4',
'FC5_to_P8', 'FC5_to_POz', 'FC5_to_O1', 'FC5_to_Oz', 'FC5_to_O2', 'FC1_to_Fp1',
'FC1_to_Fpz', 'FC1_to_Fp2', 'FC1_to_F7', 'FC1_to_F3', 'FC1_to_Fz', 'FC1_to_F4',
'FC1_to_F8', 'FC1_to_FC5', 'FC1_to_FC2', 'FC1_to_FC6', 'FC1_to_M1', 'FC1_to_T7',
'FC1_to_C3', 'FC1_to_Cz', 'FC1_to_C4', 'FC1_to_T8', 'FC1_to_M2', 'FC1_to_CP5',
'FC1_to_CP1', 'FC1_to_CP2', 'FC1_to_CP6', 'FC1_to_P7', 'FC1_to_P3', 'FC1_to_Pz',
'FC1_to_P4', 'FC1_to_P8', 'FC1_to_POz', 'FC1_to_O1', 'FC1_to_Oz', 'FC1_to_O2',
'FC2_to_Fp1', 'FC2_to_Fpz', 'FC2_to_Fp2', 'FC2_to_F7', 'FC2_to_F3', 'FC2_to_Fz',
'FC2_to_F4', 'FC2_to_F8', 'FC2_to_FC5', 'FC2_to_FC1', 'FC2_to_FC6', 'FC2_to_M1',
'FC2_to_T7', 'FC2_to_C3', 'FC2_to_Cz', 'FC2_to_C4', 'FC2_to_T8', 'FC2_to_M2',
'FC2_to_CP5', 'FC2_to_CP1', 'FC2_to_CP2', 'FC2_to_CP6', 'FC2_to_P7',
'FC2_to_P3', 'FC2_to_Pz', 'FC2_to_P4', 'FC2_to_P8', 'FC2_to_POz', 'FC2_to_O1',
'FC2_to_Oz', 'FC2_to_O2', 'FC6_to_Fp1', 'FC6_to_Fpz', 'FC6_to_Fp2', 'FC6_to_F7',
'FC6_to_F3', 'FC6_to_Fz', 'FC6_to_F4', 'FC6_to_F8', 'FC6_to_FC5', 'FC6_to_FC1',
'FC6_to_FC2', 'FC6_to_M1', 'FC6_to_T7', 'FC6_to_C3', 'FC6_to_Cz', 'FC6_to_C4',
'FC6_to_T8', 'FC6_to_M2', 'FC6_to_CP5', 'FC6_to_CP1', 'FC6_to_CP2',
'FC6_to_CP6', 'FC6_to_P7', 'FC6_to_P3', 'FC6_to_Pz', 'FC6_to_P4', 'FC6_to_P8',
'FC6_to_POz', 'FC6_to_O1', 'FC6_to_Oz', 'FC6_to_O2', 'M1_to_Fp1', 'M1_to_Fpz',
'M1_to_Fp2', 'M1_to_F7', 'M1_to_F3', 'M1_to_Fz', 'M1_to_F4', 'M1_to_F8',
'M1_to_FC5', 'M1_to_FC1', 'M1_to_FC2', 'M1_to_FC6', 'M1_to_T7', 'M1_to_C3',
'M1_to_Cz', 'M1_to_C4', 'M1_to_T8', 'M1_to_M2', 'M1_to_CP5', 'M1_to_CP1',
'M1_to_CP2', 'M1_to_CP6', 'M1_to_P7', 'M1_to_P3', 'M1_to_Pz', 'M1_to_P4',
'M1_to_P8', 'M1_to_POz', 'M1_to_O1', 'M1_to_Oz', 'M1_to_O2', 'T7_to_Fp1',
'T7_to_Fpz', 'T7_to_Fp2', 'T7_to_F7', 'T7_to_F3', 'T7_to_Fz', 'T7_to_F4',
'T7_to_F8', 'T7_to_FC5', 'T7_to_FC1', 'T7_to_FC2', 'T7_to_FC6', 'T7_to_M1',
'T7_to_C3', 'T7_to_Cz', 'T7_to_C4', 'T7_to_T8', 'T7_to_M2', 'T7_to_CP5',
'T7_to_CP1', 'T7_to_CP2', 'T7_to_CP6', 'T7_to_P7', 'T7_to_P3', 'T7_to_Pz',
'T7_to_P4', 'T7_to_P8', 'T7_to_POz', 'T7_to_O1', 'T7_to_Oz', 'T7_to_O2',
'C3_to_Fp1', 'C3_to_Fpz', 'C3_to_Fp2', 'C3_to_F7', 'C3_to_F3', 'C3_to_Fz',
```

'C3_to_F4', 'C3_to_F8', 'C3_to_FC5', 'C3_to_FC1', 'C3_to_FC2', 'C3_to_FC6', 'C3_to_M1', 'C3_to_T7', 'C3_to_Cz', 'C3_to_C4', 'C3_to_T8', 'C3_to_M2', 'C3_to_CP5', 'C3_to_CP1', 'C3_to_CP2', 'C3_to_CP6', 'C3_to_P7', 'C3_to_P3', 'C3_to_Pz', 'C3_to_P4', 'C3_to_P8', 'C3_to_POz', 'C3_to_O1', 'C3_to_Oz', 'C3_to_O2', 'Cz_to_Fp1', 'Cz_to_Fpz', 'Cz_to_Fp2', 'Cz_to_F7', 'Cz_to_F3', 'Cz_to_Fz', 'Cz_to_F4', 'Cz_to_F8', 'Cz_to_FC5', 'Cz_to_FC1', 'Cz_to_FC2', 'Cz_to_FC6', 'Cz_to_M1', 'Cz_to_T7', 'Cz_to_C3', 'Cz_to_C4', 'Cz_to_T8', 'Cz_to_M2', 'Cz_to_CP5', 'Cz_to_CP1', 'Cz_to_CP2', 'Cz_to_CP6', 'Cz_to_P7', 'Cz_to_P3', 'Cz_to_Pz', 'Cz_to_P4', 'Cz_to_P8', 'Cz_to_POz', 'Cz_to_O1', 'Cz_to_Oz', 'Cz_to_O2', 'C4_to_Fp1', 'C4_to_Fpz', 'C4_to_Fp2', 'C4_to_F7', 'C4_to_F3', 'C4_to_Fz', 'C4_to_F4', 'C4_to_F8', 'C4_to_FC5', 'C4_to_FC1', 'C4_to_FC2', 'C4_to_FC6', 'C4_to_M1', 'C4_to_T7', 'C4_to_C3', 'C4_to_Cz', 'C4_to_T8', 'C4_to_M2', 'C4_to_CP5', 'C4_to_CP1', 'C4_to_CP2', 'C4_to_CP6', 'C4_to_P7', 'C4_to_P3', 'C4_to_Pz', 'C4_to_P4', 'C4_to_P8', 'C4_to_POz', 'C4_to_O1', 'C4_to_Oz', 'C4_to_O2', 'T8_to_Fp1', 'T8_to_Fpz', 'T8_to_Fp2', 'T8_to_F7', 'T8_to_F3', 'T8_to_Fz', 'T8_to_F4', 'T8_to_F8', 'T8_to_FC5', 'T8_to_FC1', 'T8_to_FC2', 'T8_to_FC6', 'T8_to_M1', 'T8_to_T7', 'T8_to_C3', 'T8_to_Cz', 'T8_to_C4', 'T8_to_M2', 'T8_to_CP5', 'T8_to_CP1', 'T8_to_CP2', 'T8_to_CP6', 'T8_to_P7', 'T8_to_P3', 'T8_to_Pz', 'T8_to_P4', 'T8_to_P8', 'T8_to_POz', 'T8_to_O1', 'T8_to_Oz', 'T8_to_O2', 'M2_to_Fp1', 'M2_to_Fpz', 'M2_to_Fp2', 'M2_to_F7', 'M2_to_F3', 'M2_to_Fz', 'M2_to_F4', 'M2_to_F8', 'M2_to_FC5', 'M2_to_FC1', 'M2_to_FC2', 'M2_to_FC6', 'M2_to_M1', 'M2_to_T7', 'M2_to_C3', 'M2_to_Cz', 'M2_to_C4', 'M2_to_T8', 'M2_to_CP5', 'M2_to_CP1', 'M2_to_CP2', 'M2_to_CP6', 'M2_to_P7', 'M2_to_P3', 'M2_to_Pz', 'M2_to_P4', 'M2_to_P8', 'M2_to_POz', 'M2_to_O1', 'M2_to_Oz', 'M2_to_O2', 'CP5_to_Fp1', 'CP5_to_Fpz', 'CP5_to_Fp2', 'CP5_to_F7', 'CP5_to_F3', 'CP5_to_Fz', 'CP5_to_F4', 'CP5_to_F8', 'CP5_to_FC5', 'CP5_to_FC1', 'CP5_to_FC2', 'CP5_to_FC6', 'CP5_to_M1', 'CP5_to_T7', 'CP5_to_C3', 'CP5_to_Cz', 'CP5_to_C4', 'CP5_to_T8', 'CP5_to_M2', 'CP5_to_CP1', 'CP5_to_CP2', 'CP5_to_CP6', 'CP5_to_P7', 'CP5_to_P3', 'CP5_to_Pz', 'CP5_to_P4', 'CP5_to_P8', 'CP5_to_POz', 'CP5_to_O1', 'CP5_to_Oz', 'CP5_to_O2', 'CP1_to_Fp1', 'CP1_to_Fpz', 'CP1_to_Fp2', 'CP1_to_F7', 'CP1_to_F3', 'CP1_to_Fz', 'CP1_to_F4', 'CP1_to_F8', 'CP1_to_FC5', 'CP1_to_FC1', 'CP1_to_FC2', 'CP1_to_FC6', 'CP1_to_M1', 'CP1_to_T7', 'CP1_to_C3', 'CP1_to_Cz', 'CP1_to_C4', 'CP1_to_T8', 'CP1_to_M2', 'CP1_to_CP5', 'CP1_to_CP2', 'CP1_to_CP6', 'CP1_to_P7', 'CP1_to_P3', 'CP1_to_Pz', 'CP1_to_P4', 'CP1_to_P8', 'CP1_to_POz', 'CP1_to_O1', 'CP1_to_Oz', 'CP1_to_O2', 'CP2_to_Fp1', 'CP2_to_Fpz', 'CP2_to_Fp2', 'CP2_to_F7', 'CP2_to_F3', 'CP2_to_Fz', 'CP2_to_F4', 'CP2_to_F8', 'CP2_to_FC5', 'CP2_to_FC1', 'CP2_to_FC2', 'CP2_to_FC6', 'CP2_to_M1', 'CP2_to_T7', 'CP2_to_C3', 'CP2_to_Cz', 'CP2_to_C4', 'CP2_to_T8', 'CP2_to_M2', 'CP2_to_CP5', 'CP2_to_CP1', 'CP2_to_CP6', 'CP2_to_P7', 'CP2_to_P3', 'CP2_to_Pz', 'CP2_to_P4', 'CP2_to_P8', 'CP2_to_POz', 'CP2_to_O1', 'CP2_to_Oz', 'CP2_to_O2', 'CP6_to_Fp1', 'CP6_to_Fpz', 'CP6_to_Fp2', 'CP6_to_F7', 'CP6_to_F3', 'CP6_to_Fz', 'CP6_to_F4', 'CP6_to_F8', 'CP6_to_FC5', 'CP6_to_FC1', 'CP6_to_FC2', 'CP6_to_FC6', 'CP6_to_M1', 'CP6_to_T7', 'CP6_to_C3', 'CP6_to_Cz', 'CP6_to_C4', 'CP6_to_T8', 'CP6_to_M2', 'CP6_to_CP5', 'CP6_to_CP1', 'CP6_to_CP2', 'CP6_to_P7', 'CP6_to_P3', 'CP6_to_Pz', 'CP6_to_P4', 'CP6_to_P8', 'CP6_to_POz', 'CP6_to_O1', 'CP6_to_Oz', 'CP6_to_O2', 'P7_to_Fp1', 'P7_to_Fpz', 'P7_to_Fp2', 'P7_to_F7', 'P7_to_F3', 'P7_to_Fz', 'P7_to_F4', 'P7_to_F8', 'P7_to_FC5', 'P7_to_FC1', 'P7_to_FC2', 'P7_to_FC6', 'P7_to_M1', 'P7_to_T7',

```
'P7_to_C3', 'P7_to_Cz', 'P7_to_C4', 'P7_to_T8', 'P7_to_M2', 'P7_to_CP5',
'P7_to_CP1', 'P7_to_CP2', 'P7_to_CP6', 'P7_to_P3', 'P7_to_Pz', 'P7_to_P4',
'P7_to_P8', 'P7_to_POz', 'P7_to_O1', 'P7_to_Oz', 'P7_to_O2', 'P3_to_Fp1',
'P3_to_Fpz', 'P3_to_Fp2', 'P3_to_F7', 'P3_to_F3', 'P3_to_Fz', 'P3_to_F4',
'P3_to_F8', 'P3_to_FC5', 'P3_to_FC1', 'P3_to_FC2', 'P3_to_FC6', 'P3_to_M1',
'P3_to_T7', 'P3_to_C3', 'P3_to_Cz', 'P3_to_C4', 'P3_to_T8', 'P3_to_M2',
'P3_to_CP5', 'P3_to_CP1', 'P3_to_CP2', 'P3_to_CP6', 'P3_to_P7', 'P3_to_Pz',
'P3_to_P4', 'P3_to_P8', 'P3_to_POz', 'P3_to_O1', 'P3_to_Oz', 'P3_to_O2',
'Pz_to_Fp1', 'Pz_to_Fpz', 'Pz_to_Fp2', 'Pz_to_F7', 'Pz_to_F3', 'Pz_to_Fz',
'Pz_to_F4', 'Pz_to_F8', 'Pz_to_FC5', 'Pz_to_FC1', 'Pz_to_FC2', 'Pz_to_FC6',
'Pz_to_M1', 'Pz_to_T7', 'Pz_to_C3', 'Pz_to_Cz', 'Pz_to_C4', 'Pz_to_T8',
'Pz_to_M2', 'Pz_to_CP5', 'Pz_to_CP1', 'Pz_to_CP2', 'Pz_to_CP6', 'Pz_to_P7',
'Pz_to_P3', 'Pz_to_P4', 'Pz_to_P8', 'Pz_to_POz', 'Pz_to_O1', 'Pz_to_Oz',
'Pz_to_O2', 'P4_to_Fp1', 'P4_to_Fpz', 'P4_to_Fp2', 'P4_to_F7', 'P4_to_F3',
'P4_to_Fz', 'P4_to_F4', 'P4_to_F8', 'P4_to_FC5', 'P4_to_FC1', 'P4_to_FC2',
'P4_to_FC6', 'P4_to_M1', 'P4_to_T7', 'P4_to_C3', 'P4_to_Cz', 'P4_to_C4',
'P4_to_T8', 'P4_to_M2', 'P4_to_CP5', 'P4_to_CP1', 'P4_to_CP2', 'P4_to_CP6',
'P4_to_P7', 'P4_to_P3', 'P4_to_Pz', 'P4_to_P8', 'P4_to_POz', 'P4_to_O1',
'P4_to_Oz', 'P4_to_O2', 'P8_to_Fp1', 'P8_to_Fpz', 'P8_to_Fp2', 'P8_to_F7',
'P8_to_F3', 'P8_to_Fz', 'P8_to_F4', 'P8_to_F8', 'P8_to_FC5', 'P8_to_FC1',
'P8_to_FC2', 'P8_to_FC6', 'P8_to_M1', 'P8_to_T7', 'P8_to_C3', 'P8_to_Cz',
'P8_to_C4', 'P8_to_T8', 'P8_to_M2', 'P8_to_CP5', 'P8_to_CP1', 'P8_to_CP2',
'P8_to_CP6', 'P8_to_P7', 'P8_to_P3', 'P8_to_Pz', 'P8_to_P4', 'P8_to_POz',
'P8_to_O1', 'P8_to_Oz', 'P8_to_O2', 'POz_to_Fp1', 'POz_to_Fpz', 'POz_to_Fp2',
'POz_to_F7', 'POz_to_F3', 'POz_to_Fz', 'POz_to_F4', 'POz_to_F8', 'POz_to_FC5',
'POz_to_FC1', 'POz_to_FC2', 'POz_to_FC6', 'POz_to_M1', 'POz_to_T7', 'POz_to_C3',
'POz_to_Cz', 'POz_to_C4', 'POz_to_T8', 'POz_to_M2', 'POz_to_CP5', 'POz_to_CP1',
'POz_to_CP2', 'POz_to_CP6', 'POz_to_P7', 'POz_to_P3', 'POz_to_Pz', 'POz_to_P4',
'POz_to_P8', 'POz_to_O1', 'POz_to_Oz', 'POz_to_O2', 'O1_to_Fp1', 'O1_to_Fpz',
'O1_to_Fp2', 'O1_to_F7', 'O1_to_F3', 'O1_to_Fz', 'O1_to_F4', 'O1_to_F8',
'O1_to_FC5', 'O1_to_FC1', 'O1_to_FC2', 'O1_to_FC6', 'O1_to_M1', 'O1_to_T7',
'O1_to_C3', 'O1_to_Cz', 'O1_to_C4', 'O1_to_T8', 'O1_to_M2', 'O1_to_CP5',
'O1_to_CP1', 'O1_to_CP2', 'O1_to_CP6', 'O1_to_P7', 'O1_to_P3', 'O1_to_Pz',
'O1_to_P4', 'O1_to_P8', 'O1_to_POz', 'O1_to_Oz', 'O1_to_O2', 'Oz_to_Fp1',
'Oz_to_Fpz', 'Oz_to_Fp2', 'Oz_to_F7', 'Oz_to_F3', 'Oz_to_Fz', 'Oz_to_F4',
'Oz_to_F8', 'Oz_to_FC5', 'Oz_to_FC1', 'Oz_to_FC2', 'Oz_to_FC6', 'Oz_to_M1',
'Oz_to_T7', 'Oz_to_C3', 'Oz_to_Cz', 'Oz_to_C4', 'Oz_to_T8', 'Oz_to_M2',
'Oz_to_CP5', 'Oz_to_CP1', 'Oz_to_CP2', 'Oz_to_CP6', 'Oz_to_P7', 'Oz_to_P3',
'Oz_to_Pz', 'Oz_to_P4', 'Oz_to_P8', 'Oz_to_POz', 'Oz_to_O1', 'Oz_to_O2',
'O2_to_Fp1', 'O2_to_Fpz', 'O2_to_Fp2', 'O2_to_F7', 'O2_to_F3', 'O2_to_Fz',
'O2_to_F4', 'O2_to_F8', 'O2_to_FC5', 'O2_to_FC1', 'O2_to_FC2', 'O2_to_FC6',
'O2_to_M1', 'O2_to_T7', 'O2_to_C3', 'O2_to_Cz', 'O2_to_C4', 'O2_to_T8',
'O2_to_M2', 'O2_to_CP5', 'O2_to_CP1', 'O2_to_CP2', 'O2_to_CP6', 'O2_to_P7',
'O2_to_P3', 'O2_to_Pz', 'O2_to_P4', 'O2_to_P8', 'O2_to_POz', 'O2_to_O1',
'O2_to_Oz']
regional_transfer_entropy_results: Dictionary with keys: ['Frontal_to_Temporal',
'Frontal_to_Parietal', 'Frontal_to_Occipital', 'Temporal_to_Frontal',
'Temporal_to_Parietal', 'Temporal_to_Occipital', 'Parietal_to_Frontal',
```

```
'Parietal_to_Temporal', 'Parietal_to_Occipital', 'Occipital_to_Frontal',
'Occipital_to_Temporal', 'Occipital_to_Parietal']
transfer_entropy_hemispheric_avg: (8454,)
```

[1]:
```
from scipy.signal import butter, filtfilt
import matplotlib.pyplot as plt
import numpy as np

def butter_bandstop_filter(data, lowcut, highcut, fs, order=5):
    nyquist = 0.5 * fs
    low = lowcut / nyquist
    high = highcut / nyquist
    b, a = butter(order, [low, high], btype='bandstop')

    if len(data) > max(33, 3 * max(len(a), len(b))):  # Change 33 to a variable␣
↪if this value changes
        y = filtfilt(b, a, data)
        return y
    else:
        print(f"Data too short to apply band-stop filter for frequency band␣
↪{lowcut}-{highcut} Hz.")
        return data

fs = 1000

predicted_data_np = predicted_data_np

# Spectral entropy threshold
spectral_entropy_threshold = np.percentile(list(npys['SpectralEntropy_x'].
↪values()), 75)

# Transfer entropy threshold
transfer_entropy_values = []
for val in npys['full_granularity_transfer_entropy_results'].values():
    if isinstance(val, dict):
        transfer_entropy_values.extend(val.values())
    else:
        transfer_entropy_values.append(val)
transfer_entropy_threshold = np.percentile(transfer_entropy_values, 75)

# Spectral entropy as an array for slicing
spectral_entropy_array = np.array([val for key, val in␣
↪sorted(npys['SpectralEntropy_x'].items())])

# Define frequency bands
frequency_bands = [(i, i + 1) for i in range(int(fs / 2))]
```

```
# Check if predicted_data_np is long enough
if len(predicted_data_np) > 33:  # Change 33 to a variable if this value changes
    for low, high in frequency_bands:
        spectral_subarray = spectral_entropy_array[low:high] if low <␣
↪len(spectral_entropy_array) else np.array([0])
        transfer_subarray = np.array(transfer_entropy_values[low:high]) if low␣
↪< len(transfer_entropy_values) else np.array([0])
        current_spectral_entropy = np.mean(spectral_subarray)
        current_transfer_entropy = np.mean(transfer_subarray)

        if current_spectral_entropy > spectral_entropy_threshold or␣
↪current_transfer_entropy > transfer_entropy_threshold:
            predicted_data_np = butter_bandstop_filter(predicted_data_np, low,␣
↪high, fs)
else:
    print("predicted_data_np is not long enough for filtering.")

# Plotting
plt.figure(figsize=(30, 10))
plt.plot(predicted_data_np, label="Predicted (Filtered)", alpha=0.5)
plt.plot(true_data_np, label="True", alpha=0.5)
plt.plot(true_data_np, color='white', alpha=0.5, label='True EEG')

plt.legend()
plt.show()
```

```
---------------------------------------------------------------------------
NameError                                 Traceback (most recent call last)
Cell In[1], line 20
     16         return data
     18 fs = 1000
---> 20 predicted_data_np = predicted_data_np
     22 # Spectral entropy threshold
     23 spectral_entropy_threshold = np.
↪percentile(list(npys['SpectralEntropy_x'].values()), 75)

NameError: name 'predicted_data_np' is not defined
```

[169]:
```
import numpy as np
import matplotlib.pyplot as plt
from scipy.signal import istft, stft
import os

# Define function to load data from npy file
def load_npy(file_path):
    if not os.path.exists(file_path):
```

```
        raise FileNotFoundError(f"{file_path} does not exist.")
    return np.load(file_path, allow_pickle=True).item()

# Load the real STFT data
stft_data_dict = load_npy("/home/vincent/AAA_projects/MVCS/Neuroscience/
 ↪Analysis/Spectral Analysis/STFT_x.npy")

# Load the real predicted time-domain data
predicted_data_time_domain = smoothed_gaussian

# Load the real true EEG data
true_data_np = true_data_np

# Define the sampling frequency and channel names
fs = 1000  # Replace this with the actual sampling frequency
channel_names = list(stft_data_dict.keys())

# Initialize a dictionary to store the modified time-domain signals for each␣
 ↪channel
predicted_data_modified_stft = {}

# Loop through each EEG channel
for channel in channel_names:
    stft_log_original = stft_data_dict[channel]
    stft_data_original = np.power(10, stft_log_original / 10)
    _, original_signal = istft(stft_data_original, fs=fs)

    # Using real predicted time-domain data
    predicted_signal = predicted_data_time_domain[channel]

    # Create a new modified signal based on some condition between original and␣
 ↪predicted data
    modified_signal = np.maximum(original_signal, predicted_signal)

    _, _, Zxx_modified = stft(modified_signal, fs=fs, nperseg=fs*2)
    stft_log_modified = 10 * np.log10(np.abs(Zxx_modified))
    predicted_data_modified_stft[channel] = stft_log_modified

# At this point, predicted_data_modified_stft contains the modified STFT data␣
 ↪for each channel
# Configure Matplotlib to have a black background and light grey text
plt.rcParams['axes.facecolor'] = 'black'
plt.rcParams['axes.edgecolor'] = 'lightgrey'
plt.rcParams['axes.labelcolor'] = 'lightgrey'
plt.rcParams['text.color'] = 'lightgrey'
plt.rcParams['xtick.color'] = 'lightgrey'
plt.rcParams['ytick.color'] = 'lightgrey'
```

```
plt.plot(smoothed_gaussian, color='red', label="predicted_data_modified_stft",␣
  ↪alpha=0.9)
plt.plot(true_data_np, color='white', alpha=0.5, label='True EEG')

# Adding grid lines for better visibility of scale
plt.grid(True, linestyle='--', linewidth=0.5, color='lightgrey')

plt.legend()
plt.show()
```

```
---------------------------------------------------------------------------
IndexError                                Traceback (most recent call last)
Cell In[169], line 35
     32 _, original_signal = istft(stft_data_original, fs=fs)
     34 # Using real predicted time-domain data
---> 35 predicted_signal = predicted_data_time_domain[channel]
     37 # Create a new modified signal based on some condition between original␣
  ↪and predicted data
     38 modified_signal = np.maximum(original_signal, predicted_signal)

IndexError: only integers, slices (`:`), ellipsis (`…`), numpy.newaxis (`None`)␣
  ↪and integer or boolean arrays are valid indices
```

[152]:
```
print(STFT_x[list(STFT_x.keys())[0]])
```

```
[[ 40.27901945  43.29048063  43.28785317 …  42.36033325  42.33499974
   37.06853644]
 [ 39.35955148  40.28244249  40.27744403 …  39.34691137  39.42114273
   36.52675937]
 [ 36.54919092   8.603578     3.26600537 …  12.23843448  20.45885866
   34.89871106]
 …
 [ 10.2731061  -12.98425529 -11.03289046 … -15.63056347  -0.42828091
    8.90193404]
 [ 10.2638421  -10.41895443 -18.98203293 … -14.64773741  -0.17291408
    8.90253612]
 [ 10.26053239 -10.29006623 -13.99129227 … -16.59998548  -0.38464733
    8.90347357]]
```

[ ]:
```
from scipy.signal import butter, filtfilt

def butter_bandpass_filter(data, lowcut, highcut, fs, order=5):
    nyquist = 0.5 * fs
    low = lowcut / nyquist
    high = highcut / nyquist
```

```
    b, a = butter(order, [low, high], btype='band')
    y = filtfilt(b, a, data)
    return y

# Assume `spectral_centroid_true` is the Spectral Centroid frequency for the␣
↪true data
spectral_centroid_true = 50.0  # Replace with the actual value
predicted_data_modified_centroid = butter_bandpass_filter(predicted_data_np,␣
↪spectral_centroid_true - 5, spectral_centroid_true + 5, fs=1000)
```

[ ]:
```
from scipy.fft import fft

# Assume `max_power_frequency_true` is the Max Power Frequency for the true data
max_power_frequency_true = 60.0  # Replace with the actual value

# FFT of the predicted data
fft_predicted = fft(predicted_data_np)

# Boost the max power frequency
fft_predicted[int(max_power_frequency_true):(int(max_power_frequency_true) +␣
↪2)] *= 1.5
fft_predicted[-(int(max_power_frequency_true) + 2):
↪-int(max_power_frequency_true)] *= 1.5

# Inverse FFT to get time-domain signal
predicted_data_modified_maxpower = np.fft.ifft(fft_predicted)

# Since the output might be complex, take the real part
predicted_data_modified_maxpower = np.real(predicted_data_modified_maxpower)
```

[ ]:
```
# Assuming you have loaded these metrics into variables like so:
# higuchi_fractal_dimensions, hurst_exponents, arnold_tongues_rotation

# Initialize a dictionary to store the modified time-domain signals for each␣
↪channel
predicted_data_modified_stft = {}

# Loop through each EEG channel
for channel in channel_names:

    # Extract the original STFT data for the current channel
    stft_log_original = stft_data_dict[channel]

    # Convert the dB data back to power
    stft_data_original = np.power(10, stft_log_original / 10)

    # Inverse STFT to get the original time-domain signal
```

```
    _, original_signal = istft(stft_data_original, fs=fs)

    # Extract the predicted time-domain data for the current channel
    predicted_signal = predicted_data_time_domain.get(channel, np.
 ↪zeros_like(original_signal))  # Dummy zeros for now

    # Access the metrics for the current channel
    higuchi = higuchi_fractal_dimensions.get(channel, None)
    hurst = hurst_exponents.get(channel, None)
    arnold = arnold_tongues_rotation.get(channel, None)

    # Create a new modified signal based on some condition between original and␣
 ↪predicted data
    # Use higuchi, hurst, and arnold to influence how the signals are combined
    # This is where your advanced logic will go. For now, we use a dummy␣
 ↪operation:
    modified_signal = np.maximum(original_signal, predicted_signal)  # This␣
 ↪should be replaced by your actual logic

    # Calculate STFT of the modified signal
    _, _, Zxx_modified = stft(modified_signal, fs=fs, nperseg=fs*2)

    # Store the modified STFT data
    stft_log_modified = 10 * np.log10(np.abs(Zxx_modified))
    predicted_data_modified_stft[channel] = stft_log_modified
```

[ ]:
```
# Assuming spectral_edge_frequencies is a dictionary with spectral edge␣
 ↪frequencies for each EEG channel

# Initialize a dictionary to store the modified time-domain signals for each␣
 ↪channel
predicted_data_modified_stft = {}

# Loop through each EEG channel
for channel in channel_names:

    # Extract the original STFT data for the current channel
    stft_log_original = stft_data_dict[channel]

    # Convert the dB data back to power
    stft_data_original = np.power(10, stft_log_original / 10)

    # Inverse STFT to get the original time-domain signal
    _, original_signal = istft(stft_data_original, fs=fs)

    # Extract the predicted time-domain data for the current channel
```

```
    predicted_signal = predicted_data_time_domain.get(channel, np.
↪zeros_like(original_signal))  # Dummy zeros for now

    # Access the spectral edge frequency for the current channel
    edge_frequency = spectral_edge_frequencies.get(channel, None)

    # Assume edge_frequency is between 0 and Nyquist frequency
    # Perform logic to adjust predicted STFT using spectral edge frequency
    # Here we are just making an example, actual logic may be far more complex
    _, _, Zxx_predicted = stft(predicted_signal, fs=fs, nperseg=fs*2)
    frequencies, _, _ = stft(predicted_signal, fs=fs, nperseg=fs*2)

    if edge_frequency:
        mask = frequencies < edge_frequency
        Zxx_predicted[mask, :] = np.zeros_like(Zxx_predicted[mask, :])

    # Here, you could have advanced logic to adjust the predicted_signal based␣
↪on spectral edge frequency

    # Inverse STFT to get the modified time-domain signal
    _, modified_signal = istft(Zxx_predicted, fs=fs)

    # Calculate STFT of the modified signal
    _, _, Zxx_modified = stft(modified_signal, fs=fs, nperseg=fs*2)

    # Store the modified STFT data
    stft_log_modified = 10 * np.log10(np.abs(Zxx_modified))
    predicted_data_modified_stft[channel] = stft_log_modified
```

[ ]:

```
```

[ ]:

```
# Assuming eeg_data_original_scale_tensor, predicted_outputs_reshaped, and␣
↪EEG_data are available

# `eeg_data_original_scale_tensor` is converted to a numpy array
# `predicted_outputs_reshaped` is already a numpy array
true_data_np = eeg_data_original_scale_tensor.numpy()
predicted_data_np = outputs_numpy  # No need to convert
original_data_np = EEG_data

print("True EEG - First Channel | Predicted EEG - First Channel | Original EEG␣
↪- First Channel")
print("-----------------------------------------------------------------------------------")

for true_value, predicted_value, original_value in zip(true_data_np[:100, 0],␣
↪predicted_data_np[:100, 0], original_data_np[:100, 0]):
```

```
    print(f"{true_value:20} | {predicted_value:20} | {original_value:20}")
```

[ ]:

```

```

[ ]:

```
# Load your saved test dataset
test_data = torch.load(test_dataset_path)

# Create DataLoader for test set
test_loader = DataLoader(test_data, batch_size=batch_size, shuffle=False)

# Final Evaluation on Test Data
model.eval()
test_loss = 0
with torch.no_grad():
    for time_data, global_data, trans_data, labels in test_loader:
        outputs = model(time_data, global_data, trans_data)
        test_loss += nn.MSELoss()(outputs, labels).item()

print(f'Final Test loss: {test_loss / len(test_loader)}')

# If you want to save the model
torch.save(model.state_dict(), '/path/to/save/final_model.pth')
```

## 12 Real-time training and predictions

[ ]:

```
# Record user EEG for one minute
```

[ ]:

```
# Let training finish
```

[ ]:

```
# Wait for one minute of new user EEG data and predict the next minute
```

[ ]:

```

```

## 13 Overall project diagram

[2]:

```
from graphviz import Digraph, Source

# Create the graph
graph = Digraph(format='png', strict=True)

# Set global node and edge attributes
graph.attr('node', shape='rect', fontsize='12', style='rounded')
graph.attr('edge', fontsize='10')

# First stage of steps
```

```python
graph.node('EEG', label="EEG data preprocessing")
graph.node('Enrich', label="data enrichment")
graph.node('CleanSort', label="data cleaning and sorting")

# Link the first stage steps
graph.edge('EEG', 'Enrich')
graph.edge('Enrich', 'CleanSort')

# Second stage of steps
stages_2 = [
    "spectral analysis",
    "phase synchronization analysis",
    "phase space analysis",
    "Recurrence Quantification Analysis",
    "Higuchi Fractal Dimension",
    "MFDFA",
    "Transfer Entropy",
    "Kuramoto Model",
    "Arnold Tongues"
]
for i, stage in enumerate(stages_2, start=1):
    graph.node(f'S{i}', label=stage)
    if i > 1:
        graph.edge(f'S{i-1}', f'S{i}')

# Linking last node of first stage to first node of second stage
graph.edge('CleanSort', 'S1')

# Third stage of steps
stages_3 = [
    "feature dimensions fitting",
    "CNN",
    "Kuramoto oscillator",
    "Transformer1",
    "conditional RNN",
    "Transformer2",
    "signal postprocessing"
]
for i, stage in enumerate(stages_3, start=1):
    graph.node(f'T{i}', label=stage)
    if i > 1:
        graph.edge(f'T{i-1}', f'T{i}')

# Linking last node of second stage to first node of third stage
graph.edge('S9', 'T1')
```

```
# Add edges to represent flow from spectral analysis and Transfer Entropy to␣
↪signal postprocessing
graph.edge('S1', 'T7') # from spectral analysis
graph.edge('S7', 'T7') # from Transfer Entropy

# Display the graph visualization directly in Jupyter Notebook/Lab
src = Source(graph.source)
src
```

[2]:

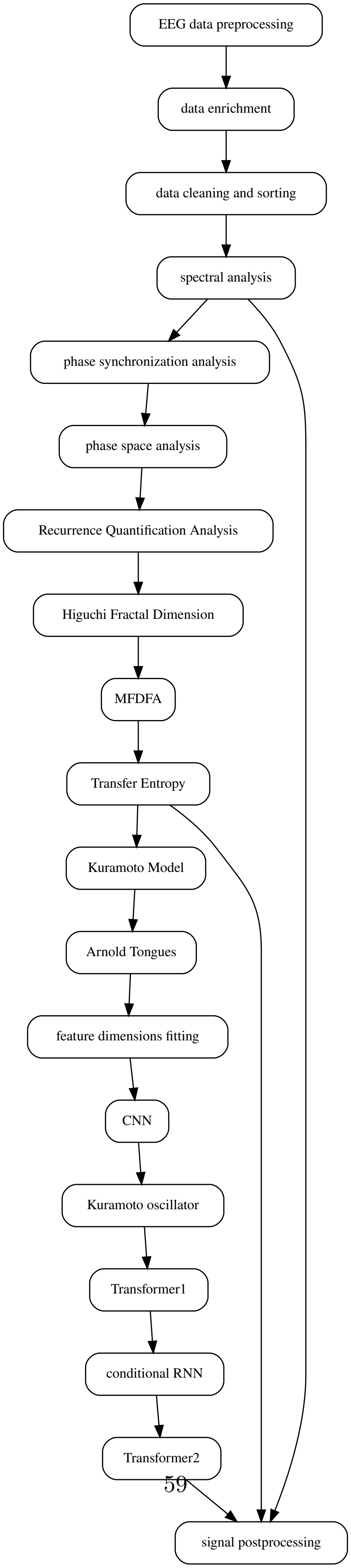

```
[ ]:
```